\documentclass[11pt,
 	a4paper,
               twoside,openright
	]{book} 

\usepackage{kpfonts} 

\usepackage[utf8]{inputenc} 

\usepackage[english]{babel}
\usepackage[T1]{fontenc}
\usepackage[binding=5mm]{layaureo}
\usepackage{float}
\usepackage{graphicx}
\usepackage{setspace}
\usepackage{wrapfig}
\usepackage{sidecap}
\usepackage{afterpage}
\usepackage{amsmath, amsthm, amssymb}
\usepackage{pspicture}
\usepackage{graphpap}
\usepackage{pifont}
\usepackage{lmodern}
\usepackage{epstopdf}
\usepackage{compactbib}
\usepackage{booktabs}
\usepackage{textgreek}

\usepackage{multicol}

\usepackage{geometry} 

\usepackage{graphicx} 

\usepackage{booktabs} 
\usepackage{array} 
\usepackage{verbatim} 
\usepackage{subfig} 

\usepackage{fancyhdr} 
\usepackage{enumitem}

\usepackage[nottoc,notlof,notlot]{tocbibind} 
\usepackage[titles,subfigure]{tocloft} 

\begin{document}

\begin{titlepage}

\begin{center}


\textsc{\LARGE Universita  degli Studi di Perugia}\\[1.5cm]

\begin{figure}[!h]
\begin{center}
\includegraphics[scale=0.8]{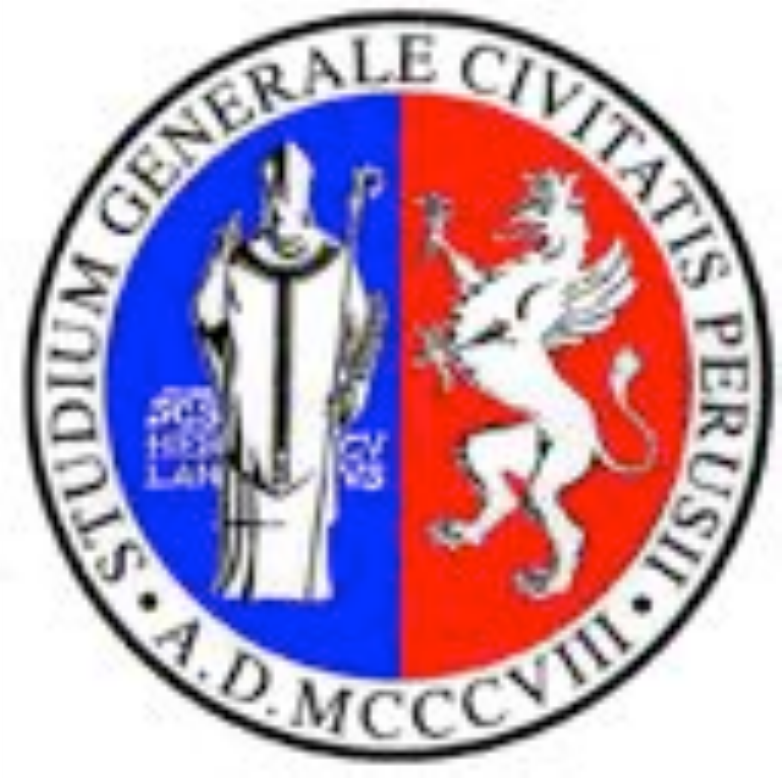}
\end{center}
\end{figure}

\textsc{\LARGE Facolta di Scienze Matematiche, Fisiche e Naturali}\\[1.5cm]
\textsc{\Large PhD Thesis in Physics}\\[0.5cm]

\vspace{1cm}
{  \huge \bfseries  Development and characterization of detectors for large area application in neutron scattering and small area application in neutron reflectometry }\\[0.4cm]

\vspace{2.5cm}

\begin{minipage}{0.4\textwidth}
\begin{flushleft} \large
\emph{Author:}\\
Giacomo \textsc{Mauri}
\end{flushleft}
\end{minipage}
\begin{minipage}{0.4\textwidth}
\begin{flushright} \large
\emph{Supervisors:} \\
Francesco \textsc{Piscitelli}\\
Alessandro \textsc{Paciaroni}\\
Francesco \textsc{Sacchetti}\\
\end{flushright}
\end{minipage}

\vfill

07 /01 /2019

\end{center}

\end{titlepage}

\shipout\null
\pagenumbering{arabic}

\tableofcontents


\chapter*{Introduction}\label{secmainInt}
\addcontentsline{toc}{chapter}{Introduction}  

Neutrons are a unique probe for investigating the structure and the dynamics of matter from the microscopic down to the atomic scale. The neutrons strongly interact with nuclei via nuclear reactions and scattering. This particle is unstable outside the nucleus and it decays through a weak interaction, the $\beta$-decay. Furthermore, neutrons undergo electromagnetic interaction because of the spin coupling with their magnetic moment. Although neutrons suffer all four fundamental interactions, they mainly interact with nuclei through the strong force. These reactions are more unusual compared with the Coulomb interactions, because of the short range of this force. Neutrons are chargeless and their electric dipole moment has a maximum upper limit of $\approx 10^{-26}\,$e$\cdot$cm. Thanks to these properties, neutrons are rather weakly interacting with matter and are more penetrating than charged particles.
\\The big advantage of neutrons is that the sample is not strongly affected by the interactions with them and the distortions can be considered as a small fluctuations from the equilibrium state. Investigations of samples under extreme conditions, e.g., high temperature, pressure and electromagnetic fields, are possible. Also, the high penetration depth of the neutrons allows the study of large or bulk samples and buried interfaces. 
\\ On the other hand, this makes them, also, difficult to detect. Both the rare interaction and limited available intensity of neutron sources lead to a poor signal generation for the neutron scattering techniques.  
This represents a challenge not only for the development of new generation, high-intensity neutron sources, but also for the design of new flexible instrumentation able to meet the needs of better performance. 
\\ At the European Spallation Source (ESS), presently under construction in Lund (Sweden), an intensity increase is foreseen of at least one order of magnitude with respect to the current neutron facilities. The peculiarity of the long pulse and low repetition rate time structure of the proton pulse, offer an unprecedented possibility to perform experiments exploiting neutron energy in the cold and thermal range (0.1 meV to 50 meV), thus focusing on slow dynamics and large-scale fluctuations of complex systems. The high brightness of the source will allow to employ smaller samples, which are typically more homogeneous, moreover it is not always possible to produce big sample and they are usually available in nature only in small volumes; the high intensity will allow as well faster measurements, increased use of polarized neutrons and detection of weaker signals. 
\\ The demands of better performances drive the development of design and operation for the instruments. The measurement of structures and dynamics over a wide length or time scales enforces a flexibility of the instrument to allow exchanges between brightness and better resolution, optimization of signal-to-noise ratio and the use of polarized neutrons as needed. In order to probe smaller sample, the instrument must be designed combining the high flux of the source and advanced neutron optics. The possibility of using the neutron polarization technique leads, instead, to improvements in different applications of physics, chemistry, biology etc.
\\ It therefore appears clear that the improvement of the detectors response is of a crucial importance to fulfil the requirement of increasing performance, since the $^3$He-based detector is at the limit of its technology. In addition to that, the $^3$He crisis of the past few years, opened up to the research for alternative solutions in order to replace the $^3$He-based neutron detector technology. 
\vspace{0.5cm}
\\This PhD project fits in this context of instrumentation development, focusing the attention on the research and the investigation of new neutron detector technologies. Two different solutions are discussed in the manuscript: a Boron-based thermal neutron gaseous detector for neutron reflectometry and a solid state Si-based thermal neutron detector coupled to a Gadolinium converter. The work regarding the first project has been carried out at the European Spallation Source (ESS) in Lund (Sweden) in the Detector Group (DG), while the second project was realized at the Department of Physics of the University of Perugia (Italy). 
\\ The Detector Group is in charge of the development of new technologies for thermal neutron detection for the instruments at ESS, and it based its research mainly on the $^{10}$Boron technology. $^{10}$B has a relative high natural abundance ($\sim 20 \%$), if compared with the one of $^3$He (1.4 parts per million), and has a large neutron absorption cross-section, even if lower than that of $^3$He. The Boron-based detector developed at ESS is in an advanced design phase, therefore not only a specific technical characterization can be performed, but also scientific cases can be studied. Further aspects can be investigated in order to have a full understanding and to validate this class of new neutron detector technology. Among others, the characterization of the background is a fundamental feature to study in order to understand the limit of the best signal-to-noise ratio available. Of particular interest, especially considering that the increase in flux leads to an increase of the background as well.
\\ The project developed at the University of Perugia is based on a prototype of a solid state thermal neutron detector started few years ago. The appealing feature of the neutron capture through Gadolinium, which has one of the highest neutron cross section, and the production of electrons instead of heavy charged particles, pushes several research programs to study possible alternatives employing the Gd. The work on the Silicon-based neutron detector, on the other hand, is in a conceptual phase. The aspects of the work mostly focus on the improvements of the different part of the design based on both basic operation measurements and simulations. 
\\ The difference on the status of the two projects reflects in the separation of the work. 
\vspace{0.2cm}
\\The Multi-Blade is a Boron-based gaseous detector, currently under development at ESS. It has been previously introduced at ILL in 2005, but never implemented until 2012. The Multi-Blade is a small area detector for neutron reflectometry applications. The most challenging requirements of the detector, set by the instruments and the research of new physics, are the spatial resolution and the counting rate capability. 
\\ Neutron reflectometry is a technique to study surfaces and interfaces on typical length-scales on the order of nano-meters. The limits are imposed by the measurements range and the instrumental resolution. Currently a typical dynamic range for reflectivities measurements is below $10^{-6}$. Strong limitations are set by the available flux. In specular reflectivity measurements, the neutron phase space is considerably reduced, indeed, the beam must be highly collimated and, monochromatized in the case of fixed wavelength spectrometers, or chopped on a time-of-flight reflectometer. Both techniques lead to an enormous waste of flux. The objective is to increase the available flux, both with the construction of new high-intensity sources, and with the development of new instruments layout able to exploit it. 
\\ From the point of view of detectors, the challenge is to achieve the better spatial resolution, about 3 times smaller than the state-of-the art, and higher counting rate capability, about 2-3 orders of magnitudes. The inclined geometry of the Multi-Blade leads to an improvement of both features and the use of a $^{10}$B$_4$C converter layer at grazing angle increases the detection efficiency as well. The neutron is absorbed in the boron layer and, in the conversion process, $\alpha$ and Li particles are emitted. This charged particles travel across the detection medium releasing their energy. The detection medium is a gas mixture of ArCO$_2$ (80/20) at atmospheric pressure. The Multi-Blade is a modular detector and each unit works as a Multi Wire Proportional Chamber (MWPC), it employs, indeed, a plane of wires (anodes) and a plane of strips (cathodes), which ensure a 2-dimensional read out.
\\ The whole process has been followed for this PhD work, from the assembling of a new demonstrator till the measurements on instruments. Several preliminary tests have been performed both at Source Testing Facility (STF) in Lund University (SE) and at the Budapest Research Center (BNC) in Budapest (HU). A complete system, mechanics, electronics and data acquisition system has been set up and finally tested on a real neutron reflectometer at ISIS in UK. The experiment concerned the technical characterization of the detector, i.e.,  uniformity and linearity, spatial resolution, stability, efficiency, but also it was proved the establishment of this technology for neutron reflectometry by performing, for the first time, science measurements using the Multi-Blade detector. 
\vspace{0.2cm}
\\To validate the effectiveness of this detector technology it has been investigated, for the first time on this class of devices, the response to fast neutrons. Previous works on $\gamma$-ray sensitivity have been carried out. The background discrimination is, indeed, a fundamental feature to investigate, in order to reduce as much as possible the effect of background events on the efficiency of the detector. 
\\A detailed study on the different types of interaction between fast neutrons and several materials is described. In order to quantify the fast neutron sensitivity of the Multi-Blade, the procedure of energy discrimination has been applied to the measurements. Different fast neutron sources has been used for this work, also some measurements with $\gamma$-ray sources has been performed, so to have a direct comparison between the $\gamma$-ray and fast neutron sensitivity. Although the measurements refer to the Multi-Blade, this detector is based on a MWPC geometry, thus the results obtained and the discussion of the underlying physical mechanisms are more general and can be extended to other gaseous-based neutron detectors. 
\\ Some measurements have been performed with an $^3$He-tube, in order to compare the two technologies. Due to the difficulty in decoupling the thermal neutron contribution from the fast neutron events, using the same experimental method adopted for the Multi-Blade, only preliminary and qualitative results can be discussed.   
\vspace{0.2cm}
\\ The Silicon-Pin detector coupled with a Gadolinium converter is a project for future application in neutron scattering, under development at the University of Perugia. The basic design is to couple a silicon microstrip sensor, well-know and widely used in particular for high-energy physics applications, together with a Gd converter. The results of the performances for this class of devices for thermal neutron detection is presented. The improvements on the design regard especially the way in which the two components, the silicon sensor and the neutron converter, can be linked. At present no optimized deposition technique for Gadolinium is available, a study of the evaporation deposition method has been performed. 
\\The requirements for large area and flexible shape detectors with submillimetric resolution could be satisfied by resorting to the technology based on solid state Si devices. The main advantage is the possibility to exploit the Integrated Circuit technology, in order to meet the demand for high-density readout electronics. Moreover, the high spatial resolution and high counting rate, typical of the Si p-n junction diode, make it a promising alternative in neutron detection, when operating under the intense neutron fluxes expected at future facilities. The development of a suitable electronics is a work in progress for the present prototypes. 
\\ For solid state Si detectors the charge signal is rather small, therefore a better data processing is relevant in the case of these kind of devices where the electronic noise becomes an important issue, in addition to the  background rejection. The ability to operate in high noise environment, where the low signal from Si detectors is a limitation, would be a major step forward for this technology. To this purpose it has been developed a new approach for a real time analysis of the detector pulses. A pulse shape analysis method is proposed and tested both on simulations and measurements. The promising performances and the capability to couple high integration electronics push forward the development of real time and single event analysis systems. Neutron scattering applications, e.g., diffraction or spectroscopy, can benefit by employing such kind of solid state Si-based neutron detectors, together with the pulse shape analysis method.
\vspace{1cm}
\\The intent of this work is to show the effectiveness of the detector technologies proposed, not only from the technical point of view, but also to demonstrate how the science can benefit by employing these new classes of devices.  

\chapter{Neutron interaction with matter}

\section{Basic properties of the neutron}\label{basic}

Neutrons together with protons are the the components of the nucleus of an atom. The neutron is made up of three quarks, (up, down, down) and it has a mass slightly larger than that of a proton ($m_p = 938.28$ MeV/c$^2$) $m_n = 939.57$ MeV/c$^2$. Neutrons are not stable outside the nucleus, they decay into a proton, electron and an electron-antineutrino via $\beta$-decay with a mean life time of $\tau_n = 885.7 \pm 0.8$ s~\cite{nakamura,Neutron-lifetime}.
\\ Neutrons are subject to all four fundamental interactions. Nuclear reactions and scattering with the nucleus are strong interactions, the $\beta$-decay of a neutron is a weak interaction. As the neutrons are uncharged particles, they are unaffected by the Coulomb potential of the electrons. The electron magnetic interaction of the neutrons is only due to the spin coupling $\Big( s = \frac{1}{2}\Big)$ with the magnetic moment. The magnetic moment of neutrons is given by:

\begin{equation}
\mu = g_s s \mu_N
\label{eq1}
\end{equation}

where $g_s$ is known as the \textit{spin g factor} and is calculated by solving a relativistic quantum mechanical equation, $s$ is the spin and the nuclear magneton is: 

\begin{equation}
\mu_N = \frac{e \hslash}{2 m_p} = 3.1525 \times 10^{-8} eV/T 
\label{eq2}
\end{equation}

where $m_p$ is the proton mass. 
The principal means of neutron interaction is through the strong force with nuclei. These reactions are much rarer in comparison with the Coulomb interactions, because of the short range of this force~\cite{DET_leo}. 
\\ Due to its neutrality the neutron is rather weakly interacting with matter, leading to important consequences. The neutron produce a small disturbance of the sample's properties, which can be treated as small fluctuations from the equilibrium state. Linear-response theory is a good approximation to extract the scattering law. The neutron has a large penetration depth, indeed it must come within $\sim 10^{-15}$m before interacting with the nuclei. The large penetration depth of the neutrons is beneficial for the investigation of materials under extreme conditions such as very low and very high temperatures, high pressures, high magnetic and electric fields. On the other hand, this is also the reason why it is rather difficult to detect these particles~\cite{furrer}.
\\ In the non-relativist limit the energy $E$ of a neutron can be described in terms of its wavelength $\lambda$ through the De Broglie relationship: 

\begin{equation}
\lambda = \frac{2 \pi \hslash}{m_n v} \quad \Rightarrow \quad E = \frac{1}{2}m_n v^2 = \frac{\pi \hslash^2}{m_n \lambda^2}
\label{eqdebroglie}
\end{equation}

where $\hslash$ is the Plank's constant and $m_n$ is the mass of the neutron. The wavevector $k$ of the neutron has a magnitude 

\begin{equation}
k = \frac{2 \pi }{\lambda}
\label{eqk}
\end{equation}

From equation~\ref{eqdebroglie} and~\ref{eqk} the energy $E$ of the neutron can be written as:

\begin{equation}
E =\frac{\hslash^2 k^2}{2m_n}
\label{eqE}
\end{equation}

The energy can be related to the temperature $T$ as well

\begin{equation}
E = k_B T
\label{eqET}
\end{equation}

where $k_B = 1.281 \cdot 10^{-23}$ J/K is the Boltzmann constant.
\\ Due to the strong energy dependence of neutron interactions, neutrons are classified according to their energy, although no specific boundaries are prescribed. In table~\ref{table1} the neutron energy regimes are also presented in terms of velocity, wavelength and temperature.

\begin{table}[htbp]
\centering
\caption{\label{table1} \footnotesize Energy classification of neutrons.}
\smallskip
\begin{tabular}{|l|l|l|l|l|}
\hline\hline
\textbf{Term}& \textbf{Energy (eV)}& \textbf{Velocity (m/s)}& \textbf{$\lambda$ (\AA)}&\textbf{Temperature (K)}\\
\hline
ultra cold & $< 2\cdot 10^{-7}$& $<6$ & $>520$ & $< 0.002$ \\
very cold & $ 10^{-7} - 10^{-5}$& $6 - 99$ & $520 - 40$ & $ 10^{-3} - 0.6$ \\
cold & $10^{-5} -  10^{-3}$& $99-10^3$ & $40-4$ & $0.6-60$ \\
thermal & $10^{-3}-0.5$& $10^3-10^4$ & $4-0.4$ & $60-6\cdot 10^{3}$ \\
epithermal &  $0.5-10^3$& $10^4-10^5$ & $0.4-10^{-2}$ &   \\
intermediate & $10^3-10^5$& $10^5-10^6$ & $10^{-2}-10^{-3}$ &   \\
fast & $10^5-10^{10}$& $10^6-10^9$ & $10^{-3}-10^{-6}$ &   \\
\hline
\end{tabular}
\end{table}

When a neutron interacts with matter a variety of nuclear processes depending on its energy may occur. Among these are:

\begin{enumerate}[label=(\roman*)]
\item Elastic scattering from nuclei, namely $A(n, n)A$, which is the main mechanism of energy loss for neutrons in the MeV region.
\item Inelastic scattering, e.g., $A(n, n')A^*$, $A(n, 2n')B$, etc. In this reaction, the nucleus is left in an excited state which may later decay by $\gamma$-ray or some other form of radiative emission. In order to occur, the neutron must have sufficient energy to excite the nucleus, usually on the order of 1 MeV or more. Below this energy threshold, only elastic scattering may occur.
\item Radiative neutron capture, i.e., $n + (Z,A)\rightarrow \gamma + (Z,A+1)$. In general, the cross-section for neutron capture goes approximately as $1/v$ with $v$ the neutron velocity. Therefore, absorption is most likely at low energies.
\item Other nuclear reactions, such as (n,p), (n,d), (n,$\alpha$), (n,t), (n,$\alpha$p), etc., in which the neutron is captured and charged particles are emitted. The energy range of these reactions is typically between eV and keV. Like the radiative capture reaction, the cross-section falls as $1/v$.
\item Fission occurs most likely at thermal energies.
\item High energy hadron shower production. This occurs at very high energy, above 100 MeV
\end{enumerate}

The total probability for a neutron to interact in matter is given by the sum of the individual cross-sections for the processes listed above:

\begin{equation}
\sigma_{tot} = \sum_i \sigma_{i}= \sigma_{elastic}+ \sigma_{inelastic}+\sigma_{capture} +... 
\label{crosssection}
\end{equation}

If we multiply $\sigma_{tot}$ by the density of atoms we obtain the macroscopic cross-section $\Sigma$, thus the mean free path length $l$ which is the inverse of $\Sigma$:

\begin{equation}
\Sigma_{tot} = \frac{1}{l} = n \cdot  \sigma_{tot} = \frac{N_A \rho}{A} \sigma_{tot}
\label{macrosigma}
\end{equation}

where $\rho$ is the material mass density, $A$ is the atomic number and $N_A$ is the Avogadro's number. If we consider a narrow beam of neutrons passing through matter, the number of detected neutrons will fall off exponentially with absorber thickness. In this case the probability for a neutron of wavelength $\lambda$ to interact with a nucleus of the matter at depth $x$ in a thickness $dx$ is given by: 

\begin{equation}
I(x,\lambda)dx = \Sigma e^{-x\Sigma(\lambda)}dx
\label{attenuation}
\end{equation}

The number of neutrons that pass through a layer of thickness $d$ is obtained by the integration of equation~\ref{attenuation}, as shown in equation~\ref{expatt}:

\begin{equation}
\frac{N(d)}{N_0} =\int_0^d I(x)dx = \int_0^d \Sigma e^{-x\Sigma}dx = 1 - e^{-d\Sigma}
\label{expatt}
\end{equation}

where $N_0$ is the initial incoming neutron flux. For a more general case of non-collimated source, equation~\ref{expatt} is no longer an adequate description. A more complex neutron transport computation is then required to predict the number of transmitted neutrons and their distribution in energy.

\section{Neutron cross section}\label{sec_crossec}

The probability per unit path length is conventionally expressed in terms of the cross-section $\sigma$ per nucleus for each type of interaction, i.e., the processes listed in section~\ref{basic}. We consider a beam of particles incident on a target with the particles in the beam uniformly distributed in space and time. We define $\Phi$ as the flux of incident particles per unit area and per unit time, and $d\Omega$ as the solid angle per unit time. Due to the randomness of the impacts, the number of particles scattered into $d\Omega$ will statistically fluctuate over different finite periods of measuring. However, in average this number will tend towards a fixed $dN_s/d\Omega$, where $N_s$ is the average number scattered per unit time. The differential cross-section is then defined as the ratio

\begin{equation}
\frac{d\sigma}{d\Omega}(E,\Omega) = \frac{1}{\Phi} \cdot \frac{dN_s}{d\Omega}
\label{diffcross}
\end{equation}

$\frac{d\sigma}{d\Omega}$ is the average fraction of the particle scattered into the considered solid angle per unit time and per unit flux. In general, the value of $\frac{d\sigma}{d\Omega}$ will vary with the energy of the reaction and the scattering angle. It is possible to calculate the total cross-section for any scattering at an energy $E$ as the integral of the differential cross-section over all solid angles as follows:

\begin{equation}
\sigma(E) = \int \frac{d\sigma}{d\Omega}(E,\Omega) d\Omega
\label{totcross}
\end{equation}

In a real situation the target is usually a slab of material containing many scattering centres. Thus we want to know how many interaction occur on the average. Assuming that the target centres are uniformly distributed and the slab is not too thick so that the likelihood of interaction is low, the number of centres per unit perpendicular area which will be seen by the beam is $n \cdot \delta x $ where $n$ is the volume density of centres and $\delta x$ is the thickness of the material along the direction of the beam. The average number of scattered particles into $d\Omega$ per unit time is defined as:

\begin{equation}
N_s(\Omega) = \Phi \cdot A \cdot n \cdot \delta x \cdot \frac{d\sigma}{d\Omega} d\Omega
\label{nscattered}
\end{equation}

where $A$ is the total perpendicular area of the target. The total number scattered into all angles is simply:

\begin{equation}
N_{tot} = \Phi \cdot A \cdot n \cdot \delta x \cdot \sigma
\label{nscatot}
\end{equation}

Considering now the more general case of any thickness $x$, we define the probability $P(x)$ for a particle not to suffer an interaction in a distance $x$ of the target. The probability to have an interaction in the interval $(x,x+dx)$ is given dividing equation~\ref{nscatot} by the total number of incident particles per unit time $(\Phi \cdot A)$. By the definition of the macroscopic cross-section (eq.~\ref{macrosigma}) we can substitute it and we obtain:  

\begin{equation}
P(x,x+dx) = n \cdot \sigma \cdot \delta x = \Sigma \cdot \delta x
\label{pscatt}
\end{equation}

The probability of not having an interaction between $x$ and $x+ dx$ is then:

\begin{align}
P(x+dx) = P(x)(1- \Sigma \cdot dx) \Rightarrow \nonumber \\
P(x) + \frac{dP}{dx} dx = P(x) - P(x) \Sigma \cdot dx \Rightarrow  \nonumber\\
P(x) = C \cdot e^{-\Sigma x} = e^{-\Sigma x}
\label{pnointeract}
\end{align}

note that the integration constant $C=1$, because it is required that $P(x=0) =1$.
\\ It is straightforward to define the probability of suffering interaction in the distance $x$ from equation~\ref{pnointeract} as:

\begin{equation}
P_{int}(x) = 1 - P(x) = 1- e^{-\Sigma x}
\label{pinteract}
\end{equation}

The mean distance travelled by a particle without suffering any collision is known as the mean free path $l$. From equation~\ref{macrosigma}, it is defined as the inverse of the macroscopic cross-section. The probability of interaction can be expressed as:

\begin{equation}
P_{int}(x) = 1- e^{-\frac{x}{l}}
\label{pinteractl}
\end{equation}

\section{Neutron sources and activity}

Some of the possible mechanisms to produce neutrons and the activity of a source will be described here, in order to define the units that will be useful for the experimental discussions (chapters~\ref{chapter4},~\ref{chapter5},~\ref{chapter6}).
\\Natural neutron emitters do not exist, although nuclei created with excitation energy greater than the neutron binding energy can decay by neutron emission. These excited states are, indeed, not produced in any convenient radioactive decay process~\cite{DET_knoll}. Neutron sources are based on either spontaneous fission or nuclear reaction.

\subsection{Spontaneous Fission}

Spontaneous fission can occur in many transuranic elements, releasing neutrons along with the fission fragments. These products can promptly decay emitting $\beta$ and $\gamma$ radiation. When used as a neutron source, the isotope is generally encapsulated in a sufficient thick container so that much of the radiation can be absorbed leaving only the fast neutrons, which are more penetrating particles.  
\\ The most common spontaneous fission source is $^{252}$Cf which has a half-life of 2.65 years. The dominant decay mechanism is the alpha decay ($\sim 97\%$) compared to spontaneous fission ($\sim 3\%$). The neutron yield is 0.116 n/s per Bq, combining both decay rate. The energy spectrum of neutrons is continuous up to about 10 MeV and exhibits a Maxwellian shape. The distribution can be described by the expression:

\begin{equation}
\frac{dN}{dE} = \sqrt{E} e^{-\frac{E}{T}}
\label{cfdistr}
\end{equation}

where the constant $T = 1.3$ MeV for the $^{252}$Cf~\cite{LORCH_n_spectra}.

\subsection{Nuclear Reactions}

A more convenient method of producing neutrons is through the nuclear reactions ($\alpha$, n) or ($\gamma$, n). Such sources are generally made by mixing a strong $\alpha$ or $\gamma$ emitter with a suitable target material. The most common target material is beryllium. It undergoes a number of reactions which lead to the production of free neutrons under bombardment by alpha-particles, such as:

\begin{equation}
\alpha + ^{9}Be \rightarrow ^{13}C^{*} \rightarrow \begin{cases} ^{12}C^{*} +n\\ ^{8}Be + \alpha + n\\ 3\alpha + n \end{cases}
\label{alphaBe}
\end{equation}

\begin{figure}[htbp]
\centering
\includegraphics[width=.9\textwidth,keepaspectratio]{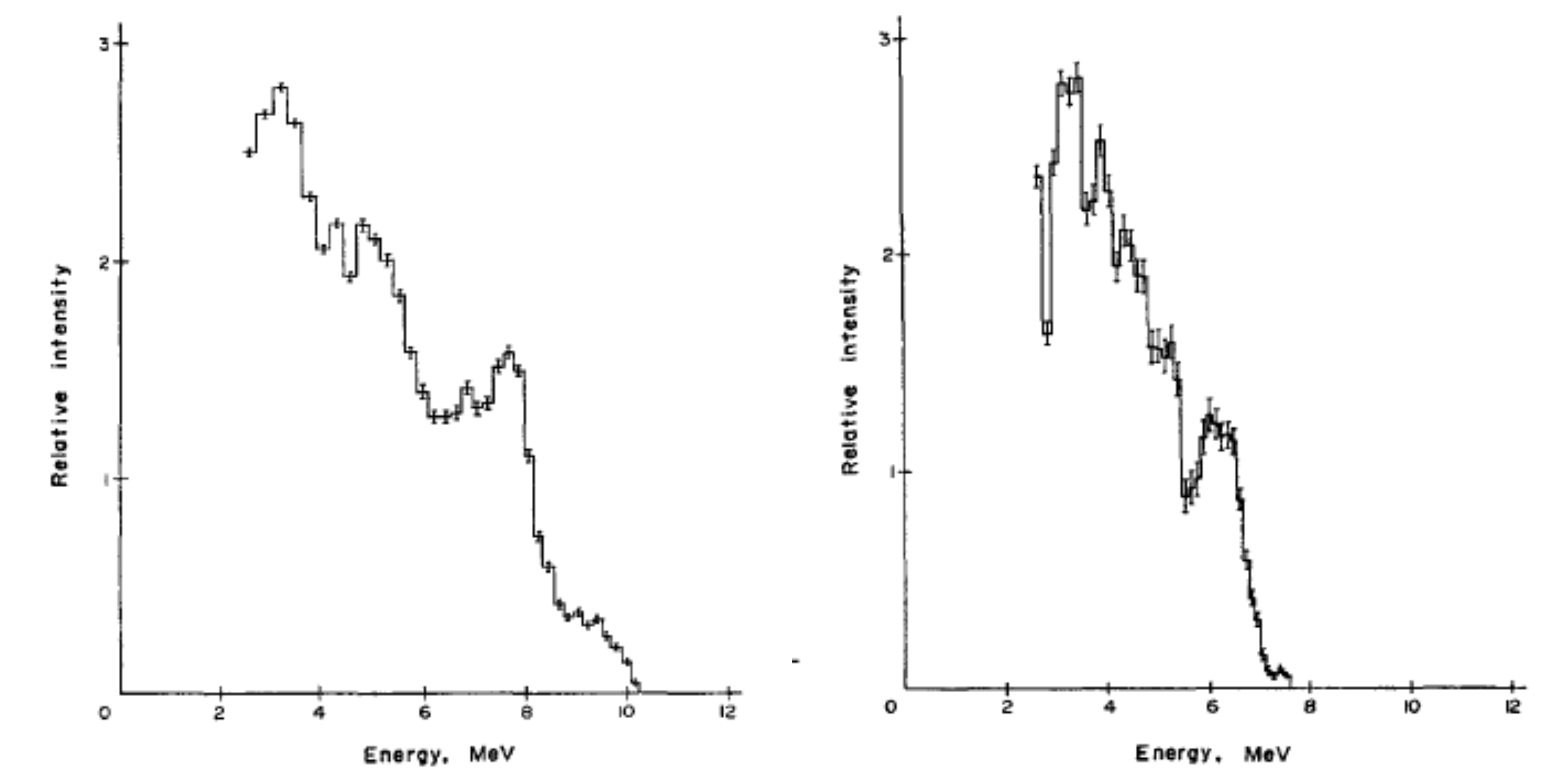}
\caption{\label{pubegaussb} \footnotesize Neutron energy spectra from $^{241}$Am/Be (left) and $^{238}$Pu/Be (right) sources (from~\cite{LORCH_n_spectra}).}
\end{figure}

The excited compound nucleus $^{13}C^{*}$ is formed, then it decays through a variety of modes depending on the excitation energy. In general, the dominant reaction is the decay to $^{12}C$. Most of the $\alpha$-particles simply are stopped in the target and only 1 in about $10^4$ reacts with a beryllium nucleus. 
\\ The actinide elements are the most diffused alpha emitters, a stable alloy can be made in the form MBe$_{13}$, where M represents the actinide metal. Some of the common choices for alpha emitters are $^{238}$Pu and $^{241}$Am. The neutron energy spectra from all such $\alpha$/Be sources are similar, any differences reflect only the small variations in the primary $\alpha$ energies. In figure~\ref{pubegaussb} the neutron energy spectra from $^{241}$Am/Be (left) and $^{238}$Pu/Be (right) source are shown~\cite{LORCH_n_spectra}.

In the case of the photo-reaction ($\gamma$, n) only two target nuclei, $^9Be$ and $^2H$, are suitable. The emission of a free neutron arises if, by the absorption of a $\gamma$-ray photon, the target nucleus is in a sufficient excitation energy state. The advantage of these kind of sources is that if the gamma-rays are mono-energetic, the neutron are, also, emitted nearly mono-energetic. On the other side the reaction yield per $\gamma$ is lower than that of the $\alpha$-type sources and the gamma-ray background is much more intense. 

\subsection{Activity}

The activity of a radioisotope source is defined as the mean number of decay processes it undergoes per unit time. Note that the activity measures the source disintegration rate, which is not necessarily synonymous with the amount of radiation emitted in its decay. The relation between radiation output and activity depends on the specific nuclear decay scheme of the isotope.  
\\ The radioactive decay law asserts that the activity of a radioactive sample decays exponentially in time. In term of quantum mechanics, this can be derived by considering that a nuclear decay process is expressed by a transition probability per unit time, $\lambda_d$, characteristic of the nuclear species. If a nuclide has more than one mode of decay, $\lambda_d$ is the sum of each constant per mode. The activity can be defined as:

\begin{equation}
A(t) = \frac{dN(t)}{dt} = -\lambda_d N(t) 
\label{activity}
\end{equation}

where $N$ is the number of nuclei and $\lambda_d$ the decay constant. Integrating equation~\ref{activity} it results in the exponential:

\begin{equation}
dN(t) = N(0) e^{-\lambda_d t}  
\label{int_act}
\end{equation}

N(0) is the number of the radioactive nuclei at the time $t=0$.  The activity can be expressed as the inverse of the decay constant $\lambda_d$, the mean lifetime $\tau = 1/ \lambda_d$, i.e., the time it takes for the sample to decay $1/e$ of its initial activity. Equally the half-life, $t_{1/2}$, can be used to define the activity. It is the time in which the source decay to one half of its original activity, thus $t_{1/2}= \tau \cdot \ln2$. It is possible to calculate the activity at the time $t$ by knowing the original activity ($A_0$) at time $t=0$ as:

\begin{equation}
A(t) = \lambda_d \cdot N(t) = \lambda_d \cdot N(0) e^{-\lambda_d t} = A_0 \cdot e^{-\lambda_d t}
\label{act_time}
\end{equation}

The traditional unit of activity has been the $Curie$ (Ci), defined as $3.7 \times 10^{10}$ disintegrations/second, which is the activity of 1 g of pure $^{226}$Ra. The laboratory-scale for radioisotope sources is usually on the order of $\mu$Ci, thus for practical aims this unit has been replaced by the Becquerel, defined as one disintegration per second. Thus 1 Bq = $2.7 \times 10^{-11}$ Ci.

\section{Neutron scattering}\label{sec_scatt}

We consider now the nuclear scattering by a general system of particles. We derive a general expression for the cross-section $\mathrm{d}^2\sigma/ \mathrm{d}\Omega \mathrm{d}E'$ for a specific transition of the scattering system from one of its quantum states to another~\cite{squires_2012}. From the general theory~\cite{cohen_QM, schiff_QM} the incoming particle can be described by a plane wave which interacts with the nucleus through a potential $V$. Suppose we have a neutron with wave-vector $\textbf{k}$ on a scattering system in a state identified by an index $\lambda$. Through the potential $V$ the particle is scattered so that its final wave-vector is $\textbf{k}^{\prime}$ and the final state of the scattering system is $\lambda'$. The resulting wave-function $\psi$ will be a superposition of the incoming wave and a spherically diffused wave. It is the solution of the Schr$\"o$dinger equation:

\begin{equation}
\bigg[- \frac{\hslash^2}{2m} \nabla^2 + V \bigg]\psi = E\psi
\label{schr_eq}
\end{equation}

$N$ is defined as the number of nuclei in the scattering system, $\textbf{R}_j$ ($j= 1,...N$) the position vector of the $j$th nucleus and $\textbf{r}$ the position vector of the neutron. 
\\ The differential cross-section $(d\sigma/d\Omega)_{\lambda \rightarrow \lambda'}$ represents the sum of all processes that involve the change of the scattering system state from $\lambda$ to $\lambda'$ and the state of neutron changes from $\textbf{k}$ to $\textbf{k}^{\prime}$.  From the equation~\ref{diffcross} is obtained:

\begin{equation}
\bigg(\frac{d\sigma}{d\Omega} \bigg)_{\lambda \rightarrow \lambda'} = \frac{1}{\Phi}\frac{1}{d\Omega} \sum_{\textbf{k}^{\prime}} W_{\textbf{k},\lambda \rightarrow \textbf{k}^{\prime},\lambda'}
\label{sum_diffcross}
\end{equation}

where $W_{\textbf{k},\lambda \rightarrow \textbf{k}^{\prime},\lambda'}$ is the number of transitions per second from the initial to the final state. To evaluate the expression~\ref{sum_diffcross} we use the \textit{Fermi's golden rule} so that

\begin{equation}
\sum_{\textbf{k}^{\prime}} W_{\textbf{k},\lambda \rightarrow \textbf{k}^{\prime},\lambda'} = \frac{2 \pi}{\hslash} \rho_{\textbf{k}^{\prime}} \vert\langle \textbf{k}^{\prime} \lambda' \vert V \vert \textbf{k}\lambda \rangle \vert^2
\label{goldenrule}
\end{equation}

$\rho_{\textbf{k}^{\prime}}$ is the number of momentum states in $d\Omega$ per unit energy range for neutron in state $\textbf{k}^{\prime}$. The matrix element can be explicit as:

\begin{equation}
\langle \textbf{k}^{\prime} \lambda' \vert V \vert \textbf{k}\lambda \rangle = \int \psi^*_{\textbf{k}^{\prime}} \chi^*_{\lambda'} V \psi_\textbf{k} \chi_\lambda\ \mathrm d \textbf{R}\ \mathrm d \textbf{r}
\label{matrix-el}
\end{equation}
 
where $\chi$ denotes the wave-function of a scattering system (i.e., $\lambda$ $\lambda'$), d$\textbf{R}_j$ is an element of volume for the $j$th nucleus and d$\textbf{r}$ is an element of volume for the neutron. 
\\ In order to calculate $\rho_{\textbf{k}^{\prime}}$, can be adopted a standard device in quantum mechanics, the \textit{box normalisation}~\cite{squires_2012}. One can imagine the neutron and the scattering system to be in a large box, thus the normalization constant of the neutron wave-function is fixed and the only allowed neutron states are those of which de Broglie waves are periodic in the box. The volume of the unit cell of the lattice is

\begin{equation}
v_k = \frac{(2\pi)^3}{A}
\label{unit-vol}
\end{equation}

where $A$ is the volume of the box. The final energy of the neutron is 

\begin{equation}
E' = \frac{\hslash^2}{2m} k'^2
\label{Efin}
\end{equation}
with
\begin{equation}
\mathrm d E' = \frac{\hslash^2}{m} k' \mathrm d k' 
\label{dE'}
\end{equation}

\begin{figure}[htbp]
\centering
\includegraphics[width=0.55\textwidth,keepaspectratio]{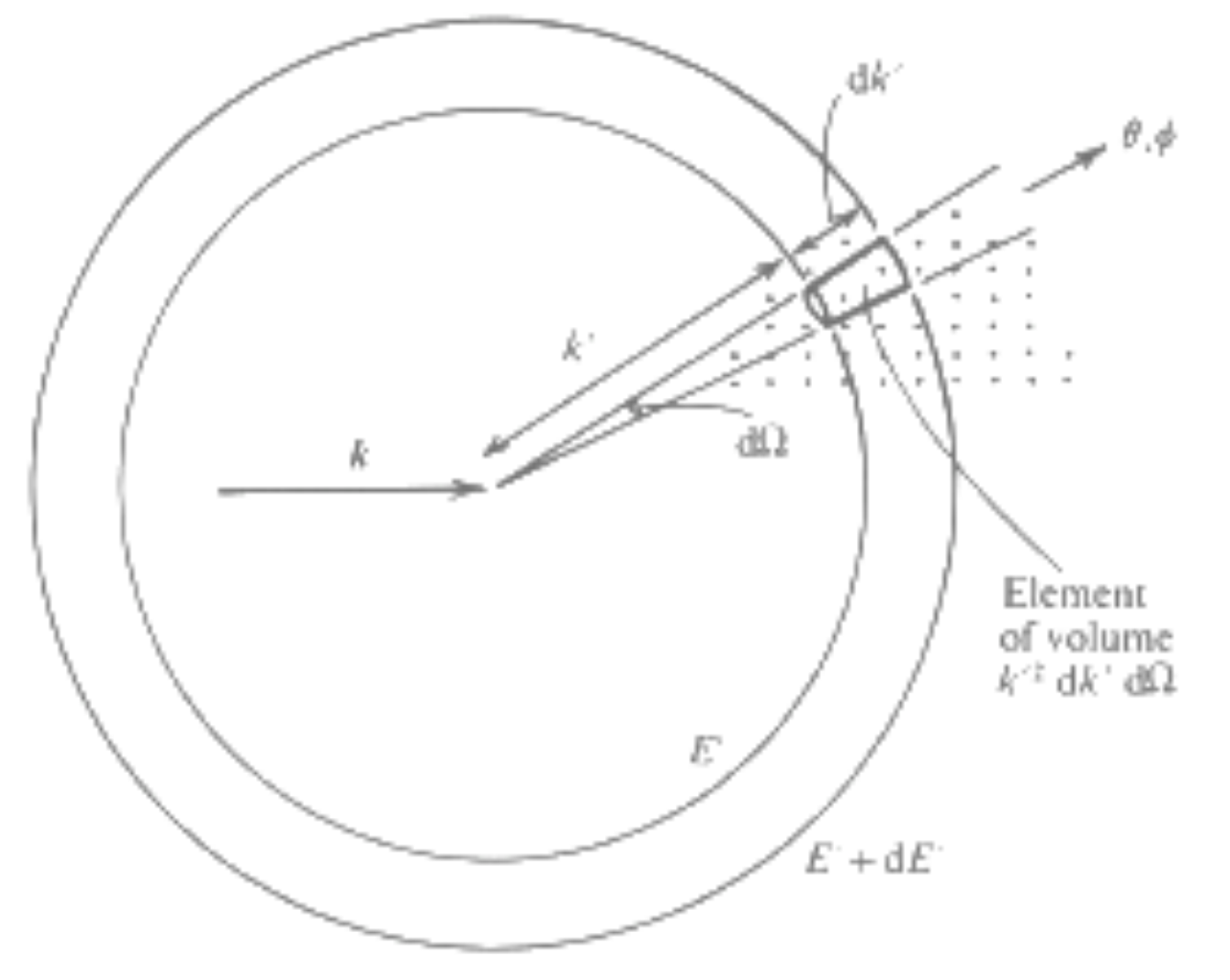}
\caption{\label{element_rok} \footnotesize The points represent $\textbf{k}^{\prime}$ values allowed by box normalization. The spheres correspond to the neutron energy $E'$ and $E'+$d$E'$. The element of volume $k'^2$d$k$d$\Omega$ lies in between the two spheres in the direction $\textbf{k}^{\prime}$. (From~\cite{squires_2012}).}
\end{figure}

The number of states in d$\Omega$ with energy between $E'$ and $E'+$d$E'$ is, by definition, $\rho_{\textbf{k}^{\prime}}$d$E'$, which can be expressed in terms of number of wave-vector points in the element of volume $k'^2$d$k$d$\Omega$. A sketch is shown is figure~\ref{element_rok}.

\begin{equation}
\rho_{\textbf{k}^{\prime}} \mathrm d E' = \frac{1}{v_k} k'^2 \mathrm{d}k' \mathrm{d}\Omega
\label{d-roE}
\end{equation}

Substituting equations~\ref{unit-vol} and~\ref{dE'} in equation~\ref{d-roE} we obtain:

\begin{equation}
\rho_{\textbf{k}^{\prime}} = \frac{A}{(2\pi)^3} k' \frac{m}{\hslash^2} \mathrm{d}\Omega
\label{rok}
\end{equation}

We now assume that there is one neutron in the box of volume $A$, so that the neutron intensity is $1/A$.  The wave-function $\psi_{\textbf{k}^{\prime}}$ is a plane wave, it can be expressed as:

\begin{equation}
\psi_{\textbf{k}} = \frac{1}{\sqrt{A}} e^{i\textbf k \textbf{r}}
\label{plane_wave}
\end{equation}

The matrix element in~\ref{matrix-el} thus can be written as follows.

\begin{equation}
\langle \textbf{k}^{\prime} \lambda' \vert V \vert \textbf{k}\lambda \rangle = \frac{1}{A} \int e^{-i \textbf{k}^{\prime} \textbf{r}} \chi^*_{\lambda'} V e^{i \textbf{k} \textbf{r}} \chi_\lambda\ \mathrm d \textbf{R}\ \mathrm d \textbf{r}
\label{matrix-el-planewave}
\end{equation}

Note that the flux of the incident neutrons is the product of their density and velocity, namely:

\begin{equation}
\Phi = \frac{1}{A} \frac{\hslash}{m} k
\label{flux}
\end{equation}

By substituting equations~\ref{goldenrule},~\ref{rok},~\ref{matrix-el-planewave} and~\ref{flux} into~\ref{sum_diffcross}, the following expression for $(\mathrm{d}\sigma/ \mathrm{d}\Omega )_{\lambda \rightarrow \lambda'}$ is obtained, which is the cross-section for neutrons scattered into $\mathrm{d}\Omega$ in the direction of $\textbf{k}^{\prime}$.

\begin{equation}
\bigg(\frac{\mathrm d \sigma}{\mathrm d \Omega} \bigg)_{\lambda \rightarrow \lambda'} = \frac{k'}{k} \bigg( \frac{m}{2 \pi \hslash^2} \bigg)^2 \vert\langle \textbf{k}^{\prime} \lambda' \vert V \vert \textbf{k}\lambda \rangle \vert^2 
\label{sum_diffcross1}
\end{equation}

Note that since $\textbf{k}$, $\lambda$ and $\lambda '$ remain constant in the element of volume considered in figure~\ref{element_rok}, all the scattered neutrons have the same energy, which is determined by the conservation of energy. The initial and final energies of the neutron are denoted by $E$ and $E'$, while $E_\lambda$ and $E_{\lambda'}$ are the initial and final energies of the scattering system. By the conservation of energy then $E +E_\lambda = E'+ E_{\lambda'}$. In mathematical terms the energy distribution of the scattered neutrons is a $\delta$-function. It is finally obtained the expression for the partial differential cross-section:

\begin{equation}
\bigg(\frac{\mathrm{d}^2\sigma}{\mathrm{d}\Omega \mathrm{d}E'} \bigg)_{\lambda \rightarrow \lambda'} = \frac{k'}{k} \bigg( \frac{m}{2 \pi \hslash^2} \bigg)^2 \vert\langle \textbf{k}^{\prime} \lambda' \vert V \vert \textbf{k}\lambda \rangle \vert^2 \delta (E_\lambda - E_{\lambda'} + E - E')
\label{partialdiffcross}
\end{equation}

We now want to evaluate the matrix element in equation~\ref{partialdiffcross}, integrating with respect to the neutron coordinate, $\textbf r$. The potential of the neutron due to the $j$th nucleus has the form $V_j (\textbf r - \textbf{R}_j)$, thus the potential for the whole scattering system is the sum of $V_j$ over $j$; we also define $\textbf{x}_j = (\textbf r - \textbf{R}_j)$. The matrix element is:

\begin{align}
\langle \textbf{k}^{\prime} \lambda' \vert V \vert \textbf{k}\lambda \rangle = \nonumber\\
& = \sum_j  \int e^{-i \textbf{k}^{\prime} \textbf{r}} \chi^*_{\lambda'} V_j (\textbf r - \textbf{R}_j) e^{i \textbf{k} \textbf{r}} \chi_\lambda\ \mathrm d \textbf{R} \mathrm d \textbf{r} \label{el-matix-calc1}\\
& = \sum_j  \int e^{-i \textbf{k}^{\prime} (\textbf{x}_j + \textbf{R}_j) } \chi^*_{\lambda'} V_j ( \textbf{x}_j) e^{i \textbf{k} (\textbf{x}_j + \textbf{R}_j) } \chi_\lambda\ \mathrm d \textbf{R} \mathrm d \textbf{x}_j \label{el-matix-calc2}\\
& = \sum_j  V_j (\kappa) \langle \lambda' \vert e^{i \kappa \textbf{R}_j) } \vert \lambda \rangle \label{el-matix-calc3}
\end{align}
The transition between equation~\ref{el-matix-calc1} and~\ref{el-matix-calc2} is made in order to simplify the calculation. For each $j$ term in the sum, integrating with respect to $\textbf r$ or to $\textbf{x}_j$ gives the same result, because both integrations are done at fixed $\textbf{R}_j$ over all space.
\\ The next step of the calculation is to insert a specific function for $V_j( \textbf{x}_j)$. In order to find a suitable solution we consider only one term, $j=1$, of the sum in equation~\ref{el-matix-calc1} and the differential cross-section $\mathrm{d}\sigma/ \mathrm{d}\Omega$ for a single fixed nucleus can be calculated. Since the nucleus is fixed at the origin, $\textbf{R}_1= 0$ and $\lambda' = \lambda$, moreover from the exponential terms in equation~\ref{el-matix-calc1}, we define $\kappa = \textbf{k} - \textbf{k'}$, which is known as the \textit{scattering vector}. In these conditions the matrix element which describe the transition is: 

\begin{align}
\langle \textbf{k}^{\prime} \lambda' \vert V \vert \textbf{k}\lambda \rangle
& = \int \chi^*_{\lambda'} \chi_\lambda \mathrm d \textbf{R} \int V(\textbf r)  e^{i \kappa \textbf{r}} \mathrm d \textbf{r} \nonumber \\
& = \int V(\textbf r) e^{i \kappa \textbf{r}}  \mathrm d \textbf{r}
\label{el-matix-calc4}
\end{align}

Inserting this result in equation~\ref{sum_diffcross1}, together with $k' = k$, gives

\begin{equation}
\frac{\mathrm d \sigma}{\mathrm d \Omega} = \bigg( \frac{m}{2 \pi \hslash^2} \bigg)^2 \ \bigg| \int V(\textbf r) e^{i \kappa \textbf{r}}  \mathrm d \textbf{r} \bigg|^2
\label{sum_diffcross2}
\end{equation}

As already mentioned, the nuclear forces for scattering have a range of about $10^{-15}$ m while the wavelength of a thermal neutron is of the order of $10^{-10}$ m. The potential is non zero only in a short range on the scale of wavelength. It can be written as a three-dimensional Dirac delta function, $\delta(\textbf r)$, of intensity $a$, which is a real constant:

\begin{equation}
V(\textbf r) = a\delta(\textbf r)
\label{V-delta}
\end{equation} 

Substituting this value in equation~\ref{sum_diffcross2} we obtain:

\begin{equation}
\int V(\textbf r) e^{i \kappa \textbf{r}}  \mathrm d \textbf{r} = a \int \delta(\textbf r) e^{i \kappa \textbf{r}}  \mathrm d \textbf{r} = a
\label{int_diffcross2}
\end{equation}

Therefore,

\begin{equation}
\frac{\mathrm d \sigma}{\mathrm d \Omega} = \bigg( \frac{m}{2 \pi \hslash^2} \bigg)^2 a^2
\label{sum_diffcross3}
\end{equation}

As we are considering the scattering of neutrons by a single fixed nucleus, the angular distribution for the wave scattering is spherically symmetric. Then the incident neutrons can be represented by a wave-function $\psi_{inc}$, and the scattered neutrons at the point $\textbf r$ is represented by the wave-function $\psi_{sc}$, given by the condition of spherical symmetry:

\begin{equation}
\psi_{inc}= e^ {ikz}, \qquad \psi_{sc}= -\frac{b}{r} e^ {ikr}
\label{wavef-inc-sc}
\end{equation}

where $z$ is the axis along the direction of $\textbf k$, $b$ is a constant and the minus sign is a standard convention which correspond to a positive value of $b$ for a repulsive potential. Note that the magnitude of the wave-vector for the incident and the scattered neutrons is the same. The energy of thermal neutrons is, indeed, too small to change the internal energy of the nucleus. Moreover, we are taking the position of the nucleus to be fixed, the neutron cannot give the nucleus kinetic energy. The scattering is elastic, so the energy of neutron, and hence the magnitude of $\textbf k$, is unchanged.
\\ Considering the expression for $\psi_{inc}$ and $\psi_{sc}$ in equation~\ref{wavef-inc-sc} and $v$ the velocity of the neutrons, the same before and after the scattering because the process is elastic, the number of neutrons passing through an area $\mathrm d S$ per second is:

\begin{equation}
v \mathrm d S |\psi_{sc}|^2 = v \mathrm d S \frac{b^2}{r^2} = v b^2 \mathrm d \Omega 
\label{vdS}
\end{equation}

The flux of incident neutron is $\Phi = v |\psi_{inc}|^2 = v$, thus from the definition of the cross-section one obtains:

\begin{equation}
\frac{\mathrm d \sigma}{\mathrm d \Omega} =  \frac{v b^2 \mathrm d \Omega}{\Phi \mathrm d \Omega} = b^2
\label{sum_diffcross4}
\end{equation}

Given equation~\ref{sum_diffcross3} and~\ref{sum_diffcross4}:

\begin{equation}
a = \frac{2 \pi \hslash^2}{m} b
\label{sc-length-b}
\end{equation}

The quantity $b$ is known as the \textit{scattering length}, it depends on the nucleus and measures the strength of the neutron-nucleus interaction. It is possible to distinguish two types of nucleus. Typically in a strict potential scattering, $b$ should be independent of incident neutron energy. This occurs if the compound nucleus formation, due to the scattering process with neutron, is not formed near an excited state. The majority of nuclei belongs to this category. In the second type the scattering length is complex and varies rapidly with the energy of the neutron. The scattering of such nuclei is a resonance phenomenon and the compound nucleus has an energy close to an excited nucleus. Since the imaginary part of the scattering length correspond to absorption, such nuclei show a strong absorption behaviour. The imaginary part of the scattering length is, therefore, small for the nuclei of the first type.
\\ Besides the particular nucleus, $b$ depends on the spin state of the nucleus-neutron system as well. As already mentioned in section~\ref{basic}, the neutron has a spin $\frac{1}{2}$. Suppose the nucleus has a spin $I$ not zero, then the spin of the system can be either $I+\frac{1}{2}$ or $I-\frac{1}{2}$. As each spin state has its own value of $b$, every nucleus with non-zero spin has two values of the scattering length. On the contrary, if the nucleus spin is zero, there is only one value of $b$ which corresponds to the nucleus-neutron system with spin $\frac{1}{2}$. The values of $b$ are determined experimentally, because of the lack of a theory able to calculate these values from other properties of the neutron.
\\ Inserting the value of $a$ of equation~\ref{sc-length-b} in equation~\ref{V-delta}, it gives:

\begin{equation}
V(\textbf r) = \frac{2 \pi \hslash^2}{m} b \: \delta (\textbf r)
\label{pseudoFermi}
\end{equation}
the potential is known as the \textit{Fermi pseudopotenial}.
\\ We come back to the expression for the cross-section for a general scattering system. If the $j$th nucleus has scattering length $b_j$ its potential is

\begin{equation}
V_j(\textbf{x}_j) = \frac{2 \pi \hslash^2}{m} b_j \: \delta (\textbf{x}_j)
\label{pseudoFermi-j}
\end{equation}

Inserting this in the expression of $V(\kappa)$, which can be derived from equation~\ref{el-matix-calc3}, gives

\begin{equation}
V_j(\kappa) = \int V_j(\textbf{x}_j) e^{i \kappa \textbf{x}_j}  \: \mathrm d (\textbf{x}_j) = \frac{2 \pi \hslash^2}{m} b_j
\label{pseudoFermi-j2}
\end{equation}

From equation~\ref{partialdiffcross},~\ref{el-matix-calc3} and~\ref{pseudoFermi-j}

\begin{equation}
\bigg(\frac{\mathrm{d}^2\sigma}{\mathrm{d}\Omega \mathrm{d}E'} \bigg)_{\lambda \rightarrow \lambda'} = \frac{k'}{k} \bigg| \sum_j \langle \lambda' \vert e^{i \kappa \textbf{R}_j} \vert \lambda \rangle \bigg|^2 \delta (E_\lambda - E_{\lambda'} + E - E')
\label{partialdiffcross2}
\end{equation}

We recall that this derivation of the cross-section is based on Fermi's golden rule. For scattering processes, it is equivalent to the Born approximation; indeed, both methods are based on first-order perturbation theory.  
\\ Consider a scattering system with a single element and where the scattering length $b$ varies from one nucleus to another due to nuclear spin or the presence of isotopes or both. The average value of $b$ for the system and the average of $b^2$ can be defined as:

\begin{equation}
\overline{b} = \sum_i \nu_i b_i , \qquad \overline{b^2}= -\sum_i \nu_i b^2_i
\label{b-average}
\end{equation}

where $\nu_i$ is the frequency with which the value $b_i$ occurs in the system. Assuming there is no correlation between the $b$ values for different nuclei. Thus whatever the value of $b$ for one nucleus, the probability that another nucleus has the value $b_i$ is simply $\nu_i$. In a scattering experiment a neutron beam hit on a target, which is a large amount of nuclei, therefore a large number of scattering systems. They are identical as regards the positions and the motions of the nuclei. The total number of each $b_i$ is the same for all the systems, but each one has a different distribution of the $b$s among the nuclei. Any combination of spins can occur, i.e., the value of $b_i$ is averaged over a large number of atoms. On this assumption of no correlation between the values of $b$ and the different nuclei, 

\begin{equation}
\begin{aligned}
  \overline{b_i b_j} = \overline{b}^2 \qquad & \mathrm{if} \ i \neq j\\
  \overline{b_i b_j} = \overline{b^2} \qquad & \mathrm{if} \ i = j,
\end{aligned}
\label{b-nocorr}
\end{equation}
it is possible to calculate the measured cross-section of a scattering process as the average over all  the systems, i.e., nuclei. This is given by:

\begin{equation}
\frac{\mathrm{d}^2\sigma}{\mathrm{d}\Omega \mathrm{d}E'} = \frac{k'}{k} \frac{1}{2 \pi \hslash} \sum_{ij} \overline{b_i b_j} \int \langle e^{- i \kappa \textbf{R}_i (0)}  e^{i \kappa \textbf{R}_j (t)} \rangle e^{-i \omega t} \mathrm d t
\label{partialdiffcross3}
\end{equation}

which is obtained expressing the $\delta$-function, in equation~\ref{partialdiffcross2}, as an integral respect to time. Moreover,  $(\mathrm{d}^2\sigma / \mathrm{d} \Omega \mathrm{d}E')_{\lambda \rightarrow \lambda'}$ is summed over all final state $\lambda'$, and then the average over all $\lambda$ is performed, for the detail see~\cite{squires_2012}. It can be shown that the cross-section defined in equation~\ref{partialdiffcross3} consists of two terms: \textit{coherent} and \textit{incoherent}. They can be derived by adding and subtracting, in equation~\ref{partialdiffcross3}, the term  $(k'/k)(1/2 \pi \hslash) \int \sum_j \langle \mathrm{exp} \{- i \kappa \textbf{R}_j (0)\} \mathrm{exp} \{i \kappa \textbf{R}_j (t)\} \rangle \mathrm{exp}\{-i \omega t\} \mathrm d t$ under the assumptions defined in equation~\ref{b-nocorr}. We obtain:

\begin{align}
\frac{\mathrm{d}^2\sigma}{\mathrm{d}\Omega \mathrm{d}E'} & = \frac{k'}{k} \frac{1}{2 \pi \hslash} (\overline{b})^2 \sum_{ij}  \int_{-\infty}^\infty \langle e^{- i \kappa \textbf{R}_i (0)}  e^{i \kappa \textbf{R}_j (t)} \rangle e^{-i \omega t} \mathrm d t \nonumber \\
& +\frac{k'}{k} \frac{1}{2 \pi \hslash} \{\overline{b^2} - (\overline{b})^2 \} \sum_{j}  \int_{-\infty}^\infty \langle e^{- i \kappa \textbf{R}_j (0)}  e^{i \kappa \textbf{R}_j (t)} \rangle e^{-i \omega t} \mathrm d t 
\label{co-incoh_term} 
\end{align}

where the first term is the coherent part and the second term is the incoherent part. As one can see from the equation~\ref{co-incoh_term}, the coherent scattering depends on the correlation between the positions of different nuclei at different times, and of the nucleus itself at different times as well. It therefore gives interference effects. Whereas the incoherent scattering depends only on the correlation between the positions of the same nucleus at different times. For both terms the scattering cross-section can be defined as:

\begin{equation}
\sigma_{coh} = 4\pi \overline{b}^2, \qquad \sigma_{inc} = 4 \pi (\overline{b^2} - \overline{b}^2)
\label{cross-sec-coh-inc}
\end{equation}

Note that it was assumed the non-correlation between the values of $b$ and any nuclei. The actual scattering system has, instead, different scattering lengths associated with different nuclei. The coherent scattering can be physically interpreted as the scattering that the same system, namely same nuclei with the same positions and motions, would give if all the values of the scattering lengths were equal to $\overline{b}$. In order to obtain the scattering that occurs from the actual system, the incoherent scattering must be taken into account. Physically it derives from a fluctuation of the scattering lengths from their mean value. As it is completely random, all interference cancel out in the incoherent part.
\\ We remind that the scattering length depends on the correlation between the neutron and its spin state. It is possible to obtain the expression for the frequencies $\nu_i$ for both $\overline{b^2}$ and $\overline{b}$. We denote $I$ as the nuclear spin of a system made up of a single isotope. Therefore, the spin for the nucleus-neutron system can be either $I + 1/2$ or $I - 1/2$. Denote the scattering lengths associated to the two spin values by $b_+$ and $b_-$. If the neutrons are unpolarized and the nuclear spins are randomly oriented, each spin state has, in principle, the same probability. So $b_+$ and $b_-$ occur with a frequency $\frac{I+1}{2I+1}$ and $\frac{I}{2I+1}$ respectively. In a more general system with several isotopes, the frequencies must be multiplied by the relative abundance of the isotope to obtain the relative frequency of the scattering length.
\\The actual total scattering cross-section is then given by the sum of the two contributions of equation~\ref{co-incoh_term}, $\sigma_s = \sigma_{coh}+\sigma_{inc}= 4 \pi \overline{b^2}$.

\section{Neutron scattering techniques}

\begin{figure}[htbp]
\centering
\includegraphics[width=1\textwidth,keepaspectratio]{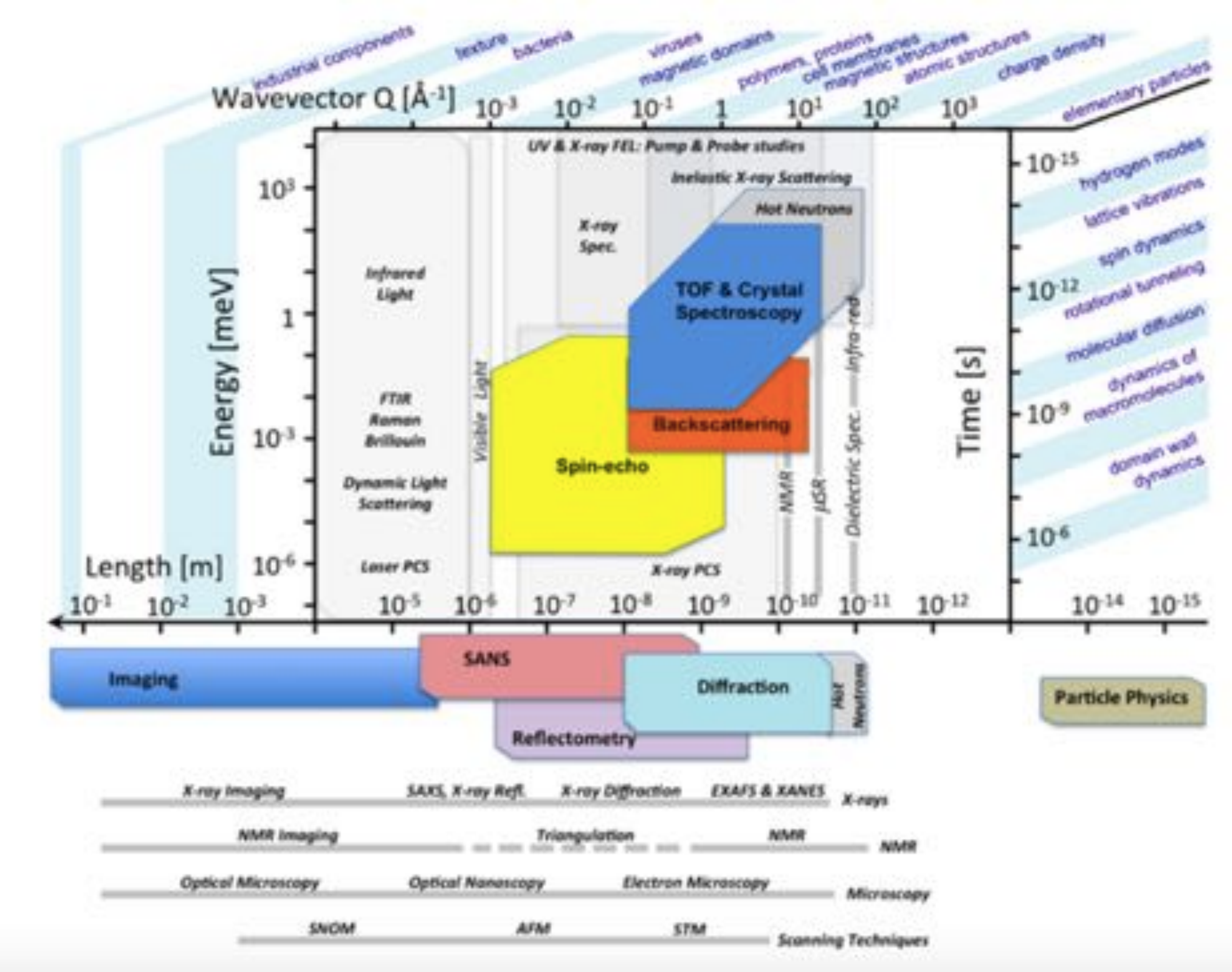}
\caption{\label{nprobe} \footnotesize Investigation of different length and time scale using neutrons as probes, compared with other techniques. The horizontal axes indicate real and reciprocal length scales, while vertical axes indicate time and energy scales. (From~\cite{ESS_TDR})}
\end{figure}

Neutron scattering can be applied to a range of scientific questions, spanning over several disciplines from physics, chemistry, geology to biology and medicine. Neutrons serve as a unique probe for revealing the structure and function of matter from the microscopic down to the atomic scale. Neutron scattering techniques enable to study the structure and dynamics of atoms and molecules over a wide range of distances and times~\cite{ESS,ESS_TDR}, as shown in figure~\ref{nprobe}. With respect to other techniques, which can provide information either within the same spatial range or the same temporal range as neutrons, neutrons scattering offers a unique combination of structural and dynamic information.

The relatively weak interaction with matter makes the neutron a high penetrating probe, which allows the study of large or bulk samples and buried interfaces. This lead to the investigation of samples under extreme conditions, e.g., very high temperature or pressure, low-temperature states without any deteriorating beam heating. Since neutrons are scattered by atomic nuclei, it is possible to discern between which element and isotope is present in a given system. This can be used to highlight particular groups of atoms in mixtures or complex biological and other hydrogen-containing materials, by substituting one isotope for another in specific regions of the molecular structure. Due to its internal magnetic moment, the neutron can be used to study microscopic magnetic structure and spin dynamics of matter. Neutrons are also useful to investigate the fundamental physics, from the creation of particles and forces right after the Big Bang, to the creation of most of the heavier elements in the explosions of massive stars.  
\\ For these reasons the research in neutron science is pushing towards understanding increasingly complex phenomena. Complexity manifests itself in investigations of multiple interrelated physical properties within materials, in studies of realistic heterogeneous samples in both extreme and natural environments, and requires, as well, experimental instruments capable of probing a wider range of length and energy scales. 

A brief overview of several techniques in neutron scattering is reported below:

\subparagraph{Small Angle Neutron Scattering:}
SANS exploits elastic neutron scattering at small angles to probe material structure on the nanometer to micrometer scale. In a SANS experiment the neutron beam is directed at a sample, either a solid, an aqueous solution, a powder or a crystal. When the plane wave $\psi= e^ {ikz}$ hits the sample, the interaction with the elementary scatters in the sample, i.e., the nuclei, gives rise to spherically symmetrical wave, $\psi= -(b/L) e^ {ikL}$; $L$ is the detector-to-sample distance and $b$ is the scattering length as defined in the previous section. The spherical waves interfere and create a pattern on the neutron detector. For the majority of the application in SANS the scattering is isotropic, so it can be expressed as a function of the modulus of the wave vector transfer:

\begin{equation}
\vert Q \vert = \frac{4\pi}{\lambda} \sin(\theta) \approx \frac{2\pi r}{\lambda L}
\end{equation}
Where $r$ is the distance of the spot of the scattered beam on the detector from that of the direct beam, $\lambda$ the neutron wavelength.

\subparagraph{Neutron reflectometry:}
It is a technique to study planar structures with a wide variety of materials from magnetic layers to biological systems. This method not only allows to investigate the material structures perpendicular to the plane, but also to observe eventual in plane correlations when the measurements are not performed in the elastic regime. A more detailed discussion of this technique is provided in section~\ref{neutrefl}, because it is the technique for which one of the two detectors, the subject of this PhD project, is designed for.

\subparagraph{Neutron diffraction:}
Neutron diffraction or elastic neutron scattering is the application of neutron scattering to determine the atomic and/or magnetic structure of a material. Through diffraction is possible to see the ordered part of systems, e.g., for ordered systems (crystals) their average structure, but also deviations from this order. With respect to the kind of interaction, two diffraction methods can be defined: nuclear diffraction when the neutrons interact with the atomic nuclei, and magnetic diffraction due to the interaction between the magnetic moments of neutrons and atoms. Two class of instruments for diffraction can be separated: powder and single crystal diffractometer.
\\ The measurement principle of this technique is based on the Bragg equation. Bragg diffraction occurs when a particle wave with wavelength comparable to atomic spacings hits a crystalline sample, it is scattered in a specular way by the atoms of the system and undergo constructive interference following Bragg's law:

\begin{equation}
n\lambda = 2d \sin(\theta)
\end{equation} 

In a ToF instrument the wavelength and the time-of-flight are related by the expression $\lambda= \frac{ht}{ m_n L}$, where $L$ is the total flight path. By using the Bragg's law we obtain the relationship between ToF and d-spacing:

\begin{equation}
t = \frac{2 m_n L}{h} d \cdot \sin(\theta)
\end{equation} 

Note that the measured d-spacing is directly proportional to ToF. For instance, considering a powder diffraction, we can express the Bragg's law for a mono-chromatic instrument as: $$\lambda = 2d_{hkl} \sin(\theta_{hkl})$$ where $\lambda$ is fixed and the measurement of the various Bragg reflections \textit{hkl} is performed by scanning in angle $\theta_{hkl}$. In a ToF instrument, instead, it is possible to measure the whole range in d-spacings at a fixed scattering angle $2\theta$ by scanning in wavelength. The Bragg equation can be rewritten as: $$\lambda_{hkl} = 2d_{hkl} \sin(\theta)$$.

\subparagraph{Neutron Spectroscopy:}
In inelastic neutron scattering, the neutrons interact with the sample changing their energy, getting either more or less energetic. While neutron diffraction investigates the structural properties of the sample, neutron spectroscopy measures the atomic and magnetic dynamics of atoms. In the inelastic neutron scattering experiment the quantity measured is the double differential cross section:

\begin{equation}
\bigg(\frac{\mathrm{d}^2\sigma}{\mathrm{d}\Omega \mathrm{d}E} \bigg) = \frac{k'}{k} b^2 S(Q,\omega)
\end{equation}

where $S(Q,\omega)$ is the neutron dynamic Structure Factor. The parameters of interest are the energy and the momentum transfer, $\hslash \omega$ and $Q$ respectively.
\\ Time-of-Flight spectrometers can be divided into two classes: direct geometry spectrometers, in which the initial energy is determined by a crystal, through Bragg scattering condition, or a chopper, which allows to select a specific wavelength, while the final energy is defined by the time-of-flight. Indirect geometry spectrometers, in which the sample is illuminated by a white incident beam, $E_i$ is determined by time-of-flight, while $E_f$ is defined by a crystal or a filter. 
\\ On a steady-state source pulsing devices to monochromatize the incident and scattered beam are required. For instance, in a triple axis spectrometer the incident and scattered wave vectors, $k_i$ and $k_f$, are selected by Bragg diffraction on the monochomator and analyser crystals respectively.

\subparagraph{Neutron Spin-echo:}
This technique is a particular form of spectroscopy that relies on the precession of a spinning neutron. Neutron spin-echo is a time-of-flight method. Polarized particles with a magnetic moment and spin $s=1/2$ behaves like a classical magnetic moment. When entering a region of magnetic field perpendicular to their magnetic moment, the particles will undergo Larmor precession. In the case of neutrons: $$\omega = \gamma B $$ where $B$ is the magnetic field, $\omega$ is the frequency of rotation and $\gamma$ is the gyromagnetic ratio of the neutrons. The Larmor precession of the neutron spin, in a zone with a magnetic field before the sample, encodes the individual velocities of neutrons in the beam into precession angles. 
\\ When the magnetic field changes its direction with respect to the neutron trajectory, it is possible to identify two limits: if this change is slow compared to the Larmor precession, the parallel component of the polarized beam will be maintained. On the contrary if the change is fast compared to the Larmor precession, the polarization will not follow the field direction. This effect is used to flip the de-phase of the incoming neutrons. A symmetric decoding zone will follow in such a way that the precession angle accumulated is compensated and all spins rephase to form the spin-echo.
\\ If an analyser is put after the second precession field at an angle $\phi$ between the polarization of a neutron and the analyser direction, the probability that a neutron is transmitted is $\cos(\phi)$. At a given $Q$, the probability of the scattering with $\omega$ energy exchange is by definition $S(Q,\omega)$. The neutron spin-echo directly measures the intermediate scattering function $ F(Q,t)$, which is the Fourier transform of the scattering function. Where: $$t = \frac{\hslash \gamma B l}{m_n v^3}$$
and it is proportional to $(1/v^3) \lambda^3$, so the resolution in time increases very rapidly with $\lambda$.

\subparagraph{Neutron imaging:}
It is a non-destructive technique, which exploits the penetration of neutrons, to investigate the internal structure of many materials and engineering components. Neutron imaging is complementary to other non-destructive imaging methods, in particular X-ray imaging. While X-rays are scattered and absorbed by electron cloud of an atom, neutrons interact with the atomic nuclei. Thus, neutrons are more sensitive than X-ray to light elements, i.e., hydrogen, lithium, boron, carbon and nitrogen. From some polycrystalline materials, a strong dependency of the attenuation is observed in the cold neutron range, because of the Bragg scattering from the crystal lattice. The instrument layout is quite simple, apart from the source, which can be either a reactor, a spallation source or a neutron emitting isotope, a collimator is needed to determine the geometric properties of the beam, and could also employ filters to modify the energy spectrum of the beam. The image resolution depends much on the collimator geometry and can be expressed by the ratio $\frac{L}{D}$, where $L$ is the collimator length and $D$ is the diameter of the aperture of the collimator on the side facing the source. The beam is transmitted through this device and recorded by a plane position sensitive detector, able to measure the 2 dimensions perpendicular to the beam direction.
\\ Neutron tomography can be obtained combining measurements at different angles. Also dynamic processes can be detected with a high neutron flux and with a neutron detector fast enough to acquire images at sufficiently high frame rate.

\subsection{Neutron reflectometry}\label{neutrefl}
In this section the attention is focused on the neutron reflectometry technique, in order to set the theoretical principles on which the experimental work illustrated in chapter~\ref{chapter4} is based on.

Neutron reflection follows the same fundamental equations as optical reflectivity, but different refractive indices. As mentioned in section~\ref{sec_scatt}, in the quantum mechanical approach, the neutron can be described by a wave-function. The dynamical theory, which takes into account the change in the incident wave within the scattering system, describes this phenomenon. The theory proceeds by solving the time-independent Schr${\"o}$dinger equation for the wave-function of the neutron, $\Psi$, which represents the neutron wave inside and outside of the reflecting surface, matching the continuity of the two wave-functions and their derivatives at the boundaries. The potential $V$ in the equation~\ref{sch_ref} represents the effect of the interaction between the neutron and the nuclei in the medium.

\begin{equation}
- \frac{\hslash^2}{2m_n} \nabla^2 \Psi + V \Psi = E\Psi
\label{sch_ref}
\end{equation}

From the discussion in the previous section, we obtained the expression for the Fermi pseudopotential (equation~\ref{pseudoFermi}). Assuming now a spread potential over all nuclei, the Fermi pseudopotential can be averaged by the density of the scattering lengths of the material, and it represents the refractive index:

\begin{equation}
V = \frac{2 \pi \hslash^2}{m_n} N_b
\label{pseudoF_average}
\end{equation}

where $N_b$ is the \textit{scattering length density} of the medium through which the neutron travels as can be defined as:

\begin{equation}
N_b = \sum_i b_in_i
\label{sc_length_dens}
\end{equation}

\begin{figure}[htbp]
\centering
\includegraphics[width=0.6\textwidth,keepaspectratio]{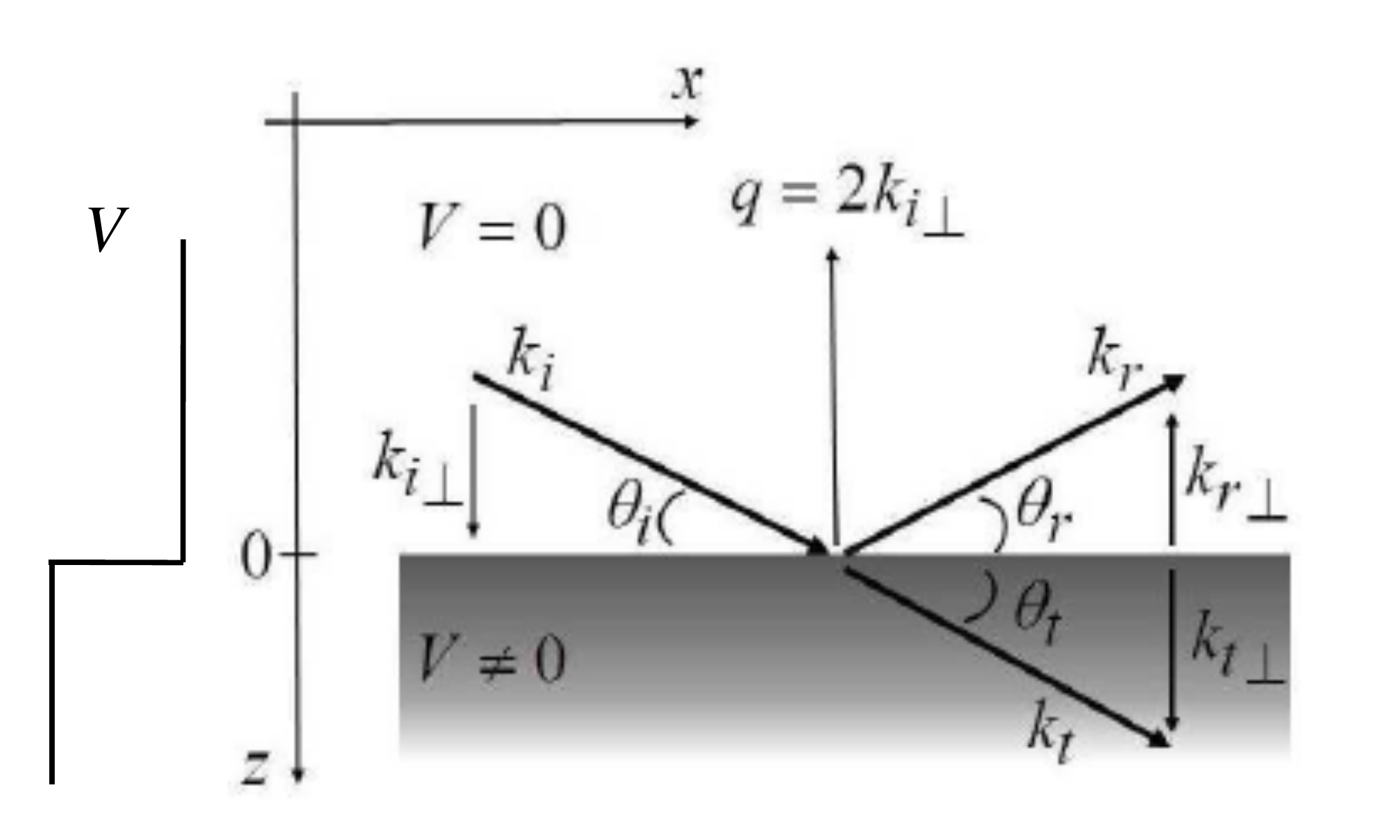}
\caption{\label{sketch_ref} \footnotesize Reflection of an incident neutron beam from an ideally flat interface. $k_i$ and $k_r$ are the incident and scattered wave vectors at angles $\theta_i$ and $\theta_r$ respectively. $q$ is the wave vector transfer and $V$ the potential.}
\end{figure}

$b_i$ is the coherent scattering length of the nucleus $i$ and $n_i$ is the number of nuclei per unit volume. 
We consider a neutron beam approaching an ideal flat surface with a bulk potential $V$, as shown in figure~\ref{sketch_ref}, the only force is perpendicular to the surface. The potential varies only the normal component of the momentum ($z$ direction), and cannot change the neutron's wave-vector parallel to the interface, i.e., in the $x$ direction. Under these conditions the solutions for the equation~\ref{sch_ref} are:

\begin{equation}
\begin{aligned}
  &\Psi_0 (z) = e^{i k_{i\perp} z} + r e^{-i k_{i\perp} z} \quad & \mathrm{if} \ z< 0\\
  &\Psi_1 (z) = t e^{i k_{t\perp} z}  & \mathrm{if} \ z>0
\end{aligned}
\label{Psi_ref}
\end{equation}

where $r$ and $t$ are the probability amplitudes for reflection and transmission. 
The perpendicular component of the incoming wave vector $k_{i\perp}$ is the normal component of the kinetic energy $E_{i\perp}$ that determines whether the neutron is totally reflected from the barrier~\cite{Cubitt_ref_principles}.

\begin{equation}
E_{i\perp} = \frac{(\hslash k_{i\perp})^2}{2 m_n} = \frac{(\hslash \, k_{i} \ \sin(\theta_i))^2}{2 m_n} = \frac{(\hslash q)^2}{8 m_n}
\label{E-perp-ref}
\end{equation}

$\theta_i$ is the incoming neutron angle with respect to the surface and $q = 2k_i \ \sin(\theta_i)$ is the momentum transfer, as depicted in figure~\ref{sketch_ref}. The total reflection occurs if $E_{i\perp} < V$, thus no neutrons penetrate into the substrate. The equivalence between the orthogonal component of the kinetic energy, equation~\ref{E-perp-ref}, and the barrier potential, equation~\ref{pseudoF_average}, identifies the critical value of the wave vector transfer $q_c$ as follows:

\begin{equation}
q_c = \sqrt{16 \pi N_b}
\label{qcritic}
\end{equation}

The angle at which it happens is called \textit{critical angle}. The reflectivity of neutrons of a given wavelength from a bulk interface is unity for angles below the critical one and it falls sharply at larger angles~\cite{Cubitt_ref_principles}.

In case of elastic scattering the momentum is conserved, $ k_{i\perp}= k_{r\perp}$, thus the incident and reflected beam angle has the same value and the reflection is specular. 
\\ On the contrary, if $E_{i\perp} > V$, the reflection is not total and the neutrons can be both reflected or transmitted into the bulk of the material. The transmitted beam, $k_t$, changes direction because the potential acts on the normal component of the kinetic energy reducing it. This change is given by $E_{t\perp} = E_{i\perp} - V$ and correspond to:

\begin{equation}
k_{t\perp}^2 = k_{i\perp}^2 - 4 \pi N_b
\label{k-transmitted}
\end{equation}

This relation allows to derive the refractive index $n$:

\begin{equation}
n = \frac{k_{t}^2}{k_{i}^2} = \frac{k_{i\parallel}^2 +(k_{i\perp}^2 - 4 \pi N_b)} {k_{i}^2}= 1- \frac{4 \pi N_b}{k_{i}^2} = 1- \frac{\lambda^2 N_b}{\pi}
\label{refract-index}
\end{equation}

where $\lambda$ is the neutron wavelength. As $N_b << 1$ for most materials, it is possible to approximate equation~\ref{refract-index}, in the thermal neutron energy range, as 

\begin{equation}
n \approx 1- \frac{\lambda^2 N_b}{2 \pi}
\label{refract-index2}
\end{equation}

The neutron refractive indices of most condensed phases are slightly less than that of air or vacuum. The total reflection, as with light, may occur when neutrons pass from a medium of higher refractive index to one of lower refractive index. Unlike that with light, where the total internal reflection is more common than the external one, with neutrons the total external reflection is mostly observed.  
\\ Measurements of the critical angle for total reflection for pure material became an important method to determine the scattering lengths of nuclei, since the neutron refractive index is related to the scattering lengths of the constituent atoms, as shown in equation~\ref{refract-index}. 
\\ Over the past few years neutron reflection has been used to investigate the inhomogeneities across the interface, either is composition~\cite{Penfold} or in magnetisation~\cite{Felcher}. Indeed, as for light, interference occurs between waves reflected at the top and at the bottom of a film at the interface. This gives rise to interference fringes in the reflectivity profile~\cite{Cubitt_ref_principles}. Informations on the structure within a surface cannot be achieved by specular reflection. Off-specular study are required~\cite{Sinha}, i.e., when the scattering is not elastic, thus the incident angle differs from the reflected angle and the wave vector transfer is not any more perpendicular to the surface, but it also has a component parallel to the surface. 
\\ Neutron reflection is now used for studies of surface chemistry (e.g., surfactants, polymers, proteins, etc.), surface magnetism (e.g., superconductors, magnetic multilayers) and solid films (e.g., Langmuir-Blodgett films, polymer films, thin solid films)~\cite{Cubitt_ref_principles}.

Referring to equation in ~\ref{Psi_ref} it is possible to derive the classical Fresnel coefficients as it is in optics, exploiting the continuity and the derivative conditions of the wave-function at the boundary. We obtain:

\begin{equation}
1+r = t \qquad k_{i\perp}(1-r)= tk_{t\perp} 
\label{cont-deriv-condition}
\end{equation}

the second relation is valid only for $E_{i\perp} > V$, thus the reflection and transmission coefficients are:

\begin{equation}
r = \frac{k_{i\perp}-k_{t\perp}}{k_{i\perp}+ k_{t\perp}}, \qquad t = \frac{2k_{i\perp}}{k_{i\perp}+ k_{t\perp}}
\label{r-t-coeff}
\end{equation}

The reflected and transmitted intensity is a function of the quantum mechanical probability amplitude squared, i.e., $R=r^2$ and $T=t^2$. By using the expression~\ref{qcritic},~\ref{k-transmitted} and~\ref{r-t-coeff} the reflectivity $R$ can be related to the wave vector transfer $q$ and $q_c$, as follows

\begin{equation}
R =r^2 =\Bigg( \frac{q-\sqrt{q^2-q_c^2}}{q+\sqrt{q^2-q_c^2}} \Bigg)^2
\label{R-intensity}
\end{equation}

When $q>>q_c$, equation~\ref{R-intensity} can be reduced to

\begin{equation}
R \approx \frac{16\pi^2 }{q^4} N_b^2
\label{R-intensity2}
\end{equation}

If we consider the wave-function within the surface $\Psi_1$, substituting equation~\ref{k-transmitted}, a real solution for $E_{i\perp} < V$ is found:

\begin{equation}
\Psi_1 (z) = t e^{+i (k_{i\perp}^2 - 4\pi N_b)^{1/2}\, z} = t e^{-\frac{1}{2} (q_{c}^2 - q^2)^{1/2}\, z}
\label{evanescent-wave}
\end{equation}

The equation shows that even when the potential barrier is higher than the particle energy orthogonal to the surface, it can still penetrate to a characteristic depth of  $(q_{c}^2 - q^2)^{1/2}$. This is an $evanescent$ wave, which decay exponentially. It travels along the surface with a wave vector $k_{\parallel}$ and after a very short time it is ejected out of the bulk in the specular direction.  
\\ The expression of the reflected intensity has been derived considering an ideal flat interface between materials, in general the interface may be rough over a large range of lengths scales. The surface roughness must be included in the reflectivity term, equation~\ref{R-intensity2}, in the approximation $q>>q_c$, becomes:

\begin{equation}
R \approx \Bigg(\frac{16\pi^2 }{q^4} N_b^2\Bigg) \cdot e^{-q^2 \sigma^2}
\label{R-intensity3}
\end{equation}
 
where $\sigma$ is the characteristic length scale of the imperfection between the layers. The result from constructive and destructive interference of waves, which are reflected from the interface, are the Kiessig fringes. The intensity oscillations in the reflectivity can be observed and the periodicity can be related to the thickness $d$ of the film via $2\pi/d$. The result can be obtained considering the condition of the Bragg diffraction in the case of diffuse neutron scattering from roughness of the interface~\cite{R_offspec1_Ott}.
\\ The aim of the specular neutron reflection experiment is to measure the reflectivity as a function of the wave-vector perpendicular to the reflecting surface, $q$. Two kind of measurements can be performed, depending on the source and on the instrument. The measurement can be performed by varying either the angle of incidence $\theta$, at constant wavelength, or measuring the time-of-flight, therefore changing wavelength, at constant angle. The incoming intensity must be also measured. The ratio between these two intensities is the reflectivity curve as a function of the transfer wave-vector, encoded by $\lambda$ and $\theta$ according to the Bragg's law: 

\begin{equation}
q = \frac{4\pi }{\lambda} \sin(\theta)
\label{q-value}
\end{equation}

\chapter{Neutron Detectors operating principles }

Both for nuclear or elementary particle physics several types of detectors have been developed. All are based on the same fundamental principle, the transfer of part or all of the radiation energy to the detector matter where it is converted into some form of electrical signal~\cite{DET_leo}. The main difference between charged and neutral particles is that the former transfer their energy to matter, primarily, through direct collision with the atomic electrons, hence inducing excitation or ionization of the atoms. On the other hand, neutral radiation must undergo some kind of reaction in the detector to produce charged particles, which successively ionized and excite the atoms. The way in which the output electrical signals are produced depends on the kind of detector: gaseous, scintillator or solid state, and their subsequent design. 
\\ Gaseous detectors are based on the direct collection of the ionization electrons and ions produced in the gas, while in scintillator detectors, the detection of ionizing radiation arises from the scintillation light produced in certain materials. When coupled to an amplifying device such as a photomultiplier, is possible to convert the light into an electrical pulse~\cite{MIO_MyThesis}. 
\\ Solid state detectors are based on semiconductor materials. The basic operational principles are similar to gas ionization devices. Instead of gas the medium is a semiconductor. The passage of an ionizing particle in an electric field produces charge carriers (electron-hole pairs) that drift and produce a signal. The semiconductor detectors benefit form a small energy gap between their valence and conduction bands, thus the energy required to create a pair is generally one order of magnitude smaller than that required for gas ionization~\cite{rancoita}.
\\ Neutrons are a powerful probes for condensed matter studies, because of the lack of charge and the weak interaction with materials. In turn these properties make difficult the construction of efficient neutron detectors. Moreover, the design is focused to the energy of the neutrons, because the energy changes the way neutrons interacts with matter. A fast neutron can transfer its energy and generate a charged particle by recoil hitting a light atom target. Thermal neutron are not enough energetic to give rise to charged products by elastic scattering. Secondary radiation is, indeed, produced by capture reactions, either $\gamma$-rays of heavy charged particles such as protons, $\alpha$, tritium or fission fragments.
\\ An overview of several detectors used in the thermal neutron enegy range applications will be presented in the next section. A general description of operation principles for both gaseous and solid state detectors will follow, the main references are~\cite{ DET_leo,DET_knoll, rancoita,Particle_Detection}.

\section{Neutron detectors}\label{secndet}

In general every type of neutron detector involves the coupling of a target material designed to carry out the neutron conversion, with a typical radiation detector as used in other disciplines, whose operation is described below. The cross section for neutron interactions strongly depends on the neutron energy as explained section~\ref{sec_crossec}. Different techniques have been developed for neutron detection in different energy regions. The attention will be focused on the slow (thermal and cold) neutron energy region, below 0.5 eV, which is the \textit{cadmium cutoff}. When looking for nuclear reactions that could be useful in neutron detection, several factors must be considered: 

\begin{enumerate}[label=(\roman*)]
\item The cross section for the reaction must be as large as possible and greater than the scattering cross section, so to keep the detector dimensions relatively small. Of particular importance for detectors in which the target material is incorporated as a gas.
\item The target nuclide should be either of high isotopic abundance in the natural element or, economically accessible in case of artificial manufacture.
\item The energy liberated in the reaction is determined by the $Q-$value. The higher the $Q-$value, the greater the energy given to the reaction products, the easier the discrimination against gamma-ray events using simple amplitude discrimination. 
\item The distance travelled by the reaction yields also concern the detector design, in terms of needful active volume to detect the released energy.  
\end{enumerate}

\begin{figure}[htbp]
\centering
\includegraphics[width=1\textwidth,keepaspectratio]{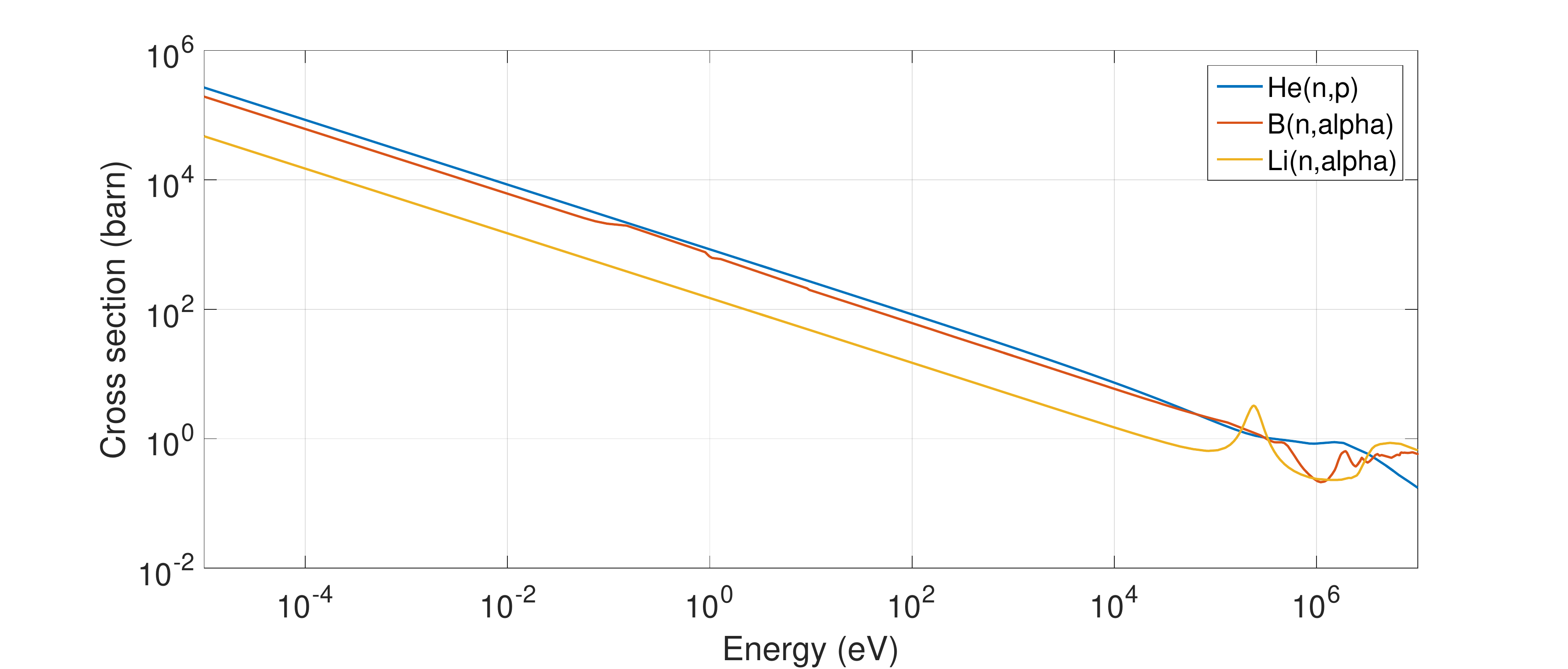}
\caption{\label{crossblihe} \footnotesize Cross section as a function of the neutron energy for $\mathrm{^{3}He(n,p)}$, $\mathrm{^{10}B(n,\alpha)}$ and $\mathrm{^{6}Li(n,\alpha)}$ reactions.}
\end{figure}

The most popular reactions for the conversion of slow neutrons are $\mathrm{^{3}He(n,p)}$, $\mathrm{^{6}Li(n,\alpha)}$ and $\mathrm{^{10}B(n,\alpha)}$. All these reactions have large positive $Q-$values and large cross sections at thermal energies. The cross sections~\cite{NIST} are shown in figure~\ref{crossblihe} as a function of the neutron energy. 

The $\mathrm{^{10}B(n,\alpha)}$ can be written as:
\begin{equation*}
n + ^{10}_{5}\mathrm{B} \rightarrow \begin{cases}^{7}_{3}\mathrm{Li} (1.02 MeV)+ \alpha ( 1.78 MeV) \quad 6\% \\ ^{7}_{3}\mathrm{Li} (0.84 MeV)+ \alpha(1.47 MeV) + \gamma (0.48 MeV)\quad 94\% \end{cases}
\label{Breact}
\end{equation*}
The two branches indicate that the reaction product $^7$Li can be left either in its ground state or in its first excited state. In this case the $^7$Li quickly returns to its ground state with the emission of a 0.48 MeV gamma ray. In either case, the $Q-$value of the reaction is very large (2.31 or 2.782 MeV) compared with the incoming energy of the slow neutron, thus the energy distributed to the reaction products is essentially the $Q-$value itself. This also means that the the incoming linear momentum is very small, and therefore the two reaction yields must show as well a total momentum of approximately zero. Namely, the products are emitted in exactly opposite directions and the energy will always be shared in the same manner between them.
\\ In the case of $\mathrm{^{6}Li(n,\alpha)}$ the reaction proceeds only to the ground state of the product and can be written as:
\begin{equation*}
n + ^{6}_{3}\mathrm{Li} \rightarrow \ ^{3}_{1}\mathrm{H} (2.73 MeV)+ \alpha ( 2.05 MeV)
\label{Lireact}
\end{equation*}
The alpha particle and the tritium must be oppositely directed when the incoming neutron energy is low. With respect to $^{10}$B the cross section is always lower until the resonance region (> 100 keV), figure~\ref{crossblihe}. The lower cross section is typically a disadvantage, but is, in part, compensate by the higher $Q-$value, i.e., the reaction products have a greater energy. 
\\ The $^3$He gas is widely used as detection medium for neutrons through the reaction:
\begin{equation*}
n + ^{3}_{2}\mathrm{He} \rightarrow \ ^{3}_{1}\mathrm{H} (0.191 MeV)+ p ( 0.573 MeV)
\label{Hereact}
\end{equation*}
The $Q-$value (764 keV) is lower than the previous two cases, but the cross section for this reaction is higher than the other two. Although $^3$He is commercially available, it is lately much less available and more expensive~\cite{HE3S_hurd,HE3S_karl}. The availability~\cite{HE3S_kouzes,HE3S_shea} and the requirements of higher performances, as explained in chapter~\ref{chapter3}, are the reason why a number of research programs are aiming to find technologies that would replace the technology based on $^3$He gas~\cite{HE3S_karl,HE3S_gao,HE3S_gebauer}.
\\ Another type of reaction exploits the neutron capture by Gadolinium which has one of the highest nuclear cross section for thermal neutron capture, about $2 \cdot 10^5\,$ barns. A more detailed description can be found in chapter~\ref{chapter6}.
\\ The fission reaction can be used as the basis of slow neutron detectors as well. The cross sections of $^{233}$U, $^{235}$U and $^{239}$Pu are, indeed, rather high at low neutron's energies. The main feature of the fission reaction is its extremely large $Q-$value ($\approx$ 200 MeV) compared with the reactions mentioned above. Detectors based on fission reaction generally give output signals orders of magnitude larger compared with the competing interactions or incident gamma rays. A clear discrimination can be achieved. Almost all fissile nuclide are naturally alpha radioactive, thus any detector will exploit this kind of reaction will also show an output signal due to the alpha decay. However, the energy of the alpha particles is always much smaller compared with the one induced by a fission reaction. Also in this case these events can be easily discriminated by pulse amplitude discrimination.

\begin{figure}[htbp]
\centering
\includegraphics[width=0.6\textwidth,keepaspectratio]{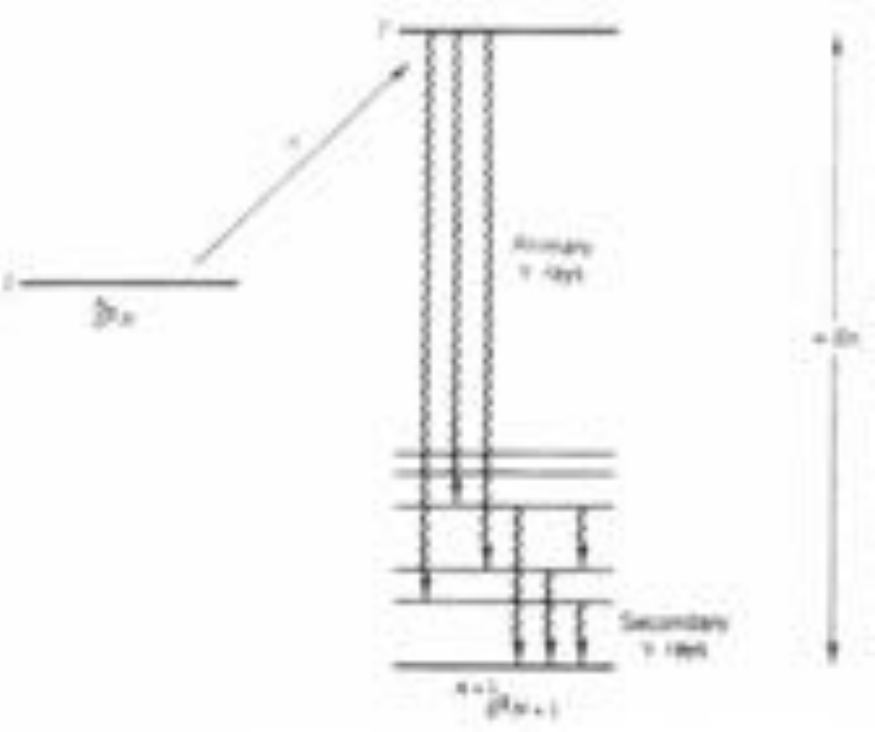}
\caption{\label{cattura} \footnotesize Sketch of a low-energy neutron capture process which leads to a state $l'$, that emits primary $\gamma$-rays followed by secondary $\gamma$-rays. (From~\cite{DET_krane}).}
\end{figure}

\paragraph{Radiative capture} Another important mechanism for low energy neutrons, is the radiative capture. The target nucleus absorbs the neutrons and goes in an excited state, the de-excitation into the ground state most probably occurs via emission of $\gamma$-ray. Although it is possible to re-emit the neutron, for heavy nuclei and very low energy incident neutrons, this mode of decay of the compound on resonant state is suppressed~\cite{DET_krane}. A sketch is shown in figure~\ref{cattura}.

The isotopic compound of the target changes because the neutron is not re-emitted, following the reaction, $$ n + A^{n+p}_{p} \rightarrow  A^{n+p+1}_{p} + \gamma \ .$$ The excitation energy $E_x$ of $A'$ is simply $S_n + E_n$, the neutron separation energy plus the energy of the incident neutron. $E_x$ is typically 5-10 MeV for low-energy neutrons.
\\ Neutron capture reactions can be used to determine the energy and spin-parity assignments of the capturing states~\cite{DET_krane}. Assuming that the nucleus had spin $I$ and parity $\pi$, thus the spin $I'$ of the capturing state is determined by the neutron orbital angular momentum $l$ and spin angular momentum $s$ added to that of the target nucleus:

\begin{equation}
I' = I + l + s
\label{spinparity}
\end{equation}
and the parities are correlated by the relation:

\begin{equation}
\pi' = \pi(-1)^ l 
\label{parityspin}
\end{equation}

For incident neutron in the thermal energies, only the $s$-wave capture will occur, for which $I' = I \pm \frac{1}{2}$ and $\pi' = \pi$. 
\\ A very interesting radiative capture process is the one of Gadolinium; the cross section for thermal neutron capture is about $2.5 \cdot 10^5\,$barns in $^{157}$Gd, one of the largest nuclear cross section in any material. This isotope is 15.7\% abundant in natural Gadolinium, and upon neutron absorption in Gd, $\gamma$-rays over a wide keV energy range, and a cascade of conversion electrons with an energy spectrum extending from 20-30 keV up to $\sim$250 keV, with a main peak at about 70 keV~\cite{Gd-capt, Gdabs1,Gdabs2}, are emitted. The secondary electron emission is characterized by a probability of $\approx$80\%~\cite{GdNeumann}. The neutron absorption cross section in natural and isotopic Gd ($^{157}$Gd) takes quite high values over the range of cold and thermal neutrons, although it decreases very sharply for energies greater than 0.1 eV. The ranges of the 70 keV conversion electrons are $\approx$ 10 $\mu$m in Gd and $\approx$ 30 $\mu$m in Si. The combination of a rather large thickness and a large absorption cross section makes Gadolinium an appropriate converter for cold and thermal neutrons. The main problem compared with the other conversion reaction, in which heavy charged particles are produced is the high $\gamma-$ray background, and an effective pulse shape discrimination techniques is needed. 
\\The description of a thermal neutron detector with employ the Gadolinium is presented in chapter~\ref{chapter6}, together with the characterization of pulse shape analysis method developed for these kind of devices. 

\subsection{Detectors based on the Helium-3 reaction}

\begin{figure}[htbp]
\centering
\includegraphics[width=.6\textwidth,keepaspectratio]{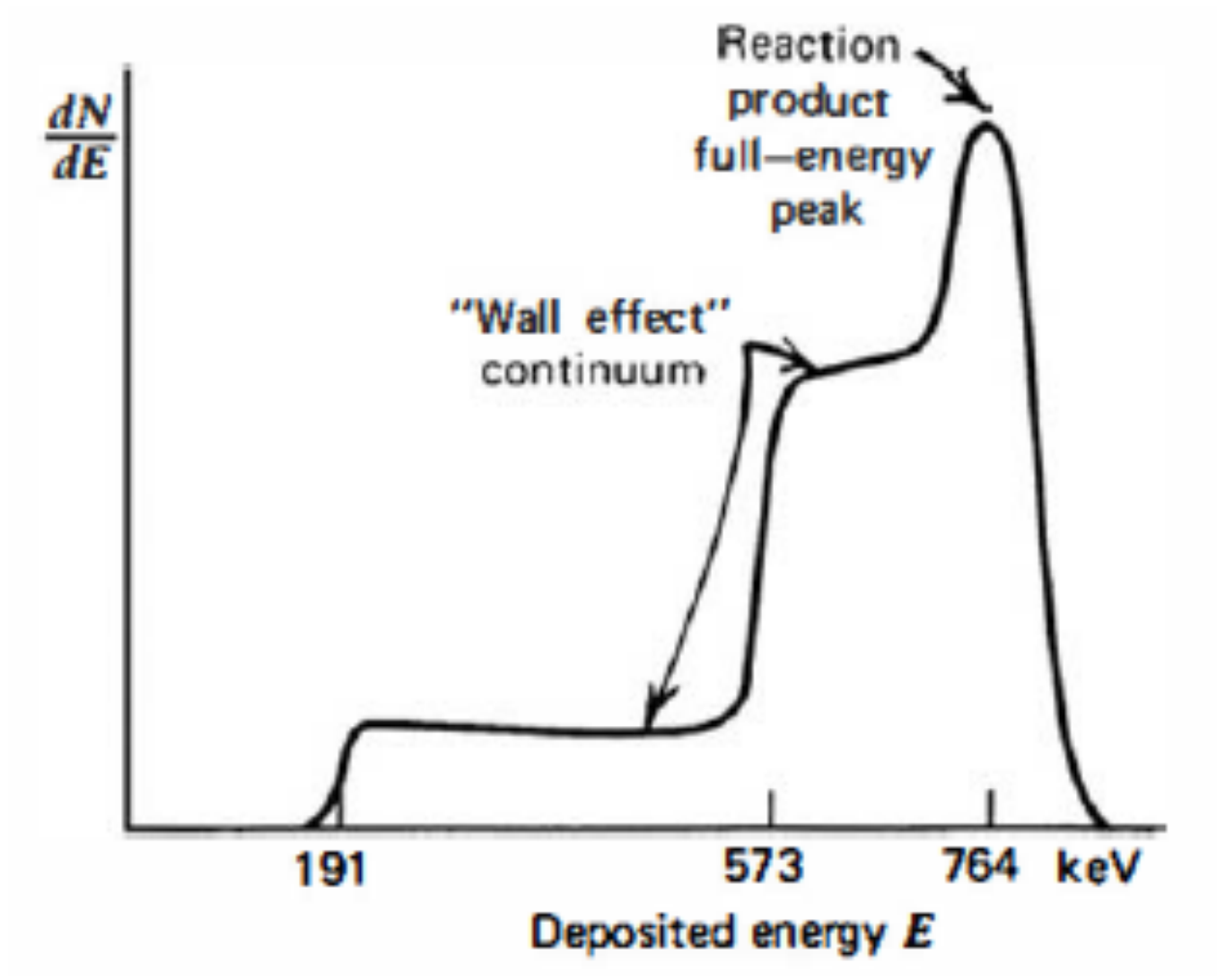}
\caption{\label{phshe} \footnotesize Sketch of the pulse height spectrum from a $^3$He tube, taking into account the wall effect. Figure from~\cite{DET_knoll}.}
\end{figure}

The $\mathrm{^{3}He(n,p)}$ reaction with the highest cross section (5330 barns) among the interactions described here, is an attractive alternative for slow neutron detection. In a large detector one would expect each thermal neutron reaction to deposit the 764 keV for the sum of both tritium and proton. However, the range of these particles is comparable with the dimensions of a typical $^3$He tube. The wall effect must be taken into account for such kind of counters. In figure~\ref{phshe} is illustrated a sketch of the pulse height spectrum for a $^3$He tube. 

The two steps correspond to the energy of the proton (573 keV) and the tritium (191 keV) respectively, while the continuum in the pulse height spectrum is due to the wall effect. Several considerations can be taken into account in the design of $^3$He tube to minimize the wall effect. The simplest one is to enlarge the counter diameter as much as possible, so that most of interactions occur far from the walls. Another way to reduce the wall effect is to increase the pressure of the $^3$He gas in order to reduce the range of the charged particles. One method of reducing the range is to add a small amount of a heavier gas to the $^3$He to achieve an enhanced stopping power. 
\\ Similar to the behaviour of boron-based detectors, the efficiency drops off with increasing neutron energy because of the decrease of the cross section. The probability of observing a pulse will be greater for those cases in which the neutron passes through the maximum distance in the gas. Often, a dead area  is observed at the ends of a tube, where the charge collection is poor. Thus, the true active length may be less the the physical length of a tube. To give an idea of typical efficiencies, a 2 cm path length through a tube at 1 atm pressure corresponds to an efficiency of 25\% for 0.025 eV neutrons. This value rise up to 76\% for a tube pressurized to 5 atm.
\\ The rise time of the output pulse observed from either a $^3$He tube or a BF$_3$ tube will depend both on the position of the neutron interaction and the orientation of the charged particle tracks with respect to the tube axis~\cite{DET_knoll}. The most significant charges that generate the output signal are produced in the avalanches around the anode wire. If the electrons have the same drift time, the avalanches are triggered at the same time. If, instead, the electrons arrivals are spread because of different drift times, the pulse rise time is slower. The slow component of the rise comes because of the slow motion of the positive ions after they leave from the proximity of the anode wire. Another factor that affect the rise time of the signals is the range of the back-to-back proton-tritium reaction products in the gas. Longer ranges correspond to greater variation in drift times of the ionization electrons, so the average pulse rise time will be longer as well. By adding a second heavy gas component (e.g., Ar or Xe) it is possible to reduce the track lengths and consequently to shorten the rise time. This will be an advantage in performances, leading to reach higher count rates with minimum dead time. 

\subsubsection{The Helium-3 crisis}

The natural abundance of $^3$He on earth is about 1.4 parts per million of all helium. It is, therefore, manufactured through nuclear decay of tritium: $$ \mathrm{^3H (t_{1/2}= 12.3 y) \rightarrow ^3He + e^- +\overline{\nu}_e }$$ The main suppliers of $^3$He are USA and Russia. The most common source comes from the US nuclear weapons program. The tritium is produced for use in nuclear warheads, over time it decays into $^3$He and must be replaced for the maintenance of the warheads. The $^3$He is a byproduct of the tritium supply.  
In the past decade the consumption of $^3$He has grown rapidly. After the terroristic attacks of September 11, 2001, the federal government installed neutron detectors at the US border for homeland security reasons. Not only the shortage of $^3$He~\cite{HE3S_shea}, but also the demands of better performances in neutron science, see chapter~\ref{chapter3}, opened to the research for alternative solutions in order to replace the $^3$He-based neutron detector technology.  

\subsection{Detectors based on the Lithium-6 reaction}

A lithium tube is not available, because a stable lithium-containing gas does not exist. Moreover the lithium is highly hygroscopic, thus it cannot be exposed to water vapor. Commercially available crystals are hermetically sealed in a thin canning material with an optical window provided on one face. Because of the high density of the material, crystal sizes need not be large for very efficient slow neutron detection. Nevertheless, the large $Q-$value of the lithium reaction offers some advantage in the case of discrimination against gamma-ray and other low-amplitude events. Moreover, the energy imparted to the yields is always the same, because the lithium reaction goes uniquely to the ground state of the product nucleus. The resulting pulse height distribution is therefore a single peak. Some applications of gas-filled detectors with solid lithium-based converters can be found~\cite{Li-joneja}, but the more common applications of this reaction use the scintillation process to detect the charged particles of the neutron-induced reaction.
\\ Scintillators operate by absorbing incident radiation that raises electrons to excited states. After the subsequent de-excitation, usually on the order of tens of ns, the scintillator emits a photon in the visible light range. This light interacts with a photocathode of a photomultiplier tube (PMT) or a photodiode, releasing electrons through the photoelectric effect. These electrons are then accelerated along the photomultiplier tube generating secondary electrons, the final amplification can be on the order of 10$^6$ or higher. Scintillators detectors include: liquid organic scintillators, crystals, plastic and scintillation fibers~\cite{Tavernier-Det}.
\\ In the case of lithium, the typical choice is crystalline lithium iodide, with a scintillation decay time of approximately 0.3 $\mu$s. Crystals of lithium iodide are usually large compared with the ranges of either of the reaction products from the neutron interaction~\cite{DET_knoll}. The pulse height will not be affected by the wall effects as in the case of a $^3$He tube or BF$_3$ tube. The range of secondary electrons produced by gamma-rays will not be large compared to the crystal dimensions as well. i.e., a 4.1 MeV electron will yield about the same light as the 4.78 MeV reaction products. The gamma-ray rejection characteristics will be less than that of a typical gas-filled neutron detector, in which a $\gamma$-ray can deposit only a small fraction of its energy. A pulse shape discrimination is needed to discriminate events~\cite{Pulse-shape-Li6}. This methods relies upon the property that the fraction of the light that appears in the slow component, often, depends on the nature of the exciting particle. 
\\ Compared with the boron, the higher $Q$-value of the neutron-lithium reaction and the more penetrating nature of the reaction products, lead to longer ranges of the yields in the converter material, thus the optimum thickness of the converter layer is larger. On the other hand, the thermal neutron cross section for the $^6$Li reaction is smaller than that for the $^{10}$B reaction (940 vs. 3840 barns), see figure~\ref{crossblihe}. 
\\ Although the ideal lithium-based converter layer would be pure $^6$Li metal, the highly reactive chemical properties of the metal limits its use. The stable compound LiF has been widely employed, model calculations~\cite{SS_mcgregor2} show that the optimal thickness of 26 $\mu$m of $^6$LiF results in a detection efficiency of 4.4\% in transmission. This value is comparable with the detection efficiency achievable with a $^{10}$B converter, with an optimal thickness about a factor 10 smaller. For both converter materials the simple case has been considered, but more complex geometries can provide higher efficiencies.  
  
\subsection{Detectors based on the Boron-10 reaction}

\begin{figure}[htbp]
\centering
\includegraphics[width=1\textwidth,keepaspectratio]{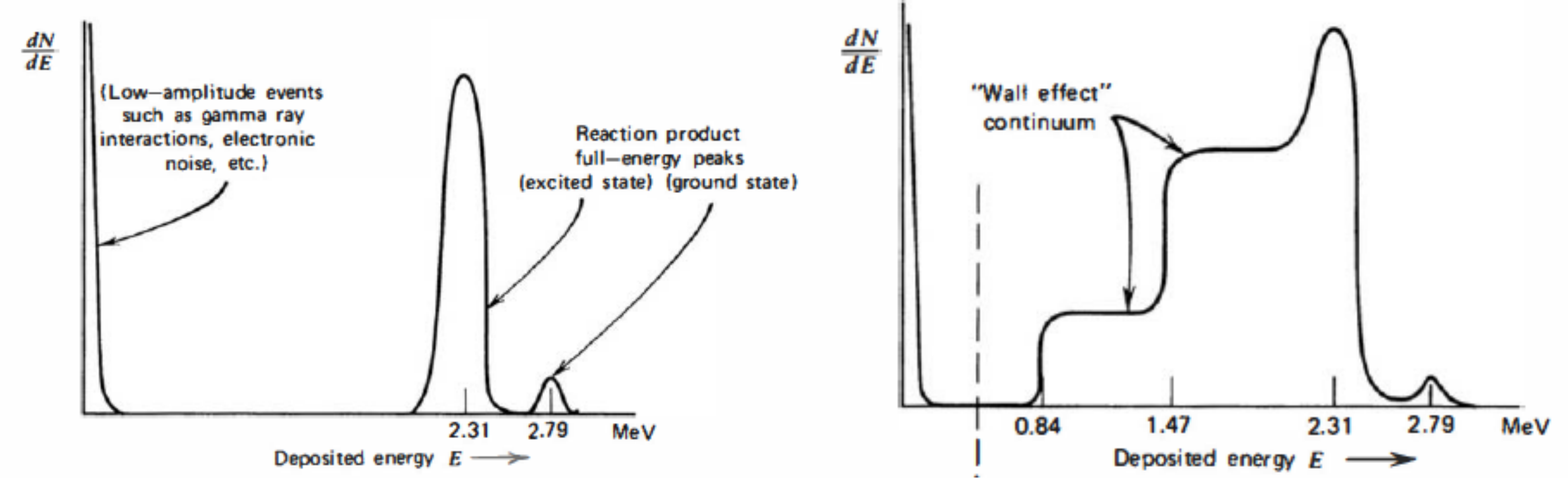}
\caption{\label{phsbf3} \footnotesize Left: Expected pulse height spectra form a large BF$_3$ tube, in which all the reaction products are fully absorbed. Right: Add of a continuum signal due to the wall effect. Figure from~\cite{DET_knoll}.}
\end{figure}

The boron-based detectors can use the boron in the gaseous form, e.g., BF$_3$ or in the form of a solid film~\cite{DET_wang,B4C_nowak}, e.g., by coating the interior walls of a tube~\cite{DET_dighe2,STRAW_lacy2011,DET_salvat} or the aluminum cathode in the case of a GEM~\cite{Bgem} or on a metal substrate as in the case of Multi-Grid~\cite{MG_IN6tests,MG_khap} and Multi-Blade~\cite{MIO_MB2014,MIO_MB2017}. The latter is the subject of study of this thesis and it will be presented in chapter~\ref{chapter4}.
\\ The approach of using solid coating has many advantages compared with the gaseous solution, indeed a more suitable gas than BF$_3$ can be chosen, and several applications, in particular for high rate and fast timing, can gain in performances by employing one of the common proportional gases, as described in section~\ref{GID}. Also the degradation problems in BF$_3$ under high fluxes can be reduced by using alternative fill gases. Moreover, the BF$_3$ is a toxic gas and the usage is now forbidden in several countries. 
\\ A detailed description of a boron-based detector in solid form is provided in chapter~\ref{chapter4}, for completeness, is briefly describe here the operation of a BF$_3$ proportional tube~\cite{DET_knoll}. In such a device, boron trifluoride acts both as a neutron converter into secondary particles and detecting medium. Like most of the proportional counters, BF$_3$ tubes are built using cylindrical outer cathodes and small-diameter central wire anodes. Generally the cathodes are made in Aluminum, because of its low neutron interaction cross section. The anode diameters are on the order of 0.1 mm or less, the operating voltage is usually about 2000-3000 V and the absolute pressure is limited to about 0.5-1.0 atm. The typical gas multiplication is on the order of 100-500. Larger diameter wires and/or higher fill gas pressures require higher applied voltages, therefore greater gas multiplication factor. BF$_3$ tubes also show significant effect of aging. This effect is related to the contamination of the anode wire and the cathode wall by molecular disassociation products produced in the avalanches~\cite{DET_knoll}. 
\\ If we consider a BF$_3$ tube large enough that all the reactions occur far from the walls of the detector, the products will deposit the full energy within the proportional gas. In this case, the pulse height spectrum would show two peaks corresponding to the $Q-$value of the two reactions of boron. The difference between the two is given by their probability ratio. A sketch is depicted in figure~\ref{phsbf3} left. 
  
The discrimination against background is convenient for these kind of devices, because of the amplitude difference between real events and background signals. For instance, gamma rays interact primarily in the wall of the counter and create secondary electrons that may produce ionization in the gas. As the stopping power for electrons in gases is quite low, in general an electron will deposit only a small fraction of its energy. Thus, one expects that most of the gamma-ray interactions will have a low amplitude, left tail in figure~\ref{phsbf3}. The amplitude discrimination can easily differentiate between neutron and gamma-ray events.     
\\ Considering now the more realistic case when the size of the tube is comparable with the range of the alpha particle and recoil lithium nucleus produced in the reaction. Some events would not deposit the full energy in the gas, on the contrary if a particle hits the chamber wall a pulse can be produced. In gas counters, this effect is known as the \textit{wall effect}. Note that the range of the $\alpha$ particle produced in the reaction is on the order of 1 cm for typical BF$_3$ tubes and it is comparable with the diameter of most of the tubes. The wall effect is, then, significant and a representation of this effect is shown in figure~\ref{phsbf3} right.  

\subsection{Fission chambers}

\begin{figure}[htbp]
\centering
\includegraphics[width=.8\textwidth,keepaspectratio]{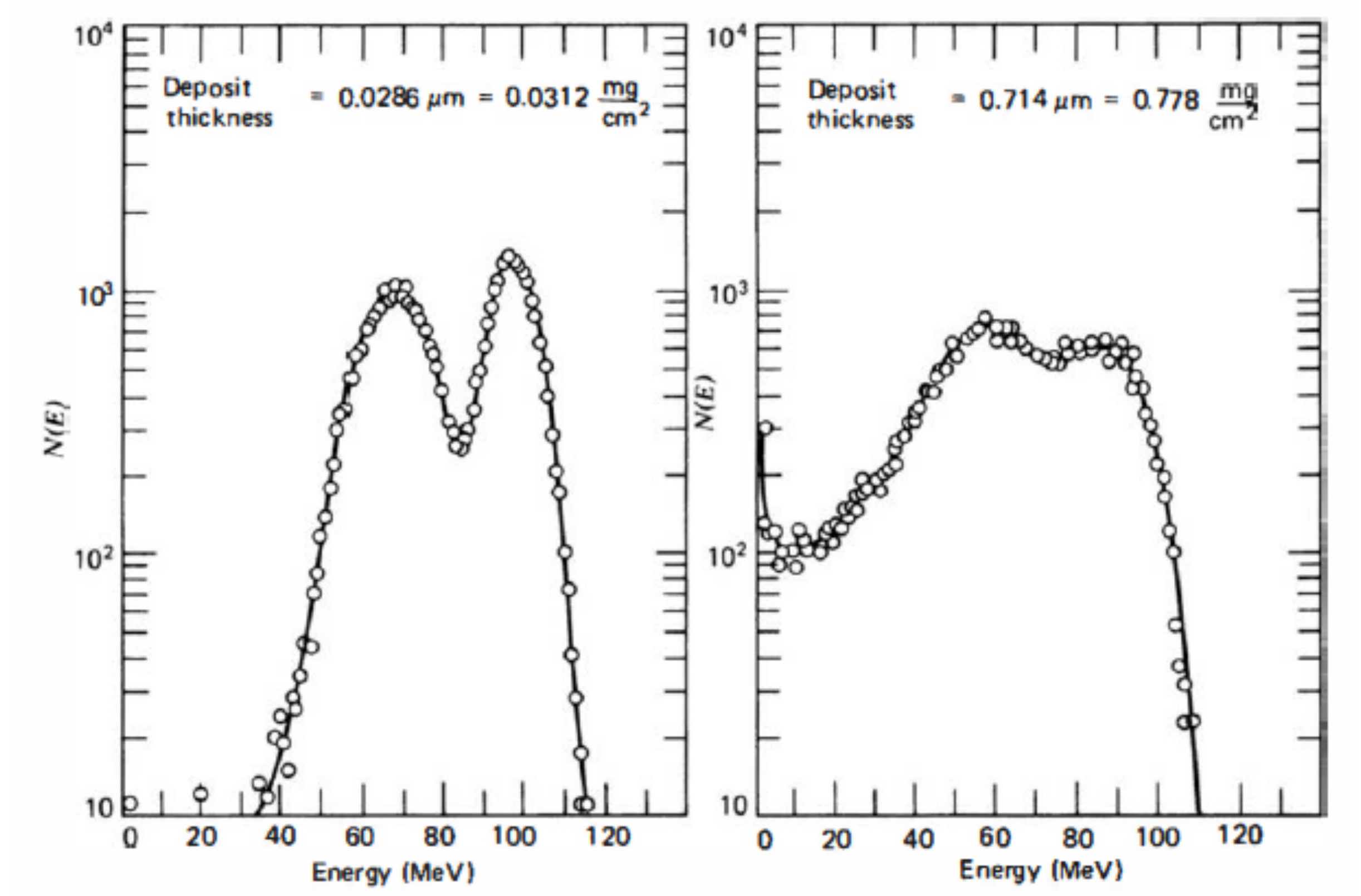}
\caption{\label{phsfission} \footnotesize Energy spectra of fission fragments from a UO$_2$ deposit of two different thickness. A 2$\pi$ detector is considered in order to get the fragments emitted from all direction by the deposit surface. Figure from~\cite{DET_knoll}.}
\end{figure}

The most popular fission detector is a ionization chamber with the inner surface coated with a fissile deposit. The pulse height spectrum expected in a fission chamber depends mainly on the fissile deposit thickness and on the geometry condition in which the fragments will be collected. If the deposit is thin compared to the fragments ranges, the fission fragment energy spectrum shows a double peak, with the light and heavy fragment distributions peaking at about 100 and 70 MeV, respectively. In case the deposit is made thicker to enhance detection efficiency, the energy loss of fragments within the deposit will reduce the average fragment energy and distort the shape of the measured distribution~\cite{DET_knoll}. Figure~\ref{phsfission} illustrates the changes in spectrum shape considering two different thickness of a deposit of UO$_2$.
\\ From a layer a highly enriched $^{235}$U of thickness 2-3 g/cm$^2$ (about 0.12 cm), the corresponding detection efficiency in 2$\pi$ counting geometry is about 0.5\% at thermal energy. Typical fission counters employ a single layer and thus are limited to an equivalent neutron detection efficiency.

The dimensions of these kind of devices is similar to that of alpha particles detectors. The fission fragments average range is approximately half the range of 5 MeV $\alpha$ particles. The energy loss of the fission fragments is at its maximum at the beginning of the track, because the fragments have a very high initial charge, and it will decrease as the yields slow down. The exact opposite happens in the case of light particles, for which the rate of energy loss peaks at the end of the track (see section~\ref{charged interaction}). In those detectors that do not fully stop the particle, fission fragments will deposit a larger fraction of their total energy than alpha particles or protons of the same range.  
\\ The two fission fragments are always emitted in opposite directions for slow-neutron-induced fission, therefore detectors with a solid coating will detect only the single fragment directed toward the active area of the chamber.   
\\ More complex geometry and design have been built to higher the detection efficiency by employing, for instance, multiple layers of fissile deposits. There have been also developed fission chambers capable to detect both fission fragments. 

\section{Principles of gaseous detectors}
A small overview of neutron detectors has been presented. The principles of operation for two classes of device will be now described: gaseous and solid state detectors. Note that the considerations on signal formation are valid for both cases. A boron-10-based detector and a Silicon-pin detector are subject of study of the thesis; we will, therefore, rely on the theoretical basis discussed in this chapter, when presenting the two detectors later.

\subsection{Charged particles interaction with matter}\label{charged interaction}
In general the passage of charge particles through matter can be characterized by two main features: a loss of energy by the particle and its change in momentum. These effects are mainly due to two processes, the electromagnetic interaction with the atomic electrons of the material and the elastic scattering from nuclei. Other processes include the nuclear reaction (in case of neutrons), the emission of Cherenkov radiation and bremsstrahlung. Compared with the atomic collision interactions, which occur almost continuously in matter, they are very rare~\cite{DET_leo}. It is necessary to separate the charged particles into two classes: electron and positron, and heavier particles. 

\subsubsection{Heavy charged particles interaction with matter}
The inelastic collisions with the electrons are mainly responsible for the energy loss of heavy particles in matters. Through these processes, energy is transferred from the particle to the atom giving rise to an ionization or excitation of the latter. Although in each collision only a small fraction of the particle's total kinetic energy is transferred, the number of collisions per unit path length is so large, that the cumulative energy loss is observed even in a relative thin layer of material. 
\\ Elastic scattering from nuclei occurs frequently even if not as often as electron collision. In general the energy transferred in these processes is very little since the masses of the nuclei of most materials are typically large compared to the incident particles. 
\\ As the inelastic collisions are statistical in nature, they occur with a certain quantum mechanical probability. Nevertheless, the fluctuations in the total energy loss are small, because of the large numbers of collisions. It is possible, hence, to work with the average energy loss per unit path length; usually one refers to it as the \textit{stopping power} or $\frac{\mathrm{d}E}{\mathrm{d}x}$. First it was derived classically  by Bohr, later by Bethe and Bloch using quantum mechanics. The obtained formula is:

\begin{equation}
-\frac{dE}{dx} = 2 \pi N_A r^2_e m_e c^2 \rho \frac{Z}{A}\frac{z^2}{\beta^2} \Bigg( \ln{\Bigg(\frac{2m_e\gamma^2v^2 W_{max}}{I^2}\Bigg)-2\beta^2 -\delta-2\frac{C}{Z}}\Bigg)
\label{bethebloch}
\end{equation}
where:
\begin{table}[h!]
\centering
\begin{tabular}{l l | l}
 $r_e$ & classical electron radius & $\rho$ \quad absorbing material density  \\
$m_e$ & electron mass & $z$ \quad charge of incident particle \\
$N_A$ & Avogadro's number & $\beta$ \quad $v/c$ of the incoming particle   \\
$I$ & mean excitation potential & $\gamma$ \quad $(1-\beta^2)^{-1/2}$  \\
$Z$ & atomic number of absorbing material & $\delta$ \quad density correction \\
$A$ & atomic weight of absorbing material & $C$ \quad shell correction  \\
$W_{max}$ & max enery transfer in a single collision &    \\
\end{tabular}
\end{table}

\begin{figure}[htbp]
\centering
\includegraphics[width=1\textwidth,keepaspectratio]{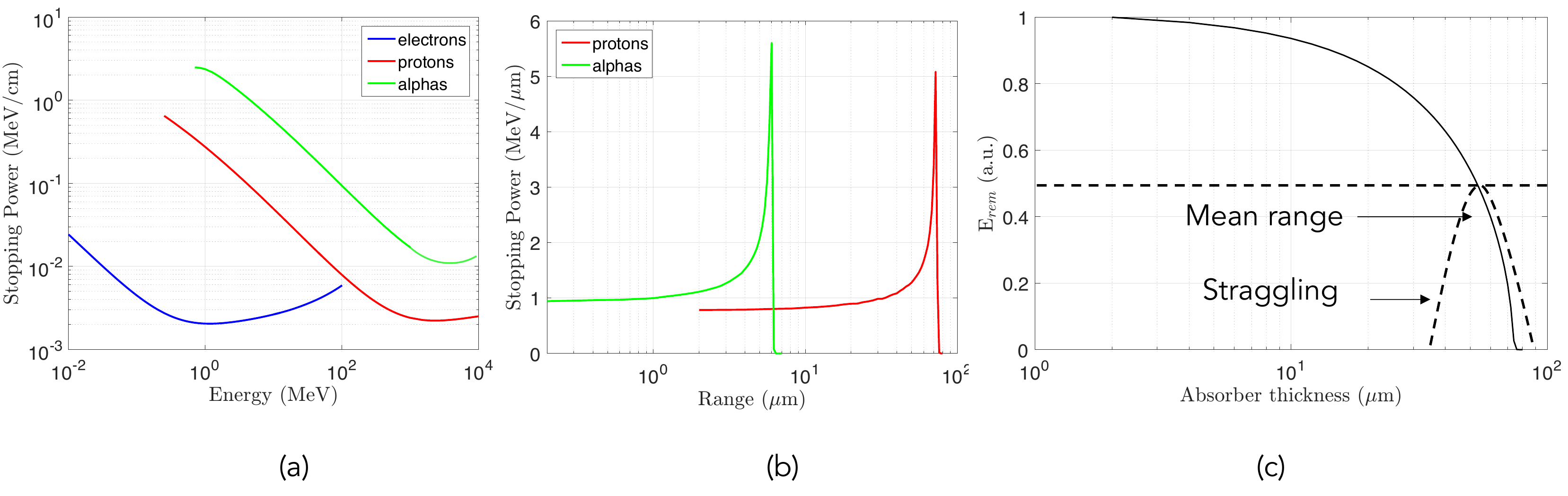}
\caption{\label{stop_pow_srim} \footnotesize (a) The stopping power $\mathrm{d}E/\mathrm{d}x$ in dry air as function of energy for different particle. (b) Bragg curve of the variation of $\mathrm{d}E/\mathrm{d}x$ as a function of the penetration depth of the particle in matter. It is shown for protons and $\alpha$s of initial energy 100 MeV in dry air. (c) Remaining energy of a proton as a function of the absorber thickness, dry air. The straggling distribution is shown and the mean value as well. It defines the thickness at which about half of the particle are stopped in the absorber.}
\end{figure}

For the full derivation see~\cite{DET_leo}. At non-relativistic energies $\mathrm{d}E/\mathrm{d}x$ varies as $1/v^2$, when comparing different charged particles of the same velocity, the only factor that may change outside the logarithmic term in equation~\ref{bethebloch}, is $z^2$. Hence, particles with greater charge will have larger specific energy loss. If we compare different material as absorbers, the stopping power depends primarily on the product between $Z$ and the number density, which represents the electron density of the absorber. High atomic number and high density materials result consequently in larger stopping power.
\\ Software packages to simulate the energy loss are freely available. It has been used either SRIM~\cite{MISC_SRIM1998,MISC_SRIM2010} or the web databases ESTAR, PSTAR and ASTAR from NIST~\cite{eloss_NIST} to calculate the stopping power. In figure~\ref{stop_pow_srim} is shown the energy loss for different charged particles. When their velocity approaches the velocity of light, at energy above hundred MeV, the different values of $dE/dx$ reach a near-constant broad minimum value. Due to this similarity, such relativistic particles are referred to as \textit{minimum ionizing particle}. From figure~\ref{stop_pow_srim}(a) it is clear that as a heavy particle slows down in matter, its rate of energy loss change as its kinetic energy changes. Indeed, more energy per unit length will be deposited towards the end of the path rather than at the beginning. This effect is depicted in figure~\ref{stop_pow_srim}(b), which shows the amount of ionization created by a heavy particle as a function of its position along its path. This feature is known as $Bragg$ curve, most of the energy is deposited near the end of the trajectory. At the very end it begins to pick up electrons and the $\mathrm{d}E/\mathrm{d}x$ falls off.

Knowing that charged particles lose their energy in matter, it is possible to define the particle \textit{range} which is the average distance a particle penetrates a material before losing all of its energy. Assuming that the energy loss is the same for all identical particles with the same initial energy in the same material, i.e., is continuous, this distance must be a well defined number. In nature the energy loss is not continuous but statistical, indeed two identical particles with the same initial energy will not, in general, undergo the same number of collisions and therefore have the same energy loss. A measurement with an ensemble of identical particles will show a statistical distribution of ranges centred about some mean value. This phenomenon is known as \textit{range straggling} and the main value of the distribution is known as the \textit{mean range}. This represent the thickness at which approximately half of the particles are stopped in the material. Referring to figure~\ref{stop_pow_srim}(c) it is possible to define the \textit{extrapolated range} as the average distance a particle can travel until it carries an energy below the minimum need to ionize an atom, $E_{th}$. This is related to the remaining energy of a particle, calculated from the stopping power function as:

\begin{equation}
E_{rem}=E_0 -\int_0^x \frac{dE}{d\epsilon} d\epsilon 
\label{Erem}
\end{equation}

where $E_0$ is the initial energy of the particle. In figure~\ref{stop_pow_srim} (c) the extrapolated range corresponds to a threshold energy of about $E_{th} \sim 0$. The distance at which the particle conserves at maximum $E_{th}$ is defined as the \textit{effective range}.

\subsubsection{Electrons and Positrons interaction with matter}
Like heavy charged particles, electrons and positrons suffer an energy loss due to the atomic inelastic collisions when passing through matter. However, because of their small mass an additional energy loss mechanism should be considered: the emission of electromagnetic radiation arising from the scattering in the electric field with nucleus (bremsstrahlung). This radiation is produced by the deceleration of an electron (or positron) when deflecting by the electrical attraction of an atomic nucleus. The total energy loss of electron and positron is composed of two parts:

\begin{equation}
\Big(\frac{dE}{dx}\Big)_{tot} = \Big(\frac{dE}{dx}\Big)_{rad}+\Big(\frac{dE}{dx}\Big)_{coll}
\label{Elossee}
\end{equation}

\begin{figure}[htbp]
\centering
\includegraphics[width=0.7\textwidth,keepaspectratio]{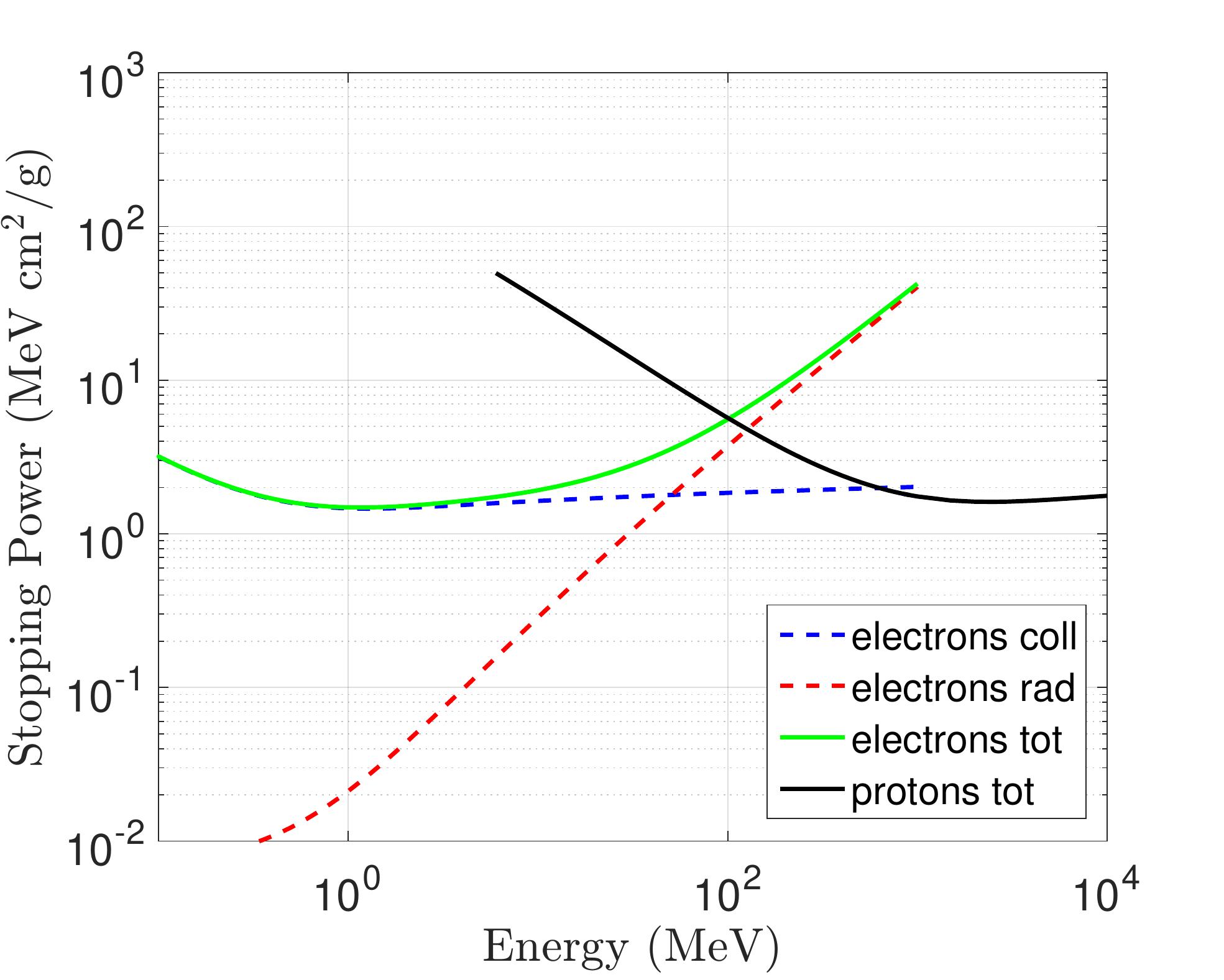}
\caption{\label{Ebremcoll} \footnotesize Radiation loss, red dashed line, vs. collision loss, blue dashed line, for electrons in Aluminium. The total energy loss for electron, green line, and protons, black line, is also shown. The calculation is performed based on the data~\cite{eloss_NIST}.}
\end{figure}

The first term in equation~\ref{Elossee} refers to the bremsstrahlung radiation energy loss, its contribution can be considered negligible below a few MeV. The loss of energy by radiation is comparable to or greater than the collision-ionization loss above a few 10's of MeV, as shown in figure~\ref{Ebremcoll}

The Bethe-Bloch formula applied for heavy charged particles (equation~\ref{bethebloch}) is in general still valid, even if must be modified for two reasons: because of their small mass, the assumption that the incident particle remains undeflected during the collision process in not valid, and for electrons the collisions are between identical particles, so the calculation must take into account their indistinguishability. These considerations change, among other terms, the maximum energy transfer in a single collision ($W_{max}$) into $W_{max} = T_e/2$ where $T_e/2$ is the kinetic energy of the incident electron or positron. The contribution of the radiation to the energy loss of the particle is notable only for electrons and positrons, indeed, the emission probability varies as the inverse square of the particle mass, i.e., $\sigma \propto r^2_e = (\frac{e^2}{mc^2})^2$. Note that radiation loss by muons, the next lightest particle, is approximately $4\cdot10^4$ times smaller than that for electrons. Since the bremsstrahlung emission depends on the strength of the electric field felt by the electron, the cross section depends also on the atomic number $Z$ of the material and the impact parameter of the incident electron.
\\ Due to the electron's greater susceptibility to multiple scattering by nuclei, the range of electrons is, in general, different from the calculated path length obtained from an integration of the stopping power function.

\subsection{Gaseous Ionization Detectors}\label{GID}

As a fast charged particle passes through a gas volume, it interacts with it by ionizing and exciting the molecules along its path. As shown in section~\ref{charged interaction}, when a neutral molecule is ionized, the resulting positive ion and free electron, the so called \textit{ion-pair}, represents the basic constituent of the electrical signal employed in a gaseous detector. The \textit{mobility} is the ability of charged particles to move through a medium under the effect of an electric field, $E$, with a velocity $v$:
\begin{equation}
\mu = v/E
\end{equation}
Due to the greater mobility of electron and ions, a gas is the obvious medium to use for the collection of ionization from radiation. Typically the electron mobility is 1000 times greater than that for ions.
\\ The basic configuration of a gas device consists of a container, a cylinder can be considered for simplicity, with conducting walls and a thin end window. The cylinder is filled with a suitable gas, usually a noble gas. Along the axis is suspended a conductive wire to which is applied a positive voltage relative to the walls. 

\begin{figure}[htbp]
\centering
\includegraphics[width=0.6\textwidth,keepaspectratio]{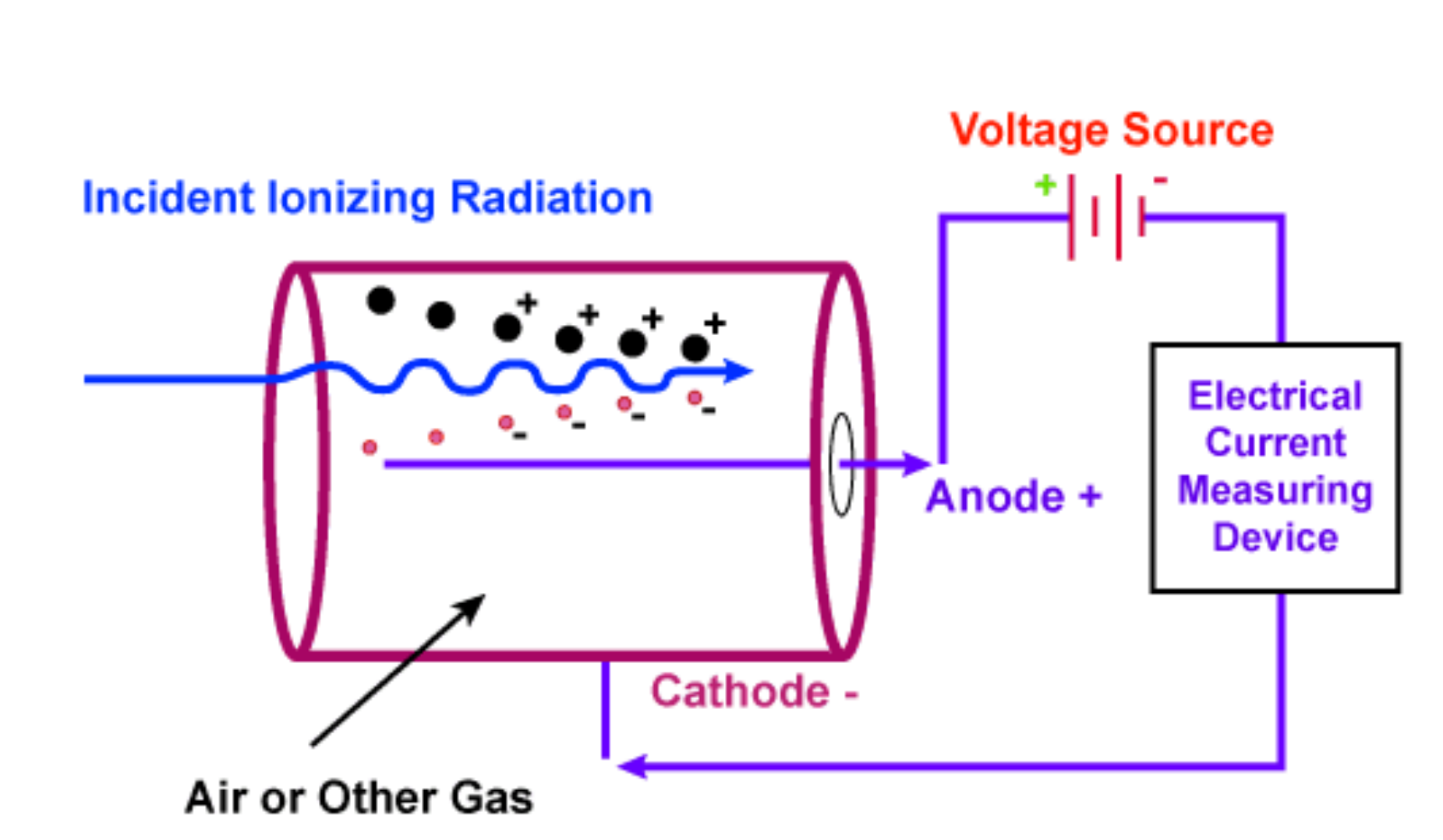}
\caption{\label{iondet} \footnotesize Sketch of a simple ionization detector.}
\end{figure} 

A radial electric field is established:

\begin{equation}
E =\frac{1}{r} \frac{V_0}{\ln{(b/a)}}
\label{E_gasch}
\end{equation}

where $r$ is the radial distance from axis, $b$ is the inside radius of the cylinder and $a$ is the radius of central wire. When a radiation penetrates the cylinder, a certain number of election-ion pairs will be created. Directly, if the radiation is a charged particle or through secondary reactions if the radiation is neutral. Regardless of the detailed processes involved, the practical quantity of interest is the total number of ion-pairs created along the track of the radiation. Both the excitation and ionization potential for several gases are listed in table~\ref{tableEnp}. The total number of ion-pair, $n_{pair}$, can be expressed by:

\begin{equation}
n_{pair} =\frac{\Delta E}{w_i}
\label{Enp}
\end{equation}

where $\Delta E$ is the charged particle energy loss, and $w_i$ is the effective average energy to produce one pair~\cite{DET_sauli1977}. This quantity is different from the ionization potential, $I_i$, because a certain amount of energy is needed to excite the particle even without creating ion-pairs. As shown in table~\ref{tableEnp}, the average energy lost by a charged particle to generate an ion-pair is about $w_i = 30$ eV, e.g., for a 3 keV particle an average of 3000/30 = 100 ion-pairs will be created. Furthermore, $w_i$ is not strongly dependent on the type of particle and slightly dependent on the type of gas.

\begin{table}[htbp]
\centering
\caption{\label{tableEnp} \footnotesize Excitation and ionization potential for several gases. The effective average energy to produce one pair is reported as well~\cite{DET_sauli1977}.}
\smallskip
\begin{tabular}{|c|c|c|c|}
\hline
 Gas & $I_{ex}$ (eV) & $I_{i}$ (eV) &$w_{i}$ (eV)  \\
 \hline
$H_2$ & 10.8 & 15.4 & 37 \\
$He$ & 19.8 & 24.5 & 41 \\
$N_2$ & 8.1 & 15.5 & 35  \\
$O_2$ & 7.9 & 12.2 & 31 \\
$Ne$ & 16.6 & 21.6 & 36  \\
$Ar$ & 11.6 & 15.8 & 26\\
$CO_2$ & 5.2 & 13.7 & 33\\
$CH_4$ & 9.8 & 13.1 & 28\\
$CF_4$ & 12.5 & 15.9 & 54\\
\hline
\end{tabular}
\end{table}

\begin{figure}[htbp]
\centering
\includegraphics[width=0.8\textwidth,keepaspectratio]{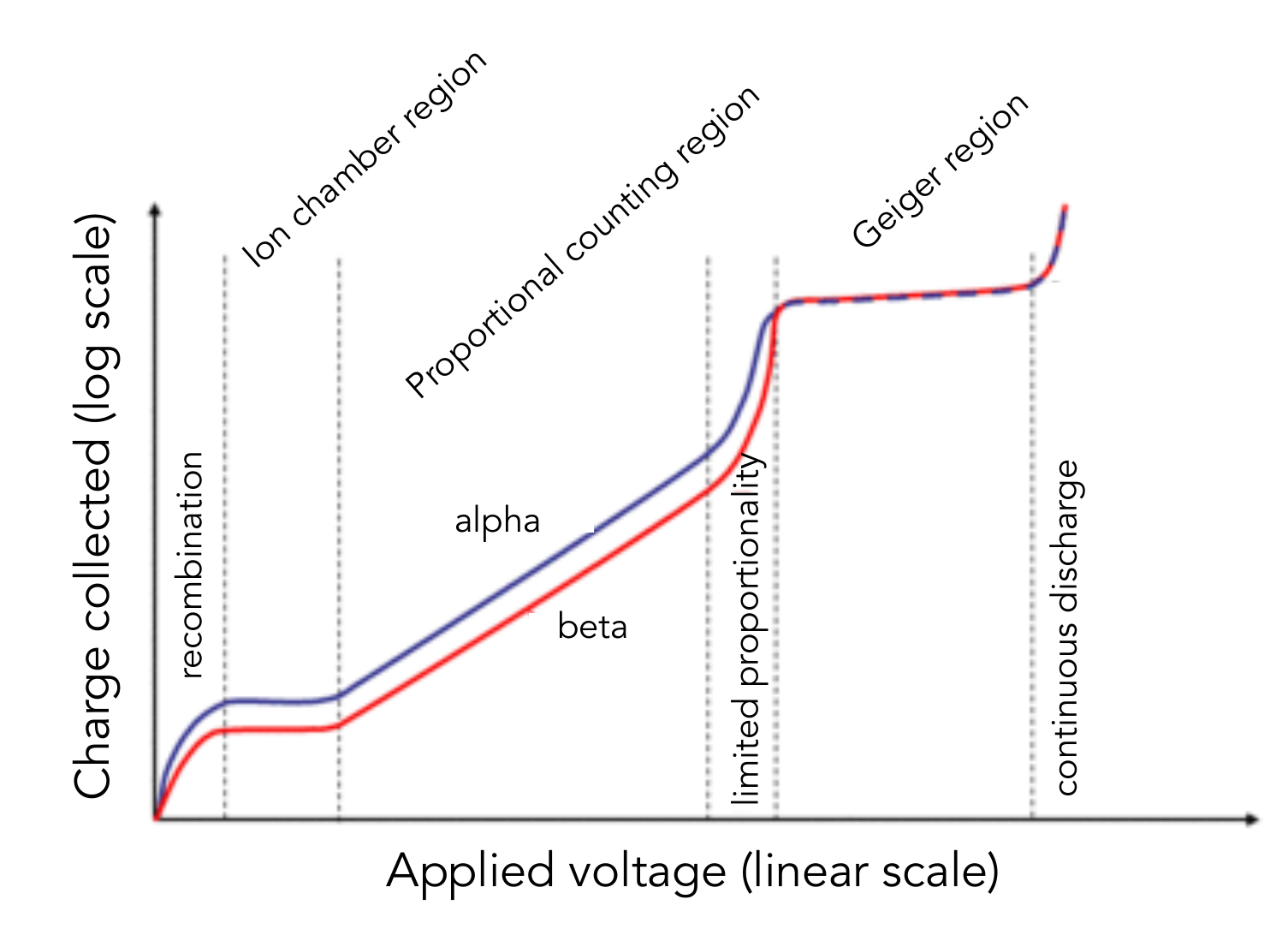}
\caption{\label{ion_region} \footnotesize Gaseous ionization detector regions. Defined with respect to the charge collected versus the applied voltage in a single wire gas chamber.}
\end{figure}

The mean number of pairs is proportional to the energy deposited in the counter. Under the effect of the electric field, the electrons will be accelerated towards the anode and the ions toward the cathode, where they can be collected. Thus, it is necessary that the ion-pairs remain in a free state long enough to be read out. As it will be shown in section~\ref{signalformation}, the signal is, indeed, induced by the motion of the charges. Positive ions and free electrons created within the gas undergo several type of collisions with the neutral gas molecules, diminishing the net charge created, i.e., by \textit{charge transfer collisions}, \textit{electron attachment} and \textit{recombination}~\cite{DET_knoll}. In the first process the collision occurs between a positive ion and a neutral gas molecule leading to an electron exchange between the two. The net positive charge tends to be transferred to the gas molecule, leaving it as a positive ion. On the other side, free electrons can be captured by electronegative atoms to form negative ions. Moreover, collision between positive ions and free electrons may result in recombination, in which the electron is captured by the positive ion and returns to the neutral state. The positive ions can also collide with a negative one, thus both ions are neutralized. 
\\ The understanding of these mechanisms is also important to choose the gas mixture. Despite the presence of electronegative gases, e.g., O$_2$ and CO$_2$, strongly diminish the amount of electron charge collected, it has been, also, observed that a small amount of CO$_2$ does not affect significantly the detector performances and helps to stop more efficiently the ionizing particles. On the other side, noble gases such as He, Ar, etc., have negative electron affinities and are a suitable choice for application in gaseous detectors~\cite{MIO_MyThesis}.
\\ The final charge collected depends on the field intensity. In absence of electric field the multiple collisions with the gas molecules lead to a thermal equilibrium between the ion-pairs and the gas. As the voltage is raised, however, the recombination mechanisms are overcome and the current begins to increase, because more electron-ion pairs are collected before they can recombine. The charge produced by ionization as a function of the voltage applied to two electrodes into a gas volume, through which the electric field is applied, is shown in figure~\ref{ion_region}, where four operating regions are identified: ionization, proportional, limited proportional and Geiger region. The recombination and continuous discharge regions are at the edges and are not used as working regions for any device.

The first flat region in figure~\ref{ion_region} is reached when all created pairs are collected after they overcome the recombination region, thus further increases in voltage does not show any effect. It is possible to distinguish between different detectors compared to the region in which they operate. A detector working in the second region is called an \textit{ionization chamber} as it collects the ionization produced directly by passing radiation. The signal is very small and this devices are generally used to measure gamma ray exposure and as monitoring instruments for large integral fluxes of radiation.
\\ As the voltage increases the electric field will be strong enough to accelerate free electrons to an energy adequate to ionize the gas molecules in the cylinder. The electrons created in these secondary ionizations are accelerated as well and are capable to produce more ionization and so on. The effect is known as ionization \textit{avalanche} or \textit{cascade}. From equation~\ref{E_gasch} the electric field is stronger near the anode, this avalanche occurs very quickly and almost completely within a few radii of the anode wire. In this region the number of ion-pairs in the avalanche is directly proportional to the number of primary electrons. The current amplification is proportional, with a multiplication factor depending on the applied voltage. A detector operating in this domain is a \textit{proportional chamber}. The output signals are much larger than that from an ionization chamber, indeed the multiplication factor can go up to $10^6$, but proportionality with the original radiation is still preserved. The proportional chamber is widely used for neutron detection or for low energy X-ray applications.
\\ A further voltage increasing leads to the loss the proportionality, indeed, the amount of ionization created through multiplication becomes large enough that the space charge effect deforms the electric field. This is defined as the region of limited proportionality. Higher voltage values produce discharge in the gas, instead of a localized avalanche along the anode wire, many multiplication processes occur and spread out along the entire length of the electrode. The secondary avalanches are generated by photons emitted by de-exciting molecules of the gas. The output signals are saturated and the information about the energy of the initial event is completely lost. The discharge can be stopped using a \textit{quenching gas}, with a low ionization potential and a complex molecular structure, that can absorb these photons. Devices working at these voltages are called \textit{Geiger-Mueller} counters. They can be employed only as simple counter of radiation-induced events and cannot be applied in direct radiation spectroscopy because all information on the amount of energy deposited by the incident radiation is lost. Above the Geiger region a continuous discharge occurs, and it is not useful for any application.
 
\subsubsection{The Multiwire Proportional Chamber (MWPC)} 
 
\begin{figure}[htbp]
\centering
\includegraphics[width=0.5\textwidth,keepaspectratio]{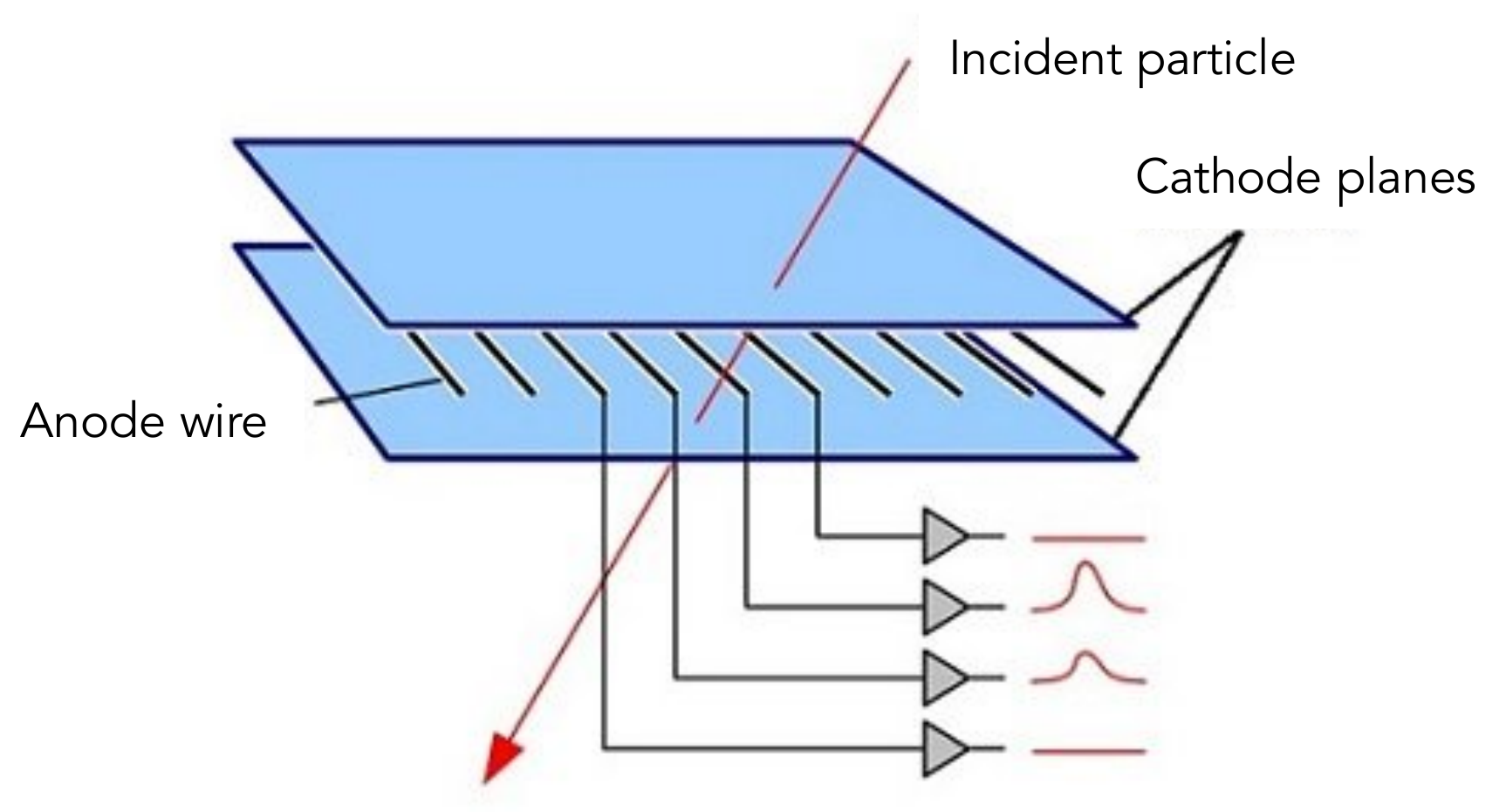}
\caption{\label{mwpc_sketch} \footnotesize Basic scheme of a multiwire proportional chamber. Each wire is independent from the others and act like a single proportional counter.}
\end{figure}

Based on the principles of a proportional counter, in 1968 Charpak~\cite{DET_charpak1968} developed the Multiwire Proportional Chamber (MWPC) which basically consist in an array of many equally spaced anode wires centred between two cathode planes, a sketch is shown in figure~\ref{mwpc_sketch}. 
 
\begin{figure}[htbp]
\centering
\includegraphics[width=0.3\textwidth,keepaspectratio]{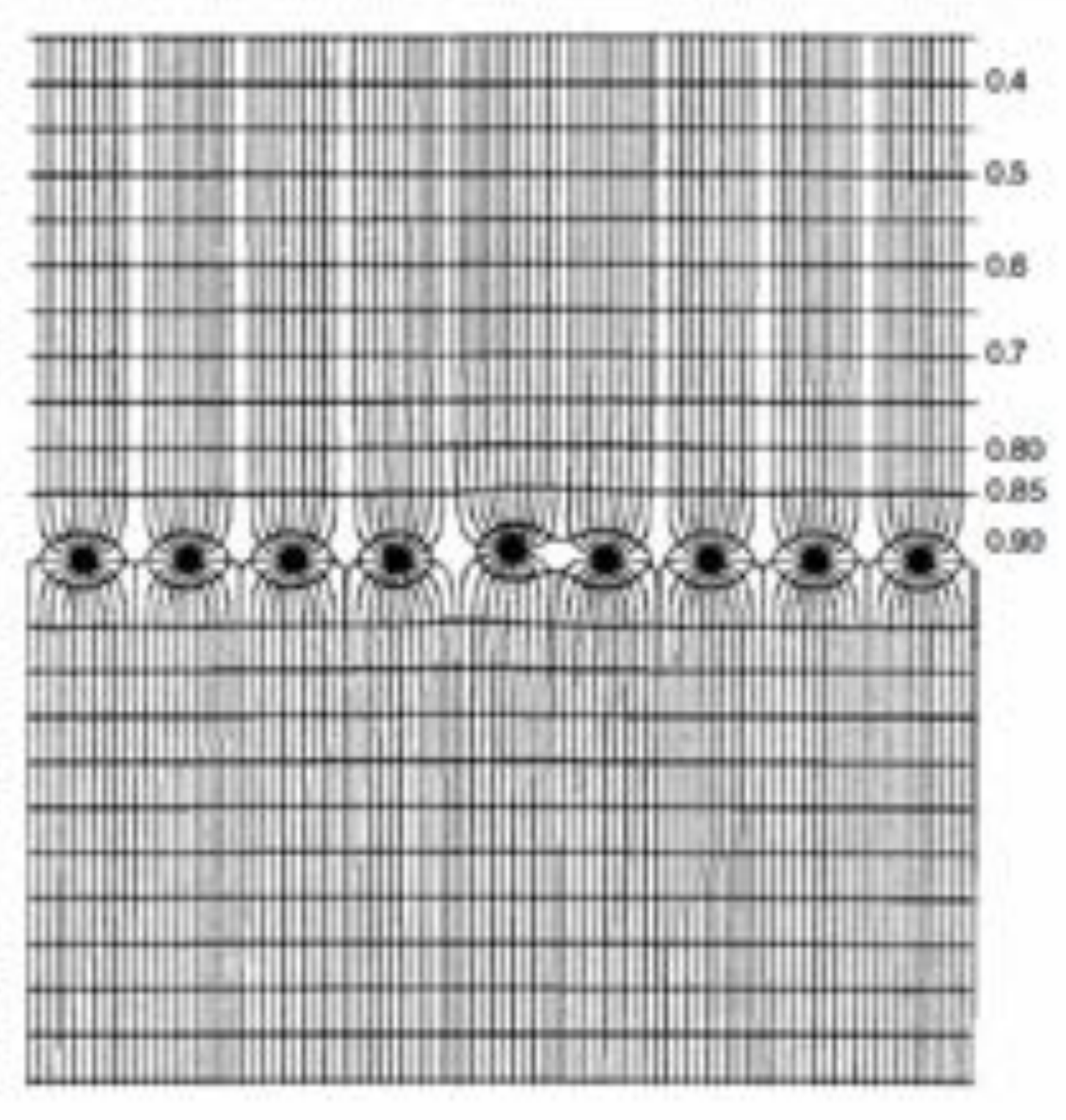}
\caption{\label{mwpc2} \footnotesize Electric field lines in a multiwire proportional chamber (from Charpak et al.~\cite{DET_charpak1984}).}
\end{figure}

If a negative voltage is applied to the cathode planes, the anodes being grounded, an electric field develops as indicated by the equipotentials and fields lines as in figure~\ref{mwpc2} arises. The field lines are practically parallel and almost constant except for the region near the anode wires. Suppose now that charges are liberated in the gas volume by an ionizing event, as in a proportional chamber, conditions are set such that electrons will drift along the field lines until they approach the high field region where avalanche multiplication occurs.

If we assume an infinite anode plane with zero diameter wires, the potential is given by~\cite{DET_ERSKINE},

\begin{equation}
V(x,y) = -\frac{CV}{4 \pi \epsilon} \ln \bigg[ 4 \bigg( \sin^2\frac{\pi x}{s} + \sinh^2\frac{\pi y}{s}\bigg) \bigg]
\label{V_mwpc}
\end{equation}

where $V$ is the applied voltage, $s$ is the wire spacing and $C$ the anode-cathode capacitance. $L$ is defined as the anode to cathode gap distance and $d$ as the anode wire diameter, if $L \gg s \gg d$, then $C$ is given by

\begin{equation}
C = \frac{2 \pi \epsilon}{\frac{\pi L}{s}-\ln \frac{\pi d}{s} } 
\label{C_mwpc}
\end{equation}

Equations~\ref{V_mwpc} and~\ref{C_mwpc} are usually a good approximation for a real MWPC, despite the assumptions are not met in a real chamber. Note that, since $d \ll s$ the value given by equation~\ref{C_mwpc} is always smaller than the capacitance of the plane condenser with the same surface $(2\epsilon s/L)$. In general it can be seen that the capacitance is quickly decreasing with the wire spacing, while it does not depend much on the wire diameter.
\\ Near the anode wires the field takes on a $1/r$ dependence similar to the single wire cylindrical chamber described above. If electrons and ions are now liberated in the constant field region they will drift along the field lines toward the closest anode wire and opposing cathode. When reaching the high field region, the electrons will be accelerated to produce an avalanche. The positive ions released in the multiplication process induce a negative signal on the anode, the neighbouring wires are also affected, however the signals induced are positive and of a small amplitude. The signal from one anode plane gives information on one coordinate, the second coordinate may be obtained by adding a second plane whose wires are perpendicular to the first, or arranging the cathode as a series of strips perpendicularly to the wires.

\section{Amplification mechanisms at the Proportional Wire}\label{sectionamplgas}

In the case of neutron detectors, to generate charged particles some conversion mechanism must occur. However, not all the neutron capture reaction can lead to yields that, carrying hundreds of keV, produce enough charge to be amplified by standard devices. In order to increase the ion-pair production it is possible to exploit the gas multiplication process, as pointed out in section~\ref{GID}. The average energy of the electron between collisions increases with the electric field, thus, it is possible to set a threshold value for the field above which this mechanism take place. In typical gases, at atmospheric pressure, this value is of the order of $10^7$ V/m.
\\ The multiplication of ionization is known as \textit{Townsend avalanche}, and is described by the first Townsend coefficient $\alpha$. The number of electrons per path d$s$ is given by 

\begin{equation}
\mathrm{d}n = n\alpha \mathrm{d}s
\label{townsend}
\end{equation}

The coefficient is determined by the excitation and ionization cross section of the electrons that have acquired sufficient energy in the field. It is zero when the electric field is below the threshold value and typically increases very quickly with the increasing of the electric field. If the gas density $\phi$ is changed while keeping the distribution of the collision energy ($\varepsilon$) fixed, then $\alpha$ changes proportionally with the density because all the linear dimensions in the gas scale with the mean free collision length. 
\\ The amplification factor on a wire is calculated by integrating equation~\ref{townsend} between the point where the field is just enough to start the avalanche ($s_{min}$) and the wire radius $a$:

\begin{equation}
\frac{n}{n_0} = \exp \int_{s_{min}}^a \alpha(s) \, \mathrm{d}s = \exp \int_{E_{min}}^{E(a)} \frac{\alpha(E)}{\mathrm{d}E/ \mathrm{d}s} \, \mathrm{d}E
\label{townsend2}
\end{equation}

where $n$ and $n_0$ are respectively the final and initial number of electrons in the avalanche; $\mathrm{d}E/ \mathrm{d}s$ is the electric field gradient. The electric field near a wire whose radius is small compared with the distance to other electrodes is given by the charge per unit length, $\lambda$, as a function of the radius:

\begin{equation}
E(r) = \frac{\lambda}{2\pi \varepsilon_0 r} 
\label{Efieldwire}
\end{equation} 

By inserting now equation~\ref{Efieldwire} into~\ref{townsend2}, one obtains:

\begin{equation}
\frac{n}{n_0} = \exp \int_{E_{min}}^{E(a)} \frac{\lambda \alpha(E)}{2\pi \varepsilon_0 E^2}\, \mathrm{d}E
\label{townsend3}
\end{equation}

The ratio $n/n_0$ is usually called $gain$, by assuming $\alpha$ to be proportional to $E$ it is possible to derive an expression for $G =n/n_0$~\cite{Diethorn}. We define $\alpha = \beta E$ and substitute it in equation~\ref{townsend3} gives:

\begin{equation}
\ln G = \frac{\beta \lambda}{2 \pi \varepsilon_0} \ln \frac{\lambda}{2 \pi \varepsilon_0 a E_{min}}
\label{gasgain}
\end{equation}

$\beta$ can be related to the average energy $e\Delta V$ required to produce one more electron. By solving the equation~\ref{townsend2}, the potential difference between $r=a$ and $r= s_{min}$ is obtained:

\begin{equation}
\Phi(a) - \Phi(s_{min}) = \int_{a}^{s_{min}} E(r)\, \mathrm{d}r = \frac{\lambda}{2 \pi \varepsilon_0} \ln \frac{s_{min}}{a} = \frac{\lambda}{2 \pi \varepsilon_0} \ln \frac{\lambda}{2 \pi \varepsilon_0 a E_{min}}
\label{potentialdiff}
\end{equation}

This gives rise to a number $Z$ of generations of doubling the electrons in the avalanche $Z=[\Phi(a) - \Phi(s_{min})]/ \Delta V$ which means $G= 2^Z$. The expression of the gas gain~\ref{gasgain} can be then written as:

\begin{equation}
\ln G = \frac{\ln 2}{\Delta V} \frac{\lambda}{2 \pi \varepsilon_0} \ln \frac{\lambda}{2 \pi \varepsilon_0 a E_{min}}
\label{gasgain2}
\end{equation}

The constant $\beta$ has a physical meaning of the inverse of the average potential required to produce one electron in the avalanche multiplied by $\ln 2$. Considering now the relation between the minimal field $E_{min}$ and the gas density $\rho$ to be $E_{min}(\rho)=E_{min}(\rho_0) (\rho/\rho_0)$, with $\rho_0$ the normal gas density. We finally obtain a relation for the gas gain proportional to the voltage and the charge density:

\begin{equation}
\ln G = \frac{\ln 2}{\Delta V} \frac{\lambda}{2 \pi \varepsilon_0} \ln \frac{\lambda}{2 \pi \varepsilon_0 a E_{min}(\rho_0) (\rho/\rho_0)}
\label{gasgain3}
\end{equation}

In a proportional counter, as the one described in~\ref{GID}, the charge density $\lambda$ is related to the voltage $V$ by 

\begin{equation}
 \frac{\lambda}{2 \pi \varepsilon_0}=  \frac{V}{\ln (b/a)}
\label{chargeD}
\end{equation}

Equation~\ref{gasgain3} can be expressed, in the so-called Diethorn's formula, by substituting this term:

\begin{equation}
\ln G = \frac{\ln 2}{\ln (b/a)}\frac{V}{\Delta V} \ln \frac{V}{\ln (b/a) a E_{min}(\rho_0) (\rho/\rho_0)}
\label{gasgain4}
\end{equation}

Experimentally $G$ can be measured varying $\rho/\rho_0$, the wire radius $a$ and $V/ \ln (b/a)$. 
\\ The gain variation with the gas density is of particular interest since typically this kind of chambers are operated at atmospheric pressure and the gas density changes proportionally to it. Indeed, from equation~\ref{gasgain3}, it can be observed that a small relative change $\mathrm{d} \rho/ \rho$ of the density results in a change of amplification of:

\begin{equation}
\frac{\mathrm{d} G}{G} = - \frac{\lambda}{\Delta V} \frac{\ln 2}{2 \pi \varepsilon_0} \frac{\mathrm{d}\rho} {\rho}
\label{gasgaindensity}
\end{equation}

In practical cases the factor that multiplies $\mathrm{d} \rho/ \rho$  ranges between 5 and 8. The fractional variation of the gas pressure causes a global variation of the gain typically 5 to 8 times larger than this. Nevertheless the gas pressure can be well monitored, and these variations can be corrected.
\\ Referring to equation~\ref{gasgain3} we denote that the gain depends on the local charge density of the wire $\lambda$. A relative change $\mathrm{d}\lambda/\lambda$ will modify the amplification by the following factor:

\begin{equation}
\frac{\mathrm{d} G}{G} =\bigg(\ln G + \frac{\lambda \ln 2}{\Delta V 2 \pi \varepsilon_0} \bigg) \frac{\mathrm{d}\lambda} {\lambda}
\label{gasgaincharge}
\end{equation}

The two multiplying factors in the parentheses are of the same order of magnitude and their sum is approximately of 10-20. The total relative variations of the charge density cause, then, a local relative variations of the gain that are in general 10-20 times larger.

\section{Signal formation}\label{signalformation}

\begin{figure}[htbp]
\centering
\includegraphics[width=0.8\textwidth,keepaspectratio]{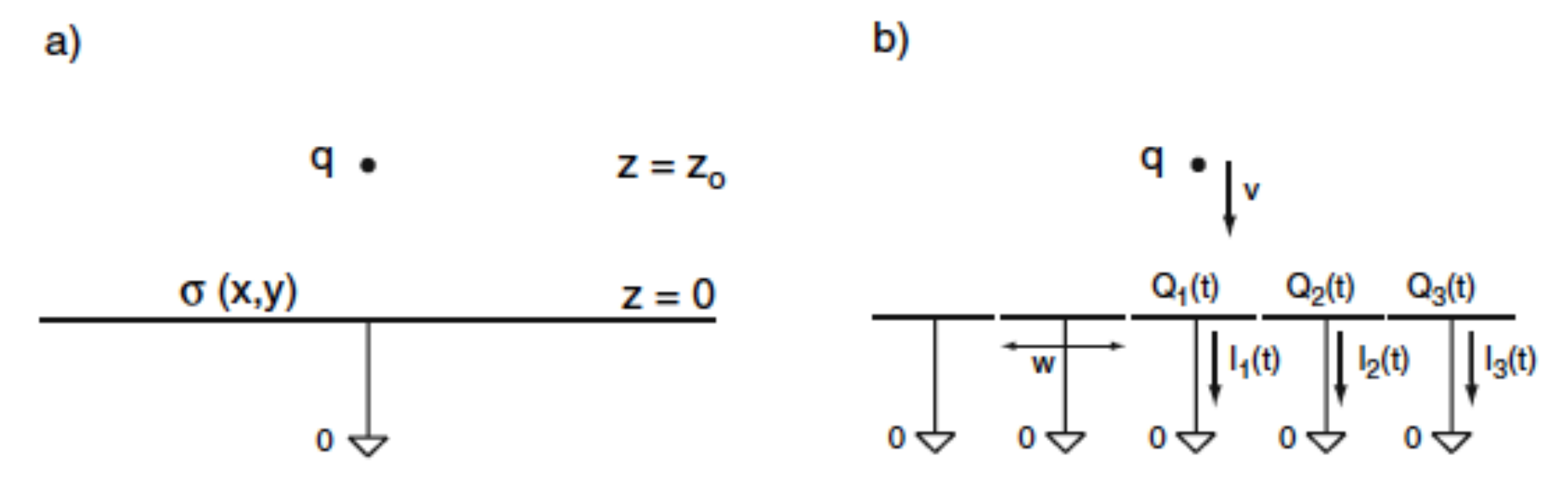}
\caption{\label{signal_form} \footnotesize (a) Sketch of a point charge $q$ inducing a charge density $\sigma(x,y)$ on the surface of a grounded plate. (b) Induced charge $Q_n$ in case of segmentation of the metal plate. The induced charges change in case $q$ is moving, and the currents flow between the strips and ground. (From Blum et al~\cite{Particle_Detection}). }
\end{figure}

The moving charges in a chamber give rise to electrical signals on the electrodes that can be read out by an amplifier. The electrons created in the avalanche close to the wire move to the wire surface within a time much less than a nanosecond, thus resulting in a short signal pulse. The ions, instead, move away from the wire with a velocity ($v \propto \mu$) about 1000 times smaller, which result in a signal with a long tail (with a duration on the order of several microseconds). The movement of these charges induce a signal both on the wire and on the cathode, which can be divided into several parts for the purpose of coordinate measurements~\cite{Particle_Detection}.
\\ Before describing how the moving charges give rise to signals on electrodes, we consider the case of a point charge $q$. As illustrated in figure~\ref{signal_form}a, in the presence of a grounded metal plate, $q$ induces a charge on the metal surface~\cite{Classical_electrodyn}. This surface charge can be calculated by solving the Poisson equation for the potential $\Phi$ with a point charge $q$ at $z= z_0$ and with the boundary condition that $\Phi=0$ at $z=0$. The resulting electric field $E = - \nabla \Phi$ on the metal surface is related to the surface charge density $\sigma$ by $\sigma(x,y)= \varepsilon_0\, E(x,y,z=0)$, following Gauss law.

The electric field on the metal surface is given by:

\begin{equation}
E_z (x,y)= - \frac{qz_0}{2 \pi \varepsilon_0 (x^2 +y^2 +z_0^2)^{3/2}} \qquad E_x =E_y =0
\label{E_field1}
\end{equation}

and the surface charge density is $\sigma(x,y)= \varepsilon_0\, E_z (x,y)$. The integration of $\sigma (x,y)$ gives the total charge induced on the metal plate: 

\begin{equation}
Q = \int_{-\infty}^{\infty} \int_{-\infty}^{\infty} \sigma(x,y) \mathrm{d}x \mathrm{d}y = -q
\label{Q_tot}
\end{equation}

and it is independent of the distance of the charge $q$ from the metal plate. We consider now the plate segmented into strips of width $w$, and each one is grounded, a scheme is shown in figure~\ref{signal_form}b. The surface charge density can be integrated over the area of the strip, from equation~\ref{E_field1}, to obtain the charge induced in one strip:

\begin{equation}
Q_1 (z_0) = \int_{-\infty}^{\infty} \int_{-w/2}^{w/2} \sigma(x,y) \mathrm{d}x \mathrm{d}y = -\frac{2q}{\pi} \arctan \Bigg(\frac{w}{2z_0} \Bigg)
\label{Q_strip}
\end{equation}

This charge depends now on the distance $z_0$ of the point charge from the metal surface. Considering now that the charge moves with a velocity $v$ according to $z_0(t)=z_0 -vt$, it is possible to find a time-dependent induced $Q_1 (z_0(t))$ and therefore an induced current of

\begin{equation}
I^{ind}_1 (t) = -\frac{d}{dt}Q_1 (z_0(t)) = -\frac{\partial Q_1 (z_0(t))}{\partial z_0} \frac{dz_0(t)}{dt} = \frac{4qw}{\pi [4z_0(t)^2 +w^2]}v
\label{I_strip}
\end{equation}

One observes that the movement of a charge induces a current which flows between the electrode and the ground. This method of calculating signals is the same for more realistic geometries, but become more complex as well. Some theorems are introduced to allow a simpler way of calculating signals induced on grounded electrodes.

\subsection{Reciprocity Theorem and Capacitance Matrix}

We refer to `metal' electrode to meet the condition that the charges on the electrode can move and that the electrode surface is equipotential. $N$ metal electrodes are set to voltages $V_n$, the potential is uniquely defined by the conditions that $\Phi (x)$ satisfies the Laplace equation and that $\Phi (x) = V_n$   on the electrode surfaces. The voltages and the charges on the electrodes are related by the capacitance matrix of the electrode system as~\cite{Classical_electrodyn}:

\begin{equation}
Q_n = \sum_{m=1}^N c_{nm}V_m
\label{Q_stripmatrix}
\end{equation}

The capacitance matrix is defined by the electrode geometry and for the capacitance matrix elements $c_{nm}$ hold the relations:

\begin{equation}
c_{nm}=c_{mn}, \quad c_{nm}\leq0 \ \mathrm{for} \ n\neq m, \quad c_{nn}\geq 0, \quad \sum_{m=1}^N c_{nm} \geq 0
\label{cmatrix_rel}
\end{equation}

A set of different voltages $\overline{V}_n$ is related to a set of different charges $\overline{Q}_n$ by the same capacitance matrix $\overline{Q}_n = \sum c_{nm}\overline{V}_n$. By inverting this relations and multiplying it by equation~\ref{Q_stripmatrix}, a relation called \textit{reciprocity theorem} is obtained:

\begin{equation}
\sum_{n=1}^N Q_n \overline{V}_n = \sum_{n=1}^N \overline{Q}_nV_n
\label{reprocity_theo}
\end{equation}

Together with this relation, we consider another one which concern the sum of all the charges of an electrode system. An electrode surrounding the $N$ metal electrodes is taken into account. By applying Gauss law to the volume between the electrodes, the sum of all charges within this volume is equal to $\varepsilon_0$ times the integral of $E(x)$ over the surface surrounding it. As this integral over a metal electrode is equal to the charge of this electrode, one finds that the sum of all charges on the metal electrodes is equal to the sum of all charges in the volume between the electrodes. In the case there are no charges in the volume between the electrodes, the sum of $Q_n$ on the electrodes must be zero for any applied voltages. Considering the capacitance matrix we find the relation

\begin{equation}
\sum_{n} Q_n =0 \ \rightarrow \ \sum_{n}\sum_{m} c_{nm}V_m = 0 \ \rightarrow \ \sum_{n} c_{nm}= 0
\label{cmatrix_rel2}
\end{equation}

Note that the capacitance matrix elements $c_{nm}$ are different from the capacitances $C_{nm}$. They are indeed related to the voltage difference between circuit nodes. The relation between the two can be expressed as

\begin{equation}
C_{nn} = \sum_{m=1}^N c_{nm}, \qquad C_{nm} = -c_{nm}\ \mathrm{if}\ n \neq m
\label{cmatrix_rel3}
\end{equation}

\subsection{The Shockley-Ramo theorem}

\begin{figure}[htbp]
\centering
\includegraphics[width=.8\textwidth,keepaspectratio]{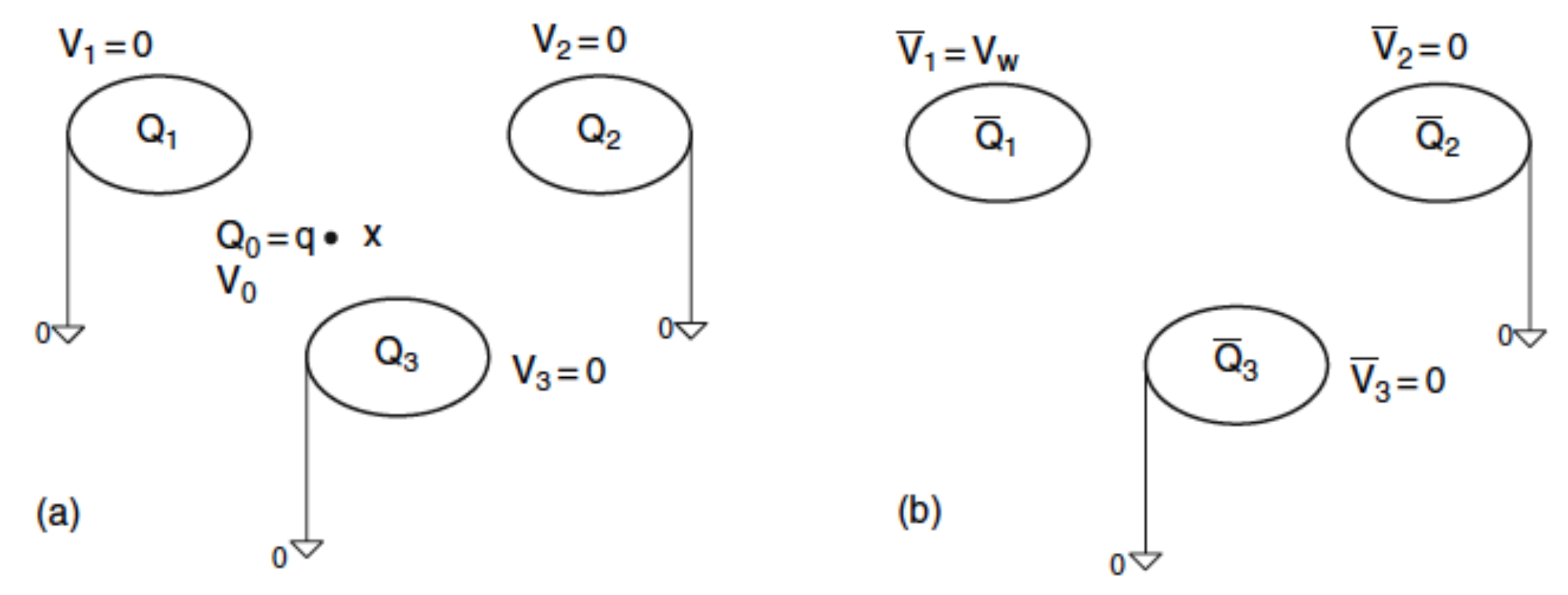}
\caption{\label{sramo1} \footnotesize (a) Three grounded electrodes and a point charge $q$ which induces charges $Q_n$ on the electrodes. (b) Configuration in which the charge $q$ is removed and the electrode 1 is set to voltage $V_w$.}
\end{figure}

The current signal read from electrodes arises from the motion of the charge carriers and not from their physical collection. This applies both for gas-filled detectors and for semiconductor detectors. The output pulse begins to form as soon as the carriers start their motion to the electrodes and it finishes once the last of the carriers reach its collecting electrode. Thus the pulse is fully developed during this motion and the time evolution of the signal is of primary importance to understand the timing properties of detectors~\cite{DET_knoll}. The method to calculate the shape of the pulse produced from the motion of charges in a detector exploits the \textit{Shockley-Ramo} theorem~\cite{DET_shockley,DET_ramo,DET_he}.
\\ We consider a set of three grounded electrodes in the presence of a point charge $Q_0=q$ at position $x$ and we imagine that the net charge is sitting on an infinitely small metal electrode. In total the system consists of four metal electrodes. A sketch is illustrated in figure~\ref{sramo1}(a).

We set $V_1 = V_2=V_3=0$ and we derive the charges $Q_1, Q_2, Q_3$ induced by the presence of charge $Q_0=q$. From equation~\ref{reprocity_theo}:

\begin{equation}
q\overline{V}_0 + Q_1\overline{V}_1 + Q_2\overline{V}_2+ Q_3\overline{V}_3 = \overline{Q}_0V_0
\label{q_ramo}
\end{equation}

Choosing another configuration, figure~\ref{sramo1}(b), where the charge $q$ is removed and the first electrode is set to voltage $V_w$ ($\overline{V}_1 = V_w$) while keeping the other two electrodes grounded, the above relation becomes:

\begin{equation}
q\overline{V}_0 + Q_1\overline{V}_w =0 \quad \rightarrow \quad Q_1 = -q \frac{\overline{V}_0}{V_w}
\label{q_ramo2}
\end{equation}

$\overline{V}_0$ is the potential of the uncharged small electrode for this configuration. Since an infinitely small uncharged electrode is the same as having no electrode, $\overline{V}_0$ is the potential at point $x$ when the point charge $q$ is removed. We refer to $\overline{V}_0= \psi_1 (x)$ as \textit{weighting potential} of electrode 1, thus the induced charge $Q_1$ is given by

\begin{equation}
 Q_1 = - \frac{q}{V_w} \psi_1 (x)
\label{q_ramo3}
\end{equation}

\begin{figure}[htbp]
\centering
\includegraphics[width=0.6\textwidth,keepaspectratio]{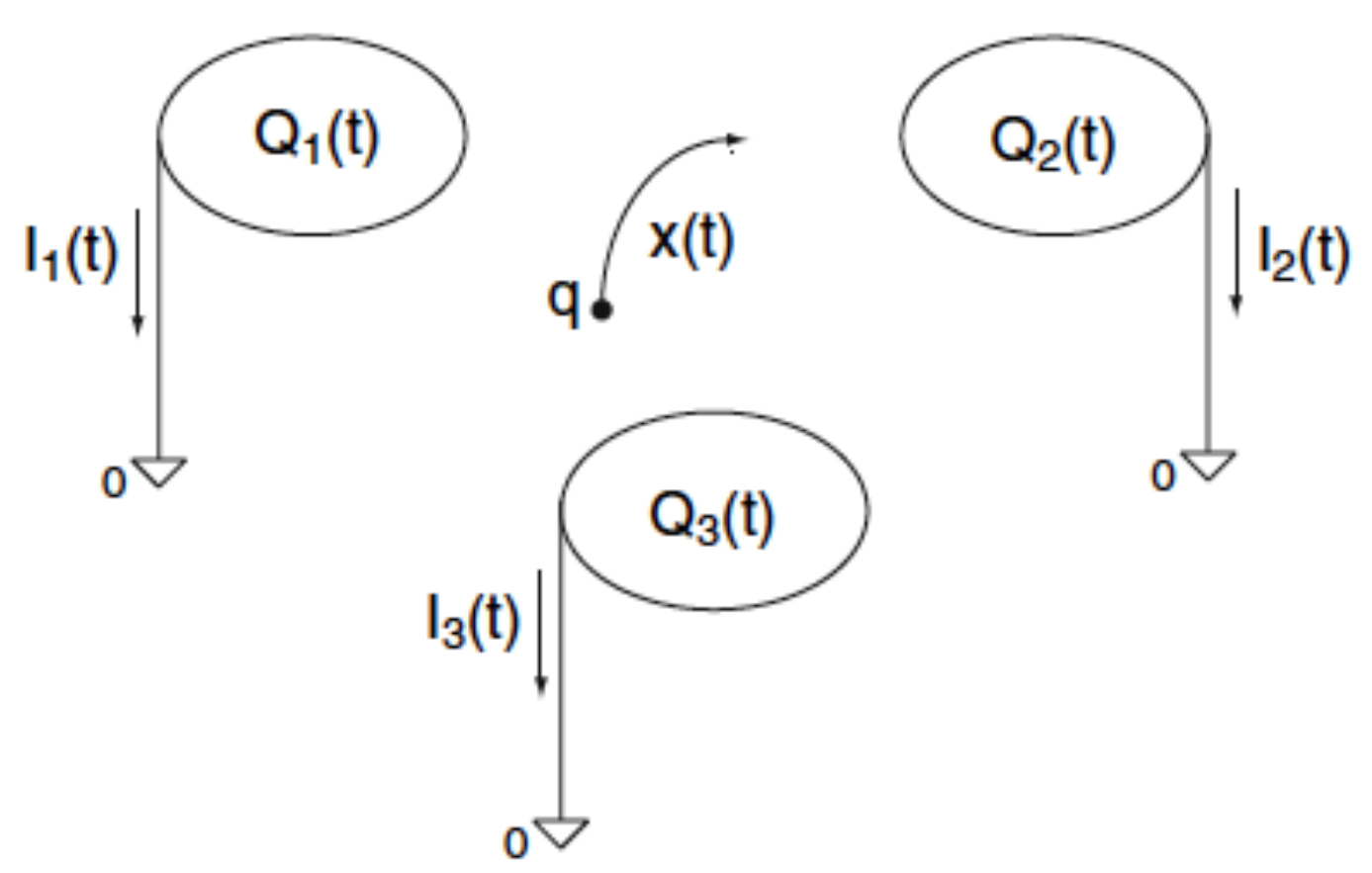}
\caption{\label{shockley-ramo1} \footnotesize Three grounded electrodes and a point charge $q$, in the case where the charge is moving there are currents flowing the electrodes and ground.}
\end{figure}

In case the point charge $q$ is moving along the trajectory $x(t)$, as shown in figure~\ref{shockley-ramo1}, a time-dependent-induced charge on electrode $n$ can be derived and, hence, a current :

\begin{equation}
I^{ind}_n (t) = -\frac{dQ_n (t)}{dt} = -\frac{q}{V_w} \nabla \psi_n (x(t)) \frac{dx(t)}{dt} = -\frac{q}{V_w}E_n [x(t)] v(t)
\label{I_ramo}
\end{equation}

We call $E_n (x) = -\nabla \psi_n (x)$ the \textit{weighting field} of electrode $n$ as stated in Ramo's theorem~\cite{DET_ramo}. Note that the sign of the induced current is not only given by the sign of the charge but also by the orientation of the particle velocity vector with respect to the orientation of the weighting fields.

If the charge $q$ is moving along a trajectory $x(t)$ from position $x_0 = x(t_0)$ to position $x_1 = x(t_1)$ the total amount of charge $Q_n^{ind}$ between its electrode and ground is equal to 

\begin{equation}
Q^{ind}_n = \int_{t_0}^{t_1} I^{ind}_n (t) \mathrm{d} t = -\frac{q}{V_w} \int_{t_0}^{t_1} E_n [x(t)] v(t) \mathrm{d} t = \frac{q}{V_w} [ \psi_n (x_1) - \psi_n (x_0) ]
\label{Q_ramo4}
\end{equation}

It is possible to conclude that this induced charge is a conservative quantity, indeed, it depends only on the end points of the trajectory and is independent of the specific path. e.g., if a pair of charges $q$, $-q$ is produced at point $x_0$ at $t= t_0$ and they arrive at two different position $x_1$, $x_2$ respectively, after a time $t_1$, the resulting charge on electrode $n$ is given by 

\begin{equation}
Q^{ind}_n = \int_{t_0}^{t_1} I^{ind}_n (t) \mathrm{d} t = \frac{q}{V_w} [ \psi_n (x_1) - \psi_n (x_2) ]
\label{Q_ramo5}
\end{equation}

\begin{figure}[htbp]
\centering
\includegraphics[width=1\textwidth,keepaspectratio]{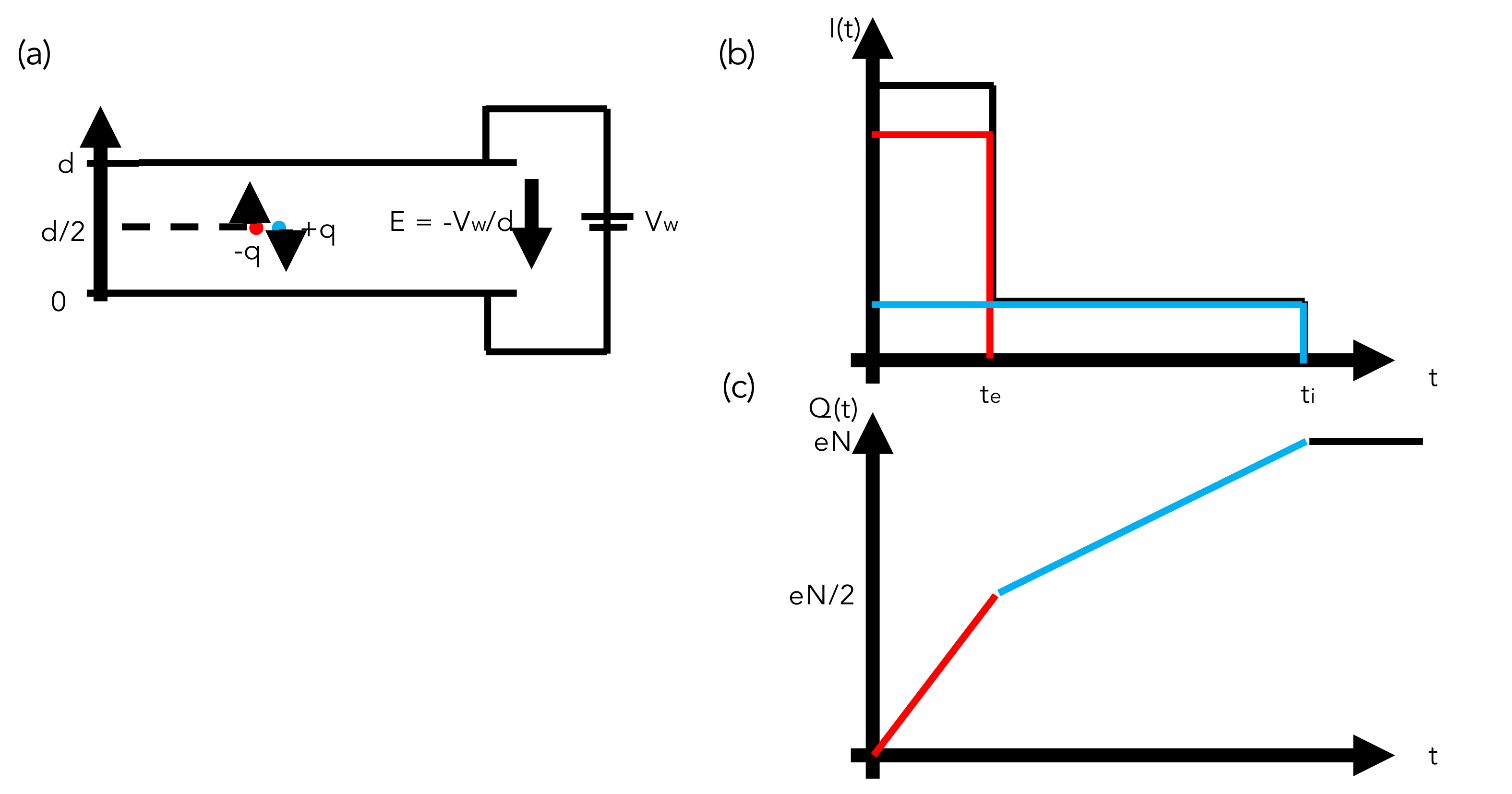}
\caption{\label{esramo1} \footnotesize (a) Sketch of a planar electrode geometry, with an electric field perpendicular to the plates, and two charges moving towards the anodes. (b) The curve of the induced current and (c) charge to one electrode are also shown.}
\end{figure}

In the case the charge $q$ moves to the surface of electrode $n$ and $-q$ moves to the surface of some other electrode, the total induced charge on electrode $n$ is equal to $q$, since $\psi_n  = V_w$ on electrode $n$ and $\psi_n = 0$ on the others electrodes. When both charges move to other electrodes, the total induced charge on electrode $n$ is zero. Therefore, it is possible to conclude that when all the charges have arrived at different electrodes, the total induced charge on electrode $n$ is equal to the charge that has arrived at that specific electrode. Note that it also leads to the conclusion that the current signals on electrodes that do not receive any charge are strictly bipolar.
\\ For simplicity we consider a planar electrode geometry operating in ionization mode, see section~\ref{GID}, the electric field inside the chamber is $E=V_w/d$. If $N$ pairs are produced by a charged particle, then the $-q$ electrons will drift to one anode, while the $+q$ ions to the opposite one, and the charge produced is $q=eN$. A scheme is shown in figure~\ref{esramo1}. 

$ v_{e}$ and $ v_{i}$ are defined as the velocity in modulus for electrons and positive ions respectively, the induced current on one electrode is obtained by equation~\ref{I_ramo}:

\begin{align}
&i_{-} = -\frac{-q}{V_w}\frac{V_w}{d}( v_{e}) = \frac{q}{d}v_{e} \ , \label{i_indes1} \\ 
&i_{+} = -\frac{+q}{V_w}\frac{V_w}{d} (-v_{i}) = \frac{q}{d}v_{i} \ , \label{i_indes2}\\ 
&i_{tot} = i_{-}+i_{+} 
\label{i_indes3}
\end{align}

The current will be the same but with opposite sign on the other electrode. Note that both the charged particles, positive and negative, give the same contribution to the induced current.  The total induced charge on the specific electrode can be calculated following equation~\ref{Q_ramo5}, in this case one obtains:

\begin{equation}
Q =  \int i_{tot} (t) \mathrm{d} t = \frac{q}{d}v_{e} t_{e} + \frac{q}{d}v_{i}t_{i}
\label{Q_indes1}
\end{equation}

\begin{figure}[htbp]
\centering
\includegraphics[width=1\textwidth,keepaspectratio]{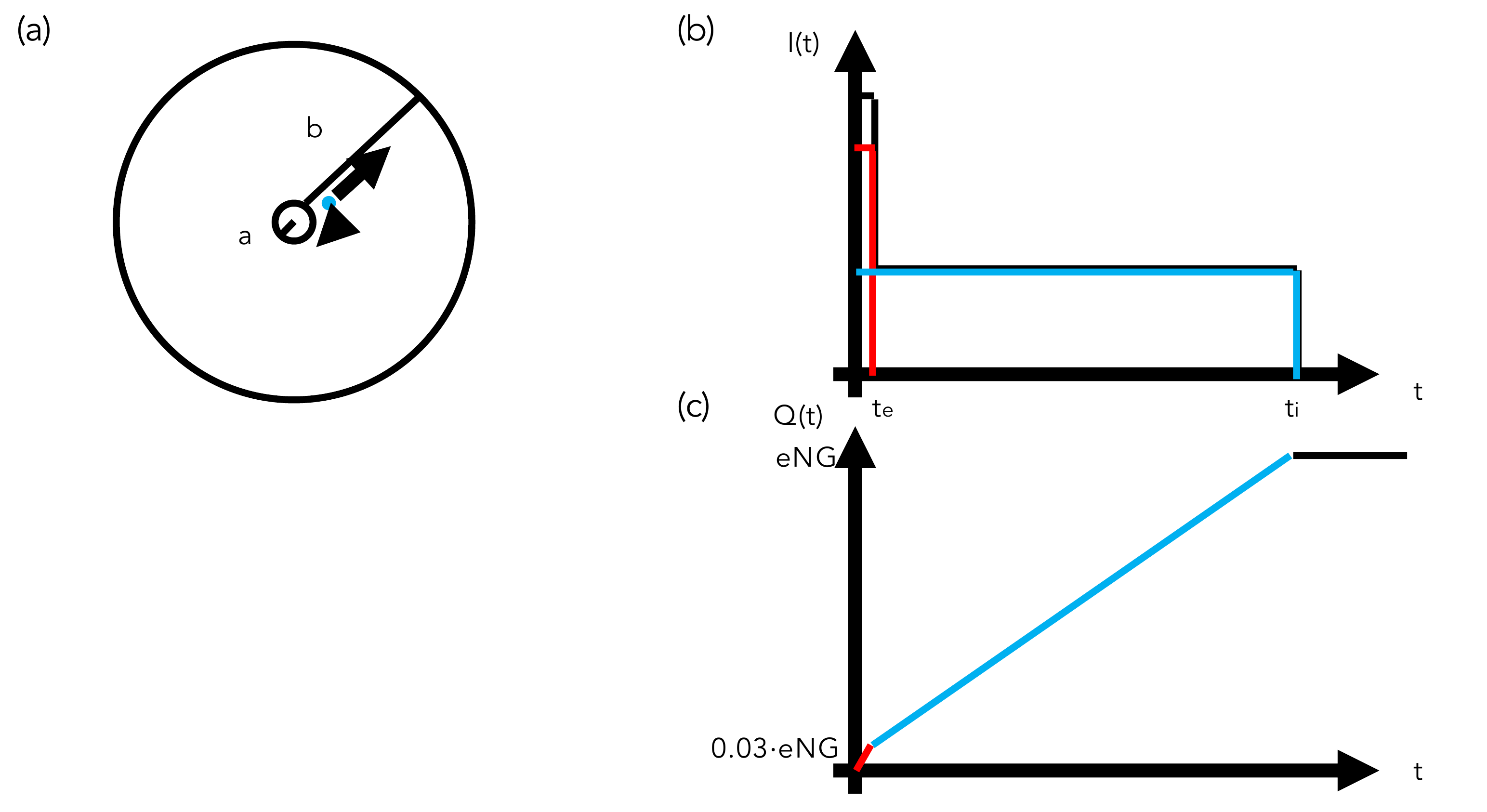}
\caption{\label{esramo2} \footnotesize (a) Sketch of a cylindrical wire chamber, a is the wire radius while b is the cathode radius. (b) The curve of the induced current and (c) charge to one electrode is also shown.}
\end{figure}

If the ion-pair is produced at distance $z_0$ from one of the two plane than $t_{e}$ and $t_i$ can be rewritten as $t_e=(d-z_0)/v_e$ and $t_i=(z_0)/v_i$. Substituting these values in equation~\ref{Q_indes1} we obtain: 

\begin{equation}
Q=q\frac{(1-z_0)}{d}+ q\frac{z_0}{d} = q = eN
\label{Q_indes11}
\end{equation}
This is the Shockley-Ramo theorem, whereby the total induced charge on the electrodes is equal to the charge that has arrived at that electrode. A sketch of both curves for induced current and induced charge as a function of time is depicted in figure~\ref{esramo1}, in the case the ion-pairs are produced in the middle of the chamber ($z_0 =d/2$). The electrons of the ion-pair have a much larger velocity ($\approx$ 1000 times higher) than that of the positive ions.  The time $t_e$ is much lower than $t_i$, therefore the electrons contribution to the current is peaked, while the positive ions contribution is more spread on time. As the integral is the same for both particles, the contribution to the induced charge is divided in half between the electrons and the positive ions, as shown in figure~\ref{esramo1}.   
\\ Considering now a wire chamber with cylindrical geometry operating in proportional mode, we will observe that the main contribution to the signal on electrodes is due to positive ions. A scheme of the geometry is illustrated in figure~\ref{esramo2}.

The electric field in a cylindrical geometry is given by equation~\ref{E_gasch}. In typical chamber $b \gg a$, so that the electric field rises in a region close to the wire, the multiplication region. Therefore, most of the ion-pairs are formed close to the anode wire. Considering a voltage $V_0 = 1000$V, a wire radius $a=10 \mu$m and a tube of radius $b = 4$mm, an electric field of 170 kV/cm is obtained. Given $q=eN\cdot G$, where $G$ is the gain due to the proportional mode operation (section~\ref{sectionamplgas}), the total charge created by the avalanche process and by considering that it take place at a distance $d=2 \mu$m from the wire anode, the charge induced on the anode can be calculated using the Schokley-Ramo theorem (eq.~\ref{Q_ramo5}):

\begin{equation}
Q_{anode}=-\frac{q_i}{\ln\Big(\frac{b}{a}\Big)} \int^{r_2}_{r_1} \frac{\mathrm{d}r}{r}= -\frac{q_i}{\ln\Big(\frac{b}{a}\Big)} \ln(r)\mid^{r_2}_{r_1}
\label{Q_indes2}
\end{equation}

\begin{figure}[htbp]
\centering
\includegraphics[width=0.6\textwidth,keepaspectratio]{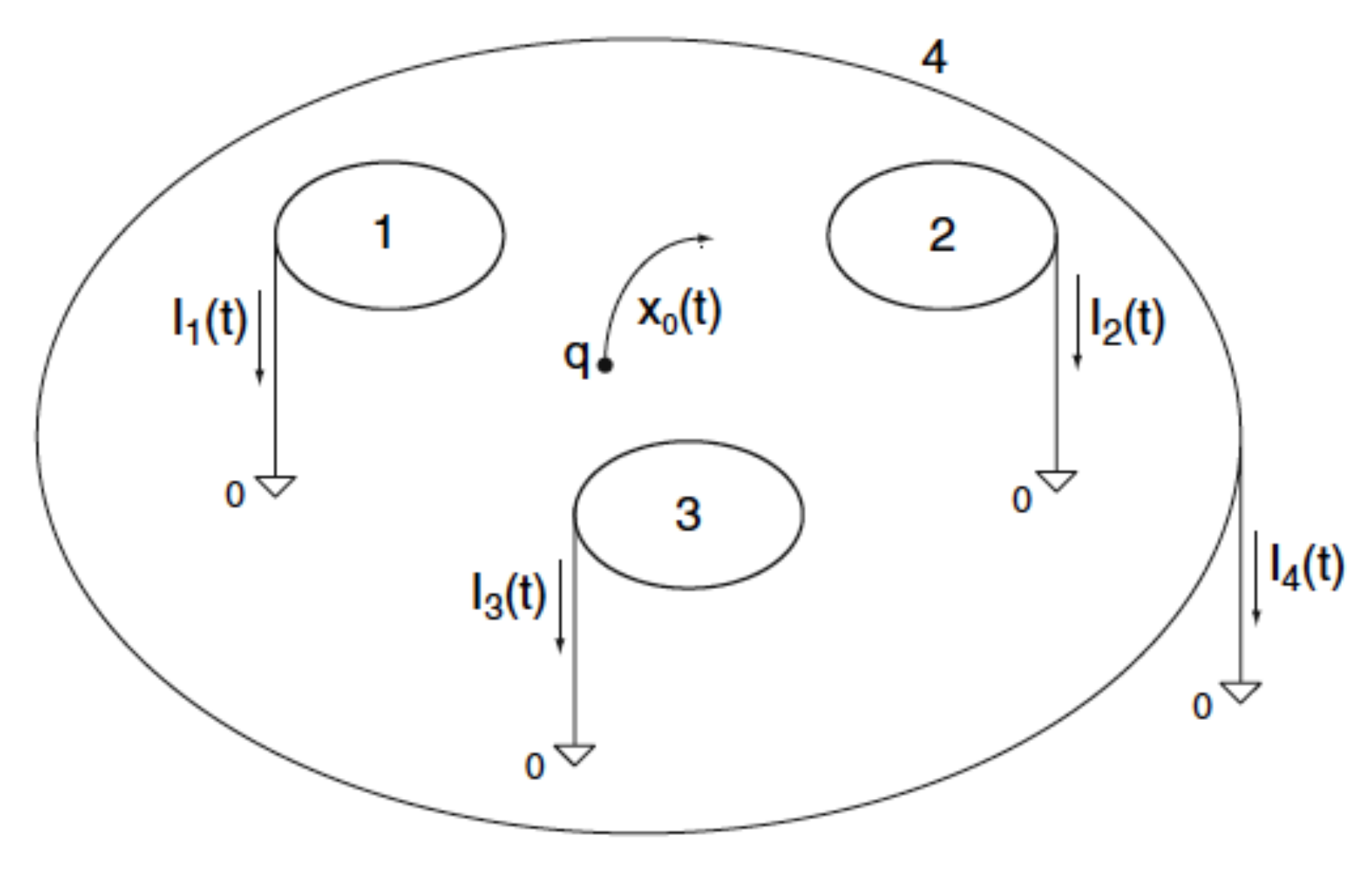}
\caption{\label{shockley-ramo} \footnotesize Three grounded electrodes enclosed by one electrode. }
\end{figure}

The contribution of electrons and ions can be addressed separately. In the first case $q_i=-q$, while for the ions the charge will be positive ($q_i= +q$). As we have chosen $d=2 \mu$m, the electrons will travel toward the anode between $r_1 =a+d$ and $r_2 =a$, and the ions toward the cathode between $r_1 =a+d$ and $r_2 =b$. By subsituting these values in equation~\ref{Q_indes2} we obtain:

\begin{align*}
Q_{e}= -\frac{-q}{\ln\Big(\frac{b}{a}\Big)} \ln\bigg(\frac{a}{a+d}\bigg) \simeq - 0.03q\\
Q_{ion}= -\frac{q}{\ln\Big(\frac{b}{a}\Big)} \ln\bigg(\frac{b}{a+d}\bigg) \simeq - 0.97q
\label{Q_indes3}
\end{align*}
As expected, almost all the signal is due to the ions contribution.
\\In a more general case, one can calculate the sum of the signals induced on several electrodes either using the weighting fields and induced currents for the individual electrodes and add the currents or the weighting fields can be calculated for the entire set of electrodes by setting all of them to voltage $V_w$ and grounding the remaining ones. Referring to figure~\ref{shockley-ramo} we assume that one electrode encloses the others.

$\psi_{tot}$, sum of all induced currents, is defined by setting all electrodes to potential $V_w$. This results in $\psi_{tot} = V_w$ in the considered volume and consequently a weighting field of $E_{tot} (x)= 0$. It is possible to conclude that the sum of all induced currents on grounded electrodes is zero at any time if there is one electrode enclosing all the others.

\subsubsection{Shockley-Ramo application on MWPC}
\begin{figure}[htbp]
\centering
\includegraphics[width=0.6\textwidth,keepaspectratio]{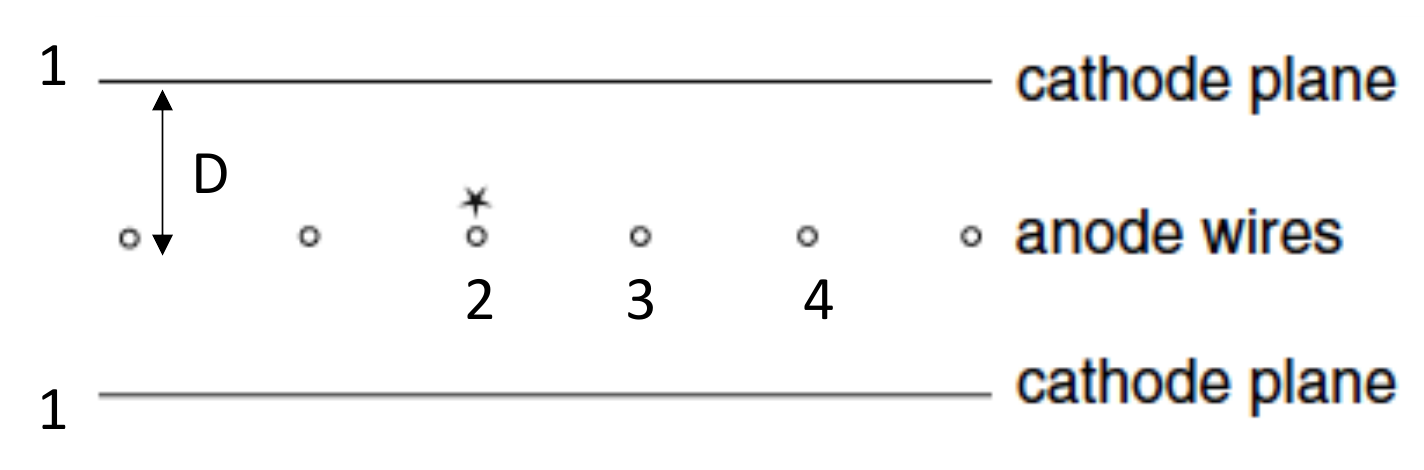}
\caption{\label{anode-cathode} \footnotesize Sketch of the electrode arrangement in a MWPC, in which the wire 2 carries the avalanche. }
\end{figure}

Let us consider now a plane of anode wires situated between two plane electrodes facing each other at distance $D$, as in case of a MWPC, the scheme is shown in figure~\ref{anode-cathode}. By applying the method described above the main characteristic for the signals of the wires and cathode can be derived.

The wire radii (in general 10 -30 $\mu$m) are small compared to the distance between the wires and the distance between the wires and the cathode, $D$, typically of a few millimetres. This results in a coaxial electric field close to the wire, as mentioned in previous sections. We set the wires to potentials $V_n$ which results in charges $q_n$ on the wires, determined by the capacitance matrix through $q_n = \sum_{m=1}^N c_{nm}V_m$. As depicted in figure~\ref{anode-cathode} we label the individual wires as 2, 3, 4, etc. and the two cathode as electrode 1. Assuming the wires to be infinitely long, the electric field at a distance $r$ from a line charge $q$ is given by

\begin{equation}
E(r) = \frac{q}{2\pi \varepsilon_0 r} \widehat{r}
\label{efield}
\end{equation} 

so at a point with distance $r_1$ and $r_2$ to two wires carrying charge $q_1$ and $q_2$, the electric field is:

\begin{equation}
E= \frac{1}{2\pi \varepsilon_0} \bigg( \frac{q_1}{r_1}\widehat{r}_1+\frac{q_2}{r_2}\widehat{r}_2 \bigg)
\label{campE}
\end{equation}

In case $r_1 \ll r_2$ the field is dominated by the charge on this wire as long as the charge $q_1$ and $q_2$ are of similar magnitude. It is possible to conclude that the electric fields at distance $r$ from wire $n$ is approximately coaxial and its magnitude is given by 

\begin{equation}
E(r) = \frac{q_n}{2\pi \varepsilon_0 r} \Bigg( \sum_{m=1}^N c_{nm}V_m \Bigg)
\label{efield2}
\end{equation} 

This is valid in case $r < D$. Under these assumptions the trajectory of the avalanche ions is:

\begin{equation}
r(t) = a \sqrt{1+\frac{t}{t_0}} \ ; \qquad \frac{1}{t_0} = \frac{\mu}{a^2 \pi \varepsilon_0 r} \Bigg( \sum_{m=1}^N c_{nm}V_m \Bigg)
\label{ion_traject}
\end{equation} 

where $a$ is the wire radius and $\mu$ is the ions mobility ($ v = \mu E$). Knowing the ion trajectory $r(t)$ we need to calculate the weighting field of the wires in order to find the induced current signals. The weighting field of the wire 2, where the avalanche takes place, is found by setting the wire to voltage $V_w$ and grounding all others wires and electrodes. That is a charge $q_2 =c_{22}V_w$ on this wire, and the magnitude of the weighting field in the coaxial approximation is 

\begin{equation}
E_2(r) = \frac{c_{22}V_w}{2\pi \varepsilon_0 r} 
\label{Ewire2}
\end{equation} 

Thus the current induced by $N_{tot}$ ions of charge $e_0$, moving away from the wire surface is 

\begin{equation}
I_2 (t) = -\frac{N_{tot}e_0}{V_w}E_2 [r(t)] \frac{\mathrm{d}r}{\mathrm{d}t} = -\frac{N_{tot}e_0}{4\pi \varepsilon_0 }\frac{c_{22}}{t +t_0}
\label{Iwire2}
\end{equation}

Note that the signal on the avalanche wire is negative because from equation~\ref{cmatrix_rel} $c_{22} > 0$. The weighting field of the closest wire (electrode 3) is given by setting wire 3 to potential $V_w$ and grounding all other wires, thus $q_3 =c_{23}V_w$ and the current signal is $I_3 = -\frac{N_{tot}e_0}{4\pi \varepsilon_0 }\frac{c_{23}}{t +t_0}$. The cross-induced signal has the same shape as the signal on the avalanche wire but it is positive because $c_{2n} < 0$ with $n \neq 2$, see equation~\ref{cmatrix_rel}. The other wires have the same polarity and shape with their relative amplitudes $c_{24}, c_{25}$, etc. Using the relations in~\ref{cmatrix_rel3} we can express the relative signal amplitudes respect to the capacitances $C_{nm}$:

\begin{equation}
\frac{I_n(t)}{I_1(t)} = \frac{c_{2n}}{c_{22}} = -\frac{C_{1n}}{\sum_{m=1}^N C_{2m}}
\label{signal_amp}
\end{equation} 

The amplitude of the cross-induced signal from the avalanche wire to wire $n$ is given by the mutual capacitance between the avalanche wire and wire $n$ divided by the sum of all mutual capacitances of the avalanche wire to the other electrodes~\cite{Particle_Detection}. It can be noticed that the capacitive coupling between wires causes a signal of same polarity on the neighbouring wires.
\\ The sum of all wire signals is given by 

\begin{equation}
I_w(t) =\sum_{n=2}^N I_n (t)= -\frac{N_{tot}e_0}{4\pi \varepsilon_0 }\frac{1}{t +t_0} \sum_{n=2}^N c_{1n} =   -\frac{N_{tot}e_0}{4\pi \varepsilon_0 }\frac{C_{12}}{t +t_0}
\label{signal_amp_tot}
\end{equation} 

As the two cathode planes completely enclose the wires the sum of all capacitance matrix elements $\sum_{m=1}^N c_{nm}$ is zero. The sum of all wires signals thus has a negative polarity, if the same voltage $U$ is set for each of them, the characteristic time constant from equation~\ref{ion_traject} becomes:

\begin{equation}
\frac{1}{t_0} = \frac{\mu}{a^2 \pi \varepsilon_0 r} \Bigg( U \sum_{m=2}^N c_{nm} \Bigg) = \frac{\mu U C_{12}}{a^2 \pi \varepsilon_0}
\label{time_tot}
\end{equation} 

When the wires are enclosed by the two cathode planes, like in figure~\ref{anode-cathode}, the sum of all chamber signals is zero at any time and the cathode signal $I_c(t)$ is equal to the negative summed wire signal $-I_w(t)$, based on the Shockley-Ramo theorem. In the coaxial approximation, the signal on a single cathode plane, considering that in our case they are symmetrically placed with respect to the wire plane, is:

\begin{equation}
I_{c1}(t) =\frac{1}{2} I_c(t) = -\frac{1}{2} I_w(t) = \frac{1}{2} \frac{N_{tot}e_0}{4\pi \varepsilon_0}\frac{C_{12}}{t +t_0}
\label{I_cathode}
\end{equation} 

As mentioned above it is possible to localize the avalanche in the direction perpendicular to the sense wire subdividing the cathode into a series of strips or pads. In this case the weighting field of the cathode strip is calculated by setting the strip to potential $V_w$ while grounding all other strips and wires. The method to derive the expression for the weighting field and induced signal can be found in detail in~\cite{GORDON}.
\\We first calculate the signal on a strip of infinitesimal width, we then integrate over the strip width $w$. It is defined, in this way, the so-called cathode charge distribution $\Gamma(\lambda)$ through

\begin{equation}
\mathrm{d} I_{c1} (t,\lambda) = I_c(t) \Gamma(\lambda) \, \mathrm{d}\lambda \ ; \qquad \int_{\infty}^{\infty} \Gamma(\lambda) \, \mathrm{d}\lambda = \frac{1}{2}
\label{chargedistr_cathode}
\end{equation}

where $\lambda = x/D$ is the distance of the infinitesimal strip from the avalanche position. The expression for $\Gamma(\lambda)$ can be calculated by investigating the weighting field of an infinitesimal cathode strip, and can be found at~\cite{GATTI_stripCathode}.

\begin{equation}
\Gamma(\lambda) = K_1 \frac{1- \tanh^2 K_2 \lambda}{1+ K_3 \tanh^2 K_2 \lambda} 
\label{chargedistr_cathode2}
\end{equation}

The parameter $K_3$ depends on the chamber geometry, while $K_1$ and $K_2$ are uniquely defined by $K_3$. Thus for a strip of finite width $w$ centred at position $\lambda$, the signal $I(t,\lambda,w)$ is obtained by integrating equation~\ref{chargedistr_cathode}:

\begin{equation}
I (t,\lambda,w) = \int_{\lambda -w/2}^{\lambda +w/2} \mathrm{d} I_{c1} (t,\lambda') \, \mathrm{d}\lambda' = I_c(t) \int_{\lambda -w/2}^{\lambda +w/2} \Gamma(\lambda') \, \mathrm{d}\lambda' = I_c(t) P_0 (\lambda)
\label{chargedistr_cathode_tot}
\end{equation}

where $P_0 (\lambda)$ is the \textit{pad response function} and depends on the strip width $w$ and on the parameters $K_1, K_2, K_3 $. It determines the fraction of the total cathode signal induced in the strip as a function of the distance between the centre of the strip and the avalanche position in the cathode plane. 
\\ The strip width $w$ and the anode-cathode distance $D$ have to be chosen such that the typical pulse height induced on two or three adjacent strips falls into the dynamic range of the readout electronics. If the limitation in position resolution along the wire is given by the electronic pulse-height measurement, the best position resolution is then given for a strip width near the cathode-anode distance, namely $w \approx D$.

\section{Semiconductor Detectors}\label{secsemisdet}

Semiconductor, or solid state, detectors are based on crystalline semiconductor materials, typically silicon and germanium. The basic operating principle of solid state detectors is analogous to gas ionization devices. However, instead of a gas, the medium is a solid semiconductor material. The passage of ionizing radiation creates electron-hole pairs, instead of electron-ion pairs, which drift and produce signal when an electric field is applied. The semiconductor benefits from a small energy gap between their valence and conduction bands~\cite{rancoita}. Therefore, the advantage is that the average energy required to create an electron-hole pair is about 10 times smaller than that required for gas ionization. For a given energy, the amount of ionization produced is an order of magnitude greater, resulting in increased energy resolution. Moreover, as their density is larger compared to that of gases, the stopping power is greater than the gas detectors. Thus, semiconductor detectors are compact in size and can have very fast response times~\cite{DET_leo}. The section mainly refers to the books~ \cite{rancoita,Rad-Det-principles,Semiconductor-spieler,Sze-semiconductor,Lutz-SolidDet} for the principles of operation discussion. 

\subsection{Basic Principles of Operation}

One of the main characteristics that identifies the semiconductor materials is the small gap between the electronic conduction and valence band. e.g., in case of silicon the amount of energy needed to excite an electron from the valence band into the conduction band is $E_g=1.12\,$eV. This is in contrast to insulators that have an energy gap of $E_g> 5\,$eV and conductors that have their valence and conduction bands in contact. The hole left by an electron in the valence band under some excitation has a positive electric charge. 
\\ If an electric field is applied, the passage of electrons from the valence into the conduction band, generates holes moving in the valence band in a opposite direction to that of electrons in the conduction band. The motion will be the combination of a random thermal velocity and a drift velocity parallel to the direction of the field. Since electrons move in an opposite direction to the electric field vector, holes move in the same direction as the electric field, as expected for a point positive charge. The charge carriers drift velocity is proportional to the applied electric field through the mobility as follows:

\begin{eqnarray}
& v_e = \mu_e \: E  \label{e-hvelocity1} \\
& v_h = \mu_h \: E
\label{e-hvelocity2}
\end{eqnarray}

where the quantity $\mu_e$ and $\mu_h$ are the electron and hole mobility, respectively. In the case of silicon hole mobility is smaller than the electron mobility ($\mu_h = 450\,$cm$^2$V$^{-1}$s$^{-1}$, $\mu_e = 1350\,$cm$^2$V$^{-1}$s$^{-1}$)~\cite{rancoita}. At higher electric field values, the drift velocity increases more slowly with the field, till a saturation velocity, independently of further increases in the electric field, is reached. Many semiconductor detectors are operated with electric field values high enough to result in saturated drift velocity for the charges carriers. These velocities are on the order of 10$^7$ cm/s, thus the time needed to collect the signals over a typical dimension of 0.1 cm will be under 10 ns.
\\ The mobility determine the current in a semiconductor, by using equation~\ref{e-hvelocity1} and~\ref{e-hvelocity2} the current density, $J_c$, for concentrations $n$ of electron carriers and $p$ of holes carriers, is given by:

\begin{equation}
J_c = q(n\mu_e +p\mu_h)E
\label{current-density}
\end{equation}

where $q$ is the electronic charge. At a given temperature, an equilibrium between generation and recombination of free electron and holes is established. 
\\ Crystals are not perfect, but present defects and various impurities, which reduce the mean free carrier lifetime. The addition of electrically active donor and acceptor impurity atoms forms $n$-type and $p$-type conductivity semiconductors. They have an excess of electron and holes, respectively. At temperature $T$, the product of concentrations of electrons and holes remains constant and is given by 

\begin{equation}
np = n^2_{int}
\label{concentration-np}
\end{equation}

where $n_{int}$ (cm$^{-3}$) is the intrinsic carrier concentration, e.g., at $T= 300\,$K in silicon $n_{int} \approx 1.45 \times 10^{10}\,$cm$^{-3}$. It can be expressed as:

\begin{equation}
n_{int}= \sqrt{N_c N_v} \ \exp{\bigg(-\frac{E_g}{2KT}\bigg)}= AT^{3/2} \ \exp{\bigg(-\frac{E_g}{2KT}\bigg)}
\label{n-int}
\end{equation} 

where $N_c$ is the number of states in the conduction band and $N_v$ is the number of states in the valence band, $E_g$ the energy gap at 0 kelvin and $K$ is the Boltzmann constant. $N_c$ and $N_v$ can be calculated from Fermi-Dirac statistics and each vary as $T^{3/2}$. $A$ is a constant independent of temperature. Since the semiconductor is neutral, the positive and negative charge densities must be equal so that $N_D +p = N_A +n$, where $N_D$ and $N_A$ are the donor and acceptor concentrations. In a n-type material, where $N_A= 0 $ and $n \gg p$, the electron density is therefore $n=N_D$ and the electrical conductivity is determined almost exclusively by the flow of electrons. The resistivity, $\rho$ or conductivity, $\sigma$, can be calculated from the dopant concentration and the mobility of the majority carrier as:

\begin{equation}
\frac{1}{\rho} = \sigma = eN_D \mu_e
\label{resitivity}
\end{equation} 

\subsubsection{The $p-n$ Junction}

\begin{figure}[htbp]
\centering
\includegraphics[width=0.6\textwidth,keepaspectratio]{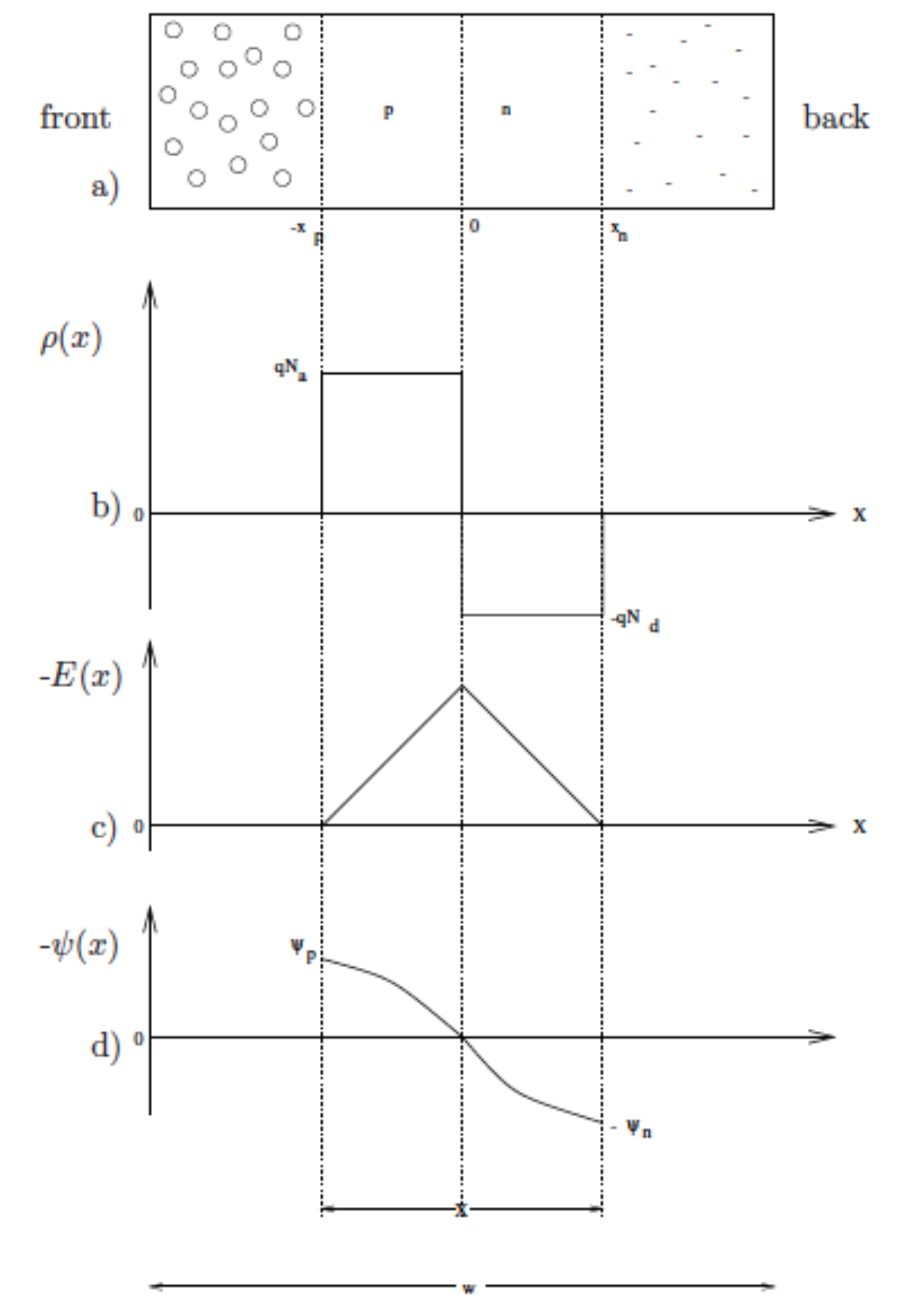}
\caption{\label{pn-junc} \footnotesize Sketch of a $p$-$n$ junction of thickness $w$. $X$ is the depletion region. (a) represents the junction, (b) the charge distribution, (c) the electric field and (d) the electrostatic potential (from~\cite{rancoita}).}
\end{figure}

When a $n$-type region in a silicon crystal is put adjacent to the $p$-type region in the same crystal, a $p-n$ junction is formed. Basically, such a junction is built by diffusing acceptor impurities into a $n$-type silicon crystal or vice versa donors in a $p$-type silicon crystal. For instance, a $p^+ -n$ junction results from an acceptor density on the $p$-type side being much larger than the donor density on the $n$-type side. The formation of a $p-n$ junction creates a special zone about the interface between the two materials. This region is created as the result of the diffusion of electron from $n$-type material into $p$-type material and diffusion of holes in the other direction. The diffusion is the consequence of the motion of carriers from regions of high concentration to regions of low concentration. Thus, the diffusing electrons fill up the hole in the $p$-region while the diffusing holes capture electrons on the $n$-region. Since the $p$-side is injected with extra electrons, it becomes negative and the $n$-region becomes positive. This create an electric field gradient across the junction. The space-charge region is called \textit{depletion region}.
\\ The shape of the electrostatic potential ($\Psi$), the electric field and the width of the depletion zone of a function can be obtained by solving the Poisson equation:

\begin{equation}
\frac{\mathrm{d}^2 \Psi}{\mathrm{d}x^2} = -\frac{\rho (x)}{\varepsilon}
\label{depletion-potential}
\end{equation}

where $\varepsilon$ is the dielectric constant. The equation~\ref{depletion-potential} can be solved in one-dimension, for an abrupt junction with a charge density $\rho (x)$ given by:

\begin{equation}
\rho(x) \rightarrow \begin{cases} q N_D \quad & \mathrm{for} \ 0 \leq x \leq x_n \\ -q N_A \quad & \mathrm{for} \  -x_p \leq x \leq 0 \end{cases}
\label{rho-cases}
\end{equation}

where $x_n$ and $x_p$ are the depletion length on the $n$-side and $p$-side, respectively, a sketch is shown in figure~\ref{pn-junc}(a). Note that the charge density is zero outside the depletion region~\ref{pn-junc}(b). The absence of net total charge in the depletion zone leads to:

\begin{equation}
N_D x_n = N_A x_p
\label{nocharge-juct}
\end{equation}

By integrating equation~\ref{depletion-potential}, one obtains the electric field $E$ generating from the charges separation, considering as boundary condition $E(x_n) = E(-x_p)= 0$:

\begin{equation}
E(x) = - \frac{\mathrm{d} \Psi}{\mathrm{d}x} = \begin{cases} E_n(x)=q\frac{N_D}{\varepsilon} (x-x_n)\quad & \mathrm{for} \ 0 \leq x \leq x_n \\ \\ E_p(x)= -q\frac{N_A}{\varepsilon} (x+x_p) \quad & \mathrm{for} \  -x_p \leq x \leq 0 \end{cases}
\label{E-cond-pn}
\end{equation}

The electric field is only due to the different concentrations of electrons and holes at the junction, when no external voltage is applied. The diffusion process will move $n$-type material electrons into $p$-type material and holes in opposite direction. This process is slowed down by the generation of an electric field due to the ionized dopants, which pushes electrons back to the $n$-type side and holes to the $p$-type side, until a dynamical equilibrium is reached.
\\ Free charges can be generated in excess of the equilibrium by ionizing particles traversing the diode in this depletion region. For instance, in silicon the energy required to produce an electron-hole pair is $E_{ion}=3.62\,$eV. The charges produced by ionization in the depletion zone are separated and induce an electron hole signal.
\\ The integration of equation~\ref{E-cond-pn} leads to the calculation of the electrostatic potential. We define the integration constant $\Psi(-x_p) = \Psi_p$ and $\Psi(x_n) = \Psi_n$, one obtains the electrostatic potential as shown in figure~\ref{pn-junc}(d):

\begin{equation}
\Psi(x) =  \begin{cases} \Psi_n(x)= \Psi_n - q\frac{N_D}{2\varepsilon} (x-x_n)^2 \quad & \mathrm{for} \ 0 \leq x \leq x_n \\ \\ \Psi_p(x) = \Psi_p + q\frac{N_A}{2\varepsilon} (x+x_p)^2 \quad & \mathrm{for} \  -x_p \leq x \leq 0 \end{cases}
\label{psi-cond-pn}
\end{equation}

The contact potential is defined by:

\begin{equation}
V_0 = - \int E(x) \mathrm{d}x = \Psi_n -\Psi_p = \frac{K_b T}{q} \ln\bigg(\frac{N_A N_D}{n^2_{int}}\bigg)
\label{juct-pot}
\end{equation}

e.g., for silicon at $T=200\,$K is $\approx 0.3 - 0.6\,$V. The depletion depths are calculated by imposing the continuity of $V_0$, the potential at $x=0 \ [\Psi_n(0)=\Psi_p(0)]$ and using equation~\ref{nocharge-juct}:

\begin{equation}
x_n = \frac{1}{N_D}\sqrt{\frac{2\varepsilon V_0}{q} \bigg(\frac{1}{N_A} + \frac{1}{N_D}\bigg)^{-1} }
\label{xn-depth}
\end{equation}

and

\begin{equation}
x_p = \frac{1}{N_A}\sqrt{\frac{2\varepsilon V_0}{q} \bigg(\frac{1}{N_A} + \frac{1}{N_D}\bigg)^{-1} }
\label{xp-depth}
\end{equation}

Summing equations~\ref{xn-depth} and~\ref{xp-depth} one obtains the total depth of the depletion region $X$:

\begin{equation}
X = \sqrt{\frac{2\varepsilon V_0}{q} \bigg(\frac{1}{N_A} + \frac{1}{N_D}\bigg)}
\label{X-depth}
\end{equation}

Note that without external polarization, the depletion depth is typically a few microns. The capability of the junction for particle detection is strongly limited, therefore an external voltage between the $n$- and $p$-regions is applied. $V_0$ will be replaced by $V_0+V_b$ in equations~\ref{xn-depth} and~\ref{xp-depth}. The depletion zone grows for increasing values of $V_b$ until it reaches the total thickness of the detector.  

\subsection{Neutron Detection with Silicon Detectors}

Silicon detectors can be used to detect neutral particles via measurement of indirect ionization for photons, or secondary radiation for neutrons. Typically the active area of the detector, the silicon component, is coupled to a neutron converter layer, in which the neutron interacts giving rise to secondary heavy charged particles either via nuclear reactions or as energy recoils from elastic scattering of neutrons on nuclei. These secondary charged particles generated in the converter must have a range larger than the distance between the interaction point in the converter and the converter layer-detector interface, in order to reach the silicon layer where they deposit the energy. Charge carriers are then generated in silicon producing a signal.
\\ Different converters means different mechanism of charged secondary particles generation, the choice depends on the neutron energy. i.e., thermal neutrons have large nuclear cross section, thus converter as $^{10}$B, $^{113}$Cd, $^{155}$Gd, $^{157}$Gd, $^6$LiF are in general the most used for neutrons of these energies. However, the neutron capture reactions have a cross section dropping strongly for higher energies of neutrons, e.g., fast neutrons. Another type of converter has to be used, and therefore, another mechanism of charged particles generation has to be exploited. Fast neutrons can be detected via energetic recoils produced from their elastic scattering on nuclei of a converter. Converters with high hydrogen content and with a restraint amount of heavier atoms must be selected.  
\\ In the case of thermal neutrons the heavy charged particles produced in the converter material have their direction of emission kinematically constrained. The reaction products are emitted in opposite directions, hence, only one can cross the converter layer-detector interface into silicon. The range, $R_i (E)$, of travelling of the secondary particles can be calculated from SRIM~\cite{MISC_SRIM1998,MISC_SRIM2010}. The probability of signal generation will depend on $R_i (E)$, on the number ($N_n$) of incident neutrons on the converter, on the atomic concentration ($N_i$) of the target isotopes in the converter and on the cross section ($\sigma_i$) of the neutron-$i$-target isotope reaction. In general, summing over all particles produced in the neutron reaction, the probability of signal generation can be expressed as:

\begin{equation}
S= \sum_{i} S_i
\label{Si-signal-gen}
\end{equation} 

where $S_i= N-n N_i \sigma_i R_i P$ and the factor $P$ represents the total or integrated geometrical probability that any charge particle generated in the converter reaches the detector and produce a signal.

\begin{figure}[htbp]
\centering
\includegraphics[width=0.6\textwidth,keepaspectratio]{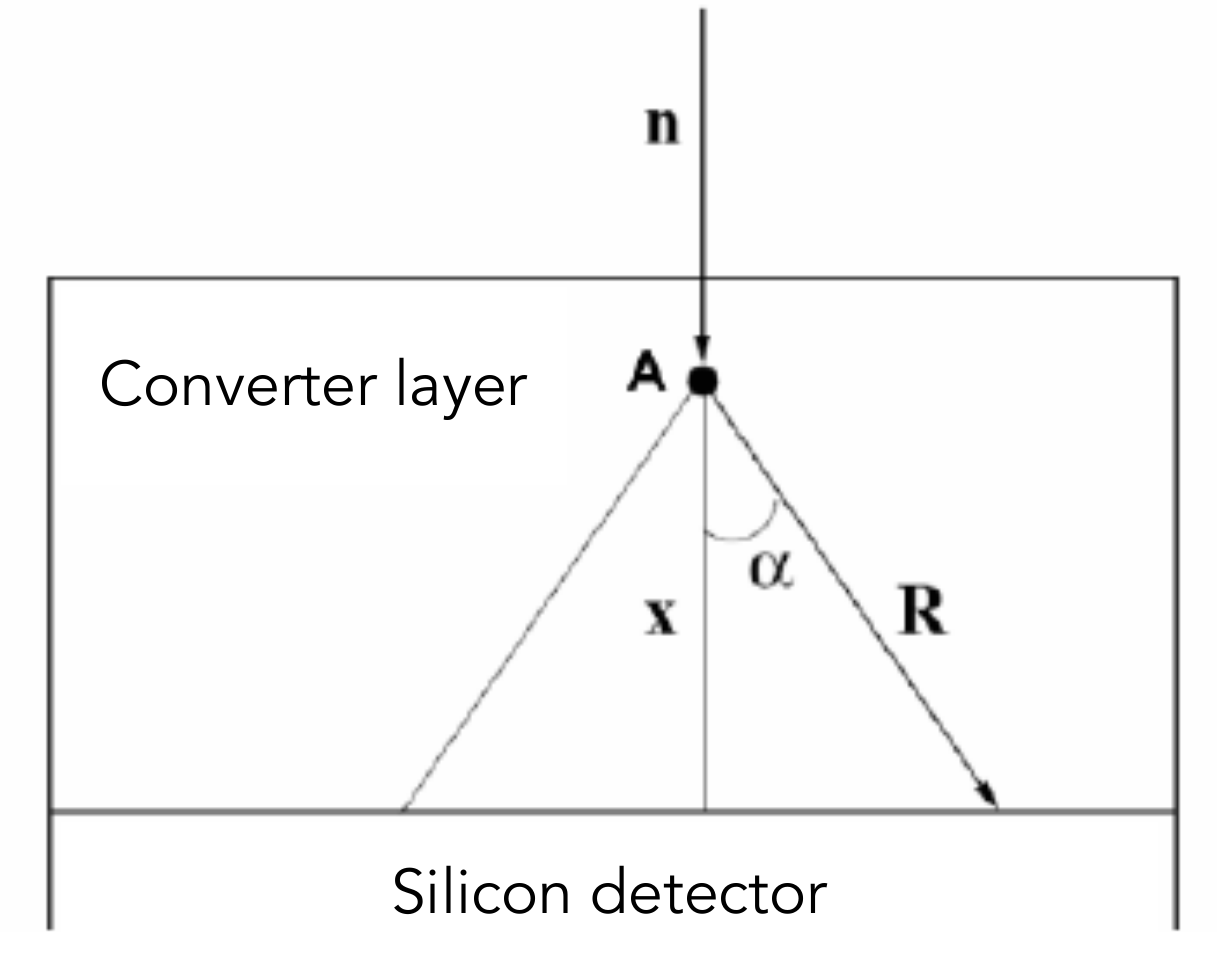}
\caption{\label{si-sketch} \footnotesize $A$ identifies the interaction point between neutrons and the target nuclei, $x$ is the distance from the interface converter-detector. The particles are produced in a cone of opening $\alpha$ with a range $R$ in that direction.}
\end{figure} 

Referring to figure~\ref{si-sketch}, the probability $P$ can be calculated. Considering $A$ the interaction point of neutron conversion, the product particles are emitted in a cone of opening angle $\alpha$ and a particle emitted at this angle stops after the range $R$ in the converter when its energy reaches zero. If the distance from the point $A$ and the cone basis is $x$, one has $\alpha = \arccos(x/R)$. The solid angle of particle emission is then:

\begin{equation}
SA = \int^{2\pi}_0 \int^{\alpha}_0 \sin\theta \, \mathrm{d} \theta \, \mathrm{d} \phi = 2\pi \, (1-\cos \alpha)
\label{Si-solidangle}
\end{equation}

The probability that a particle emitted at point $A$ with a range $R$ in the converter reaches the interface between the converter layer and the silicon detector at a distance $x$ is, then, the ratio between the solid angle, $SA$ and the total sphere solid angle, $SA_{sphere}= 4\pi$:

 \begin{equation}
P(x,R)=\frac{SA}{SA_{sphere}} = \frac{2\pi \,(1-\cos \alpha)}{4\pi} = \frac{1}{2}\bigg( 1- \frac{x}{R} \bigg)
\label{Si-probability}
\end{equation}

Introducing the variable $\eta = x/R$ and integrating from $\eta = 0$, origin of the particle at $x=0$, up to $\eta=1$, origin of the particle at $x=R$, we obtain:

\begin{equation}
P \sim \frac{1}{2} \int_0^1 P(\eta) \mathrm{d} \eta = \frac{1}{4}
\label{Si-probability2}
\end{equation}

On the contrary, the probability that any charged particle generated in the converter goes in a opposite direction of the interface can be expressed as:

 \begin{equation}
P(d-x,R)= \frac{1}{2}\bigg( 1- \frac{d-x}{R} \bigg)
\label{Si-probability3}
\end{equation}

Equation~\ref{Si-probability3} is needed in the case of a silicon detector inserted in between two converter layers.
\\ The detection efficiency increases with increasing converter thickness because more neutrons are captured in the converter material. Nevertheless, the detection efficiency starts to decrease at a value, $d_o$, of the thickness, when the range of the charged particles cannot reach the detector active area. To calculate it, we consider the detection efficiency, d$P_{det}$, of a neutron by a system as the one shown in figure~\ref{si-sketch}, (converter layer coupled to a silicon sensor):

\begin{equation}
\mathrm{d}P_{det}= e^{-\mu x} \, \frac{1}{2}\bigg( 1- \frac{d-x}{R} \bigg) \mu \, \mathrm{d}x
\label{Si-det-eff}
\end{equation} 

where $\mu \, \mathrm{d}x$ is the probability for the interaction of a neutron with  nucleus in the converter material at depth $x$. The exponential term represents the probability that a neutron will traverse a path of length $x$ without interaction. $1/2( 1- (d-x)/R)$ is the probability that the products of the neutron reaction will escape into the silicon sensor. It can be expressed as the ratio between the solid angle defined by the cone of emission and the total solid angle (equation~\ref{Si-probability3}). Integrating equation~\ref{Si-det-eff} over the converter thickness, $d$, we find

\begin{equation}
P_{det}=\frac{1}{2} \bigg[ -e^{-\mu d} + 1 - \frac{e^{-\mu d}}{\mu R} -\frac{d}{R} + \frac{1}{\mu R} \bigg]
\label{Si-det-eff2}
\end{equation}

Imposing the condition $\mathrm{d} P_{det}/\mathrm{d}d=0$ the optimal thickness $d_o$ can be found:

\begin{equation}
d_o = \frac{1}{\mu} \ln(\mu R +1)
\label{d-optimal}
\end{equation}

\chapter{Neutron Scattering Science: beyond the state-of-the-art}\label{chapter3}

The increasing complexity in science investigations driven by technological advances is reflected in the studies of neutron scattering science which enforces a diversification and an improvement of experimental tools, from the instrument design to the detector performance. It calls as well for more advanced data analysis and modelling. The operation of new generation, high-intensity neutron sources is therefore necessary; like the Spallation Neutron Source~\cite{SNS} (Oak Ridge Laboratory, TN, USA), the Japan Spallation Source~\cite{JPARC} (Tokai-mura, Japan), and the European Spallation Source (ESS, Lund, Sweden), which is presently under construction, where an intensity increase of at least an order of magnitude is expected~\cite{Mezei2007, VETTIER_ESS,ESS2011}. The high brightness opens up the possibility of studying currently not solved problems. Interesting samples are, indeed, often available only in limited quantities and may be unstable over the time. Furthermore, smaller samples tend to be more homogeneous, increasing the precision of the results obtained and making possible more advanced and conclusive investigations~\cite{ESS_TDR}. Thus the instruments must be flexible permitting exchanges between brightness and better resolution, optimization of signal-to-noise ratio, and the use of polarized neutrons when necessary. Together with the implementation of sources and instruments, the enhanced detector response is of a crucial importance to fulfil the increased performance demands.
Some of the scientific topics are described in the following section, including the current neutron application and the challenges and prospects for the future. 

\section{The European Spallation Source ESS}

\begin{figure}[htbp]
\centering
\includegraphics[width=.9\textwidth,keepaspectratio]{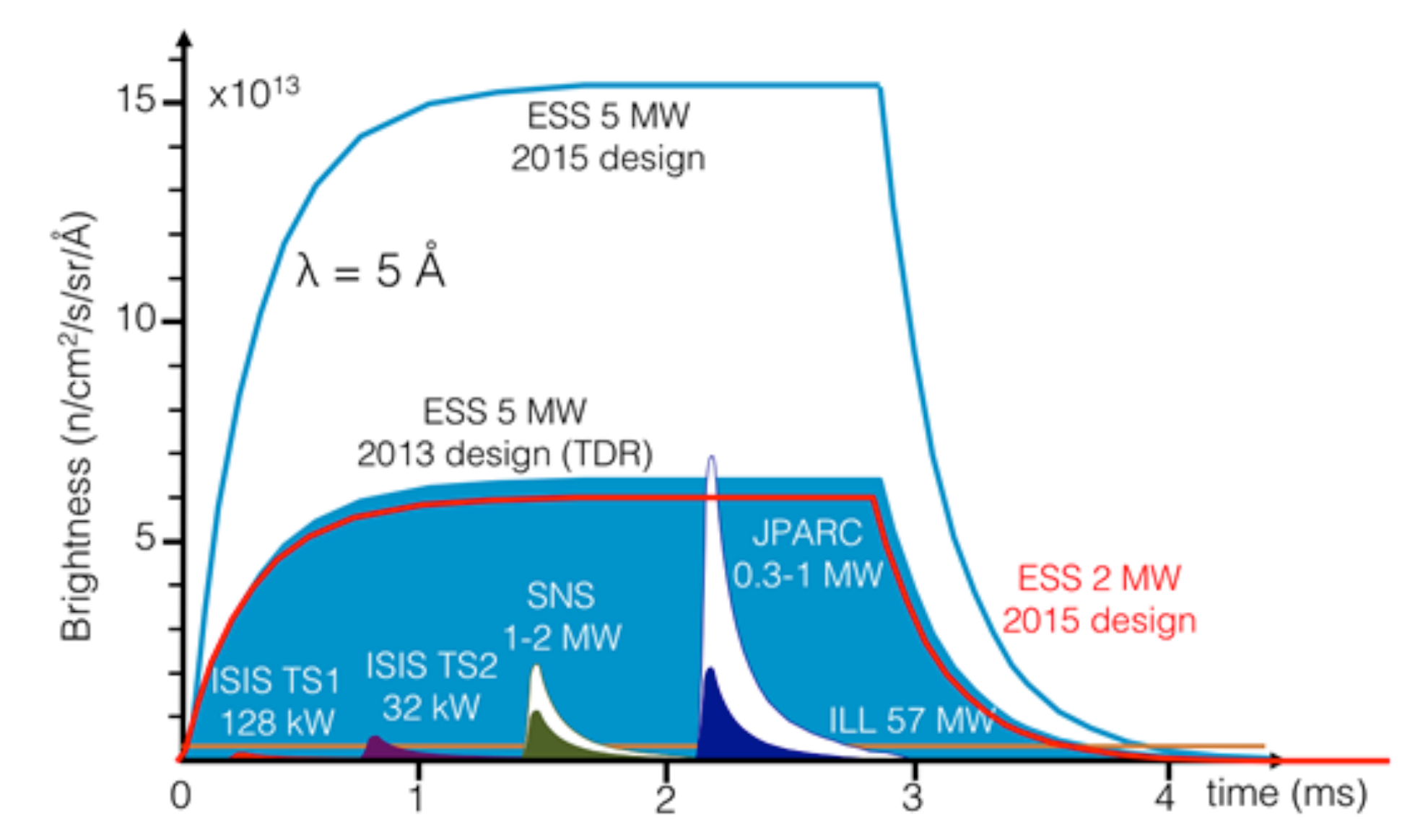}
\caption{\label{pulse-ess} \footnotesize Single pulse source brightness as a function of time at a given wavelength compared between several neutron sources: ESS, ILL, SNS, J-PARC and ISIS. Brightness at ESS is shown at both 5 MW and 2 MW accelerator power.}
\end{figure}

Neutron sources are low-brilliance particle sources compared to the electron or photon sources. The reactor technology, developed in the 1940s and 1950s, reached a plateau in neutron performance in the 1970s with the construction of ILL~\cite{ILL} in France and HFIR~\cite{HRIF} in the United States of America. Already in the late 40's the neutron production in spallation reactions has been observed, opening the opportunity to produce brighter neutron beams than reactor-based facility~\cite{Spallation-physics}. Currently high-flux spallation sources include ISIS~\cite{ISIS}, SNS~\cite{SNS} and J-PARC~\cite{JPARC}. 
\\ The European Spallation Source (ESS) will be the brightest neutron source among the others, thanks to the long pulse~\cite{mezei-longpulse,MEZEI-longpulse1,SCHOBER-longpulse} and low repetition rate time structure of the proton pulses. This unique characteristic will make ESS the ideal source for long-wavelength neutrons, with energies in the range of about 0.1 meV to 50 meV ($30 \AA -1\AA$). Existing pulsed sources deliver short pulses, because the focus at the time of their construction was on higher energy neutrons, 100 - 1000 meV ($1\AA - 0.3 \AA$). Current trends in condensed matter sciences (soft matter, magnetism, life science, engineering and chemistry materials) point up the importance of slow dynamics and large-scale fluctuations of complex systems. ESS aims to follow these trends by focusing on cold and thermal neutrons, exploiting a long-pulse time structure, as shown in figure~\ref{pulse-ess}. The neutron energy range provided by ESS will hold the same information of both hot neutron beams delivered by reactor-based sources and of short-pulse accelerator-based spallation sources~\cite{ESS_TDR,ESS-design}.

ESS will be at least 5 times brighter than the world's leading neutron sources. High fluxes will allow faster measurements, increased use of polarized neutrons, detection of weaker signals and measurements of smaller samples. Thanks to the high brightness many investigations, that are limited today, will become possible. Along with that, the neutron beams will be delivered in an unique time structure, with long pulses (2.86 ms) at low frequency (14 Hz), as shown figure~\ref{pulse-ess}, which leads to high flux. Indeed, if we denote the peak flux as $\phi_p (\lambda)$ as a function of the neutron wavelength and the pulse length as $\delta t$, the relation for the time average flux $\phi_{av}$ can be derived as described in~\cite{MEZEI-longpulse1}:

\begin{equation}
\phi_{av} (\lambda) = \nu \; \delta t \; \phi_p (\lambda)
\label{flux-pulse}
\end{equation}

where $\nu$ is the repetition rate. This structure will open the possibility of using long-wavelength neutrons with an unprecedented brightness. This feature will allow instruments to achieve a wider dynamic range, bispectral beams and tunable resolution as needed. 
\\ The investigations of real-world sample in real-world conditions often require demanding or extreme conditions, i.e., high magnetic fields, low or high temperatures and pressures. These conditions can be achieved through massive sample environment set-ups that limit the available volume for the sample itself. In order to investigate smaller samples, larger intensity is required.  
\\ ESS with the high brightness, long pulses structure and the real world capabilities, is of crucial importance to push forward the present limitation in the neutron scattering science, allowing a wider and a clearer investigation of the structure and dynamics of materials.

\section{Scientific case}\label{scicase}

\subsubsection{Soft condensed matter}

Soft condensed matter is an interdisciplinary area of research regarding system that respond to weak external stress or thermal fluctuations. It includes the study of polymers, surfactants, gels, foams and many other biological materials. During the last decades of the twentieth century, neutron scattering has played an important role in this area, based on the fact that neutrons scatter differently from hydrogen than from deuterium. This distinguishability, unique to neutrons, enable to label various part of soft matter systems so that their structure and dynamics can be investigated. Moreover, the cold neutrons provide simultaneous access to the relevant length (micro and mesoscopic scale) and time scales (10$^{-9}-$1 s) useful to soft matter research. 
\\Soft matter can be easily deformed through applications of weak external stresses or thermal fluctuations, and therefore, a non-invasive characterization is mandatory. It is important to understand the equilibrium and non-equilibrium properties of soft condensed matter not only in a well defined model system, but also in everyday life and industrial applications. The typical techniques to study soft matter are small angle neutron scattering (SANS), reflectometry, neutron spin-echo spectroscopy and quasi-elastic neutron spectroscopy. In order to make progress on this front, it is necessary to probe smaller volumes within larger and complex samples, over a wide range of length scales and with high temporal resolution.
\\ For example, time-resolved studies are needed to detect how surfactant molecules self-assemble into micelles and how such structures transform into different shape~\cite{soft_matter1, soft_matter2}. In order to be able to detect transformations on the millisecond to second time scale, a high neutron flux is required. Such study will benefit from the high flux using a polarized SANS instrument and a broadband small sample SANS instrument, able to probe small gauge volumes and cover a broad range of scattering vectors. The $Q$-range capability offered by this instrument can be exploited to investigate polymer-polymer interactions. In polymers the atomic motion define, indeed, the overall structural configuration and the macroscopic properties. At present, it is difficult to investigate dynamics at all relevant length scales. The availability of high flux in the instruments will open the possibility to study new phenomena and new materials across various length scales.
\\ As already mentioned, time-dependent surface processes and reactions on the millisecond to second time scale are important in a range of soft materials. For example, neutron reflectivity provides unique information about surface and interface structures. The actual experiments are limited by the strong attenuation of neutrons in case of liquids. The expected increasing of the flux will enable a reduction in the experimental path length, allowing time-resolved experiments in a broad $Q$-range available~\cite{INSTR_FREIA}. 
\\ Neutron reflectometry and grazing incidence SANS (GISANS) can be also used to probe thin films in which active layers are often buried deep within the system structure. Typically these samples are available only in small volumes, reducing the beam size will allow more widespread investigation, but in the same way, the intensity of the signals is very low and long time measurements are mandatory. High-intensity source and improved instrument design would permit studies of very small samples with dimensions of a millimetre or less~\cite{INSTR_ESTIA}.  

\subsubsection{Magnetic end electronic mechanisms}

The investigation of magnetic and electronic phenomena is a crucial activity for both basic and applied research. It represents, indeed, a key point to improve fundamental knowledge about quantum matter and identifies the static and dynamic properties in a broad range of materials, i.e., magnetic hetero-structure, molecular magnet and magnetic nano-particles that can be used in advanced technologies devices. The long pulse flexibility of ESS will allow to use the high neutron flux, in order to carry out higher precision measurements, and providing simultaneous coverage of length and time scales. The possibility to exploit the polarized neutron scattering with an high neutron flux, will provide quantitative information unavailable with any other measurement technique.  
\\ The direct microscopic information obtained by using neutron spin as a quantitative probe makes neutron scattering very useful for benchmarking theoretical models against experimental results~\cite{ESS_TDR}. The energy scale of many of these magnetic and electronic phenomena is in the sub-meV to tens of meV energy range. In this energy region, high resolution and precision is required to untangle the competing interactions that involve much of the physics of these kind of materials. Low energy cold neutron and the long pulse time structure of ESS will be a valuable tool in the investigation of low energy magnetic and electronic states. Thanks to the brightness of the new source will be possible to study novel materials that are synthesized in quantities too small for present-day neutron investigations. 
\\ Neutron scattering has been used to study changes in the symmetry of spin dynamics upon entry into the superconducting state~\cite{magnetic1,magnetic2}. The possibility to measure four dimensional spatial and energy neutron scattering maps, using chopper spectrometers with position sensitive detectors, simplifies probing the excitation spectrum of high temperature superconductors~\cite{magnetic1}. However, at present, the neutron spectroscopy is limited to specific energy regions, cold or thermal neutron scattering, while complex phenomena affect dynamic behaviour over broad scales. Novel instrumentation such as the bispectral chopper spectrometer will provide extended energy range in this field~\cite{ESS_TDR}.
\\ Novel concepts for electronic devices are based on thin film and multi-layer structures and oxide heterostructures. The latter is made up of metal oxide materials that shows a broad range of physical phenomena. These phenomena include large spin polarization, magnetoresistance, electronic phase ordering and charge orbital spin ordering~\cite{magnetic-ref1,magnetic-ref2,magnetic-ref3}. The interface and surface effects in thin film structure are the responsible for such phenomena, and they are probed using polarized neutron reflectometry complementary to X-ray investigations. New instrumentation, e.g., the vertical reflectometer~\cite{INSTR_ESTIA,INSTR_ESTIA1,INSTR_ESTIA2} at ESS, can profit from the high brightness available to measure the very weak scattering signals that impede the actual investigations.
\\ The miniaturization of electronic and biological devices to an extreme degree can be realized with molecular magnets, nanoparticles and excitons in confined systems. Molecular magnets are new classes of magnetic materials at the nano-scale. Their function depends on microscopic correlations revealed via single crystal neutron studies. Recent works have performed on molecular magnets exploiting inelastic neutron scattering~\cite{magnetic-ref4}. These experiments are challenging because these materials are usually available in small single crystals, leading to a limited scattering signals. The optimization of the instruments for small samples, i.e., bispectral chopper spectrometer, cold-crystal analyser spectrometer, will be a key component together with the need of high brightness neutron source. 
A variety of these nano-sized structures can be observed using vertical reflectometer or general purpose polarised SANS instrument while dynamics can be probed using cold chopper spectrometer~\cite{ESS_TDR}. The detector technology has to be developed in parallel, in order to enable mapping of wide \textit{Q}-range with high energy resolution and high spatial resolution based on the specific instrument. 

\subsubsection{Life science}

The mechanistic study of complex biological systems at atomic and molecular scale is of high impact on understanding modern life science, from healthy ageing to food security. Neutrons have many advantages as a probe for structure and dynamics in biological systems. The biological samples are typically hydrogen-rich materials, and they can be easily damaged by X-ray or electron-scattering techniques. The capability of neutrons in isotope labelling and non-destructive mechanism can be exploited for these kind of investigations. Nowadays, the most limitations are given by the relatively low source brightness and the number of appropriate neutron instruments for biological samples. A bright source, like ESS, will allow the investigation of small samples, moreover the long neutron pulses are well suited for several neutron techniques relevant for biological systems.
\\ A macromolecular diffractometer (NMX) addresses the challenge in studying systems with smaller crystals or larger unit cell volumes. Neutron crystallography offers, indeed, the possibility to determine the positions of hydrogen atoms in the complex structure of proteins. The determination of these hydrogen positions is crucial in many biological processes, i.e., enzyme mechanisms protein-ligand interactions or proton transport across membranes, not only for chemical interest but for applications in pharmacy and industrial engineering. The instrument length allows a wavelength frame of $\sim$ 1.8 \AA\, and, thanks to the Time of Flight method, it is possible to disentangle the signal from the background while making full use of the integrated intensity of the long pulse. An adequate and flexible detector system must be employed to fulfil the sub-millimetre spatial resolution requirements. The possibility to change both the sample-detector distance and the 2$\theta$ angle at the center of the detector allows the reflections from larger unit cells to be better separated on the detector plane. A suitable technology is under development at ESS~\cite{DET_doro1}.
\\ The large multi-component structure involved in the biological function of most macromolecules are usually difficult to crystallize. However, such systems can be clarified in solution using small angle neutron scattering in combination with selective deuteration. With this method it is possible to discern separate components and their motions, by changing the isotopic composition of the individual components of the complex and the solvent. Thus, it is possible to understand the complicated interactions between biological macromolecules in solution. High flux will open opportunities to perform time-resolved studies with macromolecular complexes, and a dedicated broadband small sample SANS instrument will be suitable for such experiments. The instrument will enable single pulse scattering measurements in a broad wavelength band, because of the small source-sample distance, allowing simultaneous $Q$-range.
\\ In addition to the broadband small sample SANS instrument, an horizontal reflectometer can be useful for these kind of investigations, i.e., the study of the structure across membranes. Biological membranes play a key role in mediate or regulate  many cellular functions. With neutron reflectometry and SANS, deuterium labelling allows to obtain information about internal structure, such as location, dimensions and orientation of membrane proteins under physiologically relevant conditions. The advantages of exploiting high fluxes will be, for example, to study smaller samples with better time-resolution, and elucidate biochemical mechanisms.

\subsubsection{Energy research}

Energy materials such as polymer solar cells, batteries, fuel cells and thermoelectric materials, are rich of light atoms, i.e., oxygen, lithium and hydrogen. The high penetration depth of the neutron can be used to investigate the structural and dynamic processes responsible for relevant macroscopic properties in these materials. The advantage of isotopic contrast provided by neutron-based techniques and their sensitivity to light atoms are fundamental characteristic for carrying mechanistic studies, which are difficult to perform with X-ray scattering techniques.   
\\ In the case of fuel cells, of great interest in a number of energy end-use sectors, from electric vehicles to power plants, the main challenge is to optimize their energy efficiency. A tunable resolution in space and time is required to have a better understanding of the relationship between proton diffusion and chemical structure. At ESS will be possible to exploit the long pulse structure to make versatile instruments, i.e., vertical reflectometer, thermal powder diffractometer, bispectral powder diffractometer, that can cover and extend spatial domains needed for the efficiency optimization of fuel cells. 
\\ Batteries are the most widely available solution to energy storage in a large range of applications. Many of these devices are based on the ion exchange and the ionic conductivity properties of lithium and hydrogen, suitable features for using neutron-based techniques. Time-dependent studies can be performed using pulsed monochromatic powder diffractometer and multi-purpose imaging instruments, together with the bispectral powder diffractometer. Diffusion mechanism can be directly investigated by quasi-elastic neutron scattering on the weakly scattering lithium ions. This kind of experiments can be performed using chopper spectrometers, while dynamic properties can be probed using neutron spin-echo spectrometer. 

\subsubsection{Geosciences and engineering materials}

The use of neutron scattering is widely used for engineering materials and geosciences, from application in industrial research to analysis of geomaterials. Neutron scattering can provide information about the composition, microstructure and stresses in materials, in ambient conditions and as a function of different external parameters. The deep penetration capability of neutron is of primary importance allowing to carry out non-destructive investigations of inner structure, properties and mechanisms locally within bulk materials and complex systems with high precision and sensitivity. One of the most challenging features that characterize these kind of investigations is the complex sample environment which reproduce realistic or extreme operating environments of the samples. Typically they are multi-component systems with sensitivities to different physical and thermal interactions, in contrast to the well defined and isolated samples used for fundamental studies. A key component of such research will be the possibility to combine several methods as diffraction, SANS and imaging. Moreover the high brightness of the neutron beam will enable to analyse smaller samples respect to the existing facilities, or to analyse large samples with larger spatial resolution under more extreme conditions, over shorter time scales and with higher precision.
\\ Higher spatial resolution, down to the sub-millimetre region, is needed for further progress in monitoring residual stress, i.e., high pressure and high temperatures conditions, and for detecting diffusion and reactions at interfaces in metals. High intensity will allow \textit{in situ} real time experiments, and investigation of fast, time-dependent phenomena with isotope sensitivity. The suitable instruments are multi-purpose imaging and polarized SANS instruments together with diffractometers, which allow the study of structure property relationships in mineral phases as well as the physical properties of multi-components fluids.

\section{Instruments design}

The increased performance expected for the new neutron sources and the resulting extent of scientific studies drive the development of design and operation for the instruments as well. As pointed out in the previous section, many of the problems that ESS will face require the measurement of structures and/or dynamics over several length or time scales. It is crucial to be able to adapt the resolution and bandwidth of the measurement to the sample. Such flexibility of instrument performance takes advantage of the long-pulse concept foreseen at ESS. Indeed, the resolution can be set by controlling the opening time of a pulse-shaping chopper, instead of hardwiring the resolution into the moderator line shape as it is done for instruments at a short-pulse source. Furthermore the full time-frame is accessible by using the repetition-rate and the wavelength-frame multiplication, tuning the bandwidth of the measurement according to the requirements of each experiment. The instruments must be designed to probe smaller sample volumes, by combining the high flux of the source with advanced focusing optics. ESS's source brightness and time structure are well suited to cold neutrons, providing high flux and a wide dynamic range. As shown in section~\ref{scicase}, cold neutrons are required to deeper investigate structure and dynamics of soft-matter and biological systems. Instruments using a bispectral extraction system will profit of these characteristics by accessing a dynamic range in wavelength, which is unattainable nowadays. Moreover, thanks to the high flux, instruments using the neutron polarization technique can be widely exploited leading to an enhance in several applications throughout physics, soft matter, chemistry and biology.   
\\ The unique property of the pulse provided by ESS (figure~\ref{pulse-ess}) drives the instruments design in an intermediate stage between those on a short-pulse and those on a continuous source requiring novel design. Short-pulse instruments benefit from the high peak brightness, but the time-widths are limited by the choice of the moderator. Time-of-Flight instruments at continuous sources can freely choose their time structures using chopper systems, but the peak brightness is significantly lower. The designs of instruments at ESS will be less limited by the time-structure of the long-pulse source compared with instruments at short-pulse sources. They can benefit from the higher peak brightness, while retaining much of the flexibility of continuous-source instruments. 
\\ The requirements for good wavelength resolution and bandwidth are a large part of the instruments optimizations by using appropriate combinations of instruments lengths and chopper systems. Indeed, the pulse length at the source represents the uncertainty in the emission time of neutrons. This can be reduced comparing their Time-of-Flight by making the instrument longer. The length of the instruments affect also the bandwidth, which is defined by the longest wavelength that can be measured for a given pulse, before it overlaps in time with the shortest wavelength of the following pulse. An alternative method of improving the wavelength resolution is the use of pulse-shaping choppers.  
\\ Based on the choice of one or the other option it is possible to distinguish two categories of instruments: large pulse width and small pulse width, depending on the required pulse width compared to the intrinsic length ($\tau =2.86\,$ms) of the neutron pulse. SANS, spin-echo and macromolecular crystallography instruments are well suited to the long-pulse time structure. They can use the full ESS pulse width and thus profit from the high peak and time-average brightness. Single-crystal and powder diffractometer, crystal-analyser spectrometer, reflectometers, backscattering and imaging instruments employ a pulse-shaping chopper to reduce the length of the source pulse in order to achieve the required wavelength resolution. Note that this makes them very flexible, because, if needed, they can gain flux by relaxing resolution. 
\\The pulse-shape chopper is placed as close as possible to the source (6 m in the case of ESS), when the instrument length after the chopper reaches 150 m, the bandwidth is sufficient to fill the full time frame between  following pulses. Instruments shorter that this length need wavelength-frame multiplication (WFM) in order to fill the time frame. For some instruments, it is combined with a pulse shaping double chopper, which defines a constant wavelength resolution, and a set of frame overlap choppers that prevent spurious neutrons from reaching the detector~\cite{WFM_Nekrassov}. 
There are also chopper spectrometers ranging in length from 25 m to 156 m. They will employ the repetition-rate multiplication (RRM)~\cite{RRM_Mezei} to fully use the long repetition period of ESS. The idea is to select several monochromatic wavelengths from each source pulse, so that the repetition rate of the instrument increases. In order to select different energies it is possible to use different frame lengths for each one of them, by combining several chopper systems. 
 
A short overview of the instruments class is shown below, pointing out the characteristics of each one, this part is mainly based on the ESS TDR~\cite{ESS_TDR} suite of 22 instruments, 15 of which have been chosen for the initial construction.

\subparagraph{Small angle neutron scattering:} 

SANS provides access to the largest length scales possible with neutron scattering. SANS instruments are white-beam instruments that require large pulse-width. Compared with the present-day SANS instruments the counting rate for the ESS instrument will be higher by an order of magnitude, as well as the $Q$-range covered in a single measurement. 
The general-purpose polarized SANS instrument simultaneously will cover the $Q$-range of current conventional SANS instruments, 10$^{-3}$ \AA$^{-1}$ to 0.1\AA$^{-1}$, using multiple detector banks. The instrument will have 
flexible resolution and bandwidth and will make available the option of reaching very low $Q$. While a broadband small sample SANS instrument will be optimized for very high counting rates and smaller samples, because of its shorter length it is a lower-resolution instrument, but it can cover a very large $Q$-range in a single measurement.

\subparagraph{Neutron imaging:} 

This is a real-space technique, by looking at the transmitted beam it is possible to examine the inner structure of complex systems. The imaging beam-line will push spatial resolution down to the micron range thanks to brightness of ESS, in parallel with the development of an adequate detector technology. The instrument concept takes full advantage
of the flexibility made possible by the ESS time structure, allowing wavelength resolution, bandwidth and collimation to be tuned for each application.

\subparagraph{Neutron reflectometry:} 

Neutron reflectometry is used to probe the structure of surfaces and interfaces in the \AA\ to micron range. The horizontal reflectometer uses an inclined beam to measure horizontal surfaces. It is possible to cover the full $Q$-range of interest without moving the sample. The instrument is optimized especially for very fast measurements while keeping a good resolution. The vertical reflectometer is a versatile instrument for measuring solid interfaces with high resolution and on small samples. A novel double-elliptical mirror guide will allow flexible arrangement of the beam divergence leading to fast measurements with high flux intensity. In order to fulfil these demanding requirements an appropriate detection system is mandatory.

\subparagraph{Powder diffraction:} 

ESS's pulse structure makes possible very wide dynamic range, while the long-pulse flexibility allows the resolution to be tuned with respect to the requirements of each experiment by adjusting the opening time of the pulse-shaping chopper. The thermal powder diffractometer will cover the \textit{in situ} processing science case and crystallography for thermal neutron up to a $Q \sim 25 \ \AA^{-1}$. With a tunable wavelength resolution, which in the best case can achieve a resolution < 0.01\%. The bispectral power diffractometer can access longer wavelengths and \textit{in situ} experiments in the intermediate $Q$ range ($Q_{max} < 13\ \AA^{-1}$). The pulsed monochromatic powder diffractometer will use a set of crystal monochromators with access to a wide range of angles to tune the $Q$ range, flux and instrument resolution.

\subparagraph{Single crystal diffraction:} 

The single-crystal magnetism diffractometer will exploit a polarized incoming beam, optimized for the determination of complex magnetic structure, spin density distributions, diffuse scattering. By taking advantage of the high flexibility and low background long-pulse structure, it will be possible to measure very small samples. Moreover, the choice of a 156 m long instrument will allow to access the full long source pulse. The thermal spectrum of the pulse can be adapted using the pulse-shaping chopper to provide a wavelength-dependent pulse length. The macromolecular diffractometer is dedicated to biological crystallography and it is optimized for sub-millimetre single-crystal samples with large unit cells.

\subparagraph{Time-of-Flight and crystal spectroscopy:} 

Time-of-flight and crystal spectroscopy is carried out by two types of instruments: chopper spectrometers and crystal-analyser instruments. All the chopper spectrometers are designed for optimized repetition-rate multiplication (RRM) using multiplexing chopper systems. The range of incident energies will be adapted to the range of time scales that need to be covered, and the spacing between adjacent energies will be set in order to cover the width of the inelastic scattering. The ESS peak brightness and the use of the RRM system will allow a total flux on the sample up to more than an order of magnitude larger than what is possible today. Moreover, a wider dynamic range in both $Q$ and energy can be simultaneously probed. The cold chopper spectrometer will focus in the cold neutron energy region with an energy resolution of 50 - 200 $\mu$eV. It is a low background and high resolution instrument well suited for probing weak signals and collective excitations. The background from the prompt pulse neutrons will be minimized by avoiding direct line from the moderator through the use of an S-shaped ballistic guide which provide a spatially homogeneous beam distribution at the sample position and a very clean short-wavelength cut-off. The bispectral chopper spectrometer will provide a simultaneous measuring technique that covers dynamics over a wide energy range of 0.1 - 100 meV. The bi-spectral extraction is possible either via the use of a supermirror switch to reflect cold neutrons into the guide, or by pointing the guide to the edge of the cold moderator to access a warmer spectrum. It is an instrument optimized for quasi-elastic scattering on small samples (mm$^3$ size), allowing to cover more than three order of magnitude in fluctuation times in a single measurement. The thermal chopper spectrometer is designed for broad mapping of thermal excitations with high energy and spatial resolution in hard condensed matter. Via RRM will be possible to probe in a single measurement an energy range between approximately 11 and 160 meV. The beam optics are tunable, allowing efficient focusing for small samples as well high Q-resolution single-crystal measurements.
\\ The cold crystal-analyser spectrometer is a very high intensity, indirect geometry instrument covering  wide $Q$-range (0.04 -7.8 $\AA^{-1}$). It is especially suited for studies of excitation spectra as function of wavevector transfer for several applications in geoscience and materials studying. The instrument will use graphite crystals as analysers covering an energy range from 2.5 meV and 8 meV.  

\subparagraph{Backscattering:} 

The backscattering spectrometer will provide access to a unique combination of high energy resolution, intermediate $Q$ and large dynamic range. It will employ Si analyses crystals arranged in near-backscattering geometry and a flexible chopper system to match the source time structure, in order to allow a continuous variation of the energy resolution from 2 to 300 $\mu$eV with a variable dynamic range down to $\sim$1 meV. This combination results in an order of magnitude performance increase compared to currently world-leading backscattering instruments. 

\subparagraph{Spin-echo spectroscopy:} 

Spin-echo spectroscopy reaches the longest time scales accessible with neutrons, allowing the measurement of polymer and protein dynamics, complex soft matter systems and magnetism. The high-resolution neutron spin-echo spectrometer uses the neutron spin to encode very small changes in the velocity of the neutron, thus decoupling the energy resolution and intensity. The instrument covers a broad energy, 10 ps -1 $\mu$s, and $Q$-range, 0.01 - 2 $\AA^{-1}$. It is optimised to provide maximum flux on the sample, reaching higher Q-values. Due to the large wavelength bandwidth a very large time and spatial range can be covered within one experiment. Compared with this instrument, the wide-angle spin echo spectrometer is a medium energy resolution instrument for large scattering angles and with a large detector solid angle. It will provide measurements for a very broad simultaneous $Q$-range, up to 4 $\AA^{-1}$, with a very high count rate. The spectrometer will have a longer moderator-to-sample distance compared to the high-resolution spin-echo instrument. Note that the instrument will have a smaller usable wavelength band, but it provides a better wavelength, and hence Q-resolution, improving performance for studies of crystalline samples.
Compared to currently world-leading instruments, it will lead to improvement in performance of up to an order of magnitude, especially at the shorter wavelength. A gain in performance will arise from the use of a wide detector bank as well.

\section{Detector technologies}

In this chapter a short summary of how the neutron scattering science will benefit from the operation of new generation, high-intensity neutron sources, and in particular from the construction of ESS is presented. The enhancement of available flux leads to the need to develop new instruments design able to exploit it, in order to further extend scientific investigations. The increase in performance demands represents a challenge not only for the instruments, but for the neutron detector technology as well. 
\\Besides all the possible extended investigations in neutron scattering, some of which are discussed above, the availability of new instrumentation opens the possibility to new science, which is outside the core science case, and enabled by the possibility of better measurements. Thus, better instrumentation is a crucial driver for novel scientific fields. The improvements in resolution, count rate and signal-to-background, achievable with the new instrumentations, also drive the possibility to new investigations that could not be done before. 
\\ Several detector technologies can be used with respect to the different techniques, based on the need for each one: a boron-based gaseous detector both for perpendicular of parallel neutron incidence geometry, lithium-based scintillator detectors, for instance for powder diffraction of small angle scattering, and gadolinium-based detectors where high resolution is needed and gamma rejection can be relaxed~\cite{HE3S_kirstein}.
\\A summary of the improvement factors required for the detectors for some neutron scattering techniques is given, considering the state-of-the-art for each feature as one. For instance in neutron reflectometry the actual spatial resolution is 2-3 mm, this is set as one, and the requirements are about 0.5-1.5 mm, i.e., 2-3 times larger compared with the present achievable resolution. The values are reported in table~\ref{tablech3} and it gives an idea of the challenges to face in this field.

\begin{table}[htbp]
\centering
\caption{\label{tablech3} \footnotesize Detectors improvement factors for different neutron scattering techniques: Neutron Reflectometry, Neutron Spectroscopy, Neutron Diffraction and Small Angle Neutron Scattering (SANS).}
\smallskip
\begin{tabular}{|c|c|c|}
\hline
 &Spatial Resolution & Count rate\\
\hline
Neutron Reflectometry &  2-3 & 100-1000  \\
\hline
Neutron Spectroscopy & 1-2 & 10 \\
\hline
Neutron Diffraction & 3-4 & 10  \\
\hline
SANS & 2 &10 -100 \\
\hline
\end{tabular}
\end{table}   

In the following chapters the development and the characterization of two different detector technologies is shown. The first detector presented is designed to fulfil the requirements of high spatial resolution and counting rate capability for neutron reflectometry. The second prototype is in an earlier stage of development, but it is proposed for large area applications with sub-millimetre resolution detection at high rate.

\chapter{Boron-10-based gaseous detector for neutron reflectometry}\label{chapter4}

The chapter provides the description of a Boron-10-based gaseous thermal neutron detector, the Multi-Blade detector technology~\cite{MIO_MyThesis,MIO_MB2014,MIO_MB2017,MIO_HERE}, developed to face the challenge arising in neutron reflectometry at neutron sources. The Multi-Blade detector has been designed according to the requirements given by the reflectometers at the European Spallation Source (ESS) in Sweden. 
\\ As highlighted in the previous chapter, neutron scattering science is nowadays increasing its instrument power expanding the neutron scattering technique to faster kinetics, smaller and more complex structures. Neutron reflectometers are the most challenging instruments in terms of detector counting rate requirements. The peak brightness of ESS will be without precedent but present detectors for reflectometry are at the limit and already inhibit instrument performance at today's neutron sources to some extent.
\\ A full characterization of the detector is reported both on the technical aspects and scientific measurements. This part of the work is mainly based on a campaign of measurements performed at the CRISP reflectometer~\cite{CRISP1} at ISIS (Science $\&$ Technology Facilities Council in UK~\cite{ISIS}). Together with this are presented some results of the characterization carried out at the Budapest Neutron Centre (BNC)~\cite{FAC_BNC} and at the Source Testing Facility (STF)~\cite{SF2,SF1} at the Lund University in Sweden. 
\\ My contribution involves several aspects of the project, from the assembling of the prototype used to perform these tests, to the design of improvements for the next device as more technical tasks. From the scientific point of view, I carried out the experiments mentioned above, I performed the data analysis and I wrote part of the paper published about the technical characterization of the detector~\cite{MIO_MB16CRISP_jinst} and I fully wrote the paper regarding the neutron reflectometry measurements on standard samples~\cite{MIO_ScientificMBcrisp}.

\section{Introduction}

The Multi-Blade is a gaseous thermal neutron detector based on $\mathrm{^{10}B}$ solid converters for neutron reflectometers. The design of the Multi-Blade detector has been driven by the requirements set by the two reflectometers foreseen at the European Spallation Source (ESS~\cite{ESS,ESS_TDR,ESS-design}) in Sweden: FREIA~\cite{INSTR_FREIA,INSTR_FREIA2} (horizontal reflectometer) and ESTIA~\cite{INSTR_ESTIA,INSTR_ESTIA1,INSTR_ESTIA2} (vertical reflectometer). 
\\ Neutron reflectometers are a challenging class of instruments in terms of detector requirements. The need for better performance in addition to the scarcity of $\mathrm{^3He}$~\cite{HE3S_kouzes,HE3S_kouzes2} are both driving the developments of new detector technologies for neutron scattering science in general. The key detector requirements for neutron reflectometers are the counting rate capability and the spatial resolution. These are essential features for the detectors at the ESS reflectometers, whereby the expected instantaneous local flux is about $10^{5}$/s/mm$^2$~\cite{ESS_TDR,DET_rates,HE3S_kirstein,HE3S_cooper} and a sub-mm spatial resolution (Full-Width-Half-Maximum, FWHM) is required.
\\ The state-of-the-art detector technology is reaching fundamental limits. The spatial resolution is limited to approximately $2 \times 8\,$mm$^2$, with a maximum counting rate capability of $40\,$kHz integrated over the whole beam intensity on the detector~\cite{INSTR_FIGARO}. The latter corresponds to a few hundred Hz/mm$^2$. Note that at the incident rates available at existing sources (pulsed and reactors) detectors already saturate. 
\\ At current facilities the time resolution for kinetic studies is limited by the available flux and by the detector performance. In order to open the possibility of sub-second kinetic studies, a new instrument layout, which exploits a higher neutron flux, has been presented for reactor sources~\cite{R_Cubitt1,R_Cubitt2} and it requires high spatial resolution detectors. Therefore, a more performing detector technology is needed, not only for the ESS reflectometers, but also at existing reflectometers at current facilities. The Multi-Blade detector is designed to fulfil these challenging requirements. 
\\ Table~\ref{tab1} summarizes the detector requirements for the two ESS reflectometers. 
\begin{table}[htbp]
\centering
\caption{\label{tab1} \footnotesize Detector requirements for neutron reflectometers at ESS.}
\smallskip
\begin{tabular}{|l|l|l|}
\hline
\hline
  & FREIA~\cite{INSTR_FREIA,INSTR_FREIA2} & ESTIA~\cite{INSTR_ESTIA,INSTR_ESTIA1}  \\        
\hline
\hline
wavelength range (\AA) &  2.5 - 12  & 4 - 10 \\
\hline
minimum detection efficiency  &  >40\% at 2.5\AA  & >45\% at 4\AA \\
\hline
sample-detector distance (m) & 3 & 4 \\
\hline
instantaneous local rate & & \\
on detector ($kHz/mm^2$) & $10$  & $30$ \\
\hline
sensitive area: x-direction (horizontal) ($mm$)  & 300 & 500 \\
sensitive area: y-direction (vertical) ($mm$)  & 300 & 250 \\
\hline
spatial resolution (FWHM) x ($mm$)  & 2.5 & 0.5 \\
spatial resolution (FWHM) y ($mm$)  & 0.5 & 4 \\
\hline
uniformity ($\%$)  & 5 & 5 \\
\hline
desired max window scattering  & $10^{-4}$ & $10^{-4}$ \\
 \hline
$\gamma$-ray sensitivity & $<10^{-6}$ & $<10^{-6}$ \\
 \hline
 \hline
\end{tabular}
\end{table}
\\ The Multi-Blade detector has been previously characterized and several demonstrators have been built. It has been shown that this detector technology represents a valid alternative to the state-of-the-art technology for neutron reflectometry instruments that use cold neutrons (2.5-30\AA~\cite{INSTR_FIGARO,INSTR_D17}). Most of the detector requirements have already been fulfilled by the Multi-Blade technology. A spatial resolution of $\approx 0.6\,$mm has been measured together with a detection efficiency ($\approx 44\%$ at the shortest wavelength 2.5\AA). The counting rate capability of this detector has been measured up to $\approx 17\,$kHz/ch limited by the available neutron flux at the source. The gamma-ray sensitivity~\cite{MIO_MB2017,MG_gamma} below $10^{-7}$ has been measured with the Multi-Blade detector. The study on the sensitivity to fast neutrons ($1-10\,$MeV)~\cite{MIO_fastn} has been carried out, for the first time on these kind of devices, at the Source Testing Facility (STF) at the Lund University in Sweden. A dedicated chapter is presented in the manuscript.  
\\ The technical and scientific results obtained during the test performed on the CRISP reflectometer at ISIS are shown in the following sections. The aim of this test was to make a detector technology demonstration on a neutron reflectometry instrument. These tests are crucial to validate the Multi-Blade technology to be installed at the ESS reflectometers. The reflectivity of several reference samples have also been measured with the Multi-blade detector at the instrument.

\section{The Multi-Blade detector}\label{MBtes}

\begin{figure}[htbp]
\centering
\includegraphics[width=.8\textwidth,keepaspectratio]{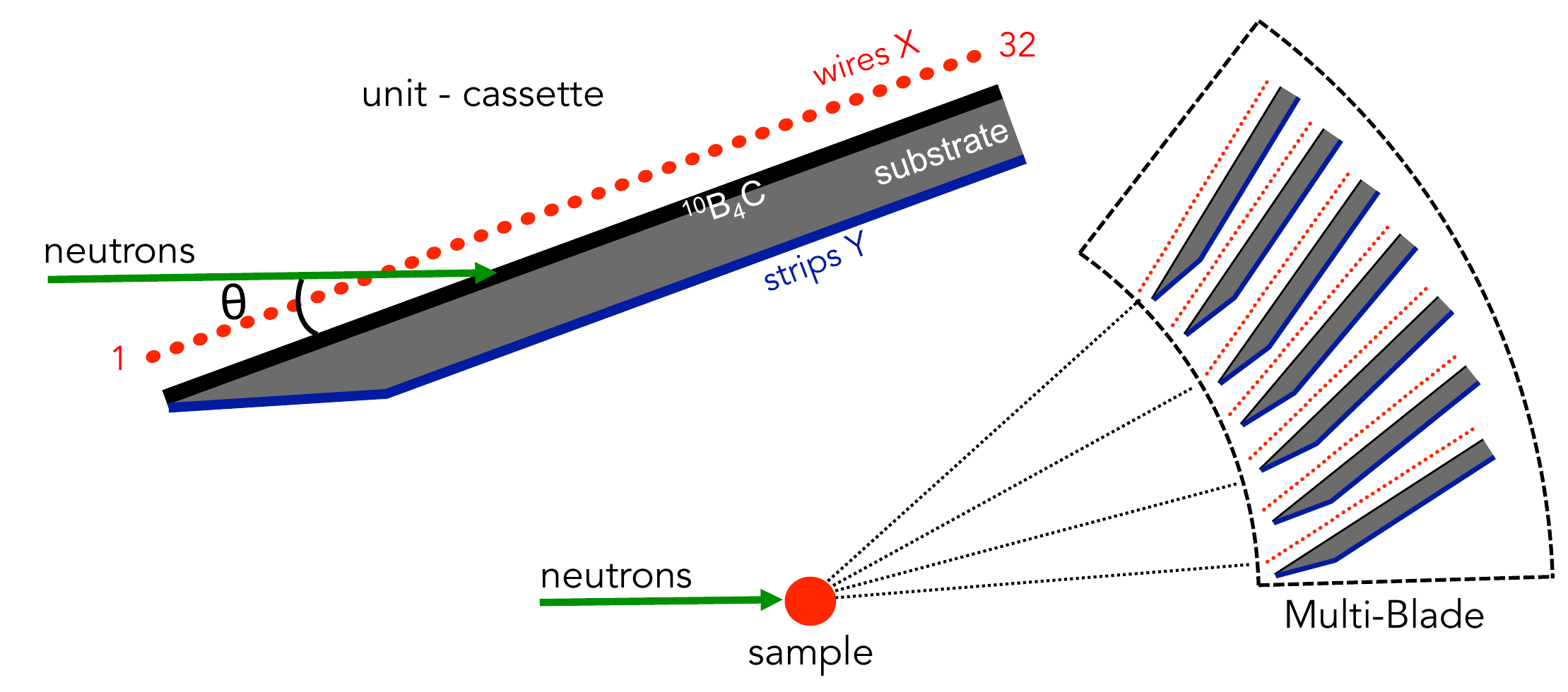}
\caption{\label{fig99} \footnotesize Schematic view of the cross-section of the Multi-Blade detector made up of identical units (cassettes) arranged adjacent to each other. Please note that the scale is exaggerated for ease of viewing. Each cassette holds a $\mathrm{^{10}B_4C}$-layer; the readout is performed through a plane of wires and a plane of strips. Figure from~\cite{MIO_MB2017}.}
\end{figure}

The Multi-Blade is a stack of Multi Wire Proportional Chambers (MWPC) operated at atmospheric pressure with a continuous gas flow ($\mathrm{Ar/CO_2}$ 80/20 mixture). A sketch of the Multi-Blade detector is shown in figure~\ref{fig99}. The Multi-Blade is made up of identical units, the so-called `cassettes'. Each cassette holds a `blade' (a flat substrate coated with $\mathrm{^{10}B_4C}$~\cite{B4C_carina,B4C_carina3,B4C_Schmidt}) and a two-dimensional readout system, which consists of a plane of wires and a plane of strips. The readout of a single converter is performed by the facing anode wire plane which mechanically belongs to the cassette and the strip plane that belongs to the adjacent cassette. Each $\mathrm{^{10}B_4C}$-converter (blade) is inclined at grazing angle ($\beta = 5$ degrees) with respect to the incoming neutron beam. The inclined geometry has two advantages: the neutron flux is shared among more wires with respect to the normal incidence (the counting rate capability is correspondingly increased) and the spatial resolution is similarly improved. Moreover, the use of the $\mathrm{^{10}B_4C}$ conversion layer at an angle also increases the detection efficiency, which is otherwise limited to a few percent at thermal energies for a single converter~\cite{MIO_analyt}. The cassettes are arranged over a circle around the sample and they have some overlap; i.e., each blade makes a small shadow over the adjacent one in order to avoid dead areas.

\begin{figure}[htbp]
\centering
\includegraphics[width=.8\textwidth,keepaspectratio]{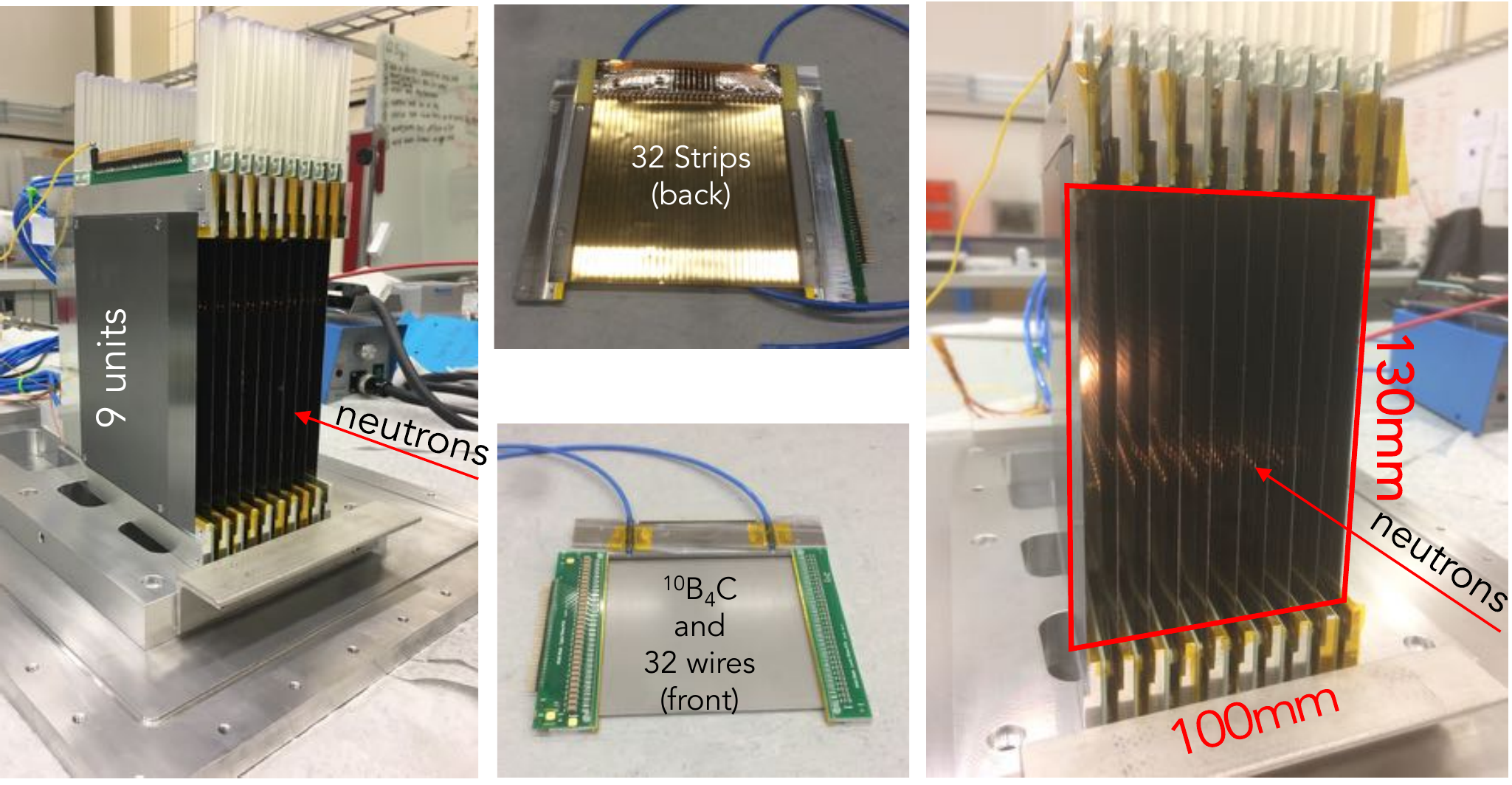}
\caption{\label{fig1} \footnotesize A picture of the Multi-Blade detector with Ti-blades made up of 9 units (cassettes), left and right. A picture of the two sides of a cassettes holding the wires, the blade with the $\mathrm{^{10}B_4C}$ layer and the strips (centre). Figure from~\cite{MIO_MB16CRISP_jinst}.}
\end{figure}

The present detector consists of 9 units (576 channels in total), figure~\ref{fig1} shows a picture of the Multi-Blade detector, and a picture of the front and back of a single cassette, holding the wires and strips. The previous demonstrator employed $2\,$mm thick Al-blades coated with the neutron converter layer ($\mathrm{^{10}B_4C}$) $7.5\,\mu$m thick. It has been shown in~\cite{MIO_MB2017}, that this thickness is needed to absorb $\approx 99\%$ of the neutrons at the shortest wavelength 2.5 \AA\, of the two ESS reflectometers (any neutron of longer wavelength is absorbed with higher probability). For deposition on a single side of the substrate (blade), a deformation of the substrates was observed. In the present detector the Al-blades have been replaced with Titanium (Ti) or Stainless Steel (SS) of thickness of $2\,$mm, and they both show a very good response in terms of mechanical stress, i.e., planarity is not an issue even if the $\mathrm{^{10}B_4C}$ layer is deposited on a single side. Note that the planarity of the substrate is crucial for the uniformity of the electric field of the MWPC and for defining the position of the neutron detection, together with the alignment and resolution. 
 
The coating thickness on the blades of this demonstrator is $4.4\,\mu$m. The detection efficiency is saturated above $3\,\mu$m and any extra film thickness will only help to absorb neutrons. A neutron that crosses the layer and reaches the substrate, can be scattered causing unwanted spurious events within the detector. 
\\ The cassettes are mounted on an array plate which gives the alignment and matches the circle of $4\,$m radius which is needed for ESTIA; this distance is measured from the centre of the sample to the first wire of each cassette (wire no. 1 in Figure~\ref{fig99}). Moreover, the cassettes are not parallel, but the relative angle between two adjacent cassettes is $0.14$ degrees. This arrangement requires the positioning of each cassette on the array plate with a precision of approximately $0.15\,$mm.  
\\ The vessel of the present Multi-Blade detector has a $1\,$mm-thick aluminium entrance window. No neutron shielding was foreseen for such a vessel, further improvements will take this into account. 

\subsection{Front-end and readout electronics}\label{secFEE}

\begin{figure}[htbp]
\centering
\includegraphics[width=.98\textwidth,keepaspectratio]{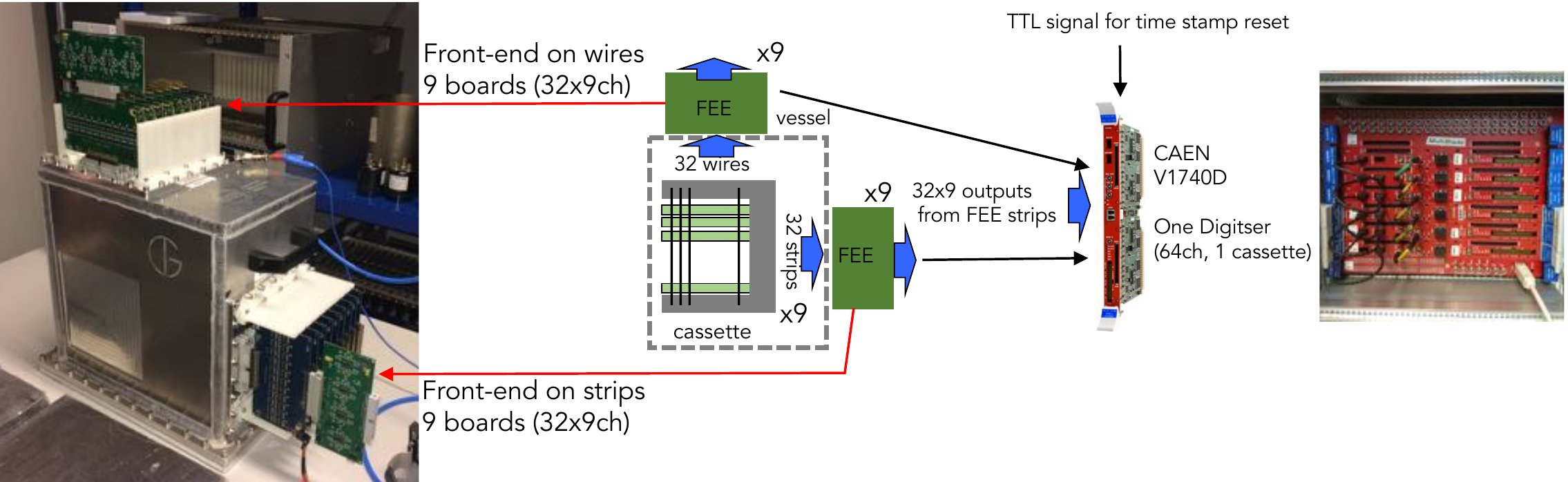}
\caption{\label{fig3} \footnotesize A picture and a sketch of the readout electronics scheme of the Multi-Blade detector. Individual readout boards (32 channels) based on FET are connected to each pane of wires and strips; each board is connected to a CAEN V1740D digitizer which allows the reset of the time-stamp.Figure from~\cite{MIO_MB16CRISP_jinst}.}
\end{figure}

With respect to the previous demonstrator, the new detector does not employ any charge division readout~\cite{DET_chargediv} and each channel (64 per cassette, 32 wires and 32 strips) is read out individually. Each channel is connected to a FET-based charge pre-amplifier and shaper with an approximate gain of $10\,$V/pC and shaping time of $\approx 2\,\mu$s. There is a front-end electronics (FEE) board (32 channels) connected to each plane of wires and each plane of strips. Figure~\ref{fig3} shows a sketch of the readout system, 9 units and 576 channels in total. The individual readout is the sole scheme that can lead to a high counting rate applications for two main reasons: global dead time is reduced since each channel is independent from the others, the amount of charge needed to perform the individual readout is generally lower than that of charge division to achieve the same signal-to-noise. 

\begin{figure}[htbp]
\centering
\includegraphics[width=.7\textwidth,keepaspectratio]{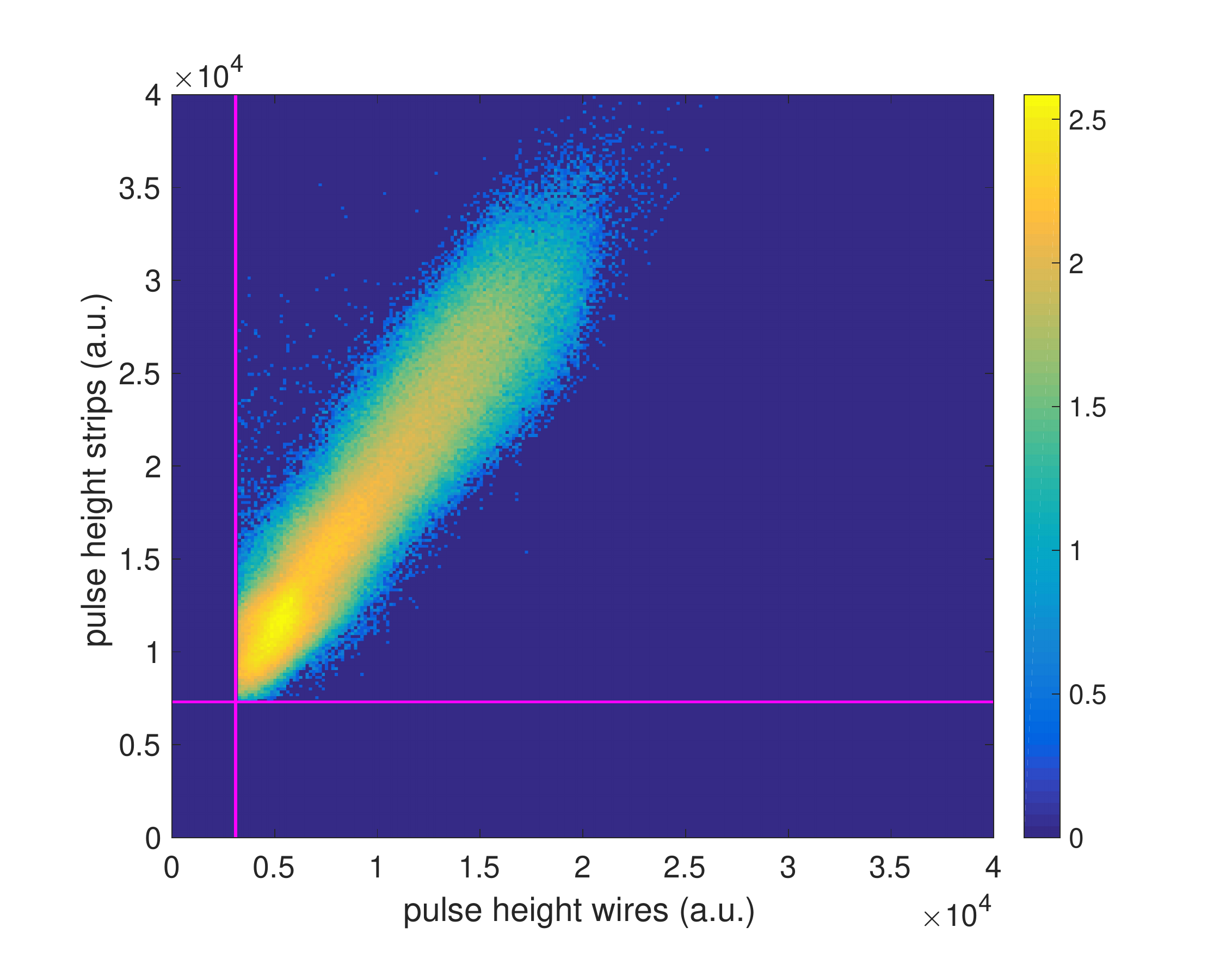}
\caption{\label{fig3334} \footnotesize Wire and strip pulse height correlation. The color scale is counts in logarithmic units. The two pink lines represent the hardware threshold applied to the readout channels to reject the electronic noise. Figure from~\cite{MIO_MB16CRISP_jinst}.}
\end{figure}

\begin{figure}[htbp]
\centering
\includegraphics[width=.7\textwidth,keepaspectratio]{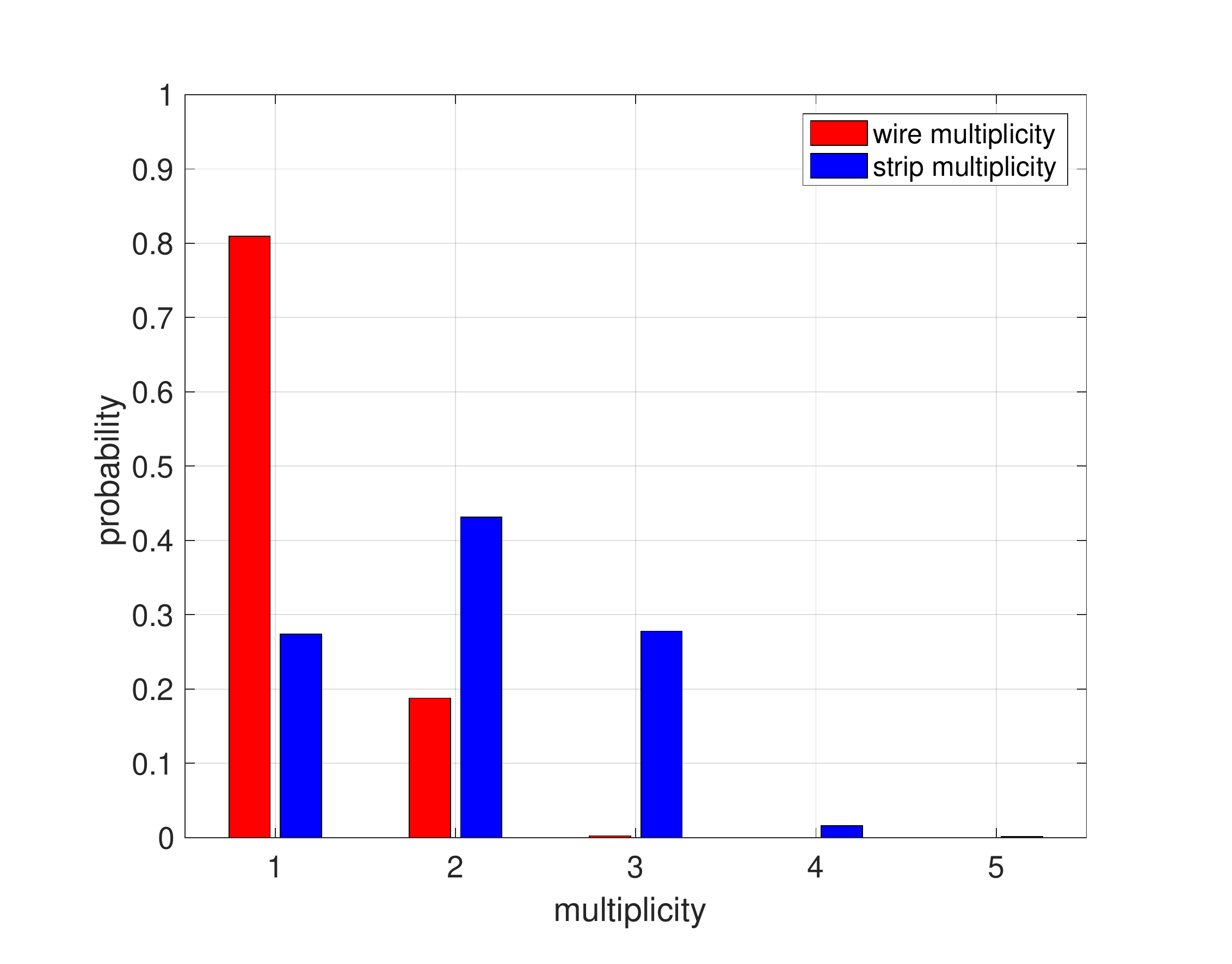}
\caption{\label{fig47} \footnotesize Probability of each multiplicity of wires and strips in a cluster for detected thermal neutron. The multiplicity is normalized to 1 and defined as the number of wires or strips respectively with a detected signal above the software threshold. Figure from~\cite{MIO_MB16CRISP_jinst}.}
\end{figure}

Moreover, less charge means smaller space charge effects~\cite{RC_Ivaniouchenkov,RC_mathieson1,RC_mathieson2} which affects the gas gain variation of the detector at high rates. Note that, at high rate operation, the individual readout (as opposed to charge division) is mandatory to disentangle hits occurring nearly at the same time (that is, unresolved due to the finite time resolution of the detector). The measured amplitudes on the wires and on the strips are strongly correlated (since they are induced by the same avalanche), therefore with sufficient dynamic range, the ambiguity might be resolved by requiring matching amplitudes. Figure~\ref{fig3334} shows the correlation between the pulse height of strips and that of wires. The ratio between the amplitudes on wires and strips depends on the amount of charge collected at the electrode and the gain of the front-end channel. In figure~\ref{fig3334} the amplitudes on wires are cut below approximately $3\times10^3$ a.u. (vertical pink line in the plot) and on strips below approximately $7.5\times10^3$ a.u. (horizontal pink line in the plot) which correspond to $5\,$mV and $15\,$mV respectively (or approximately $0.2 - 0.3\,$fC). These values are the hardware thresholds set to reject the electronic noise. 
\\ Each 32-channel board is connected to a CAEN V1740D digitizer (12 bit, $62.5\,$MS/s) \cite{EL_CAEN}. There are 6 digitizers in total and each can readout 64 channels, i.e., one cassette. Thus, out of the 9 cassettes only 6 could be used simultaneously in the tests. The 6 digitizers can be synchronized to the same clock source and a TTL logic signal can be sent to one of them and propagated to reset the time-stamp which is associated to an event. This feature is needed to perform any type of Time-of-Flight (ToF) measurement. e.g., in the case of CRISP, the reset of the time-stamp is given by the proton pulse of the ISIS source. The system is asynchronous and each time any channel is above the set hardware threshold the digitizer calculates the area of the trace in a given gate and records this event to file with the relative time-stamp. Since the signals are shaped, any value among amplitude, area of the pulse or time-over-threshold (ToT) give the same information: a value related to the energy released on a wire or a strip. 
The raw data, containing the channel number and its time-stamp, is reduced to clusters of groups of channels. A software threshold can be applied to each channel in order to reject background events. The software thresholds used in these measurements corresponds to approximately $100\,$keV~\cite{MIO_MB2017,MIO_fastn}. See chapter~\ref{chapter5} for further considerations.
\\A cluster is defined as a group of events in the file with the same time-stamp within a time window of $6\, \mu$s. This window is defined from the first active wire of a cluster since the strip signals arrive at the same time or later. The information of a cluster, or a hit, is given by a triplet such as (X,Y,ToF). As already shown in~\cite{MIO_MB2017}, each neutron event could have a multiplicity more than one; each time a neutron is converted about $75\, \%$ of the times a single wire is involved in the detection process, about $25\, \%$ two wires are firing; the probability to get three or more wires involved in a detection process is below $1\, \%$. About $25\, \%$ of the times only one or three strips are involved in a detection process whereas $50\, \%$ of the times two strips are firing at the same time. The probability to get four or more strips firing is below $1\, \%$. The most probable cluster in the Multi-Blade geometry is then a wire and two strips firing at the same time. The multiplicity can be used to discriminate against background events; for instance gamma-rays interactions and fast neutron interactions have in general higher multiplicity due to the longer range in gas of electrons and protons~\cite{MIO_fastn}. 
\\ Note that the multiplicity depends on the applied thresholds and the gas gain at which the detector is operated. Hence a systematic variation of these values are expected, without qualitative change of the confirmed behaviour. Figure~\ref{fig47} shows the probability of each multiplicity of wires and strips in a cluster obtained in these tests. 

\begin{figure}[htbp]
\centering
\includegraphics[width=.7\textwidth,keepaspectratio]{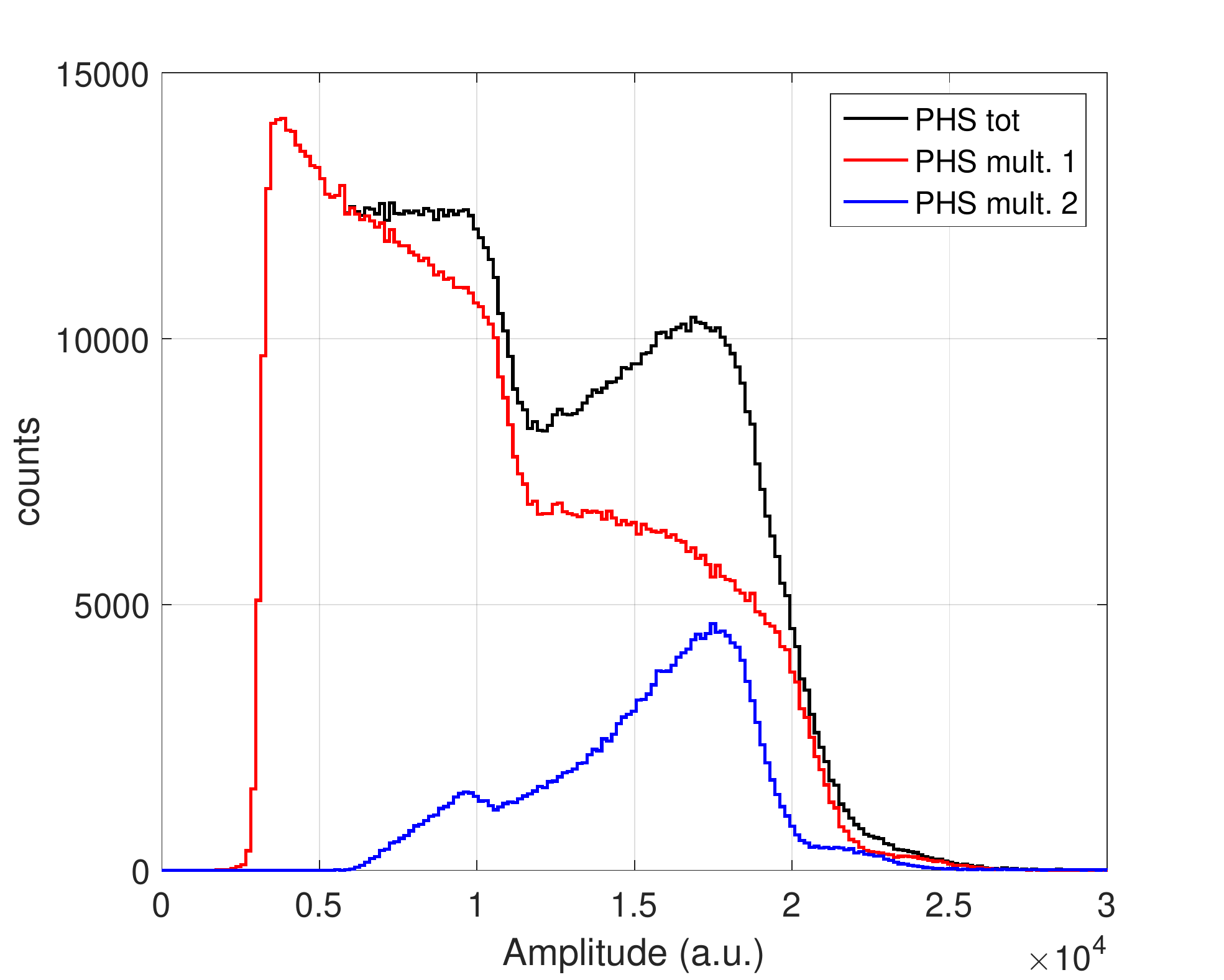}
\caption{\label{phs_multi} \footnotesize Total Pulse Height Spectrum recorded at 800 V at the BNC (black line). PHS considering multiplicity one and two on the wires, red and blue line, respectively.}
\end{figure}

The fact that wires and strips have in general different multiplicity~\cite{Particle_Detection} is due to the combination of two phenomena: the actual extension in space of the neutron capture fragment tracks ($\alpha$ and Li particles) which in our case is comparable to the wire pitch and strip width ($4\,$mm); and the physical process of induction of the signal from the avalanche at the various electrodes. The first is responsible of the spread of the charge among more than one wire; if the ionization track of one of the neutron capture fragments is shared between two wires there will be two avalanche processes, one at each wire. Note that on all the other wires, which are not taking part to the detection process, will be induced a bipolar signal with amplitude proportional to their distance to the firing wire(s). 
\\ In figure~\ref{phs_multi} is shown the Pulse Height Spectrum (PHS) measured at the Budapest Neutron Centre. The measurements were recorded at 800 V. In the total PHS are visible both the peaks corresponding to Li and $\alpha$ particles respectively (black line). When the PHS is selected for multiplicity one, red line in the plot, the contribution of Li particles is more evident than that of $\alpha$ particles. They are, indeed, heavier and less energetic than the $\alpha$ so they release their energy most probably on one wire. The blue line in figure~\ref{phs_multi} refers to the PHS calculated for multiplicity two. It is more visible the peak given by the $\alpha$ particles, which have a higher energy (1470 keV respect to 840 keV of Li particles). The probability of releasing energy in more than one wire is higher because of the larger track length of $alpha$ particles. Any contribution to the spectrum taking into account higher multiplicity values is negligible (below 1$\%$). It can be attributed both to neutrons with higher energy or $\gamma$-ray and fast neutron background events.

The other phenomenon is due to the physical induction mechanism on the electrodes: even if only one wire receives the whole charge, this does not happen for the facing strips that share the charge. It is demonstrated less than $50\,\%$ is the fraction of signal measured on an infinitely wide strip~\cite{Particle_Detection}. Moreover, for narrower strips this fraction is still quite large, e.g., in our case of strip width of $4\,$mm and comparable wire-strip distance, $\approx 28\,\%$ is collected when the avalanche is at the centre of a strip and $\approx 14\,\%$ if it is over the centre of the neighbouring strip. Note that the best position resolution is achieved when the strip width is matching the wire-strip distance~\cite{GATTI_stripCathode}, as in the Multi-Blade. See section~\ref{signalformation} for the theoretical discussion of signal formation. 
\\ When the coordinate of the avalanche is exactly in between two strips, it induces two equal signals on them. If the avalanche is produced in correspondence with the centre of one strip, it induces a large signal on it and two smaller signals on the two adjacent strips. Typically one records two signals in the first case and three in the second. Multiplicity of 1 on strips can be obtained when three signals are induced, but the two side strips do not receive enough charge to generate a signal above the electronic noise. It appears clear then how the multiplicity 1 and 3 on strips, in the Multi-Blade geometry, add up to $\approx50\, \%$ as it is for the multiplicity 2. 

\begin{figure}[htbp]
\centering
\includegraphics[width=.49\textwidth,keepaspectratio]{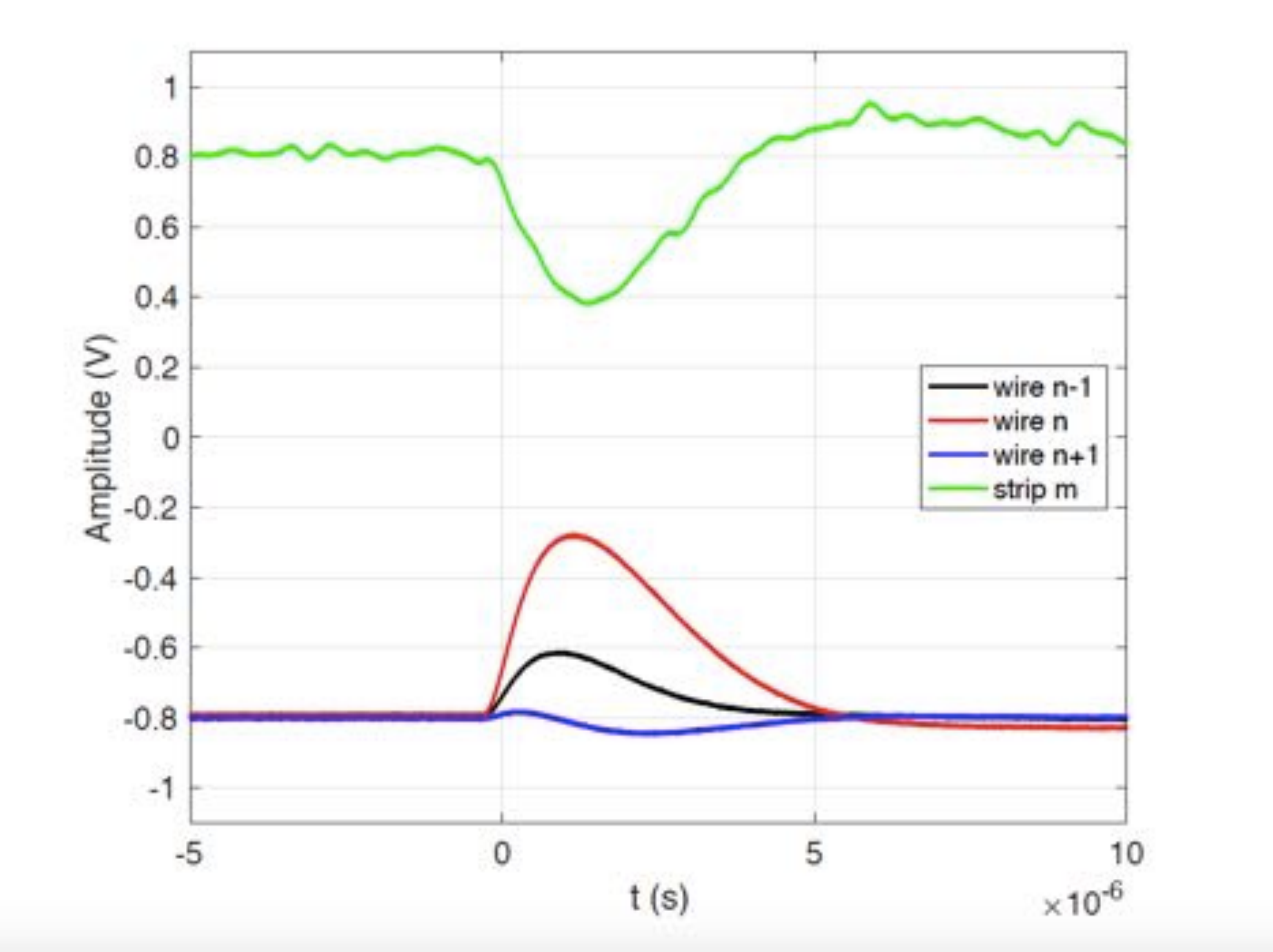}
\includegraphics[width=.49\textwidth,keepaspectratio]{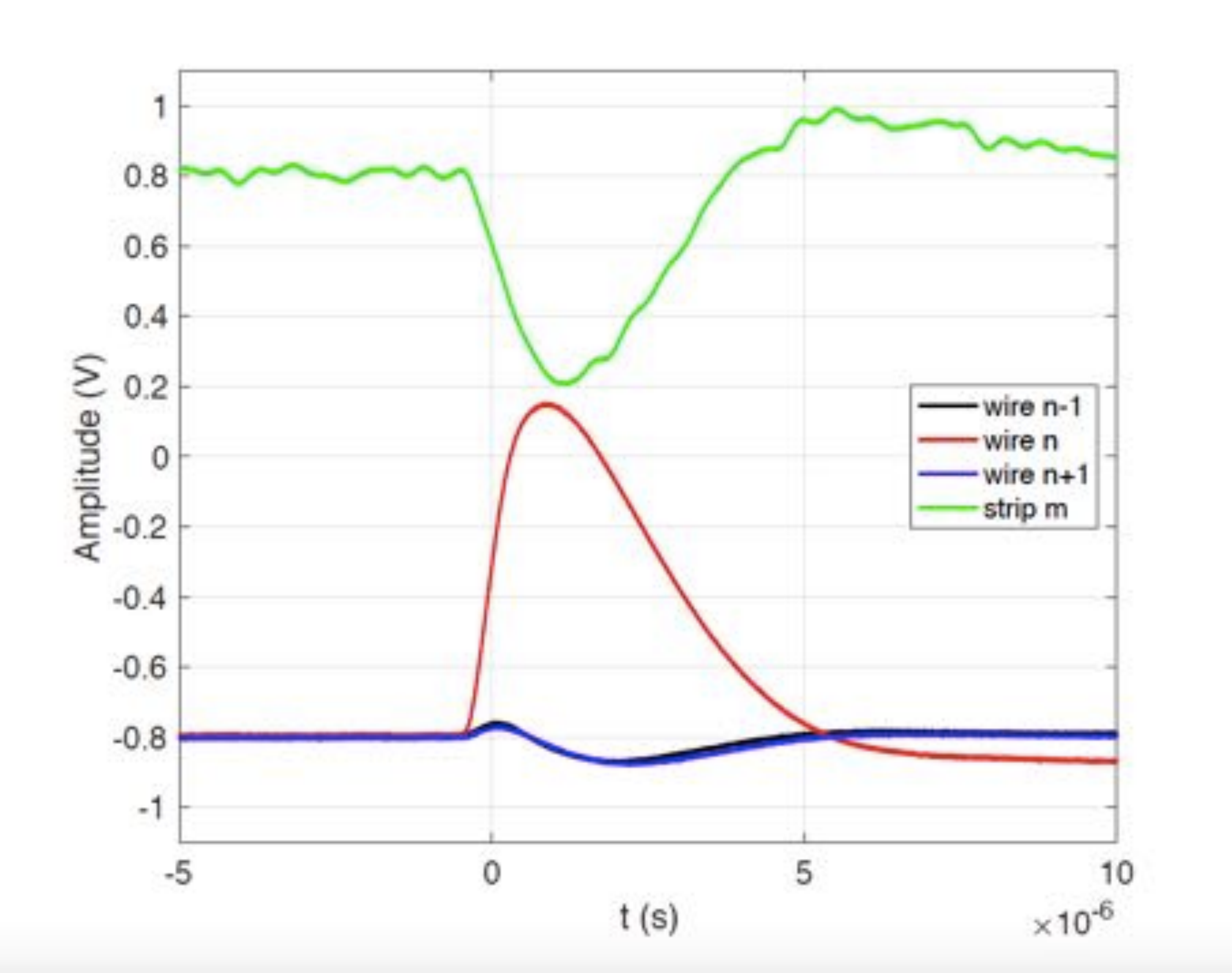}\\
\includegraphics[width=.49\textwidth,keepaspectratio]{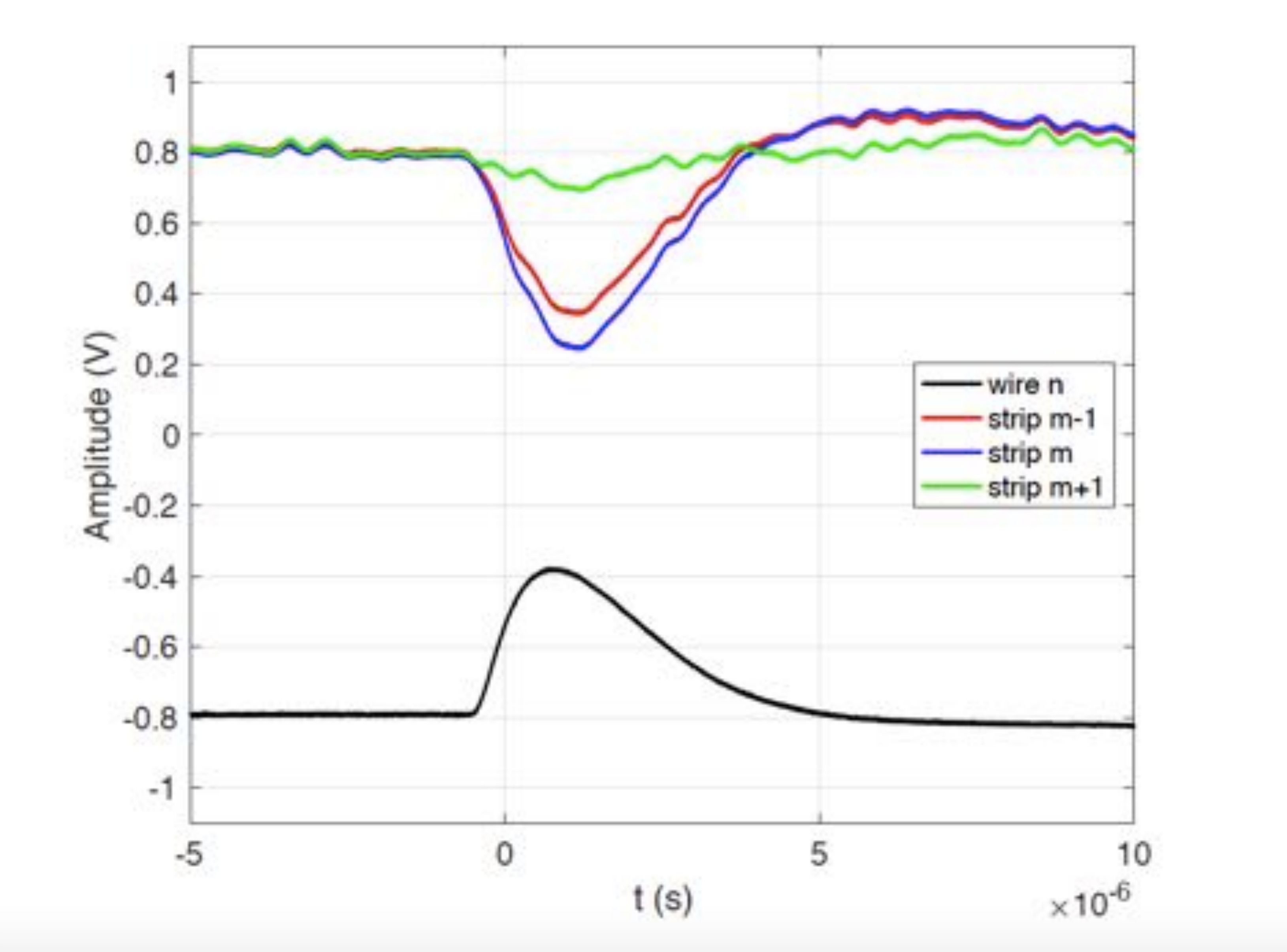}
\includegraphics[width=.49\textwidth,keepaspectratio]{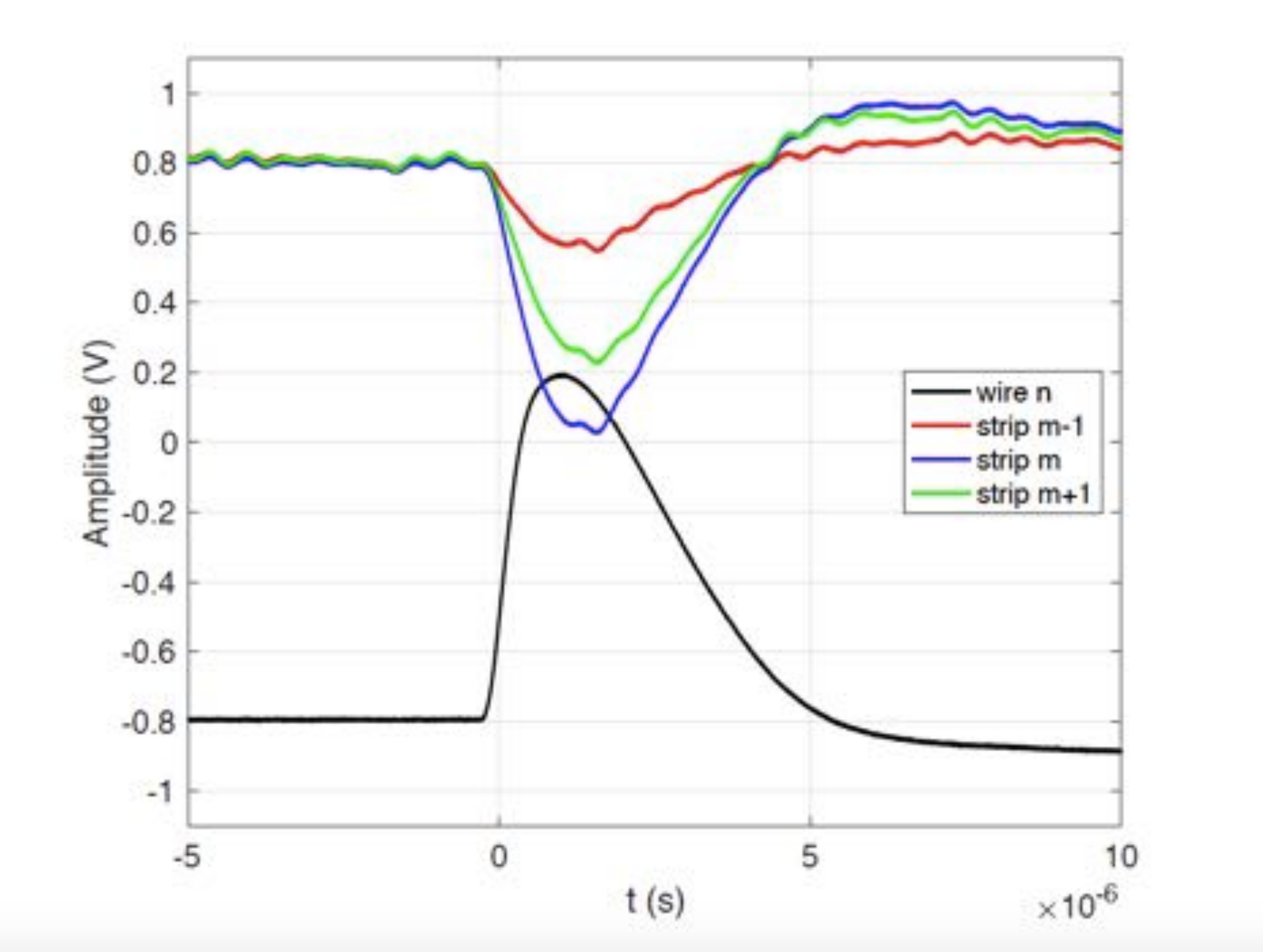}
\caption{\label{fig3b} \footnotesize Typical examples of the signals from wires and strips when firing due to a neutron interaction. Three adjacent wires and one strip with an event having multiplicity 2 (top-left) and 1 (top-right) on wires. Three adjacent strips and one wire with an event having multiplicity 2 (bottom-left) and 3 (bottom-right) on strips. The signals are all inverted due to the use of inverting amplifiers. Strip and wire signals have been shifted arbitrarily on the y-axis for viewing purposes. Figure from~\cite{MIO_MB16CRISP_jinst}.}
\end{figure} 

Figure~\ref{fig3b} shows some examples of signals from the FEE boards from adjacent wires and strips when a neutron is detected. The two top plots show the signals of a strip and three adjacent wires when two wires are firing (wire multiplicity 2) and when only one wire is firing (wire multiplicity 1). The two bottom plots show the signals of a wire and three adjacent strips when two strips are firing (strip multiplicity 2) and when three strips are firing (strip multiplicity 3). The clusters, i.e., triplets, define a three-dimensional space containing the information where the neutron was detected with associated ToF, i.e., its wavelength. Referring to figure~\ref{fig4}, the spatial coordinates, X and Y, of a triplet, reflect the physical channels in the detector: 32 strips and 32 wires, respectively. The spatial coordinates, X and Y, represent the projection over the detector entrance window ( i.e., the projection of the blades toward the neutron incoming direction). Note that the ToF of each triplet is the time of arrival of that neutron to the specific wire. If the ToF has to be encoded in neutron wavelength ($\lambda$), the physical position of each wire in depth (Z) must be taken into account. The flight path must be corrected with the distance ($Z_i$) of the wire $i-th$ of each cassette according this formula:  
\begin{equation}
\label{equadep}
D_i = D_0+Z_i = D_0+(Y_i - 1)\cdot(p\cdot\cos(\beta))
\end{equation}
where $D_0$ depends on the instrument geometry and in our case is the distance from moderator to the first wire (front wire) of the Multi-Blade corresponding to $Y_1=1$, $p=4\,$mm is the wire pitch and $\beta=5^o$ is the inclination of each blade with respect to the sample position.  

\section{Scientific interest}\label{intro}

The detector requirements are set by the two reflectometers that are being designed for the European Spallation Source (ESS~\cite{ESS}) in Sweden: FREIA~\cite{INSTR_FREIA,INSTR_FREIA2} (horizontal reflectometer) and ESTIA~\cite{INSTR_ESTIA,INSTR_ESTIA1,INSTR_ESTIA2} (vertical reflectometer). In the past few years several methods have been proposed to improve the performance of reflectometry instruments and the ESS reflectometers are based on these new concepts. 

Neutron reflectometry and off-specular scattering are powerful techniques to study depth profiles and in-plane correlations of thin film samples~\cite{MISC_pike2002scattering,R_fermi,Lauter2016}. A detailed description of this type of scattering can be found in section~\ref{neutrefl}. In a typical neutron reflection experiment the reflection of neutrons as a function of the wave-vector transfer in direction of the surface normal, $q_z$, is measured:

\begin{equation}
\label{eqaf1}
q_z=\frac{4\pi}{\lambda}\sin(\theta)
\end{equation}

where $\theta$ is the angle between the beam and sample surface (which is the same for incident and reflected beam, $\alpha_i=\alpha_f=\theta$). 
\\The optical properties of neutron propagation arise from the fact that quantum-mechanically the neutron is described by a wave-function. We remind that the potential ($V$) in the Schr\"{o}dinger equation, which is the averaged density of the scattering lengths of the material, plays the role of a refractive index:

\begin{equation}\label{eqasld}
V=\frac{2\pi\hbar^2}{m_n} N_b=\frac{2\pi\hbar^2}{m_n}\sum_i b_i n_i
\end{equation}

The reflectivity of neutrons of a given wavelength (or given $q$) from a bulk interface is unity at  angles smaller than the critical one and falls sharply at larger angles. As with light, interference can occur between waves reflected at the top and at the bottom of a thin film, which gives rise to interference fringes in the reflectivity profile~\cite{MISC_pike2002scattering}. 
\\The typical neutron wavelengths ($\lambda$) in a reflectometry experiment are in the range of 2 - 20 \AA , which corresponds to a range between 0.05 and 3 nm$^{-1}$ in the wave-vector transfer ($q_z$). In the real space this corresponds to length-scales between 2 and 100 nm~\cite{R_offspec0_Ott}. The limits are imposed both by the measurement range and the instrumental resolution.
In the case of off-specular scattering it is possible to investigate objects in the plane with a correlation length of the order of several micrometers (50 to 0.5 $\mu$m). The upper limit is set by the resolution of the intruments and the size of the direct beam. The lower limit is determined by the available neutron flux~\cite{R_offspec0_Ott} and background. 
\\ In the last two decades the reflectometers have been optimized and allow to measure reflectivities below $10^{-6}$, enough for most experiments~\cite{OTT_general}. The next step is to increase the available flux, this leads to a significant speed up of reflectivity measurements and the possibility of using smaller samples (see Chapter\ref{chapter3}). 
\\Several techniques have been recently proposed to improve the operating performance of reflectometry instruments. The methods are based on spin-space~\cite{INSTR_R_Spin2}, time-space~\cite{OTT_tiltof} or energy-space encoding~\cite{OTT_gradtof,OTT_refocus,R_Cubitt1,R_Cubitt2}. The first technique is used for off-specular measurements~\cite{INSTR_R_Spin} and encodes the incident angle by the rotation of the neutron spin in a magnetic field. The time-space encoding (TilTOF) enables an increase in the incoming flux on the sample, removing the chopper and modulating mechanically the angle of the sample to determine the time shape of the beam, and thus the wavelength. The idea of energy-space encoding is to analyse the neutron energies through a spatial spread of the reflected beam produced by an energy dispersive device, either a refractive crystal~\cite{R_Cubitt1,R_Cubitt2} or a magnetic field gradient~\cite{OTT_gradtof}. It is also possible to correlate the neutron wavelength and the incident angle, hence before the sample, using a divergent beam focused on the sample. The REFocus~\cite{OTT_refocus} technique employs an elliptical graded multilayer monochromator to focus the neutrons on the sample.
This concept has been modified and adapted to the time-of-flight instrument AMOR at PSI~\cite{PSI}, using an elliptic-shaped reflector: the \textit{Selene} guide~\cite{INSTR_ESTIA0,INSTR_ESTIA1}. A scaled-down demonstrator is implemented on AMOR at PSI~\cite{INSTR_ESTIA2} to prove the concept and to test the performances of the guide. The full-scale Selene guide will be a primary feature for ESTIA, a reflectometer instrument at European Spallation Source (ESS, Lund, Sweden) now under construction.
\\ The general aim of all these optimizations is to increase the available neutron flux on the sample; thus time resolved measurements for kinetic studies can be performed, smaller samples can be used, faster measurements scaling down from hours, typical time for present day reflectivity experiments, to minutes can be performed. This gives the possibility to probe a dynamic range of reflectivity measurements down to $10^{-7}$.
\\ These improvements represent a challenge not only for the instrument design, but also for the performance of the detector technologies to be employed. The current detector technology is reaching fundamental limits, e.g., a sub-mm spatial resolution (Full-Width-Half-Maximum, FWHM) and high counting rate capabilities are required for the new instruments and it is not achievable with the state-of-the-art technology.

\section{Experimental setup on CRISP}\label{rifle}

\begin{figure}[htbp]
\centering
\includegraphics[width=.8\textwidth,keepaspectratio]{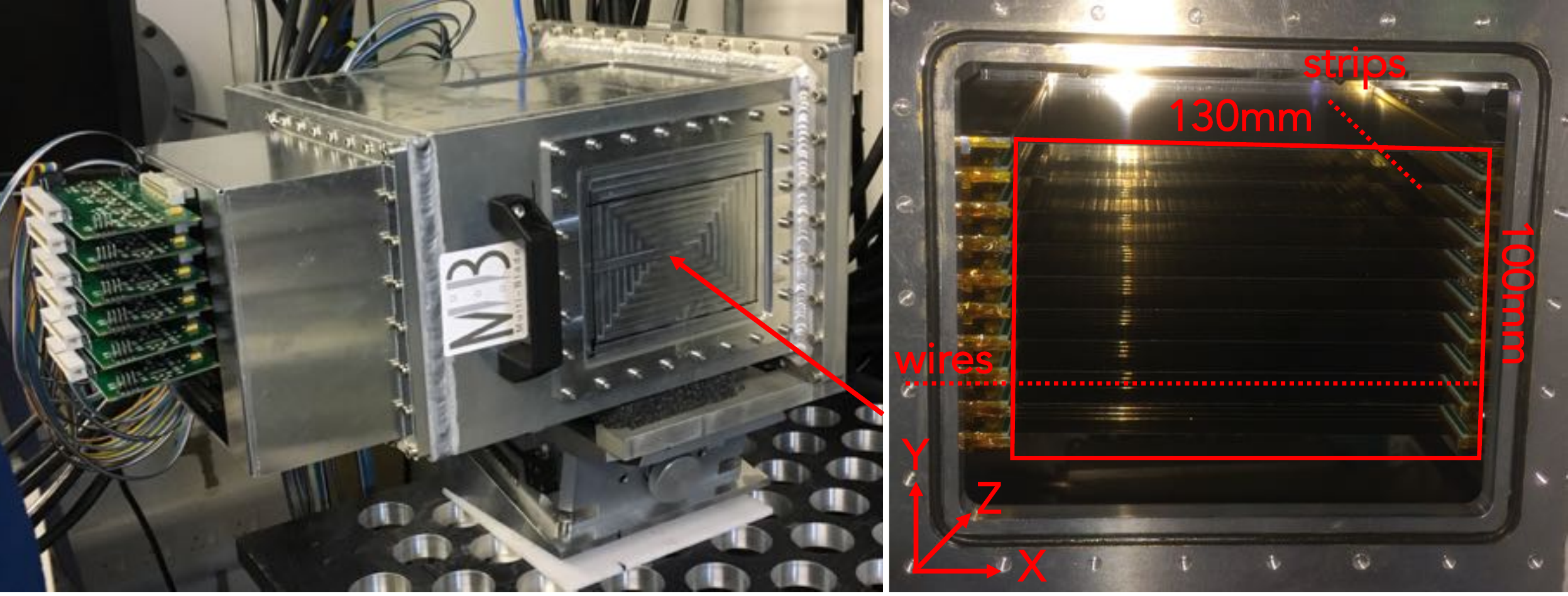}
\caption{\label{fig4} \footnotesize The Multi-Blade installed on the table of CRISP on a goniometer with the FEE boards (left). A view of the cassettes behind the vessel window to show the orientation of the wires and strips in space (right). Figure from~\cite{MIO_MB16CRISP_jinst}.}
\end{figure}

CRISP is an horizontal neutron reflectometer at ISIS, Target Station 1, that uses a broad band neutron Time-of-Flight (ToF) method for determining the wavelength, (and hence $q$), at fixed angles ($\theta$). A detailed description of the CRISP reflectometer can be found in~\cite{CRISP1}. The instrument views an hydrogen moderator giving an effective wavelength range of $0.5-6.5$\AA\, at the source frequency of $50\,$Hz. The wavelength band extends up to 13\,\AA\, if operated at $25\,$Hz. A frame overlap mirror suppresses the wavelengths above 13\,\AA. The distance from the moderator to the sample is $10.25\,$m and the sample to detector distance is $1.87\,$m. The detector is a single $\mathrm{^3He}$ tube filled with $3.5\,$bar $\mathrm{^3He}$. The Multi-Blade detector was installed at a distance of $2.3\,$m. Figure~\ref{fig4} shows the Multi-Blade installed on CRISP and the orientation of the blades behind the vessel entrance window. The Multi-Blade detector was electrically insulated from the moving stage underneath in order to decouple the detector and the motion unit grounds.

\begin{figure}[htbp]
\centering
\includegraphics[width=0.8\textwidth,keepaspectratio]{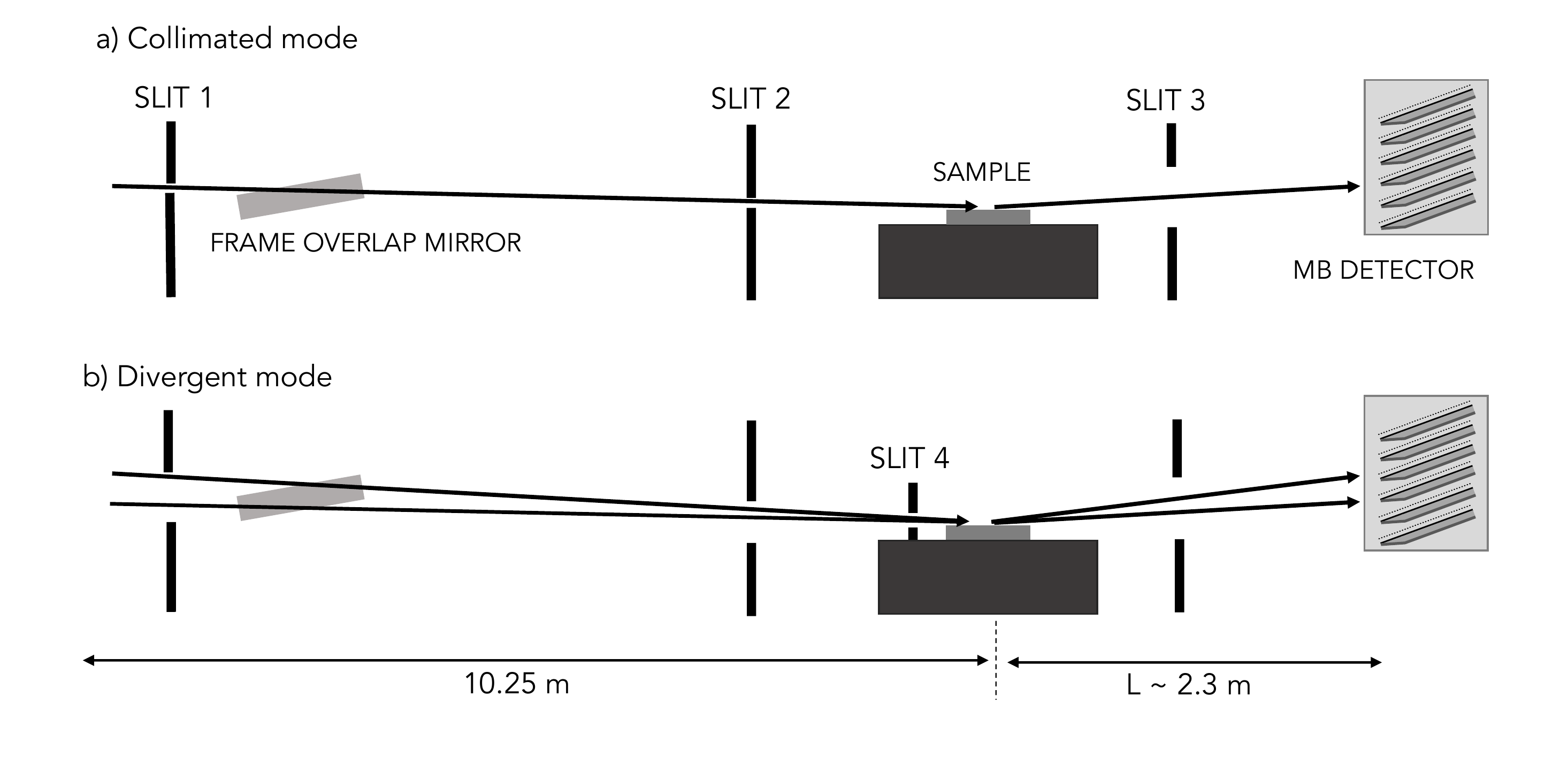}
\caption{\label{fig4bis} \footnotesize A sketch of the CRISP horizontal reflectometer and the MB detector showing the orientation of the cassettes. The beam can be collimated at the sample position either with a low divergence (collimated mode a) or allowing more divergence of the beam (divergent mode b). Figure from~\cite{MIO_ScientificMBcrisp}.}
\end{figure} 

The beam can be well-collimated using adjustable slits along the beam line, a sketch is shown in figure~\ref{fig4bis}. According to the position and the opening of the slits, the measurements have been performed in two working modes: collimated and divergent. In the collimated mode the slits are set in order to achieve a good collimation of the beam at the sample. The divergent mode is obtained by opening as much as possible the slits before the sample. According to the concept of REFocus~\cite{OTT_refocus}, proposed for ESTIA~\cite{INSTR_ESTIA1,INSTR_ESTIA0}, one more slit with a narrow opening ($\approx 1\,$mm) was added before the sample, as shown in figure~\ref{fig4bis}.
\\ The results on threshold choice and background scattering, dynamic range, detection efficiency, stability and uniformity along with the study of the reconstruction algorithms are part of the measurements performed on CRISP, that concern the technical characterization of the detector, together with the stability test performed at the STF at the Lund University.
\\ The scientific results from the test on CRISP are shown in the dedicated section~\ref{secsci}. Three standard and well-known samples have been used in the tests: an iridium (Ir) sample, slightly bent, deposited on a silicon substrate (4 $\times$ 4 cm$^2$), a bare silicon (Si) sample ($\approx$ 8 cm diameter) and a Fe/Si super-mirror ($\approx$ 4 cm diameter) which is used in neutron optics to deliver neutrons to the instruments. The Ir sample has been used to study the effect of the spatial resolution of the detector on the measured reflectivity curve and it will be shown in section~\ref{secIr}. The Si sample has been used to study the collimated and divergent modes. This will be discussed in details in section~\ref{colldivsi}. The Fe/Si super-mirror has been used to study the off-specular scattering with the Multi-Blade and it will be discussed in section~\ref{offfesi}.

\section{Technical Results}

In this section, a full characterization of the detector performance is provided. The measurements refers to the direct beam, hence no sample have been used. Comparisons with the results obtained with the $^3$He-tube installed on the instrument are presented as well. The section on stability concern a set of measurements performed at the STF at Lund University.

\subsection{Threshold choice and scattering from substrate}\label{scatt}

\begin{figure}[htbp]
\centering
\includegraphics[width=.49\textwidth,keepaspectratio]{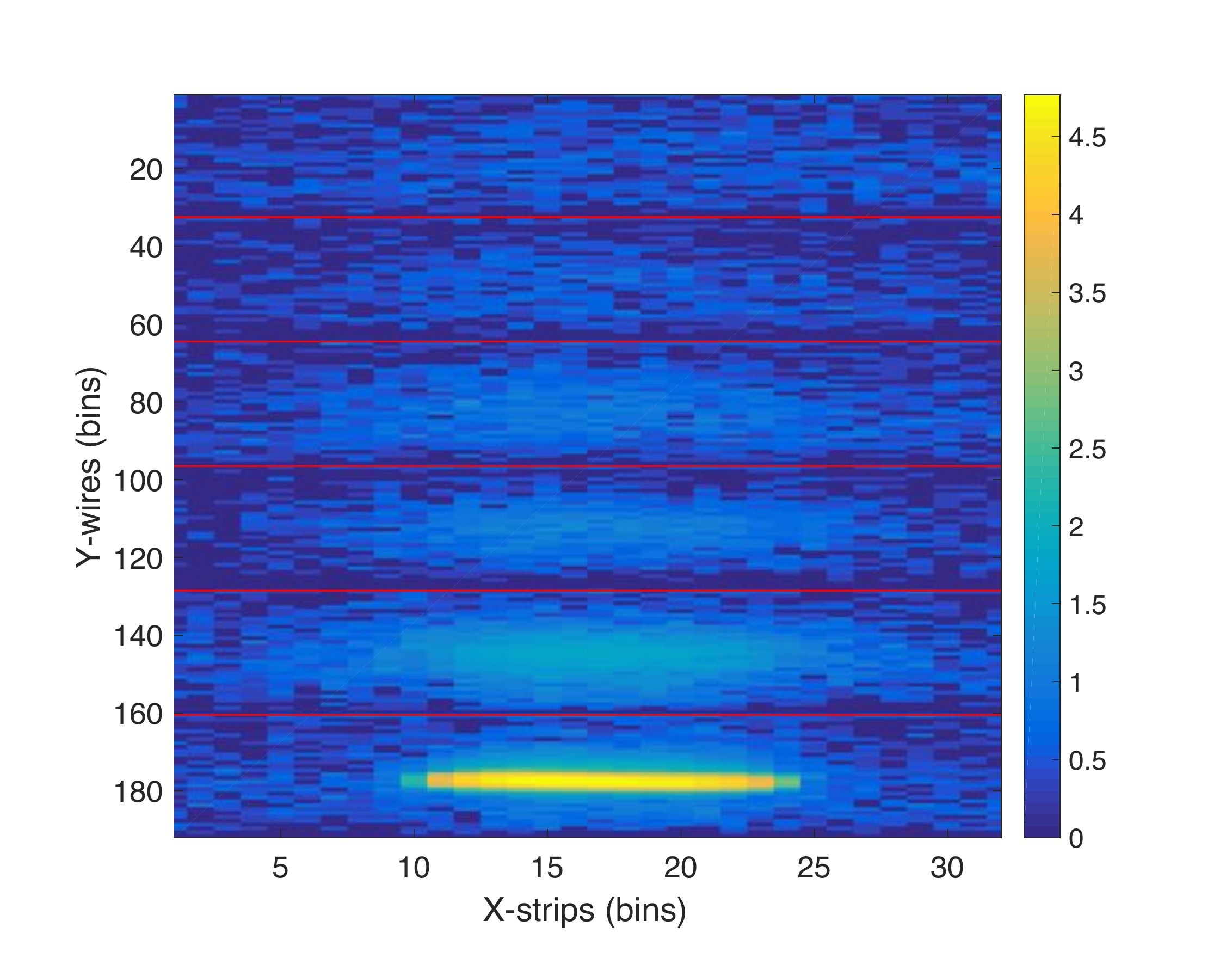}
\includegraphics[width=.49\textwidth,keepaspectratio]{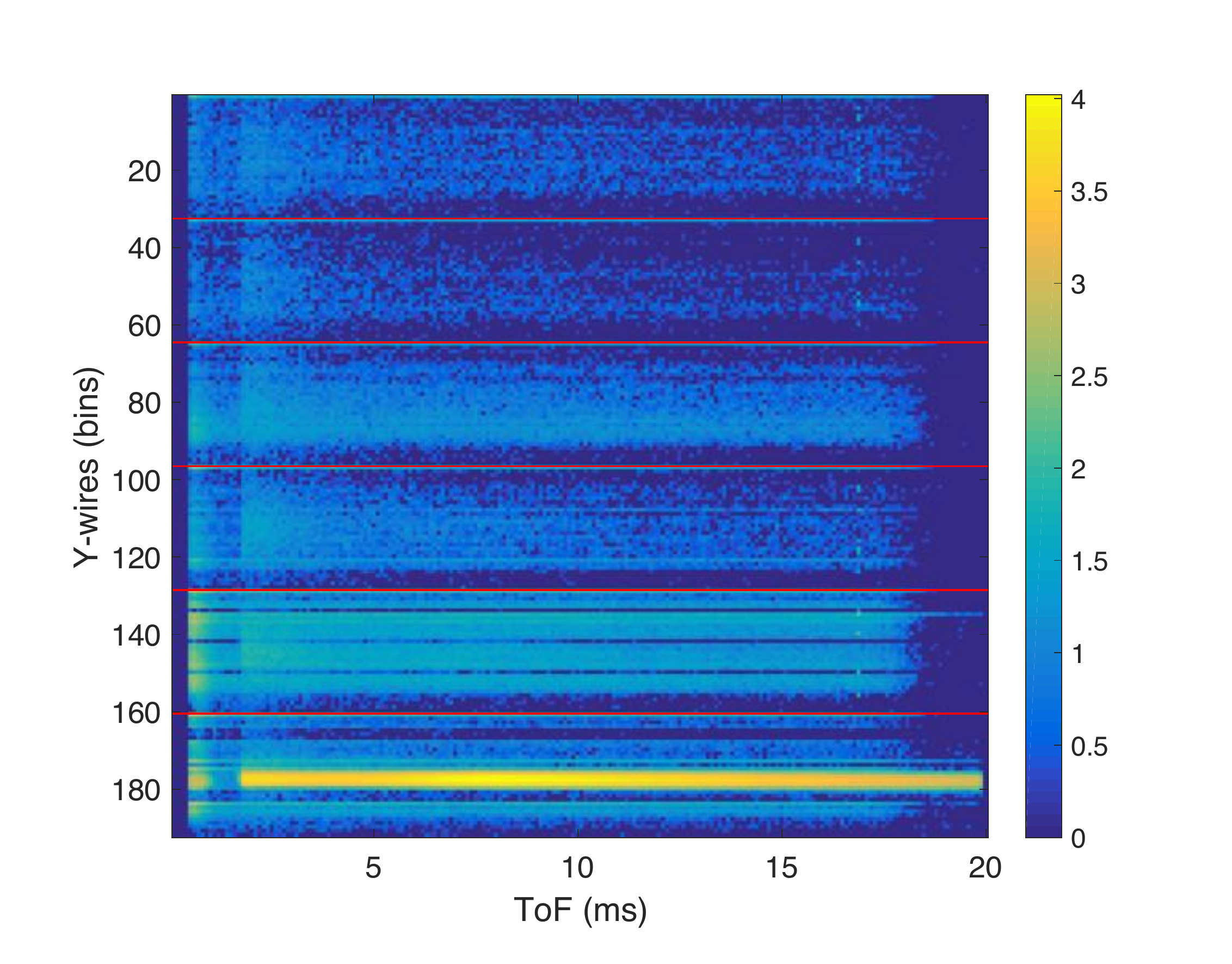}
\caption{\label{fig5} \footnotesize 2D image of six cassettes of the Multi-Blade detector (left). ToF image of the detector integrated over the X-direction (strips) (right). This is a raw uncorrected data with no software threshold applied. The color bar represents counts in logarithmic scale. Figure from~\cite{MIO_MB16CRISP_jinst}.}
\end{figure}

\begin{figure}[htbp]
\centering
\includegraphics[width=.49\textwidth,keepaspectratio]{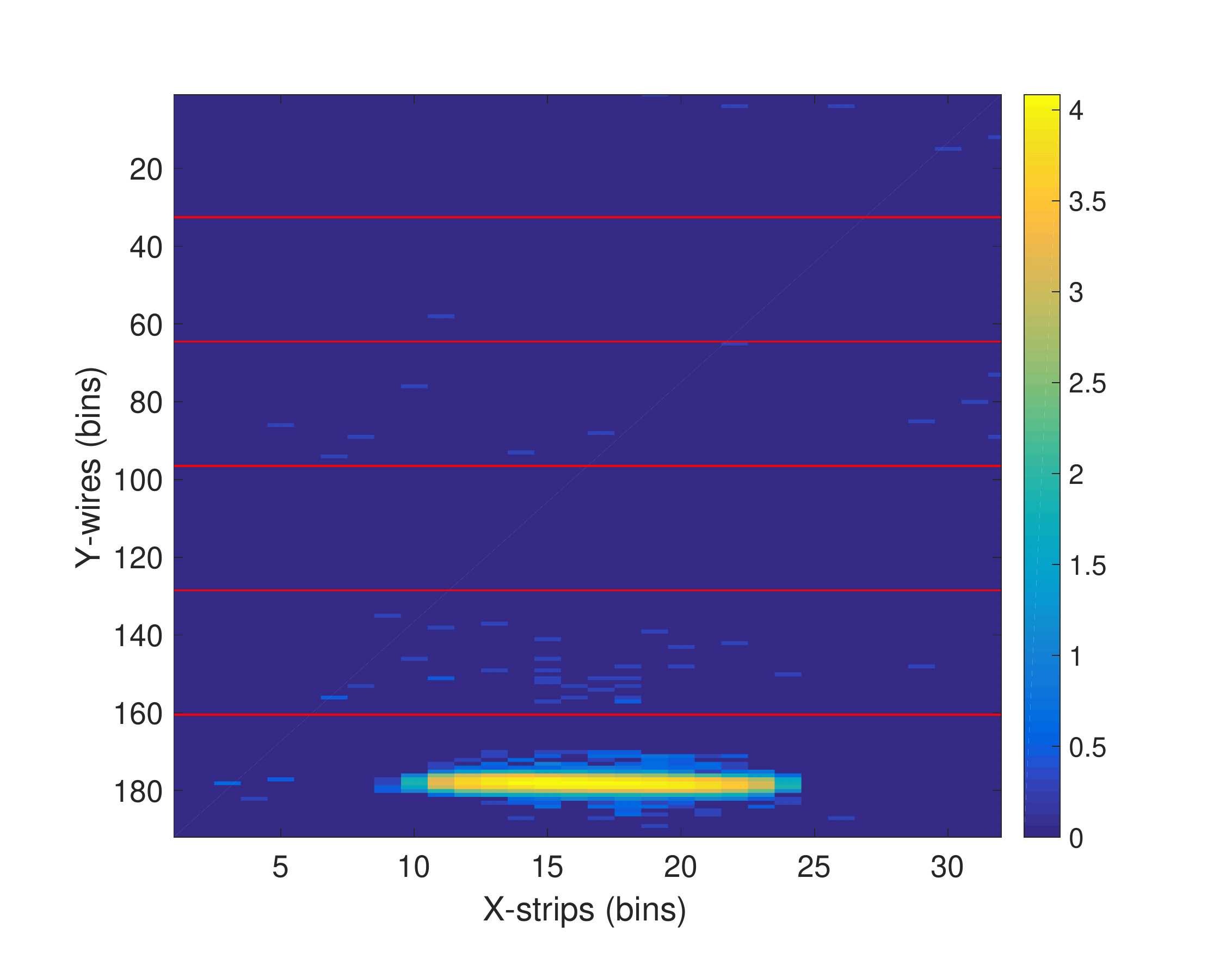}
\includegraphics[width=.49\textwidth,keepaspectratio]{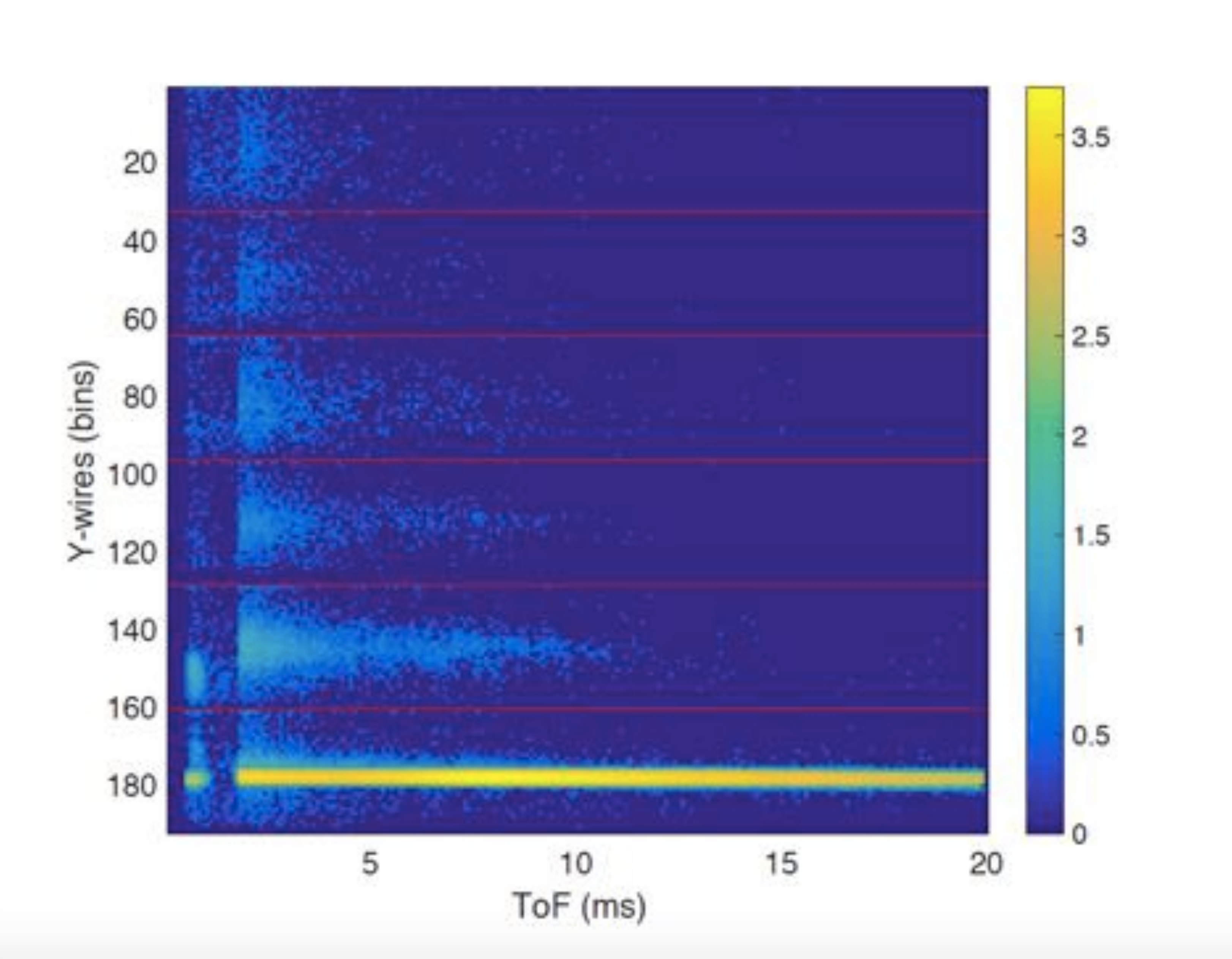}
\caption{\label{fig6} \footnotesize 2D image of the detector when software thresholds are applied to discriminate against $\gamma$-rays and the ToF is gated between $12.5\,$ms and $20\,$ms (4\AA\,- 6.5\AA) (left). ToF image of the detector when software threshold is applied (right). The color bar represents counts in logarithmic scale. Figure from~\cite{MIO_MB16CRISP_jinst}.}
\end{figure}

\begin{figure}[htbp]
\centering
\includegraphics[width=.5\textwidth,keepaspectratio]{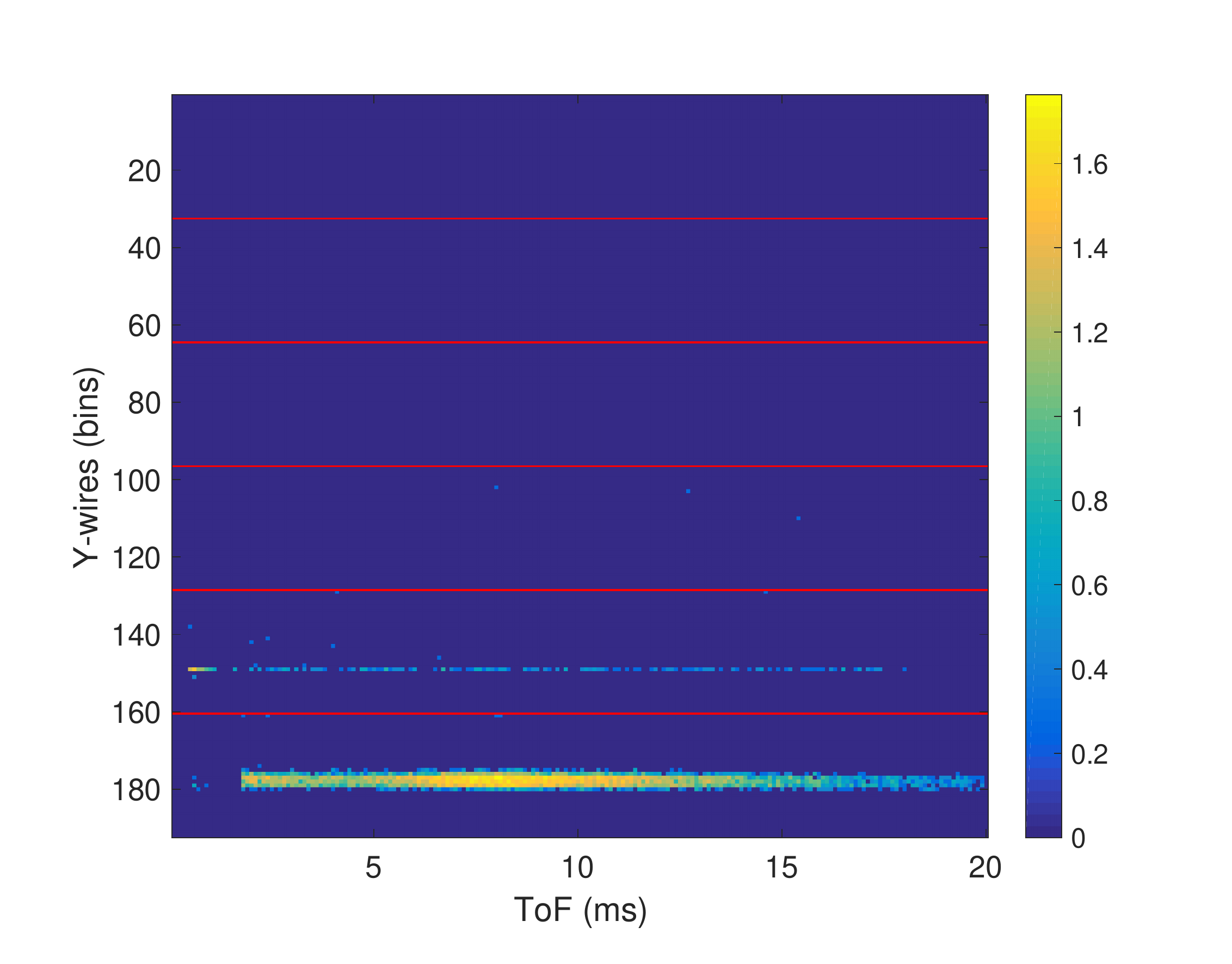}
\caption{\label{fig6fn} \footnotesize ToF image of the detector when software threshold is applied and the events below threshold are taken into account. The color bar represents counts in logarithmic scale.}
\end{figure}

We recall that the hardware thresholds are applied to the individual channels of the digitizers (either wires or strips) and it is needed to reject the electronic noise. Note that these values have been set not to fully discriminate against low energy events, e.g., background $\gamma$-rays. Therefore, a software threshold can be applied to filter these events directly in the data. 
\\ The triplets $(X,Y,ToF)$, that identify an event and were described in section~\ref{MBtes}, can be represented by two-dimensional plots: the 2D image of the detector which corresponds to the $(X,Y)$ coordinates and the ToF image of the detector which corresponds to the $(Y,ToF)$ coordinates and it is integrated over the other spatial coordinate ($X$, the strips). The $(X,ToF)$ image can also be used but it is not relevant for the purpose of this work. Moreover, the 2D image $(X,Y)$ can be either integrated over the ToF coordinate or gated in any range of time. A ToF of $6\,$ms corresponds approximately to 1.8\AA, $8\,$ms to 2.5\AA\, and $12.5\,$ms to 4\AA.  
\\ Figure~\ref{fig5} shows the 2D image and the ToF image of the part of the detector that has been read out by the six digitizers (six cassettes); the horizontal red lines indicate where each cassette starts and ends. The direct beam was directed to the lower cassette of the Multi-Blade detector without being reflected by any sample and its footprint, at the detector, was approximately 3$\,$mm$\times$60$\,$mm. The instantaneous peak rate on the whole detector area (as defined in~\cite{DET_rates}) was of $\approx10^4\,$Hz (corresponding to an instantaneous local rate of $R\approx50\,$Hz/mm$^2$ at peak).
\begin{equation}
R = \frac{Counts}{(n_f \cdot n_{ToF})\, t_b\, A} = \bigg[ \frac{counts}{s \cdot mm^2}\bigg] = \bigg[ \frac{Hz}{mm^2}\bigg]
\label{ratemb}
\end{equation}
where $n_f$ is the number of acquisitions, $n_{ToF}$ is the number of ToF per file, $t_b$ is the bin size in time and $A$ is the illuminated area.
\\ Note that if no software thresholds are applied nor the wires are requested to be in coincidence with the strips, a constant background is visible and it is mostly due to $\gamma$-rays. This background almost disappear if the coincidences are selected, although no software thresholds are applied. 
\\ When a software threshold is applied, the plot on the right in figure~\ref{fig6} is obtained. The constant background, which can be attributed to the $\gamma$-rays, vanishes. However a background at short wavelengths ($<4$\AA, $12.5\,$ms) is still visible and reproduces the shape of the direct beam. The plot on the left in Figure~\ref{fig6} depicts the 2D image of the direct beam with software thresholds applied and with the ToF gated between $12.5\,$ms and $20\,$ms (above 4\AA). Note that if the ToF is not selected in the indicated range, the background does not vanish.
\\ On the right side of figure~\ref{fig6} a bright peak at $1\,$ms in ToF is visible. The plot in figure~\ref{fig6fn} is obtained when software thresholds are applied and events below these values are taken into account. The spot disappears, and it can be attributed to fast and/or epithermal neutrons of wavelength below 0.3 \AA . Indeed, the fast neutrons generally release large amount of energy in the gas~\cite{MIO_fastn} (see chapter~\ref{chapter5}) and it is not visible if only events below the software threshold are selected.

In figure~\ref{fig6fn} is possible to distinguish a firing wire in the second cassette from the bottom. The spark may be due to some dirtiness on the wire. It is less probable that the effect arises from a deflection of the wire from the proper tension, because the non-uniformity of the electric field would produce a much higher noise~\cite{Particle_Detection}. 

\begin{figure}[htbp]
\centering
\includegraphics[width=.7\textwidth,keepaspectratio]{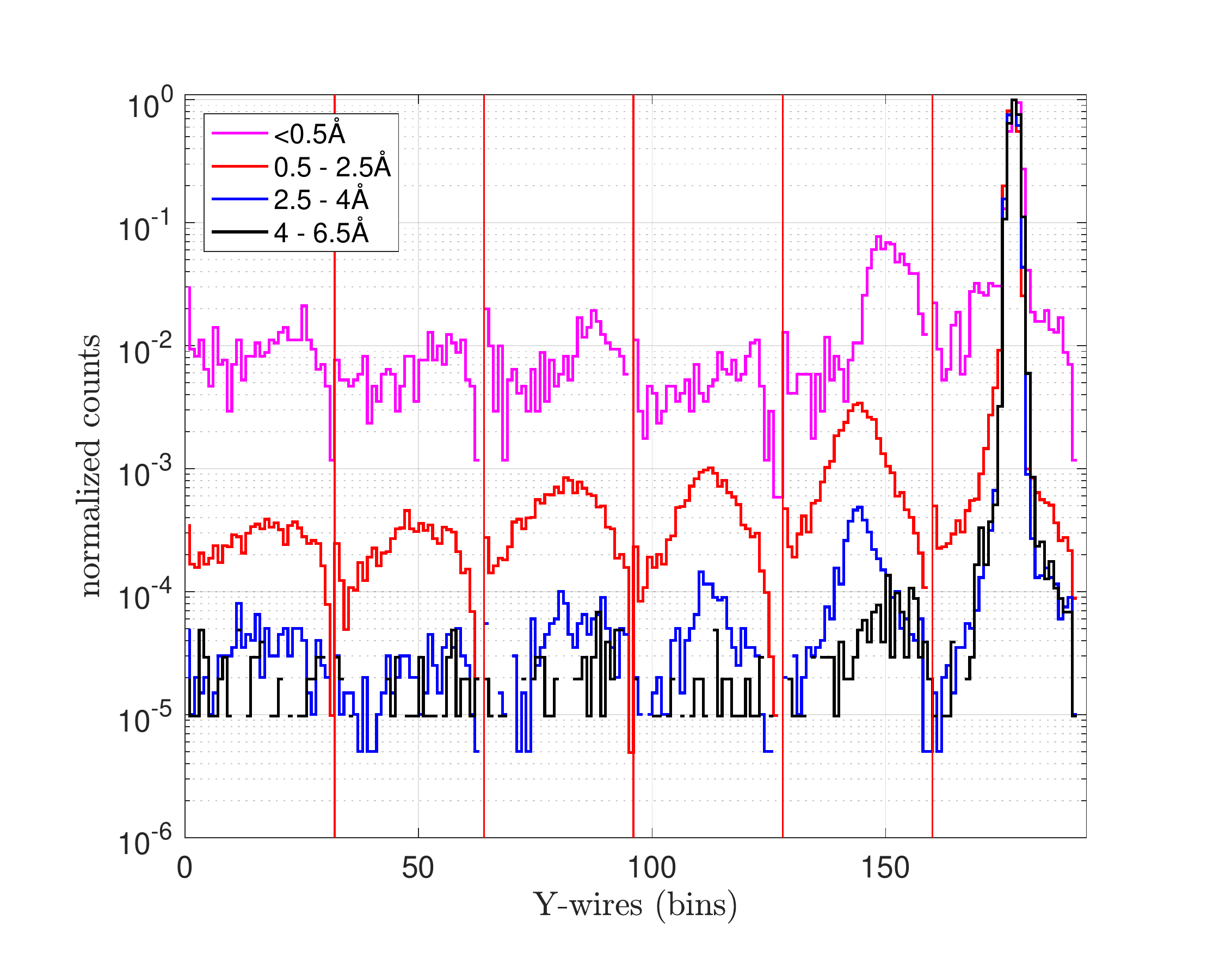}
\caption{\label{fig7} \footnotesize Normalized counts in the 6 cassettes from the 2D image in Figure~\ref{fig6} integrated over the X-direction (strips) for four different gates in the ToF: below $1.5\,$ms ($\approx 0.5$\AA), between $1.5\,$ms and $8\,$ms (0.5 \AA\,- 2.5\AA), between $8\,$ms and $12.5\,$ms (2.5 \AA\,- 4\AA) and between $12.5\,$ms and $20\,$ms (4 \AA\,- 6.5\AA). Figure from~\cite{MIO_MB16CRISP_jinst}}
\end{figure}

\begin{figure}[htbp]
\centering
\includegraphics[width=.6\textwidth,keepaspectratio]{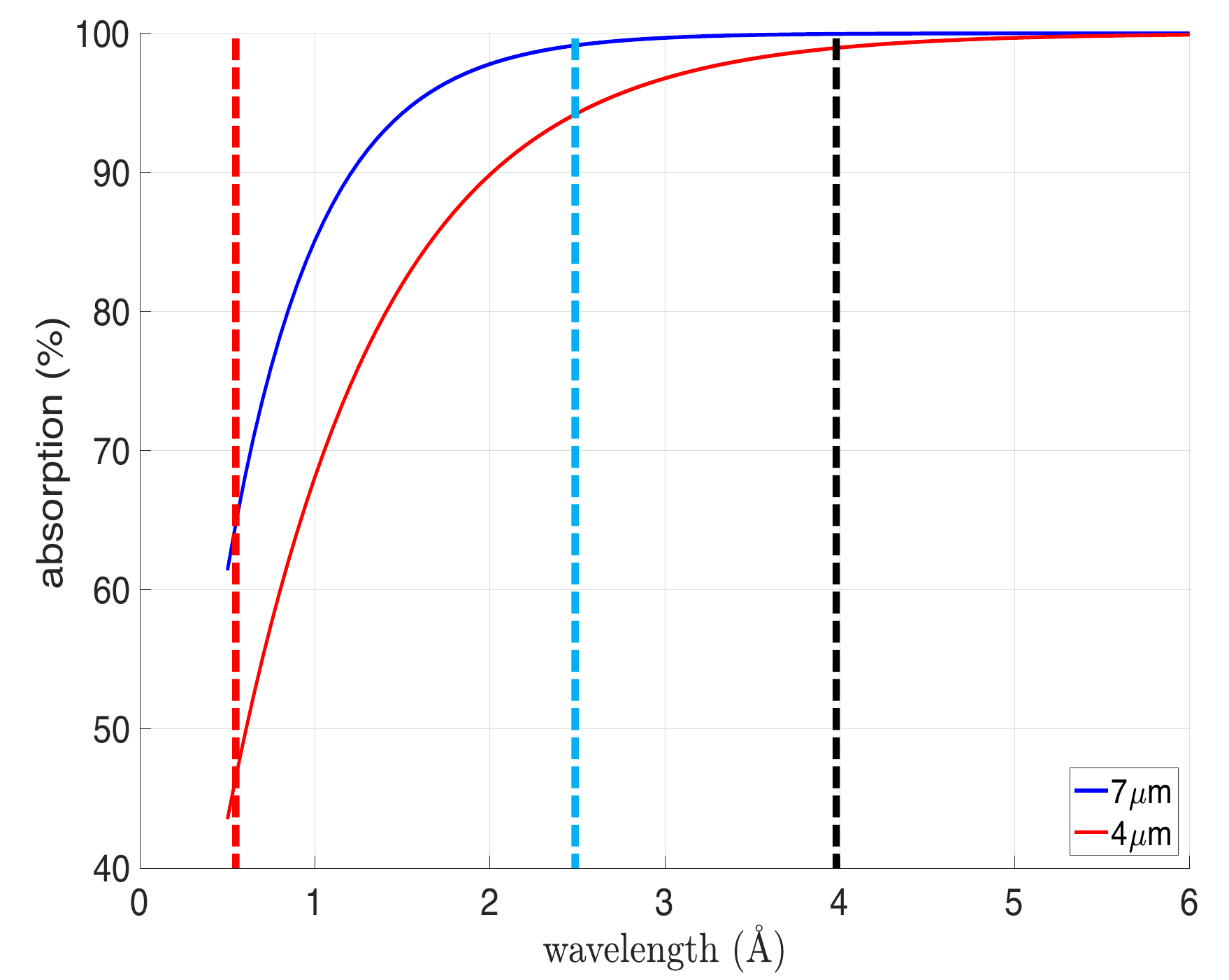}
\caption{\label{fig2} \footnotesize Calculation of the absorbed neutron flux as a function of wavelength for the nominal coating ($7.5\,\mu$m, blue) and for the coating thickness in the present detector ($4.4\,\mu$m, red) inclined at 5 degrees. The vertical dashed lines correspond to the value of the several gates applied to reject the scattering background shown in figure~\ref{fig7}. The red line corresponds to $ 0.5$\AA\ which is the lower limit of the second gated range, the blue line (2.5\AA) represents the lower limit of the third range, and the black line at 4\AA\ is the lower limit of the highest range.}
\end{figure}

Figure~\ref{fig7} shows the normalized counts in the 6 cassettes from the 2D image in figure~\ref{fig6} integrated over the X-direction (strips) for four different gates in ToF, below $1.5\,$ms ($\approx 0.5$\AA), between $1.5\,$ms and $8\,$ms (0.5 \AA\,- 2.5\AA), between $8\,$ms and $12.5\,$ms (2.5 \AA\,- 4\AA) and between $12.5\,$ms and $20\,$ms (4 \AA\,- 6.5\AA).
\\ The shape of the direct beam, centred in the lower cassette, is reproduced into the others and the intensity decreases with the distance. This effect can be attributed to the neutrons that cross the $\mathrm{^{10}B_4C}$ coating without being absorbed. They are scattered by the substrate and they are detected in other cassettes. Note that the nominal recommended thickness is $7.5\,\mu$m, but the present blades (Ti and SS) have been coated with $4.4\,\mu$m.  
Figure~\ref{fig2} shows the amount of absorption in the $\mathrm{^{10}B_4C}$ layer on the blades as a function of the neutron wavelength. The values of the gates applied are also shown (dashed vertical lines).
We expect, with the actual coating, that about $50\%$ of neutrons at the shortest wavelengths (see Figure~\ref{fig2}) are not stopped by the layer causing this background. Note that any material chosen among SS, Ti or Al for the substrate of a blade at 5 degrees, angle of incidence, correspond to about a $11$ times thicker layer. Therefore, the amount of scattered neutron flux from any $2\,$mm-thick substrate of a blade is close to unity~\cite{MIO_MB2017}. 
\\ By applying the gate above $12.5\,$ms (4\AA) the background is completely suppressed, indeed the absorption efficiency of the $\mathrm{^{10}B_4C}$-coating is always above $98\%$ for this or longer wavelengths. 
\\ From the detector requirements set by the instruments, the shortest wavelength that will be used is 2.5\AA\,(Table~\ref{tab1}). The nominal coating ($7.5\, \mu$m) at 2.5\AA\, is expected to be as efficient at absorbing neutrons as the $4.4\,\mu$m coating at 4\AA.

\subsection{Dynamic range}

In order to quantify the spatial and time dynamic range of the detector the direct beam was directed to the lower cassette of the Multi-Blade. The spatial dynamic range is related to the ability of the detector to measure in different pixels (adjacent or not) two different counting rates at the same time: the difference between the counting rates defines the actual spatial dynamic range of the detector. Equivalently, the time dynamic range defines the ability of the detector to measure in the same pixel two different counting rates in subsequent time bins.
\begin{figure}[htbp]
\centering
\includegraphics[width=.49\textwidth,keepaspectratio]{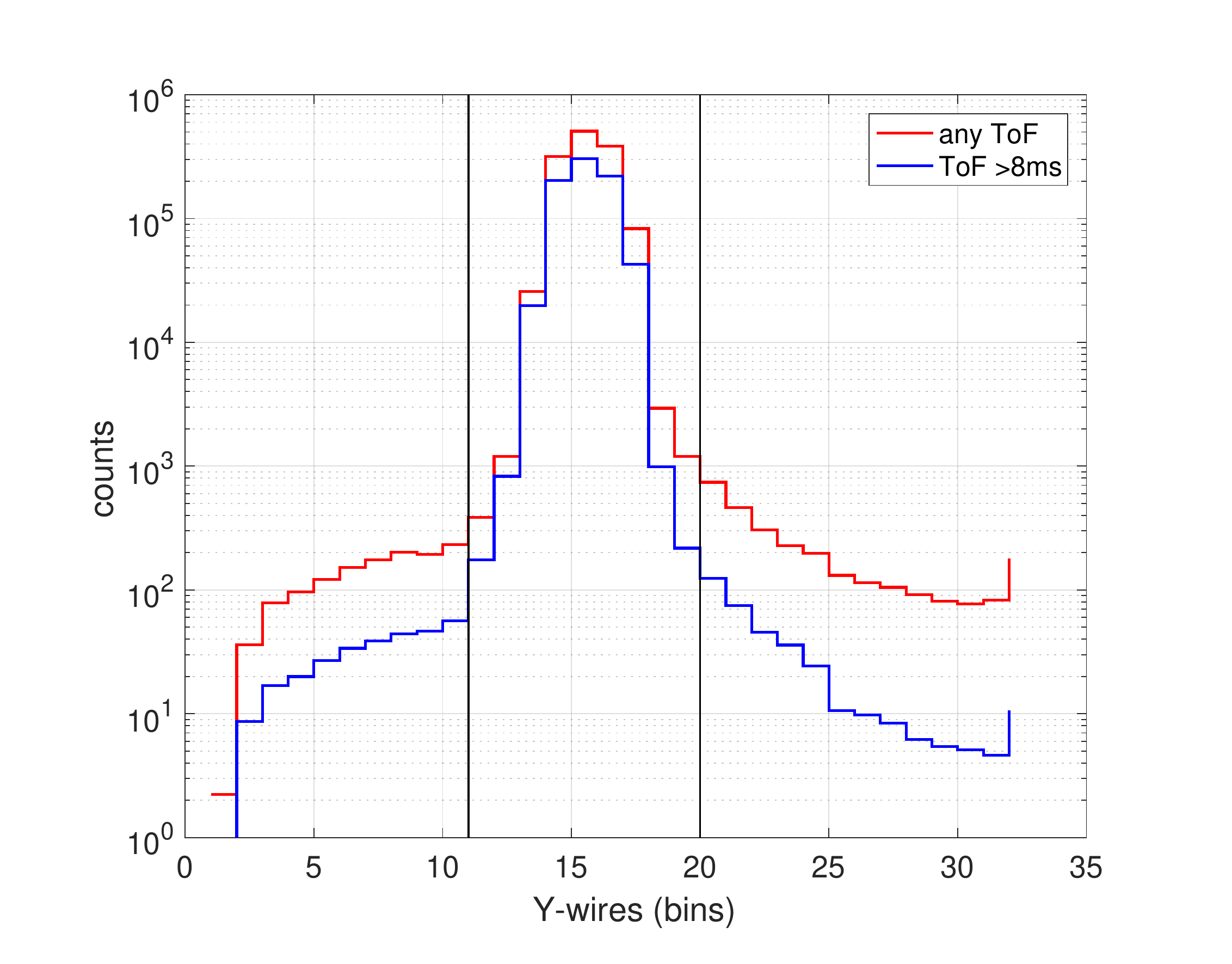}
\includegraphics[width=.49\textwidth,keepaspectratio]{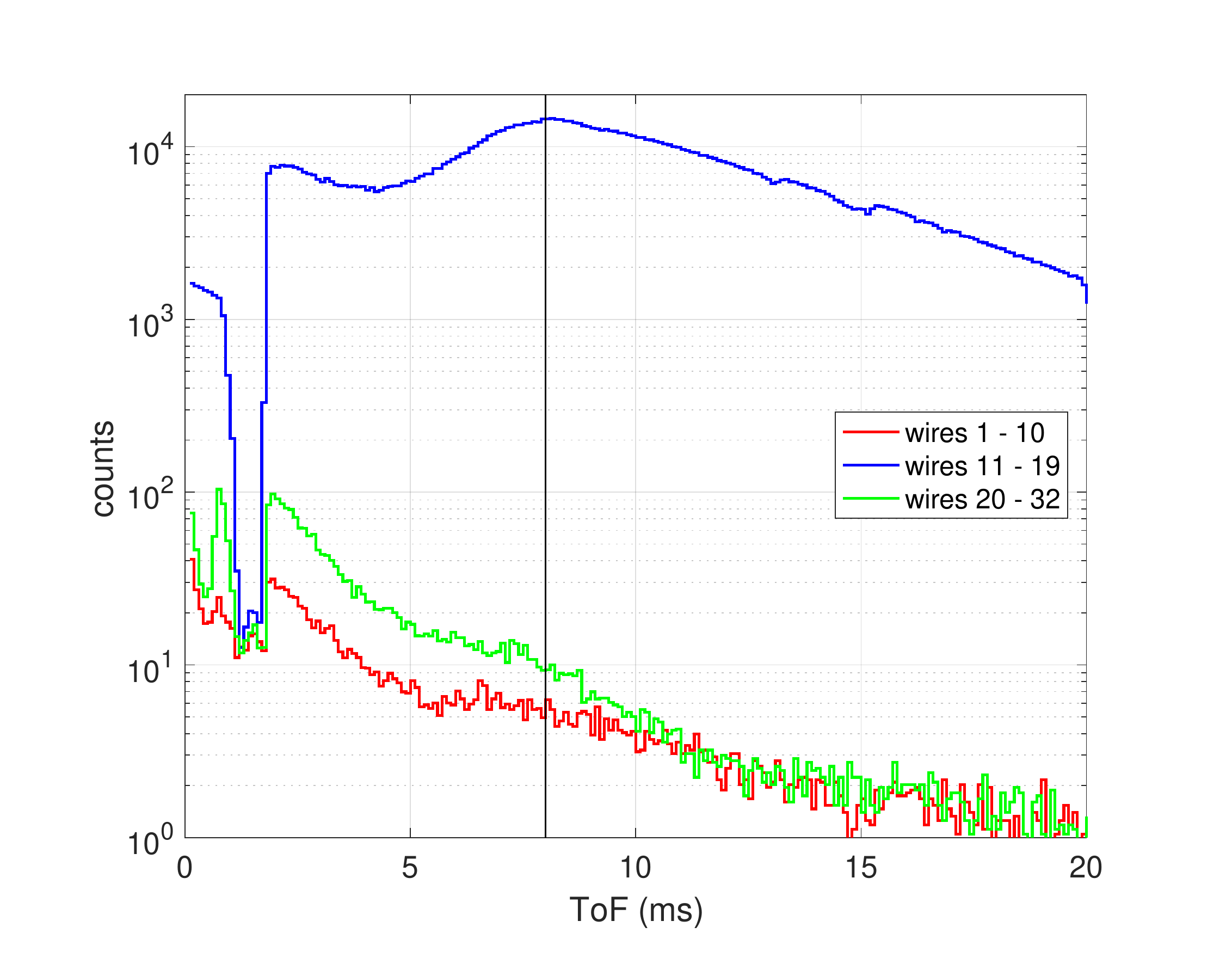}
\caption{\label{fig7bis} \footnotesize Direct beam profile on wires integrated over the strips for the full ToF spectrum and for events gated above $8\, ms$ (2.5\AA) (left). ToF spectra for wires in the direct beam (11-19) and for the wires in the tails of the direct beam (1-10 and 20-32) (right). Figure from~\cite{MIO_MB16CRISP_jinst}.}
\end{figure} 
\\ Figure~\ref{fig7bis} shows the profile of the direct beam on the wires of the illuminated cassette, integrated over the strips and the relative ToF spectra. The direct beam is comprised within wire no. 11 and no. 19. The tails of the beam extend all over the wire plane. it is shown the comparison of the beam profiles integrated over the full ToF spectrum and gated above $8\,$ms (2.5\AA) in order to decrease the contribution of the scattered neutrons within the detector as described in the previous subsection. The spatial dynamic range is about $10^4$ (peak to tail).
\\ The ToF spectra in Figure~\ref{fig7bis} is shown for the wires in the beam and for those in the tails. Above $8\,$ms, a difference of approximately 3 orders of magnitude in counting rate, in the same ToF bin, is visible between any wire in the interval 11-19 and any other in the tails (1-10 or 20-32). Moreover, the ToF spectrum varies by 3 orders of magnitude within two subsequent time bins (at $\approx 2\,$ms).  
\\ The measured dynamic range with the Multi-Blade detector is the actual dynamic range of the CRISP instrument~\cite{INSTR_OSMOND_CRISP}, a lower background environment would be required to determine the limits of the detector technology.

\subsection{Detection efficiency}\label{subeff}

\begin{figure}[htbp]
\centering
\includegraphics[width=.49\textwidth,keepaspectratio]{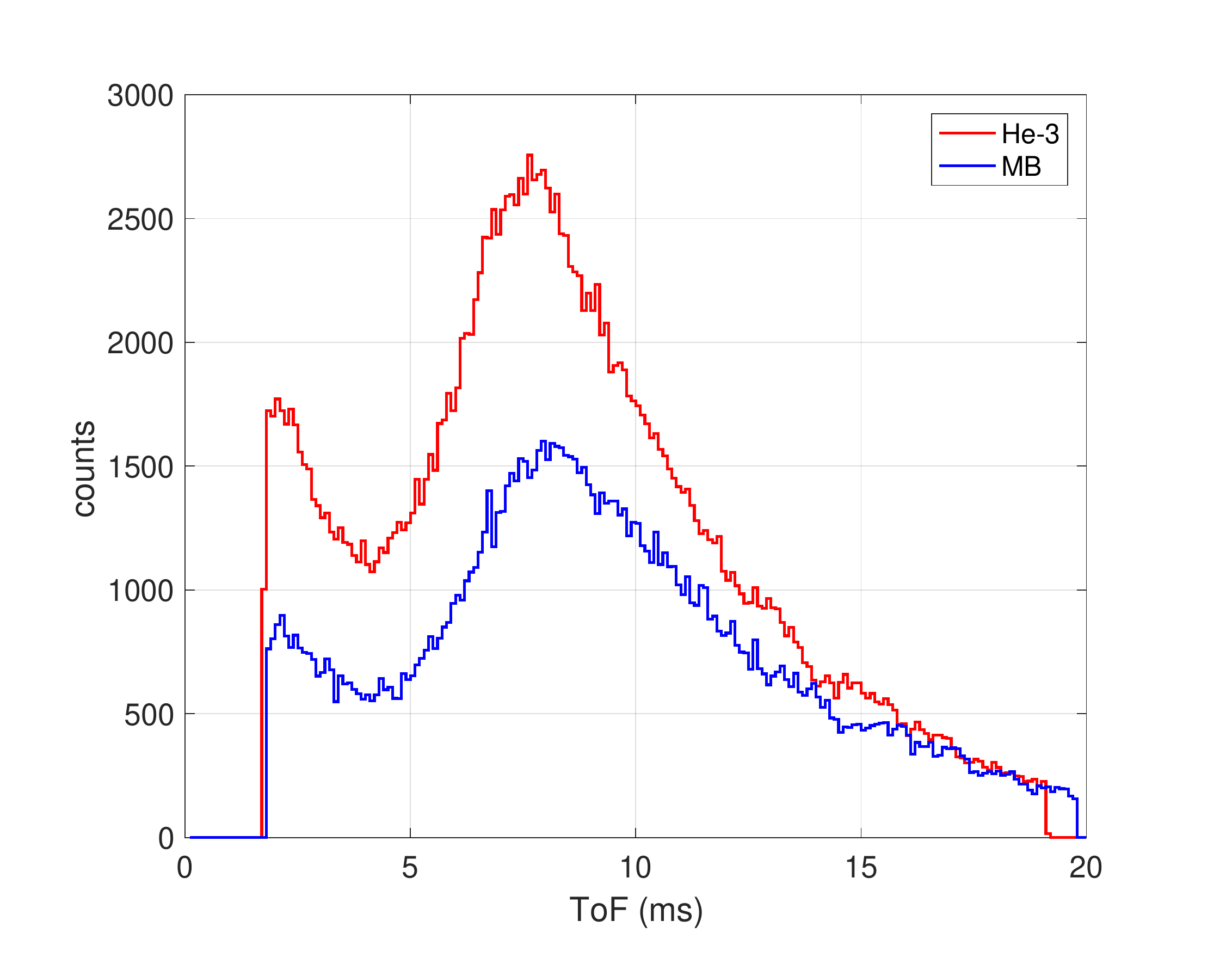}
\includegraphics[width=.49\textwidth,keepaspectratio]{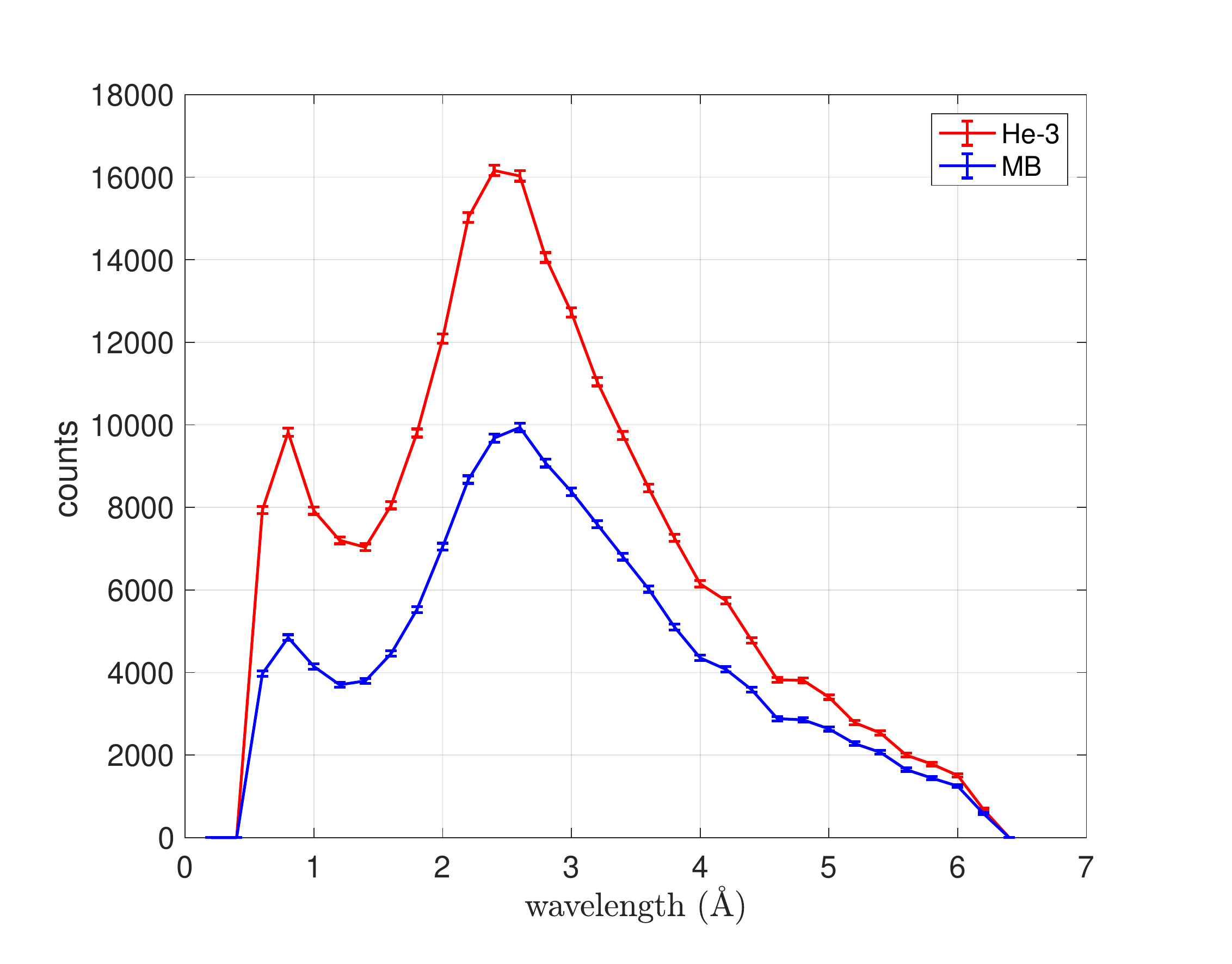}
\caption{\label{fig8} \footnotesize The raw uncorrected ToF spectra of the CRISP $\mathrm{^3He}$ detector and the Multi-Blade (left). The difference in the ToF spectrum of $0.5\,$ms for the slower neutrons is due to the fact that the $\mathrm{^3He}$ detector was $\approx 0.5\,$m closer to the sample than the Multi-Blade detector. The spectra of the CRISP $\mathrm{^3He}$ detector and the Multi-Blade as a function of the neutron wavelength (right). Note that the ToF spectrum of the Multi-Blade detector is corrected with the depth of the detector according to the equation~\ref{equadep}. Figure from~\cite{MIO_MB16CRISP_jinst}.}
\end{figure}

\begin{figure}[htbp]
\centering
\includegraphics[width=.49\textwidth,keepaspectratio]{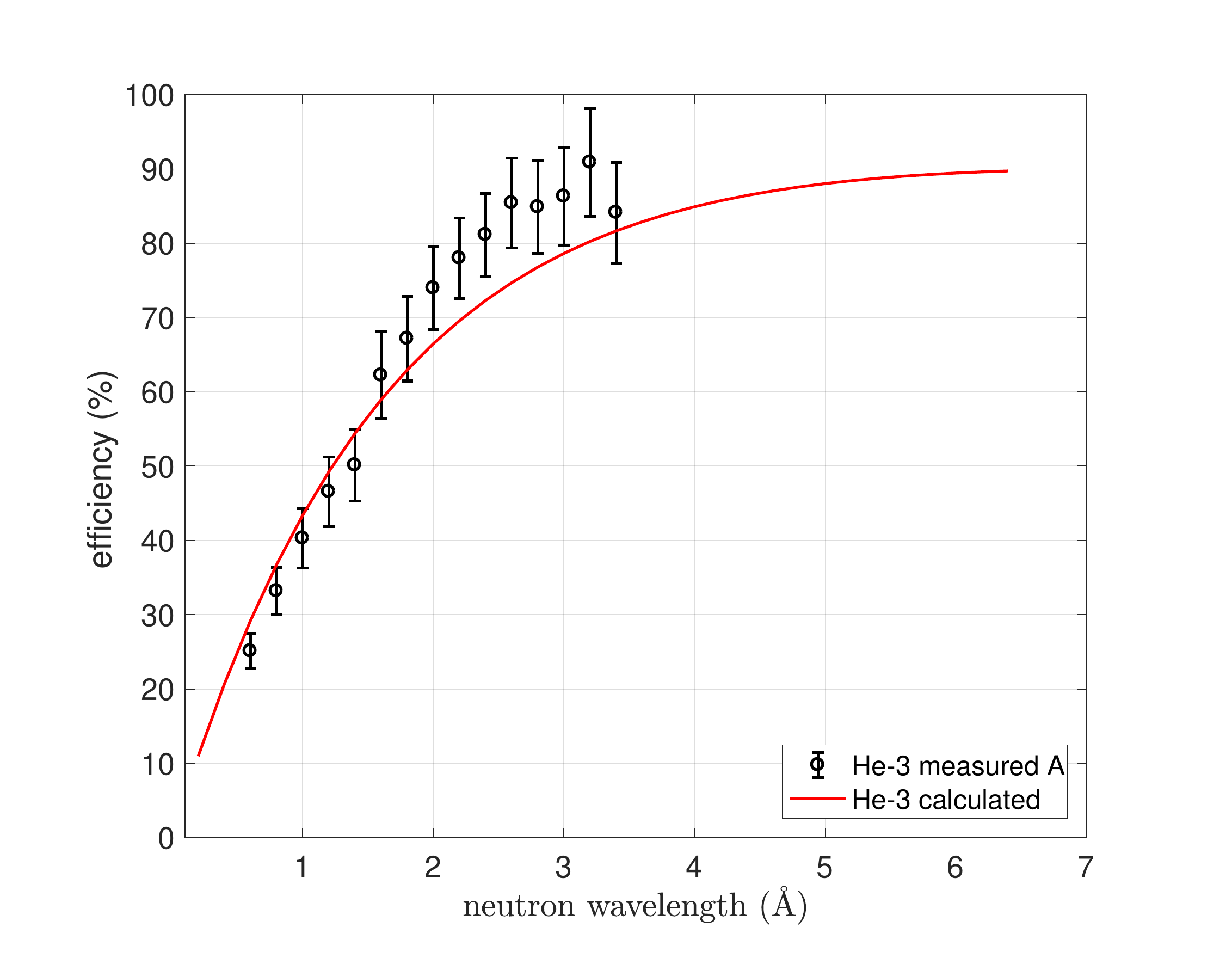}
\includegraphics[width=.49\textwidth,keepaspectratio]{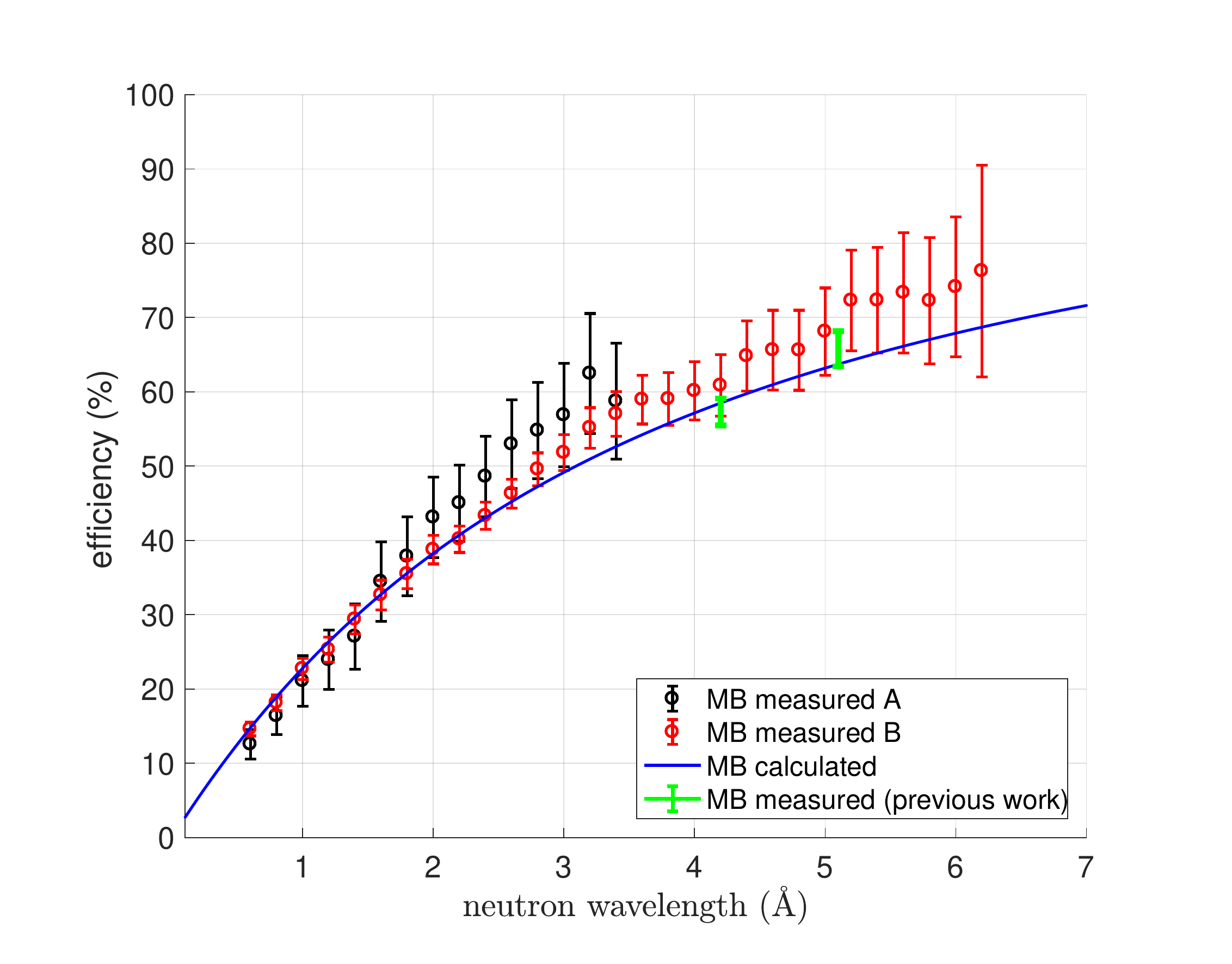}
\caption{\label{fig8bis} \footnotesize The measured and the calculated efficiency of the $\mathrm{^3He}$ detector on CRISP (left). The measured efficiency can be considered valid up to 3.5\AA\, ($11\,$ms). The efficiency is calculated assuming a $\mathrm{^3He}$ gas pressure of $3.5\,$bar. The efficiency of the Multi-Blade detector calculated according to~\cite{MIO_analyt} and measured against the $\mathrm{^3He}$ tube of CRISP (right). The efficiency is normalized assuming either the measured or calculated efficiency of the $\mathrm{^3He}$ detector, measurement A and measurement B, respectively. The two measured points from a previous detector charaterization~\cite{MIO_MB2017} are also shown. Figure from~\cite{MIO_MB16CRISP_jinst}.}
\end{figure}

The detection efficiency is defined as the ratio of detected neutrons over the incoming neutrons in the beam in a defined area. A set of data is recorded with the Multi-Blade detector using the direct beam. Software thresholds were applied to the data in order to discriminate against background events. The same configuration was used to illuminate the $\mathrm{^3He}$-detector of CRISP which was previously calibrated, thus its efficiency, as a function of the neutron wavelength, is known. The data for the $\mathrm{^3He}$-tube efficiency can be considered valid up to 3.5\AA\, ($11\,$ms) due to high background at larger wavelengths. Figure~\ref{fig8bis} (left) shows the measured efficiency of the CRISP detector up to 3.5\AA\,and the calculated efficiency for a $\mathrm{^3He}$ gas pressure of 3.5 bar. Since the absolute efficiency of the $\mathrm{^3He}$-detector is known, the absolute efficiency of the Multi-Blade can be calculated. The ratio between the ToF spectra of the two detectors defines their relative efficiency. This method is complementary and independent to the more commonly used method employing monochromatic pencil beams and it has different and independent systematic effects affecting the uncertainty on the final result.
\\ The counts in the beam are integrated over the spatial coordinates of the Multi-Blade and the resulting ToF spectrum is shown in figure~\ref{fig8} (left) along with the spectrum of the $\mathrm{^3He}$-tube. The $\mathrm{^3He}$ is physically $0.5\,$m closer to the sample than the Multi-Blade, this results into a difference in the ToF spectrum, i.e., the slower neutrons above 6\AA \, ($\approx 20\,$ms), arrive at the Multi-Blade detector with $0.5\,$ms delay with respect to the $\mathrm{^3He}$ tube. 
Hence, the two ToF spectra are slightly shifted and stretched relative to each other. Both the ToF spectra can be plotted as a function of the neutron wavelength by knowing the distance from the target: $12.6\,$m and $12.1\,$m for the Multi-Blade and the $\mathrm{^3He}$-tube respectively. Figure~\ref{fig8} (right) shows the spectra of the two detectors as a function of the neutron wavelength. Observe that the ToF spectrum of the Multi-Blade detector is corrected with the depth of the detector according to the equation~\ref{equadep}. 
\\ The ratio between the Multi-Blade and $\mathrm{^3He}$ detector spectra as a function of the neutron wavelength can be calculated and this gives us the relative efficiency of the two detectors. In order to calculate the absolute Multi-Blade efficiency the latter must be normalized to the absolute efficiency of the  $\mathrm{^3He}$ detector. The absolute Multi-Blade detector efficiency is shown in figure~\ref{fig8bis} (right) as a function of the neutron wavelength and compared to the theoretical efficiency calculated according to~\cite{MIO_analyt,MIO_decal}. The black points (measurement A) represent the calculated efficiency for the Multi-Blade by normalizing to the measured efficiency of the $\mathrm{^3He}$ detector, whereas the red points (measurement B) are the calculated efficiency for the Multi-Blade by normalizing to the $\mathrm{^3He}$ detector efficiency calculated from the $\mathrm{^3He}$ pressure ($3.5\,$bar). 
\\ The measured efficiency shifts systematically toward higher values at larger wavelengths, this is also visible in the measured $\mathrm{^3He}$ detector efficiency which propagates in the Multi-Blade efficiency normalization. The overall trend is as expected; the measurements are consistent with calculation~\cite{MIO_analyt}, with the exception of the points at the longest wavelength, where the Multi-Blade systematically tends towards the upper end of the expectation. Moreover, the obtained Multi-Blade efficiency agrees with the previously measured efficiency shown in~\cite{MIO_MB2017}. 

\subsection{Stability}

\begin{figure}[htbp]
\centering
\includegraphics[width=1\textwidth,keepaspectratio]{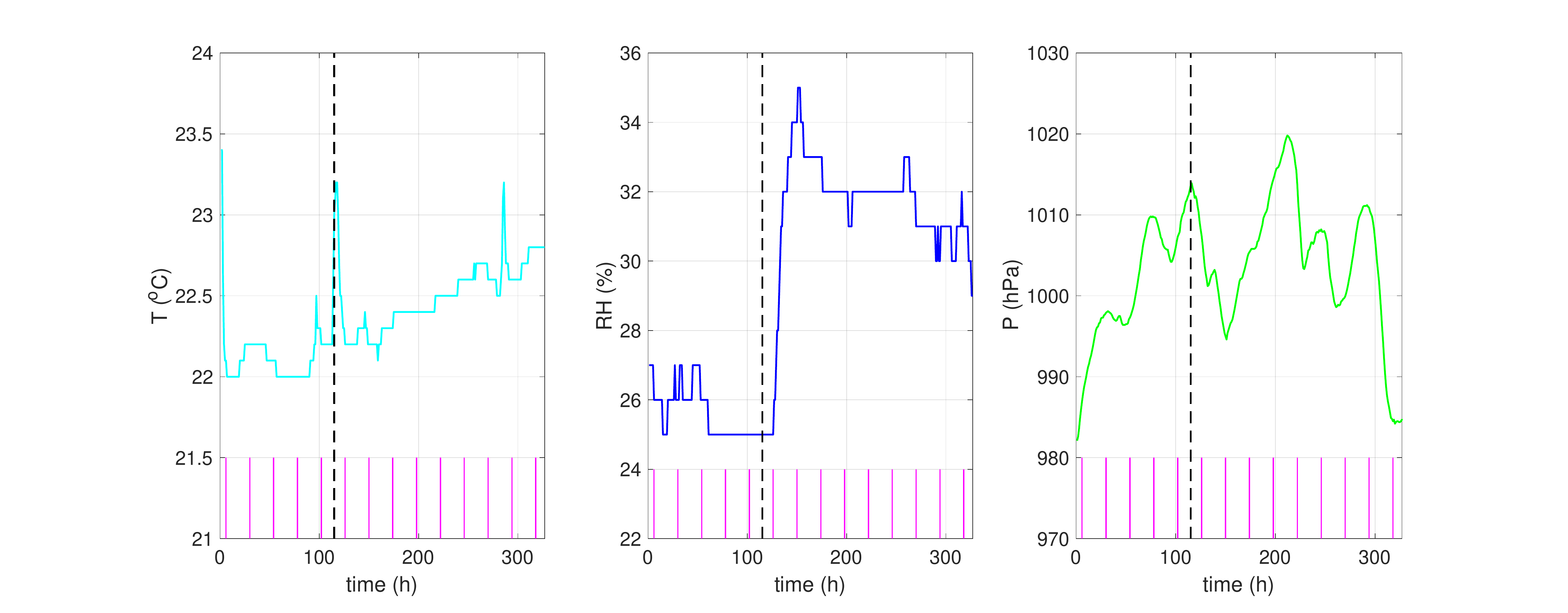}
\caption{\label{fig9bis} \footnotesize Trends of the recorded temperature (left), relative humidity (centre) and atmospheric pressure (right). The pink vertical lines indicate the midnight of each day. The vertical black dashed line indicates the change of the detector flow from $60\,$l/day to $30\,$l/day. Figure from~\cite{MIO_MB16CRISP_jinst}.}
\end{figure}

\begin{figure}[htbp]
\centering
\includegraphics[width=1\textwidth,keepaspectratio]{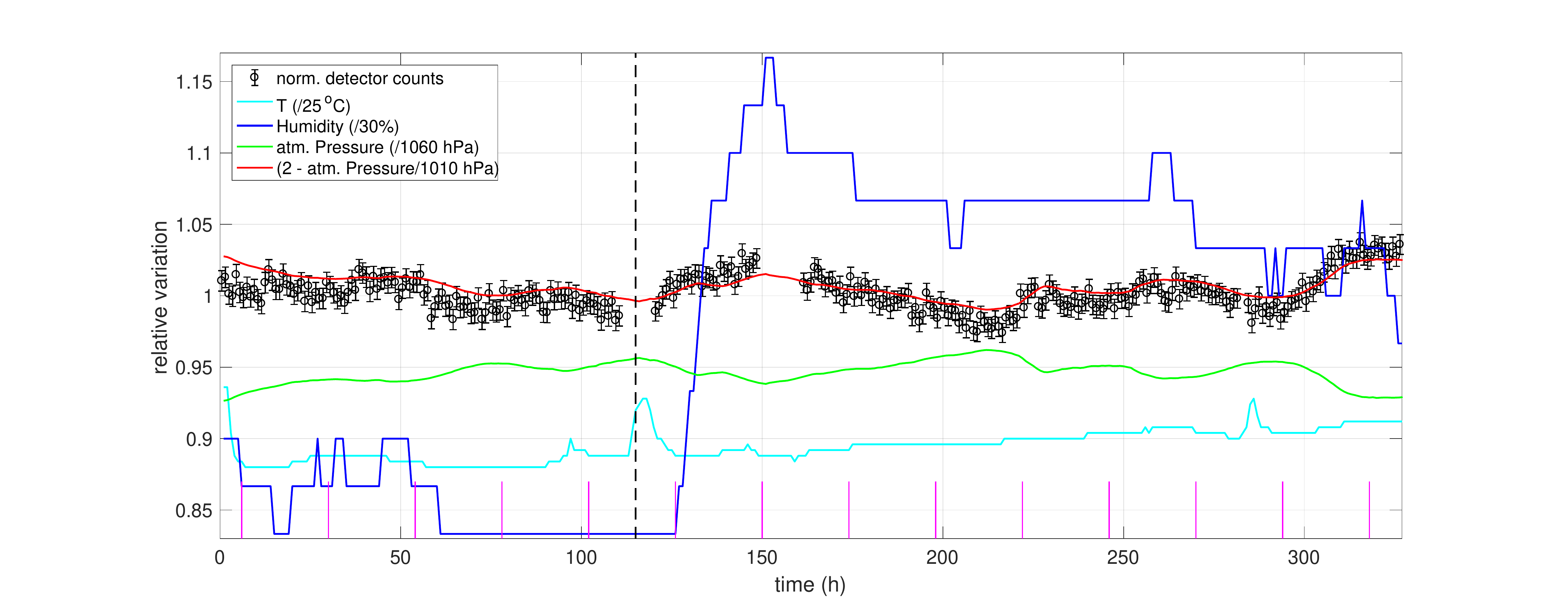}
\caption{\label{fig9} \footnotesize Normalized detector counts over 2 weeks along with the temperature, humidity and atmospheric pressure (also normalized). The atmospheric pressure is also shown as ($2-P/1010$hPa). The pink vertical lines indicate the midnight of each day. The vertical black dashed line indicates the change of the detector flow from $60\,$l/day to $30\,$l/day. Figure from~\cite{MIO_MB16CRISP_jinst}.}
\end{figure}

The gas gain of a MWPC is well described by the Diethorn's formula~\cite{Diethorn} and it is influenced by the atmospheric pressure and temperature variation as described in~\cite{Particle_Detection,DET_Stability}. A derivation of the relation can be found in section~\ref{sectionamplgas}.
\\ The Multi-Blade detector was placed in front of a moderated $\mathrm{Pu/Be}$ neutron source at the Source Testing Facility (STF) at the Lund University in Sweden. The detector was flushed with $\mathrm{Ar/CO_2}$ (80/20) at $2.4\,$l/h resulting into approximately $60\,$l/day. Since the detector vessel is approximately $30\,$l, the full gas volume is renewed twice per day. A set of data is recorded every hour for approximately two weeks. The total number of counts in the detector integrated over an hour, normalized to the average counting rate, is shown in figure~\ref{fig9} as a function of time. The temperature, humidity and atmospheric pressure were also monitored and they are shown in figure~\ref{fig9bis}. The temperature varied within $\approx0.5\,\%$ ($^o\mathrm{C}$), the relative humidity within $\approx30\,\%$ and the atmospheric pressure within $\approx4\,\%$ in the two weeks of the measurements. 
In Figure~\ref{fig9}, the trend of the atmospheric pressure is also shown as $2-P/1010$hPa) in order to better visualize the change of the detector counts with pressure. The missing points in the detector counts corresponds to periods when the Source Testing Facility (STF) was occupied with other users and background was altered, thus those points have been disregard. 
\begin{figure}[htbp]
\centering
\includegraphics[width=.7\textwidth,keepaspectratio]{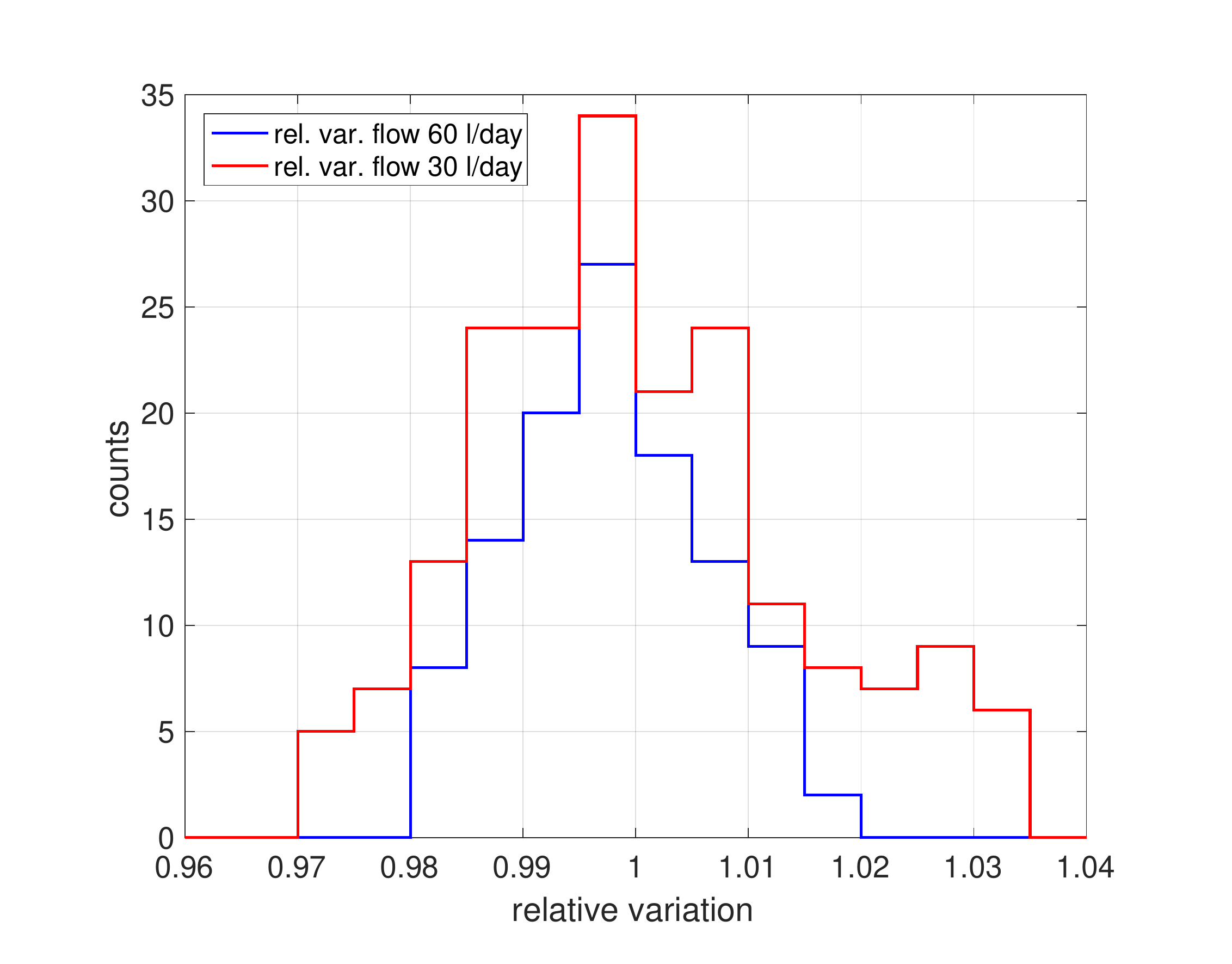}
\caption{\label{fig9bis2} \footnotesize Histogram of the relative variation of the detector counts for the first 115 hours with a gas flow of $60\,$l/day, and for the rest of the measurements with a gas flow of $30\,$l/day.}
\end{figure} 
\\ After approximately 115 hours of measurement (see the vertical black dashed line in the plot in Figure~\ref{fig9}) the detector flow was changed to one volume per day, i.e., $\approx 1.7\,$l/h ($30\,$l/day). In the first 115 hours the atmospheric pressure relative variation is $\approx \pm 1.5\%$ and the detector counts variation is $\approx \pm 1.7\%$ (expressed as a range (max-min)/2). Figure~\ref{fig9bis2} shows the histogram of the relative variation of the detector counts for the two gas flow regimes ($60\,$l/day and $30\,$l/day).
\\ When the flow is reduced, the atmospheric pressure relative variation is $\approx \pm 1.7\%$ and the detector counts variation is $\approx \pm 3\%$. 
\\ In either cases of gas renewal of 1 or 2 volumes per day, no contamination of the gas due to pollutant in the detector has been observed. A steady decrease of the counts would be visible otherwise.
\\ The trend in detector counts is clearly influenced by the atmospheric pressure. Although the atmospheric pressure variation was comparable in the two configurations of flow, the detector counts are influenced more when the flow is lower. This has to do with the higher over-pressure set in the detector to allow a larger flow. A larger over-pressure in the detector is affected by the variation of the atmospheric pressure to a minor extent. 
\\ The counting rate in the detector is stable within $\pm 1.7\%$ during several days with a flow that replace approximately 2 detector volumes per day. 
\\ In order to further improve the detector stability in time, the gas gain or thresholds must be adjusted according to the atmospheric pressure and temperature variations. Thus, an active feedback on the signal thresholds or on the high voltage as shown in~\cite{DET_Stability} or a post-processing of the data can be used. 

\subsection{Overlap, uniformity and linearity}\label{linea}

\begin{figure}[htbp]
\centering
\includegraphics[width=.8\textwidth,keepaspectratio]{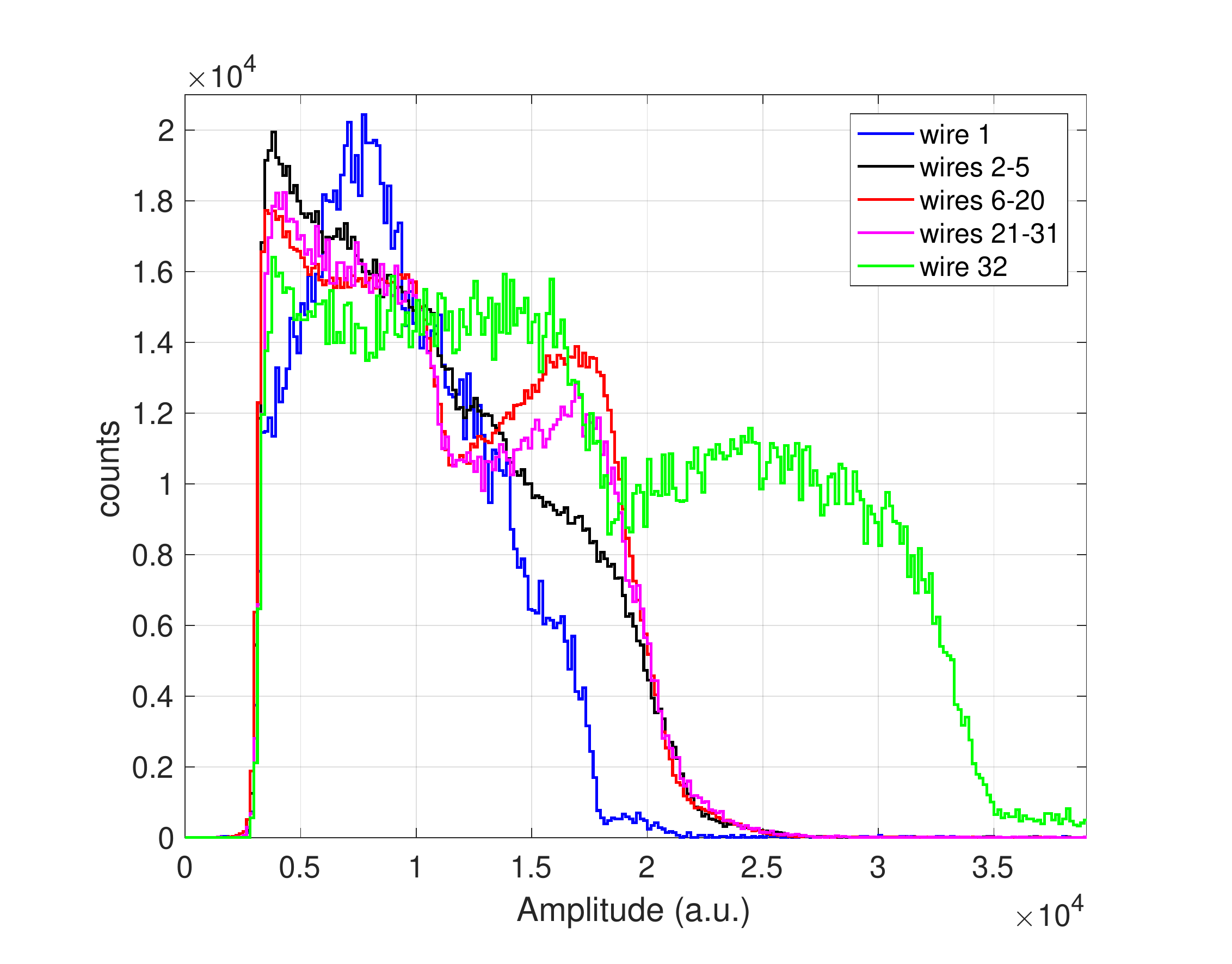}
\caption{\label{fig10} \footnotesize Pulse Height Spectrum (PHS) for individual wires or group of wires in one cassette of the Multi-Blade detector. Wire 1 is at the front of the detector and wire 32 is at the back. The gas gain varies due to the geometry of the cassette. Figure from~\cite{MIO_MB16CRISP_jinst}.}
\end{figure}

\begin{figure}[htbp]
\centering
\includegraphics[width=.49\textwidth,keepaspectratio]{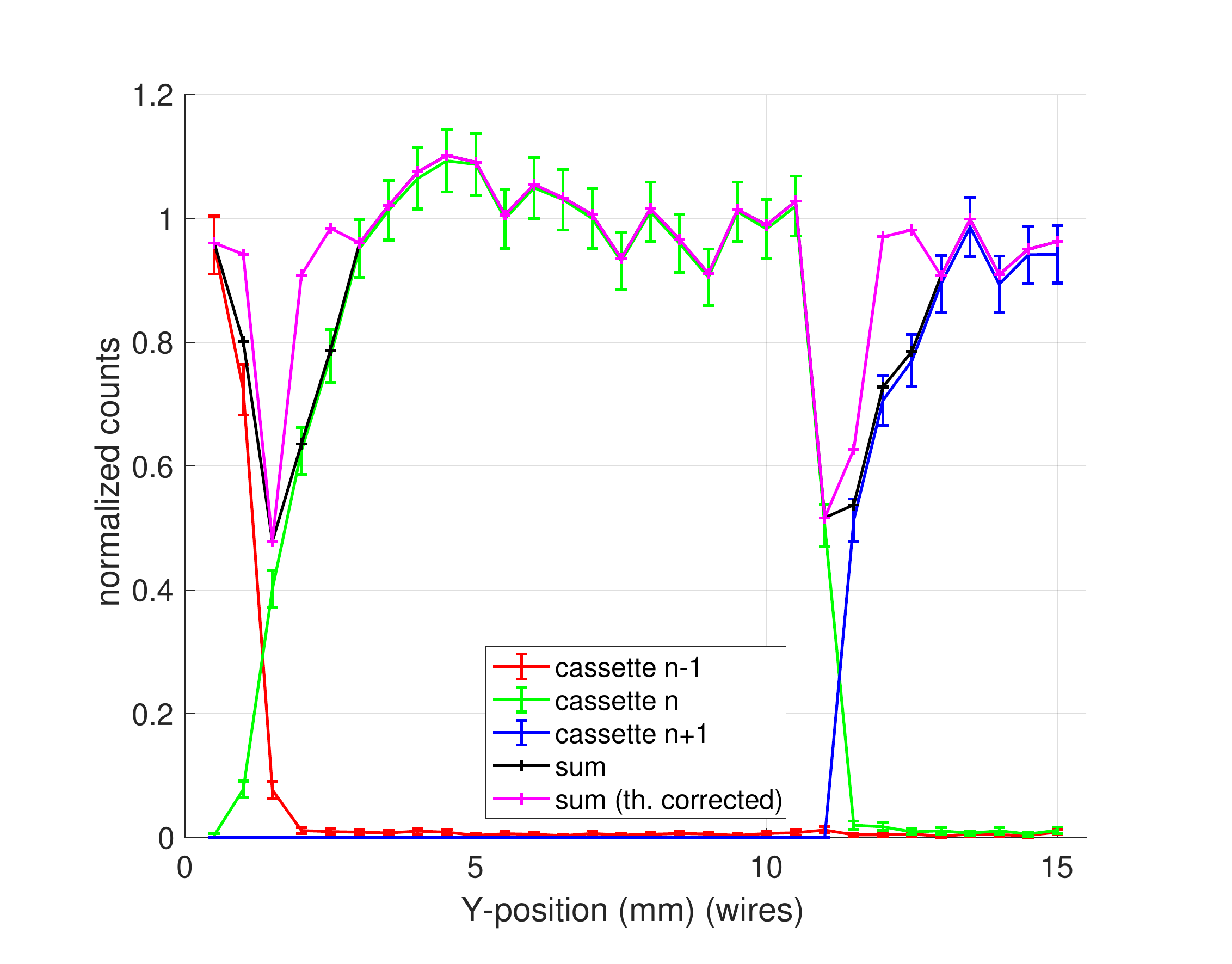}
\includegraphics[width=.49\textwidth,keepaspectratio]{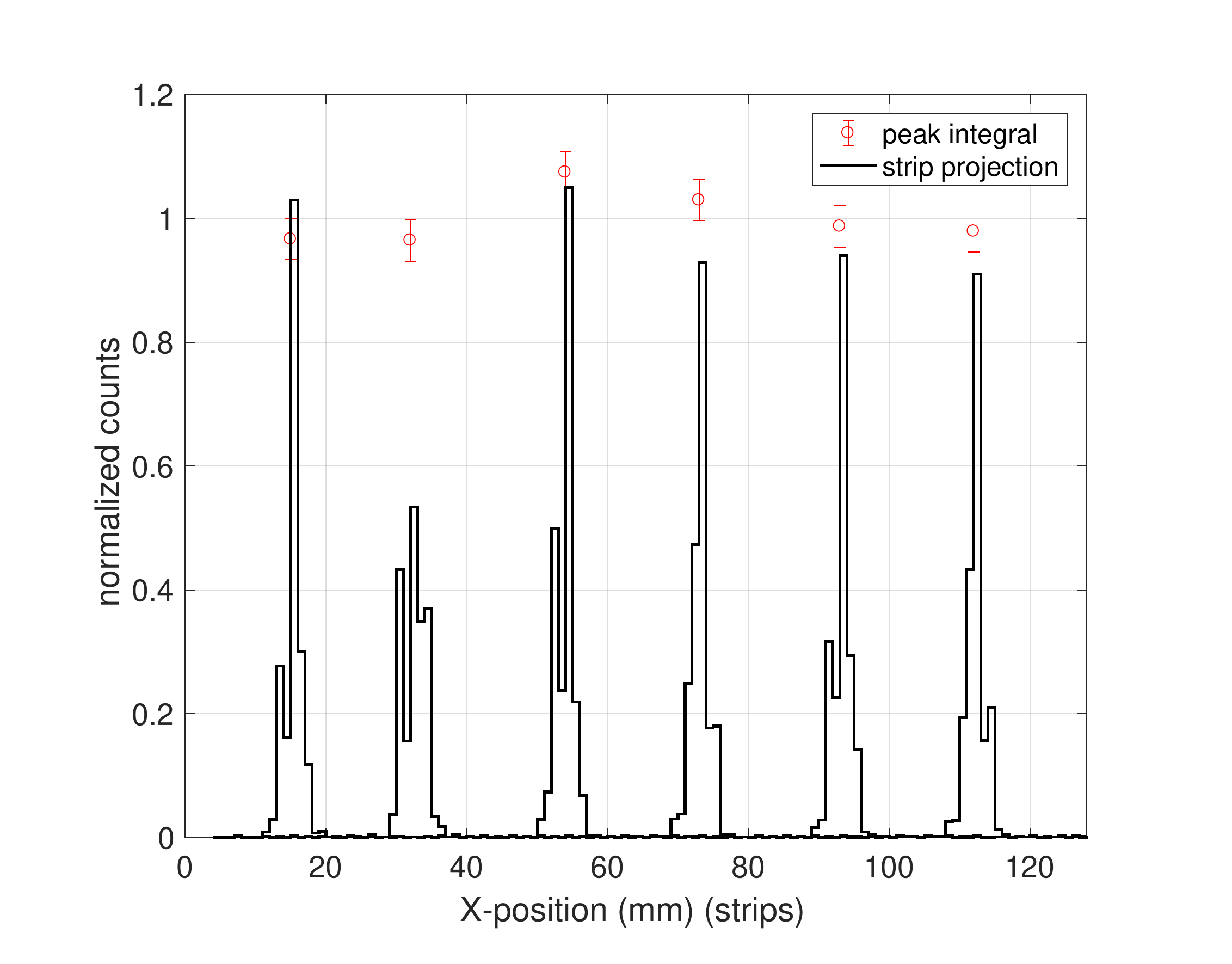}
\caption{\label{fig11} \footnotesize The normalized counts of the Multi-Blade detector scanned across the wires of three adjacent cassettes with a $0.5\,$mm step (left). In black the sum of the three cassettes and in pink the counts in the three cassettes when the threshold is adjusted for individual channels. The normalized counts of the Multi-Blade detector scanned across the strips of one cassette in steps of $\,20$mm (right). The red points are the integral of the peaks.}
\end{figure}

Due to the blade geometry the gas gain differs for different wires within a cassette. Each cassette has 32 wires and 32 strips. The wire no.1 is labelled as the one closer to the sample position (see Figure~\ref{fig99}), i.e., at the front of each cassette, and the wire no. 32 the one at the back. Electric field simulations and measurements have been carried out to investigate the gain variation at each wire due to the geometry and they have been described previously in~\cite{MIO_MB2017}. The gain is approximately constant for each wire from no. 8 to no. 31. The wire no. 32 has appropriately a double gain due to the lack of a neighbour. Each of the wires from no. 1 to no. 7 have a different and smaller gas gain with respect to those between no. 8 and no. 31. The gain drops at the first 7 wires can be compensated by adjusting individual thresholds, in hardware or software, on each channel.
Otherwise the gas gain compensation can be implemented with a separated high voltage supplies at the first wires. Figure~\ref{fig10} shows the Pulse Height Spectrum (PHS) for the wires in a cassette. The peaks from the $\mathrm{n(^{10}B, \alpha)^7Li}$ reaction can be identified for any wire apart from the front wires (from no. 1 to no. 5). Even if the threshold is adjusted according to the gain, at the very first wire there is a loss in gain which corresponds to a drop of $50\%$ with respect to the nominal efficiency. This region of reduced sensitivity, is where two cassettes overlap and it is about $0.5\,$mm wide. 
\\ The Multi-Blade detector was scanned across three cassettes and with a collimated beam. A set of data was recorded for each position in steps of $0.5\,$mm. Figure~\ref{fig11} (left plot) shows the normalized counts of the detector as for each position and for each cassette (red, green and blue curves). The black curve in the plot is the sum over the 3 cassettes with no threshold correction applied: the counts drop due to the gain variation in the two overlap regions scanned. The pink curve is then obtained if the thresholds are adjusted for individual channels and this results in the reduced sensitivity region of about $0.5\,$mm. This meets expectation but it can be improved. 
The scan was repeated across the strips in steps of about $20\,$mm. Figure~\ref{fig11} (right plot) shows the normalized counts for each position of the scan and the integral of the counts (also normalized) in each peak. The overall variation of the gain, i.e., the uniformity, in the scanned cassette in both directions (wires and strips) is $\pm 10\%$.

\subsection{Masks and reconstruction algorithms}\label{spatresd}
A set of images of BN (Boron-Nitride, HeBoSint C100~\cite{hebosint}) masks were captured with the Multi-Blade on CRISP to investigate how the the position reconstruction algorithm affects the reconstructed image. The BN masks are $5\,$mm thick and natural enriched in boron. The typical attenuation of these masks is approximately $100\,\%$ from and above 1\AA. Two algorithms can be used to reconstruct the $(X,Y)$ coordinates from the raw data: a maximum amplitude (or area) algorithm (MAX) or a Center of Gravity algorithm (CoG). The first associates the hit of a cluster to the two channels (wire and strip) which have the maximum area (i.e., energy deposition) among the events in the cluster, thus the $X $and $Y$ coordinates are 32 integers for wires and 32 integers for strips. The MAX algorithm does not exploit the information about the multiplicity of an event. On the hand, the CoG algorithm uses the multiplicity in a cluster to better position the hit across the two coordinates. If two, or more, adjacent channels are firing at the same time and belong to the same cluster, the position of the hit is calculated weighting the energies (areas or amplitudes) of the channels. i.e., if two adjacent strips are firing at the same time and they perfectly share the energy in two identical parts, the hit will be placed exactly in the middle of the two. In the case of CoG are used 128 bins instead of 32 . The CoG is an continuous quantity and the resolution improves independently from the binning but it has been chosen to bin it in 4 times more bins. In the reconstructed position of the wires there is not much difference if the MAX or the CoG algorithm is used since the multiplicity of wires is often 1. It has been shown that the spatial resolution improves from $\approx0.6\,$mm to $\approx0.55\,$mm~\cite{MIO_MB2017}. On the other hand, the most common multiplicity on strips is 2, and the CoG algorithm improves significantly the spatial resolution for the strips. 

\begin{figure}[htbp]
\centering
\includegraphics[width=.95\textwidth,keepaspectratio]{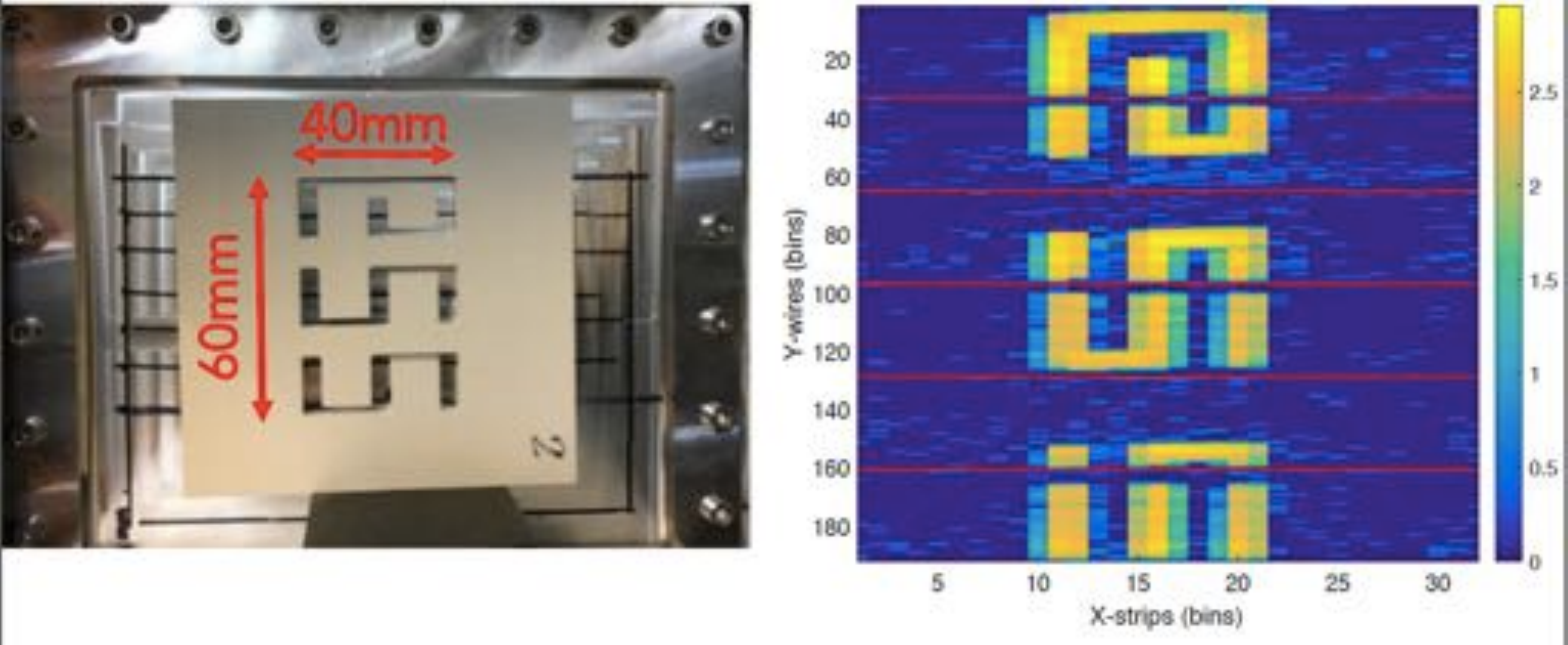}
\caption{\label{fig12} \footnotesize BN (Boron-Nitride) mask of the ESS logo positioned on the entrance window of the Multi-Blade detector (left). Raw reconstructed image of the ESS mask with the MAX algorithm showing all the channels (right). The color bar is shown in logarithmic scale and represents counts. The bin size on the X-axis is $4\,$mm and on the Y-axis is $0.35\,$mm. Figure from~\cite{MIO_MB16CRISP_jinst}.}
\end{figure}

\begin{figure}[htbp]
\centering
\includegraphics[width=.49\textwidth,keepaspectratio]{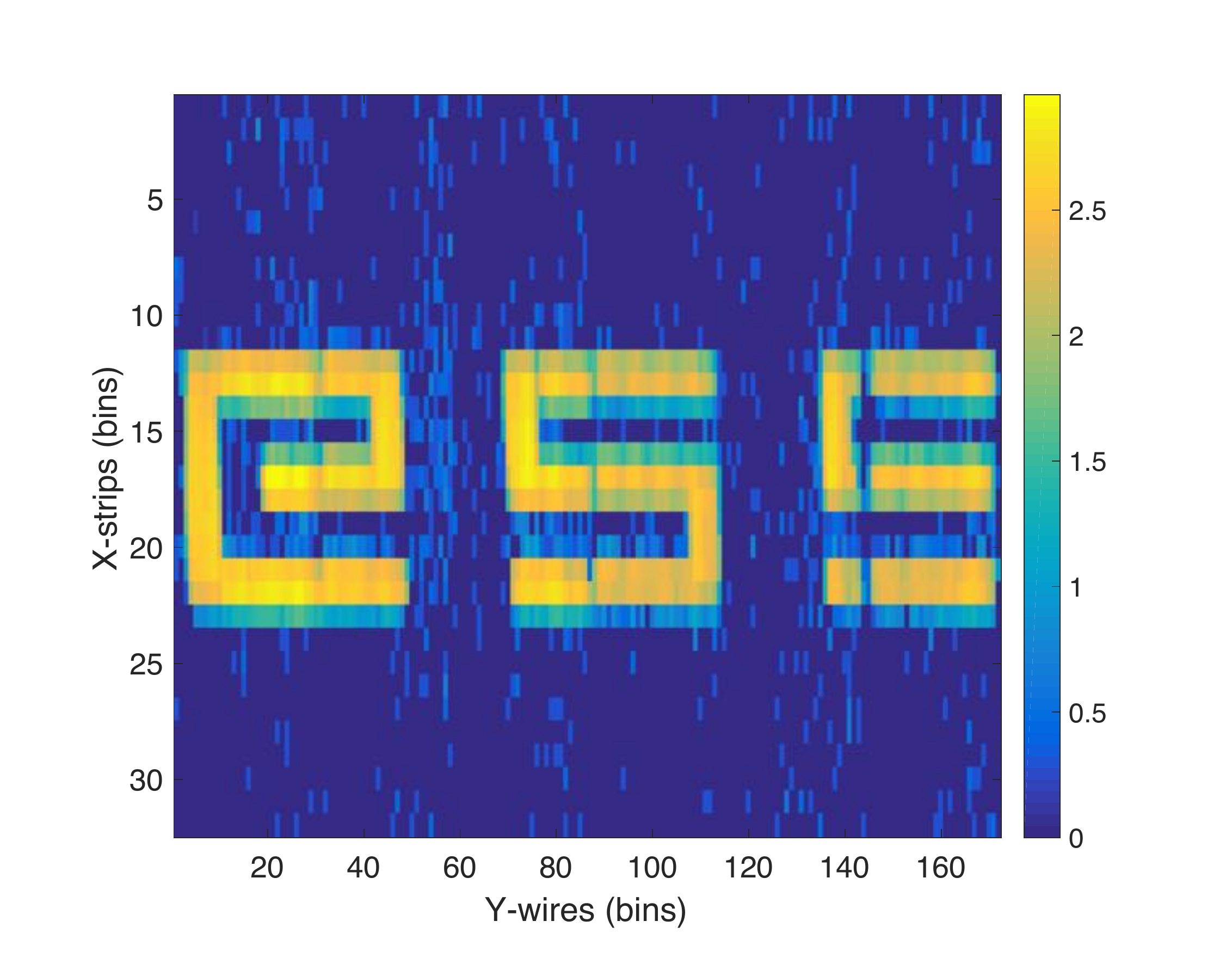}
\includegraphics[width=.49\textwidth,keepaspectratio]{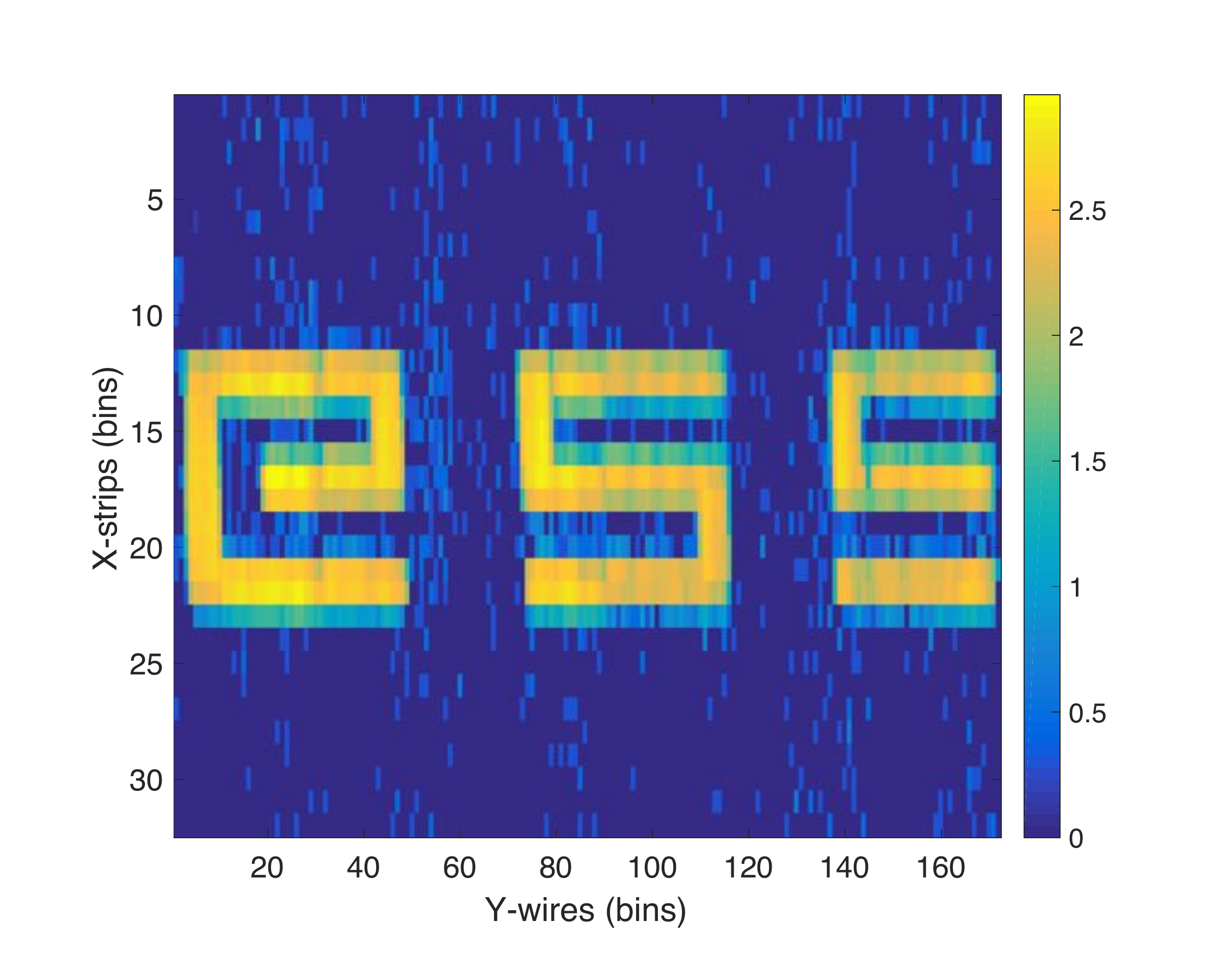}
\caption{\label{fig12bis} \footnotesize Raw reconstructed image of the ESS mask rotated by 90 degrees with shadowed channels removed. The correction is applied considering the $4\,$m (left) and $2.3\,$m (right) cassette array radius arrangement. The color bar is shown in logarithmic scale and represents counts. The bin size on the X-axis is $4\,$mm and on the Y-axis is $0.35\,$mm.Figure from~\cite{MIO_MB16CRISP_jinst}.}
\end{figure}

\begin{figure}[htbp]
\centering
\includegraphics[width=.95\textwidth,keepaspectratio]{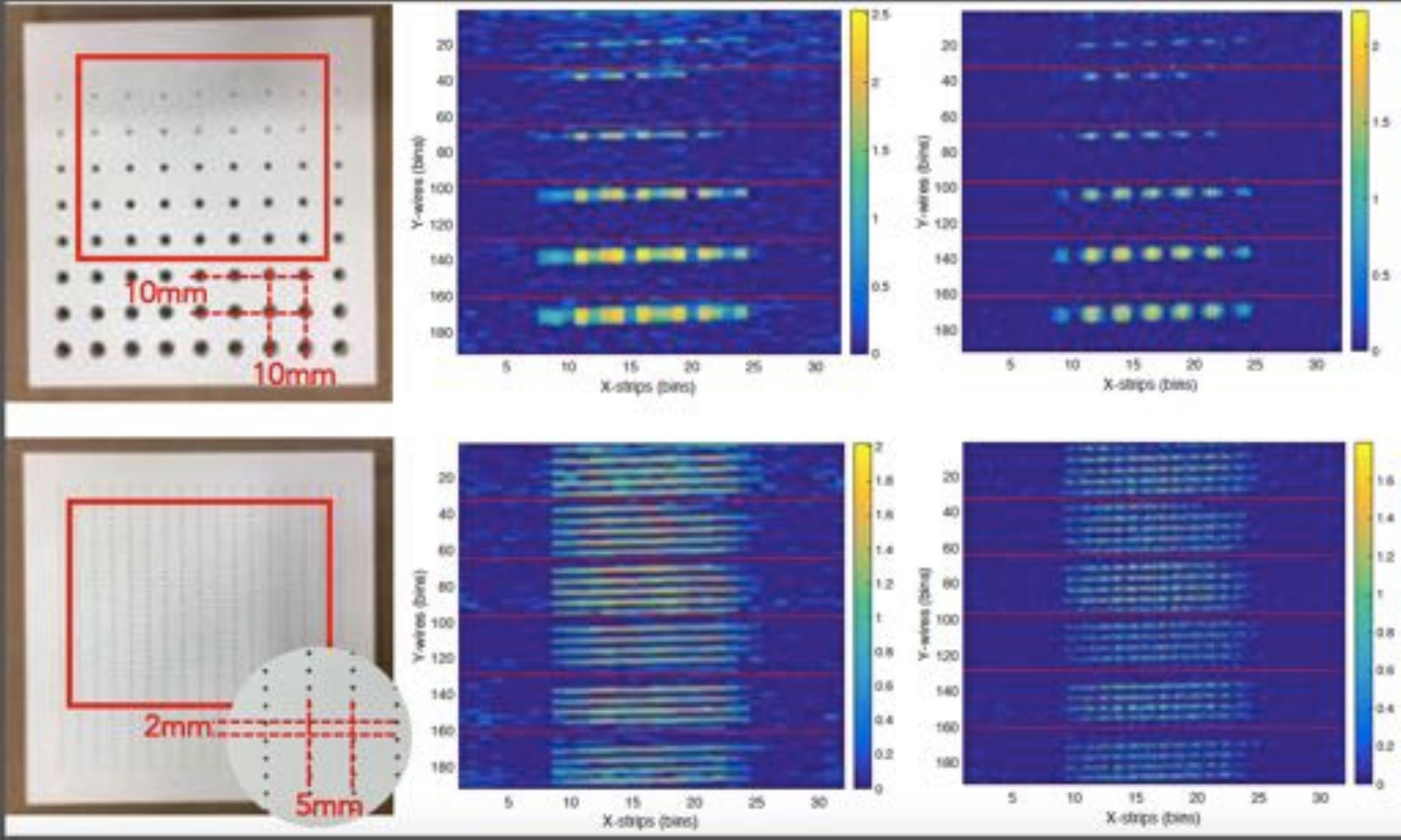}
\caption{\label{fig13} \footnotesize A picture of the BN masks (left) along with the reconstructed images with the MAX (center) and the CoG (right) algorithms. The color bar is shown in logarithmic scale and represents counts.}
\end{figure}

\begin{figure}[htbp]
\centering
\includegraphics[width=0.95\textwidth,keepaspectratio]{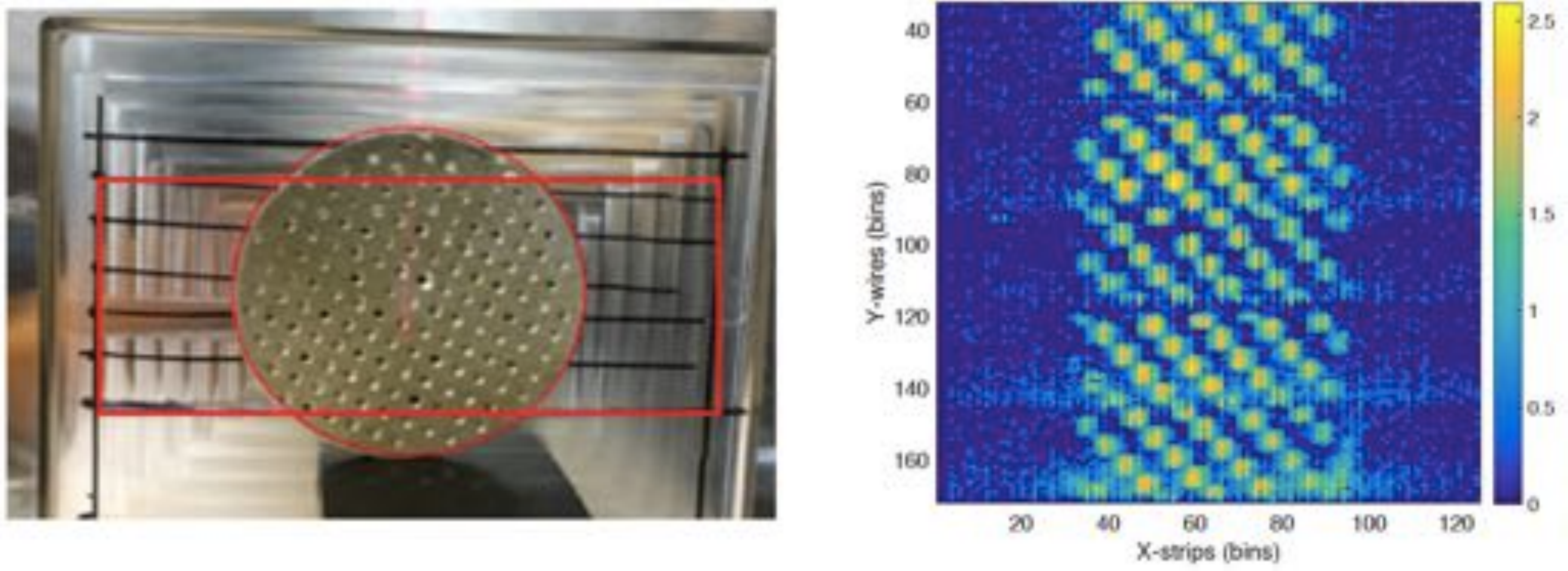}
\caption{\label{figMisis} \footnotesize A picture of the mask (left) along with the reconstructed image with the CoG (right) algorithm. The color bar is shown in logarithmic scale and represents counts.}
\end{figure}

\begin{figure}[htbp]
\centering
\includegraphics[width=1\textwidth,keepaspectratio]{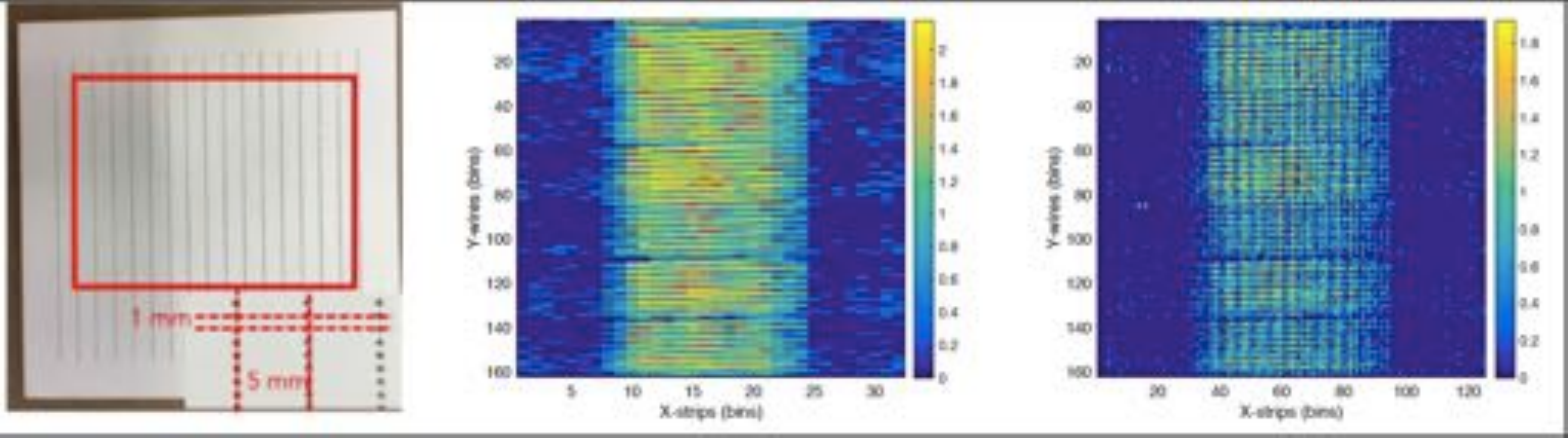}
\caption{\label{figM4} \footnotesize A picture of a BN mask with the closest gap between two adjacent rows, 1 mm (left) along with the reconstructed image the MAX (center) and the CoG (right) algorithms. The color bar is shown in logarithmic scale and represents counts.}
\end{figure}

All the images shown in this section are gated in ToF above $8\,$ms (2.5\AA). As the direct beam at the instrument is narrow and not enough to illuminate entirely the masks, a super-mirror (Fe/Si multi-layer) at the sample position was used. During the acquisition the sample angle was changed steadily in order to scan across the mask positioned at the detector window. 
\\ The first mask used in the tests is shown in figure~\ref{fig12} (left) and it reproduces the ESS logo. This mask was used to study what happens at the overlap between cassettes. The image reconstructed with the MAX algorithm is also shown in figure~\ref{fig12} (right). 
\\  In the present detector the cassettes are arranged over a circle of $4\,$m radius, as for the ESTIA configuration, and each blade is positioned with $0.14$ degrees angle with respect to the adjacent unit. In this configuration the shadowing of each cassette on the neighbour is of about 3 wires, i.e., 3 bins (of 32). The bin size on the X-axis (strips) is $4\,$mm and on the Y-axis (wires) is $0.35\,$mm: $\sin(5^o) \times 4\,$mm. The very last 3 wires of each cassette should not receive any neutron because they are physically hidden behind the neighbour blade (bins from 33 to 35, 65 to 67, 97 to 99, 129 to 131 and 161 to 163 in the image in figure~\ref{fig12}). 
\\ On CRISP the Multi-Blade is positioned at $2.3\,$m from the sample, and this distance was the maximum distance allowed by the CRISP setup. In this configuration, the sample-detector distance and the cassette array radius are not matching, and this does not allow to align all the cassettes of the detector at 5 degrees with respect to the incoming beam, but only one: the third from the top, bins from 64 to 96. Therefore, the shadowing is not expected to be constant across the cassettes. From the image in figure~\ref{fig12} (right), the shadowing is between 2 and 3 wires (or bins from 33 to 35) for the top cassette and almost 5 wires for the bottom cassette (bins from 161 to 165). It is important to note that this is not a dead area of the detector and this shadowing effect can be removed from the images without losing any information. Figure~\ref{fig12bis} shows the reconstructed images (rotated) when this effect is removed. In the left image the shadows are removed according to the $4\,$m array radius, i.e., 3 bins for each cassette. The larger the distance from the aligned cassette (bins from 64 to 96), the wider is the shadow. On the other hand, in the right image, the correction is applied according to the geometry at $2.3\,$m on CRISP, thus the number of removed bins is not constant across the detector and the shadows disappear completely.
\\ As described in section~\ref{linea}, the blades overlap with a region of reduced sensitivity due to the reduced charge collection at the very first wire of each cassette (see the pink curve in the left plot in Figure~\ref{fig11}). When removing the shadows the reduced sensitivity area is visible in figure~\ref{fig12bis} at the bin position 32, 89 and 143.
\\ Two more BN masks were used in the tests in order to compare the reconstruction algorithms. Figure~\ref{fig13} shows a picture of the masks (left) along with the reconstructed images with the MAX (center) and the CoG (right) algorithms. The red square marked on the picture of the masks represents the  area illuminated with neutrons that is reconstructed in the images. The CoG algorithm is applied only to reconstruct the strip position (X-axis in the plots) because the improvement on the wires is not significant. Note that the rows of holes in both the reconstructed images are slightly tilted because the detector was rotated by approximately 2 degrees with respect to the horizontal. 
\\ The reconstructed images in figure~\ref{fig13} show a good separation across the wires (Y axis) being the resolution of the detector much better in this direction with respect to the other direction. With the MAX algorithm the separation across the strips (X axis) is barely visible. When the CoG algorithm is applied, the good separation between the $10\,$mm holes of the top mask appears and the separation between the $5\,$mm holes of the mask at the bottom start to be visible in the image. 
\\ Note that this is not a measurement of the spatial resolution of the detector, because the actual footprint of the neutron spot on the detector is not defined due to the unknown divergence of the neutron beam through each hole of the mask. For instance, the actual $0.5\,$mm diameter of the holes of the bottom mask is instead wider, due to the absorption of BN being a significant fraction of the thickness of the mask, and unknown at the detector. 
\\ A figure of merit ($fom$) of the improvement of using the CoG algorithm with respect to the MAX algorithm can be defined as the average ratio between the counts at the peaks and the counts between the peaks. For the top image in figure~\ref{fig13} a $fom\approx3.6$ is obtained for the MAX algorithm and $fom\approx5.7$ for the CoG algorithm when this is applied to the strip signals. 
\\Other results are shown in figure~\ref{figMisis} and~\ref{figM4}.

\section{Scientific Results}\label{secsci}

A series of scientific measurements with several samples have been performed on the CRISP reflectometer. The results presented in the manuscript refer to the paper published on this topic~\cite{MIO_ScientificMBcrisp}. The aim of this test is not only to prove the capabilities of the detector in an actual instrument, but to show as well the improvements that arise from operating the CRISP reflectometer in a configuration which reproduces the ESTIA operation mode. This is exclusively possible by exploiting the features of the Multi-Blade. The measurements provide a link between science instrument and detector improvements.
\\ As described in section~\ref{scatt} the triplets $(X,Y,ToF)$ that identify a neutron event, can be represented by two-dimensional plots. Referring to figure~\ref{fig6}, it is possible to integrate the ToF over the $X$ coordinate and to gate in the $Y$ coordinated around the direct beam area. Thus, the spectrum of the direct beam in ToF is obtained and it is used to normalize the reflectivity measurements of the samples described in the following sections. The time binning of $100\,\mu$s is chosen to match the ISIS-Target Station 1 pulse length.

\begin{figure}[htbp]
\centering
\includegraphics[width=.8\textwidth,keepaspectratio]{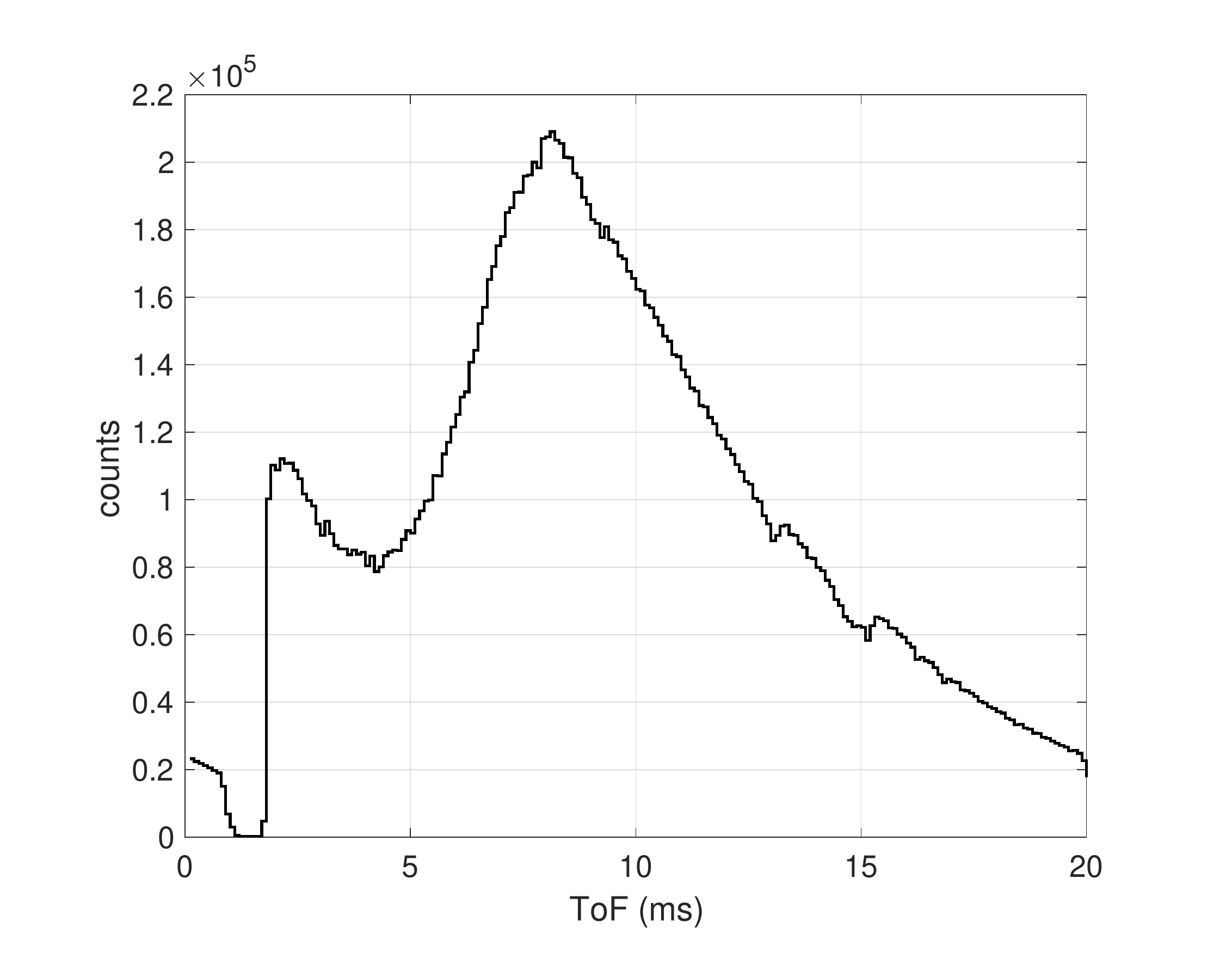}
\caption{\label{fig4sc} \footnotesize Intensity of the direct beam in ToF, integrated over the $X$-coordinate and gated in the $Y$-coordinate around the beam intensity. The bin size is 100 $\mu$s on the ToF axis. Figure from~\cite{MIO_ScientificMBcrisp}.}
\end{figure}

The gate in ToF is applied in order to reject the background arising from the spurious scattering from the substrate of the cassettes (the blades). This effect is due to the neutrons that cross the $\mathrm{^{10}B_4 C}$ layer without being absorbed. They are scattered by the substrate and detected in the other cassettes. The background is explained in detail in section~\ref{scatt}. 

In the last paragraph some results from the first test carried out with the actual demonstrator at the the Budapest Neutron Centre (BNC)~\cite{FAC_BNC} are presented. It must be recall that this was a test for the electronics, therefore the results obtained have not been published and are presented in the manuscript for the first time.

\subsection{Specular reflectometry on Ir sample: improvement of the q-resolution with the detector spatial resolution}\label{secIr}

\begin{figure}[htbp]
\centering
\includegraphics[width=.59\textwidth,keepaspectratio]{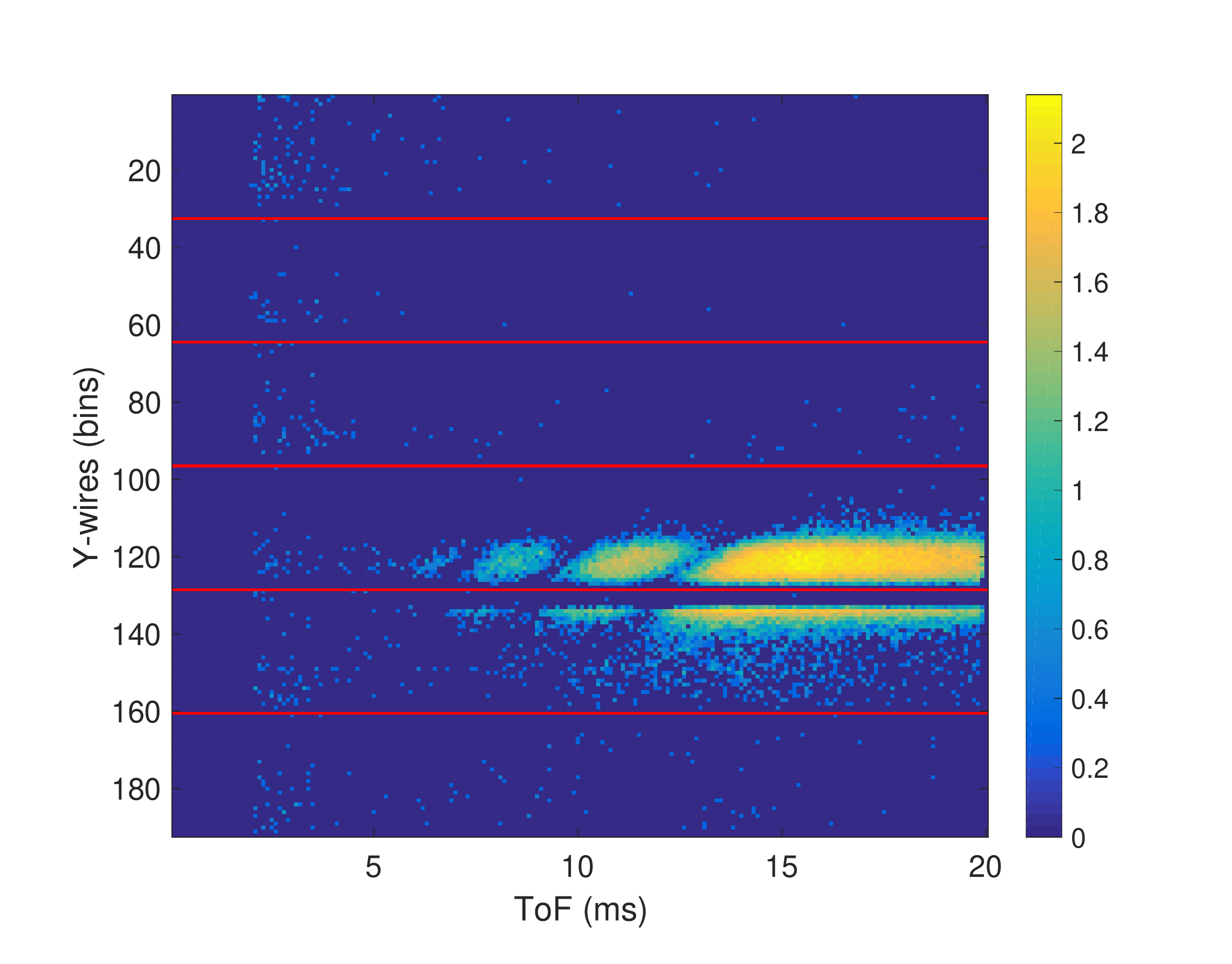}
\includegraphics[width=.59\textwidth,keepaspectratio]{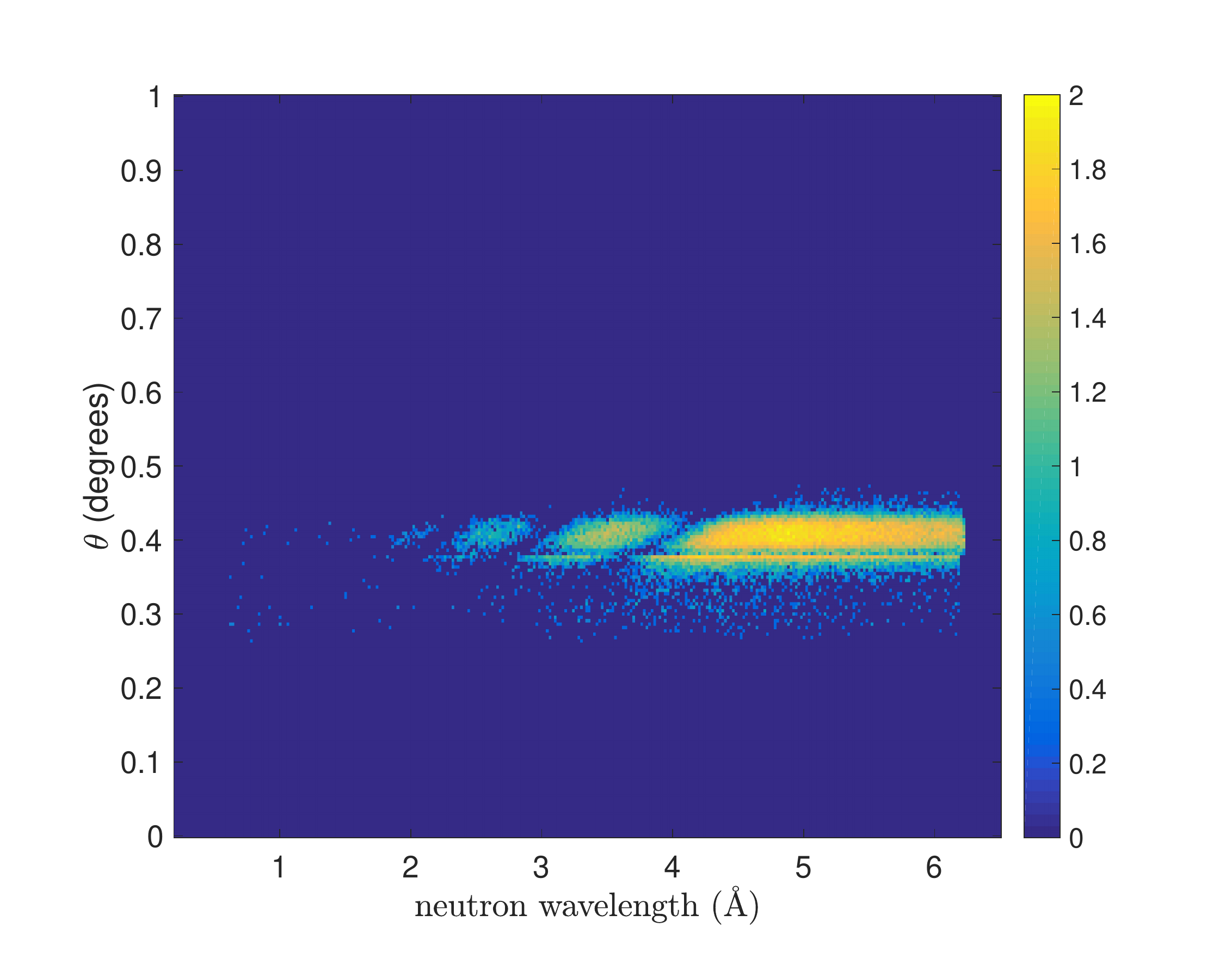}
\caption{\label{lt} \footnotesize Top: ToF spectrum of the reflected beam from the Ir sample. The bin size on the $Y$ axis is 0.35 mm and 100 $\mu$s on the $ToF$ axis. The horizontal lines depict the end of each cassette and the gap in between is the shadowing effect due to geometric properties of the detector. Bottom: ToF spectrum reduced in the $(\theta, \lambda)$ space. The gap does not represent a dead area, thus it can be removed without losing information as shown for the reduced data in $(\theta, \lambda)$ space. The color bar represents counts in logarithmic scale. Figure from~\cite{MIO_ScientificMBcrisp}.}
\end{figure}

An Iridium (Ir) sample has been used to perform measurements of specular reflectivity. The aim of this measurement was to show how the data analysis can be improved, if the detector spatial resolution is taken into account, and how a finer spatial resolution affects the quality of the results. The reflected intensity from the Ir sample in the $(Y,ToF)$ coordinates is shown on the left graph of figure~\ref{lt}.

The horizontal lines represent the boundaries of each cassette and the gap in between is a shadowing effect caused by the arrangement of the blades. Two subsequent cassettes are arranged in order to have an overlapping area, therefore the gap is not a dead area of the detector. The last firing wire of one cassette, not necessarily the last physical wire, is, in the projected space $(X,Y)$, the neighbour of the first wire of the adjacent cassette. Thus the gap can be removed without losing any information. Moreover, due to the blade geometry the gas gain differs for different wires within a cassette as shown in~\cite{MIO_MB2017, MIO_MB16CRISP_jinst}. The gain drops in the first 7 wires, but it can be compensated by adjusting individual thresholds on each channel. At the very first wire the loss in efficiency corresponds to a drop of $50\%$ with respect to the nominal efficiency. This region of reduced sensitivity is where two cassettes overlap and it is about 0.5 mm wide as shown in figure~\ref{fig99}.
\\In the bottom panel of figure~\ref{lt} the $(\theta, \lambda)$ phase space obtained from the $(Y,ToF)$ space is shown. Note that in this plot the gaps have been removed and the sole reduced sensitivity area is still visible in the plot. The neutron wavelength ($\lambda$) is calculated from the ToF corrected with the depth of the detector (equation~\ref{equadep}), thus the exact neutron wavelength can be calculated. 
\\According to equation~\ref{eqaf1}, the wave-vector transfer $q_z$ depends on $\theta$ (determined by the instrumental settings) and $\lambda$. The maximum intensity correspond to the angle between the scattered beam and the sample, $\alpha_f$, being equal to the incident angle $\alpha_f=\alpha_i=\theta$. According to the conventional analysis, for each wavelength, $q_z$ is calculated with a fixed and unique $\theta$ following the equation~\ref{eqaf1} and integrating the intensity over the full size of the beam. The width of the reflected intensity is defined in a range $\alpha_f=\theta \pm \Delta \theta$. The latter originates from the divergence of the beam. 
\\The spatial resolution of the detector can be used to include a correction over $\theta$, as for a small projected sample size this position directly correlates with the reflection angle. This can be used to correct for the increased spread of the reflected beam caused by a slight curvature of the sample surface, which would otherwise reduce the q-resolution.
In contrast to the conventional analysis, each value of $q_z$ is calculated according to its relative $\theta_i = \alpha_i + \delta \theta_i$ defined by the position on the detector. The correction is shown in equation~\ref{thetacorr}:

\begin{equation}
\theta_i = \alpha_i + \delta \theta_i = \alpha_i + f \cdot arctan \Big( \frac{(Y_i-Y_0) \cdot p_s}{L} \Big)
\label{thetacorr}
\end{equation}

where $\mathrm{Y_0} $ is the position of the bin corresponding to $\alpha_f=\alpha_i$, $\mathrm{Y_i}$ is any other position in the integration range, $L$ is the distance between the sample and the detector ($2.3\,$m) and $p_s$ is the pixel size of the detector. Note that the pixel size of the Multi-Blade is $p_s=p\cdot \sin(\beta) \approx 0.34\,$mm, where $p =4\,$mm is the wire pitch, is finer than the spatial resolution of the detector $\approx 0.6\,$mm. The factor $f =1/2$ has to be introduced, as the curvature of the sample surface acts as a change in sample angle and leads to a change in reflection angle by $2\theta$. 
Different combinations of $\lambda$ and $\theta$ correspond to the same $q_z$ in a diagonal cuts of the $(\theta, \lambda)$ space; this leads to an improvement of the resulting reflectivity profile. 
\\ Figure~\ref{lt} clearly visualizes the effect for the bent Ir sample in this manner as it is possible to distinguish three intensity minima from the thickness oscillations that are spread over an extended detector area, much larger than the direct beam. 

\begin{figure}[htbp]
\centering
\includegraphics[width=1\textwidth,keepaspectratio]{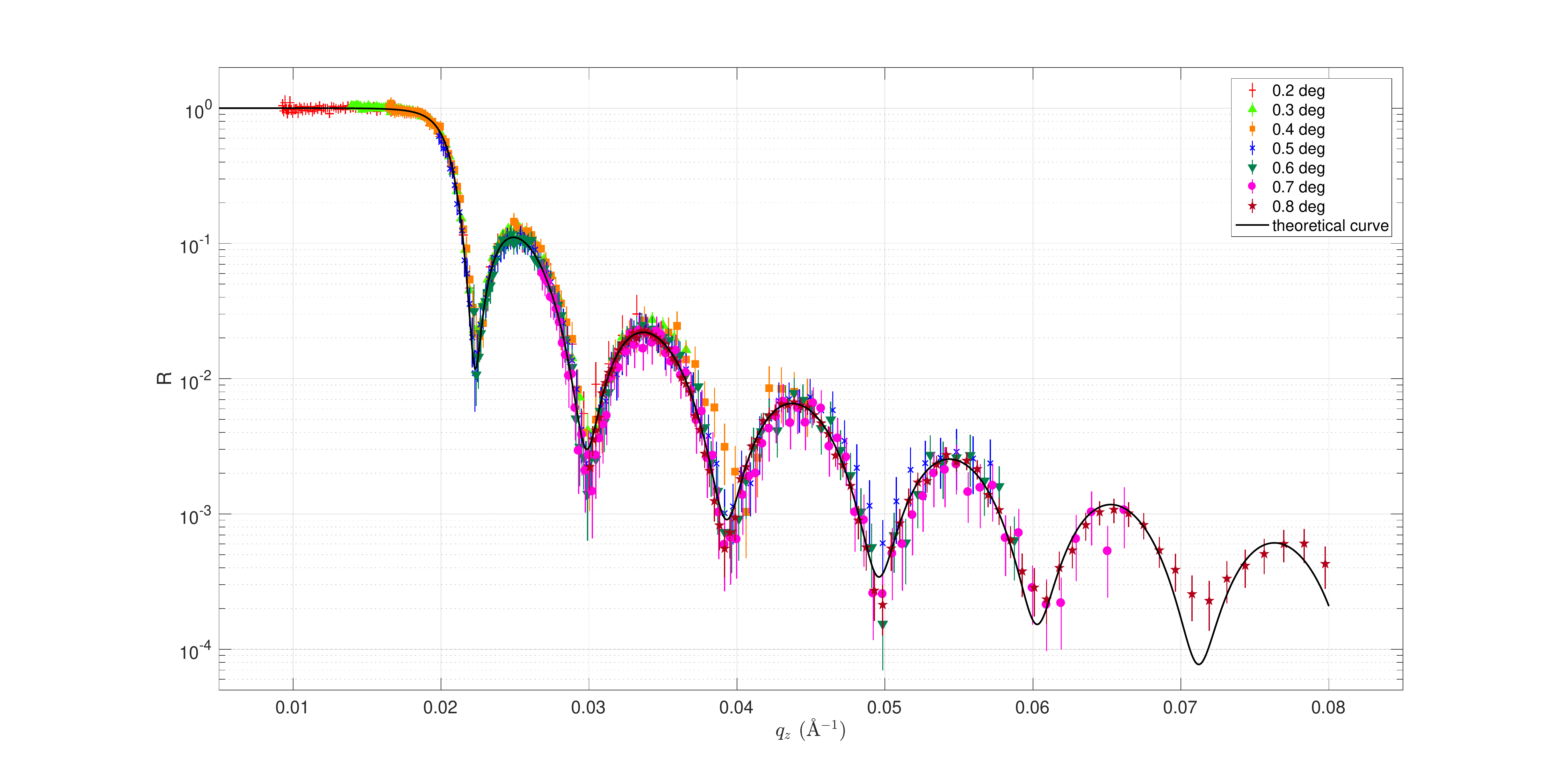}
\includegraphics[width=1\textwidth,keepaspectratio]{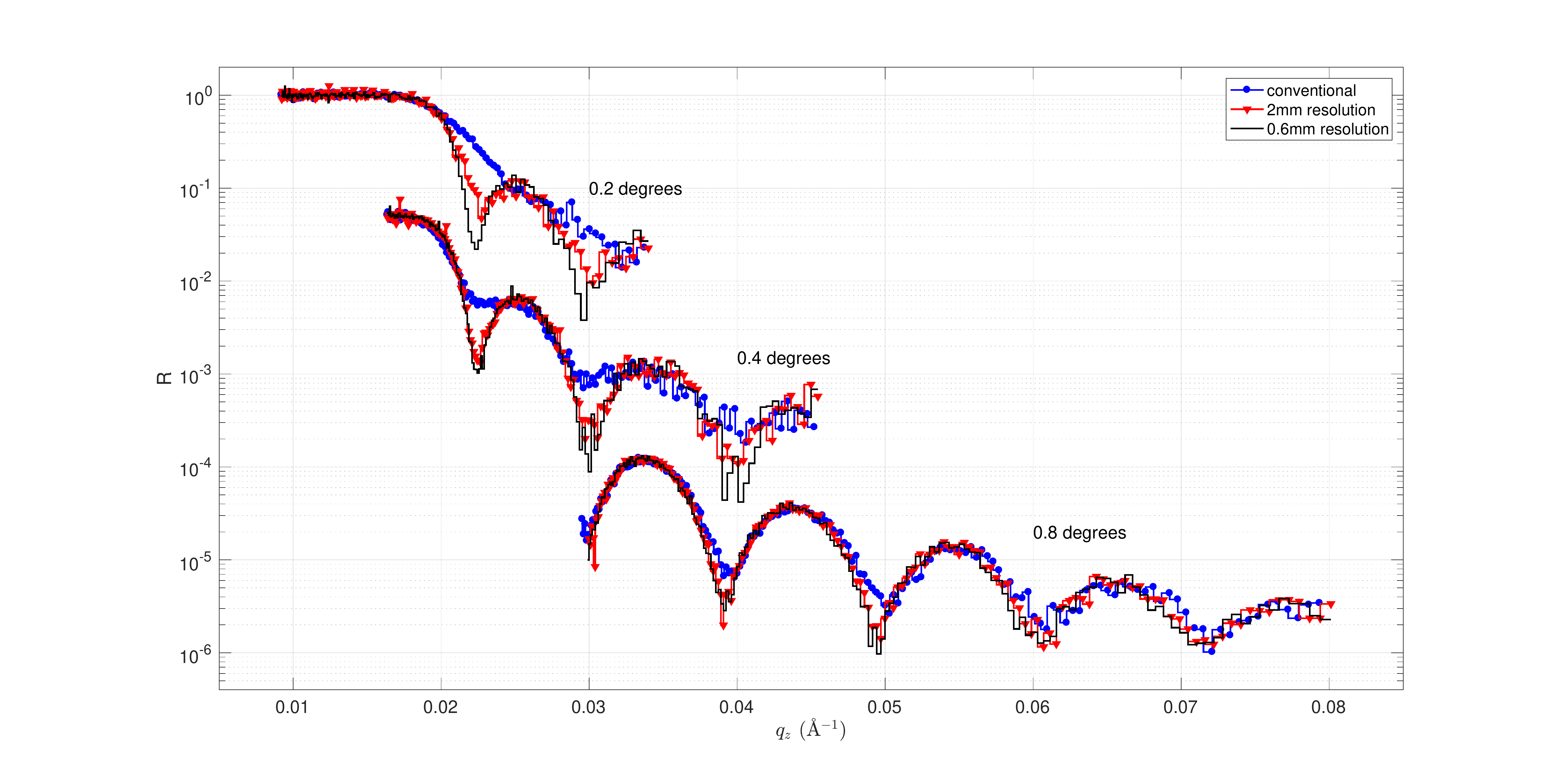}
\caption{\label{figir} \footnotesize Reflectivity curves (R) as a function of the wave-vector transfer ($q$) from an Ir-sample measured with the Multi-Blade detector at several angles and the theoretical curve (top). Reflectivity curves for the three angles 0.2, 0.4 and 0.8 degrees using the conventional analysis and the $\theta$-corrected analysis with two values for the spatial resolution of the detector, $0.6\,$mm and $2\,$mm (bottom). Figure from~\cite{MIO_ScientificMBcrisp}.}
\end{figure}

The sample is a layer of Ir of 550\,\AA{} deposited on a Si substrate. The roughness between the two interfaces is $\approx 10$\AA\ with scattering length density $N_b = 7.3\cdot10^{-6}$\AA$^{-2}$ (see equation~\ref{eqasld}). The top panel of figure~\ref{figir} shows the reflectivity curves for several angles used in the measurement, in the range 0.2-0.8 degrees, in steps of 0.1 degrees. The theoretical reflectivity is also shown and it is calculated using the Parratt formalism~\cite{Parratt} and is in good agreement with the experimental data.
  \\For a system with several interfaces, the Fresnel coefficients $r_{j,j+1}$ are related to the reflection amplitudes of each interface, e.g., $r_{0,1}$, $r_{1,2}$ etc., after a combination with an appropriate phase factor, $e^{ik_{i}2\Delta}$, where $\Delta$ is the thickness of the thin film. The period of the oscillations in figure~\ref{figir}, far from the critical edge where the dynamic effects are more pronounced, is approximately $2\pi/ \Delta$ as expected from theory (see section~\ref{neutrefl}). The amplitude of the oscillations is instead related to the constrast between the scattering length density across the interface. According to the Parratt formalism, the reflection amplitudes are calculated from the bottom interface to the next higher one, using $r_{j,j+1}$ as the reflection amplitude with the corresponding phase-shift factor. This equation is applied recursively until the top interface is reached. The ratio $X_j$ of the reflected and transmitted amplitude in the medium between interfaces $j$ and $j+1$ is:

\begin{equation}
X_j= \frac{R_j}{T_j}= \frac{r_{j,j+1} + X_{j+1}\,e^{(2ik_{z,j+1}z_j})}{1 +r_{j,j+1}\, X_{j+1}\,e^{(2ik_{z,j+1}z_j})} e^{-2ik_{z,j}z_j}
\label{parratt1}
\end{equation}    

with the Fresnel coefficients as expressed in equation~\ref{r-t-coeff}. For a n-interface system on a semi-infinite substrate, the recursion stars with $R_{n+1}=0$, and all $X_j$ can be calculated afterwards. The roughness can also be taken into account by a factor:

\begin{equation}
r_{j,j+1}^{rough} = r_{j,j+1}^{ideal} \, e^{(-2k_{z,j}\,k_{z,j+1} \, \sigma_j^2)}
\label{r-rough}
\end{equation} 

Where $\sigma_j$ is the inter-facial roughness. Note that the roughness must be much smaller than the thickness to be interpreted as an interface width and $\Delta$ as a layer thickness.
\\ A comparison between the conventional analysis and the $\theta$-corrected reduction is shown in figure~\ref{figir} on the bottom for the three angles: 0.2, 0.4 and 0.8 degrees. The $\theta$-correction was applied considering two pixel sizes, the actual Multi-Blade resolution and a reduced $\approx 2\,$mm resolution which is the current limit of state-of-art detectors used in neutron reflectometry. 

\begin{figure}[htbp]
\centering
\includegraphics[width=.8\textwidth,keepaspectratio]{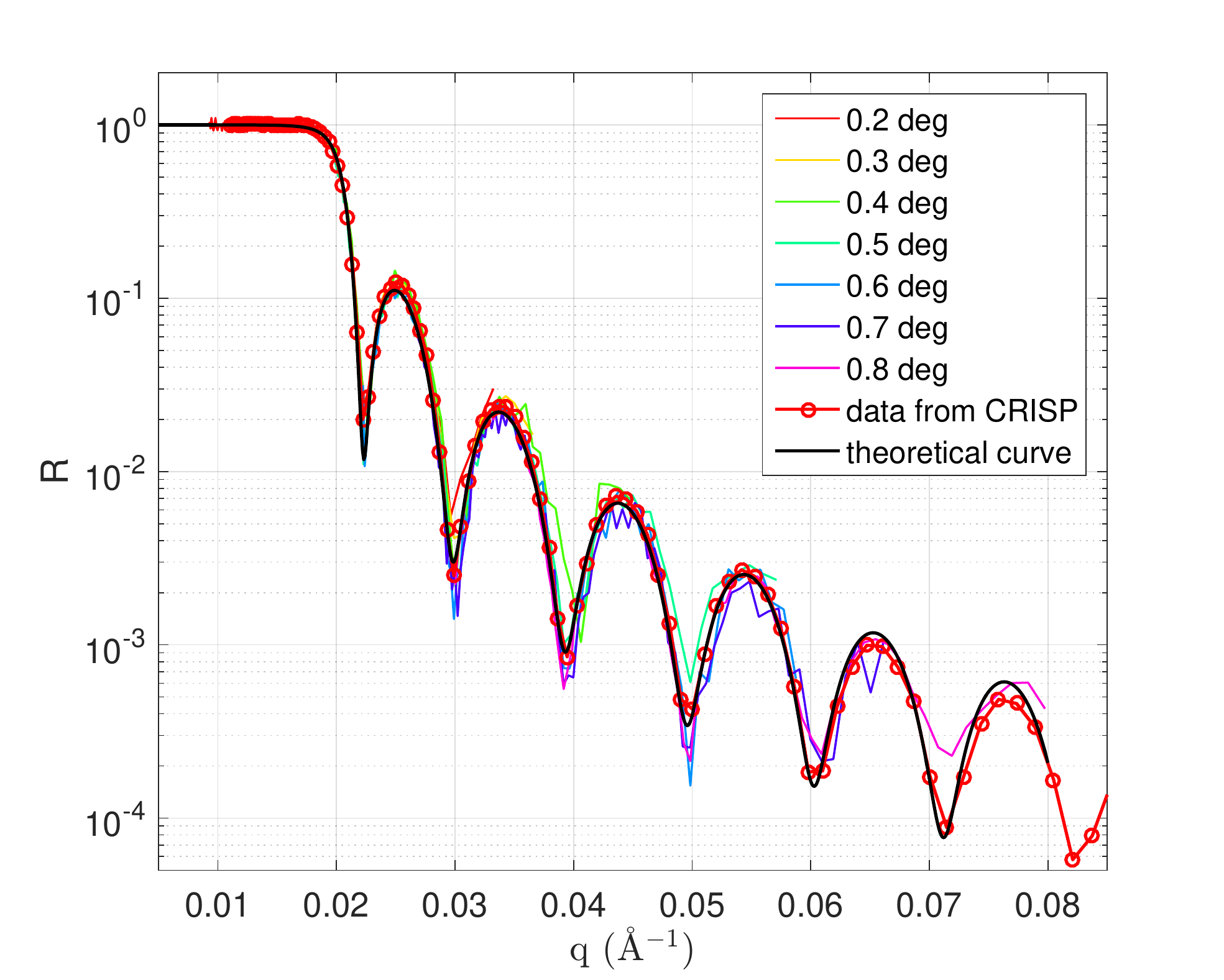}
\caption{\label{figirhe} \footnotesize Reflectivity curves (R) as a function of the wave-vector transfer ($q$) from an Ir-sample measured with the Multi-Blade detector at several angles, compared with some measurements performed on the same sample at ISIS.}
\end{figure} 

At smaller angles, the $q$-resolution depends on the detector spatial resolution to a larger extent. By applying the conventional analysis, the fringes at 0.2 and 0.4 degrees are less visible than if the $\theta$-corrected analysis is used as the better spatial resolution of the detector leads to deeper fringes. 
\\ A set of measurements were previously performed with the same sample using an $^3$He detector. In figure~\ref{figirhe} both the data collected with the $^3$He-tube and the Multi-Blade are shown. The theoretical curve is also depicted. A good agreement is obtained in both cases, further proof that the parameters used to calculate the reflectivity curve are correct. It must be noticed that the measurements performed with the $^3$He-tube were performed over a whole night with respect to about one hour of measurement with the Multi-Blade, and that the set-up with the $^3$He-tube required a high collimation of the beam.

\subsection{Specular reflectometry on Si sample: dynamic range, collimated and divergent modes}\label{colldivsi}

\begin{figure}[htbp]
\centering
\includegraphics[width=.59\textwidth,keepaspectratio]{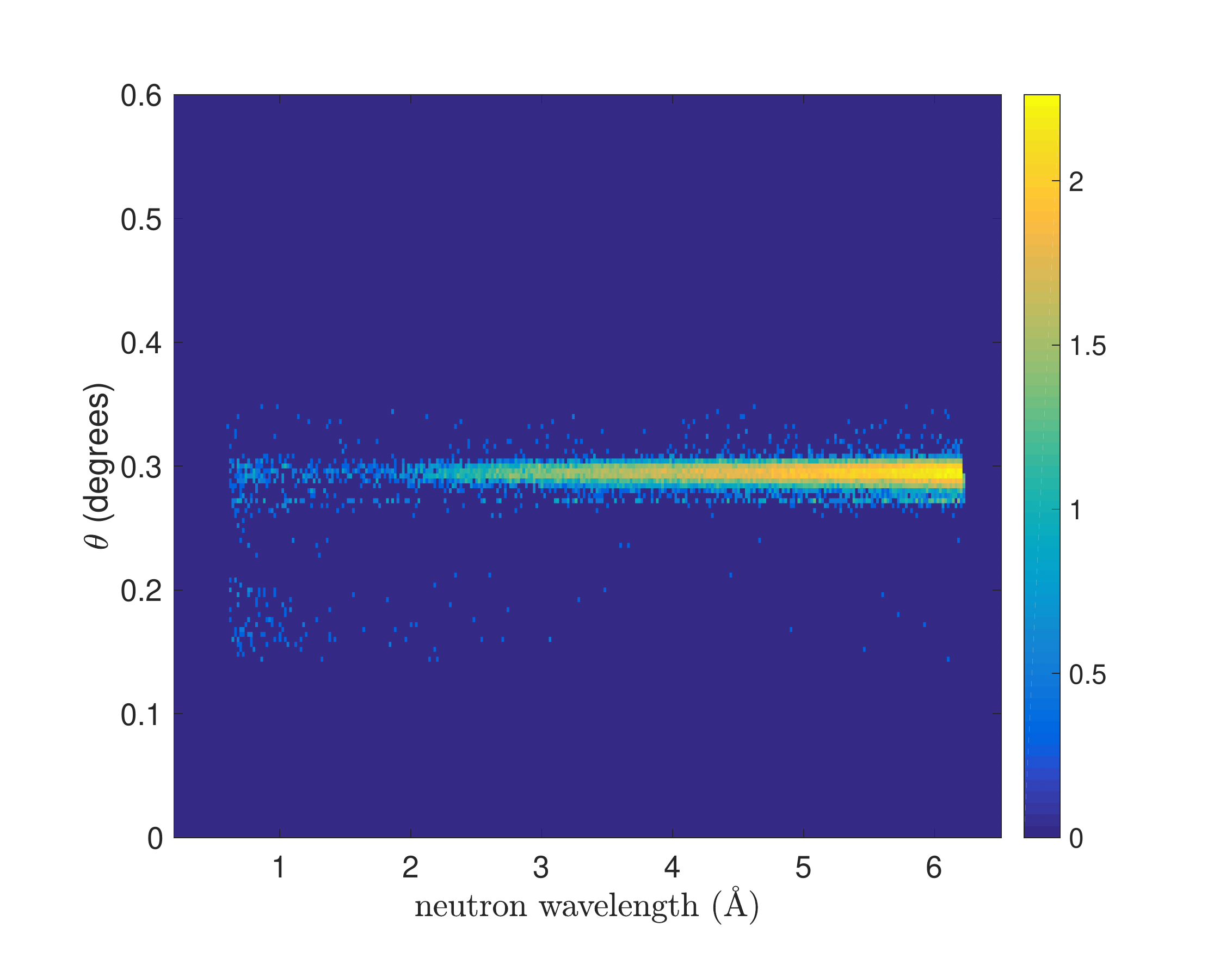}
\includegraphics[width=.59\textwidth,keepaspectratio]{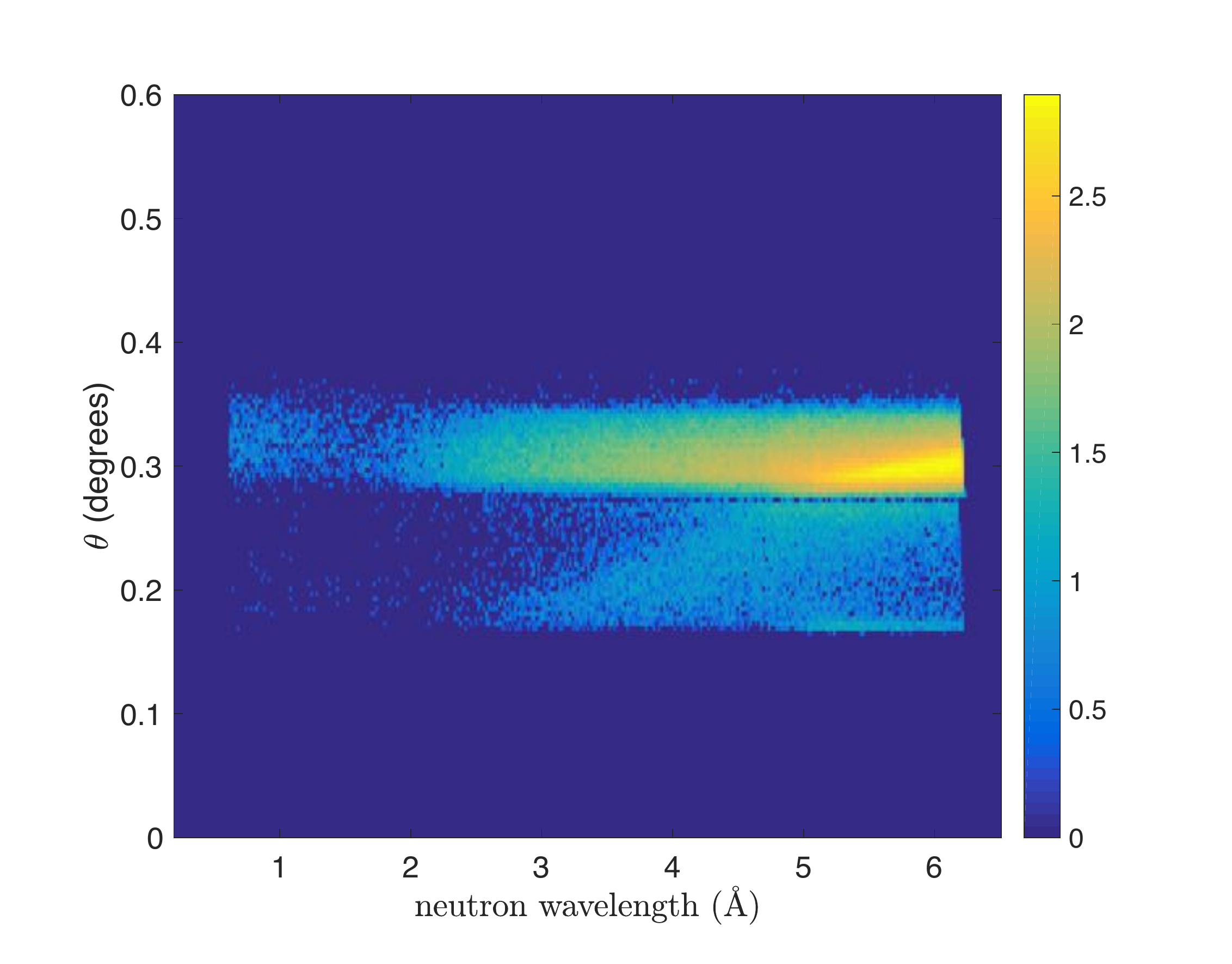}
\caption{\label{figsi1} \footnotesize The $(\theta, \lambda)$ space for the reflectivity of the Si sample at $0.3$ degrees in the collimated (top) and divergent (bottom) configuration. The color bar represents counts in logarithmic scale. Figure from~\cite{MIO_ScientificMBcrisp}.}
\end{figure}

\begin{figure}[htbp]
\centering
\includegraphics[width=1\textwidth,keepaspectratio]{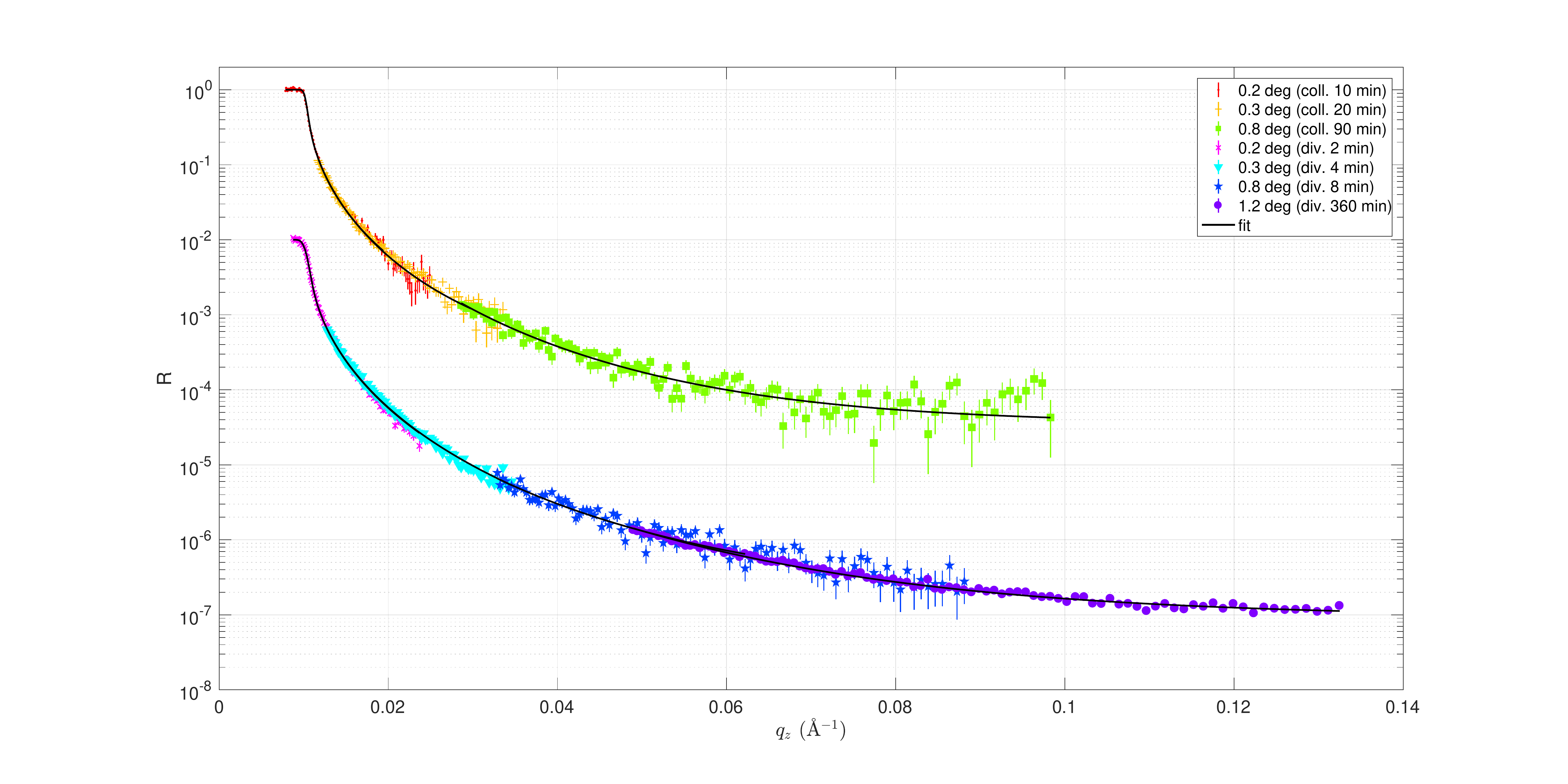}
\caption{\label{figsi2} \footnotesize Specular reflectivity (R) as a function of the wave-vector transfer ($q$) of the Si sample obtained with the collimated and divergent modes. The curves obtained with the divergent mode are shifted by 0.01 in R for clarity. Figure from~\cite{MIO_ScientificMBcrisp}.}
\end{figure}

The aim of the measurements presented in this section is to demonstrate the Multi-Blade detector capabilities in a setup as similar as possible to the ESTIA working configurations as described in section~\ref{rifle}. The instrument was operated in two configurations (collimated and divergent modes, see figure~\ref{fig4bis}) and measuring the specular reflectivity from a Si sample.
\\The collimated mode is the conventional working configuration of a reflectometer, where the divergence of the beam is limited due to the slit settings and typically its contribution to the $q$-resolution is set similar to the $\lambda$ contribution.
\\On the other hand the divergent mode exploits the full divergence available at the instrument by only constraining parts of the beam that would not impinge on the sample with the slits. The position of the neutron on the detector is used to encode $\theta$ in a similar manner as described in the previous section, according to equation~\ref{thetacorr}. Now the factor $f$ is not needed as the sample surface is flat and the change in reflection angle corresponds to the same change in incidence angle. By allowing a wider divergence of the beam, the sampled $\theta$-range is also larger; the available flux at sample increases and thus the measuring time is reduced. This method for the data reduction refers to the one that will be used with ESTIA to allow measurement from very small samples. A detailed description is reported in~\cite{INSTR_ESTIA2}.
\\Although the geometry used for these measurements on CRISP is only an approximate reproduction of the focusing concept used in ESTIA~\cite{INSTR_ESTIA2}, it is useful to test the effectiveness of the Multi-Blade detector response. Note that the focusing obtained with the slits instead of a focusing guide, leads to lower signal and a higher background as the available divergence is smaller and the sample area is strongly over illuminated~\cite{INSTR_ESTIA2}. 
\\ The measurement of specular reflectivity was performed in either configurations on a Si sample at three angles (0.2, 0.3, 0.8 degrees). A further measurement at 1.2 degrees was performed for the divergent mode to reach a wider dynamic range. 
\\ In figure~\ref{figsi1} the intensity of the beam in the $(\theta, \lambda)$ space in the collimated (left) and the divergent (right) modes are shown. The illuminated area of the detector is about 5 times larger in the divergent mode than that of the collimated mode.
\\Figure~\ref{figsi2} depicts the extracted reflectivity of the sample in the two configurations. The total acquisition time for the three angles in the collimated mode is 120 minutes. The same result is obtained in 14 minutes by performing the measurements in the divergent mode. The acquisition time is thus improved by about one order of magnitude.
\\Despite the high background due to the poor shielding of the Multi-Blade setup on CRISP, a dynamic range of $\approx$ 4 orders of magnitude with the three angles was achieved. With a further measurement at 1.2 degrees, one extra order of magnitude in the dynamic range was achieved, which is shown in figure~\ref{figsi2}. Five orders of magnitude is the dynamic range typically reached on this instrument~\cite{INSTR_OSMOND_CRISP}.
\\It is expected with the Multi-Blade to measure a deeper dynamic range in a better shielded instrument operating environment.

\subsection{Off-specular scattering: Fe/Si supermirror sample}\label{offfesi}
\begin{figure}[htbp]
\centering
\includegraphics[width=.45\textwidth,keepaspectratio]{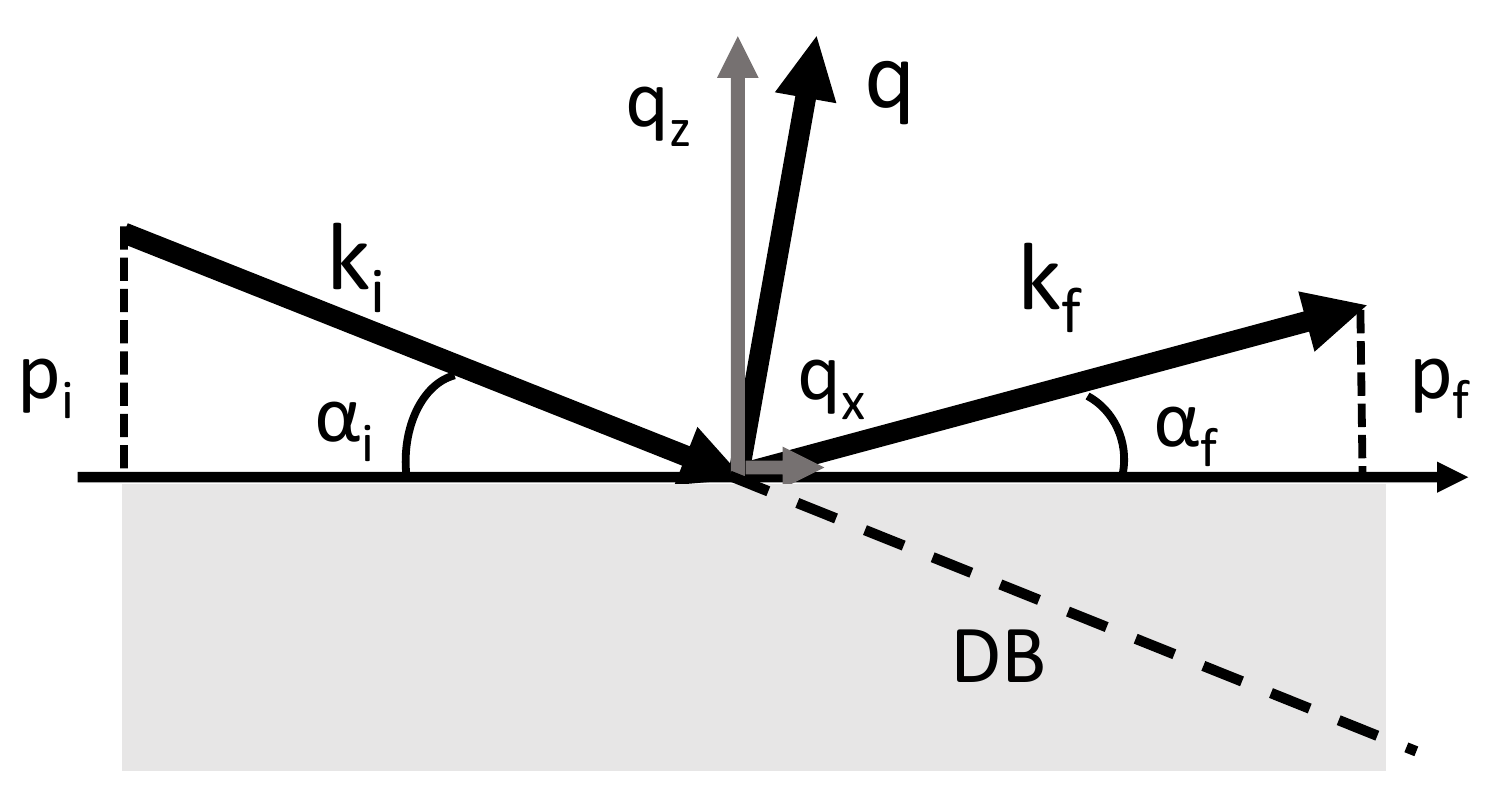}
\caption{\label{fig70} \footnotesize Sketch of the wave vectors definition used in the off-specular scattering.}
\end{figure}

The specular reflectivity allows to probe the structure of a sample across the depth, indeed the scattering vector $q$ is perpendicular to the sample surface. It is possible to probe the in-plane structure of a sample introducing a small parallel component of the scattering vector~\cite{R_offspec0_Ott}; a sketch is shown in figure~\ref{fig70}. The parameter used to reproduce the results of the off-specular scattering are the components of $q$ and the projections of the initial and final wave vectors, they are reported in the equation~\ref{Qcomp}. 

\begin{align}
\label{Qcomp}
&p_i = \frac{2 \pi}{\lambda} \sin \alpha_i & \nonumber \\
&p_f = \frac{2 \pi}{\lambda} \sin \alpha_f & \\
&q_x = \frac{2 \pi}{\lambda} (\cos \alpha_f - \cos \alpha_i)& \nonumber\\
&q_z = \frac{2 \pi}{\lambda} (\sin \alpha_f + \sin \alpha_i)& \nonumber
\end{align}

Neutron off-specular scattering probes the in-plane structure at the $\mu$m length scale. The limitation of this technique is set by both the limited available neutron flux and the small scattering probability. Similarly correlations at the nm length scale can be reached with a collimated beam in both directions, so called grazing incidence small-angle scattering (GISANS), which is described in detail in~\cite{Lauter2016,INSTR_GISANS1,INSTR_GISANS2,INSTR_GISANS3}.
On magnetic samples the off-specular technique allows the depth resolved measurement of correlations from magnetic domains as used in~\cite{Josten2010,Nickel2001}.
\\ Several specific areas can be identified in the off-specular scattering, based on the direction of the final wave vector determined by the reflected angle~\cite{R_offspec1_Ott}. The horizon is defined as $\alpha_f = 0$, when the neutron beam is parallel to the surface of the sample. The specular reflection is found at  $\alpha_i = \alpha_f$ and all other areas above the horizon mark the off-specular scattering region. The direct beam, DB in figure~\ref{fig70}, meets the condition $\alpha_f = - \alpha_i$. When the incident angle is close to the critical angle $\alpha_c$, the transmitted beam is also refracted and thus this equality does not hold for small $\alpha_i$. 
Finally, at $\alpha_i=\alpha_c$ and $\alpha_f=\alpha_c$ one finds the so called Yoneda wings, which are results of dynamic effects mostly produced from surface roughness and magnetic domains.

\begin{figure}[htbp]
\centering
\includegraphics[width=.9\textwidth,keepaspectratio]{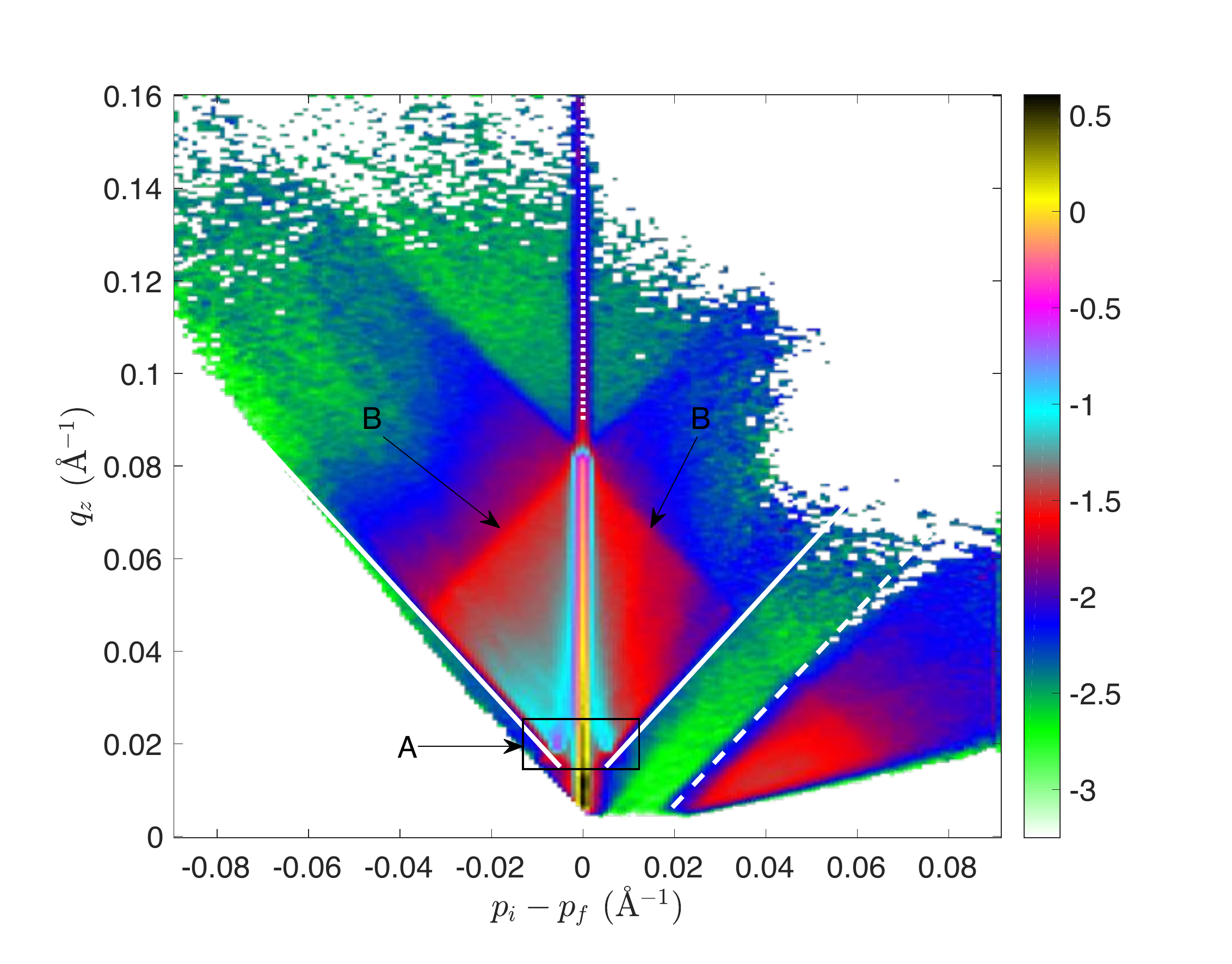}
\caption{\label{figoffsp} \footnotesize Off-specular scattering, expressed as $q_z$ as a function of $p_i - p_f$, from Fe/Si super-mirror: the solid lines corresponds to the two Yoneda wings, the dashed line denotes the beginning of a region of scattered neutrons in the transmission direction (anti-Yoneda), while the dotted line marks the specular reflectivity. The spin flip signal from the layers is highlighted with the rectangle indicated with A. The correlated domains from the sample layers correspond to the blue and red rhombus shaped area (B). The colour scale is counts in logarithmic units. Figure from~\cite{MIO_ScientificMBcrisp}. }
\end{figure} 

The sample employed to carry out the measurements was a super-mirror Fe/Si (m = 3.8). It shows a strong off-specular scattering when un-magnetized due to complex magnetic domain structures. The off-specular measurements are used to test the detector performances. Good uniformity and spatial resolution as well as large dynamic range are needed to fully characterize the features of the off-specular scattering on the sample.
Some measurements were performed using the collimated beam to scan the sample in angle in order to get a fine tuning in uniformity and reach a wider q-space. The sample position was tilted in step of 0.01 degree in the 0.2-0.8 degrees range. The measurements were performed over the whole night. 
The data are presented in the ($p_i - p_f$, $q_z$) coordinates in figure~\ref{figoffsp} and the typical features of the sample are well reproduced. In figure~\ref{figoffsp} the solid lines corresponds to the two Yoneda wings, the dashed line denotes the beginning of a region of scattered neutrons in the transmission direction (anti-Yoneda), while the dotted line marks the specular reflectivity. In the specular reflectivity direction both Silicon and the super-mirror edge are identified, (black spot and the crossing point on the top of the line respectively). The correlated domains from the sample layers correspond to the red and light blue rhombus shaped area. 
This almost featureless area corresponds to magnetic spin-flip scattering within the super-mirror that can be separated by polarization-analysis into two asymmetric components as is demonstrated in~\cite{KLAUSER2016}.  
Neither beam polarization nor magnetic field has been employed, therefore the magnetic scattering of all spin-states are summed together, resulting in the rhombus area. Nevertheless this test demonstrates the establishment of the Multi-Blade detector technology for neutron reflectometry application. 

\subsection{BNC tests: Bragg-Edge Transmission on graphite sample}

\begin{figure}[htbp]
\centering
\includegraphics[width=.6\textwidth,keepaspectratio]{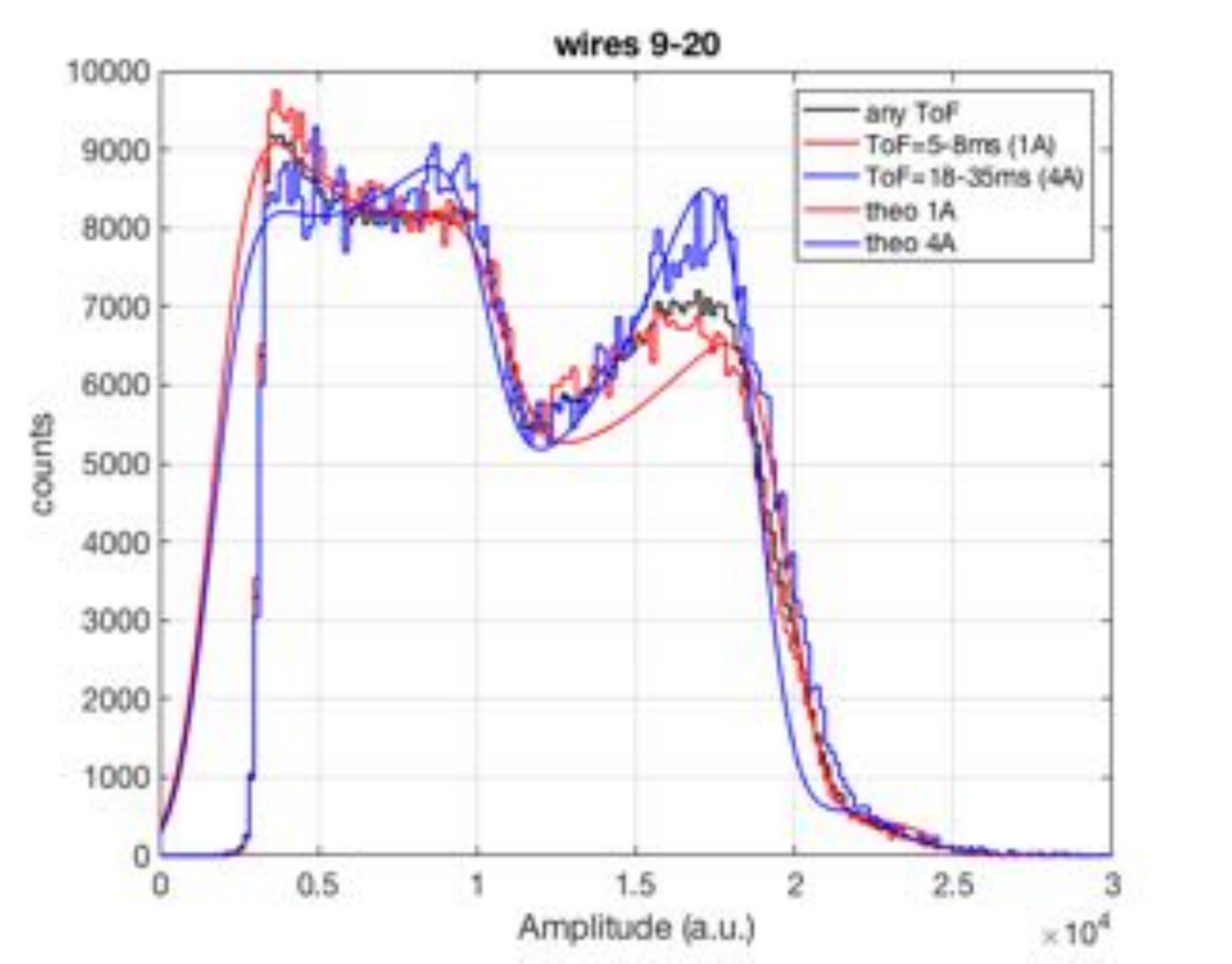}
\caption{\label{spettrook}\footnotesize Spectrum of wires between 9 and 20 (same gas gain) for any ToF, ToF = 5-8 ms and ToF = 18-35 ms, black red and blue line respectively. The theoretical spectrum is also shown for $\lambda =$ 1 and 4 \AA.}
\end{figure}

The collaboration between the ESS Detector Group and the Wigner Research Centre for Physics in Budapest allows for beam time at the Budapest Neutron Centre (BNC) to perform tests on the detector prototypes. In the case of the Multi-Blade this collaboration is mainly based on the development of electronics.
\\ The measurements performed at the Budapest Neutron Centre served to test the operation of the detector in all its aspects, from the mechanics to the electronics.  Indeed, a new prototype have been assembled with several modifications with respect to the previous version~\cite{MIO_MB2014}. The major improvements were focused on the new front end electronics, with the corresponding software acquisition system. Therefore, the main aim of this test was to prove the new front end electronics and the relative software acquisition system employed on the latest version of the Multi-Blade detector, which is described in section~\ref{secFEE}. The majority of tests were dedicated to prove the effectiveness of the full read out system and to understand its limitations. Thus to lead improvements in the next design iterations. 
\\ Together with these tests, some qualitative measurements were performed on a graphite sample. The beam line (TOF) is the only one that exploits the Time of Flight technique. All the measurements were compared with the $^3$He-tube installed on the instrument. 
\\The energy spectrum of the conversion particles due to the interaction between neutrons and a boron converter layer is well known. Therefore it is an easy way to verify if the detector is working properly, in figure~\ref{spettrook} is shown the spectrum recorded with the Multi-Blade detector. A comparison with the theoretical curve is depicted and a good agreement in achieved. The spectrum is plotted for two different ToF regions, namely considering two different neutron wavelengths for the theoretical calculation. The spectrum in the total ToF range is shown as well. As discussed in section~\ref{linea}, because of the geometry, the gas gain along the wires is not uniform. The spectra are analysed considering the wires with the same gain. A detailed discussion and the implementation to reduce this effect is described in section~\ref{secfuture}. 

A series of measurements in transmission through a sample, a graphite slab, have been carried out. In general, by neutron diffraction is possible to determine the crystal structure of a sample under some conditions on the angles with respect to the direction of the incident neutron beam and the neutron wavelengths. Indeed, at certain wavelengths, strong intensity signals are detected, i.e., the Bragg peaks. The neutrons are scattered over the full solid angle, 4$\pi$, while only a small fraction of the solid angle is covered by the detectors. In such a set-up a large amount of neutron is unused. In the transmitted direction, the yield is much higher and the experimental set-up for such measurements is straightforward, because sample and detector are on the same axis, which corresponds to the one of the incident neutron beam.  
\begin{figure}[htbp]
\centering
\includegraphics[width=.49\textwidth,keepaspectratio]{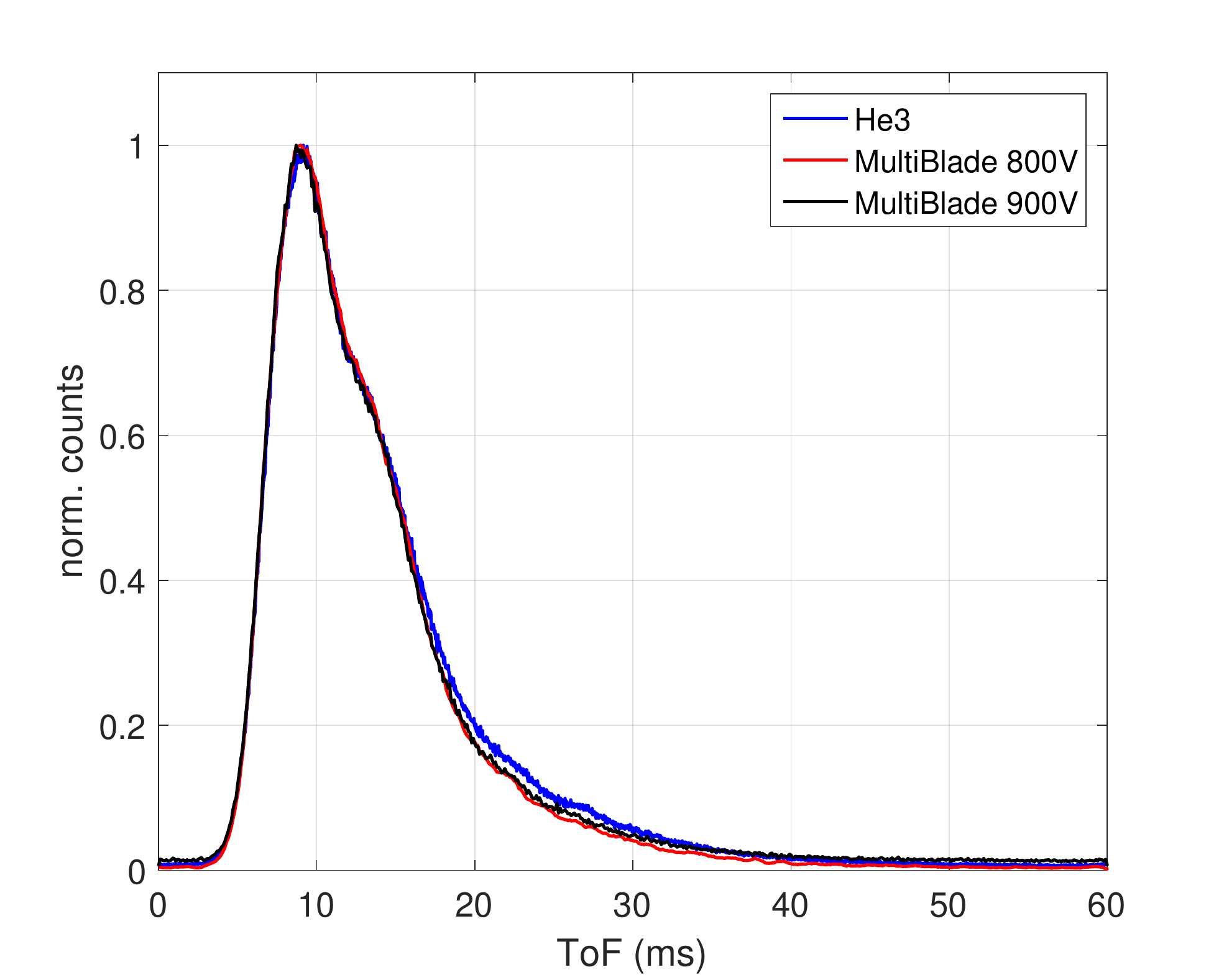}
\includegraphics[width=.49\textwidth,keepaspectratio]{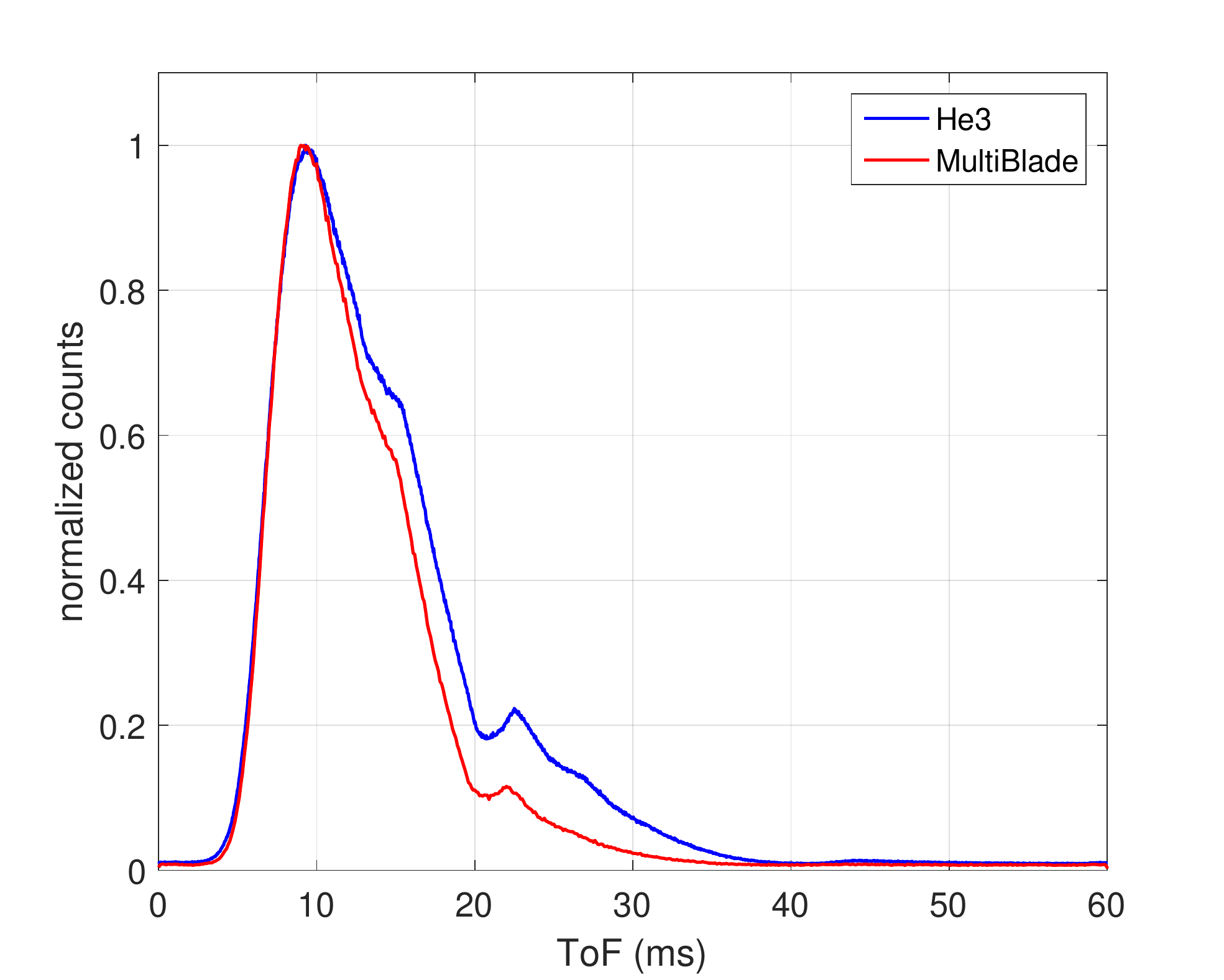}
\caption{\label{spectofbnc} \footnotesize Left: ToF spectra from direct beam measurement normalized to unity for the $^3$He-tube and the Multi-Blade detector applying two different high voltage values (800 V and 900 V). Right: ToF spectra from measurement in transmission through the graphite sample normalized to unity  at the peak value of counts for the $^3$He-tube and the Multi-Blade detector.}
\end{figure}    
\\ The transmission spectrum of thermal neutrons through a polycristalline sample displays sharp, well-defined increases in intensity as a funcion of the neutron wavelength~\cite{bragg_edg1}, as governed by Bragg's law. These Bragg edges occur because for a given $\{h\, k\, l\}$ reflection, the Bragg angle increases as the wavelength increases until $2\theta =180^{\circ}$. At greater values of $\lambda$ than this critical one, no scattering by this particular lattice spacing can occur, thus a sudden increase in the transmitted intensity is seen~\cite{bragg_edg1}. From Bragg's law, this condition is verified for $\lambda = 2d_{hkl}$, when a measure of the $\{h\, k\, l\}$ $d$-spacing in the direction of the direct beam is performed. When the neutron wavelength is shorter than or equal to $2d_{hkl}$, neutrons are scattered by the $\{h\, k\, l\}$ plane, thus the transmitted intensity is reduced~\cite{bragg_edg3}. 
\\ Measurements of the direct beam are needed to obtain the transmitted intensity and to normalize the two set of measurements. Figure~\ref{spectofbnc} shown the spectra in ToF for both detectors in the case of direct beam (left) and after the scattering from the graphite sample (right). All the spectra are normalized in time. For view purpose, the pulse heights are also normalized to unity. In both acquisitions with the Multi-Blade the spectrum fully follows the one obtained with the reference device. Although the shape of the transmitted intensity is well reproduce by the Multi-Blade with respect to the one obtained with the $^3$He-detector, the counts decrease at higher ToF values. It must be taken into account that the total illuminated area of the two detectors is different. Moreover, the measurements with the $^3$He have been provided by the local contact in order to have a reference. An accurate comparison was not performed because the intent of these tests is purely demonstrative and the investigation of the sample does not have scientific research purposes. However, it allows a qualitative proof of the detector operation.

\begin{figure}[htbp]
\centering
\includegraphics[width=.49\textwidth,keepaspectratio]{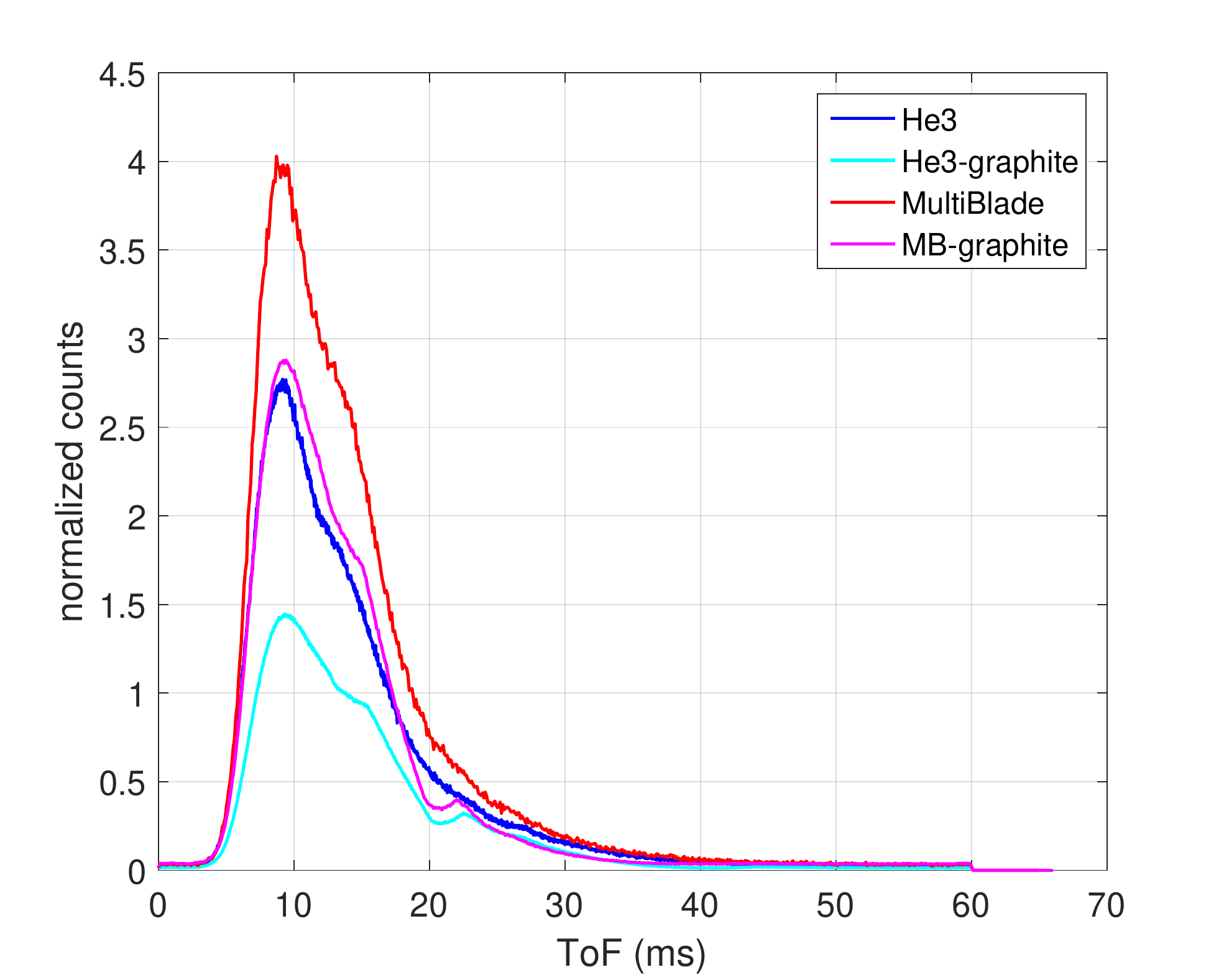}
\includegraphics[width=.49\textwidth,keepaspectratio]{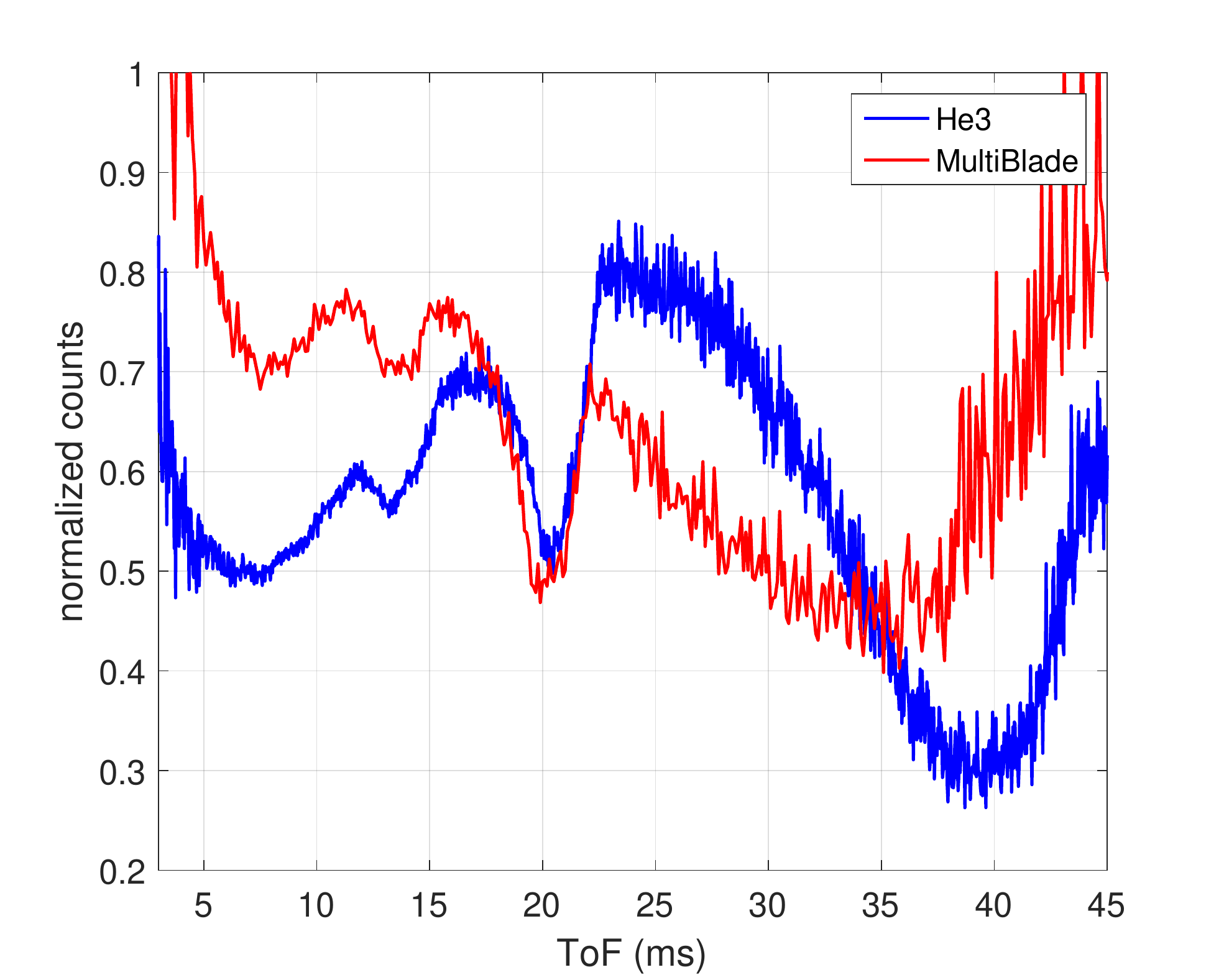}
\caption{\label{braggedge} \footnotesize Left: ToF spectra of both detectors recorded with the direct beam and the transmission after the graphite sample measurements. The normalization is different in the case of the two devices, because of difference in the illuminated area. Right: Transmitted intensity, the Bragg edges of the sample diffracted structure are visible in both cases.}
\end{figure} 

In figure~\ref{braggedge} left is shown the PHS in ToF of both detectors recorded in the two measurements set-up: direct beam and transmission after the graphite sample. Note that the direct beam and the transmission measurements are normalized between them, but no normalization is applied between the two detectors, because of the difference in active area. The transmitted intensity is shown in figure~\ref{braggedge} right. It is the ratio between the transmitted and the direct beam intensity. Although the normalization is not the same, the pattern of the sample diffracted structure is reproduced in both cases. The typical decreases in intensity which refer to the particular diffractions of the lattice space are in agreement with literature~\cite{bragg_edg2}.

\section{ Future improvements}\label{secfuture}

The several tests performed either at the Source Testing Facility in Lund, at the Budapest Research Centre and the experiment carried out at ISIS provided a full overview of the status of the project. This part of the work is, instead, dedicated to some of the design implementations arisen from these tests, in order to improve the detector performances. 
\\ One of the main issues of the present device is the non-uniform response of the wires due to the geometrical effects. The wires are placed at 4 mm distance in between two adjacent blades. In the edge of the blade (knife) this distance is not constant any more. The ions drifting to the cathode have to travel more, this leads to a decreasing of the collected charge, because the electric field is not uniform and as discussed in section~\ref{sectionamplgas}, these effects reflect on the gas gain amplification. 

\begin{figure}[htbp]
\centering
\includegraphics[width=1\textwidth,keepaspectratio]{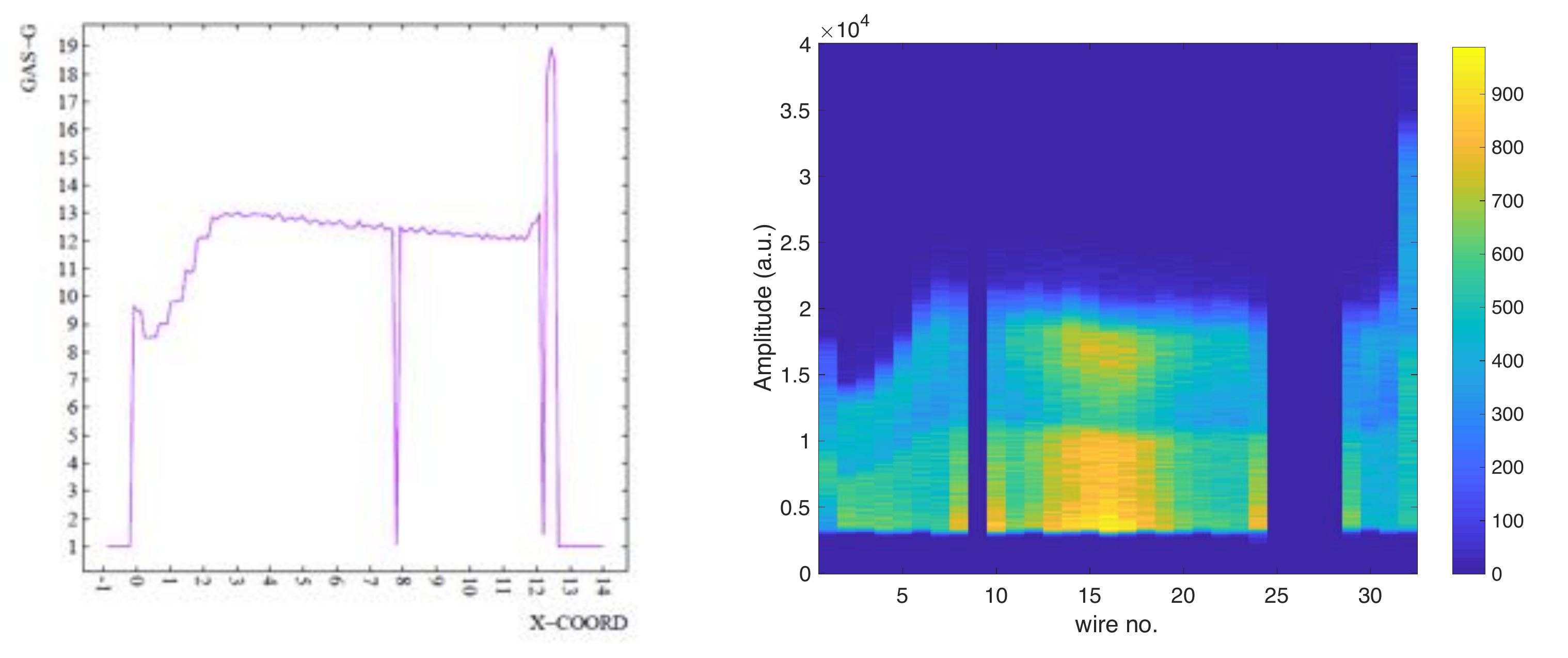}
\caption{\label{specali} \footnotesize Gas gain map across the wire plane obtained with Garfield (left), and spectrum of $\alpha$ and Lithium particles obtained by the measurements of the direct beam (right). }
\end{figure} 

\begin{figure}[htbp]
\centering
\subfloat[]{\includegraphics[width=1\textwidth,keepaspectratio]{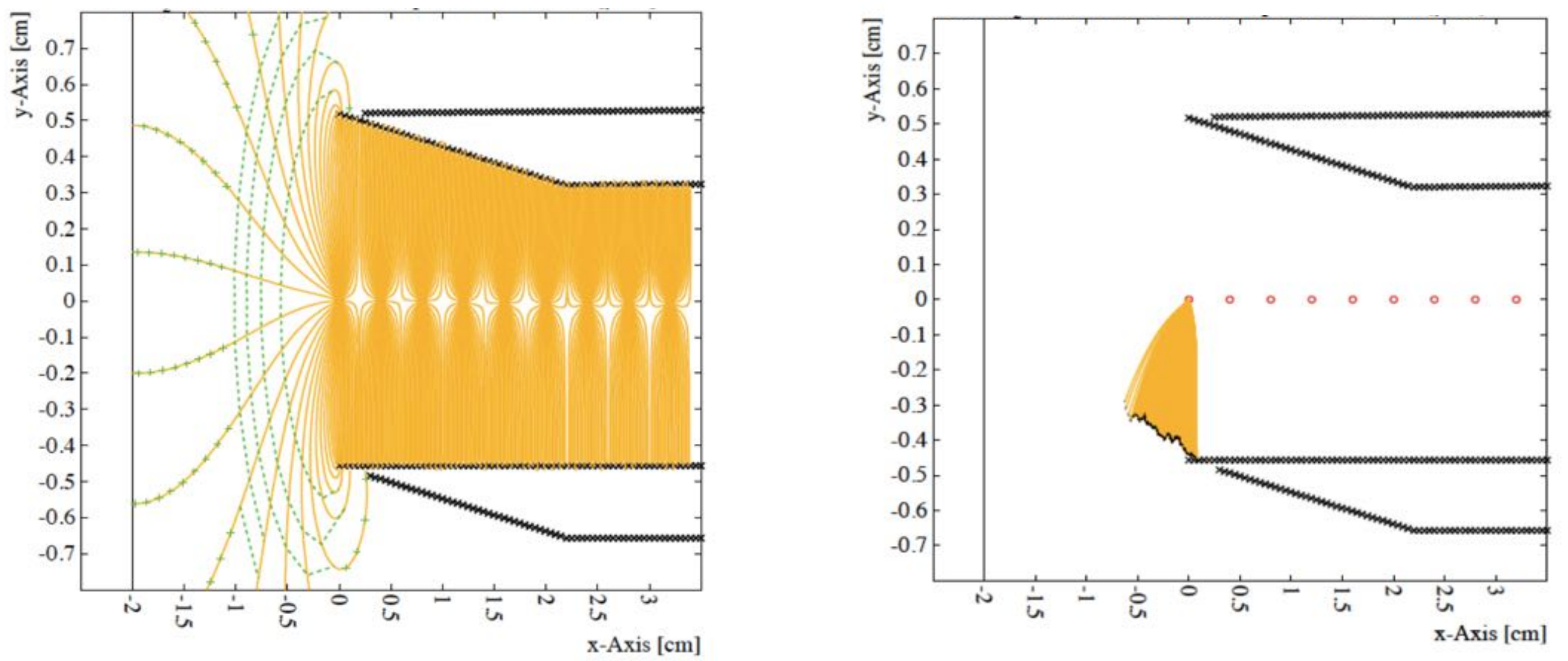}}\\
\subfloat[]{\includegraphics[width=.49\textwidth,keepaspectratio]{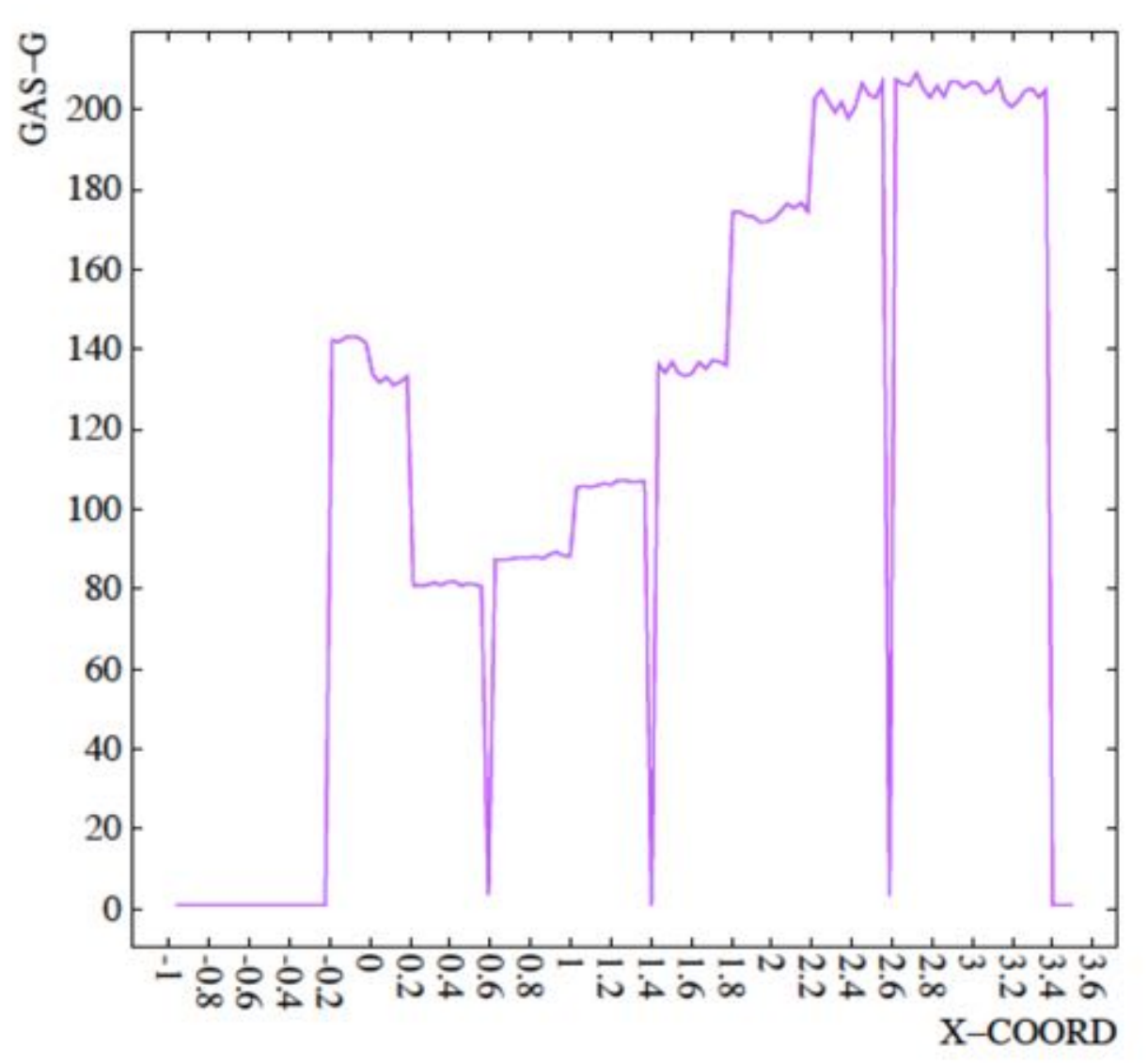}}
\caption{\label{garw0} \footnotesize (a) Electric field lines when the first wire is place at the edge of the blade X=0 mm (left). Simulation of the charge collection of an $\alpha$ particle escaping the boron layer towards the edge of the blade. (b) Gas gain on the first 8 wires.}
\end{figure}

Referring to figure~\ref{specali}, the spectrum of $\alpha$ and Lithium particles across the wires (x-axis) obtained by the measurements of the direct beam, is shown together with the simulated gas gain map. The measurement has been performed during the test at the BNC, while the simulations have been carried out with Garfield~\cite{GARFI}, which is a software for detailed simulation of drift chambers. In the measured plot there are two spots clearly visible. The highest intensity of the two, refers to the corresponding peak of the Li and $\alpha$ particles  generate from the neutron conversion with the B$_4$C layer. The missing parts are dead channels. The slight slope between wire 9 and 31 is expected. Remember that the cassettes are arranged over a circle around the sample, thus they are placed at a relative angle of 0.14 degrees from each others. Also, the last wire has approximately a double gain compared with the rest of the wires. The higher gain is foreseen because there is no neighbour wire that shield the electric field. 
\\ It must be noticed that the first wire has a double gain compared to the last one. Moreover, even if the track of the yield particle is oriented toward the outside of the wire plane, the total charge is collected by the wire. The simulation is shown on the right side of figure~\ref{garw0}~(a), all the charge is collected by the wire. As shown is figure~\ref{garw0}~(a) left, the first wire is placed at $\mathrm{x} = 0$, i.e., at the edge of the blade. In this scenario about half of the field lines does not close on the strips cathode, this reduce the charge collected on the strips. 

\begin{figure}[htbp]
\centering
\subfloat[]{\includegraphics[width=0.9\textwidth,keepaspectratio]{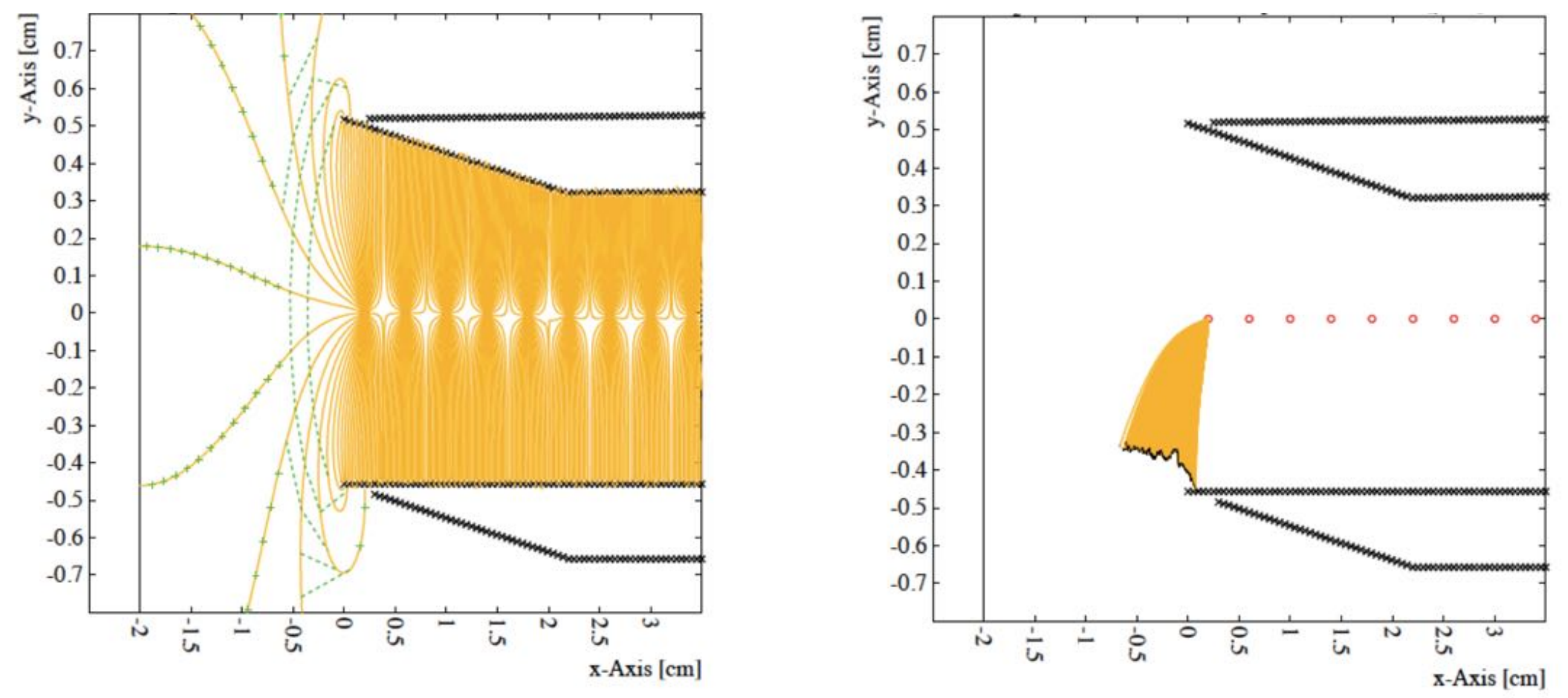}}\\
\subfloat[]{\includegraphics[width=.49\textwidth,keepaspectratio]{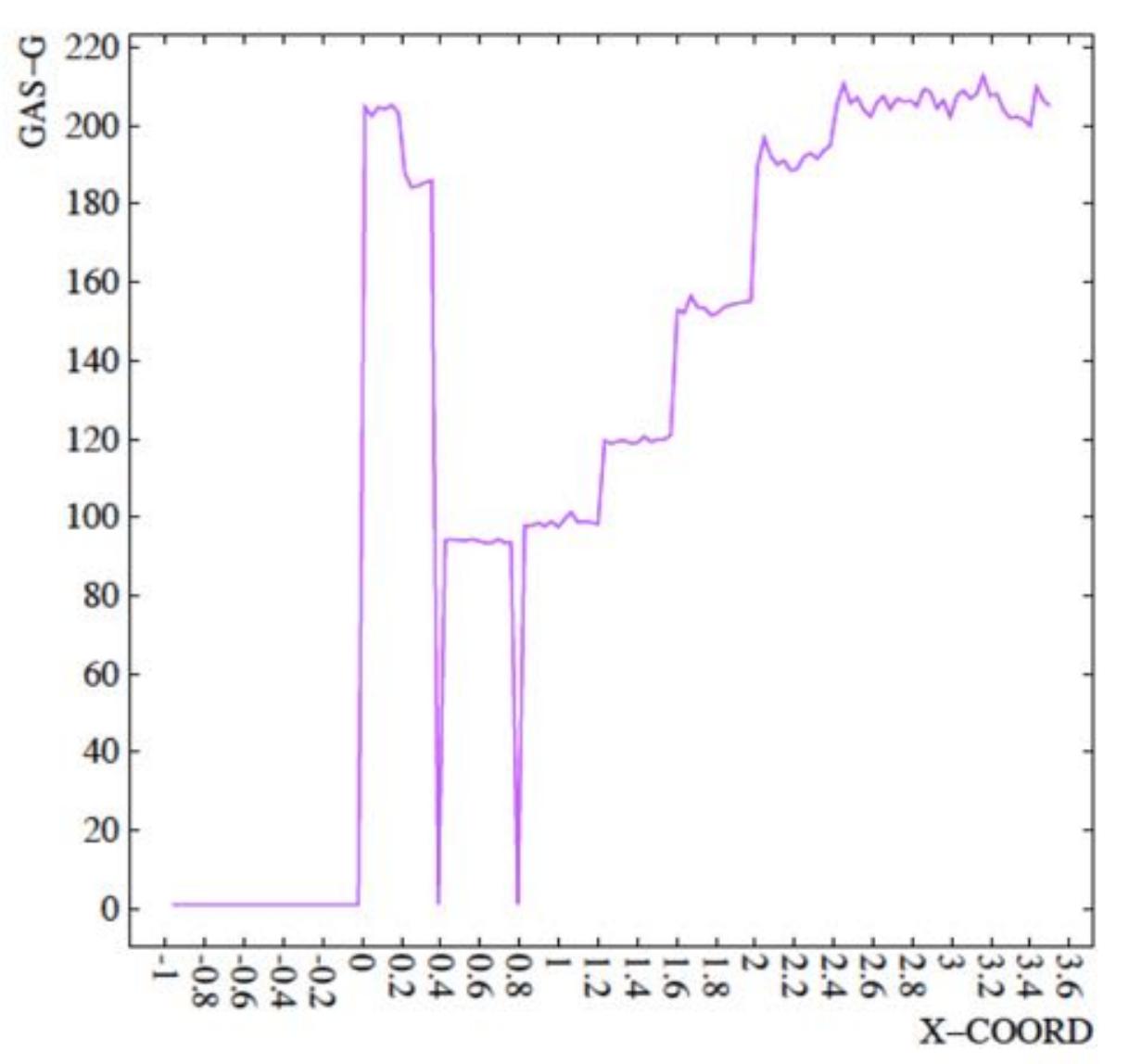}}
\caption{\label{garw2} \footnotesize (a) Electric field lines when the first wire is shifted X=0.2 mm toward the inside of the cassette (left). Simulation of the charge collection of an $\alpha$ particle escaping the boron layer towards the edge of the blade. (b) Gas gain on the first 8 wires.}
\end{figure}

A drop in efficiency is observed when the coincidence wires-strips, which identifies a neutron event in 2-dimensions, is taken into account. The gain variation on the first 8 wires is depicted in figure~\ref{garw0}~(b), both this and the effect of the knife is reproduced by the simulation. The drops in the plot is a binning effect and does not have any physical meaning.   
\\ In order to decrease the loss of energy the wire plane has been shifted toward the inside of the cassette. The simulations have been performed from 0 to 4 mm, i.e., between the starting position to the x-coordinate of the second wire. It has been found that for $\mathrm{x} = 2\,$mm almost all the field lines close on the strips. Over a total of 25 lines going out from the wire toward the cathode plane, 14 reaches it when $\mathrm{x} = 0\,$mm, 18 when $\mathrm{x} = 2\,$mm and 21 when $\mathrm{x} = 4\,$mm, namely the second wire. The plots of the electric field lines, the charge released by a product particle and the gas gain variation are shown in figure~\ref{garw2}. In this case the coincidence with the strips is ensured and the gas gain of the first wire is aligned with the rest, but the reduced sensitivity given the knife is not solved yet.

\begin{figure}[htbp]
\centering
\includegraphics[width=.6\textwidth,keepaspectratio]{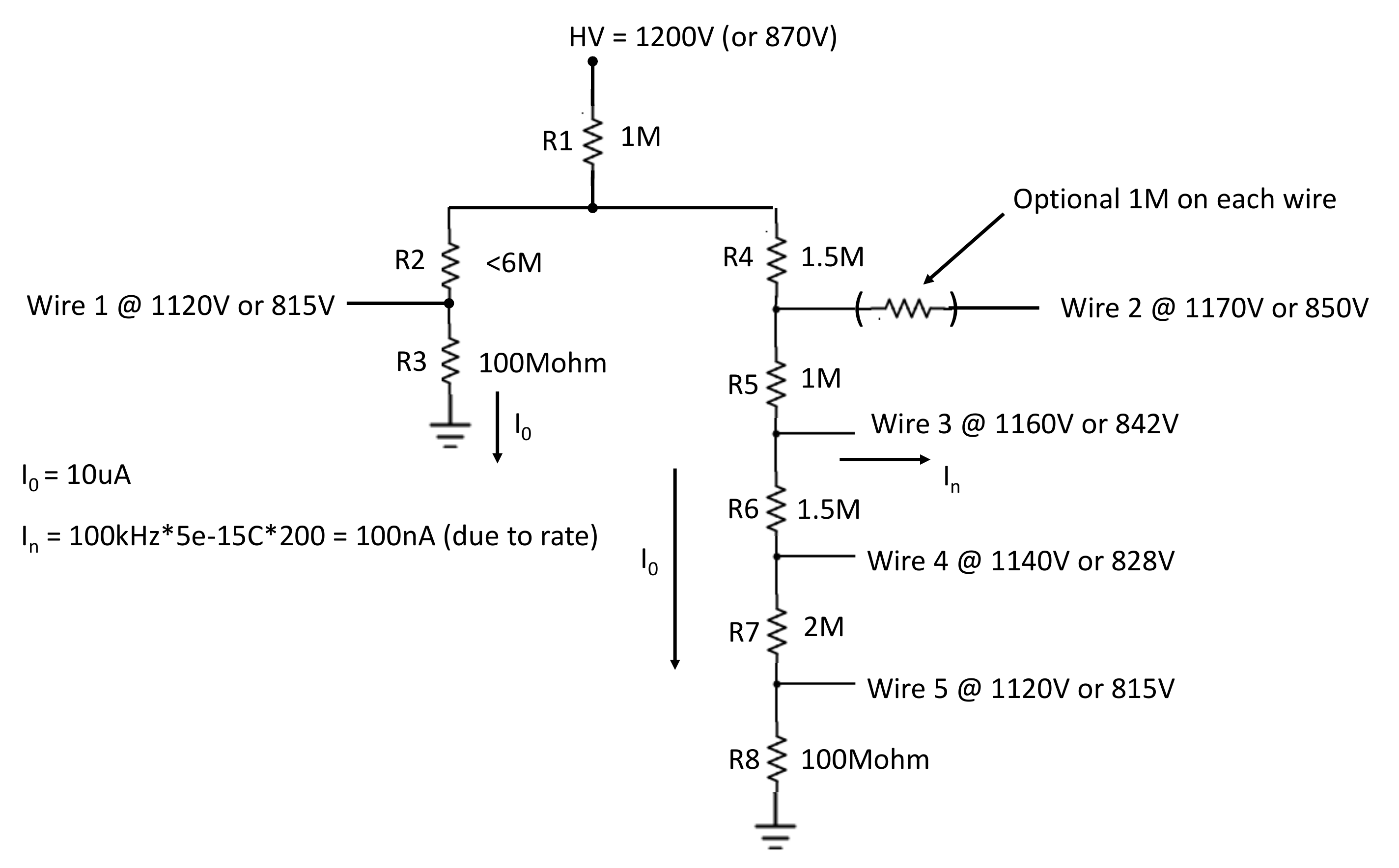}
\caption{\label{voltdiv} \footnotesize Sketch of the voltage divider used to compensate the gas gain of the first 5 wires.}
\end{figure}

By Diethorn's formula (eq.~\ref{gasgaincharge}), described in section~\ref{sectionamplgas}, it is known that the gain amplification is related to the applied voltage. Therefore, a voltage divider has been designed to avoid the decreasing of the charge collection on the first wires. A sketch is depicted in figure~\ref{voltdiv}, knowing the initial high voltage, the resistance and the current wire to ground, it is possible to calculate the value of the resistances to add in order to module the voltage on the wires of interest. The values of the resistances are independent from the absolute value of the initial voltage, thus with this divider is feasible to work at different initial high voltages. 

\begin{figure}[htbp]
\centering
\includegraphics[width=.6\textwidth,keepaspectratio]{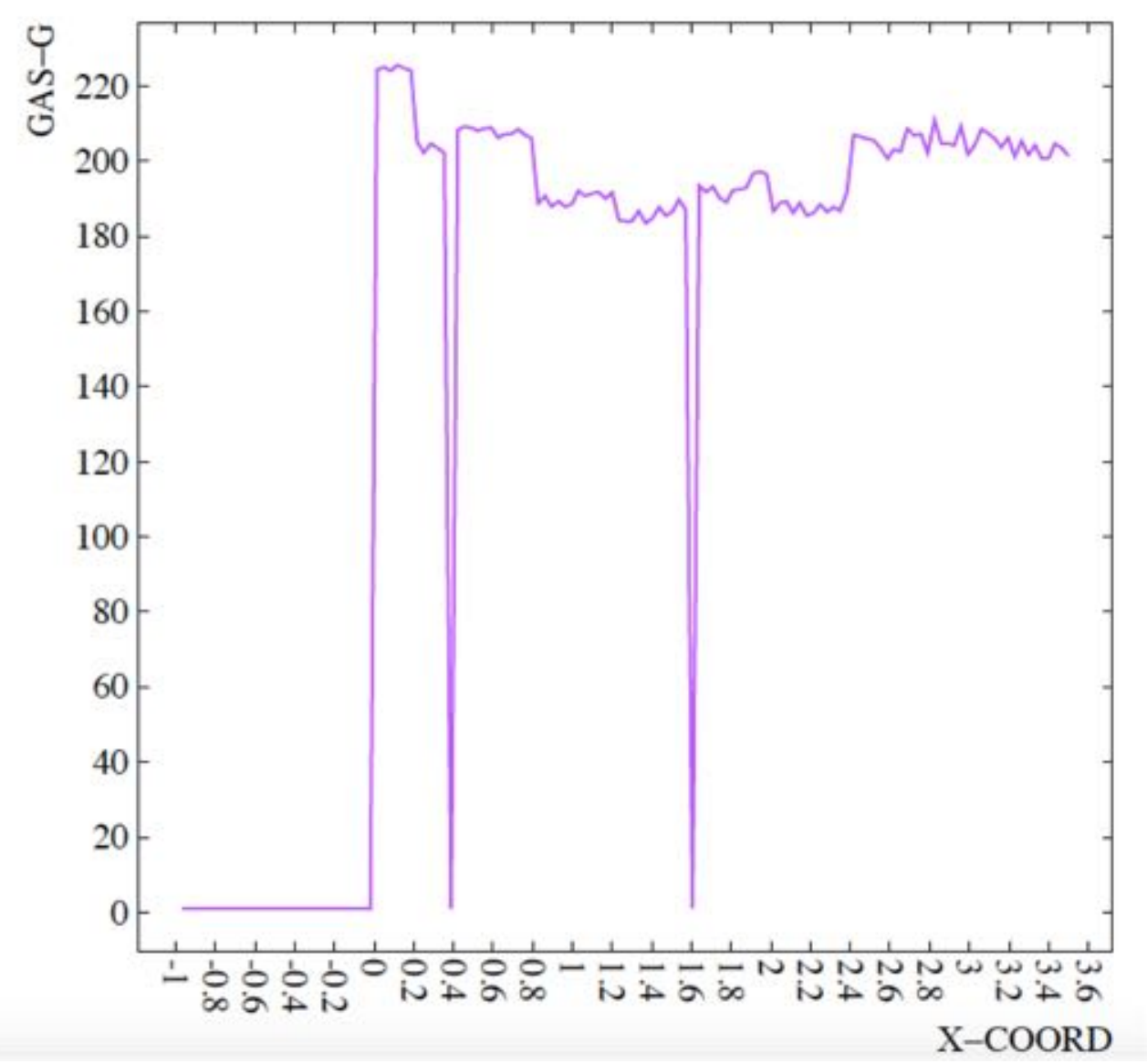}
\caption{\label{gargasfin} \footnotesize Gas gain of the first wires after the voltage divider compensation.}
\end{figure}

The electronic circuit has been successfully tested and the implemented version is now ready for the future detectors. The simulated gas gain variation is shown in figure~\ref{gargasfin}. A good uniformity is now achieved. The remaining variation in uniformity can be corrected by detailed calibration.

\begin{figure}[htbp]
\centering
\includegraphics[width=1\textwidth,keepaspectratio]{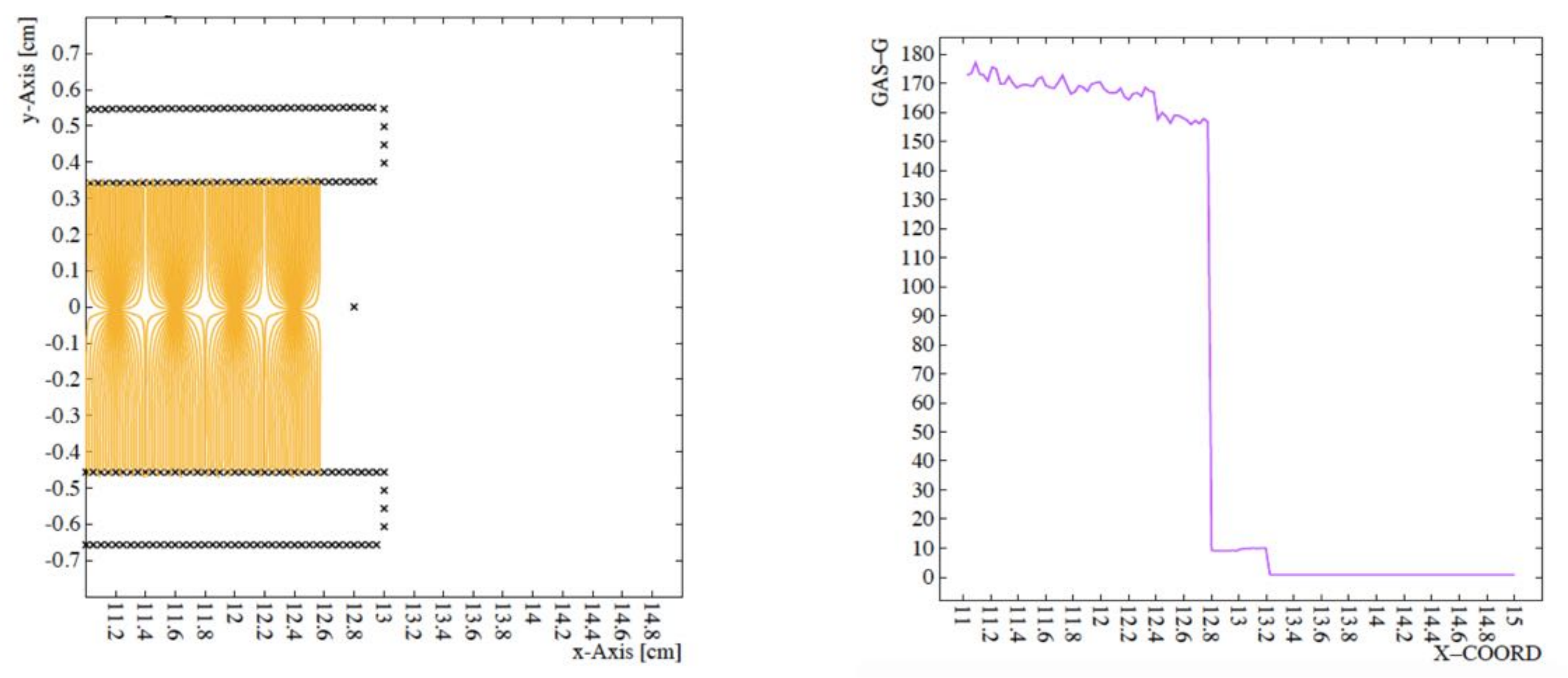}
\caption{\label{garguardia} \footnotesize Sketch of the guard wire, 50 $\mu$m diameter, added at the end of the cassette, in order to align the gas gain of the last read wire (left). Gas gain variation in the last 5 wires, approximately same gas gain, and the guard wire, which gas gain is almost zero (right).}
\end{figure}

The last point to solve is the double gain of the last wire. The simple idea is to add an extra wire, a so-called guard wire. In order not to add extra high voltage channels it is kept at the same high voltage of the rest of the wires. Once this quantity is fixed, several configurations have been simulated by varying the values of wire diameters. With the guard wire of the same dimension of the read out wires the gain is uniform. Even if the added wire is not read out, it may generate some spurious events on the strip, because of its double gain. To mitigate this it is noted that the higher is the diameter the lower is the electric field. Therefore, a guard wire of 50 $\mu$m has been simulated, compared with the 15 $\mu$m diameter of the read out wires. In this case there is a drop in gain with respect to the others. The final configuration is depicted in figure~\ref{garguardia}, where both the configuration and the gas variation is shown.

\begin{figure}[htbp]
\centering
\includegraphics[width=1\textwidth,keepaspectratio]{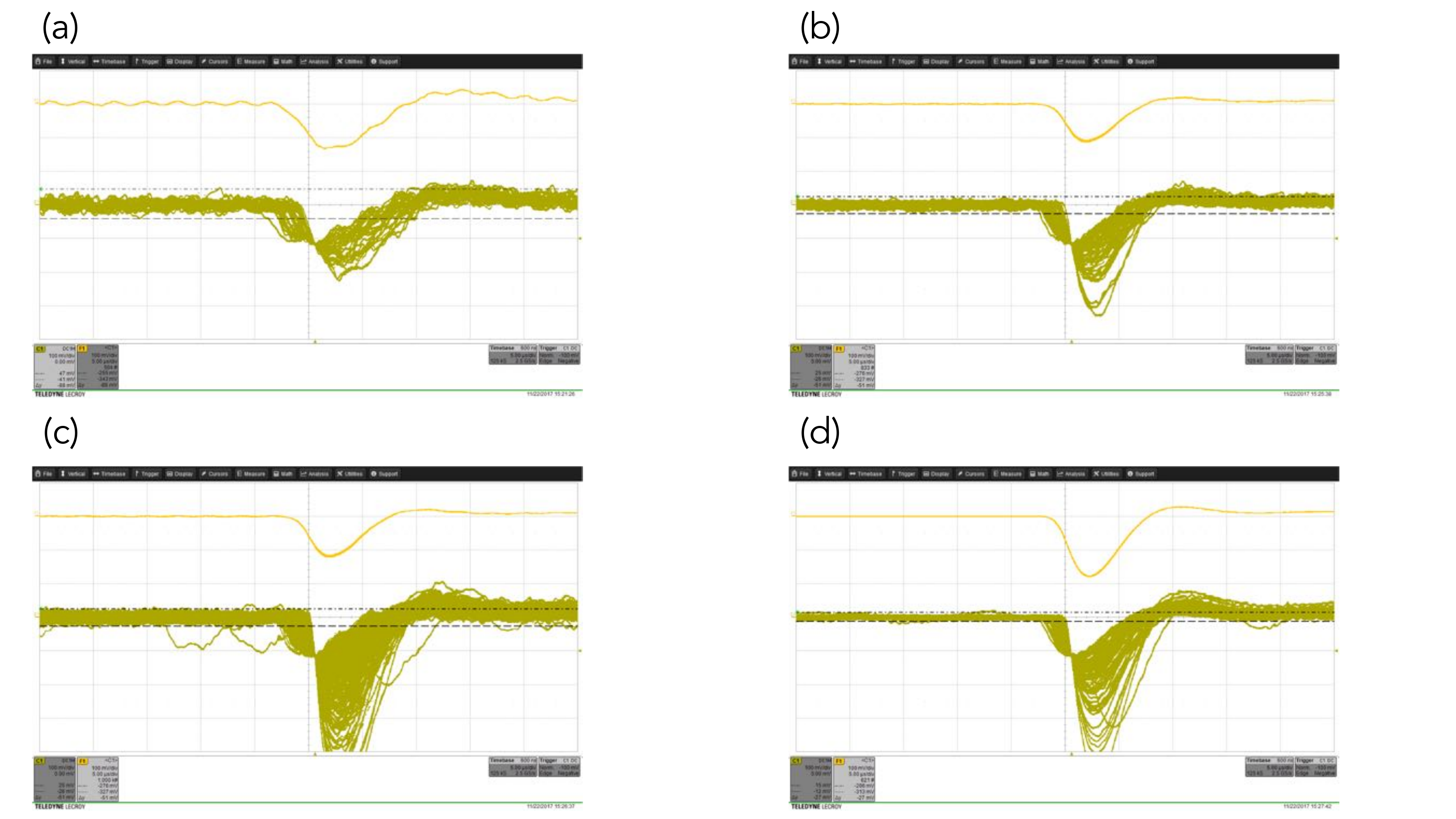}
\caption{\label{kapthick} \footnotesize Configurations with different polyimide (Kapton) thickness in order to better decouple the capacitance of the strips and the blade. A nominal substrate of 75$\mu$m is employed, panel (a). The signals with an extra layer of 75 $\mu$m, panel (b), two (c) and three (d) extra layers of the same thickness are shown as well.}
\end{figure}

Another issue experienced in the tests has been the difficulty of reading out the signals from the strips with respect to the wires signals. Due to this problem, it is necessary to operate the detector at higher values of high voltage. This reduces the performance in speed and rate capability, because of a larger space charge effect and because the higher the voltage, the higher the gas gain and the less is the proportionality, see section~\ref{GID}. The nominal voltage is around 800-900 V, but most of the measurements have been performed at 1100-1200 V. The gas gain in the second case is one order of magnitude larger than the nominal scenario.  
\\ Electronics consideration apart, it is possible to reduce the noise-to-signal ratio (NSR) insulating the blade substrate from the strip cathodes. Indeed, they form a capacitor that reduce the amplitude of the collected signals on the strips, therefore the capacitive coupling must be smaller, i.e., by insulating strips and substrate.
In the present detector, a layer of polymide (Kapton) 100 $\mu$m thick is placed between the blade and the strip. A picture of real signals is shown in figure~\ref{kapthick}~(a). The measurements have been performed at the STF in Lund. In this case the signals amplitude is around 200 mV and a noise-to-signal ratio = 0.44 is measured. For this test the cassette has been modified adding one, two and three layers of Kapton 75 $\mu$m thick, panel (b), (c) and (d) of figure~\ref{kapthick}, respectively. A clear improvement is visible already when one layer is added (b), not only the amplitude values are higher, around 300 mV, but also the noise is diminished from $\approx$ 90 mV to $\approx$ 50 mV. The noise-to-signal ratio is equal to 0.15, three times better than the actual configuration. In the case of two more layers the noise level is unaffected but the signals have an amplitude of more than 400 mV. The noise-to-signal ratio is slightly improved, it goes down to 0.11. If three layers are joint together, the noise level is below 30 mV with amplitudes more than 400 mV for the signals. A further factor two from the previous scenario is gained, noise-to-signal ratio = 0.06.  
\\ The last configuration has been chosen for the production of the next strips. They will be coupled with a Kapton layer of about 300 $\mu$m to ensure a good insulation from the metal blade, but 100 $\mu$m at the blade edge in order not to cover the adjacent blade, because of the 5 degrees inclination the thickness is 10 times higher.
  
The results reported in this chapter show that the detector technology is well established and the implementation discussed here are detailed refinements. The effectiveness of the detector has been widely proved, as discussed in this chapter. Besides this, a dedicated study on the background response for this kind of detector technologies has been evaluated. This will be the topic of next chapter.

\chapter{Fast neutron sensitivity of neutron detector based on boron-10 converter layers}\label{chapter5}

This part of the work is based on a campaign of measurements performed at the Source Testing Facility (STF)~\cite{SF2,SF1} at the Lund University in Sweden. The study is focused on the fast neutron sensitivity for the class of boron-10-based neutron detectors. The experiment discussed in this chapter has been performed with the Multi-Blade detector and the text is based on the paper published by the author of this thesis in Reference~\cite{MIO_fastn}. This was the first time fast neutron sensitivity has been looked at for B-10 based thermal neutron detectors. Moreover, one of the figure-of-merit for most neutron scattering instruments is strongly related to the signal-to-background ratio. The signal is defined by the source, and a better background rejection can improve this figure-of-merit, leading to significant impact on instrument's operation, especially at the new high intensity sources. Further measurements with another boron-10-based detector (Multi-Grid~\cite{MG_2017}) and with a $^3$He-tube have been carried out. A brief discussion on the results obtained with the $^3$He-tube is presented together with a qualitative analysis derived from a specific set of measurements performed at the CRISP reflectometer.

\section{Introduction}

The science progress in neutron physics is made possible thanks to the development of both sources and instrumentations, including neutron detectors. The European Spallation Source (ESS) will be a prominent infrastructure in this regard~\cite{ESS, ESS_TDR,ESS-design}. As highlighted is chapter~\ref{chapter3}, ESS is designed to be the world's brightest neutron source, thus the instantaneous flux of the instruments will be the highest of any other existing neutron source.
\\ The $\mathrm{^3He}$-based techonologies have been the most used detectors for thermal neutrons. Both the availability and the requirements of higher performance are the reasons why a number of research programs are now aiming to find technologies that would replace the $\mathrm{^3He}$~\cite{HE3S_karl}. A promising technique is based on solid converter layers ($\mathrm{^{10}B}$, Gd) and gas proportional counter as the sensing medium. Some examples are the Multi-Grid~\cite{MG_IN6tests,MG_2017,MG_patent,MG_joni} for large area applications in inelastic scattering, the Multi-Blade for neutron reflectometry~\cite{MIO_MB2014,MIO_MB2017}, the Jalousie detector~\cite{DET_jalousie} for diffractometers, BandGEM~\cite{Bgem,MPGD_GEMcroci} and the Boron-coated straw-tubes~\cite{STRAW_lacy2011} for SANS and Gd-GEM~\cite{DET_doro1,gdgem} for neutron macromolecular diffractometer and CASCADE~\cite{DET_kohli} for neutron resonance spin echo. 
\\ Among these, the Multi-Blade, described in detail in Chapter~\ref{chapter4}, was used to perform the measurements presented subsequently. As it is a relative new technology in neutron science, all the aspects have to be investigated in order to be installed it on a real instrument; from the mechanics to the electronics, from the response to neutrons to the background characterization.   
\\ The characterization of the background is, indeed, a fundamental feature to study in order to understand the limit of the best signal-to-background ratio achievable. The main sources of background, which affect the detectors and instrument performance, are $\gamma$-rays, fast neutrons (1-10 MeV), environmental neutron counts (thermal and epithermal) and electronic noise~\cite{BG_cherkashyna2014}.
\\ The fast neutron energy range (1-10 MeV) considered for this experiment is set by the sources available in the laboratory for the measurements. Note that this range closely represents the energy spectrum of the fast neutron background at reactor facilities, these being based on fission; whereas that of spallation sources is much broader.
\\ For the first time the fast neutron sensitivity of $^{10}$B-based thermal neutron detectors has been investigated. A detailed description of the nuclear processes involved in the neutron interaction with the different materials of the detector is, then, reported. 
\\ All the measurements have been performed at the Source Testing Facility of Lund University~\cite{SF2,SF1} with the Multi-Blade detector. A comparison with the $\gamma$-ray sensitivity~\cite{MIO_MB2017,MG_gamma, MPGD_CrociGamma} follows the main analysis on fast neutron sensitivity.
\\ Theoretical analysis and Monte Carlo simulations are used to interpret the experimental results obtained with a Boron-10-based detector.
 
\section{Description of the setup}\label{mbdescr}

\begin{figure}[htbp]
\centering
\includegraphics[width=.45\textwidth,keepaspectratio]{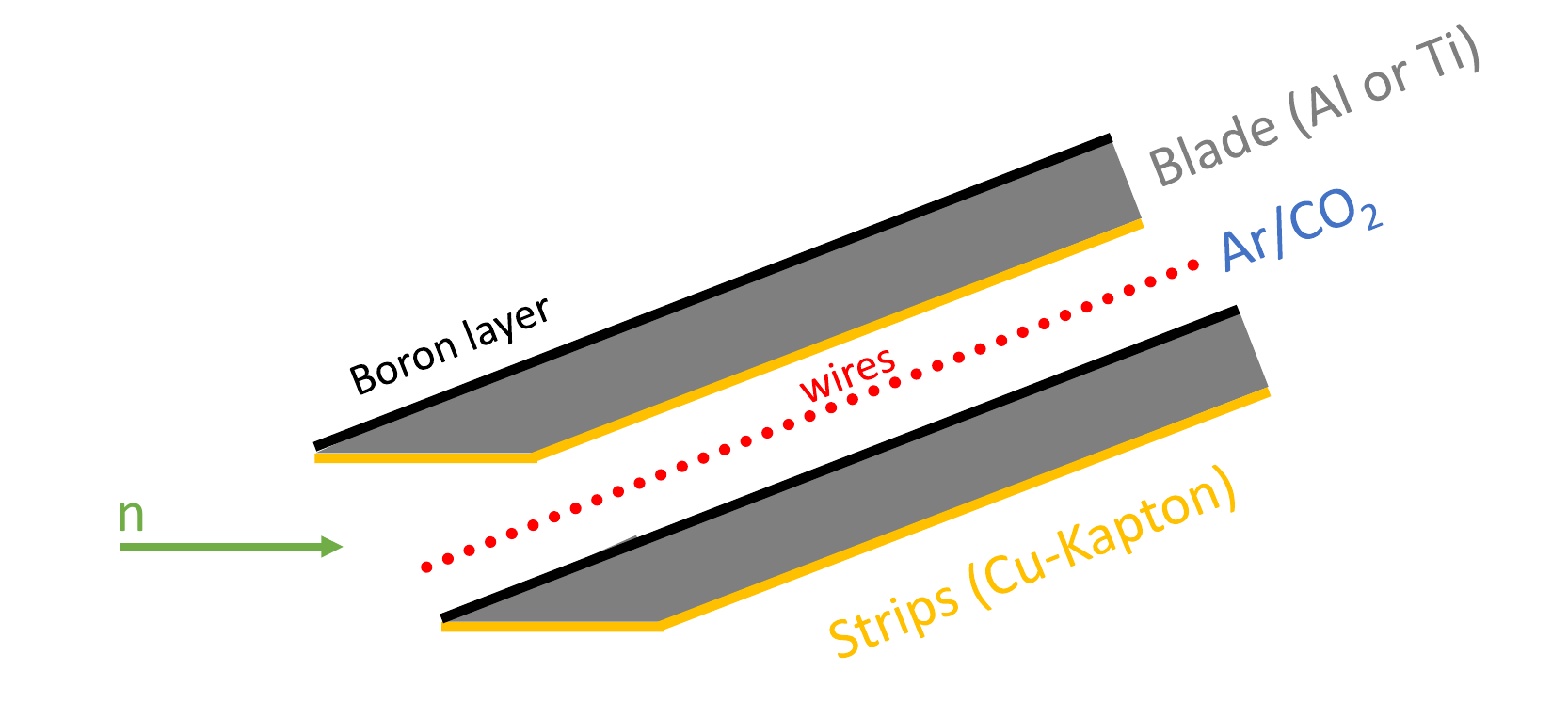}
\caption{\label{cstmb} \footnotesize Schematic view of the Multi-Blade cassettes. Figure from~\cite{MIO_fastn}.}
\end{figure}

It is recalled that the Multi-Blade is a stack of Multi Wire Proportional Chambers (MWPC) operated at atmospheric pressure with a continuous gas flow (Ar/CO$_2$ 80/20 mixture in volume). A sketch of the cassette is shown in figure~\ref{cstmb}. The purpose of the picture is to visualize the materials in the detector that are more relevant for our study, i.e., \textit{Ar , CO$_2$} (gases) ,\textit{ Al , Ti} (blade materials) , \textit{Cu , Kapton} (strip materials) , \textit{$^{10}$B$_{4}$C} (neutron converter).

\begin{figure}[htbp]
\centering
\includegraphics[width=1\textwidth,keepaspectratio]{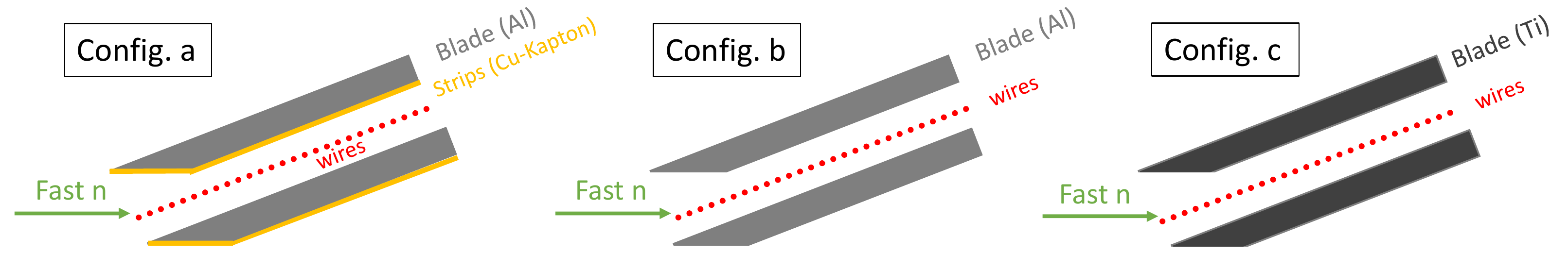}
\caption{\label{cassettesc} \footnotesize Schematic view of the Multi Blade cassettes for three different configurations. In all of them the $^{10}$B$_{4}$C layer was removed. a- Aluminium blade and strip-plane in Cu and Kapton, b- Aluminium blade and strip-plane removed, c- Titanium blade and strip-plane removed. Figure from~\cite{MIO_fastn}.}
\end{figure}

In order not to be sensitive to thermal neutrons, all the measurements were performed without the $^{10}$B layer, while keeping all the rest of the detector unchanged. Three different detector configurations have been investigated, as shown in figure~\ref{cassettesc}. {\it Configuration a} is the standard configuration which is a full cassette (without the $\mathrm{^{10}B_{4}C}$ layer see figure~\ref{cstmb} for comparison), {\it Configuration b} is the configuration with Aluminum blades without strips, without Cu nor Kapton, and {\it Configuration c} is the configuration with Titanium blades and no strips.  

\begin{figure}[htbp]
\centering
\includegraphics[width=1\textwidth,keepaspectratio]{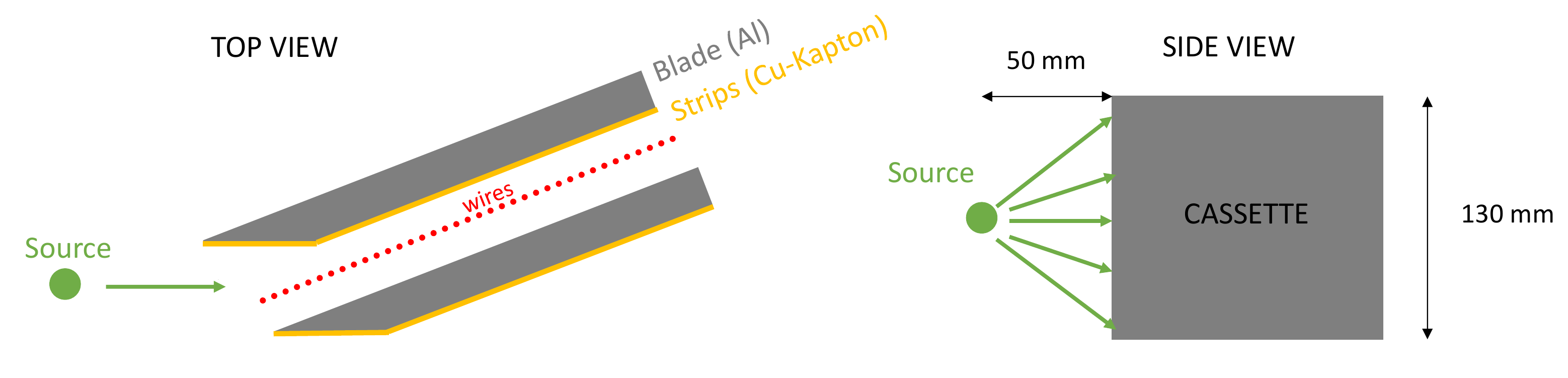}
\caption{\label{setup} \footnotesize Schematic representation of the experimental setup used to perform the measurements showing the position of the source with respect to the cassettes of the Multi-Blade detector. Figure from~\cite{MIO_fastn}.}
\end{figure}

All the measurements have been performed with the same experimental setup, a sketch of which is shown in figure~\ref{setup}. 

In the tests the response of one cassette has been measured and this is representative of the full detector because the design is modular and an event in a cassette does not influence neighbour cassettes. The cassette was equipped with an individual readout, one amplifier on each wire or strip (64 channels)~\cite{MIO_MB2017}. The pre-amplifiers were CREMAT~\cite{EL_cremat} CR-110 charge sensitive pre-amplifiers and the signals are shaped with a CREMAT~\cite{EL_cremat} CR-200-500 ns Gaussian shaping amplifier. The pre-amplifiers have a gain of 1.4 V/pC and the shaping amplifiers a gain of 10 with 500 ns shaping time. A CAEN~\cite{EL_CAEN} Digitizer was used to record the traces and the amplitude of the signals (Mod. V1740D, 12bit, 62.5 MS/s).
\\ A software threshold for each wire and strip was set to reject the electronic noise and the detector was operated at 800 V (gas gain $\approx$15).

\subsection{Sources} \label{subsources}

Three neutron sources: $^{252}$Cf, $^{241}$Am/Be, $^{238}$Pu/Be, were used to perform the measurements. A further measurement was performed with a $^{60}$Co gamma source, in order to have a comparison with the previous work on gamma sensitivity~\cite{MIO_MB2017,MG_gamma}.
\\Actinide/Be-based radioactive sources produce neutrons when the actinide decay by emission of an $\alpha$-particle which interacts with the $^9$Be nucleus. $^{252}$Cf decays by both $\alpha$-particle emission ($\sim$ 97\%) and spontaneous fission ($\sim$ 3\%), the latter giving rise to fast neutrons directly. The activity of the Actinide/Be sources used in this work is about one order of magnitude higher than that of the $^{252}$Cf.
\\The mean of the incident $\alpha$-particle energies is 5.48 MeV for the $^{241}$Am source and 5.49 MeV for the $^{238}$Pu source. The energy spectra of emitted neutrons are similar~\cite{PuBe_AmBe}. The fast neutron emission spectrum of $^{252}$Cf has an average energy of 2.1 MeV, with a most probable energy of 0.7 MeV~\cite{Cf}.
\\ Besides the differences between the energy spectra of the Actinide/Be-based radioactive sources and of the $^{252}$Cf, the energy distribution range considered is $1-10$ MeV for the three sources.
\\ The solid angle has been calculate on a rectangular surface of the detector cassette under test (130x10)mm$^2$ for a fixed distance, d, between the source and the detector surface.  Considering a point-like source, all the measurements were performed with d = 50 mm, see figure~\ref{setup}. It results in a solid angle of $\approx 0.1$ sr ($\approx 1\%$ of a sphere). It has been estimated that the measurement of the distance between the source and the sensitive region can lead to an uncertainty which is less than a factor two variation for a misplacement of a $\pm 15$ mm from the fixed location of the source.
\\ All the measurements are normalized to the flux, number of neutrons per unit time and area, multiplying the activity by the solid angle. 

\section{Physical processes}\label{theocalc}

The thermal reaction mediated by a $\mathrm{^{10}B}$ layer is $\mathrm{^{10}B(n,\alpha)}$Li : 

\begin{equation}
n + ^{10}_{5}B \rightarrow \begin{cases}^{7}_{3}Li (1.02 MeV)+ \alpha ( 1.78 MeV) \quad 6\% \\ ^{7}_{3}Li (0.84 MeV)+ \alpha(1.47 MeV) + \gamma (0.48 MeV)\quad 94\% \end{cases}
\label{tras}
\end{equation}

The range of the product particles are of the order of a few $\mu$m in solids and a few mm in gasses at atmospheric pressure.

\subsection{Cross section and interaction probability calculation}\label{sp}

A preliminary study was performed on the nuclear processes that may occur between neutrons of energy range 1-10 MeV and nuclei present in the detector material. This investigation gives a qualitative indication which reactions are significant for the fast neutrons sensitivity of the detector. 
\\For this study the main materials in the Multi-Blade detector~\cite{MIO_MB2017} listed in section~\ref{mbdescr} have been taken into account. The probability of interaction in a medium, $P(x)$, and the probability of deposition in the gas, \textit{P(r)}, have been calculated as follows:

\begin{equation}
P(x) = 1 - e^{-\Sigma x}
\label{inter}
\end{equation}

\begin{equation}
P(r) = 1 - e^{-\Sigma r}
\label{escape}
\end{equation}

where \textit{x} is the path length and \textit{r} is the range of the particle travelling in the medium and $\Sigma = N \sigma$ is the macroscopic cross section. $\Sigma$ defines the probability per unit path length for a specific process, described by the microscopic cross section $\sigma$; $N$ is the number of nuclei per unit volume. The number of interactions decreases exponentially with the absorber thickness. The relation of attenuation is described as $\frac{I}{I_{0}} = e^{-\Sigma x}$. \textit{P(r) = P(x)} for the gas.

The macroscopic cross section has been calculated from the number density $n_p$ and the microscopic cross section. In the case of a gas, the calculation starts from the ideal gas law $PV = nRT$. The number of moles $n$ can be written as $N_p / N_A$, where $N_p$ is the number of particles and $N_A$ is the Avogadro constant. By knowing that $R/N_A = K_B$, which is the Boltzmann constant, per volume unit it is obtained: $n_p = \frac{P}{K_B T}$ where $P$ is the pressure, $T$ is the temperature of the gas. For the solids $n_p = \frac{N_A}{M} \cdot \rho$, where $M$ (g/mol) is the molar mass and $\rho$ (g/cm$^3$) is the density of the material.
\\ In a proportional chamber the gas is the detection medium, for the Multi-Blade detector, this is an 80/20 mixture of Ar/CO$_2$ operating at atmospheric pressure and room temperature. Considering the weight $w$ and the fraction $f$ of the two compounds. The number density for the Ar/CO$_2$ is shown in table~\ref{gasnp}, while for pure materials the $n_p$ is listed in table~\ref{sol}.

\begin{table}[htbp]
\centering
\caption{\label{gasnp} \footnotesize Atomic density value for the gas Ar/CO$_2$ (80/20) in the detector.}
\smallskip
\begin{tabular}{|c|c|c|c|}
\hline
  & f & w & $n_p$ ($10^{22}/cm^3$) \\
\hline
Ar & 1 &  0.8 & 0.0212 \\
\hline
C & 1/3 & 0.2 & 0.00177 \\
\hline
O$_2$ & 2/3 & 0.2 &  0.00354\\ 
\hline
\end{tabular}
\end{table}

\begin{table}[htbp]
\centering
\caption{\label{sol} \footnotesize Atomic density value for solid pure materials in the detector.}
\smallskip
\begin{tabular}{|c|c|c|c|}
\hline
 &  $\rho $&  $M $& $n_p $ ($10^{22}/cm^3$)\\
\hline
Al  & 2.7 &  26.98 & 6.026 \\
\hline
Ti & 4.506 & 47.87 & 5.669 \\
\hline
Cu & 8.96 &63.55 & 8.491 \\ 
\hline
\end{tabular}
\end{table}

The Kapton ($\mathrm \rho$ = 1.42 $\mathrm g/cm^3$) is composed of several elements, the macroscopic cross section is  $\Sigma_{mix} = \sum_i N_i \sigma_i$, the $\mathrm{^{10}B_{4}C}$ layer ($\mathrm \rho$ = 2.45 $\mathrm g/cm^3$) as well. This density differs from the material data and refers to a measured value for the films employed in this detector~\cite{MIO_B10refl}. The $n_p $ calculation for both compounds is reported in table~\ref{kap}.

\begin{table}[htbp]
\centering
\caption{\label{kap} \footnotesize Atomic density value for the kapton and the $\mathrm{^{10}B_{4}C}$ elements.}
\smallskip
\begin{tabular}{|c|c|c|c|c|c|}
\hline
 & & $w$ & $M $ & $\rho $  & $n_p $ ($10^{22}/cm^3$) \\
\hline
& H & 0.0264 &  1.01 & 0.037 &  2.23 \\
\cline{2-6}
Kapton & C & 0.6911 & 12.01 & 0.981 &  4.92 \\
\cline{2-6}
& N & 0.0733 & 14.01  & 0.104 & 4.47 \\
\cline{2-6}
& O & 0.2092 &15.99 &0.297 & 1.12  \\ 
\hline
\hline
$\mathrm{^{10}B_{4}C}$ & B & 0.7826 & 10.81 & 1.972 &  10.98 \\
\cline{2-6}
& C & 0.2174 & 12.011 & 0.548 &  2.75 \\
\hline
\end{tabular}
\end{table}

\begin{figure}[htbp]
\centering
\includegraphics[width=1\textwidth,keepaspectratio]{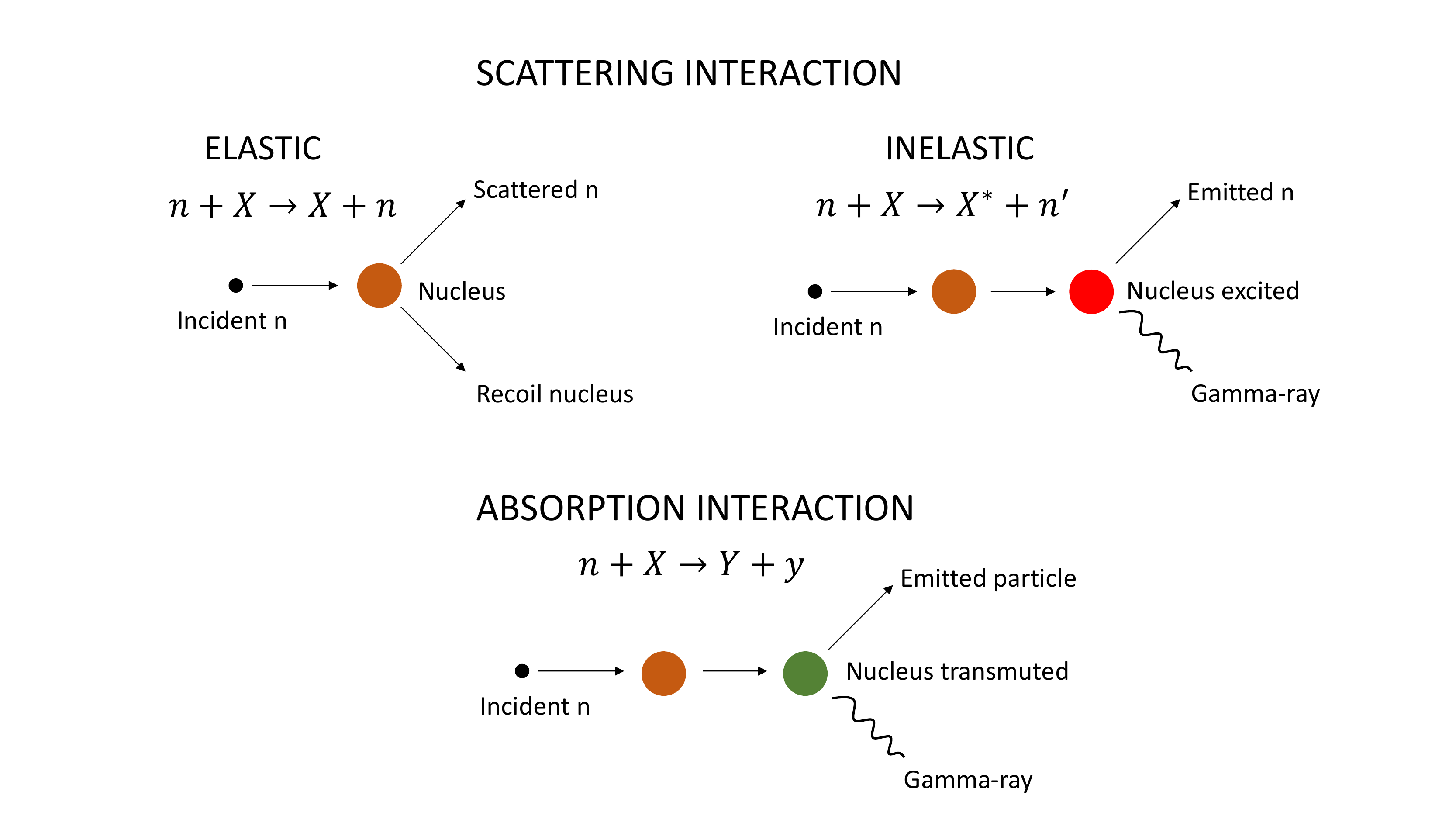}
\caption{\label{interaction} \footnotesize  A schematic view of the interactions of fast neutrons with matter. Elastic scattering interaction (top-left): the recoil nucleus and a scattered neutron is emitted. Inelastic scattering (top-right): the nucleus goes into an excited state and emits one or more gamma-rays. Absorption interaction (bottom) includes all the processes with heavy particles or gamma-rays as a yield. Figure from~\cite{MIO_fastn}.}
\end{figure}

A neutron can have many types of interaction with a nucleus. In the detector these can give rise to charged products that can release an amount of energy in the detection medium and can be interpreted as real thermal neutron events, instead of being rejected. For this work the fast neutron energy range between 1 and 10 MeV is taken into account, see subsection~\ref{subsources}. In this range several interactions between neutrons and nucleus may occur. Our focus is on two main types: \textit{Scattering} and \textit{Absorption}. 
\\ When a neutron is scattered by a nucleus, the scattering interaction transfers some portion of the neutron kinetic energy to the target nucleus resulting in a \textit{recoil nucleus}. The total kinetic energy of the neutron-nucleus system remains unchanged by the interaction, i.e., the Q-value of the reaction is zero.
In the considered energy range the maximum possible recoil energy of the nucleus is 

\begin{equation}
E_R = \frac{4A}{(1+A)^2}  E_n
\label{e_recoil}
\end{equation}

where A is the mass of the target nucleus , $E_R$ the recoil nucleus kinetic energy and $E_n$ the incoming neutron kinetic energy~\cite{DET_knoll}. 
\\ Referring to figure~\ref{interaction}, the scattering interaction also includes the \textit{inelastic scattering} in which the recoil nucleus goes to one of its excited states during the collision and then de-excites emitting one or more gamma rays or internal conversion electrons. 
\\ If the neutron is absorbed or captured, instead of being scattered, a variety of processes may occur. The nucleus may rearrange its internal structure and release gamma rays (n,$\gamma$) or charged particles with the more common ones being protons (n,p), alpha particles (n,$\alpha$) and deuterons. For this energy range only p and $\alpha$ are relevant. All these kind of interactions are included in absorption processes~\cite{n_interaction_matter}. A sketch of these reactions is shown in figure~\ref{interaction}.

\begin{figure}[htbp]
\centering
\includegraphics[width=.95\textwidth,keepaspectratio]{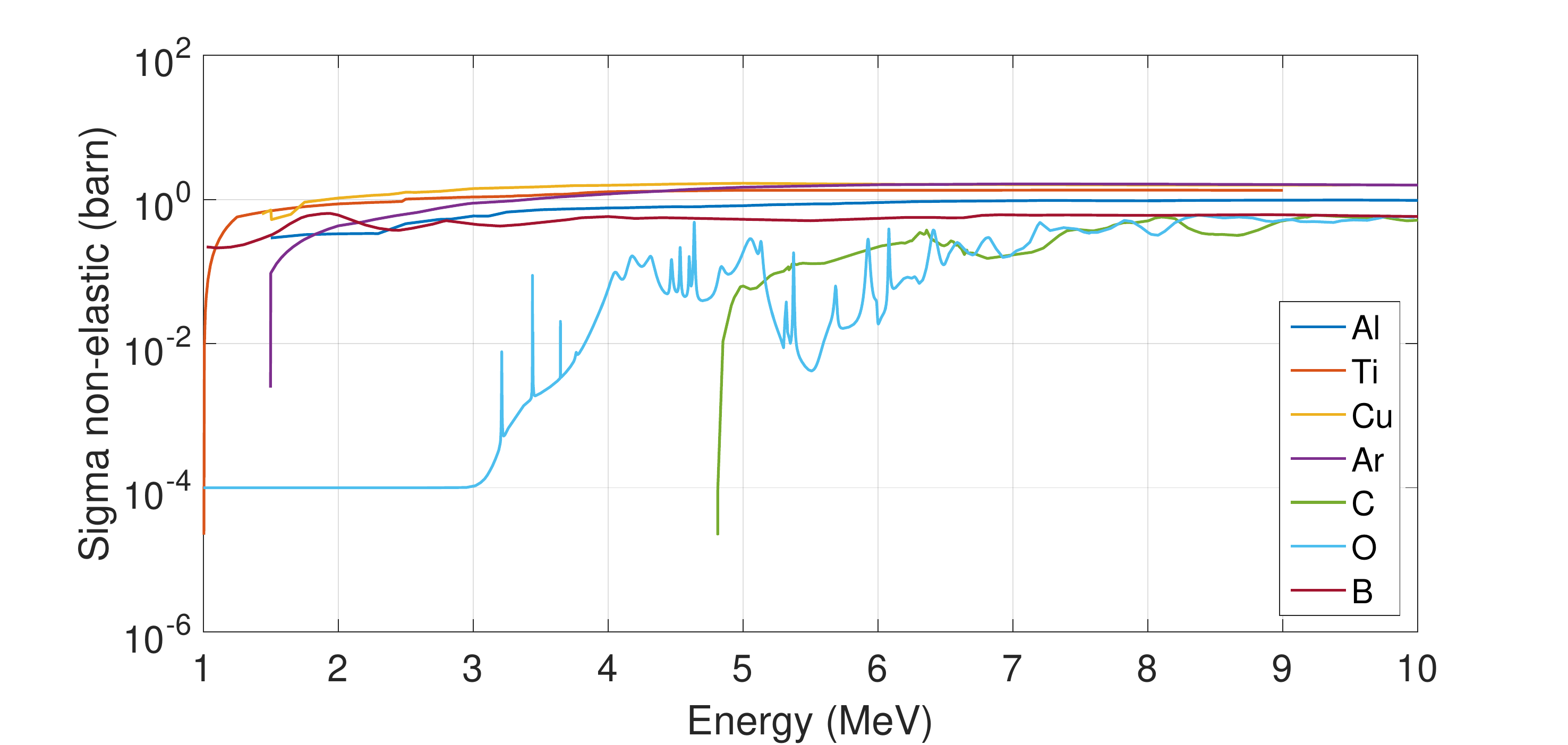}
\includegraphics[width=.95\textwidth,keepaspectratio]{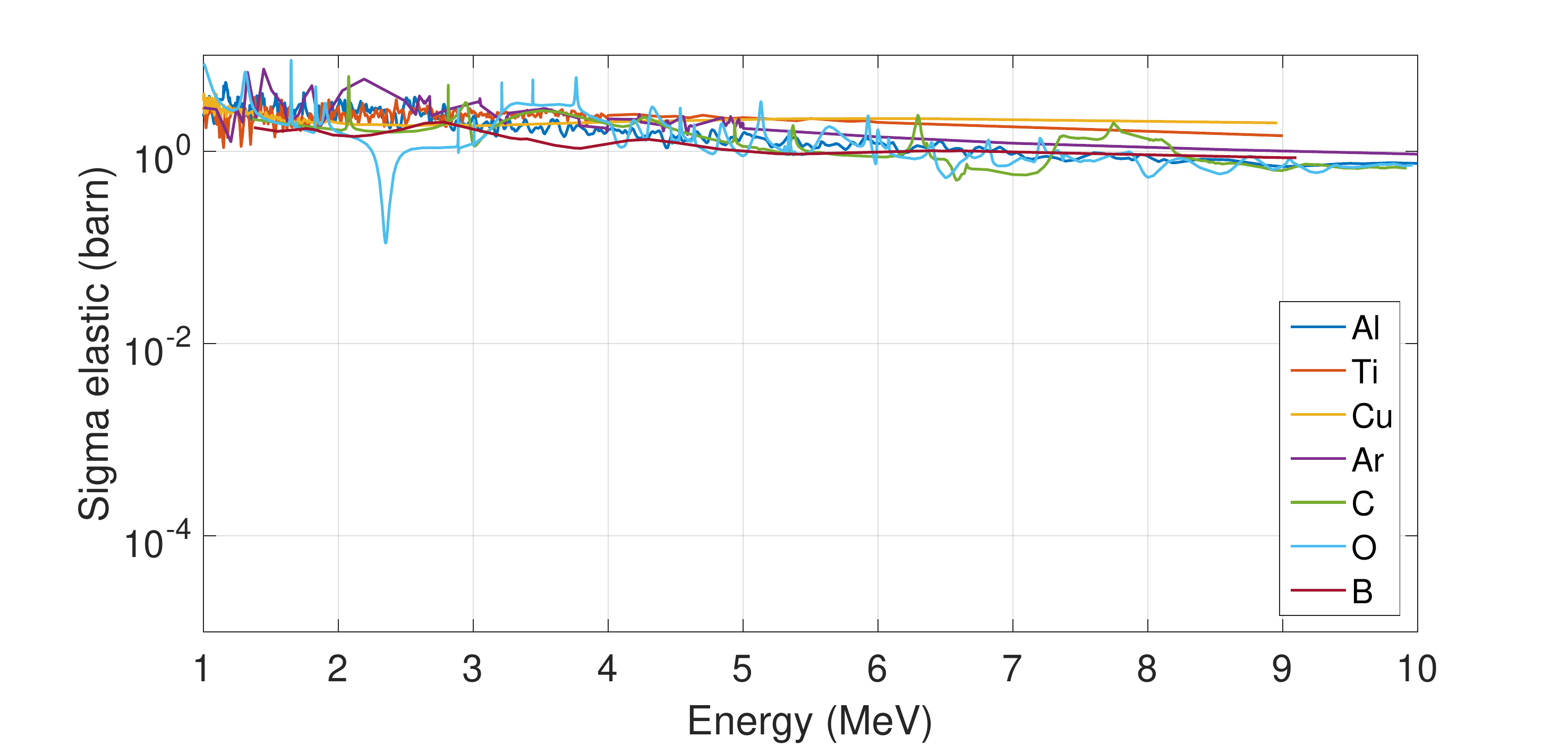}
\caption{\label{snoel} \footnotesize  Non-elastic (top) and elastic (bottom) cross sections~\cite{NIST} for the materials of the Multi-Blade detector. The considered energy range is 1-10 MeV, matching the emission spectrum of the sources used in the measurements. Figure from~\cite{MIO_fastn}.}
\end{figure}

An important difference between the fast neutron interaction and the $\gamma$-rays interaction is the final products. When a $\gamma$-ray interacts with a medium, it can generate a detectable signal, via an electron, typically in a photoelectric or Compton interaction, while for fast neutrons protons, $\alpha$ and heavier particles can also be emitted.
\\The range of protons and heavier particles in a medium is typically much shorter than that of electrons for a given energy. Therefore the amount of energy released per unit path in the gas is much larger for heavy particles than for electrons~\cite{MG_gamma}. 
\\ The fast neutron interaction can give rise to $\gamma$-rays as well which undergo the process as described above. In $\mathrm{^{10}B}$-based detector, it has been denoted that the detection probability of $\gamma$-rays is low of the order of $10^{-8}$, as reported in~\cite{MIO_MB2017,MG_gamma}. Therefore a larger contribution to the sensitivity to fast neutron is mainly expected from (n, particles) reactions and the elastic scattering. The contribution from the inelastic scattering and (n,$\gamma$) interactions is negligible. 
\\ The \textit{Non-Elastic} interaction is defined as the sum of inelastic and absorption processes. Figure~\ref{snoel} shows several non-elastic (top) and elastic (bottom) cross sections, in the energy range 1-10 MeV~\cite{NIST}.

In the considered energy range, the elastic scattering interaction has the largest cross sections and the higher probability of deposition in the gas $P(r)$ (eq.~\ref{escape}). The inelastic scattering and (n,$\gamma$) processes are comparable for each considered material and they have the lowest probability. 
\\The most relevant absorption processes taken into account are~\cite{Al_np, Ti_np, O_nalpha}: 

\begin{equation}\label{al_p}
^{27}Al(n,p)^{27}Mg \quad Q = -1.8MeV
\end{equation}

\begin{equation}\label{ti_p}
^{48}Ti(n,p)^{48}Sc \quad Q = -3.19 MeV
\end{equation}

\begin{equation}\label{o_alpha}
^{16}O(n,\alpha)^{13}C \qquad Q = -2.2 MeV
\end{equation}

In general these absorption processes are:

\begin{equation}
n + X \rightarrow Y + y
\label{reaction}
\end{equation}

\begin{figure}[htbp]
\centering
\includegraphics[width=1\textwidth,keepaspectratio]{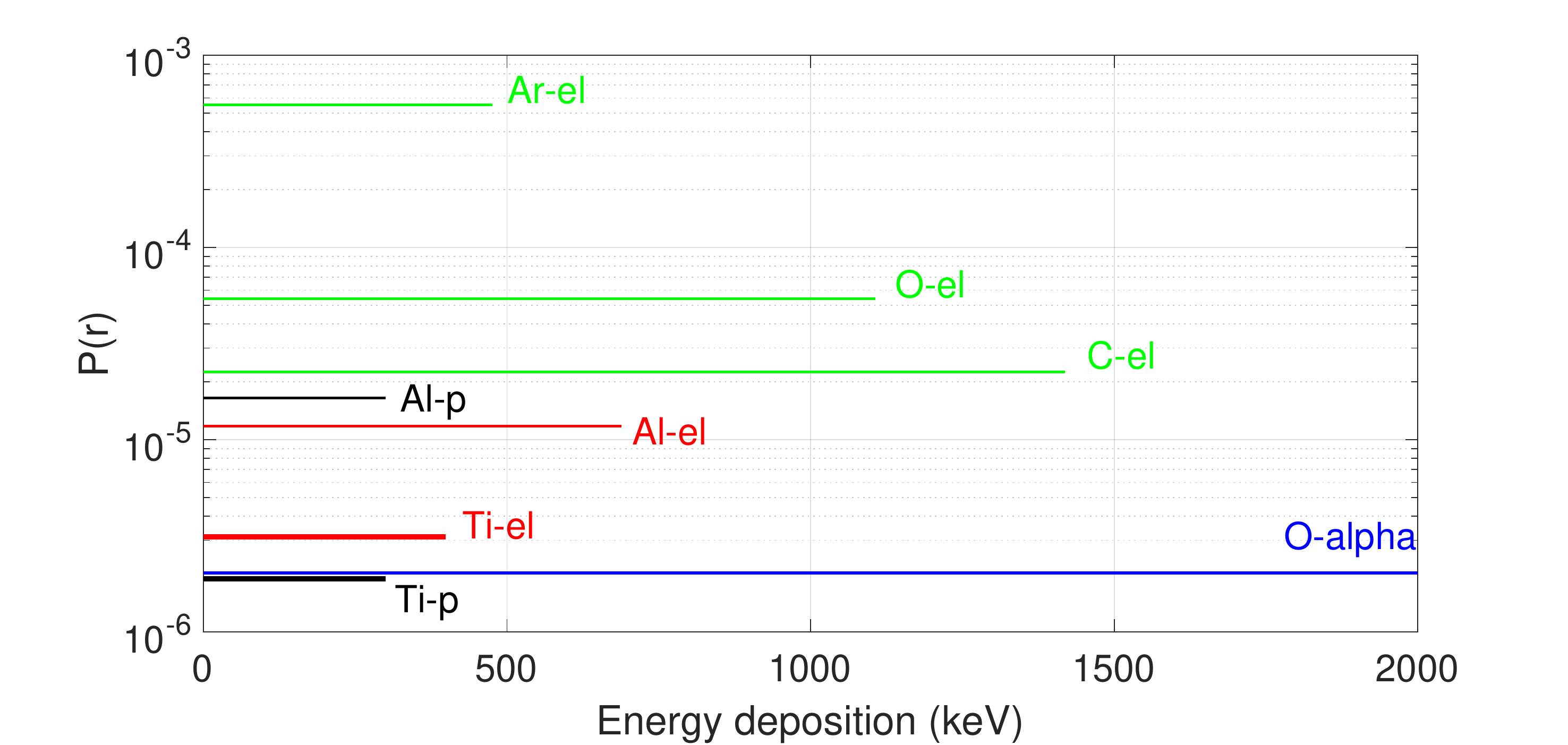}
\caption{\label{theo_el} \footnotesize Theoretical calculation of the maximum energy deposition in the gas and its probability for the different components of the detector. The contribution of the gas is dominant with respect to that of solids. The processes describes are elastic (-el), (n,p) (-p) and (n,$\alpha$) (-alpha). Figure from~\cite{MIO_fastn}.}
\end{figure}

For each reaction the energy threshold ($\mathrm E^{th}_n$) is calculated for the incoming neutron to generate the reaction as 

\begin{equation}
E^{th}_n = - \Big(1+ \frac{m_n}{M_X} \Big)\cdot Q
\label{e_th}
\end{equation}

where $m_n$ is the neutron mass and $M_X$ is the mass of the element $X$ in equation~\ref{reaction}. The energy of the produced particles ($\mathrm E_y$) is calculated from the definition of the Q-value and the conservation of energy and momentum as 

\begin{equation}
E_y =  \Big( \frac{M_Y}{M_Y + m_y}\Big)\cdot Q_n
\label{e_y}
\end{equation}

where $M_Y$ and $m_y$ are respectively the mass of the elements $Y$ and $y$ in equation~\ref{reaction}. $Q_n = E_n +Q$ is the sum of the incoming energy of neutrons and the Q-value.
\\ In order to compare these results to a theoretical prediction, an incoming neutron energy of $E_n = 5$ MeV has been chosen for the calculations, because it is the average value of the energy range of interest and is approximately the average of the emission spectrum of the fast neutron sources used for the measurements (see section~\ref{subsources}).
The probability of deposition $P(r)$ (eq.~\ref{escape}) summarized in table~\ref{tabsp} has been calculated considering the average of the cross sections $\sigma$ for each interaction in the given energy range and the number density $n_p$ for each material, in the detector. 
\\ For the elastic scattering and the absorption interactions $E_R$ and $E_y$ were calculated respectively, as described above. These energies have been used as an input in a SRIM simulation~\cite{ MISC_SRIM2010, MISC_SRIM1998} in order to estimate the maximum range $r$. Considering, instead, the nominal thickness $x$ of the medium for the inelastic scattering interactions, because the yields of this process are $\gamma$-rays. 
\\ The values are shown in table~\ref{tabsp} and the individual contribution of the elements for the elastic and the absorption interactions are shown in figure~\ref{theo_el}.

\begin{table}[htbp]
\centering
\caption{\label{tabsp} \footnotesize Macroscopic cross sections and probability of deposition in the gas values for the maximum range, r, and the thickness, x, given by the detector geometry.}
\smallskip
\begin{tabular}{|c|c|c|c|}
\hline
\multicolumn{4}{|c|}{\textbf{Elastic Interaction}}\\
\hline
 &  $\Sigma (10^{-3}/cm)$ & $r (cm)$ & $P(r)(\%)$  \\
\hline
 Al & 139.2 & 0.0000874  & 0.00117 \\
 \hline
Ti & 134.9 & 0.0000232 &  0.000313\\
\hline
Cu & 224.2 &  0.0000089 &  0.000199 \\
\hline
$\mathrm{Ar/CO_2} $ & 0.79 & 0.803 & 0.0629 \\ 
\hline
$\mathrm{^{10}B_{4}C}$ & 202.2 & 0.000148 & 0.0029\\
\hline\hline
\multicolumn{4}{|c|}{\textbf{Inelastic Interaction}}\\
\hline
 &  $\Sigma (10^{-3}/cm)$ & $x (cm)$ & $P(x)(\%)$  \\
\hline
 Al & 41.83 & 2.295  & 9.48 \\
 \hline
Ti & 44.67 & 2.295 &  10.04\\
\hline
Cu & 94.14 &  0.021 &  0.197 \\
\hline
Kapton & 8.259 &  0.046 &  0.038 \\
\hline
$\mathrm{Ar/CO_2} $ & 0.188 & 8.032 & 0.0168 \\ 
\hline
$\mathrm{^{10}B_{4}C}$ & 58.95 & 0.008 & 0.046\\
\hline\hline
\multicolumn{4}{|c|}{\textbf{Absorption Interaction}}\\
\hline
 &  $\Sigma (10^{-3}/cm)$ & $r (cm)$ & $P(r)(\%)$  \\
\hline
 Al(n,p) & 44.09 & 0.0081  & 0.0016 \\
\hline
Ti(n,p)& 44.67 & 0.0026 &  0.00018\\
\hline
O(n,$\alpha$) & 103.1 &  0.803 &  0.0002 \\
\hline
\end{tabular}
\end{table}

A direct comparison between the materials is possible because the probability of deposition in the gas is normalized by the density. An interaction in the solid generates charged particles that have to escape the surface and reach the gas to be detected. With the SRIM calculation has been quantified the portion of surface involved in the process (maximum range r). A direct comparison is then possible between solids and gas components as well.
\\ Referring to figure~\ref{theo_el}, the energy released by the elastic process of the gas elements is dominant ($\approx$ 40 times larger) compared to the same contribution of the several solid components. The energy deposited by the various elastic reactions is 1-2 orders of magnitude larger than the contribution of the absorption interactions.
\\ Although the $\mathrm{^{10}B_{4}C}$ layer is not present in the detector or this experimental setup, from the calculation has been understood that its contribution would be approximately 1/20 of that of the $\mathrm{Ar/CO_2} $.
\\ The comparison between Aluminium and Titanium is crucial for the Multi-Blade design, because it concerns the choice of the blade material, aiming at an improved detector performance. A difference emerges from the calculation of the probability of deposition for elastic scattering. For Aluminium, P(r) is approximately 4 times larger than for Titanium and about 6 times higher than for Copper. Cu has one of the largest microscopic cross sections among the studied materials (section~\ref{mbdescr}) in the energy range (subsection~\ref{subsources}), but it is the heaviest element among them. It is then less probable for a particle to escape the surface and release energy in the gas. Referring to the absorption processes described above, the deposition probability of the (n,p) process for Aluminium is 1 order of magnitude larger than for Titanium. 

\begin{figure}[htbp]
\centering
\includegraphics[width=1\textwidth,keepaspectratio]{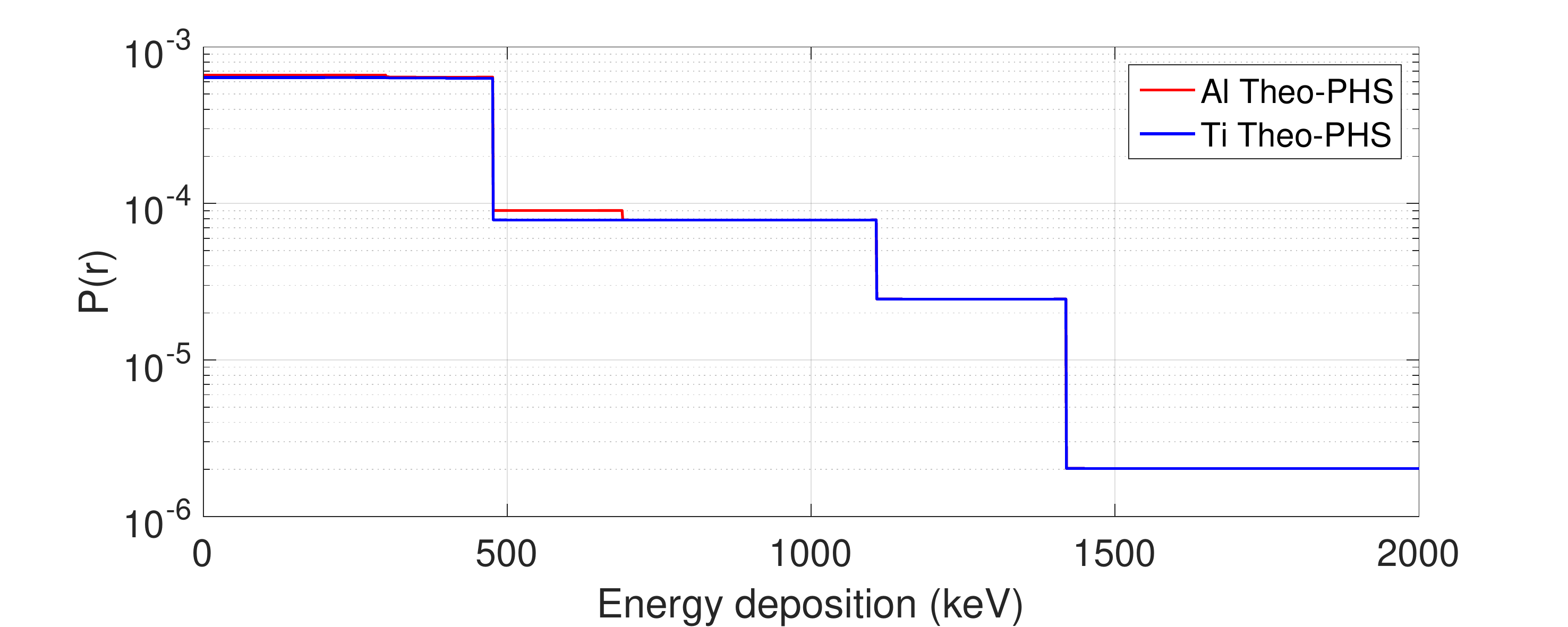}
\caption{\label{theo_phs} \footnotesize Theoretical representation of the energy deposition in gas in case of Aluminium and Titanium blades assuming the same probability, $P(r)$, in all the possible energy range defined by SRIM calculation. Figure from~\cite{MIO_fastn}.}
\end{figure}  

\begin{figure}[htbp]
\centering
\includegraphics[width=0.95\textwidth,keepaspectratio]{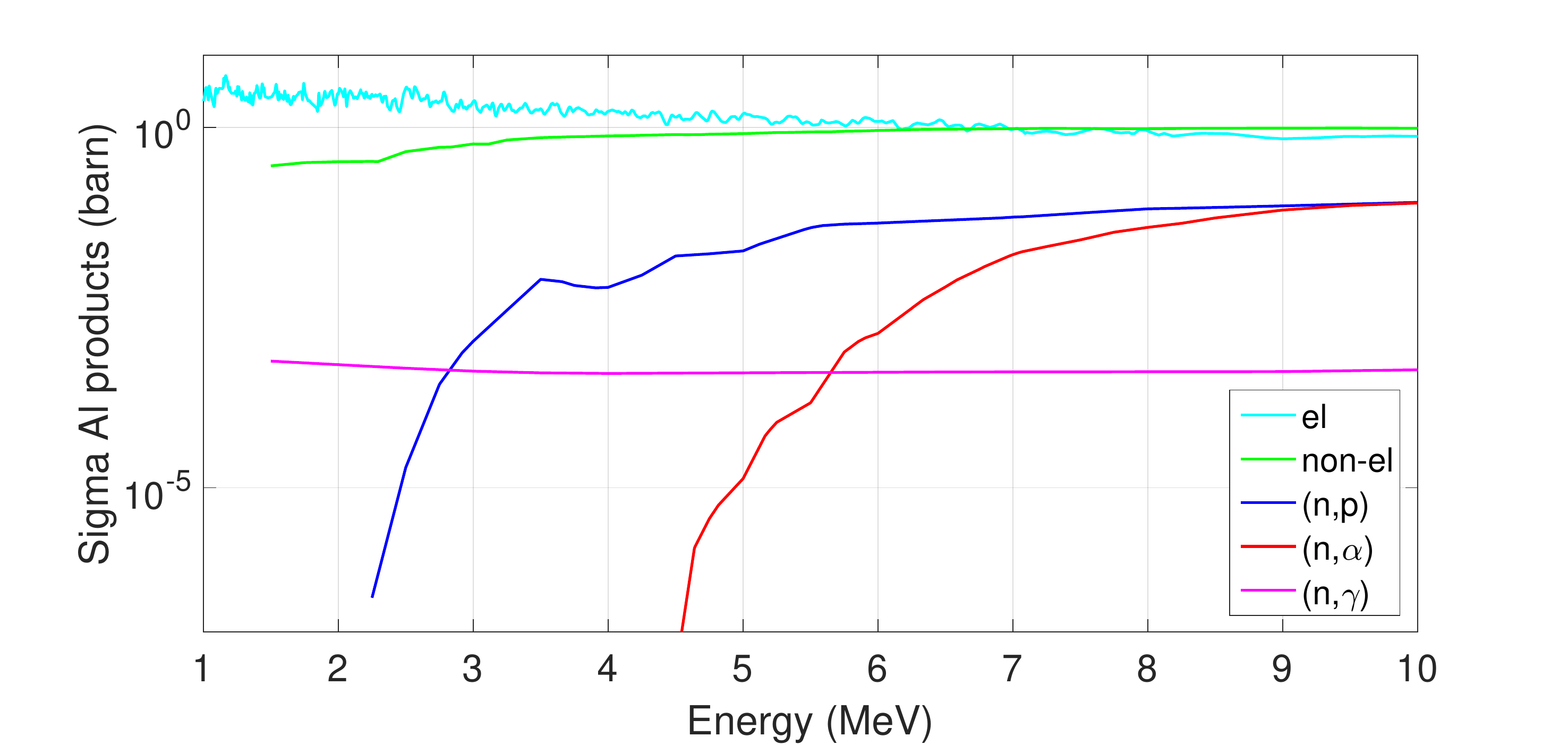}
\includegraphics[width=0.95\textwidth,keepaspectratio]{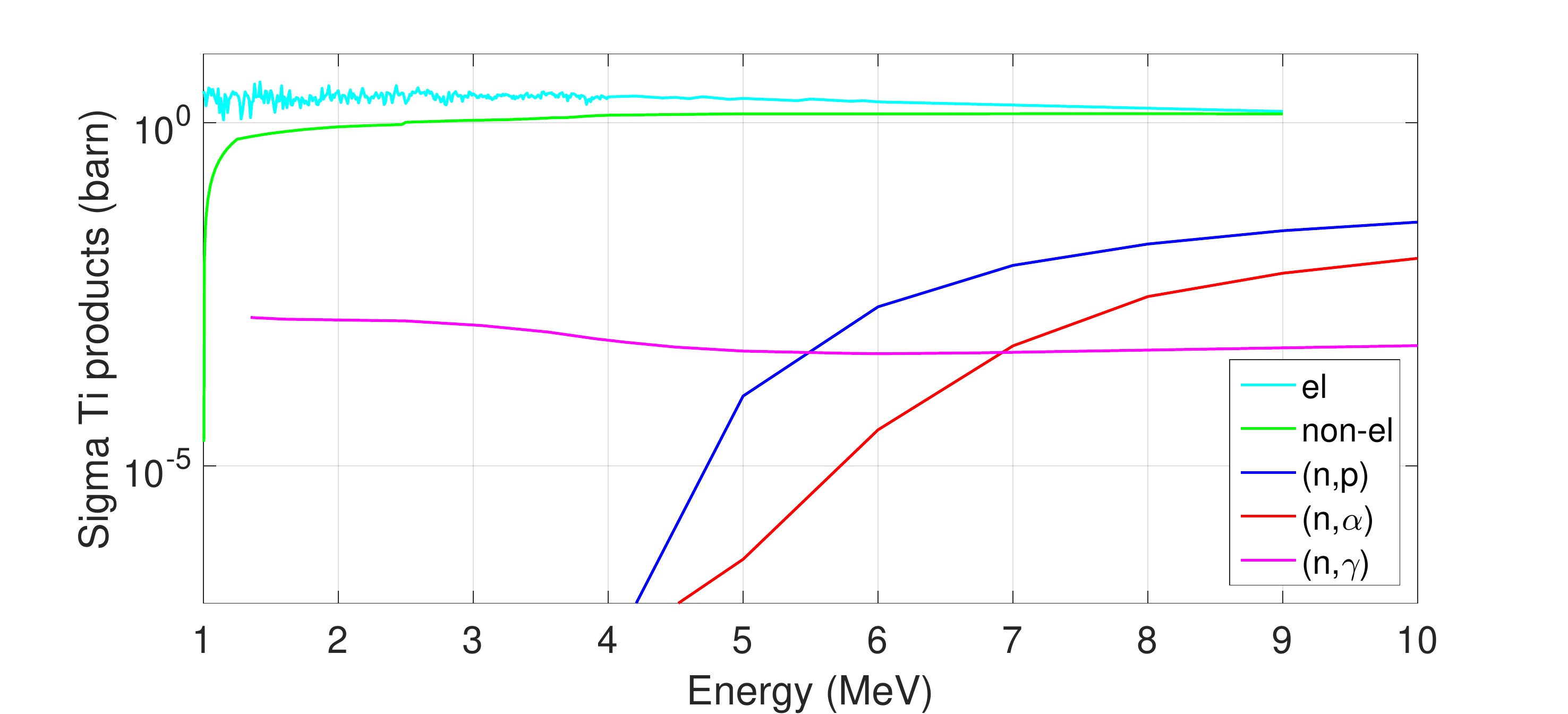}
\caption{\label{sigmaAlTi} \footnotesize Comparison between elastic and total non-elastic cross sections. The components of the non-elastic cross section are also shown:  $(n,p)$, $(n,\alpha)$, $(n,\gamma)$ processes both for Aluminium (top) and Titanium (bottom). In Aluminium the $(n,p)$ process occurs at 3 MeV, while in Titanium it has the same probability at 6 MeV. Figure from~\cite{MIO_fastn}.}
\end{figure}

Figure~\ref{theo_phs} depicts the sum of the individual energy deposition in the gas according to figure~\ref{theo_el} distinguishing between Aluminium and Titanium, as described in section~\ref{mbdescr}. The plot is a theoretical representation of a Pulse Height Spectrum (PHS) assuming that a particle has the same probability, P(r), to deposit any energy, in all the possible energy range defined by SRIM calculation.
\\The cross sections for the three most relevant absorption processes: $(n,p)$, $(n,\alpha)$, $(n,\gamma)$ are compared together with the total non-elastic cross section of two materials, figure \ref{sigmaAlTi}.
\\ The $(n,p)$ process in Al occurs approximately at $\approx$3 MeV with a probability  $\sim10^{-3}$, while for Ti the cross section of this process has the same order of magnitude at $\approx$6 MeV. Note that the neutron sources used for the measurements have an intensity of the emission spectrum which falls off at larger energies up to 10 MeV.

\section{GEANT4 simulations}\label{simu}

\begin{figure}[htbp]
\centering
\includegraphics[width=0.6\textwidth,keepaspectratio]{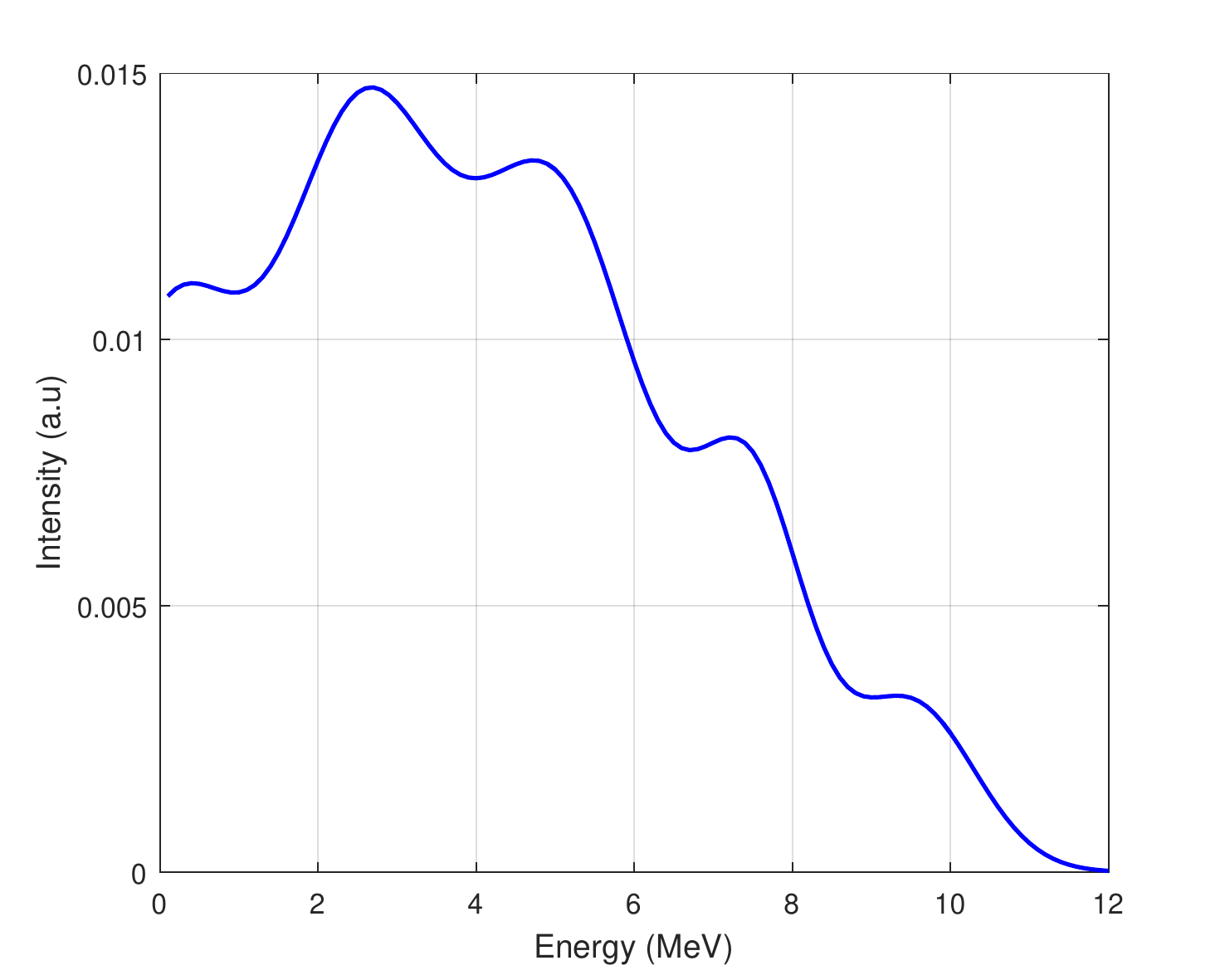}
\caption{\label{pubegauss} \footnotesize $^{238}$Pu/Be-like fast neutron energy spectrum
  used as input for the Geant4 simulations. Figure from~\cite{MIO_fastn}.}
\end{figure}

As a qualitative cross validation of the theoretical conclusions on the shape
of the PHS, a GEANT4~\cite{g1,g2,g3} simulation is
performed with a ``realistic'' energy spectrum of the neutron
source. In particular, an approximate neutron energy spectrum of a $^{238}$Pu/Be
source is used as input (see figure~\ref{pubegauss}). It is modelled using several Gaussian distributions, referring to the work on the characterization of the $^{238}$Pu/Be and $^{241}$Am/Be sources~\cite{PuBe_AmBe}.

\begin{figure}[htbp]
\centering
\includegraphics[width=0.5\textwidth,keepaspectratio]{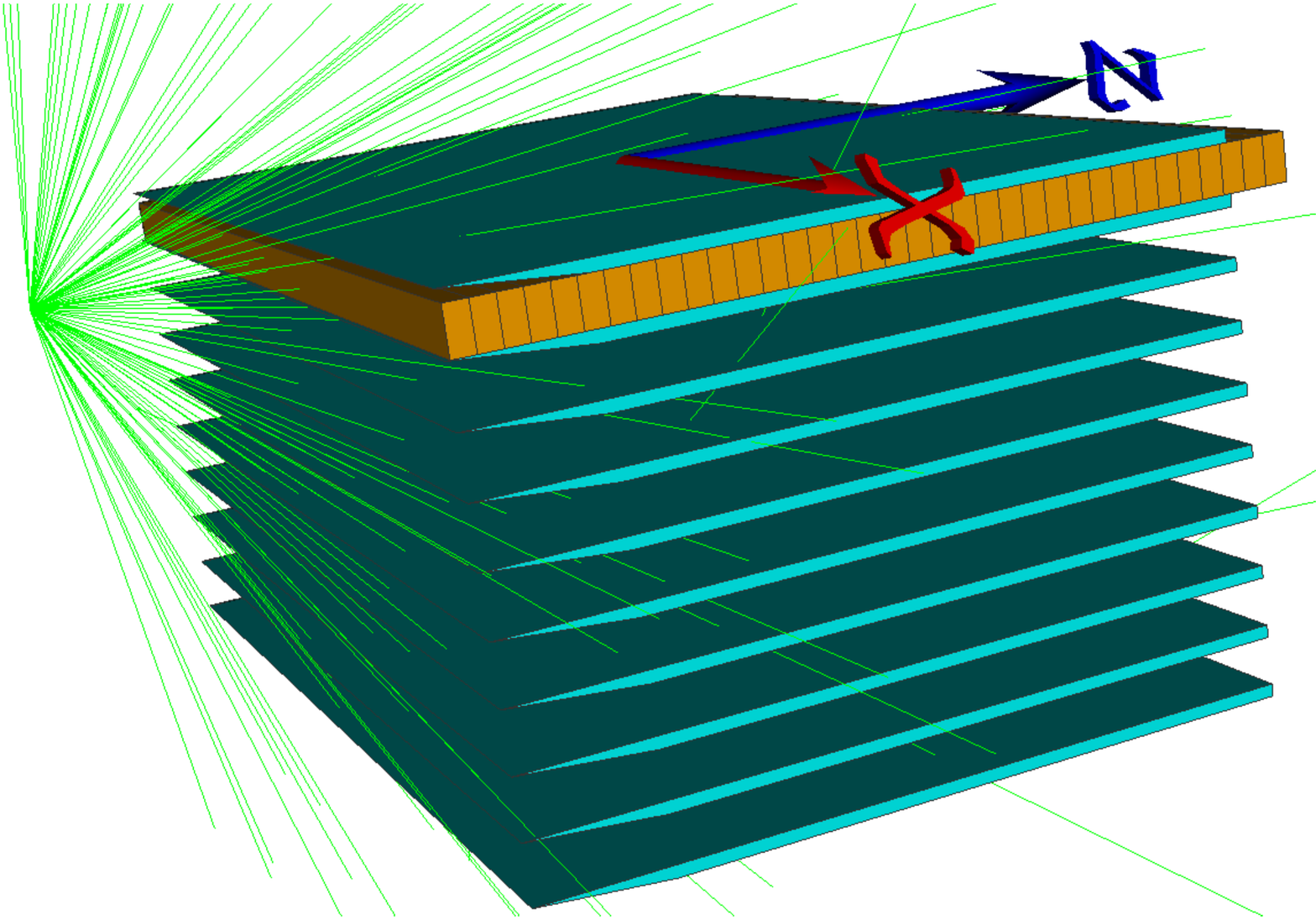}
\caption{\label{geoMB} \footnotesize Approximated Multi-Blade geometry
  with source placement as in the experimental setup. The orange volumes represent the wire
  voxels. In green are the tracks of the primary neutrons generated at the
  source. Figure from~\cite{MIO_fastn}.}
\end{figure}

The implemented detector geometry and the sources placement match the experimental setup and are
depicted in figure~\ref{geoMB}. The detector consists of ten blades
arranged in parallel, with a segmented gas region between them. The
segmentation resembles the
sensitive volumes around the wires and readout of the experimental conditions. The model
is an approximation of the actual detector in the sense that the
blades are parallel instead of having the fan-like arrangement used
for thermal neutron detection, a compromise
that facilitates the implementation of the gas volume segmentation without an impact on the
fast neutron detector response. 

The neutron source emits isotropically and is placed 5~cm from the
edge of the readout wire plane as shown in figure~\ref{geoMB}. The
particle generator produces only neutrons, without photons or other
sources of background included. No detector vessel, kapton, tungsten wires or copper strips are present in the particular geometry
model either. Their impact has been studied in a separate set of
GEANT4 simulations and as the results have fully matched the
experimental findings, a simplified detector geometry model is
utilised for the results presented here.

The material chosen for the gas composition is a mixture of Ar and
CO$_2$ with an 80:20 by mass ratio. The material of the blades is
either pure Al or Ti of natural isotopic composition with a
poly-crystalline structure. The latter is enabled via the use of
the NCrystal library~\cite{ncrystal} and will properly treat the interactions
of low energy neutrons with the crystalline material of the blades, in
case thermalization occurs in the setup. In terms of physics
list, the standard QGPS\_BIC\_HP covers the processes of interest.

Assuming perfect energy resolution, the energy deposited by any particle in the gas volume is collected by
applying the experimental condition of a 100~keV threshold per wire
volume. All wire volumes are included in the formation of the
simulated PHS to gain a statistical advantage, as the gas
amplification stage is excluded, thus equalizing their response. 

\subsection{Individual contributions to the PHS}\label{sim_contribute}

\begin{figure}[htbp]
\centering
\includegraphics[width=1\textwidth,keepaspectratio]{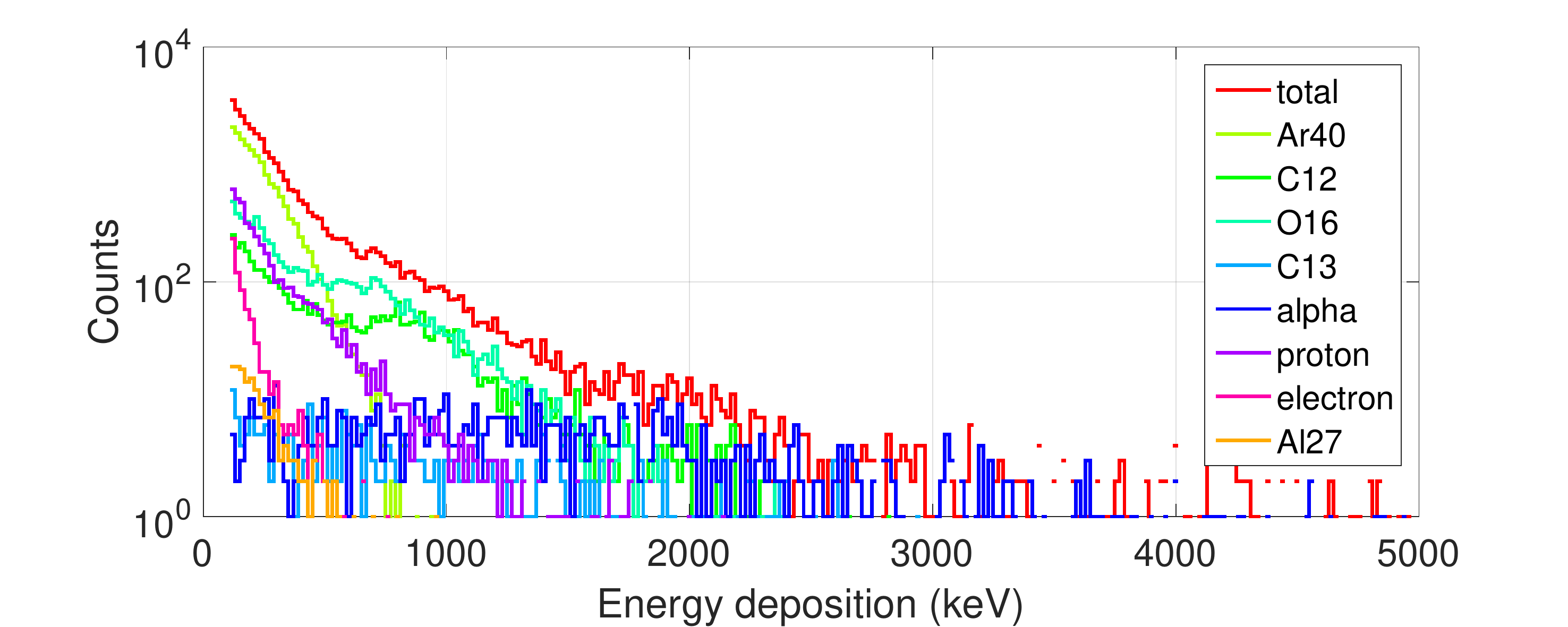}
\includegraphics[width=1\textwidth,keepaspectratio]{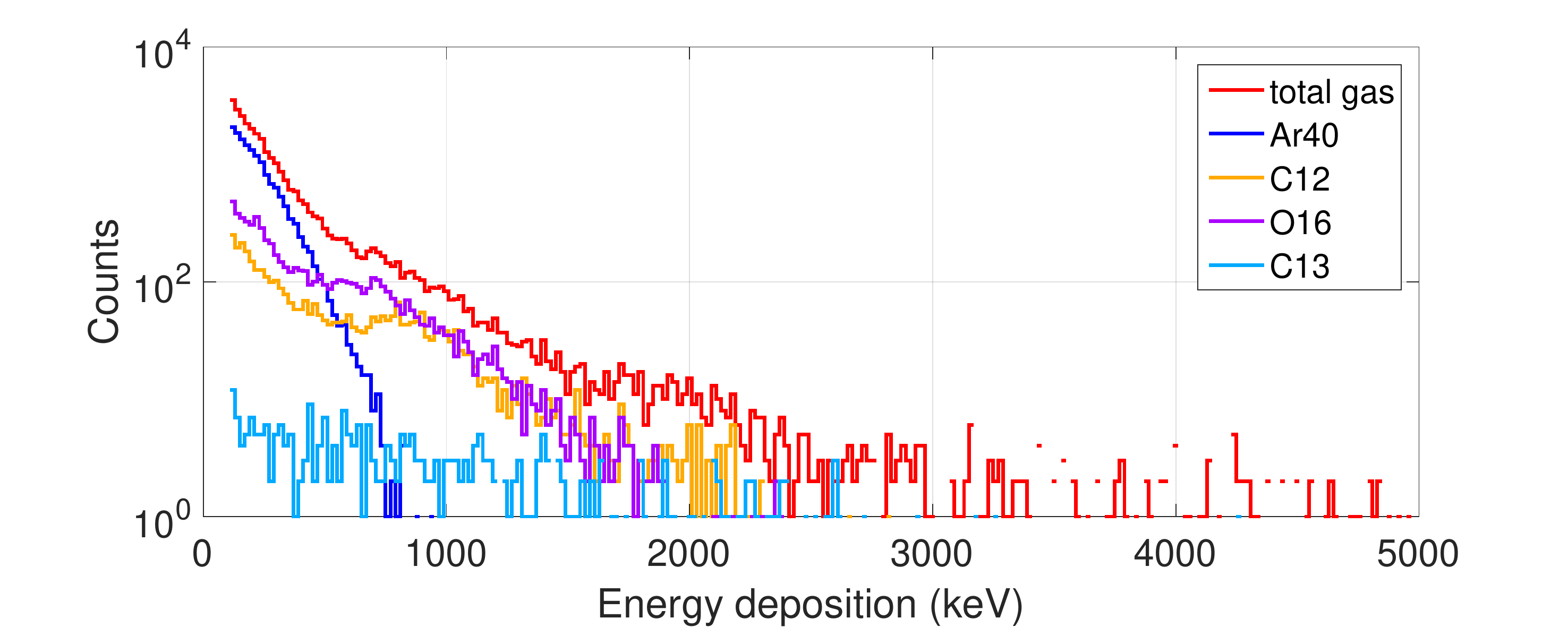}
\caption{\label{sim_gas_edep} \footnotesize Simulated energy
  deposition of the main detector components (top) and the gas components (bottom). The spectrum spreads up to 5000~keV due to the contribution of $\alpha$-particles and protons from any material considered in the detector. The energy depositions from $^{12}$C and $^{16}$O, mainly from the elastic and absorption processes, are the reason for the spectrum bump around 800~keV. Figure from~\cite{MIO_fastn}.}
\end{figure}

The individual contributions to the deposited energy in the gas are
identified and the shape of the PHS can be explained by them. Figure~\ref{sim_gas_edep} demonstrates in a quantitative way that the
energy depositions from $^{12}$C and $^{16}$O are the reason for the
spectrum bump around 800~keV. This feature arises mainly from the elastic and absorption processes of these two elements. The $^{40}$Ar contribution drops
relatively sharply without structure until 1000~keV, while the
$^{13}$C spectrum is flatter and extends up to 3000~keV. 
Beyond 1000~keV, the spectrum spreads up to 5000~keV due to the contribution of $\alpha$-particles and protons from any material considered in the detector. Note that the main source of $\alpha$-particles is the O(n,$\alpha$) process.
The elastic interactions contribute to the
energy deposition spectrum in a wide energy region, while all inelastic processes can be attributed to the electron energy deposition in the gas, whose contribution does not extend above 500 keV.
These findings are consistent with and validate the
prior theoretical analysis of cross sections in section~\ref{theocalc}. 
The contributions from the blade atoms are minimal and sharply drop in
the low values of the spectrum, as seen in figure~\ref{sim_gas_edep}. 

\begin{figure}[htbp]\centering
\includegraphics[width=1\textwidth,keepaspectratio]{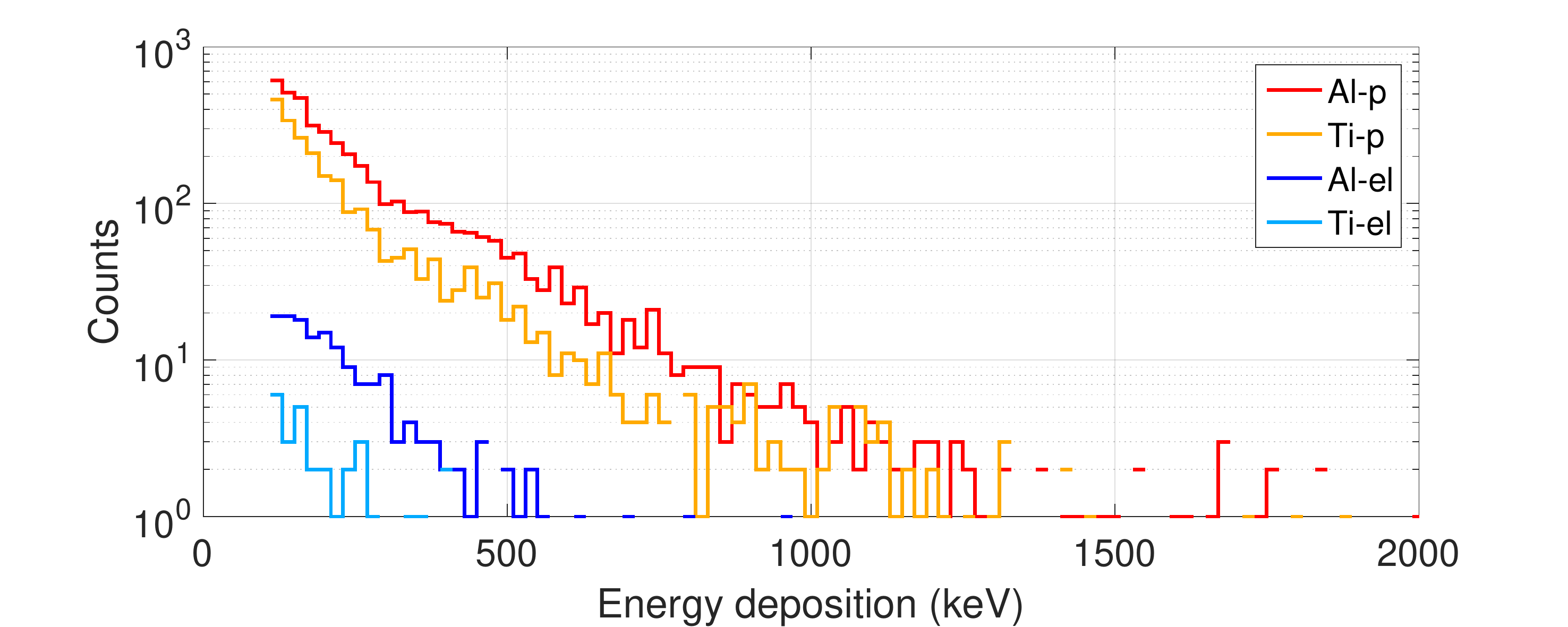}
\caption{\label{proton} \footnotesize Proton energy loss in the gas and nucleus recoil energy from elastic scattering for Aluminium and Titanium blade material. Figure from~\cite{MIO_fastn}.}
\end{figure}

A closer study at the individual components between the spectra for Al and Ti blades leads to the conclusion that the difference is mostly attributed to
the proton energy loss in the gas and the elastic reaction (see figure~\ref{proton}), in agreement with the analysis presented in section~\ref{theocalc}. 
Moreover for the same number of initial events, the ratio between the PHS obtained with Al and Ti is (2$\pm 1$), in agreement with the calculation shown in subsection~\ref{sp}. 
\\It will be shown, in subsection~\ref{materiali}, that this result is also in agreement with the measurements.  

\newpage
\section{Measurements}\label{measure}

\subsection{Verification of lack of thermal neutron sensitivity}\label{lacknsens}

\begin{figure}[htbp]
\centering
\includegraphics[width=1\textwidth,keepaspectratio]{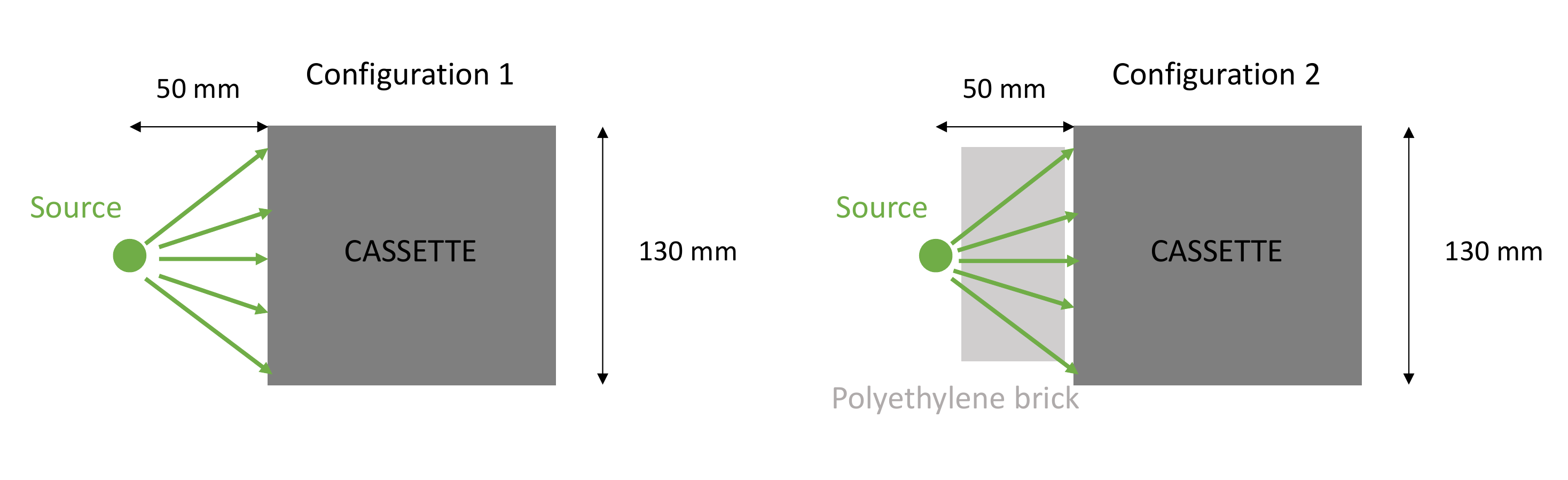}
\caption{\label{setup_brick} \footnotesize Sketch of the experimental setup with (right) and without (left) polyethylene brick in between. Figure from~\cite{MIO_fastn}.}
\end{figure}

Before proceeding with fast neutron measurements, a study of the thermal neutron contribution was performed to ensure that it does not affect the measurements. The measurements with and without a polyethylene brick between the source and the detector are presented in order to compare thermalized and fast neutron flux. A sketch of these configurations is shown in figure~\ref{setup_brick}, while all other measurements were performed in the {\it Configuration 1}.

\begin{figure}[htbp]
\centering
\includegraphics[width=1\textwidth,keepaspectratio]{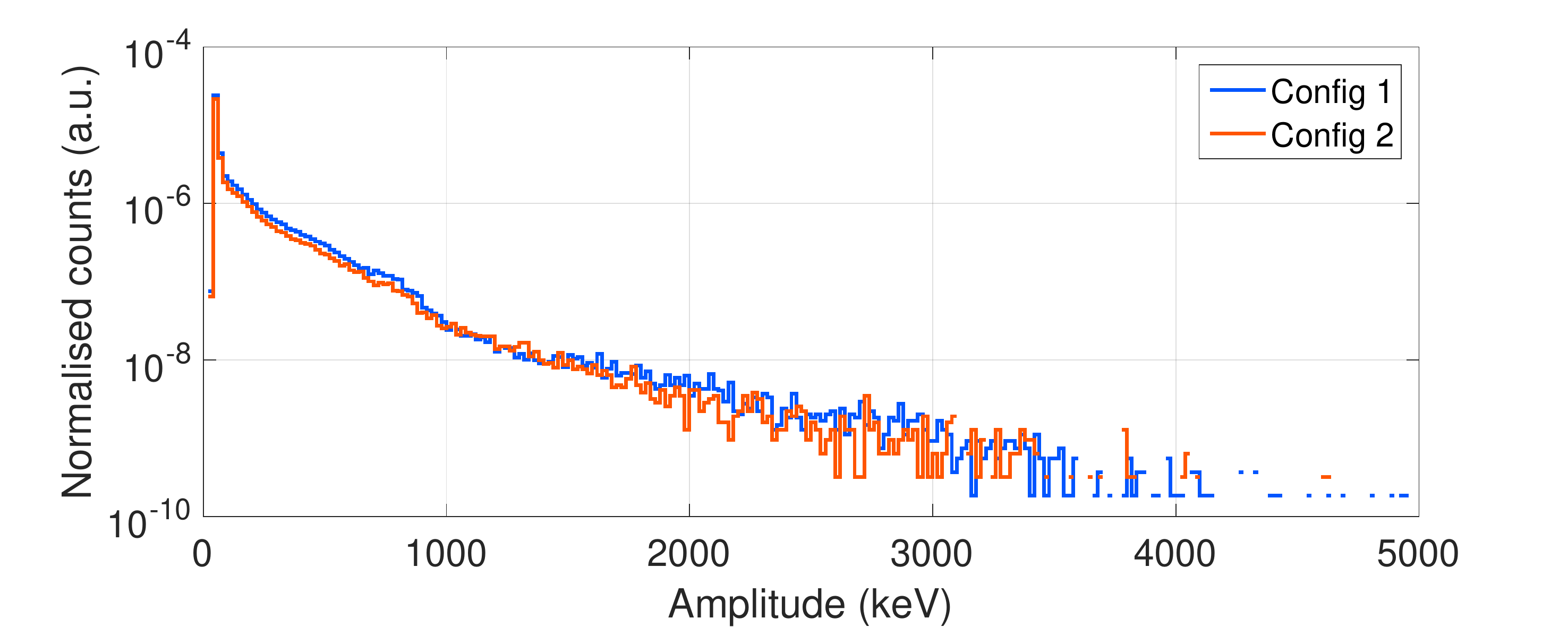}
\caption{\label{config} \footnotesize PHS acquired in the {\it Configuration 1} (blue curve) and with a polyethylene brick between the source and the detector, {\it Configuration 2} (red curve). For comparison an arbitrary scaling factor of $\approx$2 to the counts of the PHS in {\it Configuration 2} is applied. Figure from~\cite{MIO_fastn}.}
\end{figure}

A PHS for the two configurations was recorded (see figure~\ref{config}). For these measurements the $^{238}$Pu/Be source is used. To get the PHS the counts recorded with the four wires in the middle of the cassette are summed together in order to increase the statistics and to take into account only wires with approximately the same gas gain. Due to the geometry, the detector has a variable gain across the wire plane as described in chapter~\ref{chapter4}. 
\\ In order to calibrate the experimental spectra, a measurement with another cassette with the Boron layer was performed in order to measure the alpha peak of the 94\% branching ratio of the neutron capture reaction, corresponding to an energy of 1470 keV and the center of the peak is used to convert the PHS X-axis from ADC levels to energy. This calibration method is applied in all the following plots.
\\ In order to understand if one is sensitive to the thermal neutron contribution with no $\mathrm{^{10}B_{4}C}$ layer on the blade, the PHS obtained in the two considered cases are compared in figure~\ref{setup_brick}. Fewer neutrons reach the detector in the {\it Configuration 2} (figure~\ref{setup_brick}) as they are scattered and thermalized by the polyethylene brick. An arbitrary scaling factor of $\approx$2 is applied to the counts of the PHS obtained, in this case to compare with the PHS acquired in the {\it Configuration 1} (see figure~\ref{config}). No difference emerges from this comparison, proving that this setup is not sensitive to thermal neutrons. As expected by removing the $^{10}$B layer no other thermal contributions are involved in the measurements.

A further test to identify the thermal and the fast neutron signal signature is the multiplicity. When a neutron is converted into charged particles, they travel in the gas releasing energy on their path. This energy can be collected on wires and strips. The number of wires or strips involved in the detection of a single event defines the multiplicity of such event. When a neutron is converted in the $^{10}$B$_4$C layer the event is recorded in $99.5 \%$ of cases on no more than 2 wires and 3 strips, see section~\ref{secFEE} for further details. Multiplicity for fast neutrons and gammas is in general higher, a characteristic that can be used to discriminate against background. A summary of multiplicity values is depicted in table~\ref{multiplicity}. These values are specifically for the Multi-Blade detector and they are not general because they are strongly dependent on the detector geometry. Moreover, these values are for a wide beam detector illumination: such as the granularity of the readout system (wires and strips) is averaged out.

\begin{table}[htbp]
\centering
\caption{\label{multiplicity} \footnotesize Multiplicity in percentage recorded on wires for thermal neutron, gammas and fast neutron events.}
\smallskip
\begin{tabular}{|c|c|c|c|c|c|}
\hline
 & \multicolumn{5}{|c|}{Number of wires} \\
\cline{2-6}
 & 1 & 2 & 3 & 4  & 5 \\
\hline
Thermal n & 75\% & 25\% &  0\% & 0\% &  0\% \\
\hline
Gammas & 20\% & 20\% & 50\% & 10\% &  0\% \\
\hline
Fast n & 15\% & 30\% & 40\%  & 10\% & 5\% \\
\hline
\end{tabular}
\end{table}

\subsection{Fast neutron measurements with different sources}\label{sources}

\begin{figure}[htbp]
\centering
\includegraphics[width=1\textwidth,keepaspectratio]{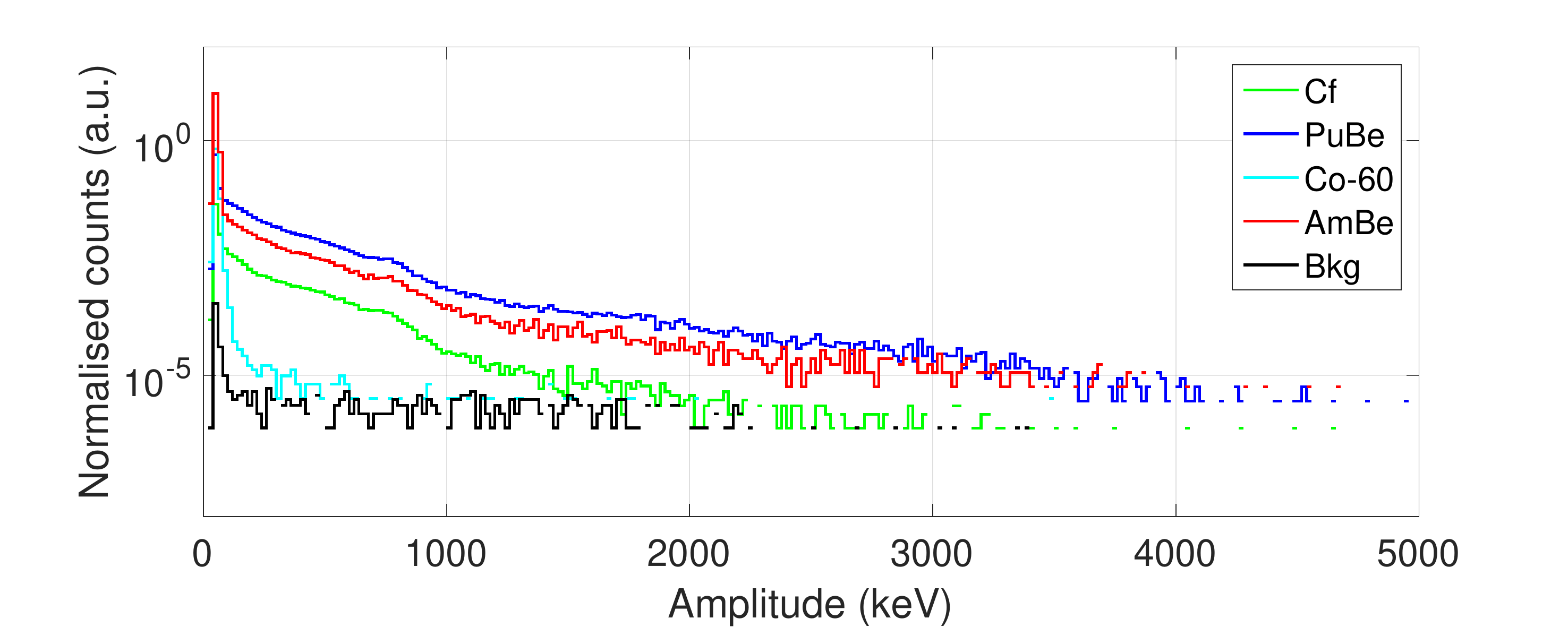}
\includegraphics[width=1\textwidth,keepaspectratio]{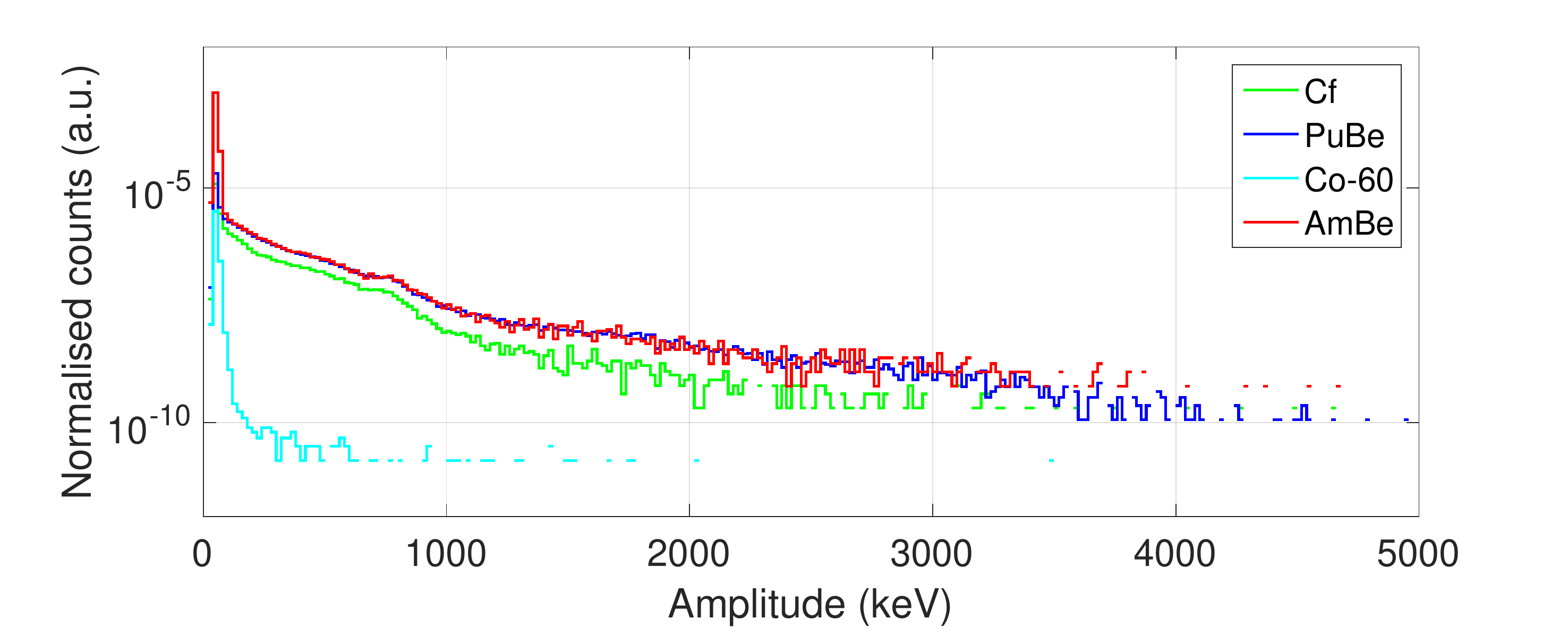}
\caption{\label{phs_sources} \footnotesize PHS of one wire for different sources and the background. Each spectrum is normalized over time (top). PHS of all the sources normalized over time, solid angle and activity (bottom). The center of the $\alpha$ peak of the neutron capture reaction is used to convert the PHS X-axis from ADC levels to energy. Figure from~\cite{MIO_fastn}.}
\end{figure}

A set of measurements have been performed with three neutron sources: $^{252}$Cf, $^{241}$Am/Be, $^{238}$Pu/Be, and with a gamma source, $^{60}$Co. The latter measurement work as a comparison with the study concerning the gamma sensitivity~\cite{MIO_MB2017, MG_gamma}. A background measurement was performed without any sources. 
\\ These measurements provide information about the possible differences on the PHS caused by the choice of a source  (section~\ref{subsources}). While the Am/Be and Pu/Be sources spectra are similar both in shape and in energy, the distribution of a Cf source has a quite different shape in the same range of energy. Figure~\ref{phs_sources} shows the PHS for each source. All the spectra refer to one wire, the plot at the top is normalized over the total time of acquisition for each set of measurements, while the plot at the bottom is normalized for activity and solid angle as described in subsection~\ref{subsources}. 

Once normalized, the PHS of the $^{241}$Am/Be, $^{238}$Pu/Be are in a good agreement because their emission energy spectra are similar both in shape and energy. The difference peak at low energy ($<$100 keV) is expected because the $^{241}$Am has a higher gamma emission ($\sim$60 keV) than that of $^{238}$Pu (figure~\ref{phs_sources}). Moreover the shape of both PHS is similar to that of $^{252}$Cf, which differs from the former by a factor in intensity, and a variation in shape at high energy, above 1 MeV, due to the difference between the emission energy spectrum of the sources, see figure~\ref{phs_sources}. As described in section~\ref{subsources} the emission distribution of the Cf peaks at lower energy compared to the Actinide/Be sources. Both from calculations and simulations it emerges that the main contribution to the energy deposition spectrum in this region is attributed to the absorption interactions which require more energy to occur, see equations~\ref{e_recoil} and~\ref{e_y}.
\\It can be concluded that although the sources differ in energy distributions, they do not produce any relevant effect on the spectrum. Moreover, a common feature is visible around 800 keV, predicted by the theoretical calculation, figure~\ref{theo_el}, and verified by the simulations shown in figure~\ref{sim_gas_edep}. This feature can be identified as the contribution of the elastic and the absorption interactions of the gas elements to the PHS. 
\\ For the following measurements the $^{238}$Pu/Be source has been used, because it has the highest activity and a lower gamma contribution among the three.

\subsection{Materials comparison}\label{materiali}

\begin{figure}[htbp]
\centering
\includegraphics[width=1\textwidth,keepaspectratio]{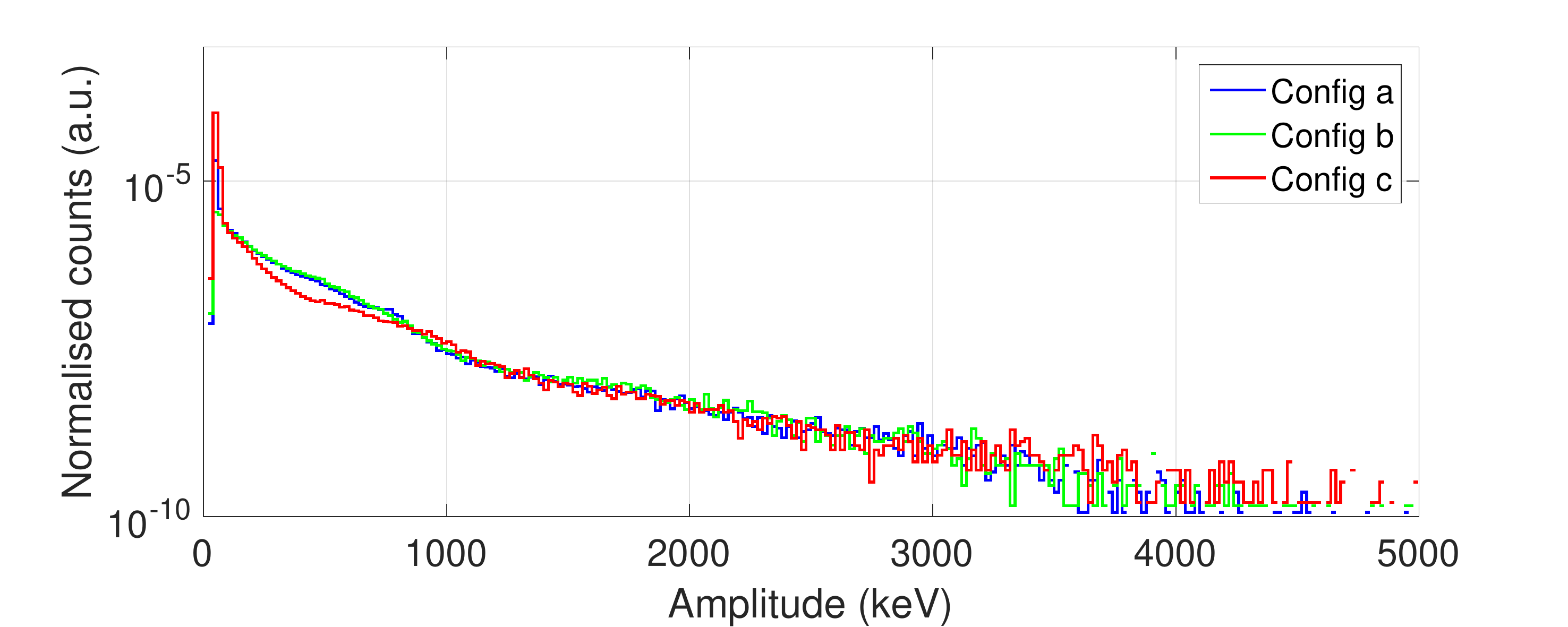}
\caption{\label{phs_material} \footnotesize The PHS are normalized by time, activity and solid angle and they refer to one single wire. The blue plot refers to the {\it Configuration a}, the green to the {\it Configuration b} and the red curve refers to the {\it Configuration c} in figure~\ref{cassettesc}. Figure from~\cite{MIO_fastn}.}
\end{figure}

Referring to section~\ref{sp}, a significant difference between the materials of the detector is expected. The cassette configuration has been changed according to figure~\ref{cassettesc}, performing the measurements for the configurations a, b and c. The PHS for each one is showed in figure~\ref{phs_material}.

From the comparison between \textit{a} and \textit{b} one can conclude that the Copper and Kapton of the strips do not contribute at our sensitivity level. The shape of the pulse height spectrum with or without the strips is similar, as expected from the theoretical calculation of the probability of deposition, see subsection~\ref{sp} and the simulations confirm it as well, see subsection~\ref{sim_contribute}. 
\\ Note that the PHS obtained with the \textit{Configuration c} (titanium blade) is different from the others below 1 MeV. The shape of the tail at higher energy is indeed unchanged between the three different setups. 
\\ A possible explanation for this behaviour can be obtained by separating solid and gas contribution to signal detection as described in section~\ref{sp}. After the neutron conversion into a charged particle, the latter has to escape and reach the gas to generate a signal. This effect in the solid is mostly significant in the low energy region of the PHS, as shown in figure~\ref{theo_el}. The gas component is the sensing medium of the detector so the conversion products release their energy directly into the gas generating a signal.
\\ From the theoretical calculation of P(r) the probability of energy deposition in the gas, see~\ref{sp} table~\ref{tabsp}, the most relevant difference is shown for the elastic scattering interaction and the (n,p) absorption reaction. The recoil nucleus of Al, from equation~\ref{al_p} and~\ref{e_y}, can release in the gas at most 690 keV for an incoming neutron energy of 5 MeV. The maximum energy released in the same condition for Ti is instead 300 keV with a probability about 4 times lower than that of Al. Moreover the P(r) of (n,p) reaction is about 1 order of magnitude higher for Al than Ti, while the energy fraction released in the gas of the emitted proton is approximately the same in both cases. The cross section of this reaction in Aluminium, as shown in figure~\ref{sigmaAlTi}, occurs at $\approx$ 3 MeV; the same reaction in Ti occurs at $\approx$ 6 MeV with roughly the same probability. The other two most significant processes take into account in the absorption interaction, namely (n,$\alpha$) and (n,$\gamma$), have approximately the same energy distribution. 
\\From this study the difference on the PHS shape can be attributed to the elastic interaction and the absorption (n,p) process. Indeed the difference is visible in the same energy range obtained from the calculation, see figure~\ref{theo_phs}, and the ratio between the two spectra is (1.5 $\pm 0.5$) in agreement with the simulations result discuss in section~\ref{sim_contribute}.  One can conclude that Titanium is less sensitive to fast neutron than Aluminium, while the background contribution of Copper and Kapton for fast neutron can be considered negligible, as expected from the theoretical considerations (subsection ~\ref{sp}) and the simulations (subsection~\ref{sim_contribute}).

\section{Fast neutron sensitivity}

\begin{figure}[htbp]
\centering
\includegraphics[width=1\textwidth,keepaspectratio]{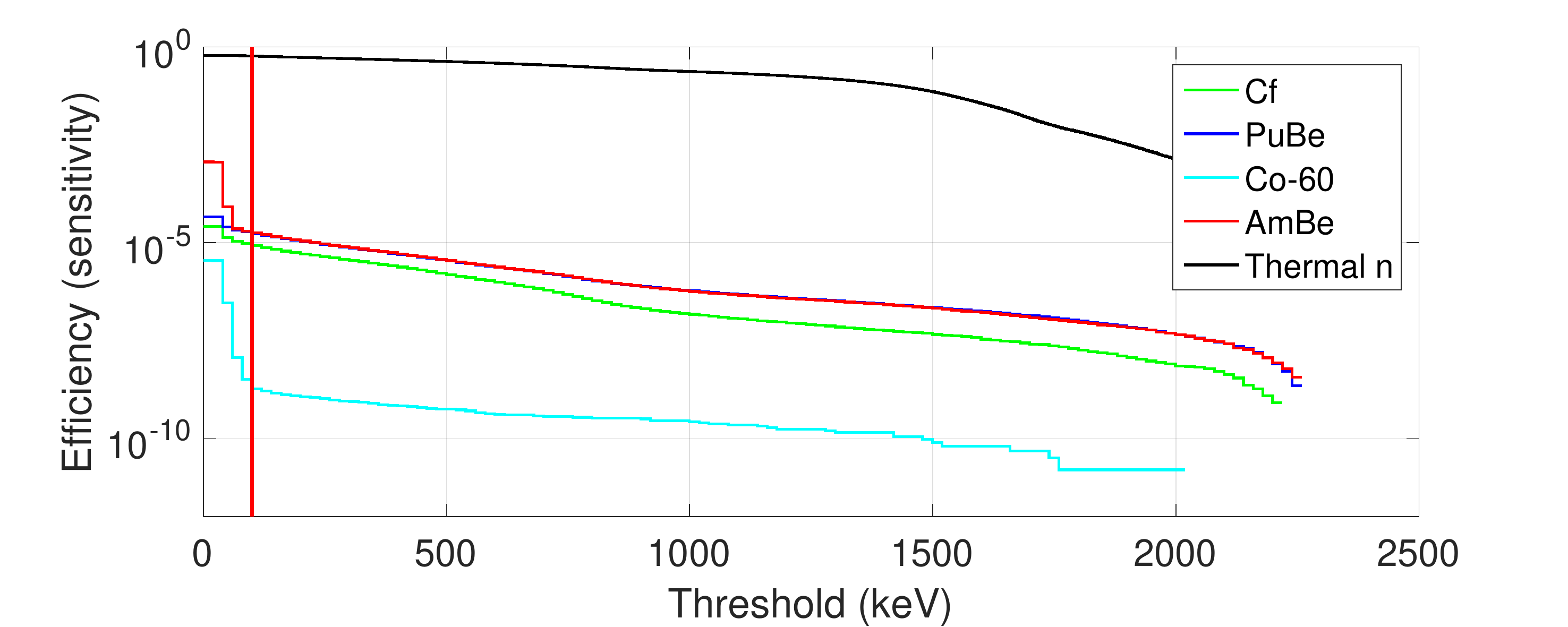}
\includegraphics[width=1\textwidth,keepaspectratio]{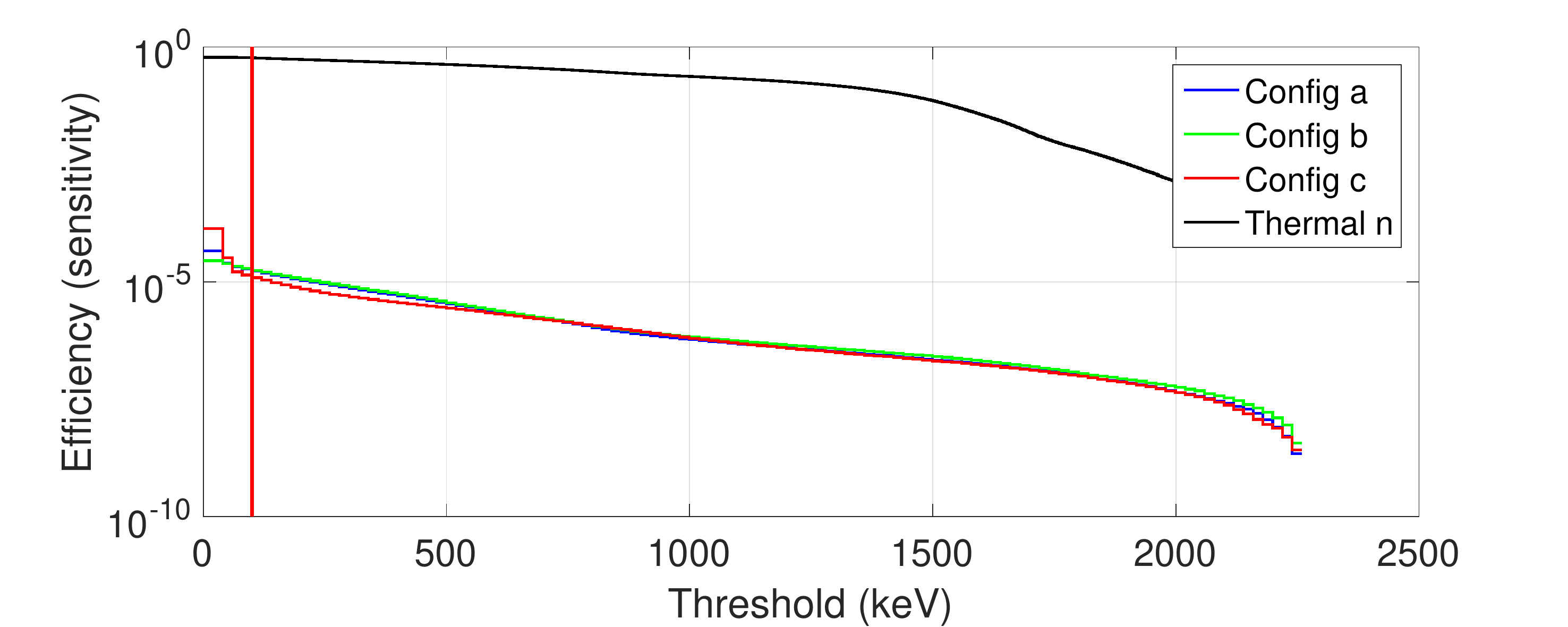}
\caption{\label{sensitivity} \footnotesize Counts in the PHS (normalized to the subtended solid angle and the activity of the sources) as a function of the applied threshold for the several sources (top). For the three configurations a, b and c (bottom). The vertical line refers to a software threshold of 100 keV applied to achieve the optimal signal-to-background ratio. Figure from~\cite{MIO_fastn}.}
\end{figure}

The fast neutron sensitivity of a thermal neutron detector is defined as the probability for a fast neutron to generate a false count in a thermal/cold neutron measurement. Along with the $\gamma$-ray sensitivity~\cite{MIO_MB2017,MG_gamma} the fast neutron sensitivity defines the best achievable signal-to-background ratio, for a given flux of each component. 
\\The interactions with fast neutrons that can give rise to background signals are described in the sections above as well as the measurements performed. When the energy for one of these processes exceed a set threshold, it results in an event. The number of events that exceed these thresholds is then normalized to the activity of the source and the solid angle. 
\\ Each source was placed close to the entrance window of the Multi-Blade, directly outside of the cassette equipped with the individual readout electronics, as showed in the sketch~\ref{setup}; the solid angle acceptance has been calculated as described in subsection~\ref{subsources}.  Note that the statistical uncertainties are almost negligible, the systematic uncertainty on the distance from the source to the sensitive element can lead to a deviation on the measurements of no more than a factor 2, as discussed in subsection~\ref{subsources}.
\\ Figure~\ref{sensitivity} (top) shows the total counts in the PHS which is normalized to the activity of the source and the solid angle as a function of the threshold for all the sources used for the experiment. The bottom plot in figure~\ref{sensitivity}, shows the sensitivity for the different cassette configurations described in section~\ref{materiali}.
In table~\ref{fastn-gamma-thr} are reported the values of the efficiency for thermal neutrons measured with a neutron wavelength $4.2\,$\AA , for the fast neutron and gamma sources used to perform the measurements described in section~\ref{measure}. The values in table~\ref{fastn-gamma-thr} refer to the {\it Configuration a} described in section~\ref{materiali}. The same value can be considered for the {\it Configuration b}, the value for the {\it Configuration c} is slightly smaller. However, this difference does not affect the consideration about the choice of the software threshold. The different sensitivities calculated for configurations {\it a} and {\it c} will be discussed later in this section. The values of the sensitivity are reported at two different software threshold of 20 keV and 100 keV, respectively.
\\ The threshold is fixed at 100 keV to achieve the optimal signal-to-background ratio, see vertical line in figure~\ref{sensitivity}. Indeed the efficiency calculated for the two Actinide/Be-based radioactive sources is different at low energies where the gamma contribution of $^{241}$Am/Be is more intense than that of $^{238}$Pu/Be. This effect is visible applying the software threshold of 20 keV while it is cut by the $\mathrm{100\,keV}$ threshold. The gamma sensitivity decreases by more than a factor 100 between the two values of threshold as well, see table~\ref{fastn-gamma-thr}. 

\begin{table}[htbp]
\centering
\caption{\label{fastn-gamma-thr} \footnotesize Efficiency (or sensitivity) for thermal neutrons, fast neutrons and gamma-rays applying two values of the threshold (20 and 100 keV) for the {\it Configuration a}. The efficiency (or sensitivity) is normalized to 1.}
\smallskip
\begin{tabular}{|c|c|c|c|c|c|}
\hline
Threshold & Th. n & $^{252}$Cf  & $^{238}$Pu/Be  & $^{238}$Am/Be & $^{60}$Co \\
\hline\hline
20 keV & 0.60 & $2.6\cdot10^{-5}$ &  $4.5\cdot10^{-5}$ & $1.1\cdot10^{-3}$ &  $3.5\cdot10^{-6}$\\
100 keV & 0.56 & $9.4 \cdot10^{-6}$ & $1.9\cdot10^{-5}$ & $1.9\cdot10^{-5}$ &  $<1\cdot10^{-8}$ \\
\hline
\end{tabular}
\end{table}

The efficiency for thermal neutrons measured at $4.2\,$\AA , is $\approx 56\%$ as described in~\cite{MIO_MB2017}. The sensitivity at the same threshold (100~keV) to the fast neutron is about 4 orders of magnitude lower. Furthermore the gamma sensitivity measurement agrees with the previous dedicated studies~\cite{MIO_MB2017,MG_gamma} and it is $\approx 10^{-8}$, about 3 orders of magnitude less intense than the fast neutron sensitivity, see figure~\ref{sensitivity}.
\\ The evaluation of gamma and fast neutron sensitivities described above is performed taking into account a pulse height cut-off through an energy threshold (100~keV). A further improvement of the background rejection can be achieved by using additional discrimination methods. i.e., the different multiplicity of thermal neutrons than that of gammas and fast neutrons can be exploited. Referring to table~\ref{multiplicity}, thermal neutrons very rarely exceed multiplicity 2 on wires, hence if all events with multiplicity equal to 3 or higher are disregarded the sensitivity to gammas and fast neutrons improves. In particular an additional rejection of 60\% for $\gamma$-rays and of 55\% for fast neutrons is achieved.
\\ The study of the various interactions that can occur between incident neutrons and the materials inside the detector is presented in this paper. Information of these reactions in Aluminium and Titanium emerges from this analysis. Focusing the attention on the elastic scattering and on two most significant absorption processes, (n,p) and (n,$\alpha$). The cross section of (n,p) process in Aluminium is about 3 orders of magnitude higher than that of Titanium in the energy range of the neutron sources used for the measurements. The probability of the same mechanism in Titanium is almost out of the energy range that can be covered by the sources as discussed in section~\ref{theocalc}. 
\\ The sensitivity obtained with the Titanium blade is slightly different compared with the one calculated with the Aluminium blade at low values of amplitude, as expected by the study of the PHS discussed in section~\ref{materiali}. The sensitivity obtained with the Ti is $1.39 \cdot 10^{-5}$, the one obtained with the Al, for the same threshold ($\mathrm{100\,keV}$), is $1.88 \cdot 10^{-5}$.
\\ By choosing, for the Multi-Blade detector, Titanium blades instead of Aluminium blades, it is possible to slightly improve the signal-to-background ratio. Both simulations and measurements show an improvement factor between 1 and 2 at the fixed software threshold of 100 keV. This change also brings mechanical advantage, as reported in~\cite{MIO_MB2017}.

\section{Further evidence of fast neutron sensitivity and comparison between the Multi-Blade detector and the $^3$He-tube}

\subsection{Fast neutron sensitivity of the Helium-3 detector}

\begin{figure}[htbp]
\centering
\includegraphics[width=.9\textwidth,keepaspectratio]{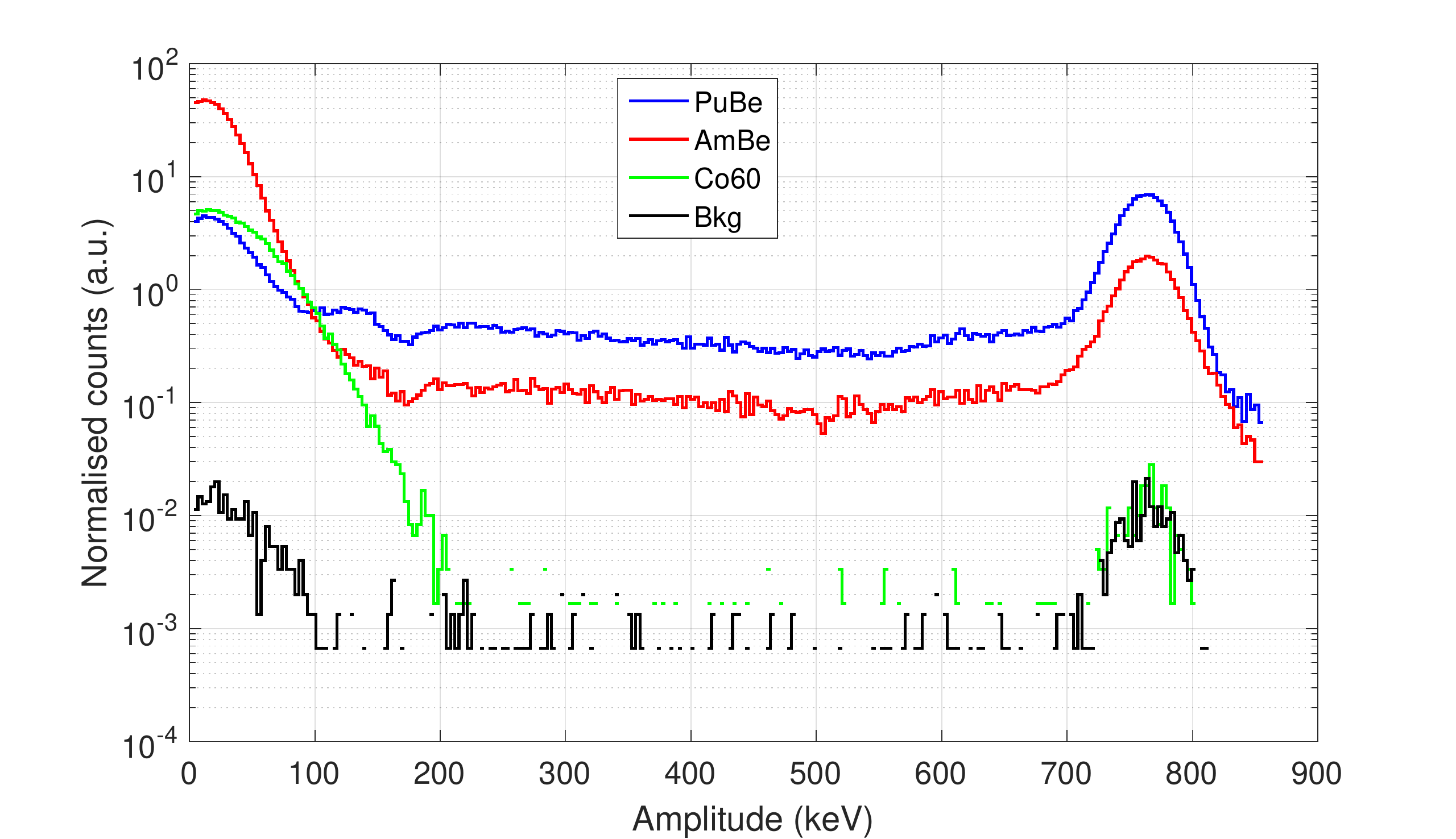}
\caption{\label{he-gen} \footnotesize Pulse Height Spectrum (PHS) normalized over time for different sources: Am/Be and Pu/Be for the fast neutrons and Co-60 for $\gamma$-rays, and the background.}
\end{figure}

Following the same proceedings used for the fast neutron sensitivity measurements performed with the Multi-Blade detector, a set of measurements have been carried out with a $^3$He-tube. 
\\ As for the experiment performed with the Multi-Blade detector, several sources have been used: Am/Be and Pu/Be for the fast neutrons and Co-60 for $\gamma$-rays. A background measurements was performed as well. The pulse height spectra, normalized over time, are depicted in figure~\ref{he-gen}; the center of the peak corresponding to an energy of 764 keV is used to convert X-axis from ADC values to energy. The same method is applied to each pulse height spectrum reported below. As described in section~\ref{sources} the Am/Be has a larger $\gamma$ emission with respect to the Pu/Be source. The effect is clearly visible in the picture, where the gammas only contribute to the PHS below 200 keV (green curve). In the region (<100 keV) the PHS measured with the Am/Be source is, indeed, about one order of magnitude more intense than the spectrum obtained with the Pu/Be source. Because of this and for consistency with the previous work, it has been chosen to carry out the measurements with the Pu/Be source.   

\begin{figure}[htbp]
\centering
\includegraphics[width=1\textwidth,keepaspectratio]{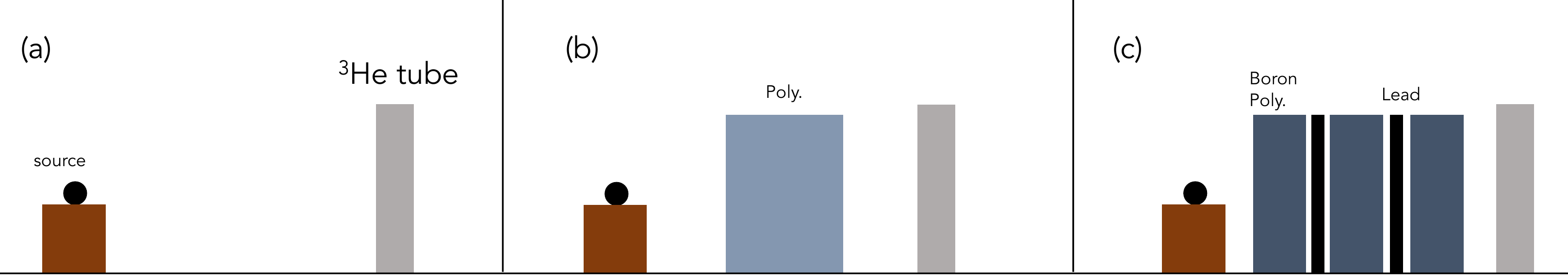}
\caption{\label{he-config} \footnotesize Different measurements configurations with the $^3$He-tube: (a) the total flux reaches the detector, (b) flux thermalized through a polyethylene block, (c) flux highly reduced through borated polyethylene and lead bricks, background measurement.}
\end{figure}

In the case of the measurements with the Multi-Blade detector, it was possible to ensure that the thermal neutron contribution was negligible by using blades without the boron coating, as described in section~\ref{lacknsens}. On the other hand, it is not possible to have a good separation between the different neutrons contribution, when using the $^3$He-tube. 
For this reason the measurements were performed in several configurations, a sketch of them is shown in figure~\ref{he-config}: in configuration (a) the total flux of the source reaches the detector, i.e., fast and thermal contribution, in (b) the flux is thermalized through a polyethylene brick, while in (c) a shielding of borated polyethylene and lead is placed between the source and the detector, in order to stop as far as possible the neutrons and to measure the background with the source.

\begin{figure}[htbp]
\centering
\includegraphics[width=.9\textwidth,keepaspectratio]{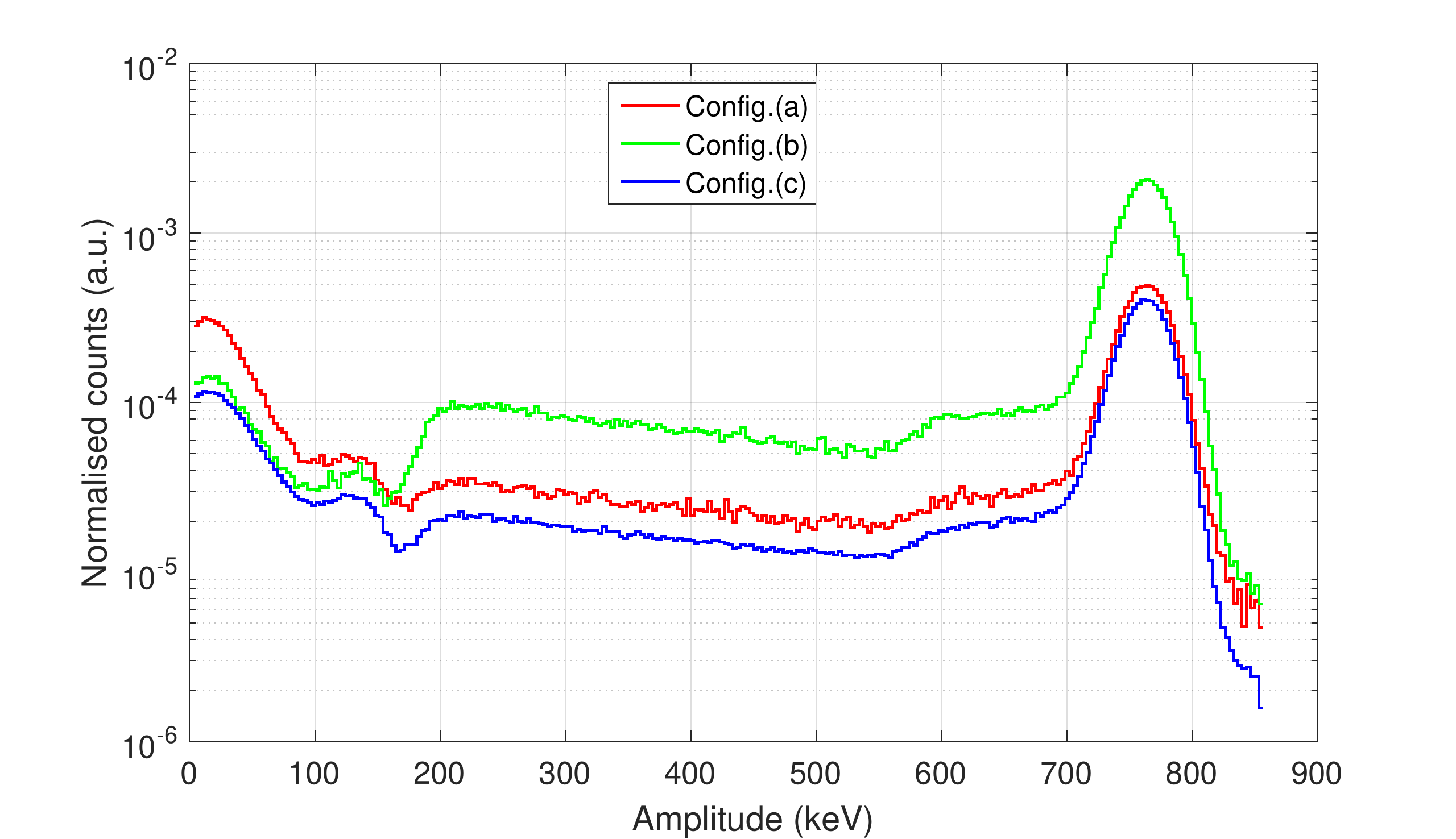}
\caption{\label{he-pube} \footnotesize PHS normalized by time, activity and solid angle, obtained in configuration (a), (b) and (c).}
\end{figure}

The measurement in Config.(c) has been performed to disentangle the different neutron contribution, i.e., fast and thermal. One assumes that this background contribution affects the measurements of fast neutrons (a) and thermal neutrons (b) in the same way, thus it can be removed in the analysis.
\\In figure~\ref{he-pube} the PHS measured in the configurations (a), (b) and (c) is shown. The background counts have been subtracted in all data. The spectra are normalized by time, activity and solid angle. Because of the high detection efficiency of the $^3$He to thermal neutrons and the approximate set-up, even with a shield in between source and detector (Config.(c)) the reaction product peak (764 keV) is only one order of magnitude less intense than in case of the thermalized flux (Config.(b)). Which confirms how difficult is to decouple fast and thermal neutrons contributions, in these measurements.

\begin{figure}[htbp]
\centering
\includegraphics[width=.9\textwidth,keepaspectratio]{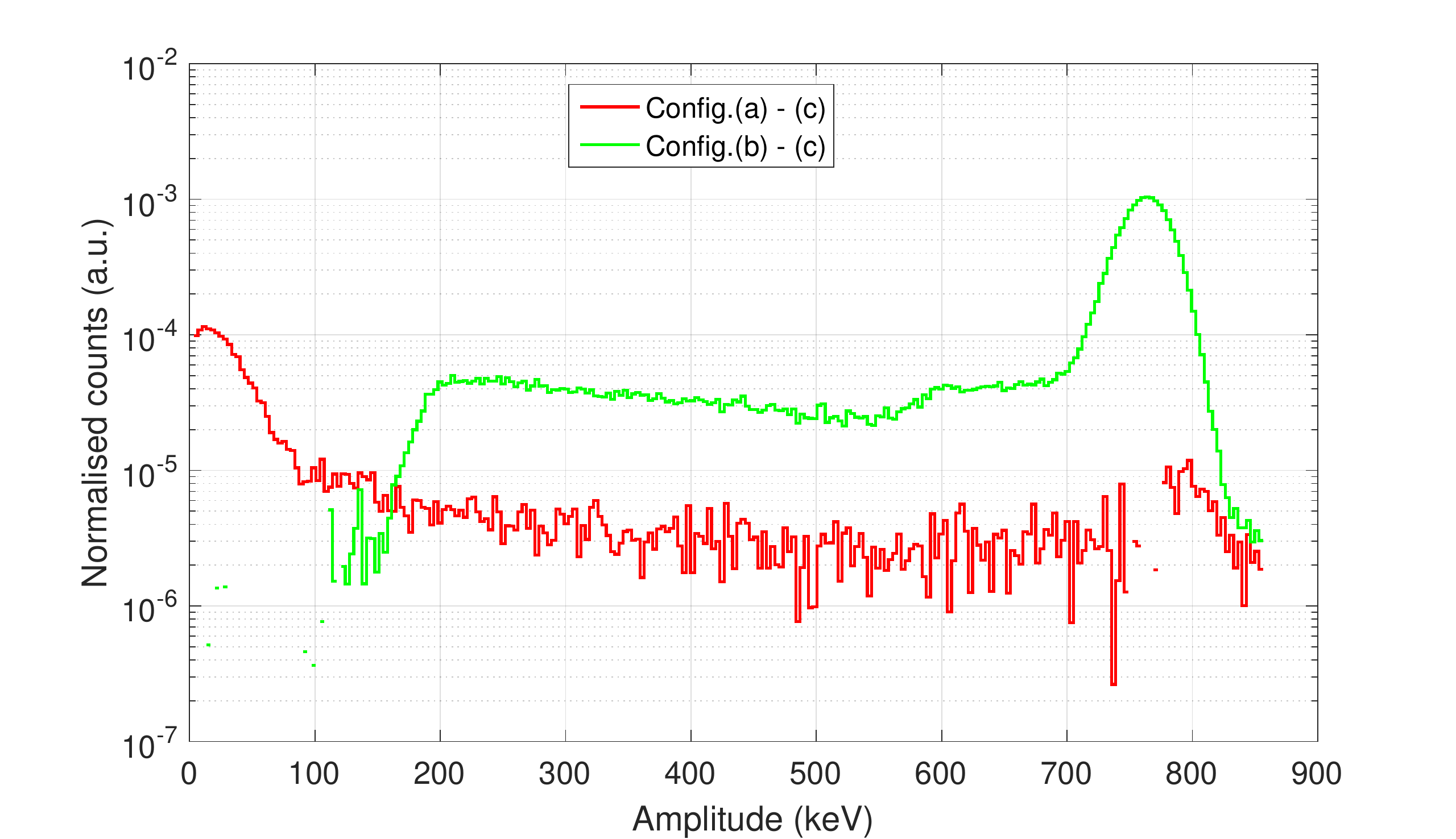}
\caption{\label{he-fth} \footnotesize PHS of thermal neutrons (blue) and fast neutrons (red), calculated subtracting the background measured with the shielding between the source and the detector.}
\end{figure}

\begin{figure}[htbp]
\centering
\includegraphics[width=.89\textwidth,keepaspectratio]{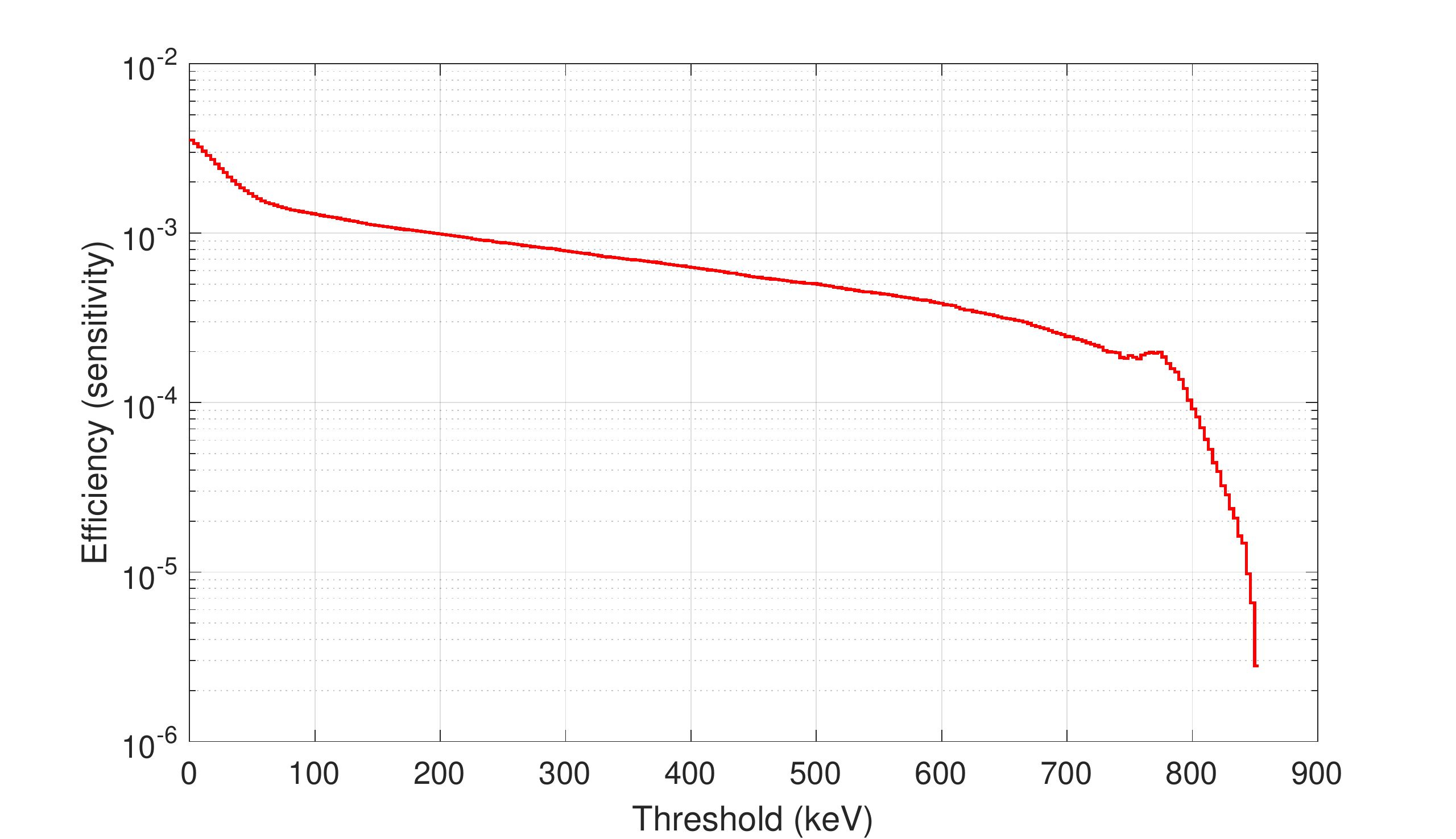}
\caption{\label{he-sens} \footnotesize Fast neutron sensitivity of the $^3$He-detector as a function of the applied threshold. For a software threshold of 100 keV a fast neutron sensitivity on the order of $10^{-3}$ is obtained.}
\end{figure}

Figure~\ref{he-fth} illustrates the thermal and the fast neutron contribution calculated subtracting the events measured in Config.(c), i.e., the background measured with the source placed in front of the $^3$He-tube. 
\\Following the same method applied in the case of the Multi-Blade detector, it is reported in figure~\ref{he-sens} the counts in PHS as a function of the applied threshold for the Config.(a) after all the subtractions. The threshold is taken from the tail of the peak and the cumulative sum of the counts is calculated. A sensitivity of  $\approx 1.3 \cdot 10^{-3}$ is observed. The value is in agreement with the one obtained at the CRISP reflectometer, as discussed in the following section~\ref{25hz}. 
\\ It is the first time this kind of investigation has been performed on a thermal neutron $^3$He-detector, and although this is a preliminary study, one can conclude that the $^3$He-detector technology is more sensitive (about 2 orders of magnitudes) to the fast neutrons compared with the boron-10-based detectors.  Moreover, the analysis of the background performed on the data from CRISP when operated at 25 Hz, confirms this experimental value. 
\\A more detailed study is foreseen by performing measurements also with an $^4$He-detector, where the thermal neutron contribution can be neglected, and performing simulations and analytical calculations as it has been done in the experiment with the Multi-Blade.    

\subsection{Background considerations at CRISP between Multi-Blade and Helium-3 detector: 25 Hz measurements}\label{25hz}

\begin{figure}[htbp]
\centering
\includegraphics[width=0.9\textwidth,keepaspectratio]{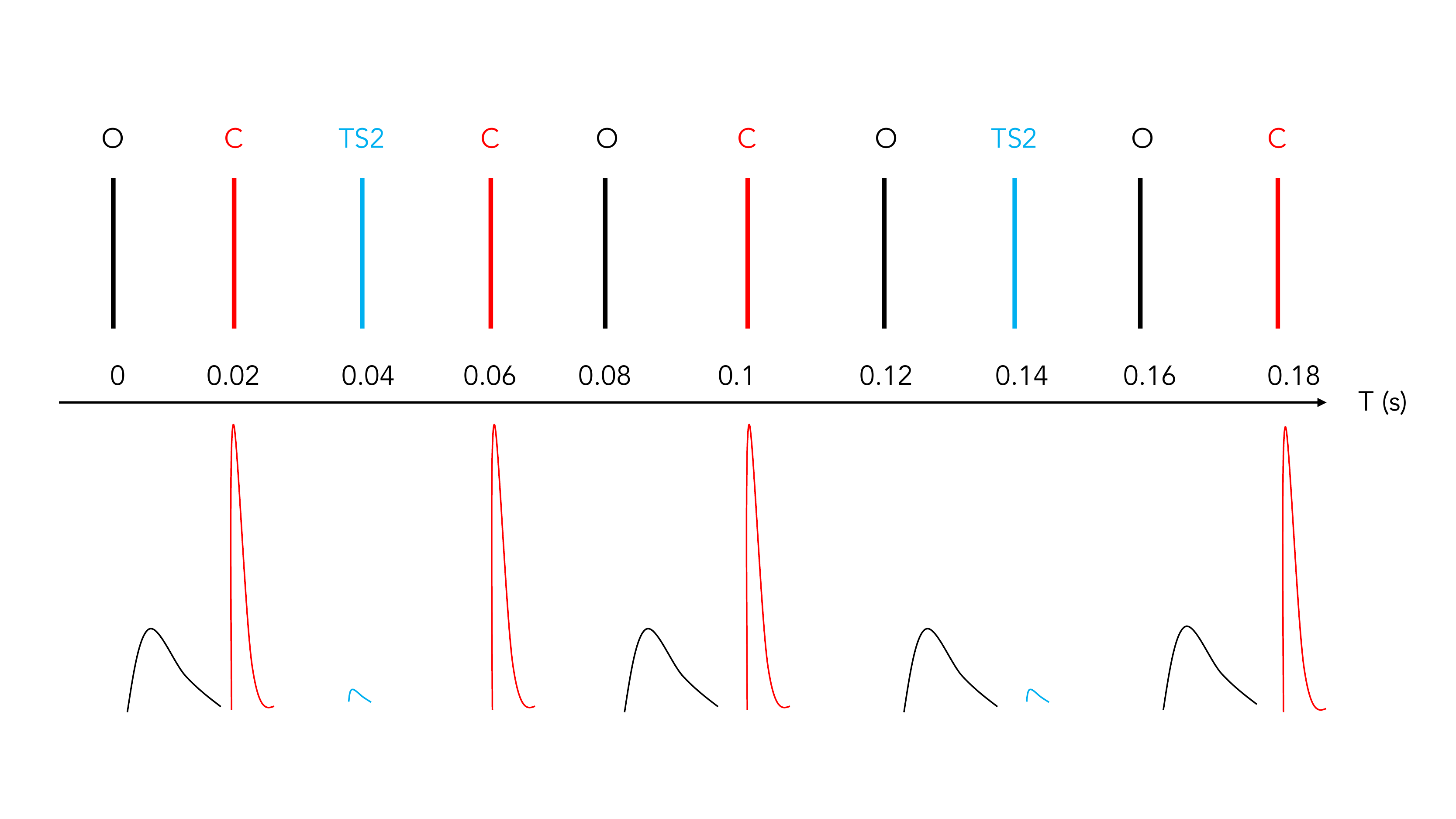}
\caption{\label{fig25hz} \footnotesize Sketch of the arrival signals in a window of 0.18 ms, when the chopper works at 25 Hz. The black identifies the signal passing throw the chopper (O = open), the red colour is used when the neutron beam hits the chopper without passing through (C = close). In this case a strong spike is expected at the detector. One every five pulses is sent to the Target Station 2 (TS2), blue in figure. A low background can be detected.}
\end{figure}

The measurements presented here refers to the test performed at the CRISP reflectometer, which is described in section~\ref{rifle}. In particular these results are obtained with the same set-up described for the efficiency measurements reported in subsection~\ref{subeff}, namely by placing both the Multi-Blade detector and the $^3$He-tube in the direction of the direct beam. The only difference is the working speed of the chopper, which in this case is half of the nominal speed, i.e., 25 Hz instead of 50 Hz.
\\In this configuration the range of wavelength is extended up to 13 $\AA$, compared to 6.5 $\AA$ at 50 Hz. A sketch of the spectrum in a time window of 0.18 ms is shown in figure~\ref{fig25hz}. There are three possible scenarios: the chopper is in phase with the proton pulse, thus the beam passes through it (O in the picture), the chopper is not in phase with the frequency of the proton pulse (C in the picture), so the neutron beam hits against the chopper producing a high background spike. Moreover, one every five pulses is sent to the Target Station 2 (TS2 in picture), a negligible background is then expected in this case. 

\begin{figure}[htbp]
\centering
\includegraphics[width=.9\textwidth,keepaspectratio]{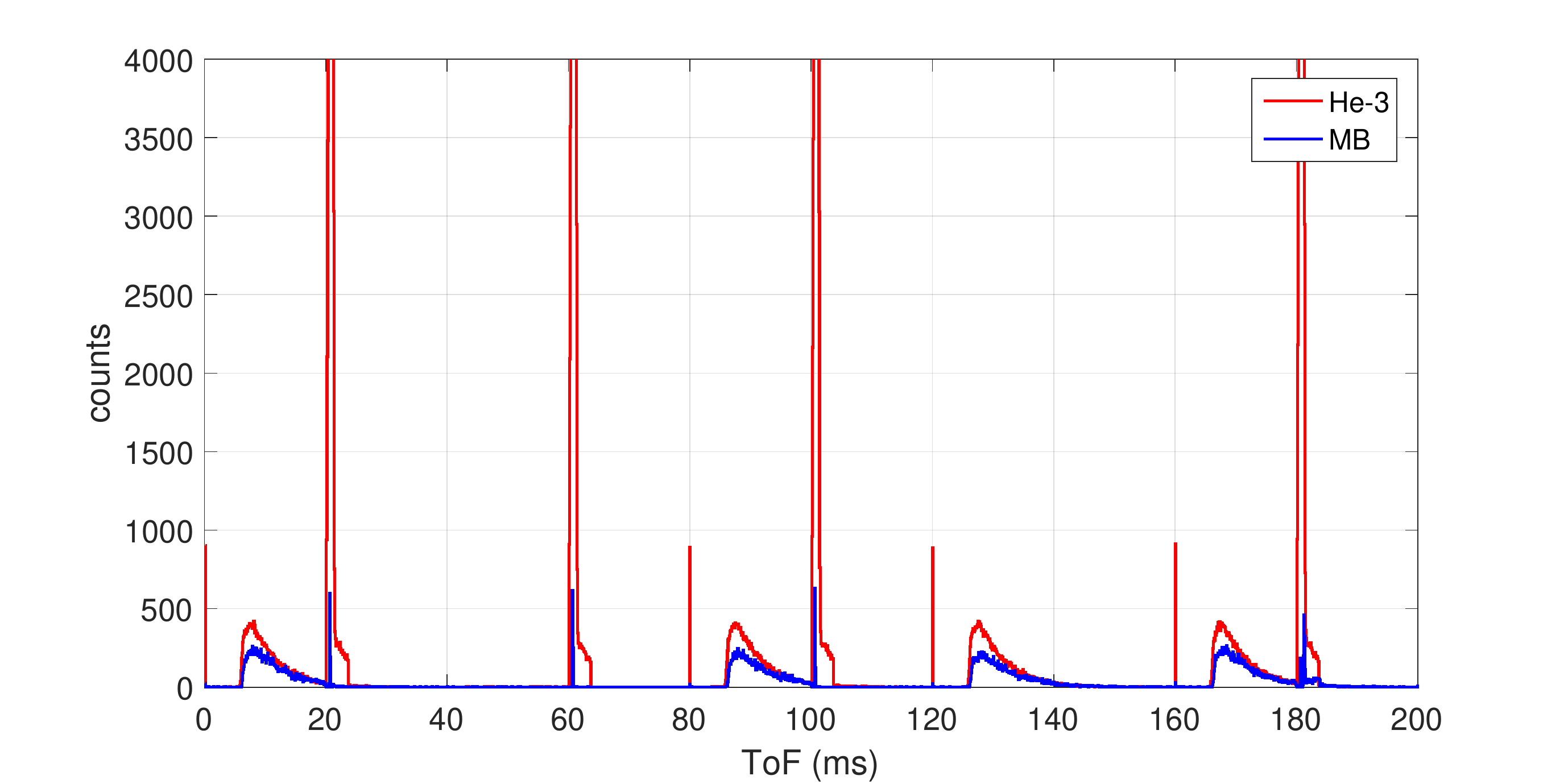}
\caption{\label{spec25hz} \footnotesize ToF spectrum obtained with the Multi-Blade detector (blue line) and the $^3$He-tube (red line) in a time window of 200 ms.}
\end{figure}

\begin{figure}[htbp]
\centering
\includegraphics[width=.9\textwidth,keepaspectratio]{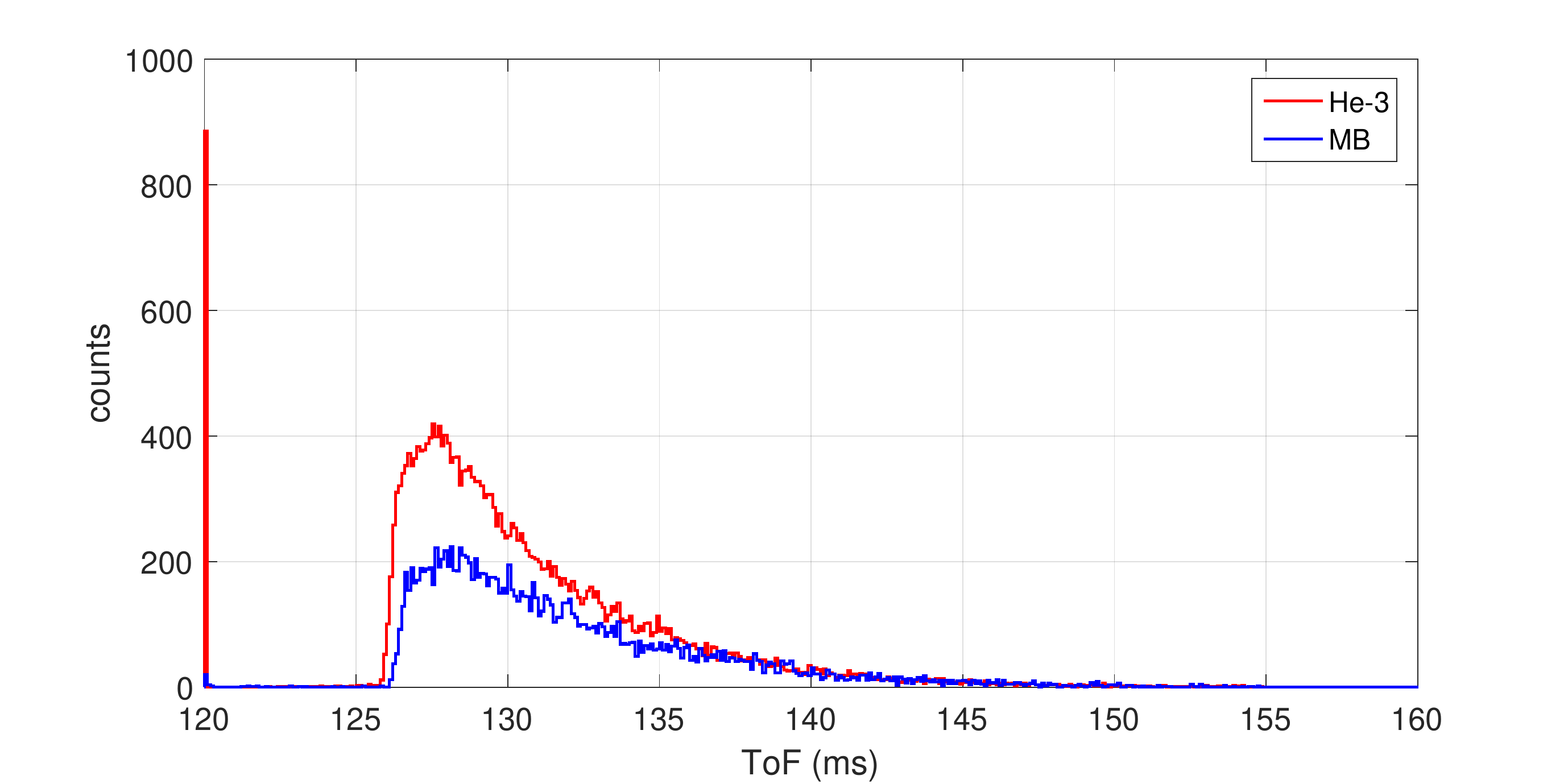}
\caption{\label{specbeam} \footnotesize ToF spectrum obtained with the Multi-Blade detector (blue line) and the $^3$He-tube (red line) for ToF = 120 - 160 ms.}
\end{figure}

A prompt pulse due to some background events is also expected in the case the chopper is open. All these events are mainly due to $\gamma$-rays, fast neutrons, environmental neutron counts (thermal and epithermal) and spurious scattering.
\\ A set of measurements have been performed both with the Multi-Blade detector and the $^3$He-tube installed at the CRISP reflectometer. A comparison between the spectra in Time of Flight obtained with both detectors is shown in figure~\ref{spec25hz}. 
\\A much higher background is recorded in the case of the $^3$He-tube, red line, compared with the Multi-Blade, blue line. In order to have a general understanding of the background, the attention has been focused in two region of the spectrum: when the chopper is open (ToF = 0.12 - 0.16 s) and when it is closed (ToF = 0.1 - 0.105 s). In figure~\ref{specbeam} is shown the spectrum when the neutrons pass through the chopper and reach the detector. The spike at low ToF is significantly higher in the case of the $^3$He-tube than the one recorded with the Multi-Blade detector.

Considering now the second region of interest, a first discrimination between the $\gamma$-rays contribution and the rest of the background signals can be investigated by analysing the events above and below the applied software threshold. These values are, indeed, set in order to discriminate the real events from the spurious ones (see subsection~\ref{scatt}). The events below threshold are less energetic, thus most probably are mainly gammas. The fast neutrons have a much broader energy spectrum, this leads to events with higher energy. A detailed discussion on the fast neutron contribution for boron-10-based detectors is reported in chapter~\ref{chapter5}. 

\begin{figure}[htbp]
\centering
\includegraphics[width=.9\textwidth,keepaspectratio]{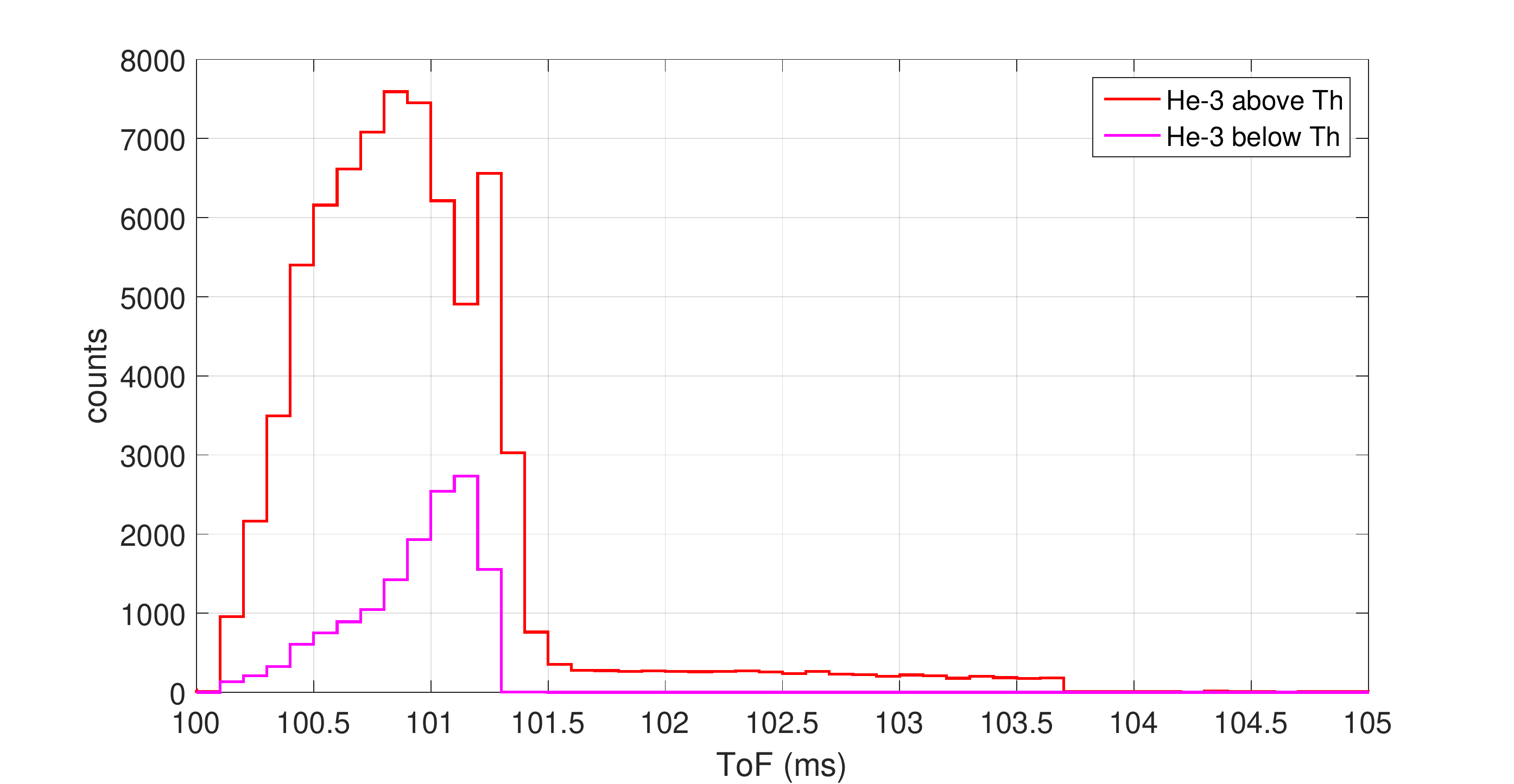}
\includegraphics[width=.9\textwidth,keepaspectratio]{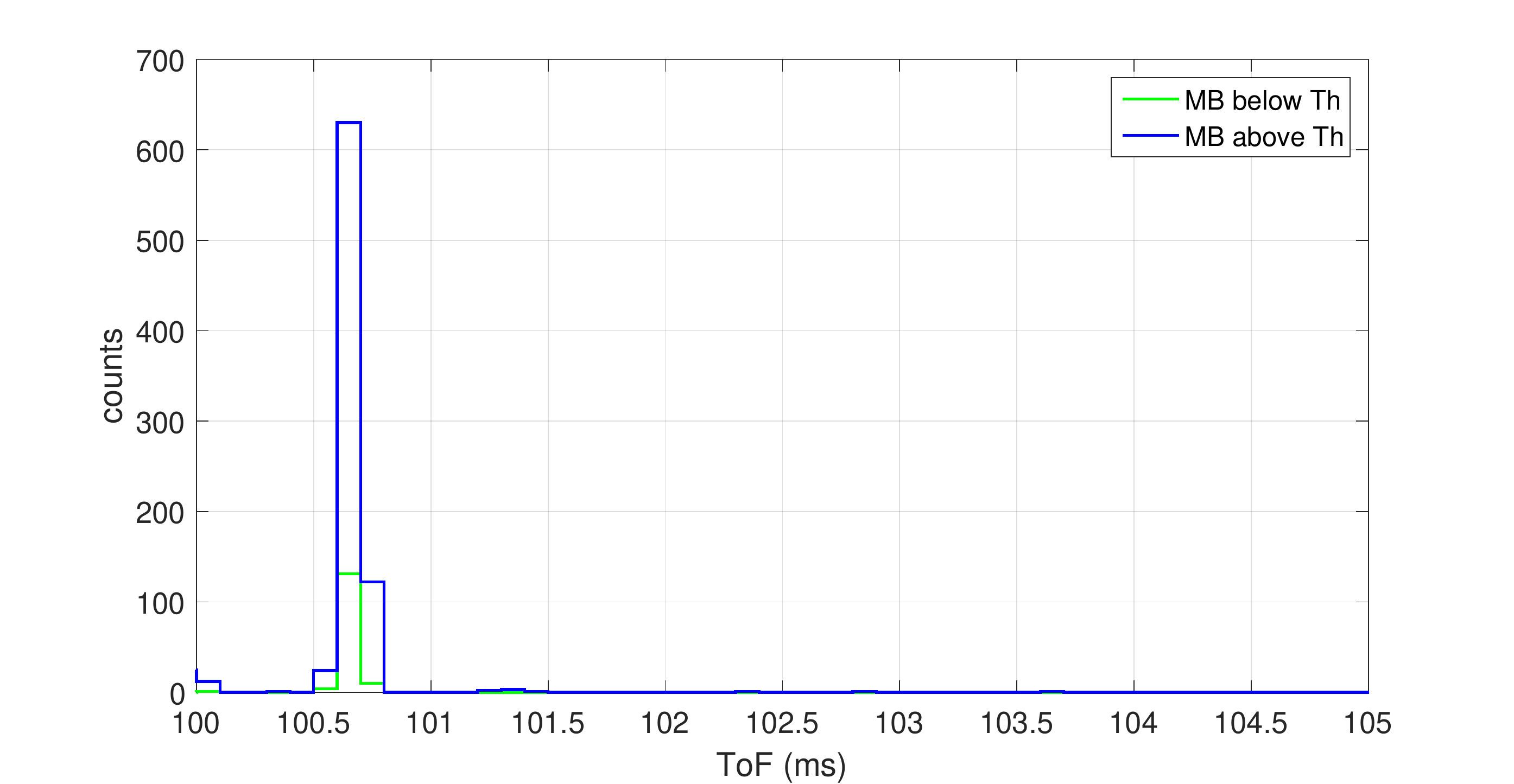}
\caption{\label{specmbhe} \footnotesize Top: ToF spectrum obtained with the $^3$He-tube below (purple line) and above (red line) threshold. Bottom: ToF spectrum obtained with the the Multi-Blade detector below (green line) and above (blue line) threshold. Both are recorded in ToF range between 100 to 105 ms, i.e., when the neutron beam hits against the chopper.}
\end{figure}

\begin{figure}[htbp]
\centering
\subfloat[]{\includegraphics[width=.49\textwidth,keepaspectratio]{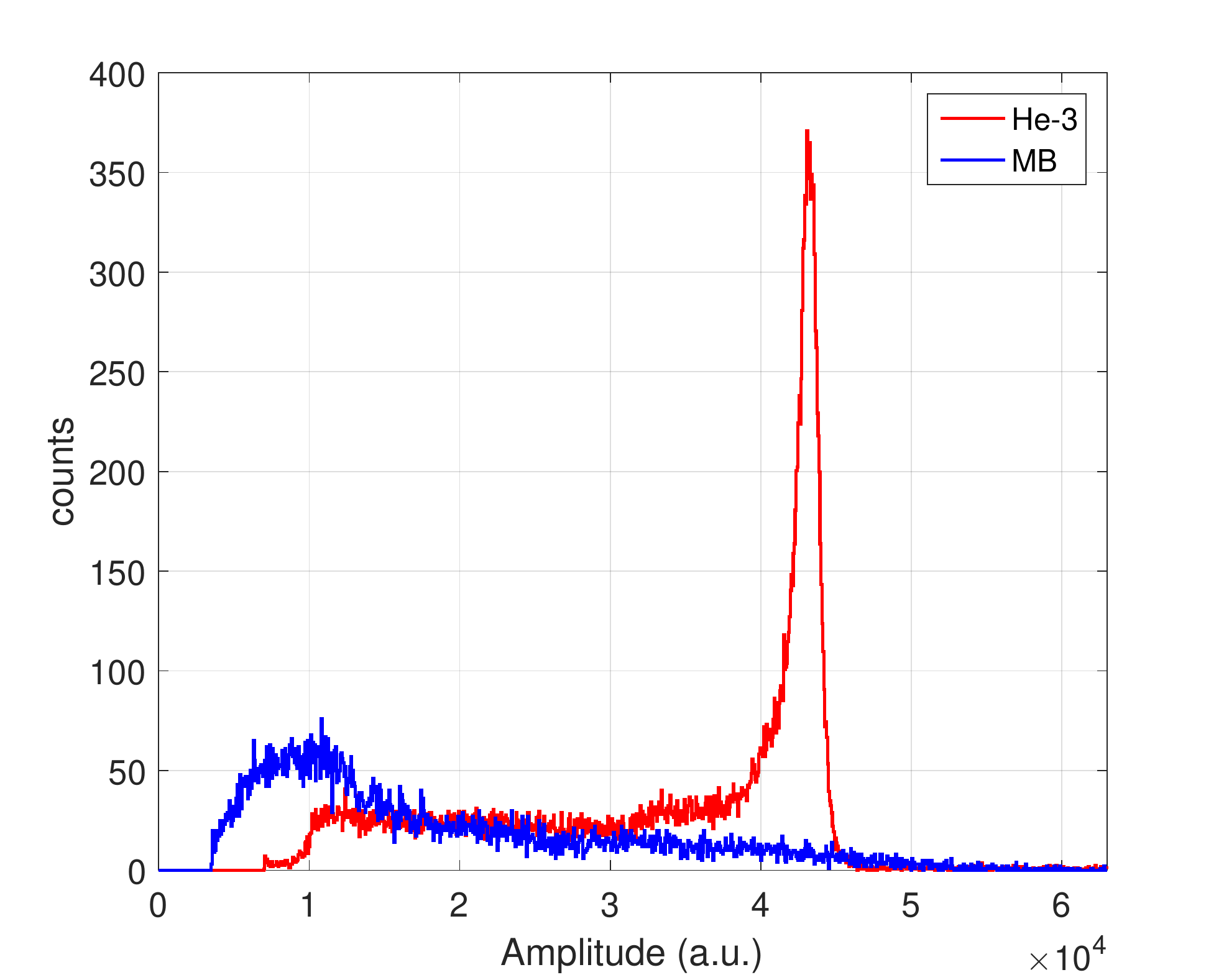}}
\subfloat[]{\includegraphics[width=.49\textwidth,keepaspectratio]{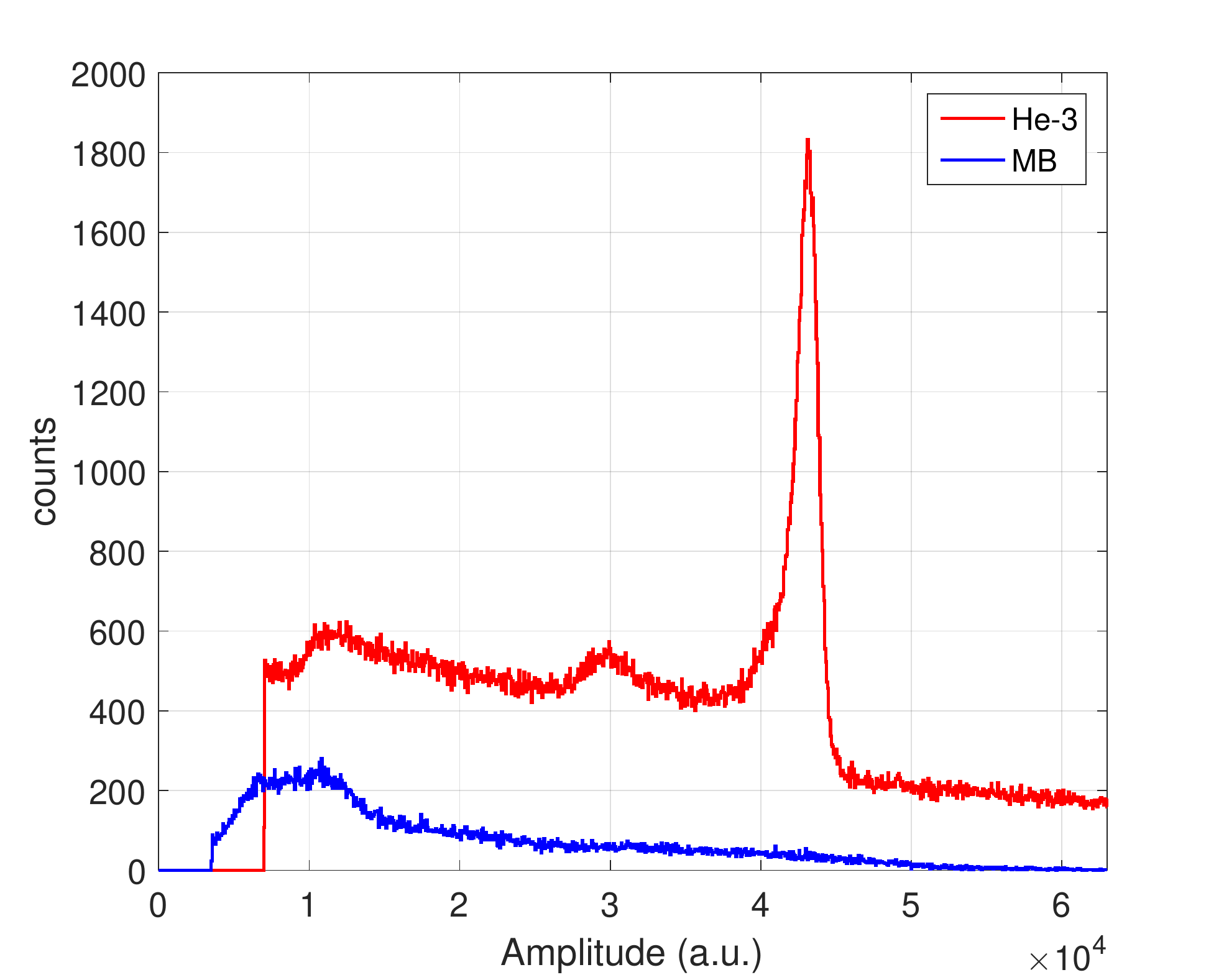}}\\
\subfloat[]{\includegraphics[width=.49\textwidth,keepaspectratio]{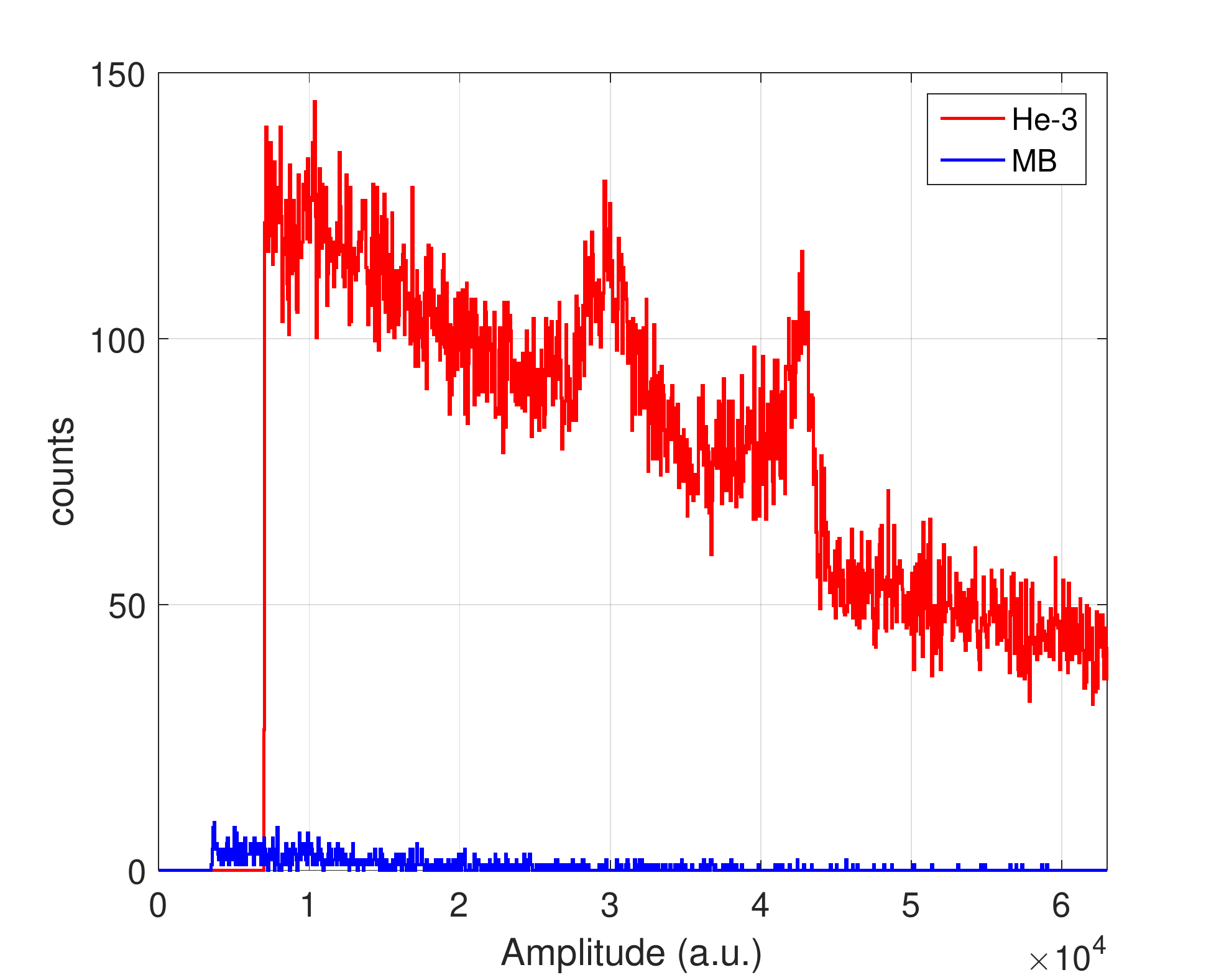}}
\caption{\label{phs25hz} \footnotesize (a) PHS in ToF = 0 - 20 ms (chopper O), (b) PHS in ToF = 100 - 105 ms and (c) PHS in the full time window 200 ms.}
\end{figure}

The ToF spectra below and above threshold are depicted in figure~\ref{specmbhe}, for the $^3$He-tube and the Multi-Blade detector separately. The time structure of the pulse is a further indication that the events below threshold correspond to the $\gamma$-rays. The gamma flash induces prompt events, in the case of the $^3$He-tube, top panel of figure~\ref{specmbhe}, is the purple pulse. When events below threshold are taking into account it is in between 1 ms. The red peak shows the signals above threshold, it is more intense because together with $\gamma$-rays of higher energy also fast neutrons interaction, thermalization of epithermal neutron and other kind of spurious scattering with the material may occur. Moreover, the peak is broader and it extends up to approximately 4 ms. In the case of the Multi-Blade the difference between the two analysis is less than in the case of the $^3$He-tube, but still significant. 
\\ In figure~\ref{phs25hz} the Pulse Height Spectrum, normalized over the time, for both detectors is shown in three different configuration: (a) PHS in ToF = 0 - 20 ms (chopper O), (b) PHS in ToF = 100 - 105 ms and (c) PHS in the full time window 200 ms. In each plot the red line identifies the $^3$He-tube while the blue line the Multi-Blade. 

The $^3$He-detector counts much more background events than the Multi-Blade. This is clearly visible in figure~\ref{specmbhe}. The integral of the PHS shown in figure~\ref{phs25hz}(a), which is the total flux on the detector, has been calculated when the beam reaches the detector and the integral of the total PHS, figure~\ref{phs25hz}(b). By subtracting the two we obtain the total flux given by any background effect. A further confirmation can be obtained by the integration of the PHS in figure~\ref{phs25hz}(c) which is recorded only at the high background peak. A factor of $\sim$ 2 orders of magnitude between the Multi-Blade and the $^3$He is  obtained in favour of the Multi-Blade. The sensitivity to the background, mainly fast neutrons and $\gamma$-rays is well characterized for the Multi-Blade detector~\cite{MIO_MB2017, MIO_fastn}. Knowing that we can estimate that $^3$He-detectors have a sensitivity to the fast neutrons on the order of 10$^{-3}$.

\chapter{Solid state Si-Gd detector for neutron scattering}\label{chapter6}

In this chapter the description of a solid state thermal neutron detector as an alternative to $^3$He-based thermal detectors is presented. This class of detectors have been developed in the early 1960s~\cite{FEIGL_sigd,RAUCH_sigd}, since then progress in both ion-implantation techniques and large scale integrated circuit (VLSI) for readout electronics made possible the design and construction of large alrea silicon detectors, either in pixel-array and microstrip form~\cite{PETRILLO-solidstate}. These devices are already widely used in high-energy physics application~\cite{Sidet-cern1,SIdet-cern2,CMS_collaboration}, and in the last decades several studies have been performed to examine the use of solid state detectors for neutron detection~\cite{PETRILLO-solidstate,SCHULTE_sigd,Mireshghi_sigd,PHLIPS-large-silicon}. The detection of neutrons with a silicon detector exploits the interactions of these neutrons with the nuclei of a converter adjacent to the detector diode, as described in section~\ref{secsemisdet}. Converters such as $^6$LiF, $^{10}$B, $^{155}$Gd and $^{157}$Gd are of standard use for thermal neutron applications. The device presented here is a PIN silicon diode coupled to $^{157}$Gd$_2$O$_3$ converters. This is one of the solid state detector prototypes under development at the Department of Physics at University of Perugia. 
\\The chapter is mainly based on a paper~\cite{Mauri_psa}, published by the author of this thesis, on the latest results obtained in several tests performed at ILL~\cite{ILL}. Apart from these results, a description of the Pulse Shape analysis method is proposed, for this class of devices, to optimize the discrimination of the neutron signals from the noise and background radiation is reported. Also, some design improvements performed during the PhD are presented, in particular the test on the deposition of Gadolinium. A paper on the implementation of the pulse shape analysis method employing an FPGA for real time application will be presented in a subsequent publication~\cite{bmac}. 

\section{Introduction}

The enormous growth of the experimental opportunities~\cite{iop} at neutron sources, despite the limited increase of the source brightness since the 1960s, has been enabled by the development of instrumentation like large position-sensitive detectors, new focusing optics and innovative exploitation of neutron polarization. The operation of new generation, high-intensity neutron sources like the Spallation Neutron Source~\cite{SNS} (US), the Japan Spallation Source~\cite{JPARC} at J-PARC, and the presently under construction European Spallation Source-ESS (Sweden), where an intensity increase of at least an order of magnitude is expected~\cite{Mezei2007,VETTIER_ESS,ESS2011}, represents a challenge because of the increased performance demands on neutron detectors, as pointed out in Chapter~\ref{chapter3}. For example, the typical process of charge collection in the widely-used $^3$He gas detectors is in the $\mu$s range, which introduces appreciable systematic uncertainties on the detector dead time and signal pile-up or errors at detection rates higher than 100 kHz. This is not of primary concern for most applications at the present installations, whereas it can become rather important in the near future when the new high rate instrumentation of the ESS will start operation. There, new technologies will be relevant to achieve better performances in neutron detection time, below 1 $\mu$s and possibly tens of ns~\cite{MIO_MB2017,DET_rates} for some specific applications~\cite{Mauri_psa}.
\\ Overall reviews of the progress made in neutron detectors are available~\cite{DET_rates,Caruso_Detsolid,Sac2015} with the search of new solutions is continuous, driven by the needs of new instruments and the anticipated shortage of $^3$He. Alternative approaches, exploiting $^{10}$B-based solutions, have been recently proposed~\cite{MIO_HERE} as large area thermal neutron detectors for high counting rates applications, and the latest results are discussed in the previous chapters. Lately, the development of new devices based on solid boron nitride crystals has been reported~\cite{Maity_bn} and their proof of concept proven. Although the implementation of boron nitride detectors needs further developments, they are quite promising devices because of the anticipated good efficiency and good spatial resolution as a consequence of the very short range of the charged particles produced by the neutron capture in $^{10}$B ($^{10}$B + n $\rightarrow ^4$He + $^7$Li) and the reduced detector thickness.
\\It is reasonable to expect novel concepts to translate into new detector applications in the near future, coupled to improved data collection techniques as prompted by an extended use of time of flight techniques. The requirements for large area, and simultaneously flexible shape, detectors with sub-millimeter resolution could be satisfied by resorting to the technology based on solid state silicon devices. Prototypes of solid state neutron detectors were proposed in the past decades~\cite{PETRILLO-solidstate,PETRILLO-TAR,PETRILLO1999,PSD_SiGd} and the possibility of reaching sub millimeter resolution~\cite{PETRILLO1999} with proportionate detection efficiency~\cite{PETRILLO-solidstate,PETRILLO-TAR,PETRILLO1999} was proven~\cite{Caruso_Detsolid,Sac2015}. Detailed investigations of the spatial resolution of microstrip detectors, based on test experiments and simulations, have been performed previously~\cite{PETRILLO1999} showing that a resolution down to 0.1 mm is possible. More recently, a 128 channels, 1-d Si microstrip detector with 0.5 mm space resolution has been tested~\cite{CASININI2012} for a local (single strip) counting rate of 500 kHz with about 10\% dead time (200 ns).
\\ Implementation of silicon detectors for neutron applications has the advantage of exploiting Integrated Circuit (IC) technology, which makes possible to meet the demand for quite high-density readout electronics. Silicon detectors can be considered as a robust and mature alternative, already available on the market, characterized by a detection efficiency comparable with current solutions~\cite{MIO_HERE} at least over the thermal neutron region. Indeed, detection efficiencies larger than 50\% can be achieved by an optimized packaging of two Si sensors with a middle layer of Gd~\cite{Mireshghi_sigd}. The high space resolution at high counting rate, typical of the Si p-n junction diode, together with the operability under vacuum and in high magnetic fields, are most appealing features of a Silicon detector for operation under the intense neutron fluxes expected at the future installations. The coupled use of Si detectors and VLSI fast electronics can push the limits of the maximum instantaneous data rate ordinarily enabled by neutron counters. Operating the detector as a counter under an intense and pulsed neutron beam requires a dead time notably less than 1 $\mu$s and the ability of managing the time information contained in the incoming neutron pulse. Readout of a large number of channels, at high speed and in parallel mode, is therefore a fundamental requisite for the construction of a large area detector.
\\The currently available VLSI (Very-Large-Scale Integration) front-end electronics can reach peaking times as short as 50 ns with still high gain ($\sim$30 mV/fC) and low noise (less than 500 e$^-$) as it is the case for instance of the ICs produced by Ideas~\cite{ideas}. It is therefore possible to get an instantaneous count rate of about 10$^6$ neutrons per channel. Nonetheless, flexibility and reliability of the data collection need to be implemented at the highest possible level. Over the last years, large effort was spent in designing and testing new neutron detectors and in developing better approaches to discriminate neutron signals against the background radiation, as well as to analyze the detector and electronic noise. In particular, several pulse shape and real time analysis approaches were proposed~\cite{DET_doro1,ps1,Mindur_hrate,ps2,ps3,psb4,MPGD_KleinCASCADE,psr2,psr1}, including the use of Field Programmable Gate Array (FPGA) to perform real time operations on the detected signals. In a recent paper~\cite{CASININI2012}, from the group of Perugia, is reported on the use of an FPGA-based system for single-event time of flight (ToF) applications as a practical tool to eliminate the time uncertainty on the collection time at the detector.

The measurements of neutron scattering experiments on standard samples performed at the spallation source ISIS (UK), by using the Si-microstrip/Gd-converter detector~\cite{CASININI2012}, is presented in~\cite{Mauri_psa}. These measurements provide a raw reference for the typical performances that can be obtained by using a Si-based device for neutron scattering applications. To optimize the Si detector performances, a new approach to signal processing has been proposed, namely a pulse shape analysis designed to extract the neutron signal from high noise, in real time mode, which will be discussed in section~\ref{secpsa}. By {\it real time} operation it is meant an acquisition system able to process the relevant information in a time short enough as compared to the rate of events per detector channel. The pulse shape analysis, efficient and simple enough for real time applications using FPGAs, is applied to the neutron data collected by two prototype detectors, namely a PIN diode/Gd-converter and a SiPM coupled to a neutron scintillator, mounted on the direct beam of the IN3 spectrometer at the ILL (France)~\cite{ILL-yb}.

\subsection{PIN diode/Gd-converter detector}\label{secpinsigd}

\begin{figure}[htbp]
\centering
\subfloat[]{\includegraphics[width=0.6\textwidth,keepaspectratio]{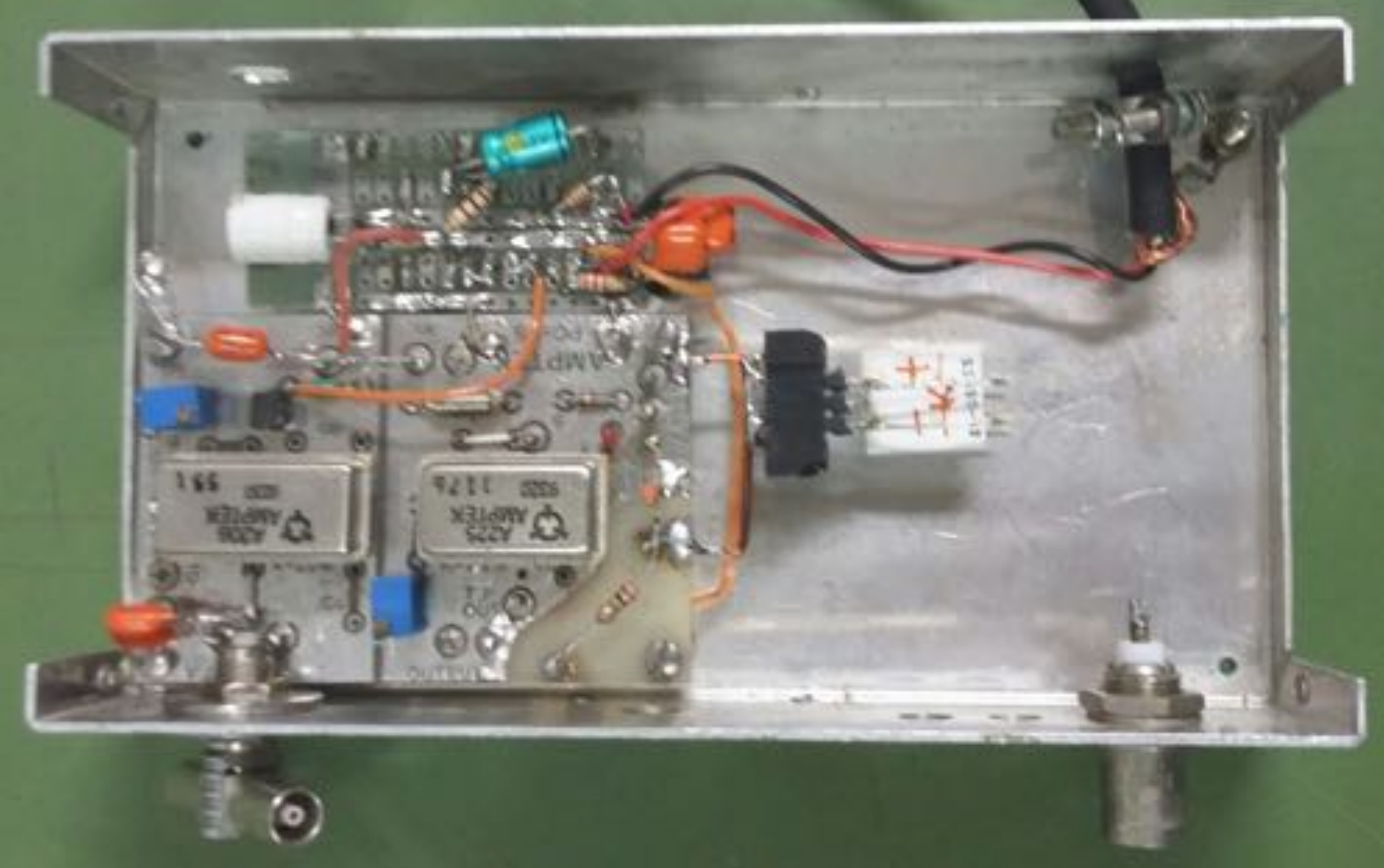}}\\
\subfloat[]{\includegraphics[width=0.62\textwidth,keepaspectratio]{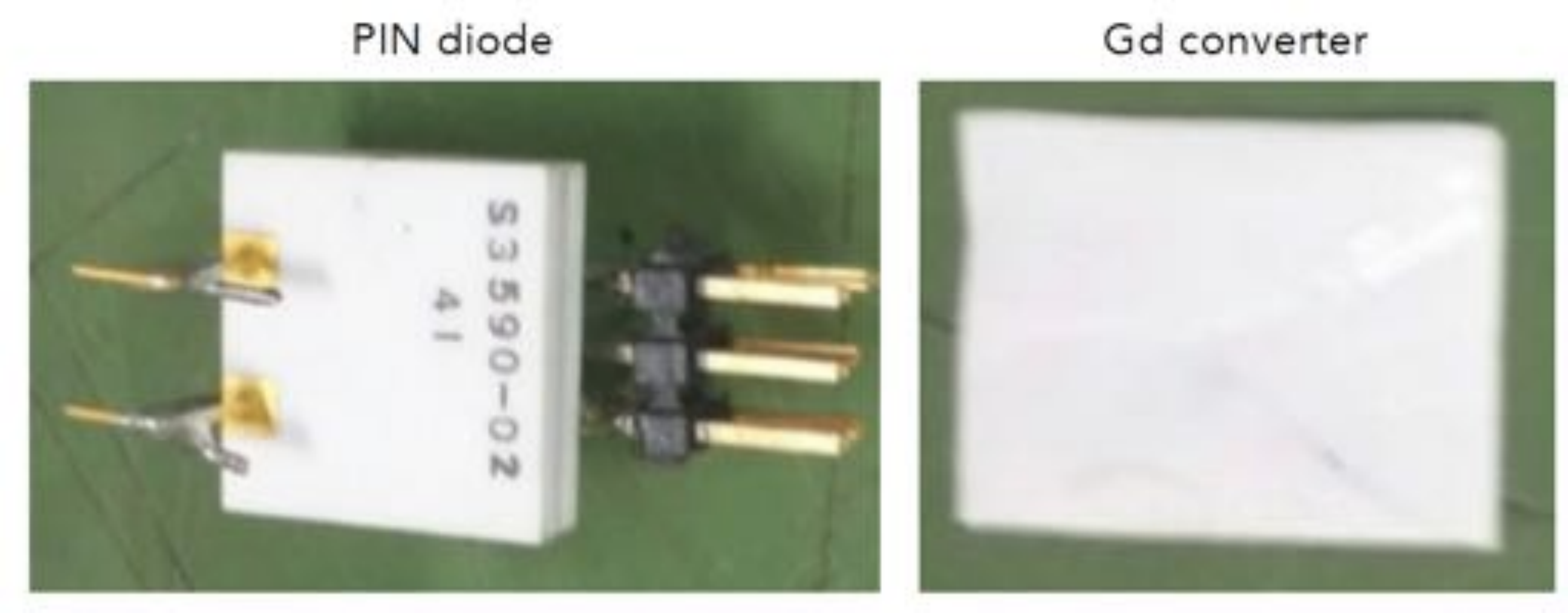}}
\caption{\label{detsigd} \footnotesize (a) Picture of the PIN diode/Gd-converter detector. (b) Detail of the PIN diode and of the Gadolinium converter deposited on a mylar foil.}
\end{figure}

The PIN diode (Hamamatsu, S3590-09), the principles of operation for this class of device is described in section~\ref{secsemisdet}, 0.3 mm thick and 10 x 10 mm$^2$ area, was coupled to a $^{157}$Gd$_2$O$_3$ converter layer, 3 $\mu$m thick, coated on a 2 $\mu$m thick mylar foil to get a removable converter.  As described in section~\ref{secndet}, the Gadolinium is a suitable converter for cold and thermal neutrons, because of its large absorption cross section. Through the radiative capture process, $\gamma$-rays over a wide keV energy range, and a cascade of conversion electrons are emitted~\cite{Gdabs1,Gdabs2}, with a probability of $\approx$80\%~\cite{GdNeumann} for the emission of secondary electrons.
\\ The main peak is at about 70 keV, the ranges of the 70 keV conversion electrons are $\approx$ 10 $\mu$m in Gd and $\approx$ 30 $\mu$m in Si, this leads to the choice of the PIN diode and the $^{157}$Gd$_2$O$_3$ converter layer thickness. In figure~\ref{detsigd}(a) the prototype is shown, in~\ref{detsigd}(b) both the PIN diode and the Gd converter are depicted. 

\begin{figure}[htbp]
\centering
\includegraphics[width=1\textwidth,keepaspectratio]{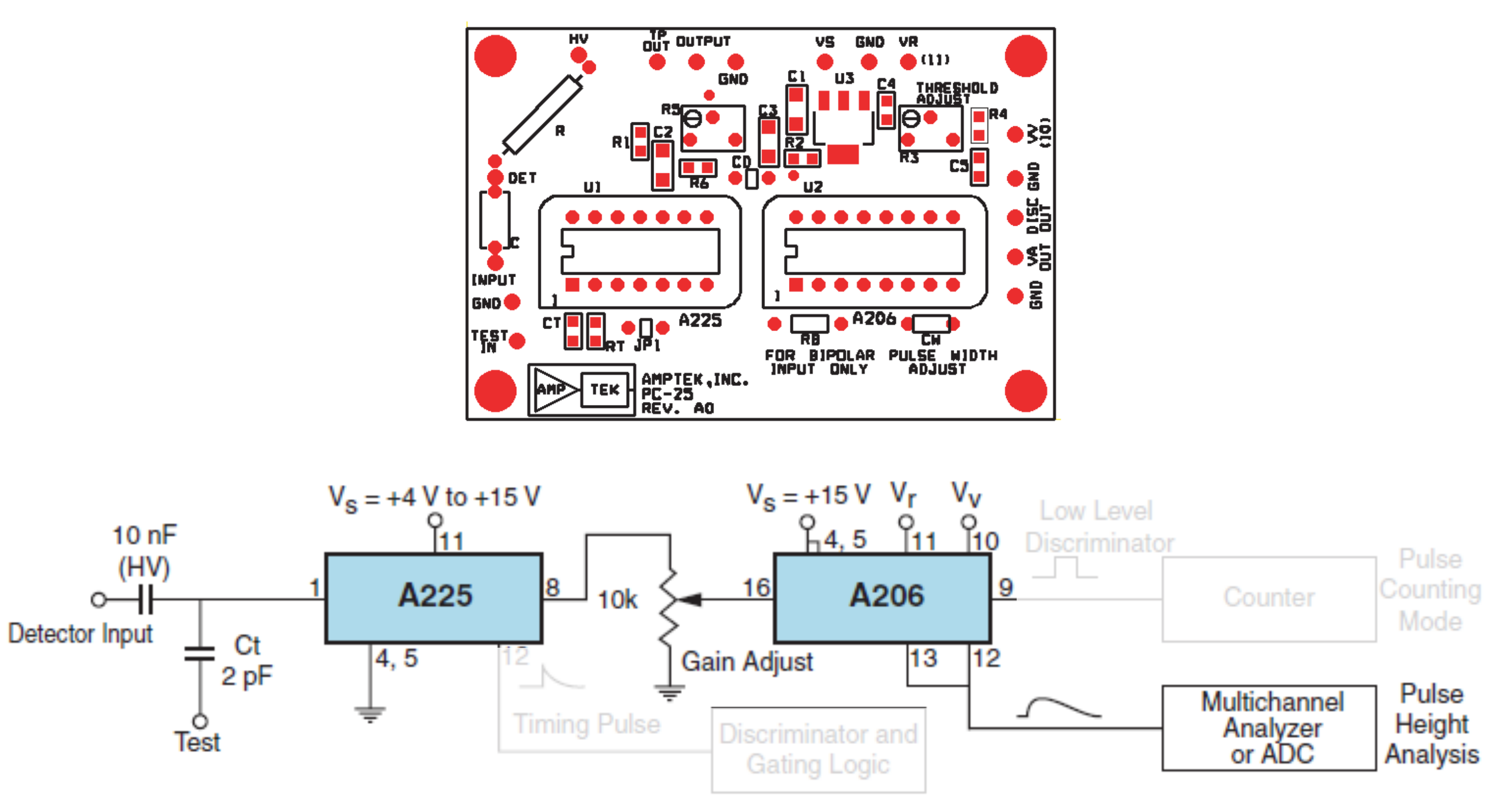}
\caption{\label{electsigd} \footnotesize Sketch of the electronics board PC25 and the two components: the preamplifier A225 and the voltage amplifier/low level discriminator A206 as they are employed in the detector. The chain of connection from the input signal to the output is depicted.}
\end{figure}

The front-end electronics of the PIN diode consisted of a low noise A225 preamplifier (Amptek, USA) followed by a linear amplification stage (voltage gain = 100), a sketch of the electronics is shown in figure~\ref{electsigd}. The signal enters the pin 1 of the preamplifier and the output 8 is selected so to have a unipolar pulse, useful to operate the pulse shape analysis method. The preamplifier is coupled with a capacitance to the voltage amplifier/low level discriminator A206 through the pin 16. To connect the amplifier and the low level discriminator pin 13 and pin 12 are linked together, as shown in figure~\ref{electsigd}. The discriminator requires a reference voltage in a range, $V_r = 4.5 - 5.5\,$V and a variable voltage $V_v < V_r$. The difference $V_v - V_r$ is the discrimination threshold. The shaping time is on the order of few $\mu$s.
\\ The input of the board is connected to the PIN diode through a six pin connector, see figure~\ref{detsigd}(b), in order to be removable and easily interchangeable to test different PIN-converter configurations. A detailed study on several combination is presented in~\cite{mauri_tesi}.

\begin{figure}[htbp]
\centering
\subfloat{\includegraphics[width=0.3\textwidth,keepaspectratio]{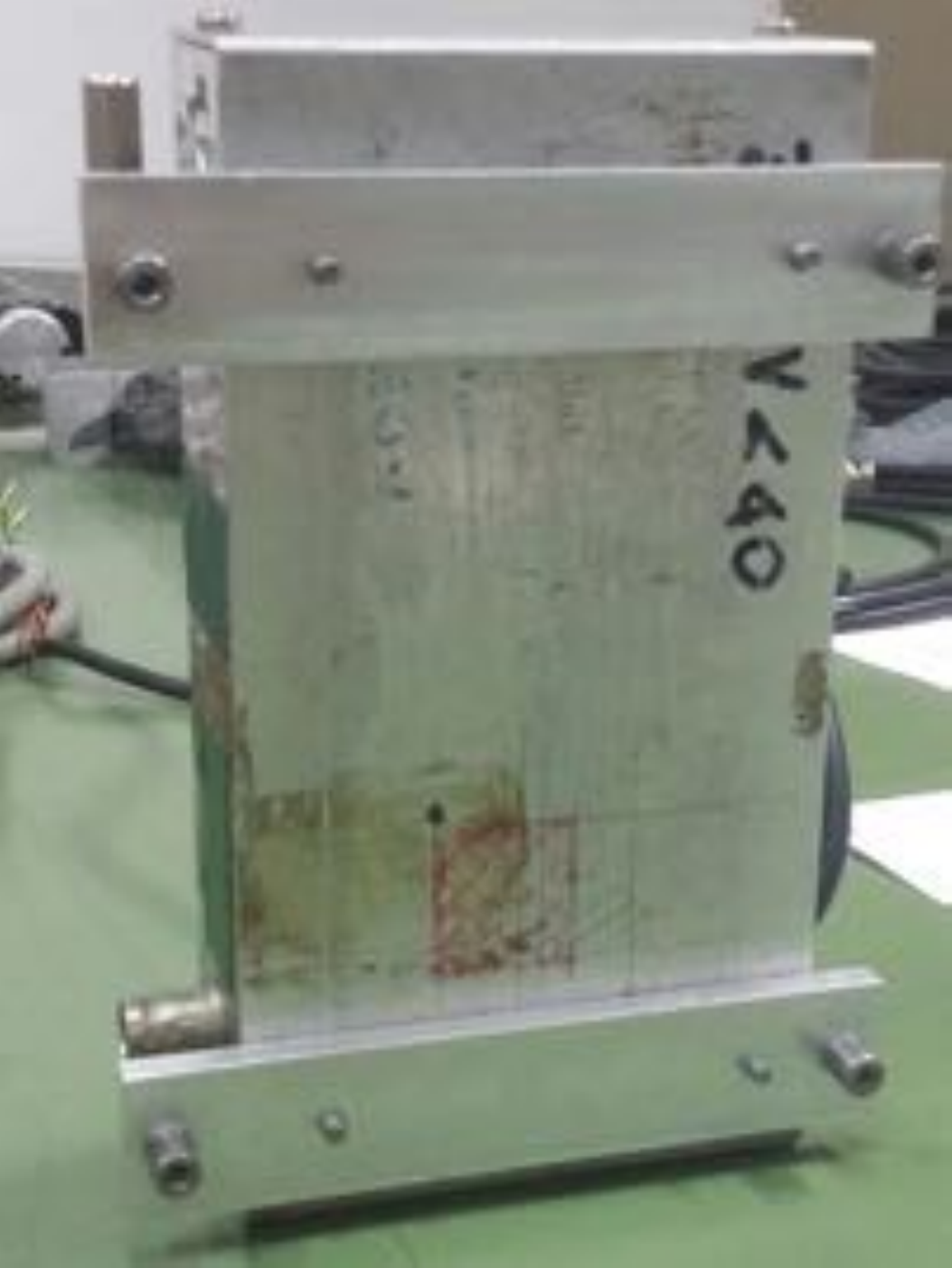}} \quad
\subfloat{\includegraphics[width=0.315\textwidth,keepaspectratio]{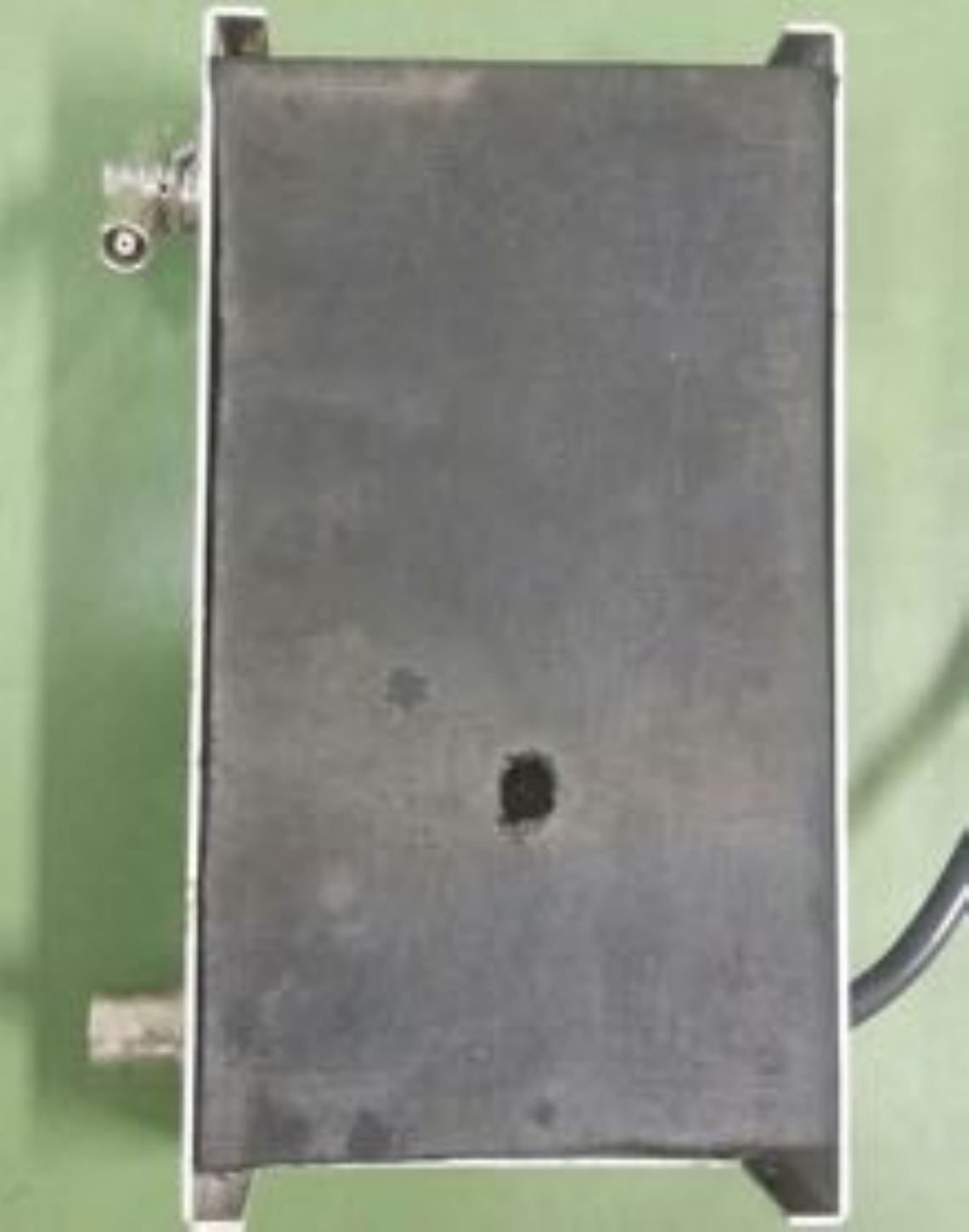}}
\caption{\label{boxsigd} \footnotesize Aluminium box to prevent the light exposure of the silicon sensor. The internal front side is covered with a boron mask with a hole in the direction of the PIN, in order to shield the electronics.}
\end{figure}

The output signal was digitized using a pico scope 2000 (Pico Technology, UK) with a bandwidth of 100 MHz and selecting a sampling time $T = $100 ns. Note that the front-end electronics coupled to the PIN diode is relatively slow. This choice was dictated by the selection of the isotopic converter $^{157}$Gd$_2$O$_3$ whose non-uniform thickness, due to a layer production procedure aimed at minimizing the loss of the expensive isotopic material, introduces a further source of noise. An alternative deposition technique has been tested and it will be presented later in section~\ref{secfutsigd}.
\\By coupling a faster electronics, which is possible, the noise will be higher and a more powerful discriminating factor would be necessary. Indeed, electron-hole pairs produced inside the Si depletion region by the secondary electrons are swept to the electrodes by the applied electric field. The so-generated electric signal can be transformed into a useful signal for analog or digital processing to discriminate against noise and to define the neutron arrival time. The collected charge can be either negative or positive with a collection time shorter for electrons and longer for holes, typically below 20 ns for a standard 300 $\mu$m thick $n$-type high-resistivity Si sensor operated at $\sim$100 V reverse bias. Due to the relatively low energy of the conversion electrons, charge collected in silicon is rather small, ranging from 1 fC to 10 fC, and therefore a low noise front-end electronics must be employed. 
\\Differently from the use of Si sensors as particle trackers in high energy physics experiments, where an external trigger is normally available and charge collection takes place over a short period corresponding to the expected occurrence of the event, in neutron scattering applications the detector needs to be {\it self-triggered}, i.e., it must be able to discriminate the neutron signal against noise and background radiation.
\\ In figure~\ref{boxsigd} is depicted the box in which the detector is placed in order to avoid the PIN diode light exposure. Moreover, the front side is covered with a boron mask with a hole in the direction of the PIN diode, in order to shield the electronics from the neutron beam and to have the full beam on the sensor. 
  
\section{Pulse Shape Analysis method}\label{secpsa}

In a standard acquisition system the front-end of the detector is a charge preamplifier, with appropriate filters, followed by a shaper amplifier and a simple threshold discriminator to produce a short pulse used to determine the arrival time of a neutron, as showed in the previous section. In presence of noise, the discriminator threshold is set higher and part of the neutron counts can go lost while some noise pulses can be counted as neutrons. These kinds of problems are more severe for solid state Si detectors with a Gd converter as the charge signal is small. For a more accurate identification of neutrons, it is in general preferable to employ a window discriminator as most of $\gamma$-rays produce a signal much higher than the neutrons. Indeed, as described in section~\ref{secndet}, the maximum energy of the electrons emitted by Gd on capturing a neutron is lower than 250 keV, while $\gamma$-rays, normally present in the neutron source environment, have energies in excess of 500 keV, like, for instance, those emitted by $^{10}$B present inside the shielding. Considering the low thickness (typically 0.3 mm) of Si sensors, the general sensitivity to $\gamma$-rays is not very high~\cite{Mireshghi_sigd}. Assuming for simplicity that the average energy release from gamma-rays in Si is that of a Compton process at 90 degs, energies of the order of 200-300 keV are obtained, consequently, such an unwanted contribution can be reduced using an upper level discrimination.

The general principle of the acquisition system turns out to be completely different when using an FPGA connected to the very front-end of the detector, a description can be found in~\cite{CASININI2012}. The basic design, under development, consists of a front-end wide band and low noise amplifier plus an Analog to Digital Converter (ADC) with a frequency appropriate for a proper sampling of the signal produced by the detector in use. All the detector channels are independent so that the digital output of the $i$-th ADC, corresponding to the $i$-th channel or pixel of the detector, is sent directly to the $i$-th digital I/O of the FPGA. The requirements of the present proposal are that many strips, or pixels, belonging to a position sensitive detector (PSD) are analysed using a single FPGA.

For an optimal discrimination of the real neutron signals from the whole of the other signals arising from either electronic or ambient noise or due to $\gamma$ radiation, some analysis of the signal is necessary. A shape analysis is a very useful tool to control the effect of noise and to reject the $\gamma$ contribution by a full pulse height analysis when the shape of the $\gamma$-ray signal were too close to that of neutron.
Indeed, a more specialised acquisition electronics would be desirable for the specific needs of a neutron counter for low energy scattering applications. In particular, some flexibility in optimizing the real time signal analysis can be crucially relevant. The ability to operate also in a high noise environment, where the low signal from Si detectors is a limitation, would be a major step forward for this technology. To this purpose a new approach for a real time analysis of the detector pulses has been developed .
\\It is briefly recalled here the basic approaches, either analogical or digital, currently used to single out the true neutron signals and to define the neutron arrival time with enough accuracy.
\\ A first analysis can be based on a filtering of the signal to get a properly shaped pulse, as it is done using standard hardware front-end electronics. The pulse amplitude is proportional to the energy released after neutron capture inside the detector. Filtering, usually accomplished by the detector preamplifier, can be obtained through a digital elaboration of the sampled signal. For example, a simple low-pass filter is formally obtained in the time domain by the following integral:

\begin{equation}
v_u(t) = {1 \over \tau} \, \int_{-\infty}^{t} \, v_i(t') \, \exp\Big[{t' - t \over \tau}\Big] \, d t'
\end{equation}

\noindent
where $v_i(t)$ and $v_u(t)$ are the input and output signals respectively and $\tau$ is the decay time of the filter. A general linear filter is simply given by:

\begin{equation}
v_u(t) = \int_{-\infty}^{t} \, v_i(t') \, T_f(t - t') d t'
\end{equation}

\noindent
where $T_f(t)$ is the transfer function of the filter. If the signals are sampled with a period $T$, the integral can be transformed into a series that, in practical cases, reduces to a finite sum:

\begin{equation}
v_u(t_k) = \sum_{i=-\infty}^\infty \, T_f^d(t_k,t_i) \, v_i(t_i)
\end{equation}

\noindent
where $t_k$ is the running sampling time and $T_f^d(t_k,t_i)$ is the transfer matrix. In the case of a low-pass filter, the transfer matrix is readily obtained:

\begin{equation}
T_f^d (t_k,t_i) = \begin{cases}\exp \Big[\frac{t_i - t_k}{\tau} \Big] \Big[1 - \exp\Big(\frac{-T}{\tau} \Big)\Big], & \mathrm{if \ t_i \le t_k} \\ 0, & \mathrm{otherwise} \end{cases}
\end{equation}

\noindent
where $T$ is the sampling time. In real cases the series must be substituted by a finite sum in a way that $t_k - t_i \gg \tau$. This equation is useful for practical purposes. Interestingly, low order filters are relatively simple to be implemented in hardware and numerical solutions.
\\ Note that all the analog linear electronic filters are infinite impulse response (IIR), namely the output signal lasts ideally forever after the arrival of a short pulse at the input. On the contrary, almost all the digital filters are finite impulse response (FIR), that is the output signal lasts for a given period after the arrival of a short pulse at the input. A real time digital analysis can use FIRs of practically any shape. Of course, IIR filters can be implemented on FPGAs for digital signal processing, but FIR filters can be more accurate when fast signals are analysed, while the (real time) computational cost is similar.
\\ For real applications, it can be important to introduce more complex data treatments to extract the cleanest information from the signal produced by the detector. A better data processing is certainly more relevant in the case of solid state detectors coupled to Gd converters where charge collection is fairly low and the electronic noise becomes an important issue, in addition to the point of $\gamma$-ray rejection. Also a specifically tailored data treatment can be valuable in the case of detectors where scintillators are coupled to SiPMs for light collection. 
\\SiPMs~\cite{SiPM-LiF} are promising devices with a potential to replace vacuum phototubes because of their very small size, good performances under vacuum and high magnetic fields, the negligible heat release and the low cost for large production. Their limitation is, however, the very high thermal noise that can mask the neutron signal, particularly when low light emission lithium glass scintillators are employed. A prototype has been tested together with the PIN diode detector at ILL and the same pulse shape analysis has been used, the results will be shown later in section~\ref{secmeassigd}.
\begin{figure}[htbp]
\centering
\includegraphics[width=0.65\textwidth,keepaspectratio]{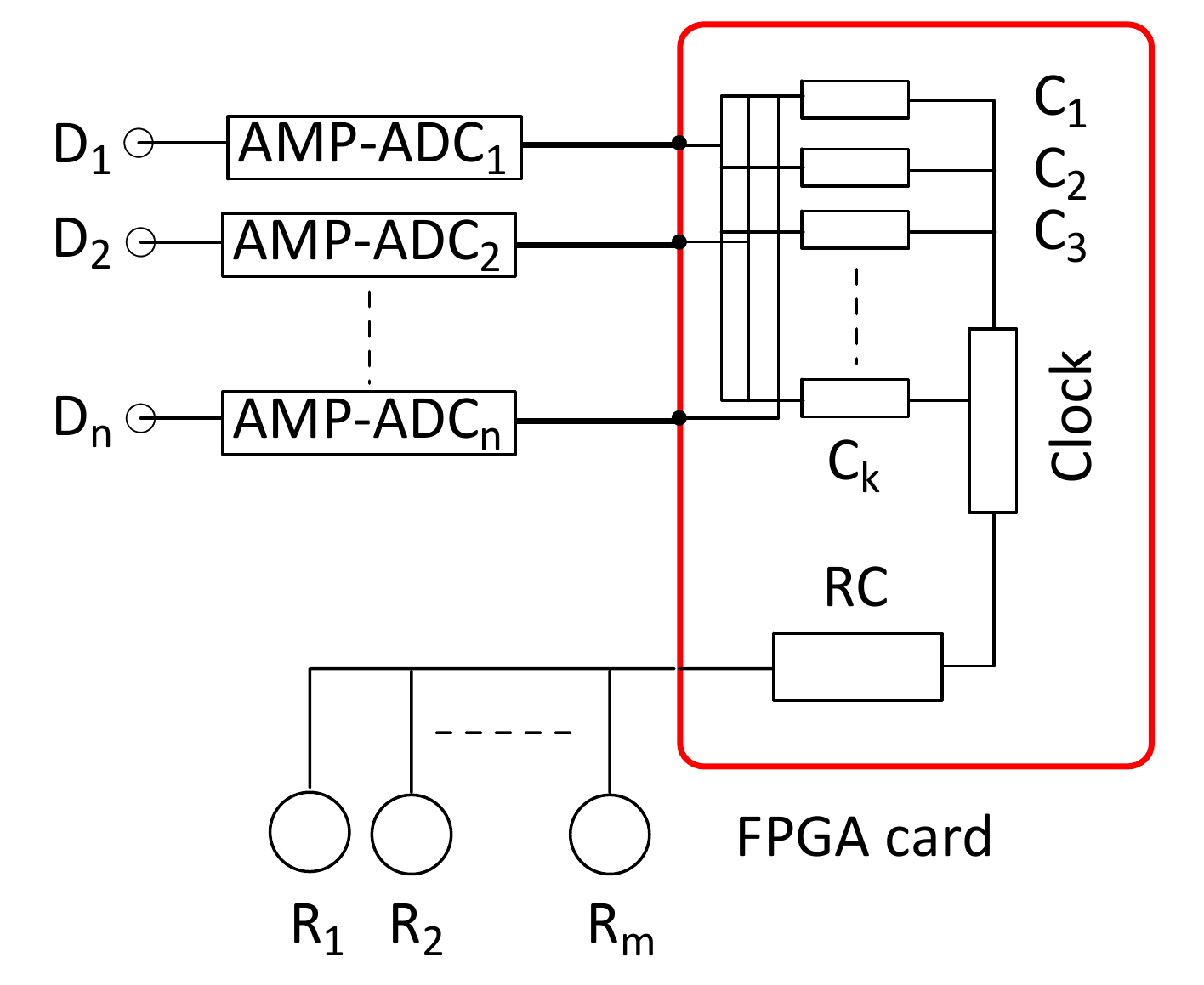}
\caption{\footnotesize Block description of the proposed acquisition system for real time pulse shape analysis and single event acquisition, and control of additional components like ToF systems. The $n$ blocks AMP-ADC are wide band amplifiers followed by the analog to digital converters. The FPGA is programmed in such a way to have specific calculation elements $C_1 ... C_k$ to perform the different steps of the pulse shape analysis. The RC is the rotation control system used to drive different components. The Clock is the same for all elements. Figure from~\cite{Mauri_psa}.}
\label{fig6sigd}
\end{figure}
\\To increase flexibility, it has been developed a procedure intended to perform all the data treatment by special programming in an FPGA. The block diagram of the proposed system is shown in figure~\ref{fig6sigd}. A wide band, low noise linear amplifier is used to amplify the incoming voltage signal. The amplified output signal is used to feed a fast enough ADC so that the parallel 8 bits output can be sent to the FPGA that provides also the clock of the ADC. A full analysis of the sampled detector signal is then performed by the FPGA to obtain a real time detection of the true neutron signals together with their arrival time stamp using algorithms described in the next section. Implementing these algorithms in real time will require a new FPGA architecture, which it is also developing by the group at the Dipartment of Physics of Perugia. This system, whose simulation and operation will  be presented in detail in a subsequent publication~\cite{bmac}, is based on the implementation of several digital processors on the FPGA. In comparison with the usual processors enabling standard numerical manipulation in FPGAs, the processors for the present application are designed to increase flexibility and they can be further programmed by the user with minor performance reduction~\cite{sitebmac}.

\subsection{Computer simulations}\label{secpcsim}

The effectiveness of the proposed pulse shape analysis was first tested by computer simulation. The present approach relies on a full pulse shape analysis, instead of turning to a filter to identify the correct signals. The correct shape of the reference neutron signal $V_r(t)$ is defined by comparison with the measured signal $V_m(t)$, sampled at $N$ time points $t_i$. The comparison can be carried out in real time if the sequence $V_m(t_i), t_i, i = 1,...,N$ is stored into an appropriate buffer. This is easily carried out in the simulation phase but, for real time applications, a proper procedure has to be implemented on an FPGA. Of course, with the goal of real time applications on measured data, a preliminary decimation protocol is considered: the data to be analysed starts when $V_m(t) - V_m(t - T) > V_{tl}$, where $V_{tl}$ is a threshold to be defined according to the noise level. Each time this condition is fulfilled, the procedure starts and the sequence is tested to check if the shape is the correct one. A high threshold was also introduced in the procedure to ignore the case of large signals coming from some background radiation, e.g., $\gamma$ radiation in the case of Si/Gd detectors. The high threshold $V_{th}$ enables starting the analysis only if the sequence, after crossing the low threshold, verifies $V(t_i) \le V_{th} \forall$ $i$.
\\ To establish if the sequence $V_m(t_i), t_i, i = 1,...,N$ refers to a genuine neutron signal, one can search for sequences such that:

\begin{equation}
S = \sum_{i=1}^N \, [ V_m(t_i) - A_s \, V_r(t_i - t_c)]^2 < S_o
\end{equation}
\noindent
where $t_c$ is a shift with respect to the starting point and $A_s$ is the normalization constant determined from the reference and the measured signals using the following simple equation:

\begin{equation}
A_s = {\sum_{i=1}^N \, V_m(t_i) \over \sum_{i=1}^N  \, V_r(t_i)}
\end{equation}

Obtaining $A_s$ implies just two simple operations, once the fixed reference signal $V_r(t)$ is stored in a given buffer. One has to calculate $\sum_i \, V_m(t_i)$ and then the ratio to the fixed sum $\sum_i \, V_r(t_i)$ which is also stored. A real time application can be simplified by a proper choice of $V_r(t)$ such that $\sum_i \, V_r(t_i) = 2^n$, and the calculation of the ratio $\sum_i \, V_m(t_i) \over \sum_i \, V_r(t_i)$ is much faster. Indeed, dividing a binary number by $2^n$ is equivalent to the shift of all the bits of the numerator by $n$ position to the {\it right}, an operation much faster than the true division. The determination of $t_c$ is performed by repeating in {\it parallel} the test by using a small set of values $t_{c_1},...,t_{c_K}$ and using the minimum value of $S$.
\\Holding the above relationships, it can be assumed that the sequence starting at $t_1$ indicates the collection of a neutron in the detector at the time $t_c$. After the procedure finds a neutron event in the detector at time $t_c$, the threshold search routine is re-activated to begin anew the search for the next neutron in the data.

Once $V_r(t)$ and $N$ are properly chosen, the whole procedure is much more efficient than a linear filtering technique, even implemented on a FPGA after digitizing the data. By using available low cost FPGA industrial cards, it is already possible to implement a real time acquisition. Preliminary simulation tests demonstrate that the complete process above described can be performed in about 20 clock cycles. If the clock is running at 200 MHz the dead time is of the order of 100 ns, therefore it is possible to work with a local peak counting rate of 1 MHz with an average 10\% loss~\cite{Mauri_psa}.

The outlined procedure was tested using simulated signals, randomly generated as to arrival time and amplitude and slightly fluctuating pulse length. The following equation has been employed to generate the pulses with random values for the arrival time $t_o$ and $b$ and multiplying $V_r(t)$ by a random amplitude.

\begin{equation}
V_r(t) = {1 \over 2} \Big \{1 + \tanh\Big[{t - t_o \over \tau}\Big] \Big\} \cdot \exp[-b (t - t_o)^2]
\label{eqpeak}
\end{equation}

\begin{figure}[htbp]
\centering
\includegraphics[width=0.9\textwidth,keepaspectratio]{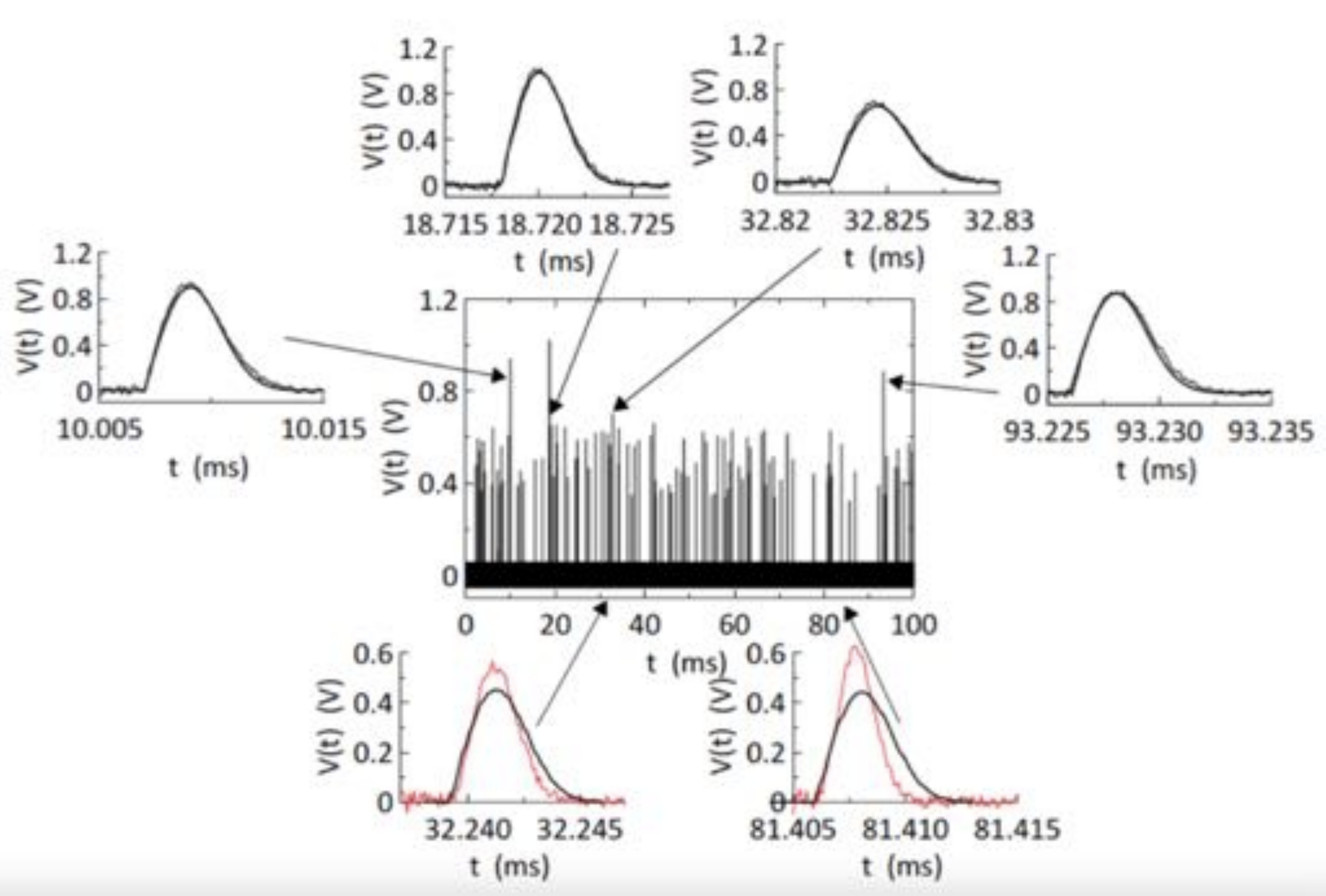}
\caption{\footnotesize Typical signal produced in the simulation, lasting 100 ms. The four neutron peaks present in the signal are shown in the expanded plots (thin line) in comparison with the model fitting curves (thick line). A couple of background signals are shown in red.}
\label{fig7sigd}
\end{figure}

A Gaussian random noise, whose amplitude is about 20\% of the average signals amplitude, was added to the so generated signals to simulate a real output from the present detector system. Additional peaks with fluctuating length were added to the data to simulate signals of similar shape and comparable amplitude produced by possible background radiation or other sources. The simulation enabled to identify the error level of the analysis based on the described approach. Indeed, the correct neutron signals were located at the known time positions and the false signals were identified as well. The results of the simulation are shown in figure~\ref{fig7sigd} where a typical sampled signal is shown together with the neutron peaks fitted by using equation~\ref{eqpeak} with $b = 170000$ ms$^{-2}$ and $\tau = 0.001$ ms. These values are chosen based on the previous work~\cite{mauri_tesi}. A couple of false data chosen at random are also shown, the value of $b$ varies between $3\cdot 10^5$ ms$^{-2}$ and $9\cdot 10^5$ ms$^{-2}$, while $\tau$ is fixed as for the neutron signals. The simulated sequences lasted 0.1 s and the sampling frequency was 10 MHz. A total number of 4 neutrons is detected in the typical data shown in figure~\ref{fig7sigd}. These 4 neutron peaks correspond to a 100\% efficiency of the analysis. No false signal was picked up by the analysis procedure even though the number of spurious peaks was 104, that is much higher than the number of neutrons. 

\begin{figure}[htbp]
\centering
\includegraphics[width=0.7\textwidth,keepaspectratio]{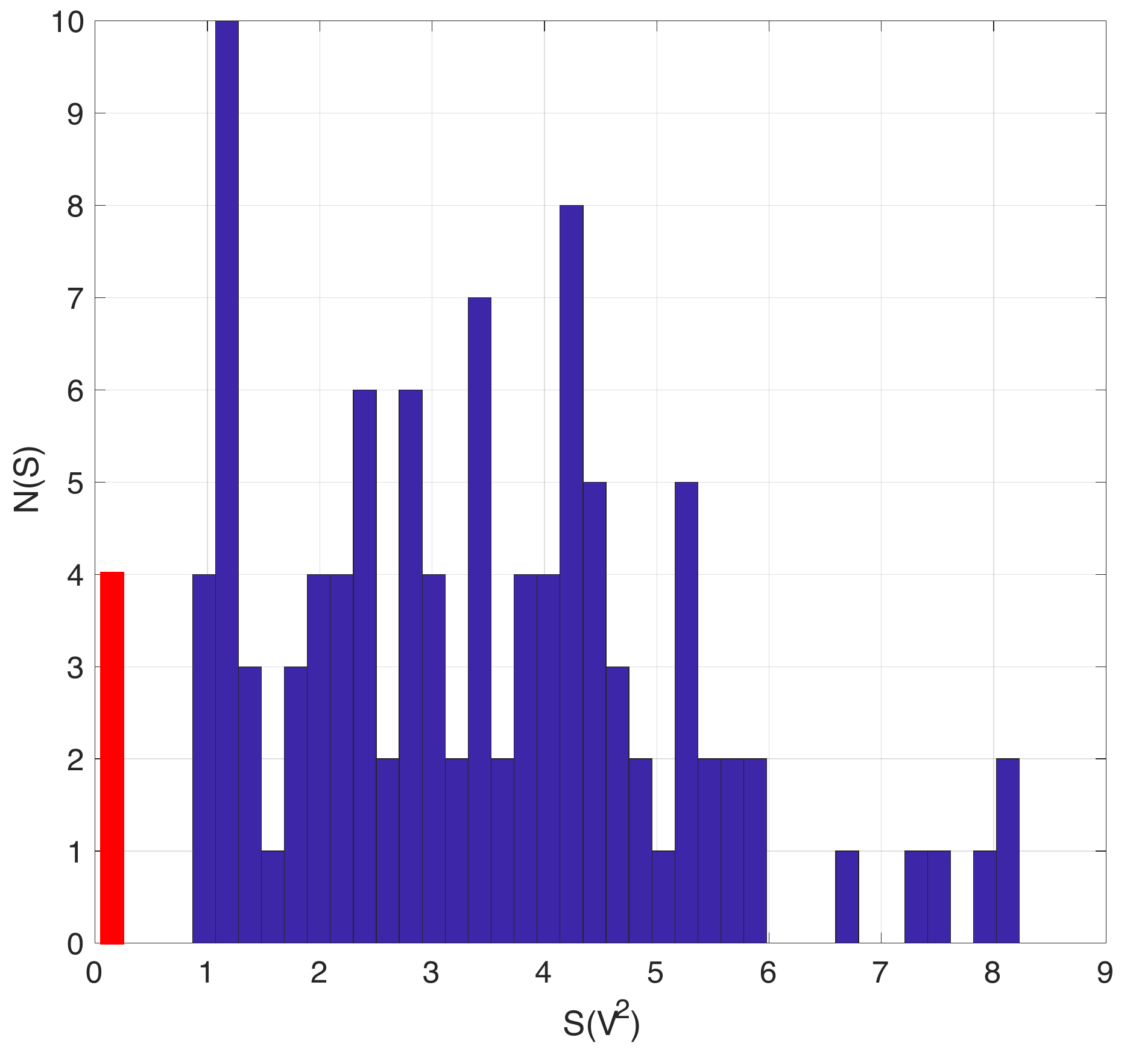}
\caption{\footnotesize Histogram of $S$ values between 1-10. The four neutron peaks have $S$ values in a small range and they correspond to the red bar, the false peaks do not show any $S$ value below 0.94.}
\label{histSsigd}
\end{figure}

The histogram of the $S$ values of the false peaks is shown in figure~\ref{histSsigd}, for those peaks exceeding 0.1 V peak height. None of the false peaks presents $S$ below 0.94 while for the 4 neutron peaks $S$ is equal to 0.069, 0.049, 0.117 and 0.069 respectively. Referring to figure~\ref{histSsigd}, indeed, it is highlighted in red the bar corresponding to the $S$ value of the four neutron peaks, while before $S=0.94$ no false signals are analysed. 
\\ To complete the simulation, several different sequences were produced in order to get better statistics. The values for the low and high threshold, $V_{tl}$ and $V_{th}$, are 0.1 V and 2 V respectively, while it has been set $S=0.6$. The error level estimated with this approach is $\approx 7 \cdot 10^{-5}$ with only 5 false signals when more than 72400 peaks were analysed. Further simulations were carried out also varying the pulse length. A set of at least 10$^5$ randomly generated pulses, 50 \% of which are true pulses with an average amplitude equal to 70 \% of the false pulses, plus a Gaussian noise of the order of 20 \% of the true pulses, were analysed. It was found that the selection of peaks characterized by $S < S_o$ with amplitude $A_s$ smaller than 0.8 is very good at all values of the pulse length, even when the pulse length of true and false pulses is the same. Note that the present selection efficiency is not the detection efficiency of the device, as here the output signal from the amplifier is analysed. 
\\ In conclusion, according to the results of the simulations, one can be quite confident that the outlined procedure can help in obtaining a good real time discrimination of the neutrons even in the worse environment and working conditions of the detector.

\section{Measurements}\label{secmeassigd}

\begin{figure}[htbp]
\centering
\includegraphics[width=0.7\textwidth,keepaspectratio]{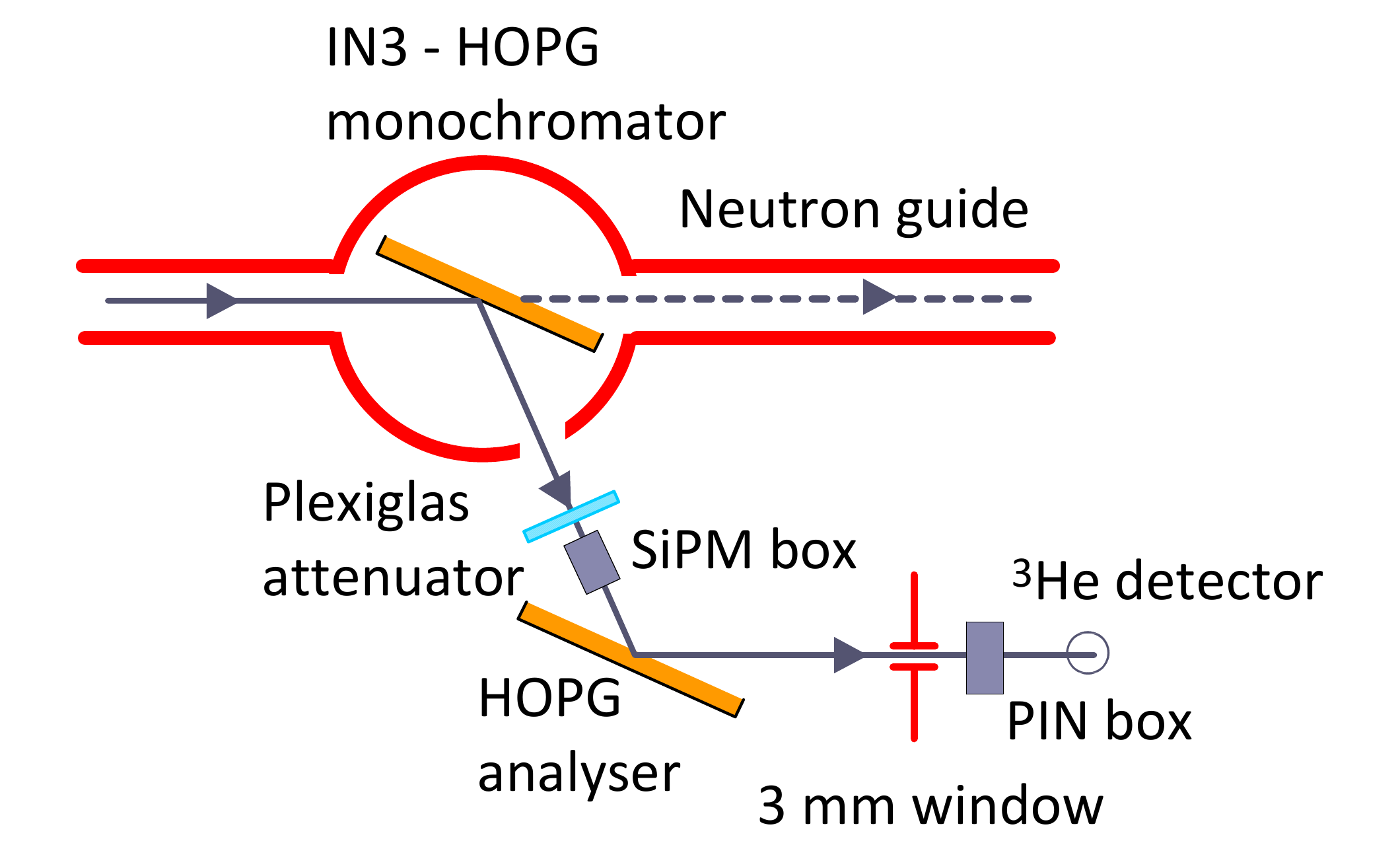}
\caption{\footnotesize Schematics of the neutron experiment set-up on 3-axis spectrometer IN3 at ILL for the test measurements on the Si PIN diode and the SiPM. The Si PIN and the SiPM were mounted not simultaneously in the sketched positions. The 50 mm diameter $^3$He instrument detector was used to measure the beam flux. Figure from~\cite{Mauri_psa}.}
\label{fig1sigd}
\end{figure}

The same method as checked in the simulation was applied to the data collected in neutron test measurements using two different devices, namely a Silicon PIN diode, described above, and a SiPM. The data were collected on the three-axis spectrometer IN3, at the ILL~\cite{ILL}, operated at 2.4 {\AA} neutron wavelength. The test configuration is sketched in figure~\ref{fig1sigd}, where both PIN and SiPM configurations are outlined. The two test experiments were not performed at the same time, because the SiPM was tested using the higher monochromatic neutron flux available just after the instrument monochromator, and the beam was attenuated by a 8 mm thick plexiglas slab (attenuation factor $\sim 10^2$), while the PIN was tested using the instrument analyser crystal to install the detector in front of the instrument $^3$He detector for a direct determination of the beam flux, which was defined by a 3 mm hole. The experimental results for both PIN and SiPM were analysed using the same procedure.

\begin{figure}[htbp]
\centering
\includegraphics[width=1\textwidth,keepaspectratio]{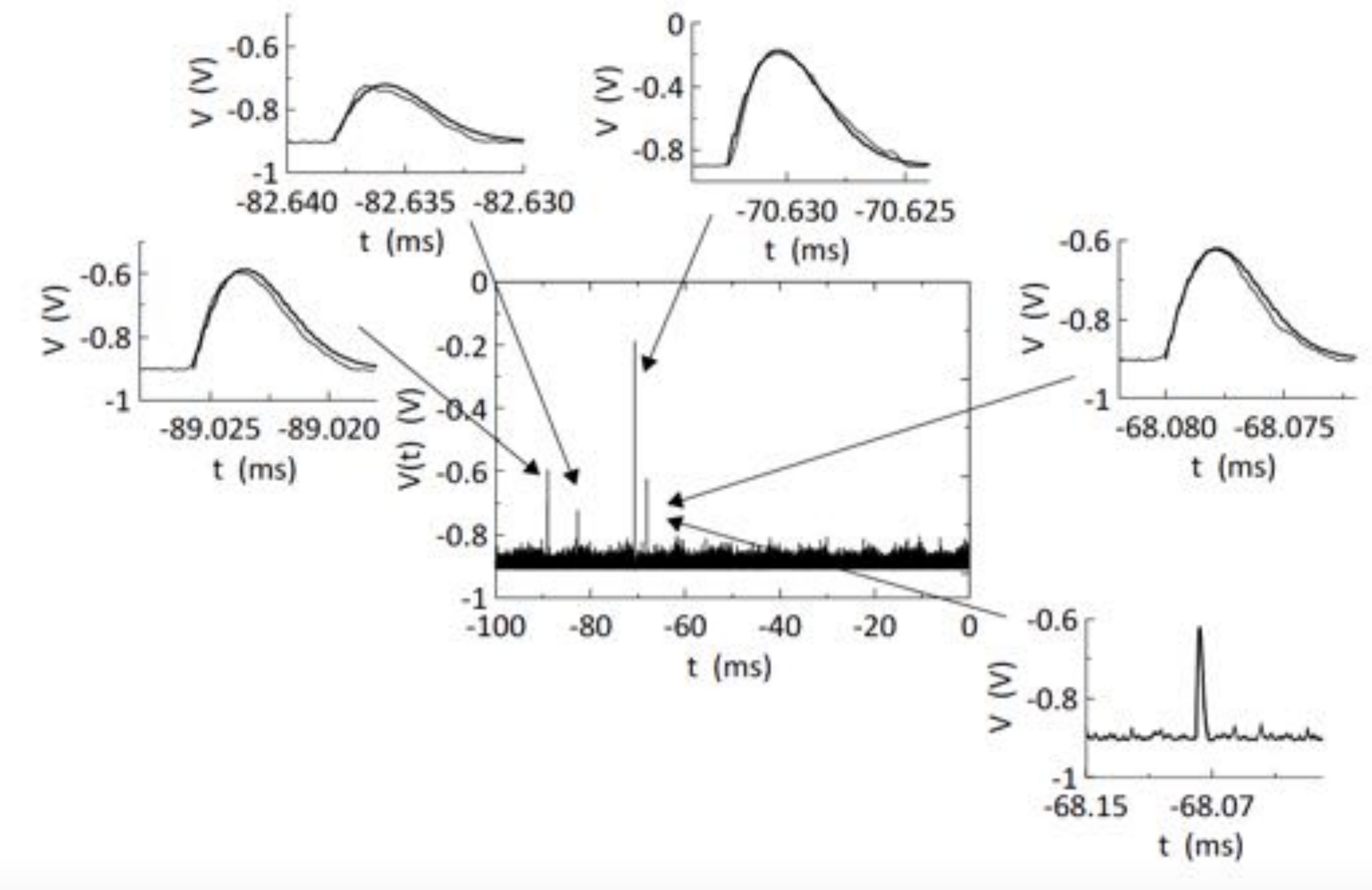}
\caption{\footnotesize Typical experimental trace collected by a Si PIN diode coupled to $^{157}$Gd$_2$O$_3$ converter. The signal contains four peaks identified as neutrons clearly visible above random noise. The four neutron peaks present in the spectrum are shown on the expanded plots (thin line) in comparison with the model fitting functions (thick line).}
\label{fig8sigd}
\end{figure}

The efficiency and space resolution of a single Si sensor coupled to a Gd converter, either natural or isotopically enriched by $^{157}$Gd, were obtained by a Monte Carlo~\cite{PETRILLO1999} simulation of the electron energy loss inside both the Gd converter and the Si sensor. Based on these simulations a neutron detection efficiency of $\sim$30\% is expected for the present PIN detector, described in section~\ref{secpinsigd}. The value is confirmed by the measurements and the detailed description is reported in~\cite{mauri_tesi}. 
\\ Improvements are possible especially considering that the converter deposition technique is not optimised yet. Indeed, thickness uniformity of the converter layer is a primary request for the detector to work properly and a non-uniform deposition layer can cause a decrease in neutron detection efficiency, a study has been performed during the PhD and it is presented in section~\ref{secfutsigd}. 
\\ Figure~\ref{fig8sigd} shows the neutron spectra versus time as measured by the Si PIN diode coupled to  $^{157}$Gd$_2$O$_3$ converter layer. The pulse shape analysis was carried out offline using $N$ = 200, a rather high value necessary to sample most part of the neutron signal in both PIN diode and SiPM cases. Increasing the sampling time, a much smaller value of $N$ can be efficiently used. Actually, $N$ of the order of 10 is a value useful for real time applications.

\begin{figure}[htbp]
\centering
\includegraphics[width=.6\textwidth,keepaspectratio]{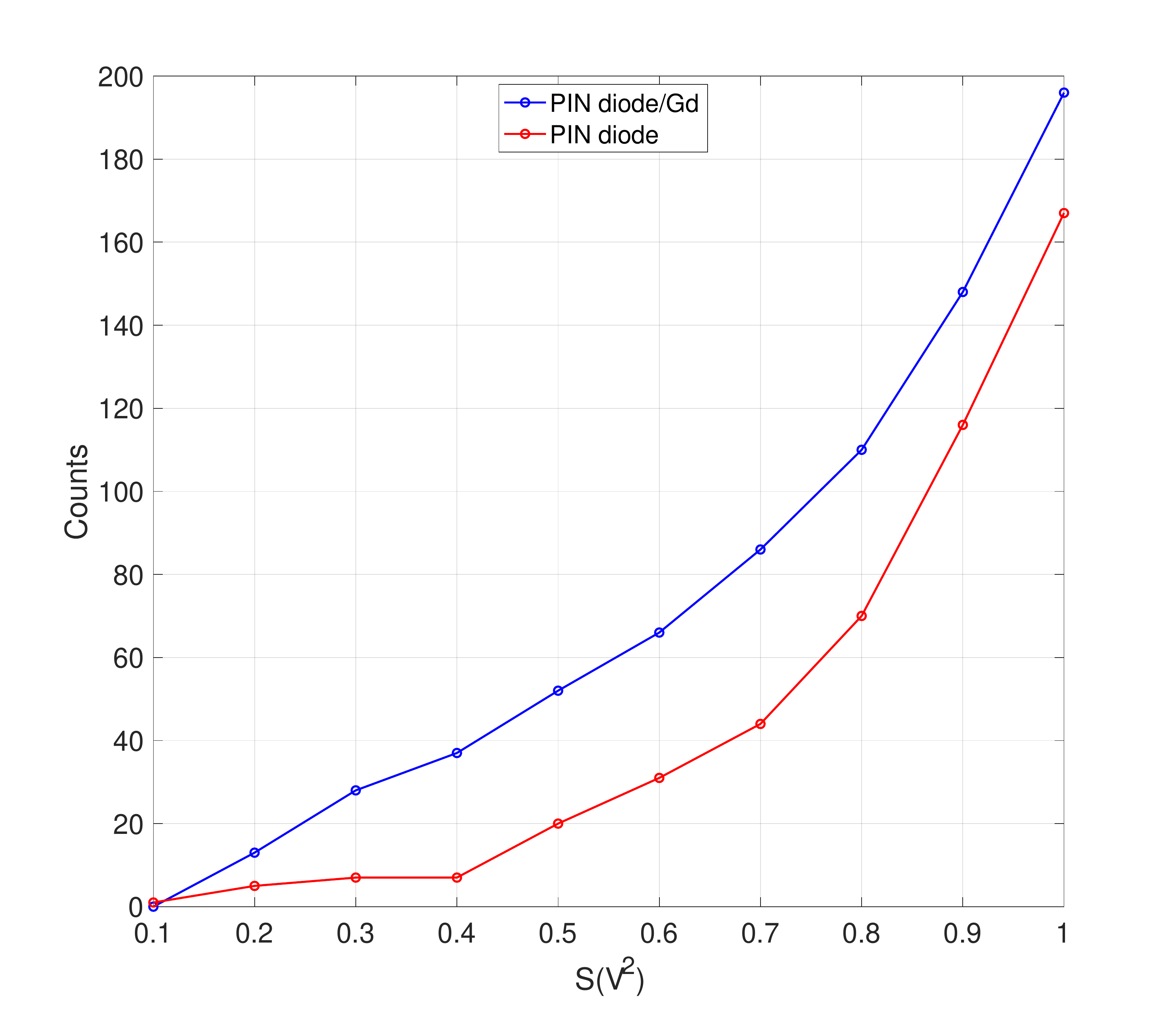}
\caption{\footnotesize Number of counts analysed with the PIN diode coupled to the $^{157}$Gd$_2$O$_3$ (blu line) and when only the PIN diode is used (red line), as a function of $S= 0.1-1$.}
\label{rap_Scounts}
\end{figure}

It can be observed that the findings of the experiment on the Si PIN, which also aimed at testing the efficiency of the isotopic converter~\cite{mauri_tesi}, are encouraging although in this case, because of the rather slow front-end electronics, the pulse shape analysis does not provide a result much better than a simple threshold discriminator. 
\\ Based on the computer simulation the same value for $S$ has been adopted and for the thresholds level. The study on the measurements is reported here. In figure~\ref{rap_Scounts} is shown the analysed counts as a function of different $S$ values, from 0.1 to 1, compared between the detector configuration with the PIN diode coupled to the $^{157}$Gd$_2$O$_3$ converter layer (PIN diode/Gd, blue line) and when the converter is removed (PIN diode, red line), which is the background configuration. 

The choice of $S=0.6$ is a reasonable compromise between the amount of counts collected when the Gd neutron converter is employed and the background configuration, i.e., when it is removed from the PIN surface. The value is in agreement with the computer simulations, as described in subsection~\ref{secpcsim}.

\begin{figure}[htbp]
\centering
\includegraphics[width=.7\textwidth,keepaspectratio]{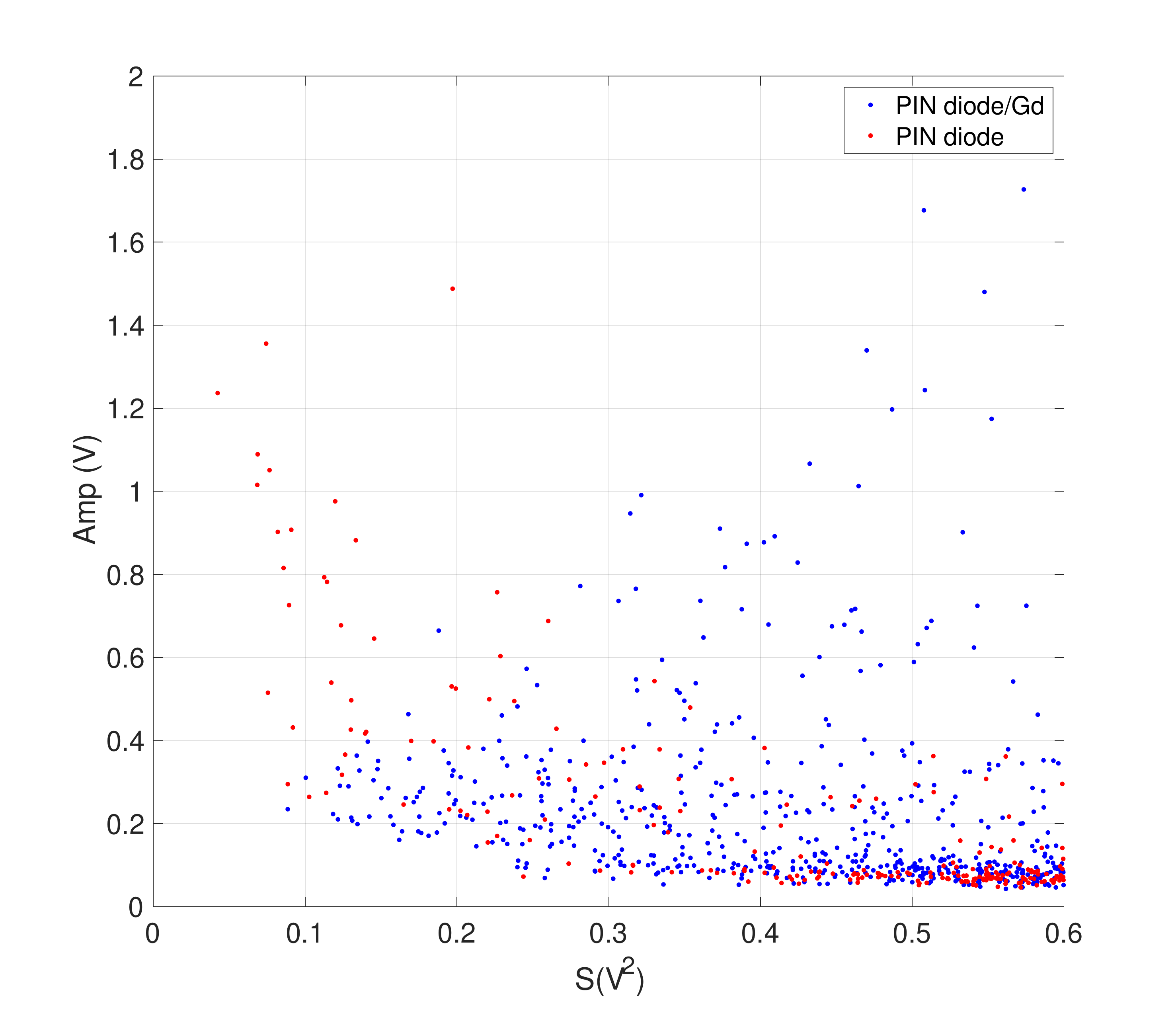}
\caption{\footnotesize Amplitude distribution of the analysed data with the PIN diode coupled to the $^{157}$Gd$_2$O$_3$ (blue dots) and when only the PIN diode is used (red dots), as a function of $S$.}
\label{S_counts}
\end{figure}

\begin{figure}[htbp]
\centering
\includegraphics[width=.85\textwidth,keepaspectratio]{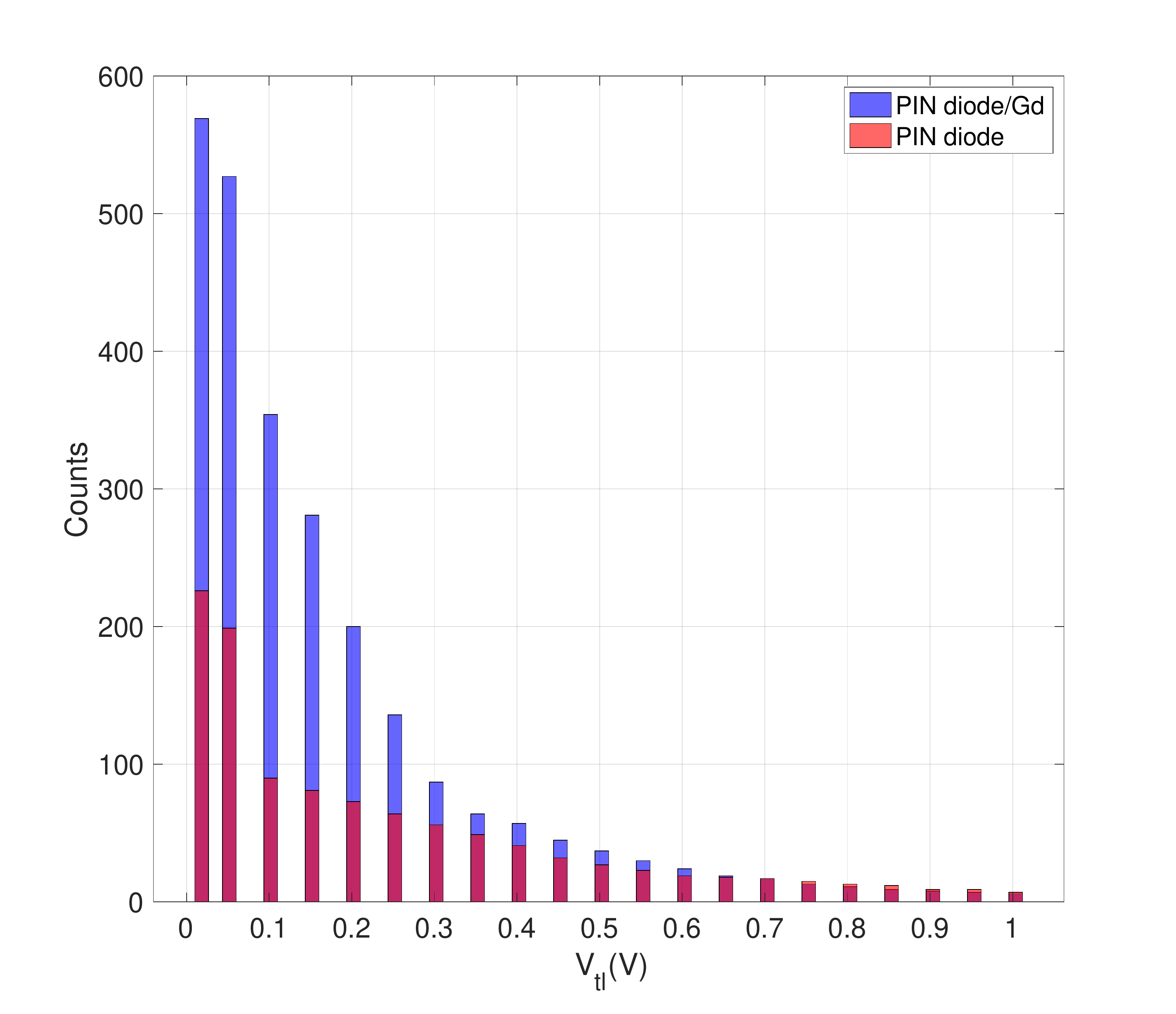}
\caption{\footnotesize Histogram of the low threshold, $V_{tl}$, values from 0.01 V to 1 V in step of 0.05 V. In blue the PIN diode/Gd configuration, in red the sole silicon PIN configuration}
\label{Th_counts}
\end{figure}

The amplitude distribution, as a function of $S$, is shown in figure~\ref{S_counts} for both cases, PIN diode/Gd and PIN diode without converter. From this is possible to identify a value for the low threshold, $V_{tl}$ around 0.1 V, in order to maximise the rejection of the background events without overly decrease the neutron counts in the analysis. To obtain a quantitative result, a series of analysis were performed, by varying the low threshold value from 0.01 V to 1 V in step of 0.05 V. The histogram is depicted in figure~\ref{Th_counts}: in blue the PIN diode/Gd configuration, in red when the sole silicon PIN is used. 

For $V_{tl} = 0.01$ V the total counts analysed in the PIN diode/Gd set-up is 569 while the only PIN diode counts 226 events, so the signal-to-background ratio is about 2.5. If is taken $V_{tl} = 0.1$ V, the neutron counts are 354, and the background counts are 90. Although a lot of neutron signal are discarded with respect to the previous case, the signal-to-background ratio = 3.93 is highly improved.
\\ The pulse shape analysis procedure described in section~\ref{secpsa} and tested via computer simulations, was applied also to the measurements. As shown above, a good agreement between the simulation and the measurements has been verified. Therefore, the same analysis method has been applied to the SiPM.

\begin{figure}[htbp]
\centering
\includegraphics[width=0.7\textwidth,keepaspectratio]{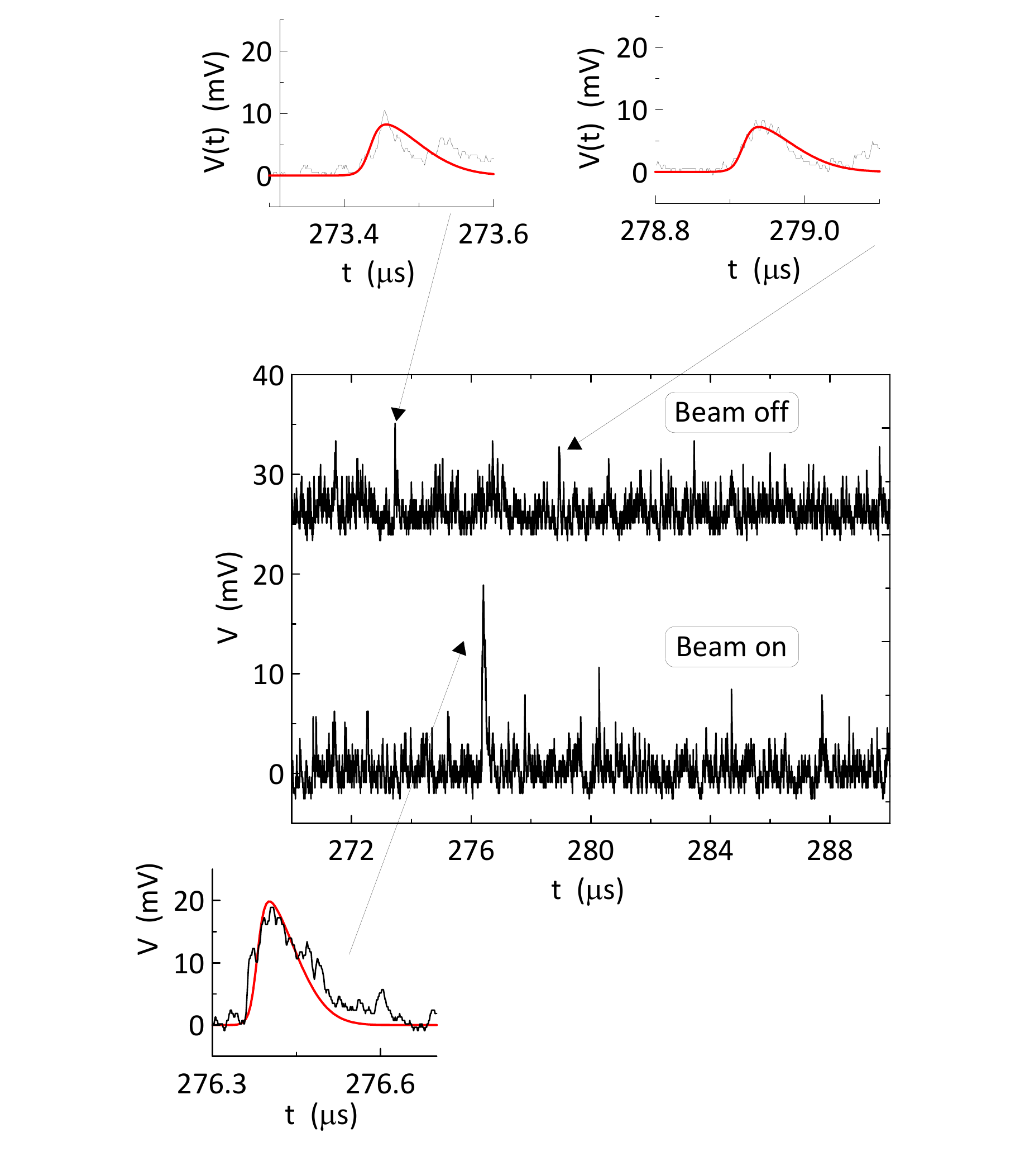}
\caption{\footnotesize Typical experimental trace collected from the SiPM coupled to a GS20 scintillator with beam on (bottom black line) and beam off (top black line). A fraction of the total acquisition duration is shown, to enhance the presence of a peak that can attributed to the neutron signal. The enlarged plot on the bottom shows the peak attributed to the neutron in comparison with the fitted peak (red thick line). Two peaks of the beam-off spectrum are also shown as compared to the reference signal. Figure from~\cite{Mauri_psa}.}
\label{fig9sigd}
\end{figure}

The SiPMs are quite fast devices adequate for applications over the ns time scale~\cite{tsipm}. The present device, produced at FBK (Italy)~\cite{sipmfbk}, was 3 x 3 mm$^2$ in size, 50 x 50 $\mu$m$^2$ cell size, coupled to a 1.5 mm thick GS20 lithium glass scintillator with the same lateral size as the SiPM. Coupling of the GS20 scintillator to the SiPM, where wire bonding is necessary for collecting the output signal, was produced by means of a 3 mm thick Plexiglas block that was used as a light guide. A teflon tape, wrapped around the scintillator and the Plexiglas block, worked as a light reflector.
\\ The SiPM had no front-end electronics as the signal was provided by the voltage on a couple of 50 $\Omega$ resistors mounted at the start and at the end of the coaxial cable that transmits the signal to the acquisition system 15 m away. The output of the SiPM (breakdown voltage $V_{BD}$ 28.5 V) was digitized using the same device as the Si PIN diode (pico scope 2000, 100 MHz bandwidth), but applying a sampling time $T = $1 ns, instead of 100 ns. Application of this digital acquisition system is convenient for the testing phase, because the sampled neutron signal can be stored for offline analysis that simulates different form of real time acquisition using the same data. 
\\ The SiPM is a rather new device for application as a neutron detector~\cite{SiPM-LiF} and a much more demanding case in terms of signal analysis because of the very high noise (close to 1 MHz of single photon excitations at room temperature) and the rather low light output of the GS20 scintillator. We remind once more that neutron detectors must be self-triggered and simple filtering techniques are hardly effective in reducing the noise. Therefore, in the case of a SiPM, a preliminary pulse shape analysis on simulated data was instrumental to understand the effectiveness of such a device for real time neutron detection applications. 
\\ The experimental configuration of the SiPM on the neutron beamline (IN3) was carefully studied since placing the detector on the primary beam, although providing an intensity high enough for all the test measurements, introduces a higher background as well. The detector was shielded by a 5 mm thick elastobore, which has a nominal attenuation by a factor 10$^3$ at thermal wavelength. The expected intensity impinging the detector was about 10$^7$ n/cm$^2$ s, corresponding to about 10$^6$ n/s on the detector area. As this rate is too high, the incoming beam was also attenuated by a plexiglas plate, 8 mm thick, which gives an intensity of the order of 4 $\cdot 10^4$ n/s on the SiPM detector. This intensity is appropriate for testing this detector. Several tests were carried out by changing the SiPM overvoltage (from 4 V to 8 V, typically 5.5 V), and the position of the detector in the beam to evaluate the effect of the non-uniform beam intensity distribution, a feature that cannot be ignored when testing a detector as small as this one. The present SiPM has an intrinsic response time with 50 $\Omega$ termination, shorter than the rise time of the acquisition system that is 3.5 ns. The decay time of the signal arising from the neutron capture event is dominated by the response of the scintillator and a typical fit to the data provides a decay time equal to 85 $\pm$ 5 ns, which is compatible with the GS20 characteristics (average decay time of about 70 ns)~\cite{Spowart-gs20}.
\\ The experimental results on the SiPM were analysed following the same procedure as the PIN diode, however, because of the relatively high environmental background, for the SiPM, and the intrinsic background sources, the shape analysis was carefully performed.
\\ A typical signal at the SiPM load resistor is shown in figure~\ref{fig9sigd}. The effect of noise can be noted as well as the appearance of some peaks. A neutron event can be attributed to peaks exceeding 12-13 mV, as, indeed, they are quite rare when the neutron beam is switched off (top black line in figure~\ref{fig9sigd}). From several similar acquisitions, all lasting 1 ms, some of the highest peaks have been selected. It was then possible to obtain reasonably good fits of the peaks attributed to neutrons, using the same simple function of equation~\ref{eqpeak} as employed for the PIN diode. 

\begin{figure}[htbp]
\centering
\includegraphics[width=0.8\textwidth,keepaspectratio]{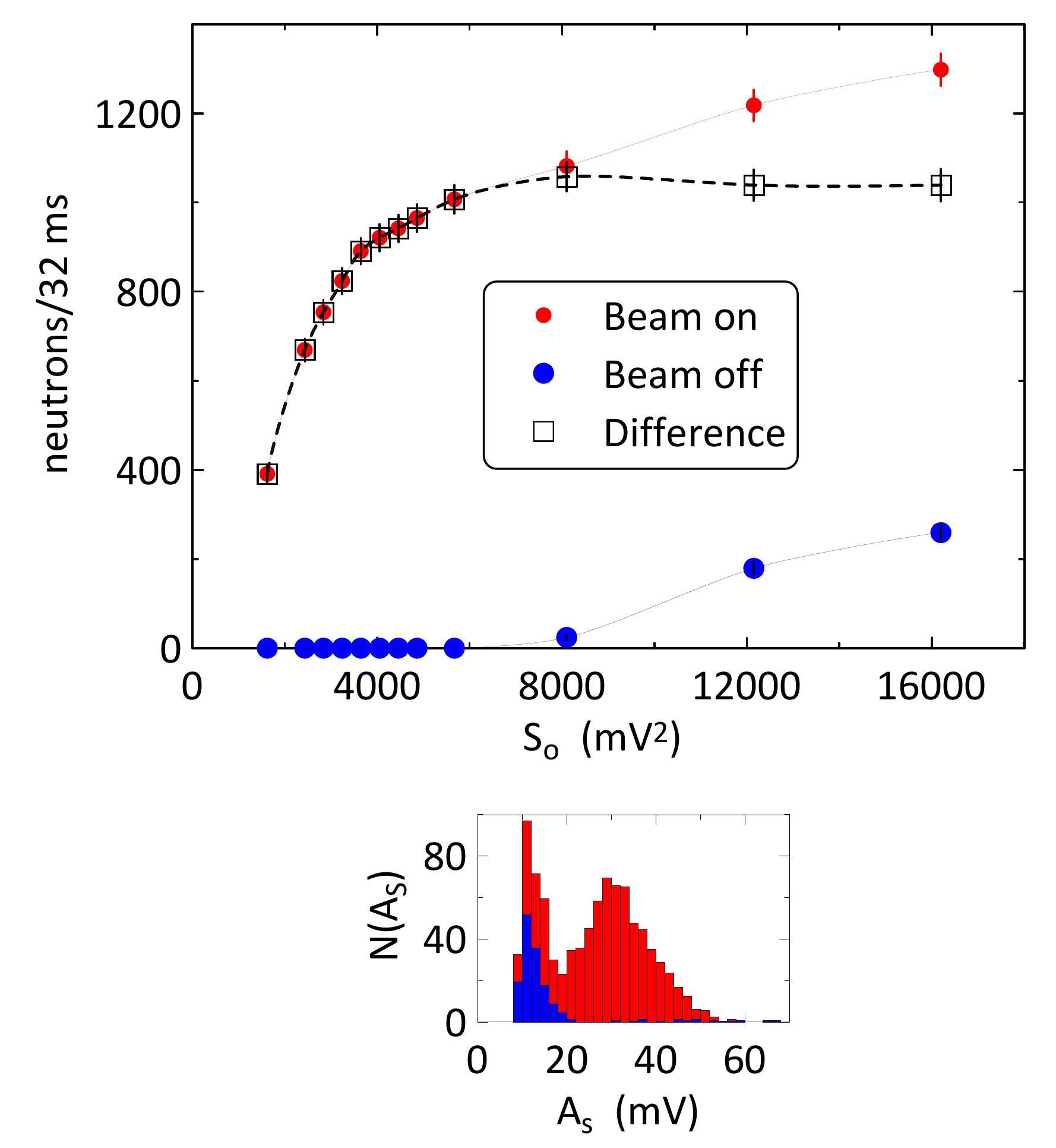}
\caption{\footnotesize Results of the full pulse shape analysis. Number of neutron signals detected in 32 ms versus  $S_o$. Red dots: beam-on data. Blue dots: beam-off data. Black empty squares: difference between beam-on and beam-off data. The lines are guide to the eye. On the bottom side of the figure, the histogram of beam-on (red) and beam-off (blue) events is shown versus the amplitude $A_s$. Figure from~\cite{Mauri_psa}.}
\label{fig10sigd}
\end{figure}

The procedure to analyse the data worked smoothly and the data were analysed by applying a low- and a high-level threshold and by varying $S_o$, as in the case of the Si PIN diode. After various trials, the low-level threshold is set at 9 mV and at 40 mV the high-level threshold. Finally, a set of 32 acquisitions, lasting 1 ms each, was stored with both \lq beam open\rq \,and \lq beam closed\rq \,operation. Both sets of data were analysed applying the same procedure, to isolate the contributions coming from noise and environmental background. The results are shown in figure~\ref{fig10sigd}, where are plotted the number of peaks identified as neutrons versus $S_o$. It is expected that a too low $S_o$, while rejecting all the background, produces also a rejection of neutron counts. By subtracting the beam-off background from the beam-on data, it was easy to identify the best choice of $S_o$ to get almost no background counts and about $3\cdot 10^4$ n/s with the beam on. 
\\ One observes that the data in figure~\ref{fig10sigd} resemble a standard counter plateau when the discriminator threshold is varied. In the present analysis $S_o$ has the same role as the threshold even though more information is carried by this quantity, as it measures the peak shape in comparison to a reference one.
\\Figure~\ref{fig10sigd} shows that the choice $S_o$ ranging between 7200 mV$^2$ and 8000 mV$^2$ gives a neutron flux corresponding to the plateau after subtraction of the background obtained with the beam-off condition (instrument local shutter closed). On increasing $S_o$ the count rate increases and the increase is completely due to the beam-off contribution that produces peaks characterized by a definitively lower intensity. This behaviour is pointed out by the histogram shown on the bottom of figure~\ref{fig10sigd}.
\\The above intensity (about 3.3 10$^5$ n/cm$^2$s) is consistent with the expected intensity, but there is no means to check it on an absolute scale. One can only observe that the scintillator has a nominal efficiency close to 100\% and the experimental results suggest a light selection efficiency also close to 100\%. $\gamma$-rays produce a smaller signal than neutrons in the GS20 scintillator, therefore an analysis of the peak amplitudes, as measured by $A_s$, could be used to minimize this background.

To state that a SiPM based system is adequate as a neutron detector on a real neutron instrument, more tests are necessary. Also, every potential use of this prototype on a larger scale requires new effort in designing and testing proper readout electronics.

\section{Design improvements}\label{secfutsigd}

As mentioned in the main Introduction, the project on the Silicon-based neutron detectors is under development at the Department of Physics in Perugia. In the past few years, several studies have been performed and different designs have been implemented and tested~\cite{PETRILLO-solidstate,PETRILLO1999,CASININI2012}. Nevertheless, various implementations can be realized before employing these devices on a real instrument. For this reason, during the PhD, part of the work has been also dedicated to investigate some design implementations. In particular, focusing the attention on some alternatives for the deposition technique of the Gadolinium on a silicon substrate.
\\ Remind that what has been performed was a feasibility investigation of the devices. The aim of the work was to understand if is possible to grown better, i.e., more uniform, samples compared with the one was used for the measurements with the Si PIN diode presented in section~\ref{secmeassigd}, which affect significantly the efficiency of the device, and with a low-cost effectiveness solution. The goal is to customize the production of Gd samples for the testing phase of the detector, in order to have Gd converters available and at a relative low price. 
\\ Several studies on deposition techniques empoying the Gadolinium have been performed; some literature references are reported in~\cite{cloruroGd,GD-ALDdep,GD-Depfilm,GD-Depion,Carina_depGd}.

\subsection{Gadolinium Thermal Evaporation feasibility tests}

Many deposition techniques have been developed to grow thin film in a controlled way. A widely used method is the sputter deposition, which is a Physical Vapour Deposition (PVD) by ejecting material from a target, due to the bombardment of the target by energetic particles, onto a substrate. Several type of deposition exploit the sputtering, such as ion-beam sputtering, reactive sputtering and magretron sputtering. 
\\ Another technique is the sol-gel deposition, where a sol is a dispersion of a solid particles in a liquid and a gel is a state where both liquid and solid are dispersed in each other, presenting a solid network with liquid components. This process usually consists of four step: a colloidal solution is formed and it is coated on a substrate by spraying, dipping or spinning; the particles in sol are then polymerised and produce a gel in a state of a continuous network, finally the gel is heated to separate the two phases and form an amorphous or crystalline coating~\cite{SOL-GEL}.

\begin{figure}[htbp]
\centering
\includegraphics[width=0.7\textwidth,keepaspectratio]{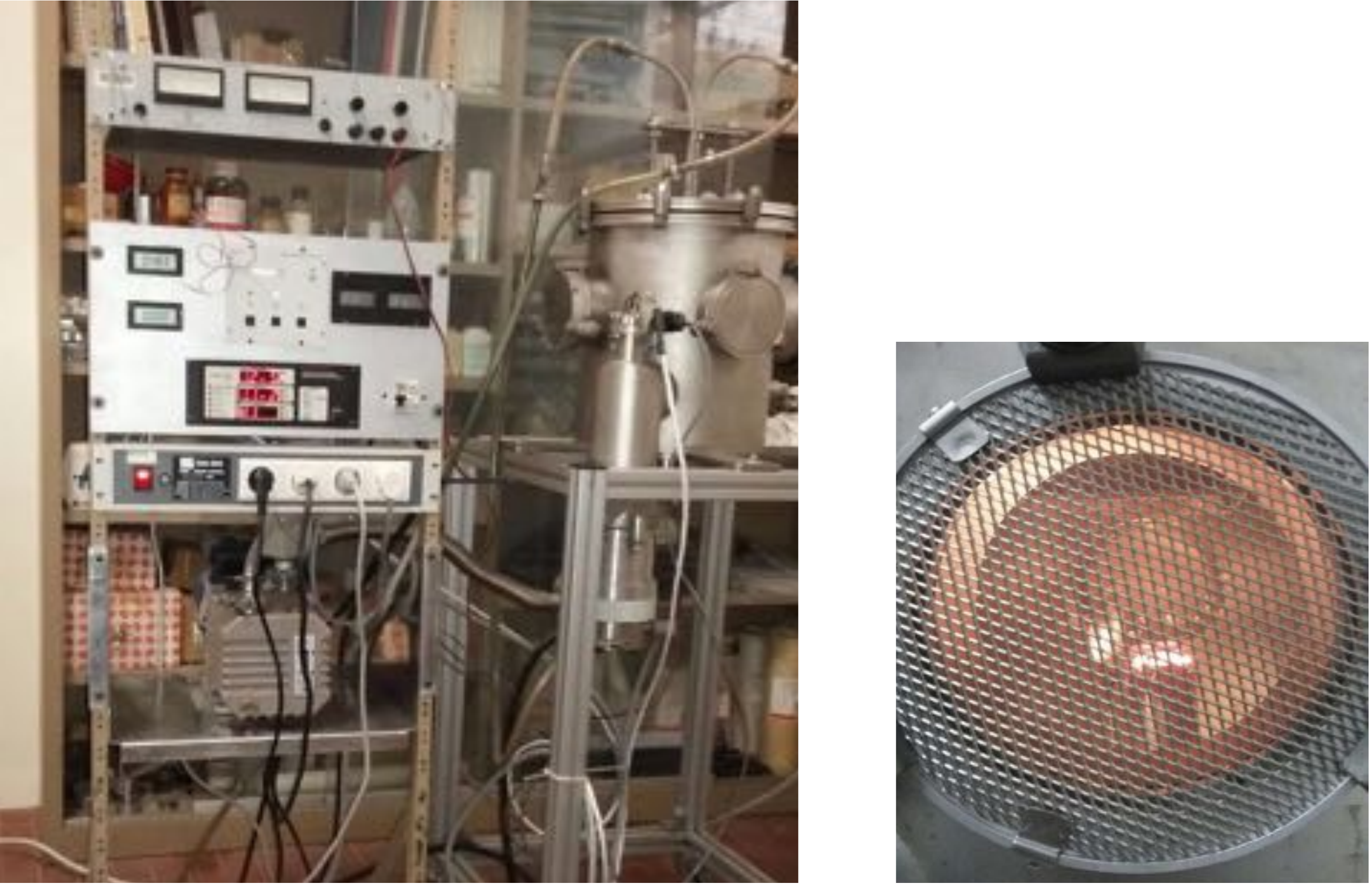}
\caption{\label{depsetup}\footnotesize Thermal Evaporation deposition set-up. Left: the vacuum chamber, the vacuum pumps and the power supply system. Right: picture of the crucible inside the chamber when heated.}
\end{figure}

One of the common methods of physical vapour deposition is the Thermal Evaporation, which is a method of thin-film deposition of pure materials to the surface of various objects. The coatings are usually in the thickness range of angstroms to microns. The thermal evaporation involves heating a solid material inside a high vacuum chamber, the temperature must be high enough to produce some vapour pressure, which is related to the evaporation rate of a liquid or a solid. Inside the vacuum, indeed, even a relatively low vapour pressure is sufficient to raise a vapour cloud inside the chamber. This evaporated material stream crosses the chamber and hits the substrate, sticking to it as a coating or film~\cite{Sze-semiconductor}.
\\ The equipment for an evaporation system includes basically a vacuum chamber and a vacuum pump, due to the simplicity of the mechanism and the availability of the tools, an experiment in our laboratory has been set. In figure~\ref{depsetup} is shown the experimental set-up; the vacuum chamber on the right, two vacuum chambers, a rotary vane pump and a turbomolacular pump, to obtain and maintain high vacuum, and a power supply, which provides the energy to heat the crucible located in the bottom of the chamber, where the source material is placed, a picture is shown on the right side in figure~\ref{depsetup}. The heating process is obtained through a simple electrical resistive heat element, in this case a filament, which allows to use low voltage (max 30 V), but with a quite high current values, up to 10 A in our set-up.

\begin{figure}[htbp]
\centering
\includegraphics[width=0.55\textwidth,keepaspectratio]{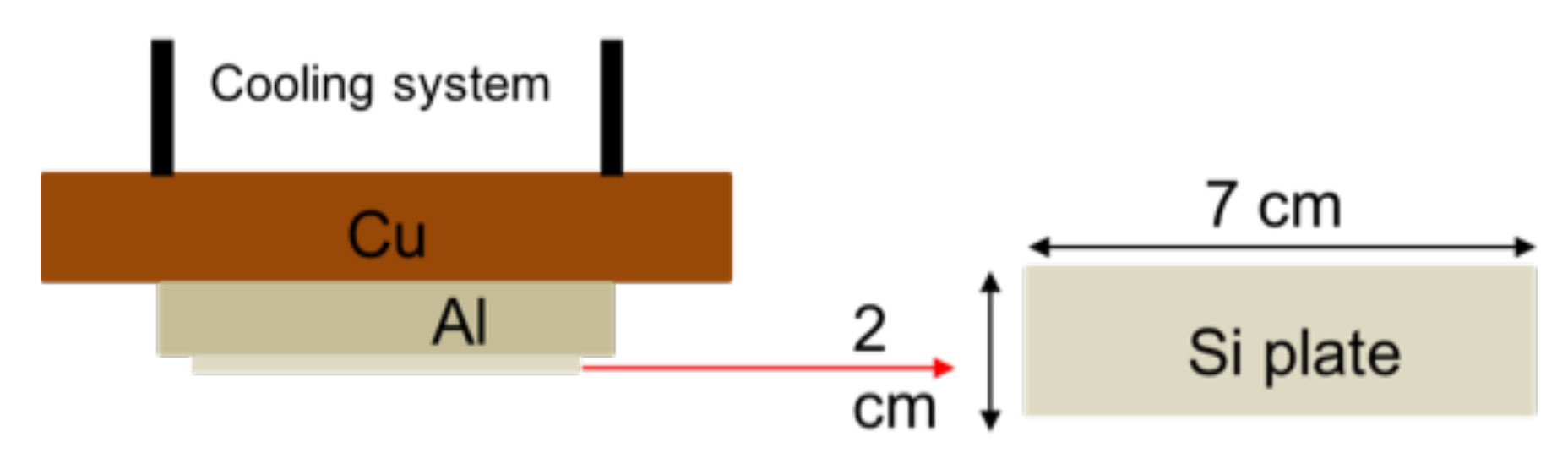}
\caption{\label{depsketch}\footnotesize Sketch of the substrate fixed at the top of the chamber with its relative cooling system, in order to keep the substrate at room temperature, preventing it from melting.}
\end{figure}

Since the material is located in the bottom of the chamber, the vapour rises above and the substrate is held inverted in appropriate fixtures at the top of the chamber. The surface to be coated is facing down toward the heated source material to receive the coating, a sketch is depicted in figure~\ref{depsketch}. The cooling system is needed to prevent the substrate from melting and to allow vapour to condense on it, therefore it is kept at room temperature. A series of thermocouple are used to monitor the different temperatures.

\begin{figure}[htbp]
\centering
\includegraphics[width=0.55\textwidth,keepaspectratio]{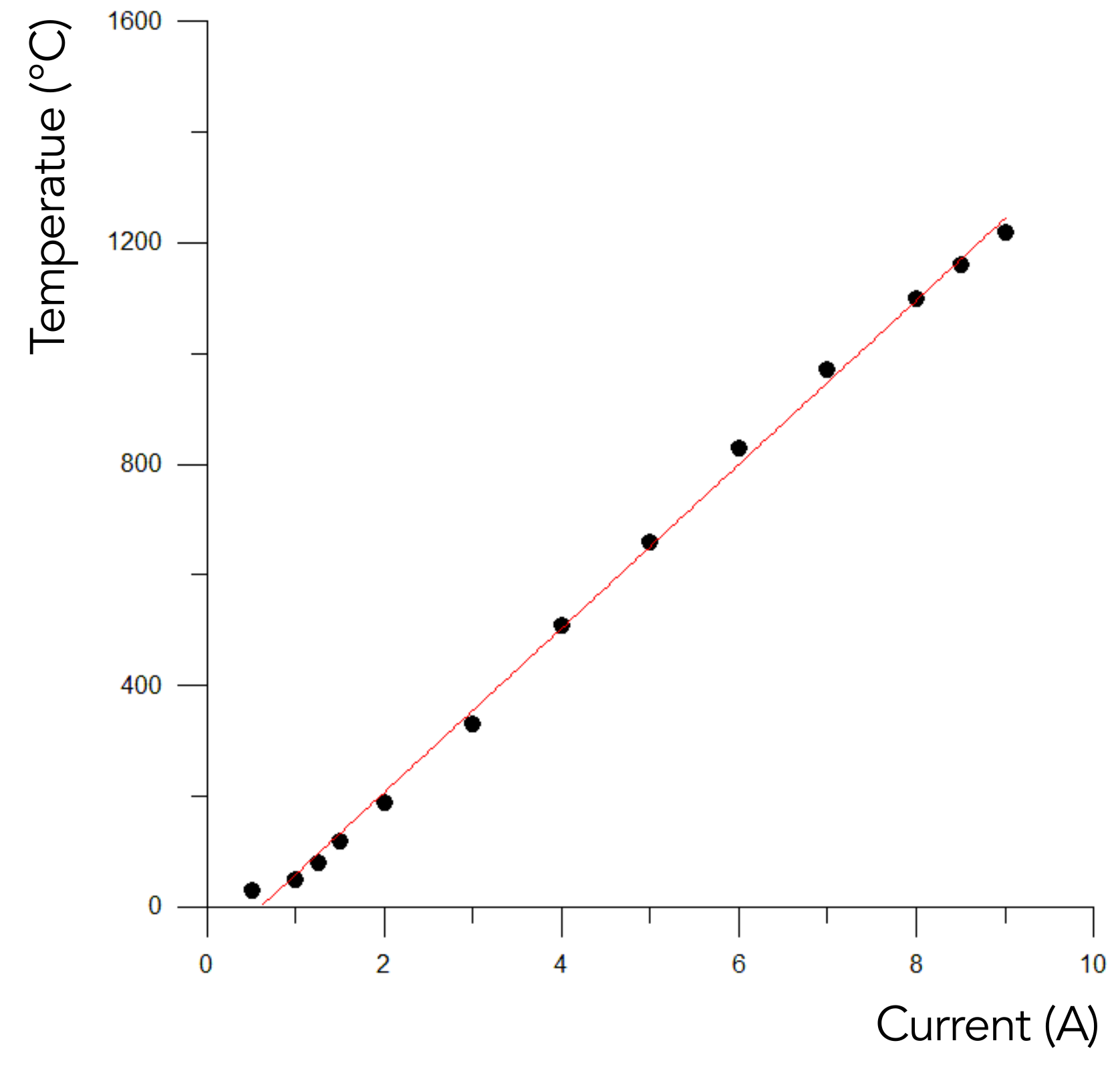}
\caption{\label{CT-prop}\footnotesize Relation between the temperature of the crucible is $^{\circ}$C and the current given by the power supply in A. A linear relation is observed.}
\end{figure}

With this set-up the crucible can reach a temperature of 1250$^{\circ}$C and it is directly proportional to the current provided by the power supply, as shown in figure~\ref{CT-prop}. 

\begin{figure}[htbp]
\centering
\includegraphics[width=1\textwidth,keepaspectratio]{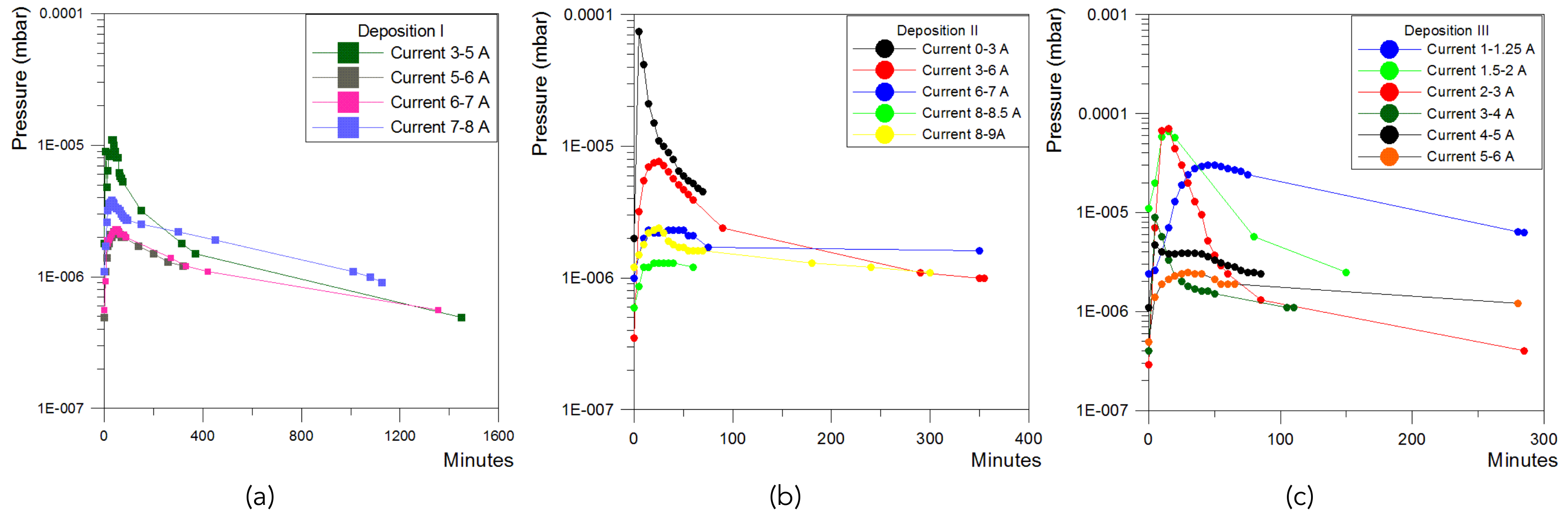}
\caption{\label{presstrend}\footnotesize Pressure trend for the different step of heating the crucible. The current and the heating temperature are linearly proportional. After each increasing, the pressure takes about one hour to stabilize again at the initial value before the increase, usually it was changed around $10^{-6}\,$mbar. (a), (b) and (c) refers to the first and second test with the Gd and the test with the GdCl$_3$, respectively}
\end{figure}

Three measurements have been performed, two of which using the natural gadolinium (Deposition\,I and II), because it is less expensive than the isotopic gadolinium, and one with a gadolinium trichloride as a source material (Deposition\,III). The melting point of natural gadolinium is $T_m \approx 1300^{\circ}$C , while $T_m \approx 600^{\circ}$C in the case of the GdCl$_3$. The vacuum chamber was at a pressure on the order of $10^{-6}\,$mbar for all the three deposition tests, the pressure trend is shown in figure~\ref{presstrend}.

In Deposition\,I a temperature of the crucible $T_c = 1100^{\circ}$C was reached with a current $i=8$A, in Deposition\,II $T_c = 1220^{\circ}$C ($i=9$A) while in Deposition\,III $T_c = 830^{\circ}$C ($i=6$A). Note that only in the case of GdCl$_3$ the temperature of the crucible is above the melting point of the material. Due to the feature of the set-up it is not possible to reach $T_c$ of the Gd, the crucible is at about 200$^{\circ}$C and 100$^{\circ}$C below the melting point in Deposition\,I and II respectively. 
\\ Although the GdCl$_3$ melting temperature is suitable for our experimental equipment, it is a hygroscopic material. Therefore, once in air the coated layer absorbs water and it separates from the substrate. On the other hand, even if was not reached the exact $T_c$ value with the Gd, some evaporation is still possible. In order to verify the coating the substrate were weighted, which is a silicon slab ($2 \times 7\,$cm$^2$) as sketched in figure~\ref{depsketch}, before ($M_{Si}$) and after the coating ($M_{SiGd}$). The thickness of the coated film ($Th$) can be easily calculated knowing the density of the material ($\rho$), the two weights and the area of the substrate $S$. The thickness is calculated by equation~\ref{dep-eq} below:

\begin{equation}
Th = \frac{\Big(\frac{M_{SiGd}-M_{Si}}{\rho}\Big)}{S} = \frac{V_{dep}}{S}
\label{dep-eq}
\end{equation}   
where $V_{dep}$ is the volume of the deposited material, $\rho=7.901\,$g/cm$^3$ in the case of Gd and $S=21\,$cm$^2$ is the area of the silicon slab. In table~\ref{dep-tab} is reported the calculated thickness in Deposition\,I and II.

\begin{table}[htbp]
\centering
\caption{\label{dep-tab} \footnotesize Thickness of the Gd coated layers on the Si substrate, calculated in the Deposition\,I and Deposition\,II measurements.}
\smallskip
\begin{tabular}{|l|c|c|c|c|}
\hline
& $M_{Si}$ (g)& $M_{SiGd}$ (g)& $V_{dep}$ (cm$^3$)& $Th$ (nm)\\
\hline
Dep.\,I & 1.5026$\pm 1 \cdot 10^{-4}$ & 1.5028$\pm 1 \cdot 10^{-4}$ & $0.25\cdot10^{-4} \pm 0.17 \cdot 10^{-4}$ & 12 $\pm 8.4$ \\
Dep.\,II & 1.5066$\pm1 \cdot 10^{-4}$ & 1.5077$\pm1 \cdot 10^{-4}$ & $1.39\cdot10^{-4}\pm 0.17 \cdot 10^{-4}$ & 66 $\pm 8.4$\\
\hline
\end{tabular}
\end{table}

The Deposition\,I lasted approximately 24 hours, while Deposition\,2 measurement run for about three days. Due to the increased temperature and the duration of the measurement, the second test shows a thickness about 6 times higher than the value calculated in the first experiment. To ensure that the deposition has occurred, an X-Ray Fluorescence (XRF) experiment was performed. This is one of the most common technique used to determine the materials in a sample. 
\\ The X-Ray fluorescence is the emission of secondary X-ray from a material, when its atoms are externally excited by high-energy short-wavelength radiation, such as X-rays. This may lead to the ejection of one or more electrons from an inner orbital, thus the atomic structure becomes unstable, leading to electrons from higher orbitals to transition into the lower orbital to fill the hole left by the ejected electron. These high-energy electrons must release energy to fill the lower energy vacancies, and this energy is emitted as photons with an energy/wavelength characteristic of the difference between the initial and the final energy state of the electrons in the two orbitals. The wavelength of emitted wave can be calculated using Plank's law:

\begin{equation}
\lambda = \frac{h \cdot c}{\Delta E}
\label{dep-plank}
\end{equation}

\begin{figure}[htbp]
\centering
\includegraphics[width=1\textwidth,keepaspectratio]{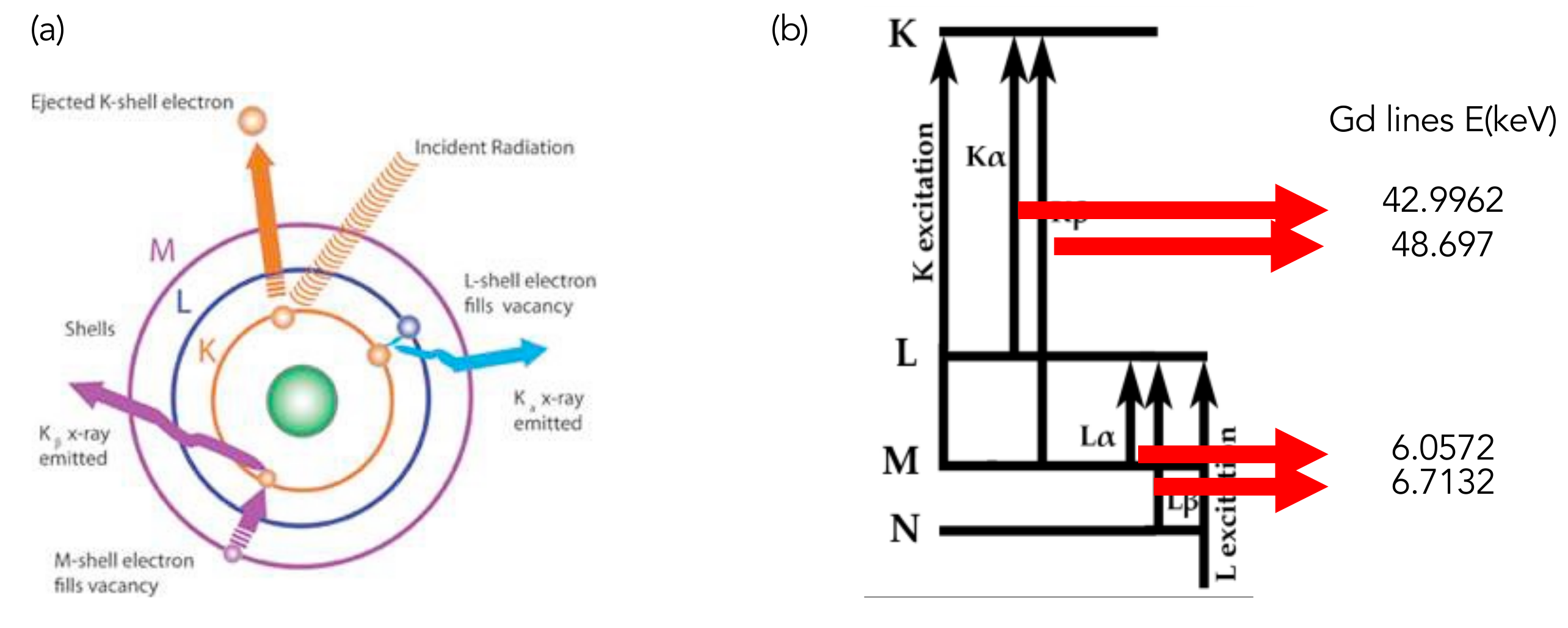}
\caption{\label{xraygd}\footnotesize (a) Schematic representation of the atomic model, showing the first three shells, and some of the possible transition between the energy levels. (b) Sketch of the spectral lines K$_\alpha$, K$_\beta$, L$_\alpha$ and L$_\beta$. For the transitions respectively: L$\rightarrow$K, M$\rightarrow$K, M$\rightarrow$L and N$\rightarrow$L. The corresponding values in the case of Gadolinium are also shown.}
\end{figure}

The electrons of an atom are distributed in energetic levels, shell, which are named corresponding to the quantum number $n =1,2,3, 4...$, respectively K, L, M, N, ..., a sketch is shown in figure~\ref{xraygd}(a). The K shell is the lowest energy electron shell and it is the closest to the nucleus. The substituting electron must fulfil the energy requirement of the K shell in order to maintain the atom electrical stable. 
\\ Different spectral lines can be defined, when an electron vacancy in the K shell is filled by an electron from the L shell, the characteristic energy/wavelength of the emitted photon is called the K-alpha (K$_\alpha$) spectral line. If the vacancy is filled by an electron from the M shell the spectral line is identified as K-beta (K$_\beta$). The transition of en electron of higher shell, M or N, in the L shell is defined as L-alpha (L$_\alpha$) and L-beta (L$_\beta$) respectively, a sketch of the spectral lines is illustrated in figure~\ref{xraygd}(b). Due to the substructure of the shells, the transitions can be also identified as $_{1,2,...}$ with respect to which energy level of a shell undergoes the transition.

\begin{figure}[htbp]
\centering
\includegraphics[width=0.6\textwidth,keepaspectratio]{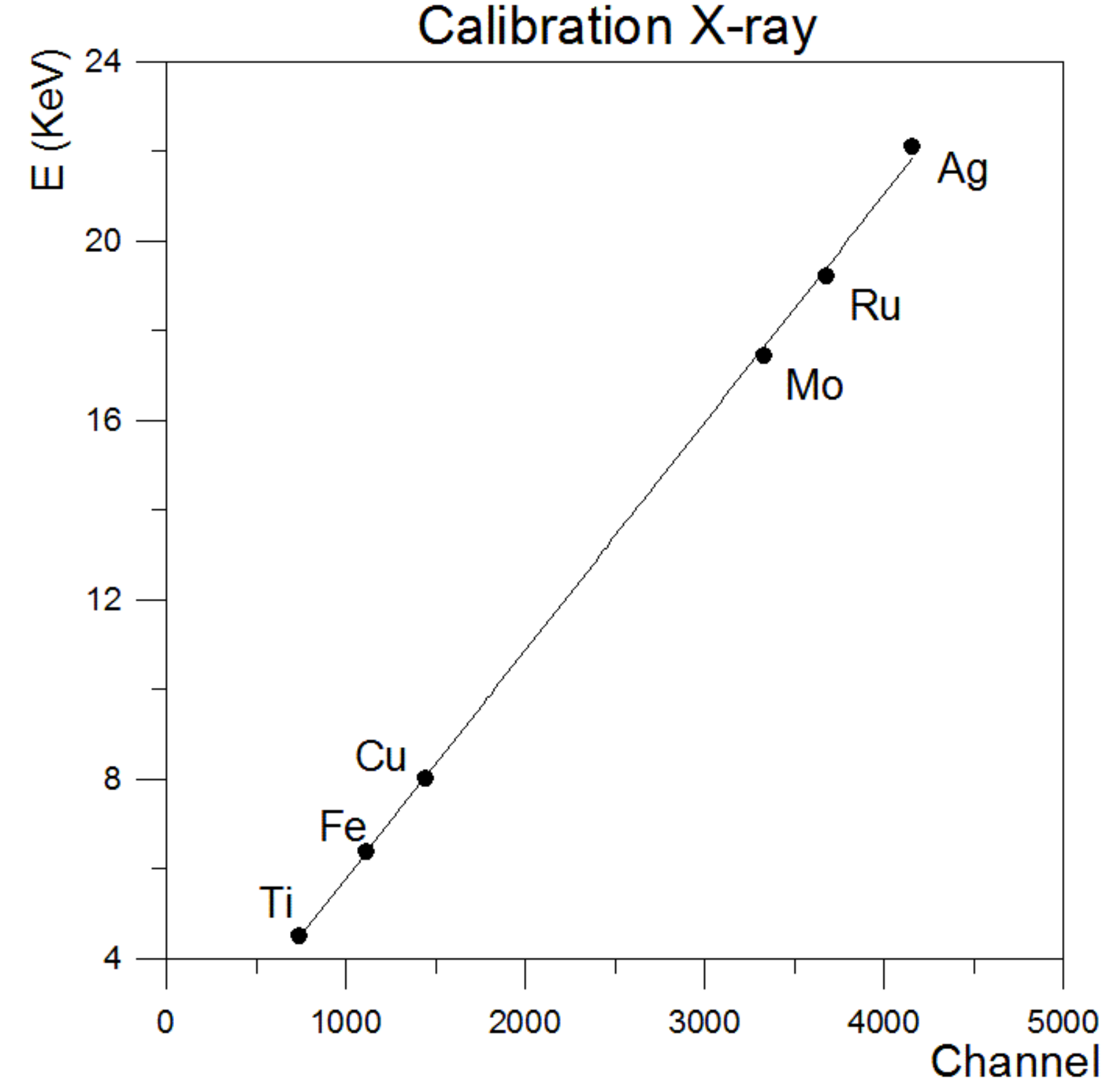}
\caption{\label{xraycalib}\footnotesize Spectrometer calibration using well know sample to relate the channel number to the energy in keV.}
\end{figure}

The experiment has been performed with the samples obtained after the gadolinium deposition and with the only silicon slab without the coating. The spectrometer has been calibrated using well know sample materials, thanks to the relation $\lambda (\AA) \simeq \frac{12.4}{E(keV)}$, the energy corresponding to the well known wavelength of the K$_\alpha$ transitions of the materials listed in table~\ref{tabxraycalib} has been calculated. Since the K$_{\alpha1}$ peak has always twice the intensity of the K$_{\alpha2}$, one can calculate the wavelength, $\lambda$, as the weighted average of the two. 
\\The relation channel-energy is shown in figure~\ref{xraycalib}. 

\begin{table}[htbp]
\centering
\caption{\label{tabxraycalib} \footnotesize Thickness of the Gd coated layers on the Si substrate, calculated in the Deposition\,I and Deposition\,II measurements.}
\smallskip
\begin{tabular}{|c|c|c|c|c|c|}
\hline
& K$_{\alpha1}$ & K$_{\alpha2}$ &$\lambda (\AA)$ & Ch.& E (keV)\\
\hline
Cu & 1.540 & 1.544 & 1.541 & 1444 & 8.042 \\
Fe & 1.935 & 1.939 & 1.937 & 1107 & 6.399 \\
Ti & 2.748 & 2.752 & 2.749 & 737 & 4.508 \\
Mo & 0.709 & 0.713 & 0.710 & 3324 & 17.445 \\
Ru & 0.643 & 0.647 & 0.645& 3676 & 19.219 \\
Ag & 0.559 & 0.563 & 0.560 & 4153 & 22.106 \\
\hline
\end{tabular}
\end{table}

\begin{figure}[htbp]
\centering
\subfloat[]{\includegraphics[width=.55\textwidth,keepaspectratio]{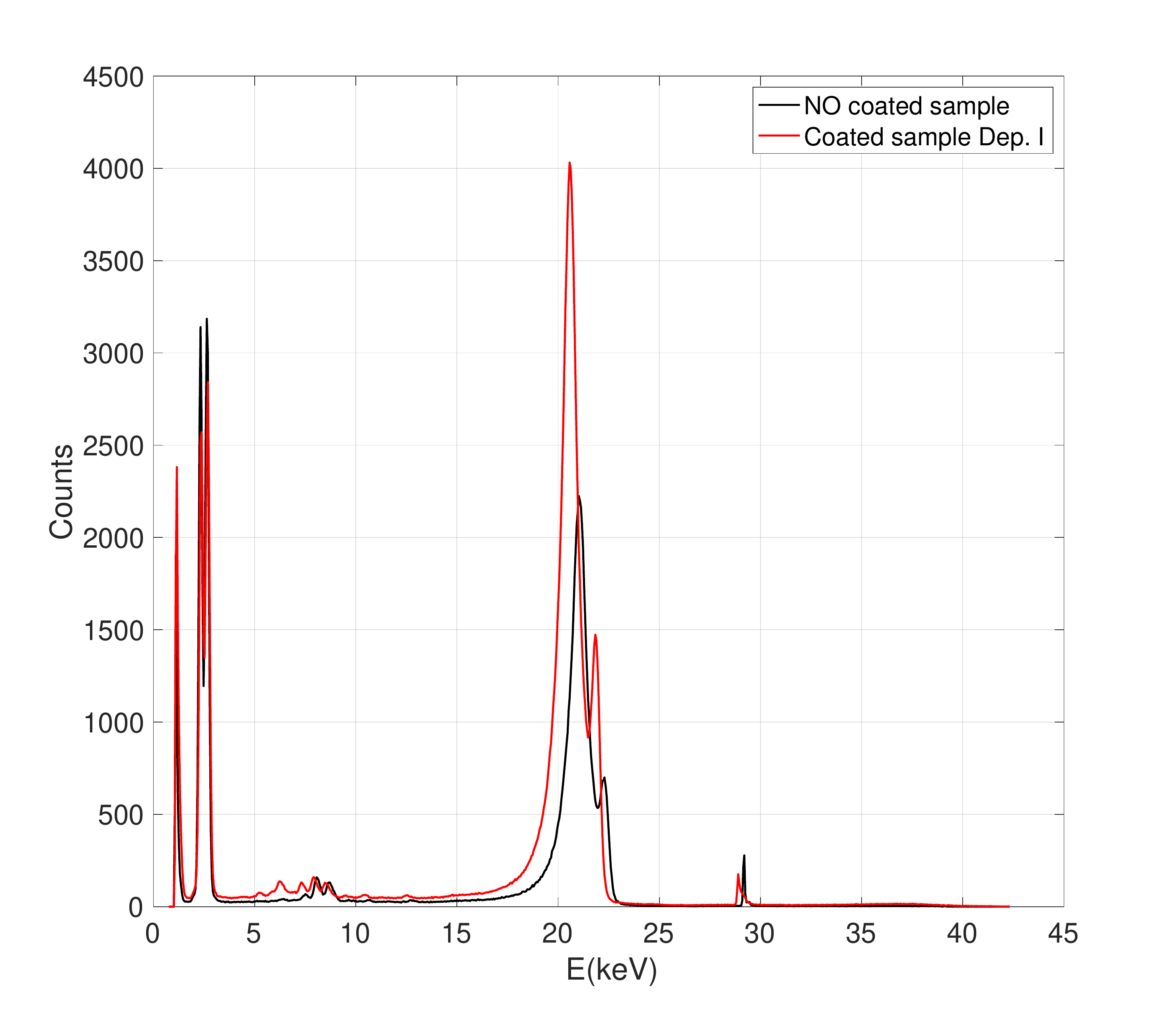}}
\subfloat[]{\includegraphics[width=.55\textwidth,keepaspectratio]{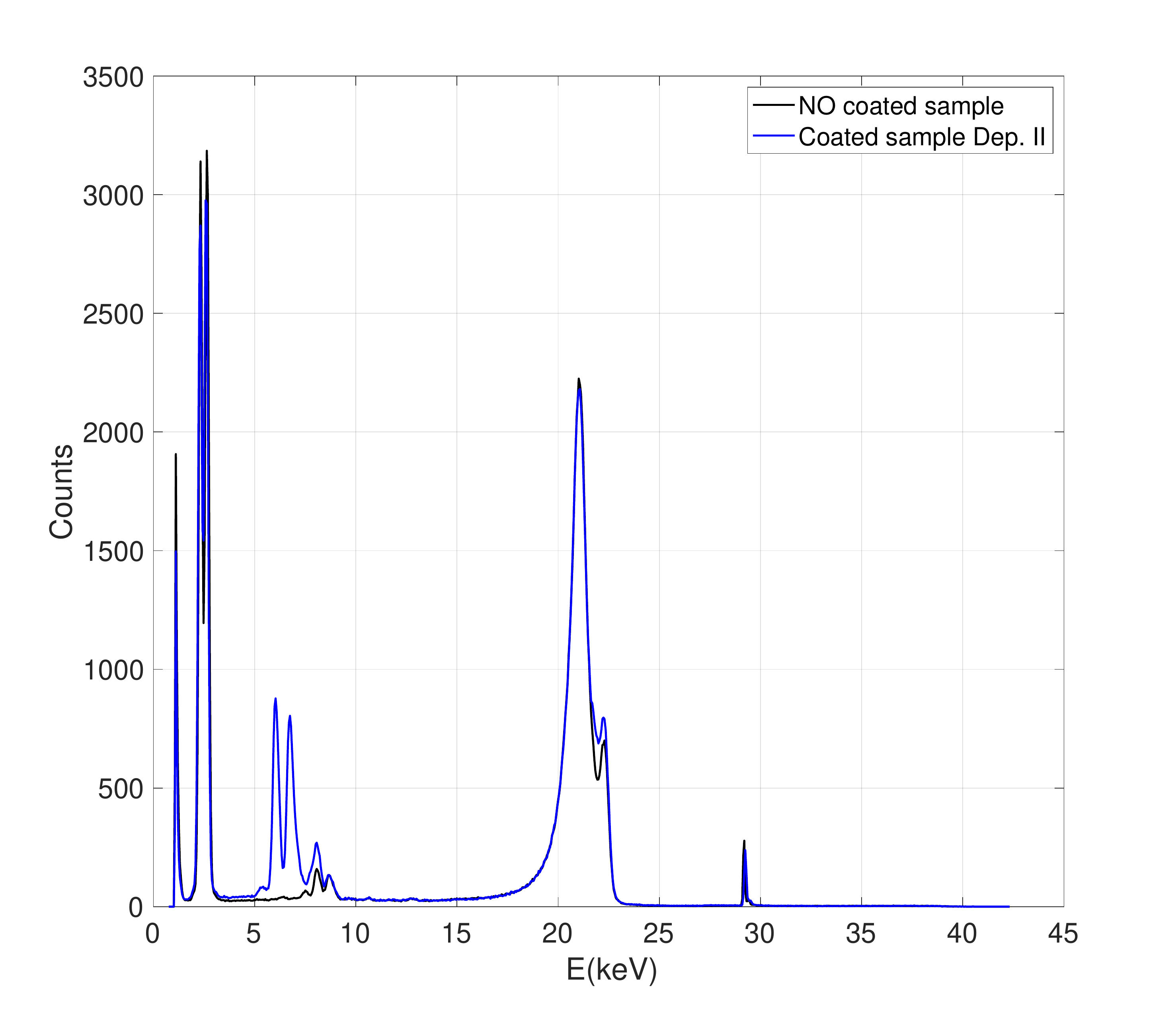}}\\
\subfloat[]{\includegraphics[width=.55\textwidth,keepaspectratio]{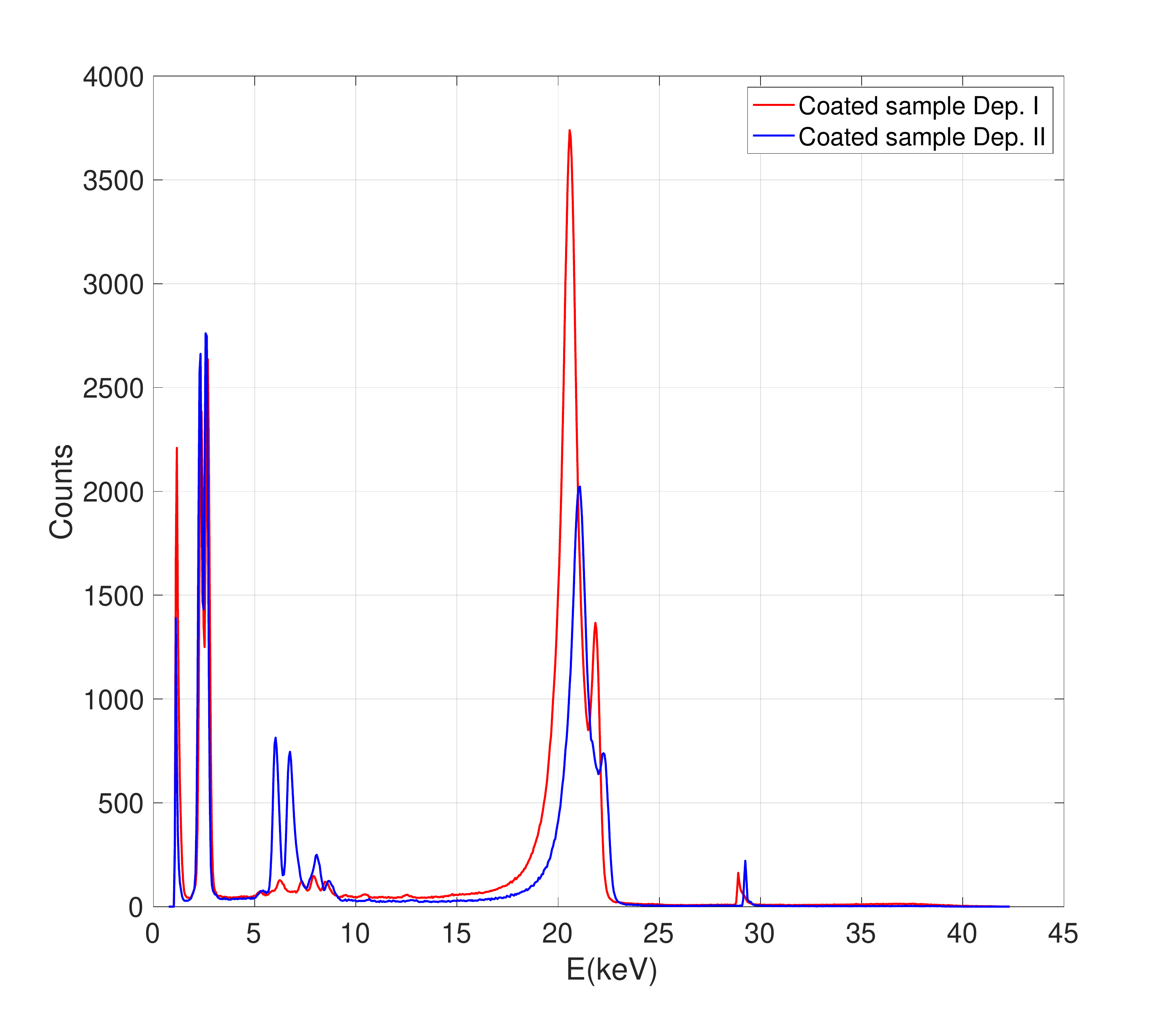}}
\caption{\label{xraygdmeas} \footnotesize (a) X-ray fluoresce spectrum of the coated sample obtained in the Deposition\,I (red line) and of the sample without the Gd coating (black line) (b) X-ray fluoresce spectrum of the coated sample obtained in the Deposition\,II (blue line) compared with the spectrum measured with the silicon slab without the Gd coating (black line). (c) Comparison between the spectra of Deposition\,I and II, red and blue line respectively.}
\end{figure}

Three set of measurements have been performed, with the two silicon sample coated respectively in the Deposition\,I and Deposition\,II test, and the relative reference measurement with a silicon sample without the coating. Figure~\ref{xraygdmeas} shows the comparison between the reference measurement and the Deposition\,I sample in panel (a), the comparison between the reference measurement and the Deposition\,II sample in panel (b) and the comparison between the spectra measured with the two coated samples in panel (c). 
\\ The two peaks L$_\alpha= 6.057\,$keV and L$_\beta = 6.713\,$keV of gadolinium are clearly visible in the second measurement, while are almost negligible in the case of Deposition\,I, as shown in figure~\ref{xraygdmeas}(c). The result is in agreement with the calculation of the thickness discussed above. Indeed, for the first deposition test it has been obtained a thickness $Th = 12\,$nm with an error of 70\%; the thickness calculated for the second deposition is, instead, $Th = 66\,$nm with an error of about 13\%.       

As mentioned before, this was a feasibility test to investigate the possibility of growing gadolinium coatings on silicon substrate. Due to the limitation of the equipment operation, especially the maximum temperature achievable by the crucible, it was not possible to grow samples in a controlled way. Nevertheless, this preliminary study provides a promising result, even if further and more detailed studies are needed. Of particular importance will be to measure, with an electron microscope, the actual thickness of the gadolinium layer and the flatness of the deposition on the substrate. A dedicated study must be performed, and it is not the topic of the present work. However, from the X-ray fluorescence measurements, it has been verified that the gadolinium has been deposited on the silicon substrate, and that thermal vapour deposition is a viable technique. 

\chapter*{Conclusion}
\addcontentsline{toc}{chapter}{Conclusion}  

Neutron scattering techniques offer a unique combination of structural and the dynamic information of atomic and molecular systems over a wide range of distances and times. The increasing complexity in science investigations driven by technological advances is reflected in the studies of neutron scattering science, which enforces a diversification and an improvement of experimental tools, from the instrument design to the detector performance. It calls as well for more advanced data analysis and modelling. The operation of new generation, high-intensity neutron sources, is therefore necessary. Among all, the European Spallation Source (ESS), which is presently under construction in Lund (Sweden). The high brightness opens up the possibility of studying currently not solved problems and to do new science, which is outside the core science case, enabled by the possibility of better measurements.
\\ The improvements in resolution, count rate and signal-to-background ratio, achievable with the new instrumentation, also drives the research of alternative technologies to replace the $^3$He-based detector technology. In addition to the $^3$He storage crisis of the past few years, this technology is, indeed, unable to fulfil the requirement of increasing performance. 
\\ Two different alternatives have been presented in this manuscript: a boron-based, and a solid state silicon-based thermal neutron detector technologies. 
\\The boron-based gaseous detector, the Multi-Blade, is currently under design at ESS, but it has been introduced at ILL in 2005. The Multi-Blade is a small area detector for neutron reflectometry application and the most challenging requirements for the detector are the spatial resolution and the count rate capability. Several tests have been performed to validate the demands established by the experimental technique. The technical aspects and characterization of the detector have been mainly explored at the Source Testing Facility at Lund university and at the Budapest Research Centre. Whereas a scientific campaign of measurements have been carried out in a real reflectometry environment at the CRISP reflectometer at ISIS, for the first time. The sole improvement in resolution of the detector are not enough to ensure the better performance of the instruments, the signal-to-background ratio affects the figure-of-merit for most instruments. Therefore, the study of background is fundamental to reach a better background rejection, which can improve the figure-of-merit, leading to significant impact on an instrument's operation, especially at the new high-intensity sources. A full characterization of the Multi-Blade detector is reported, technical and scientific, together with a dedicated study on fast neutron sensitivity performed for the first time on a boron-based thermal neutron detector.
\\ The solid state silicon-based detector is a relative new technology in the neutron field, these devices are already widely used in high-energy physics application, and in the last decades several studies have been performed to examine the use of solid state detectors for neutron detection. A Si-PIN diode couple to a Gadolinium converter layer has been described. It is under development at the University of Perugia, together with a description of the Pulse Shape analysis method proposed, for this class of devices, to optimize the discrimination of the neutron signals from the noise and background radiation. This analysis has been applied to the measurements and a set of simulations has been built up to reproduce a realistic data-set to test the effectiveness of the algorithm. The method was also applied to a SiPM, tested together with the PIN diode at ILL, which is very demanding in terms of signal analysis, because of the very high noise and the low light output of the scintillator. Both the devices are potentially suitable in application of large area detector in neutron scattering with a sub millimeter resolution and high counting rate.  
\vspace{0.5cm}
\\One of the most challenging technique in neutron scattering is the neutron reflectometry, it is particularly demanding in terms of instrument design and detection performances. Nowadays several methods have been proposed to increase the incoming flux leading to improvements for specular neutron reflectivity measurements. Along with the instruments operation the detectors response must be refined. The current detector technology is limited mainly as regards the spatial resolution, 3 times better than the state-of-the-art, and counting rate capability, $10^2-10^3$ times better than the state-of-the-art.
\\ A Multi-Blade detector has been built and it has been tested on the CRISP~\cite{CRISP1} reflectometer at ISIS (Sciente \& Technology Facilities Council in UK~\cite{ISIS}). The aim of this test was to get a full technology demonstration in a reflectometry environment. Some characterization measurements on the technical aspects of the detector have been carried out. Moreover the reflectivity of several reference sample have been measured operating the instrument in various configurations to reproduce the set-up that will be used at the ESS reflectometers.
\\ The technical characterization of the detector is needed to verify the current limitation of the design and to understand which improvements can be done, towards the final device. The detection efficiency, uniformity, linearity, stability and the spurious scattering have been characterized for the last Multi-Blade prototype. Both the spatial and time dynamic range have been measured and the actual dynamic range of the instrument was reproduced. The spatial dynamic range between pixels is about four orders of magnitude (peak to tail) and the time dynamic range between subsequent time bins is approximately 3 orders of magnitude. This was limited by the dynamic range of the instrument where the test was performed~\cite{INSTR_OSMOND_CRISP}.
The measured detection efficiency of the Multi-Blade detector is in good agreement with the previous results~\cite{MIO_MB2017} and with the theoretical model~\cite{MIO_analyt}. It is approximately $45\,\%$ at the shortest wavelength (2.5\AA) that will be used at the ESS reflectometers.
\\ A spurious scattering effect was observed in the measurements and it was attributed to those low wavelength neutrons that are not absorbed by the $\mathrm{^{10}B_4C}$-coating, and being scattered by the substrate, they are detected in other cassettes. This effect vanishes if neutrons above 4\AA\, are selected. The actual coating thickness in the detector was $4.4\,\mu$m rather than the recommended thickness of $7.5\,\mu$m. At the shortest wavelengths ($\approx 1$\AA) the $4.4\,\mu$m coating is only $50\,\%$ efficient at absorbing neutrons. The shortest wavelength that will be used at the ESS reflectometers is 2.5\AA\, the nominal coating ($7.5\, \mu$m) at 2.5\AA\, is expected to be efficient ($>98\,\%$) at reducing the scattering as the $4.4\,\mu$m coating used in these tests at 4\AA.
\\ Stability measurement of the Multi-Blade over two weeks have been carried out at the Source Testing Facility (STF)~\cite{SF2,SF1} at the Lund University in Sweden. The counting rate in the detector is stable within $\pm 1.7\,\%$ during several days with a flow that replace approximately two detector volumes ($\approx 60\,$l) per day. A clear correlation with the atmospheric pressure was found. In order to further improve the detector stability in time, the gas gain or thresholds would need to be adjusted according to the atmospheric pressure and temperature variations. An active feedback on the signal thresholds or on the high voltage as shown in~\cite{DET_Stability} or an off-line post-processing of the data can be considered. 
\\ The overall variation of the gain, i.e., the uniformity, in the scanned cassette in both directions (wires and strips) is $10\,\%$. It shows improvements with respect to the previous detector described in~\cite{MIO_MB2017} because of the new substrate used for the blades (Titanium instead of Aluminium), and the individual readout which allows to operate the detector at lower gas gain with respect to charge division.
\\ The campaign of scientific measurements not only provide a validation of the Multi-Blade as a mature technology for neutron reflectometry experiments, but it has been shown as well that the instrument operation was improved using the Multi-Blade. The spatial resolution of a detector is, indeed, deeply connected to the achievable $q$-resolution of the instrument. The calculated $q_z$ is a combination of the neutron wavelength and the scattering angle, the latter can be corrected taking into account the spatial resolution of the detector and thus a higher $q$-resolution is achieved. When measuring the specular reflectivity from a sample which shows interference fringes in $q_z$, such as the Iridium on Silicon, the fringes get more visible as the spatial resolution of the detector improves. The result has been compared to a conventional non-position sensitive detector and with a state-of-the-art detector with $2\,$mm resolution. It is possible to obtain the same results only using a very fine collimated beam, which leads to a considerable increasing in the measurement time.
\\It has been shown that the CRISP instrument can be operated, thanks to the Multi-Blade, in the REFocus mode~\cite{OTT_refocus} (divergent mode) which is one of the standard mode foreseen for the ESTIA reflectometer~\cite{INSTR_ESTIA2}. In this configuration, the correction of the scattering angle for calculating $q_z$ is mandatory and the spatial resolution and the counting rate capability of the detector is a key feature. Moreover, from the measurements of the Silicon sample, the $q$-range was measured to five orders of magnitude, reaching the limits of the instrument, despite the high background at the CRISP instrument and the poor shielding of the Multi-Blade detector. 
\\ An off-specular scattering measurement was also performed on a super-mirror Fe/Si multi-layer sample. Neither beam polarization nor magnetic field has been used in order to have a strong off-specular scattering from the sample. The ability of the Multi-Blade to measure, not only specular, but also off-specular scattering was shown. 
\\ The overall test on the CRISP reflectometer demonstrates that the detector is matching the requirements to perform neutron reflectometry measurements, and that the Multi-Blade detector technology is mature, and ready for implementation on neutron reflectometers. Not only the ESS reflectometers, but also other reflectometers at other facilities in the world, can profit from the Multi-Blade detector technology. Two papers have been published on this work~\cite{MIO_MB16CRISP_jinst,MIO_ScientificMBcrisp}.
\vspace{0.3cm}
\\The new generation high-intensity sources lead to benefit in neutron science, because of the high brightness available. Higher intensities mean higher signals, but higher background as well. The signal-to-background ratio is an important feature to study, in order to achieve the best background rejection. Fast neutrons and $\gamma-$rays are among the major sources of background. In particular, in a pulsed source, which exhibits a strong component of epithermal and fast neutrons. The $\gamma-$ray sensitivity for the boron-based neutron detector has been previously studied~\cite{MIO_MB2017,MG_gamma}, while the investigation of fast neutron sensitivity had not been studied yet, and it has not been studied in detail for any thermal neutron detector.
\\ The physical effects that affect the fast neutron sensitivity in boron-10-based gaseous detectors for thermal neutrons which are being developed at ESS have been investigated. The investigation of the physical mechanisms, cross section and the theoretical calculation of the probability of interaction in several materials, together with the simulations of the same processes provide an overview and an overall understanding of the presented work. The fast neutron sensitivity of the Multi-Blade has been measured for three fast neutron sources ($^{252}$Cf, $^{241}$Am/Be, $^{238}$Pu/Be) and it was found that for a fixed software threshold of 100 keV, which leads to the optimal neutron efficiency, it is approximately 10$^{-5}$, with an uncertainty no more than a factor two dominated by the calculation of the solid angle. A comparison between the $\gamma$-ray sensitivity and the fast neutron sensitivity has been performed. It has been used for the $\gamma$-ray measurements a $^{60}$Co source. A value for the $\gamma$-ray sensitivity (below 10$^{-8}$) was obtained, about 3 orders of magnitude lower than the fast neutron sensitivity for the same threshold. The value for gamma sensitivity comes from both a low interaction cross section and from a low probability of detection of a signal over a threshold. For the fast neutron sensitivity, only the low interaction probability helps keep the sensitivity low.
\\ The measurements presented here, and published in~\cite{MIO_fastn}, were performed for a thermal neutron detector for the first time. The Multi-Blade is based on a MWPC geometry, the results obtained and the discussion of the underlying physical processes responsible for fast neutron events in a neutron detector are more general and can be extended other gaseous-based neutron detectors. 
\\ Moreover, a preliminary study on the fast neutron sensitivity of $^3$He-based detector is presented, with a sensitivity on the order of $10^{-3}$, it is two orders of magnitude higher than the sensitivity measured for the boron-based gaseous detector. A further proof of the effectiveness of this technology compared with the $^3$He detector technology.  
\vspace{0.3cm}
\\Solid state Si-based detector technology is nowadays a viable alternative in neutron detection which can find application at the instrumentation of future sources. The promising performances and the capability to couple to high integration electronics push forward the development of real time and single event analysis systems. This is a significant advance bringing in the potential to achieve higher performances and to meet the requirements of future applications in high brilliance neutron beams. 
\\Here the results of neutron test measurements on two different Si-based devices have been presented, and a pulse shape analysis approach that was applied to both simulated and measured neutron data has been discussed. The final goal of this work is to propose a solution to the use of Si-based detectors as neutron counters with on-line and real time, pulse shape analysis system.
\\ The Si PIN diode couple to a Gadolinium converter have a potential to improve detection efficiency beyond $\approx 30$\% by optimising the deposition technique of the converter and by coupling two diodes to the same in-the-middle converter. A set of simulations has been built up based on the data acquired with this detector. The pulse shape method has been applied both to simulation and measurements. It has been shown that this method works in both cases, and a good agreement on the different parameters of the algorithm has been achieved. The measured efficiency of about 30\% is reached with the present prototype, further implementations include a better deposition of the gadolinium layer, a discussion of a feasibility test is reported in this work, and the use of integrated circuit technology for high density readout electronics, which can lead to improve the detection efficiency and the detector performance.
\\ The pulse shape analysis has been also applied to the SiPM device, which has a spatial resolution of $\approx$ 3 x 3 mm$^2$ and resolutions of $\approx 1 $ mm$^2$ can be achieved without difficulty. A system composed of many such devices, with individual plexiglass light guides and reflectors (for example, teflon tape as in the present case) does not have cross talk. Local instantaneous counting rates up to at least 500 kHz, per cm$^2$, with 10\% of dead time are already accepted as the pulse length is below 200 ns. 
\\ The pulse shape method here proposed responds well to the requirements of the signal analysis necessary for this detector technology when applied to neutron counting, and the implementation of this method leads to improved detector performances. The approach can be implemented on FPGAs, adopting simplified numerical calculation as well. A paper has been published on this work~\cite{Mauri_psa}.
\vspace{0.5cm}
\\ Each of the presented results represents a step forward and a point for further developments, mandatory for the final installation of these new neutron detector technologies in a real instrument. 
\\Two suitable alternatives for neutron detection have been presented, which are able to fulfil the increased performance demands on neutron detectors, driven by the operation of new generation high-intensity sources with improved instrument design. Better instrumentation is a core driver of novel scientific fields, it has been shown, indeed, how the science and the technical aspects are strongly correlated and the improvements deriving from it, in real experimental conditions.

\newpage
\bibliographystyle{ieeetr}
{\footnotesize\bibliography{BIBLIODB}}
\end{document}